\documentclass[manuscript,screen,nonacm]{acmart}
\setcopyright{none}
\usepackage{graphicx} 
\usepackage{algorithmic}
\usepackage{textcomp}
\usepackage{mathtools}
\usepackage{amsmath,amsthm}
\usepackage{color}
\usepackage{enumitem}
\usepackage{booktabs}
\usepackage{longtable}
\usepackage{stmaryrd}
\input{Ustmry.fd}
\DeclareFontShape{U}{stmry}{m}{n}{%
  <-6>stmary5 <6-7>stmary6 <7-8>stmary7
  <8-9>stmary8 <9-10>stmary9 <10->stmary10}{}
\usepackage{todonotes}

\usepackage{rotating}
\usepackage{multirow}

\usepackage{thmtools,thm-restate}

\hypersetup{colorlinks=true, linkcolor=red,citecolor=blue,urlcolor=blue}
\makeatother

\newcommand{\Pub}{\mathsf{P}}
\newcommand{\Cnf}{\mathsf{C}}
\newcommand{\Sec}{\mathsf{S}}
\newcommand{\Tsec}{\mathsf{T}}

\usepackage{tikz}
\usetikzlibrary{arrows,chains,matrix,positioning,scopes}
\def\BibTeX{{\rm B\kern-.05em{\sc i\kern-.025em b}\kern-.08em
    T\kern-.1667em\lower.7ex\hbox{E}\kern-.125emX}}

\makeatletter
\tikzset{join/.code=\tikzset{after node path={%
\ifx\tikzchainprevious\pgfutil@empty\else(\tikzchainprevious)%
edge[every join]#1(\tikzchaincurrent)\fi}}}
\makeatother

\tikzset{>=stealth',every on chain/.append style={join},
         every join/.style={->}}
\tikzstyle{labeled}=[execute at begin node=$\scriptstyle,
   execute at end node=$]

\newcommand{\supp}{\mathrm{supp}}

\providecommand{\dotdiv}{
  \mathbin{
    \vphantom{+}
    \text{
      \mathsurround=0pt 
      \ooalign{
        \noalign{\kern-.35ex}
        \hidewidth$\smash{\cdot}$\hidewidth\cr 
        \noalign{\kern.35ex}
        $-$\cr 
      }%
    }%
  }%
}

\newcommand{\commentout}[1]{}

\newtheorem{Thm}{Theorem}
\newtheorem{Def}[Thm]{Definition}

\newtheorem{Rmk}[Thm]{Remark}
\newtheorem{Conv}[Thm]{Convention}
\newtheorem{Exm}[Thm]{Example}

\newtheorem{Lem}[Thm]{Lemma}
\newtheorem{Pro}[Thm]{Proposition}

\usepackage{color,dsfont}
\def\gb#1{\textcolor{magenta}{}}
\def\vt#1{\textcolor{green}{}}
\def\ali#1{\textcolor{orange}{}}
\def\hide#1{}

\title{Containment of Conjunctive Queries with Negative Information over Semirings}

\author{Guillermo Badia}
\affiliation{%
  \institution{University of Queensland}
  \city{Brisbane}
  \country{Australia}
}
\email{g.badia@uq.edu.au}

\author{Carles Noguera}
\affiliation{%
  \institution{University of Siena}
  \city{Siena}
  \country{Italy}
}
\email{carles.noguera@unisi.it}

\author{Gaia Petreni}
\affiliation{%
  \institution{University of Siena}
  \city{Siena}
  \country{Italy}
}
\email{gai.petreni@student.unisi.it}

\author{Alice Portone}
\affiliation{%
  \institution{University of Siena}
  \city{Siena}
  \country{Italy}
}
\email{alice.portone@student.unisi.it}

\author{Val Tannen}
\affiliation{%
  \institution{University of Pennsylvania}
  \city{Philadelphia}
  \country{USA}
}
\email{val@seas.upenn.edu}

\begin{document}


\begin{abstract}
 In this paper, we study the containment problem for (unions of) conjunctive queries ((U)CQs) with negative information on semiring-annotated databases. Our approach to negative information follows the treatment of negation proposed by Gr\"adel and Tannen (2017, 2019, 2021 and 2024) for first-order logic and fixed point logic which applies productively to large classes of semirings, including bounded distributive lattices, as well as the important provenance semirings studied for positive CQ containment by Green (2009).
 
 Negation features in (U)CQs along two dimensions: negated equalities (disequations), and negated relational atoms. We study complexity for each of them separately, as well as jointly. For disequations, we extend a methodology that proves the same complexity bounds as the ones already known for the standard 
 Boolean semiring
 (in the second level of the \emph{polynomial hierarchy}) for most semirings of interest. Two prominent exceptions are  $\mathbb{N}[X]$ and Trio(X) where the complexity bound for containment is in the second level of the \emph{counting hierarchy}. For CQs with negated relational atoms 
 we show that the complexity, surprisingly, is the same as that of the containment of  CQs \emph{without negation} when one considers model-defining and provenance-tracking interpretations (namely, nondeterministic polynomial time).  Interestingly, among bounded distributive lattices, the Boolean two-element case is the only one where the complexity of the problem continues to be hard for the second level of the polynomial hierarchy for model-defining interpretations. Furthermore, we show that the complexity increases again for distributive lattices if we focus on other classes of interpretations. Finally, we consider queries that have both negated equalities and negated relational atoms and show that the complexity of containment continues to be that of the case with just negated equalities.

\end{abstract}

\keywords{query containment, conjunctive queries, semiring semantics, negative information, provenance}
\maketitle
\renewcommand{\shortauthors}{Badia et al.}

\tableofcontents

\section{Introduction}

\noindent{\bf Motivation~~} The study of \emph{containment} and \emph{equivalence} for various classes of queries is motivated by the practical interest in query optimization and, more generally, in static analyses applicable to query languages.
The study of the decidability and complexity of the containment and equivalence problems for \emph{conjunctive queries} (CQs) was pioneered by the seminal paper~\cite{Chandra}. The authors established the NP-completeness of both problems, using the fruitful method of containment mappings. Subsequently, these decidability results (still staying NP-complete) were extended to \emph{unions of conjunctive queries} (UCQs)~\cite{Sagiv}. These early papers assume the classical
set semantics.
The situation for bag semantics\footnote{A good survey on these issues for bag semantics is~\cite{Green2}.} was analyzed first in~\cite{Chad} where equivalence between CQs was shown to be GI-complete. This was extended to equivalence of UCQs in~\cite{Nutt2} (still GI-complete). 
In contrast, bag containment of UCQs is undecidable~\cite{Ioannidis}. The decidability of bag containment for CQs remains a long-standing open problem despite important recent developments (e.g.~\cite{Abo, mar}). These results are noted in the first and last row of Table~\ref{tab:summary_cq_ucq}.


\noindent{\bf Negative Information~~} By themselves, CQs and UCQs do not contain any negative information. 
One initial way of increasing expressive power by adding a limited form of such information is to permit negated equality atoms. We will refer to these as \emph{disequations}\footnote{For this terminology see~\cite{Kolaitis2}.}, and we denote conjunctive queries with disequations CQ$^\neq$s and UCQ$^\neq$s. For \emph{set semantics}
(see the first row of Table~\ref{tab:summary_cqneq_ucqneq}), the containment problem  for  CQ$^\neq$s is in $\Pi^p_2$~\cite{Klug}.
A matching $\Pi^p_2$ lower bound was obtained in~\cite{VdM} and refined in~\cite{Kolaitis1}. 
For set semantics, containment is inter-reducible with equivalence so the same bounds hold for CQ$^\neq$ equivalence.
The results in~\cite{Klug,VdM} also imply straightforwardly $\Pi^p_2$ upper and lower bounds for the containment and equivalence of UCQ$^\neq$s. 
For \emph{bag semantics} (see the last row of Table~\ref{tab:summary_cqneq_ucqneq}), UCQ$^\neq$ equivalence had a PSPACE upper bound~\cite{Nutt, Nutt2} that was later improved to coNP$^{\# \text{P}}$~\cite{Chen} (and thus the same applies to CQ$^\neq$s). In contrast, bag containment of CQ$^\neq$s (hence also UCQ$^\neq$) is undecidable~\cite{Kolaitis2}.

Another way of adding a limited form of negation is to allow in the body of (U)CQs \emph{negated relation atoms} rather than disequations (we call these queries (U)CQNs). Again for classical set semantics, CQN containment is known to be $\Pi^p_2$-complete \cite{Chein} if we have a fixed arity bound $k\geq 2$ on the relations appearing in our queries, and coNEXP-complete for unbounded arity~\cite{Geck}. Equivalence has exactly the same complexity  because we can express containment and equivalence in terms of each other for CQNs as for CQ$^\neq$s. Furthermore, similarly to the case of UCQ$^\neq$s, we can observe that containment and equivalence for UCQNs must also be $\Pi^p_2$-complete if we have a fixed arity bound $k\geq 2$ on the relations  in the database schema of our queries, and coNEXP-complete for unbounded arity. 

It is an interesting problem how to make sense of negated relation atoms in bag semantics (we will discuss some options below e.g.~\cite{GT, GT2, GT3, GT2a, Ba}). 
Regardless of the chosen semantics, bag containment of (U)CQNs should naturally generalize bag containment of (U)CQs and there should be a reduction from the latter problem to the former. Thus, if the bag containment problem for CQs ends up being undecidable, so would the bag containment for CQNs. Similarly, due to the undecidability of the bag containment problem for UCQs, we already have that the bag containment problem for UCQNs is undecidable. Moreover, the GI-completeness of (U)CQ bag equivalence  yields immediately a lower bound for (U)CQN bag equivalence. In fact, we show below that for a certain way of dealing with negation, equivalence of CQNs is GI-complete. For the same model-defining semantics, UCQN bag equivalence is in coNP$^{\mathrm{PP}}$.


\noindent{\bf Semiring Semantics~~} 
The framework of semiring-annotated databases was introduced in~\cite{Green0}, and has attracted considerable interest in database theory (some recent examples include~\cite{Fan, Hir, Abo, Ram}). It generalizes, for queries, both set and bag semantics (which correspond to the semirings $\mathbb{B}$ and $\mathbb{N}$, respectively) to databases whose tuples are annotated with elements from an arbitrary commutative semiring (see Section~\ref{sec:semiring-databases}).

The principal reason the paper~\cite{Green0} introduced semiring semantics was the capture of \emph{data provenance}. To that end it uses semirings of polynomials, most general of which is $\mathbb{N}[X]$ (see Section~\ref{sec:semirings}). The tuples of the input database are annotated with polynomials consisting of a single indeterminate. 
An output of a query is annotated with a polynomial that captures its provenance. For example, the annotation $2y^2 + xy$  can be ``read'' as follows: the output is computed by the query
in three different ways; two of them use the input tuple annotated with $y$ twice; the third uses the input tuples annotated with $x$, respectively $y$, once each. For provenance information that is less detailed than that given by $\mathbb{N}[X]$,
yet still valuable depending on the usage context, we can use simplified (quotiented) versions of these polynomials, see the semirings $\mathbb{B}[X]$, $\mathrm{Trio}(X)$, $\mathrm{Sorp}(X)$, $\mathrm{Why}(X)$, $\mathrm{PosBool}(X)$, and $\mathrm{Which}(X)$
\footnote{This semiring captures provenance as the set of ``usable'' input tuples~\cite{CuiWidom}.
See remarks in Section~\ref{sec:semirings} where we explain the naming.}. We review the definitions of these semirings
in Section~\ref{sec:semirings}.

Decidability and complexity of containment and equivalence for (U)CQs
under semantics based on these semirings of interest, as well as bounded distributive lattices, has been studied first in~\cite{Green} and then refined and improved in~\cite{Kostylev}, with the exception of~$\mathrm{Sorp}(X)$, the semiring of absorptive polynomials, which was introduced later~\cite{DeutchEtAl14}.
It was shown that containment for CQs over all these semirings is NP-complete while containment of UCQs is NP-complete in all cases except for $\mathbb{N}[X]$ and $\mathrm{Trio}(X)$, where the problem has a PSPACE upper bound (in the case of $\mathbb{N}[X]$, this bound was improved to coNP$^{\#P}$ --which is the same as  coNP$^\text{PP}$~\cite{Tor, Mon}-- in~\cite{Kostylev}, while in the case of $\mathrm{Trio}(X)$ a similar improvement follows by item (6) of  Theorem~\ref{ucqneqconatinmentsummary}). Equivalence for CQs is GI-complete for $\mathbb N[X]$, $\mathbb B[X]$, $\mathrm{Trio}(X)$, and $\mathrm{Why}(X)$, and NP-complete for bounded distributive lattices (including $\mathrm{PosBool}(X)$) and $\mathrm{Which}(X)$, while equivalence for UCQs is NP-complete in all cases except for $\mathbb{N}[X]$ and $\mathrm{Trio}(X)$, where it is GI-complete. 
For the  semiring $\mathrm{Sorp}(X)$, NP-completeness of containment for both CQs and UCQs follows from~\cite[Thms. 4.6 and 5.6, respectively]{Kostylev}. Membership in NP of the equivalence problem for CQs follows immediately. NP-completeness of equivalence for UCQs follows using~\cite[Prop. 7.2]{Green}. These results are noted in Table~\ref{tab:summary_cq_ucq}, rows 1-7. It is not difficult to observe that the NP upper bound on the equivalence problem for CQs in $\mathrm{Sorp}(X)$ can be improved to GI-completeness
by using the proof of Theorem~\ref{equivalcqns} (item (3)) below since equivalence would amount to isomorphism between the queries. We do not know how to improve any of the remaining bounds so we leave them open.
\vt{DISCUSS this paragraph is MAYBE too detailed for intro but OK for RELATED WORK. On second thought we can leave it like this in marXiv but we must rearrange in PODS submission.}

\begin{table}[ht!]
    \centering
    \scriptsize
    \caption{Complexity summary of containment and equivalence for CQs and UCQs from \cite{Green, Kostylev}}
    \label{tab:summary_cq_ucq}
    \medskip

    \resizebox{\textwidth}{!}{%
    \begin{tabular}{|c|cc|cc|}
        \hline
        \multirow{2}{*}{\textbf{Semirings}}
        & \multicolumn{2}{c|}{\textbf{CQs}}
        & \multicolumn{2}{c|}{\textbf{UCQs}} \\
        \cline{2-5}
        & \textbf{Containment} & \textbf{Equivalence}
        & \textbf{Containment} & \textbf{Equivalence} \\
        \hline

        Distr. Latt. (e.g. $\mathbb{B}$, $\mathrm{PosBool}(X)$)
        & NP-c 
        & NP-c 
        & NP-c 
        & NP-c \\
        \hline

        $\mathrm{Which}(X)$
        & NP-c
        & NP-c
        & NP-c
        & NP-c \\
        \hline

             $\mathrm{Why}(X)$
        & NP-c
        & GI-c
        & NP-c
        & NP-c \\
        \hline

          $\mathrm{Sorp}(X)$
        & NP-c
        & in NP
        & NP-c
        & NP-c \\
        \hline

        $\mathrm{Trio}(X)$
        & NP-c
        & GI-c
        &  in PSPACE
        & GI-c \\
        \hline

        $\mathbb{N}[X]$
        & NP-c
        & GI-c
        &  in $\text{coNP}^{\# \text{P}}$
        & GI-c \\
        \hline

        $\mathbb{B}[X]$
        & NP-c
        & GI-c
        & NP-c
        & NP-c \\
        \hline

 $\mathbb{N}$
        & \emph{open}
        & GI-c
        & undecidable
        & GI-c \\
        \hline

    \end{tabular}%
    }
\end{table}

While extending semiring semantics from (U)CQ to (U)CQ$^\neq$ is straightforward,
dealing with negated relational atoms is less so. A survey of approaches to
this issue appears in~\cite{GT} and even further possibilities are considered in~\cite{Suciu}. An early approach was based on modeling directly relational algebra subtraction using \emph{semirings with monus}~\cite{Geerts} (see also~\cite{Ba}). This approach nicely unifies Boolean and natural numbers (proper) subtraction but the free structures that capture provenance expressions are less perspicuous than the provenance polynomials. 

For this paper, we adopt the semantic setting developed by Gr\"adel and Tannen (GT) for first-order logic~\cite{GT, GT2, GT3} (further extended to fixed-point logic in~\cite{GT4}; the finite model theory of this approach continues to be extensively studied, e.g.~\cite{Bri1,Bri2,Bri3}). With this approach, in an arbitrary semiring, the additive identity, 0, stands for \emph{false}, while all the other elements of the semiring stand for ``shades'' or ``flavors'' of \emph{true}. Via a transformation of logical formulas into \emph{negation normal form (nnf)}, the semantics then uses \emph{interpretations}, which assign semiring values to the \emph{literals}, namely plain and negated relational atoms built from schema relations and constants corresponding to the elements of the universe of a finite first-order model. 
This is directly applicable to (U)CQs with negated relational atoms which are already in nnf. 

Next, we have to resolve the difficulty caused by the fact that
the GT semantics uses interpretations while containment/equivalence uses database instances. We do this by considering only interpretations \emph{compatible}  with instances
via their \emph{active domain} (cf. \cite{Kolaitis3}),
while maintaining the spirit of \emph{closed world} semantics. The default choice of
interpretation is \emph{model-defining (md)} (where exactly one of a literal or its negation is mapped to $0$). We also consider
\emph{dually model-defining (dmd)} interpretations (where exactly one of a literal or its negation is mapped to $1$) (see Section~\ref{section2.6}). 
\vt{pointer forward to sec 2.4 and to sec 2.6. I wish that were not spread over two sections.}

The GT approach to negation applies to many semirings and, notably, it accommodates a perspicuous paradigm for provenance tracking, the \emph{dual-indeterminate provenance polynomials}.
For example, $\mathbb{N}[X,\overline{X}]$ generalizes 
$\mathbb{N}[X]$, by adding the indeterminates in $\overline{X}$
for the purpose of annotating tuples (still made of elements of the active domain) whose \emph{absence} from the instance we wish to track. Note that there is no dual-indeterminate counterpart of 
$\mathrm{Which}(X)$ (we explain why in \vt{pointer forward}).
For our results pertaining to these dual-indeterminate semirings we use a more general \emph{provenance-tracking (pt)}
interpretation (where a literal and its negation can also be mapped to $x$, respectively $\overline{x}$ for tracking presence, respectively absence). \vt{pointer forward} \vt{With apologies, I now thisnk that maybe we should put md and pt back in the tables.}

\vt{pointer forward again}



\noindent{\bf Contributions} Our paper investigates containment and equivalence focusing on (U)CQ$^{\neq}$s, (U)CQNs and the queries obtained by simultaneously adding both forms of negative information, which we call (U)CQ$^{\neq}$Ns. We consider the semirings 
mentioned above 
and we establish that all these problems are decidable, by extending the methods of~\cite{Green} involving containment mappings (homomorphisms) and canonical databases to handle queries with negative information. Surprisingly, in the case of UCQNs, the complexity ends up depending dramatically on the class of interpretations under consideration (we look at model-defining and provenance-tracking interpretations, but future research might consider other options of interest).  \vt{In the detailed list and the tables we talk about three kinds of interpretations but the intro does not explain what they are. DISCUSS} \vt{I added interpretation in the intro above but see my mea culpa we should put md, dmd and pt in the tables after all!!}
One notable difference between the CQ$^{\neq}$ and  CQ cases is that containment on  $\mathbb{B}[X]$ and $\mathbb{N}[X]$ no longer coincide. 

Our results have multiple dimensions: choice of semiring for the semantics, disequations vs.~negated (relational) atoms vs.~both, CQ vs.~UCQ, and containment vs.~equivalence.
For queries with negated relational atoms we consider different kinds of interpretations.
For the dually model-defining results below, no query (in the union) on the left
contains an atom and its negation; for equivalence, this restriction applies to both queries. These restrictions also apply to Table~\ref{tab:summary_cqn_ucqn}.

More specifically, we show the following:
\begin{enumerate}
\item Containment of CQ$^\neq$s is $\Pi^p_2$-complete for bounded distributive lattices, $\mathrm{Why}(X)$, $\mathrm{Which}(X)$, $\mathrm{Sorp}(X)$ and $\mathbb{B}[X]$, while the problem has a coNP$^\text{PP}$ upper bound for $\mathbb{N}[X]$ and $\mathrm{Trio}(X)$ (but we leave the exact complexity of these two cases open).

\item Equivalence of CQ$^\neq$s is $\Pi^p_2$-complete for bounded distributive lattices, $\mathrm{Why}(X)$, $\mathrm{Which}(X)$, $\mathrm{Sorp}(X)$ $\mathbb{B}[X]$, while for $\mathbb{N}[X]$ and $\mathrm{Trio}(X)$ the problem has a coNP$^\text{PP}$ upper bound (but we leave the exact complexity of these cases open).

\item Containment of UCQ$^\neq$s is $\Pi^p_2$-complete for bounded distributive lattices, as well as $\mathrm{Why}(X)$, $\mathrm{Which}(X)$, $\mathrm{Sorp}(X)$, and $\mathbb{B}[X]$. We give coNP$^\text{PP}$ upper bounds for $\mathbb{N}[X]$ and $\mathrm{Trio}(X)$ (but we leave the exact complexity of these two cases open).

\item Equivalence of UCQ$^\neq$s is $\Pi^p_2$-complete for bounded distributive lattices, $\mathrm{Why}(X)$, $\mathrm{Which}(X)$, $\mathrm{Sorp}(X)$, and $\mathbb{B}[X]$, while for $\mathrm{Trio}(X)$ and $\mathbb{N}[X]$, it has coNP$^\text{PP}$ upper bounds (but we leave the exact complexity of these two cases open).

\item 
Containment of CQNs  is NP-complete  for bounded distributive lattices with more than two elements over model-defining interpretations, as well as for  $\mathrm{Why}(X, \overline{X})$, $\mathrm{Trio}(X, \overline{X})$, $\mathrm{PosBool}(X, \overline{X})$, $\mathbb{N}[X, \overline{X}]$, $\mathbb{B}[X, \overline{X}]$ and $\mathrm{Sorp}(X, \overline{X})$ over provenance-tracking (and model-defining) interpretations, and the problem is $\Pi^p_2$-complete for bounded distributive lattices over dual model-defining interpretations with a fixed arity bound $k\geq 2$ on relations, and coNEXP-complete for unbounded arity, while  for $\mathrm{Which}(X)$ over model-defining interpretations with a fixed arity bound $k\geq 2$ on relations (respectively, unbounded arity) the problem is  $\Pi^p_2$-complete (respectively, coNEXP-complete).

\item Equivalence of CQNs  is NP-complete  for bounded distributive lattices with more than two elements over model-defining interpretations and for  $\mathrm{PosBool}(X, \overline{X})$ over provenance-tracking interpretations, it is GI-complete  
for  $\mathrm{Why}(X, \overline{X})$,  $\mathrm{Trio}(X, \overline{X})$, $\mathbb{N}[X, \overline{X}]$, $\mathbb{B}[X, \overline{X}]$ and $\mathrm{Sorp}(X, \overline{X})$ over provenance-tracking (and model-defining) interpretations and for $\mathbb{N}$ with model-defining interpretations, while the problem is $\Pi^p_2$-complete for bounded distributive lattices over dual model-defining interpretations with a fixed arity bound $k\geq 2$ on relations, and coNEXP-complete for unbounded arity,  and for $\mathrm{Which}(X)$ over model-defining interpretations with a fixed arity bound $k\geq 2$ on relations (respectively, unbounded arity) the problem is $\Pi^p_2$-complete (respectively, coNEXP-complete).

\item Containment of UCQNs  is NP-complete  for bounded distributive lattices with more than two elements over model-defining interpretations, as well as for  $\mathrm{Why}(X, \overline{X})$,  $\mathrm{PosBool}(X, \overline{X})$,  $\mathbb{B}[X, \overline{X}]$ and $\mathrm{Sorp}(X, \overline{X})$ over provenance-tracking interpretations, 
while the problem is $\Pi^p_2$-complete for bounded distributive lattices over dual model-defining interpretations with a fixed arity bound $k\geq 2$ on relations, and coNEXP-complete for unbounded arity, and has coNP$^\text{PP}$ upper bounds for $\mathrm{Trio}(X, \overline{X})$ and $\mathbb{N}[X, \overline{X}]$ (we leave the exact complexity of these two cases open) while  for $\mathrm{Which}(X)$ over model-defining interpretations with a fixed arity bound $k\geq 2$ on relations (respectively, unbounded arity) the problem is  $\Pi^p_2$-complete (respectively,  coNEXP-complete).

\item Equivalence of UCQNs   over  bounded distributive lattices with more than two elements with model-defining interpretations, and over provenance-tracking interpretations for $\mathrm{Sorp}(X, \overline{X})$, $\mathrm{Why}(X, \overline{X})$, $\mathrm{PosBool}(X, \overline{X})$, and $\mathbb{B}[X, \overline{X}]$ is NP-complete, while the problem is $\Pi^p_2$-complete over bounded distributive lattices with dual model-defining interpretations with a fixed arity bound $k\geq 2$ on relations, and coNEXP-complete for unbounded arity, and the equivalence problems over   $\mathbb{N}[X, \overline{X}]$ and $\mathrm{Trio}(X, \overline{X})$ are in coNP$^\text{PP}$ (we leave the exact complexity of these cases open) while  for $\mathrm{Which}(X)$ over model-defining interpretations with a fixed arity bound $k\geq 2$ on relations (respectively, unbounded arity) the problem is  $\Pi^p_2$-complete (respectively,  coNEXP-complete).

\item For (U)CQ$^{\neq}$Ns, both containment and equivalence on model-defining interpretations and bounded distributive lattices with more than two elements are $\Pi^p_2$-complete, and for $\mathrm{Which}(X)$ over model-defining interpretations with a fixed arity bound $k\geq 2$ on relations (respectively, unbounded arity) both containment and equivalence of (U)CQ$^{\neq}$Ns are $\Pi^p_2$-complete (respectively, coNEXP-complete); for (U)CQ$^{\neq}$Ns and provenance-tracking interpretations on the remaining semirings ($\mathrm{Sorp}(X, \overline{X})$, $\mathrm{Why}(X, \overline{X})$, $\mathrm{Trio}(X, \overline{X})$, $\mathrm{PosBool}(X, \overline{X})$,  $\mathbb{B}[X, \overline{X}]$, and $\mathbb{N}[X, \overline{X}]$), our complexity results for both containment and equivalence problems  coincide exactly  with those for (U)CQ$^{\neq}$s.
\end{enumerate}

 In the summarizing tables we have some rows group multiple semirings when the complexity is the same. However, among semirings that share a row, while some proofs are similar, they are not quite the same. As a consequence, this paper has more than 100 theorems with distinct proofs. In what follows, we will relegate many proofs to the appendices, particularly when they are simple modifications of proofs already in the main body of the paper.

\begin{table}[ht!]
    \centering
    \scriptsize
    \caption{Complexity summary of containment and equivalence for CQ$^\neq$s and UCQ$^\neq$s}
    \label{tab:summary_cqneq_ucqneq}
    \medskip

    \resizebox{\textwidth}{!}{%
    \begin{tabular}{|c|cc|cc|}
        \hline
        \multirow{2}{*}{\textbf{Semirings}}
        & \multicolumn{2}{c|}{\textbf{CQ$^\neq$s}}
        & \multicolumn{2}{c|}{\textbf{UCQ$^\neq$s}} \\
        \cline{2-5}
        & \textbf{Containment} & \textbf{Equivalence}
        & \textbf{Containment} & \textbf{Equivalence} \\
        \hline
        
          $\mathbb{B}$
        & $\Pi^p_2$-c~\cite{Klug, VdM}
        & $\Pi^p_2$-c~\cite{Klug, VdM}
        & $\Pi^p_2$-c~\cite{Klug, VdM}
        & $\Pi^p_2$-c~\cite{Klug, VdM}\\
        \hline

          Distr. Latt. (e.g., $\mathrm{PosBool}(X)$)
        & $\Pi^p_2$-c 
        & $\Pi^p_2$-c 
        & $\Pi^p_2$-c 
        & $\Pi^p_2$-c \\
        \hline

        $\mathrm{Which}(X)$, $\mathrm{Sorp}(X)$, $\mathrm{Why}(X)$, $\mathbb{B}[X]$
        & $\Pi^p_2$-c
        & $\Pi^p_2$-c
        & $\Pi^p_2$-c
        & $\Pi^p_2$-c \\
        \hline

        $\mathbb{N}[X]$, $\mathrm{Trio}(X)$
        & in coNP$^\text{PP}$
        & in coNP$^\text{PP}$
        &  in coNP$^\text{PP}$
        &  in coNP$^\text{PP}$ \\
        \hline

 $\mathbb{N}$
        & undecidable~\cite{Kolaitis2}
        & in PSPACE~\cite{Nutt, Nutt2} (improved to coNP$^{\#\text{P}}$~\cite{Chen})
        & undecidable~\cite{Kolaitis2}
        & in PSPACE~\cite{Nutt, Nutt2} (improved to  coNP$^{\#\text{P}}$~\cite{Chen}) \\
        \hline

    \end{tabular}%
    }
\end{table}

\begin{table}[ht!]
    \centering
    \scriptsize
    \caption{Complexity summary of containment and equivalence for CQNs and UCQNs 
    }
    \label{tab:summary_cqn_ucqn}
    \medskip

    \resizebox{\textwidth}{!}{%
    \begin{tabular}{|c|cc|cc|}
        \hline
        \multirow{2}{*}{\textbf{Semirings}}
        & \multicolumn{2}{c|}{\textbf{CQNs}}
        & \multicolumn{2}{c|}{\textbf{UCQNs}} \\
        \cline{2-5}
        & \textbf{Containment} & \textbf{Equivalence}
        & \textbf{Containment} & \textbf{Equivalence} \\
        \hline

 $\mathbb{B}$ (\emph{md})
        & 
        & 
        & 
        & \\
        
        (1) fixed arity bound $k\geq 2$ on relations
        & $\Pi^p_2$-c (\cite{Chein, Geck} for $\mathbb{B}$)
        & $\Pi^p_2$-c (\cite{Chein, Geck} for $\mathbb{B}$)
        & $\Pi^p_2$-c (\cite{Chein, Geck} for $\mathbb{B}$)
        & $\Pi^p_2$-c  (\cite{Chein, Geck} for $\mathbb{B}$)\\

       (2) unbounded arity & coNEXP-c (\cite{Geck} for $\mathbb{B}$)
        & coNEXP-c (\cite{Geck} for $\mathbb{B}$)
        & coNEXP-c (\cite{Geck} for $\mathbb{B}$)
        & coNEXP-c (\cite{Geck} for $\mathbb{B}$)\\
  \hline

  Distr. Latt. (\emph{dmd})
        & 
        & 
        & 
        & \\
        
        (1) fixed arity bound $k\geq 2$ on relations
        & $\Pi^p_2$-c 
        & $\Pi^p_2$-c 
        & $\Pi^p_2$-c 
        & $\Pi^p_2$-c  \\

       (2) unbounded arity 
        & coNEXP-c 
        & coNEXP-c 
        & coNEXP-c 
        & coNEXP-c \\
  \hline

         Distr.Latt. $>2$ elements (\emph{md})
        & NP-c
        & NP-c
        & NP-c
        & NP-c \\
        \hline

            $\mathrm{PosBool}(X, \overline{X})$ (\emph{pt}) 
        & NP-c
        & NP-c
        & NP-c
        & NP-c \\
        \hline

        $\mathrm{Which}(X)$ (\emph{md})
        &  
        & 
        & 
        &  \\

          (1) fixed arity bound $k\geq 2$ on relations
        &  $\Pi^p_2$-c
        & $\Pi^p_2$-c
        & $\Pi^p_2$-c
        & $\Pi^p_2$-c \\
        
         (2) unbounded arity
        &  coNEXP-c
        & coNEXP-c
        &  coNEXP-c
        &  coNEXP-c \\
        \hline

        $\mathrm{Sorp}(X, \overline{X})$ (\emph{pt})
        & NP-c
        & GI-c
        & NP-c
        & NP-c \\
        \hline

        $\mathrm{Why}(X, \overline{X})$,  $\mathbb{B}[X,\overline{X}]$ (\emph{pt})
        & NP-c
        & GI-c
        & NP-c
        & NP-c \\
        \hline

        $\mathrm{Trio}(X, \overline{X})$, $\mathbb{N}[X,\overline{X}]$ (\emph{pt})
        & NP-c
        & GI-c
        &  in coNP$^\text{PP}$
        &  in coNP$^\text{PP}$ \\
        \hline

 $\mathbb{N}$ (\emph{md})
        & \emph{open}
        & GI-c
        & undecidable~\cite{Ioannidis}
        & in coNP$^\text{PP}$ \\
        \hline
    \end{tabular}%
    }
\end{table}

\begin{table}[ht!]
    \centering
    \scriptsize
    \caption{Complexity summary of containment and equivalence for CQ$^\neq$Ns and UCQ$^\neq$Ns 
    }
    \label{tab:summary_cqnneq_ucqnneq}
    \medskip

    \resizebox{\textwidth}{!}{%
    \begin{tabular}{|c|cc|cc|}
        \hline
        \multirow{2}{*}{\textbf{Semirings}}
        & \multicolumn{2}{c|}{\textbf{CQ$^\neq$Ns}}
        & \multicolumn{2}{c|}{\textbf{UCQ$^\neq$Ns}} \\
        \cline{2-5}
        & \textbf{Containment} & \textbf{Equivalence}
        & \textbf{Containment} & \textbf{Equivalence} \\
        \hline

        $\mathbb{B}$ (\emph{md})
        & 
        & 
        & 
        &  \\
        
        (1) fixed arity bound on relations
        & $\Pi^p_2$-c~\cite{Geck}
        & $\Pi^p_2$-c~\cite{Geck}
        & $\Pi^p_2$-c~\cite{Geck}
        & $\Pi^p_2$-c~\cite{Geck} \\
        
       (2) unbounded arity
        & coNEXP-c~\cite{Geck}
        & coNEXP-c~\cite{Geck}
        & coNEXP-c~\cite{Geck}
        & coNEXP-c~\cite{Geck} \\
        \hline

         Distr.Latt. $>2$ elements (\emph{md})
        & $\Pi^p_2$-c
        & $\Pi^p_2$-c
        & $\Pi^p_2$-c
        & $\Pi^p_2$-c \\
  \hline

       $\mathrm{PosBool}(X, \overline{X})$ (\emph{pt})
        & $\Pi^p_2$-c
        & $\Pi^p_2$-c
        & $\Pi^p_2$-c
        & $\Pi^p_2$-c \\
  \hline

        $\mathrm{Which}(X)$ (\emph{md}) 
        & 
        & 
        & 
        &  \\
        
        (1) fixed arity bound on relations
        & $\Pi^p_2$-c
        & $\Pi^p_2$-c
        & $\Pi^p_2$-c
        & $\Pi^p_2$-c\\
        
       (2) unbounded arity
        & coNEXP-c
        & coNEXP-c
        & coNEXP-c
        & coNEXP-c\\
        \hline

        $\mathrm{Sorp}(X, \overline{X})$,$\mathrm{Why}(X, \overline{X})$,$\mathbb{B}[X,\overline{X}]$ (\emph{pt})
        & $\Pi^p_2$-c
        & $\Pi^p_2$-c
        & $\Pi^p_2$-c
        & $\Pi^p_2$-c \\
        \hline

        $\mathbb{N}[X,\overline{X}]$,$\mathrm{Trio}(X, \overline{X})$ (\emph{pt})
        & in coNP$^\text{PP}$
        & in coNP$^\text{PP}$
        & in coNP$^\text{PP}$
        & in coNP$^\text{PP}$ \\
        \hline

 $\mathbb{N}$ (\emph{md})
        & undecidable~\cite{Kolaitis2}
        & in coNP$^\text{PP}$
        & undecidable~\cite{Kolaitis2}
        & in coNP$^\text{PP}$ \\
          \hline
    \end{tabular}%
    }
\end{table}

\section{Preliminaries}\label{preliminaries}

In this section, we introduce the notion of a commutative semiring and that of a database instance annotated with elements from a commutative semiring. Further, we give the syntax of the classes of queries we consider and their semantics on semiring-annotated database instances, including suitable notions of containment and equivalence between queries under such semantics.

The paper studies containment and equivalence for various semirings whose definitions are also detailed in this section. Our methodology relies on canonical databases associated with conjunctive queries and various notions of containment mapping, also included here for the reader's benefit.

\subsection{Semirings}

We start with the basic algebraic structure we need.

\label{sec:semirings}
\begin{Def}[Semiring] A \emph{semiring} is an algebraic structure $\mathbb{S}=(S, +, \cdot,0,1)$, where  addition $+$ and multiplication $\cdot$ are binary operations on $S$, and\/  $0$, $1$ are elements of $S$ such that the following properties hold: 
\begin{itemize}
\item $(S, +, 0)$  is a commutative monoid,  $(S, \cdot, 1)$ is a monoid, and $0\neq 1$;
\item (distributivity): for all elements $a,b,c \in S$, we have that 
$a\cdot (b+c) = a\cdot b + a \cdot c$;
\item (annihilation): for every element $a \in S$, we have that $a \cdot 0 = 0 \cdot a = 0$. 
\end{itemize}
$\mathbb{S}$ is said to be \emph{commutative} if $(S, \cdot, 1)$ is commutative.\footnote{All the semirings we consider below are commutative since, in the database application, multiplication is meant to be an abstraction of conjunction.} $\mathbb{S}$ is said to be \emph{idempotent} if $(S, +, 0)$ is idempotent, and \emph{multiplicatively idempotent} if $(S, \cdot, 1)$ is idempotent. If\/ $\mathbb{S}$ is both idempotent and multiplicatively idempotent, we say that $\mathbb{S}$ is \emph{fully idempotent}.
\end{Def}

\begin{Def}[cf.~\cite{GT2}]
Let \(S=(S,+,\cdot,0,1)\) be a semiring. We say that \(S\) has \emph{zero divisors} if there are non-zero elements \(s,t\in S\) such that \(s\cdot t=0\). Moreover, \(S\) is called \emph{\(+\)-positive} if \(s+t=0\) implies \(s=0\) and \(t=0\), for all \(s,t\in S\). Finally, \(S\) is called \emph{positive} if it is both \(+\)-positive and has no zero divisors.
\end{Def}
\begin{Def}
Let \(S=(S,+,\cdot,0,1)\) be a semiring. The \emph{natural preorder} on \(S\) is the binary relation \(\leq\) defined by $s\leq t$ iff there exists a $u\in S$ such that $s+u=t$. We say that \(S\) is \emph{naturally ordered} if \(\leq\) is a partial order.  If $S$ is naturally ordered, we denote its natural order by 
$\leq_S$.
\end{Def}

Examples of useful semirings include the following:

\begin{itemize}

\item
The Boolean semiring $\mathbb{B}=(\{\bot,\top\},\vee,\wedge,\bot,\top)$ is the standard habitat of 
logical truth.
\item A totally ordered set $(S, \leq)$ with least element $s$ and greatest element $t$ induces the so-called \emph{min-max semiring} $(S, \max, \min, s, t)$. 
Particularly important examples include the Boolean semiring, the fuzzy semiring $\mathbb{F}=([0,1],\text{max},\text{min},0,1)$, and the  
\emph{access control semiring}, also called the \emph{security semiring} \cite{Foster}, which
is a min-max semiring with elements $0 < \Tsec < \Sec < \Cnf < \Pub=1$ where 0 stands for ``inaccessible'' (or ``false''),
$\Tsec$ is ``top secret'', $\Sec$ is ``secret'', $\Cnf$ is ``confidential'', and
$\Pub$ is ``public''. It is used for reasoning about access restrictions to atomic facts, and
the clearance levels that are necessary to verify the truth of a sentence under such restrictions.
\item A more general class (than min-max semirings) is the class of  \emph{lattice semirings} $\mathbb{L} =(L,\lor,\land,s,t)$ induced by
a bounded distributive lattice $(L,\leq)$.\footnote{An idempotent semiring is also a  semilattice
and its natural preorder is the same as the semilattice order,  hence it is naturally ordered.} Clearly, lattice semirings are commutative and fully idempotent (both addition and multiplication are idempotent). They are also naturally ordered, and their natural order coincides with the lattice order. Moreover, they are \(+\)-positive, since \(a\vee b=\bot\) implies \(a=\bot=b\). However, lattice semirings are not necessarily positive, since they may have zero divisors (e.g.\ the Boolean algebra $\mathcal{P}(\mathbb{B})$).

\item The \emph{bag semiring} of the natural numbers $\mathbb{N} = (\mathbb{N}, +, \cdot, 0, 1)$ is used to count the number of proof trees or evaluation strategies that establish the truth of a sentence. It is also important for bag semantics in databases. 

\end{itemize}

In addition to the above, in this paper we will use the following  \emph{provenance} semirings  (cf.~\cite{Green}). Here, \(X\) denotes a
countable set of indeterminates, called \emph{provenance tokens} in~\cite{Green0}, and used to annotate the tuples of the database given as input to a query.  All the semirings below are commutative,
positive and naturally ordered (just as $\mathbb{N}$).

\begin{itemize}

    \item In general, if \(\mathbb{K}=(K,+,\cdot,0,1)\) is a commutative semiring, we denote by \(\mathbb{K}[X]\) the semiring of polynomials over indeterminates in \(X\) with coefficients in \(K\). Its elements are finite sums of monomials with coefficients in \(K\), and addition and multiplication are defined in the usual way, using the operations of \(\mathbb{K}\) on coefficients. In particular, the semirings \(\mathbb{N}[X]\) and \(\mathbb{B}[X]\) below are obtained by taking coefficients respectively in \(\mathbb{N}\) and in \(\mathbb{B}\).
    
    \item \(\mathbb{N}[X]=(\mathbb{N}[X],+,\cdot,0,1)\) is the semiring of provenance polynomials over \(X\), with coefficients in \(\mathbb{N}\) and with the usual operations.  Its natural order is the coefficientwise order: $A\leq_{\mathbb{N}[X]} B$ iff every monomial occuring in $A$ also occurs in $B$ with a coefficient at least as large as that in $A$.

    \item \(\mathbb{B}[X]=(\mathbb{B}[X],+,\cdot,0,1)\) 
    is constructed like $\mathbb{N}[X]$, except we use 
    $\mathbb{B}$ for coefficients.
%
%
    $\mathbb{B}[X]$ is idempotent and hence naturally ordered. Its order can also be characterized by the inclusion of monomials: $A\leq_{\mathbb{B}[X]} B$ iff every monomial occurring in $A$ also occurs in $B$.

    \item \(\mathrm{Trio}(X)=(\mathrm{Trio}(X),+,\cdot,0,1)\) is the quotient of \(\mathbb{N}[X]\) by the congruence that identifies polynomials after dropping exponents (that is, two polynomials are identified whenever they become equal after replacing every positive exponent by \(1\)). \vt{Think of more clear def}  To see that this relation is indeed a congruence, let $f\colon \mathbb{N}[X]\longrightarrow \mathbb{N}[X]$ be the operation that sends each monomial to its square-free\footnote{By the term `square-free', we simply mean that the exponents of the indeterminates are 1.} version, by replacing every positive exponent by \(1\), and then collects equal monomials by summing their coefficients. Thus, for instance, $f(2x^2y+3xy^4+z^3)=5xy+z.$ We define \(P\equiv_{\mathrm{Trio}} Q\) iff \(f(P)=f(Q)\). This is clearly an equivalence relation.\footnote{We show that it is compatible with the semiring operations. Let \(A\equiv_{\mathrm{Trio}}A'\) and \(B\equiv_{\mathrm{Trio}}B'\). Then, \(f(A)=f(A')\) and \(f(B)=f(B')\). We have $f(A+B)=f(f(A)+f(B)).$ Hence $f(A+B) = f(f(A)+f(B)) = f(f(A')+f(B')) = f(A'+B'),$ and therefore $A+B\equiv_{\mathrm{Trio}}A'+B'$. For example, if $A=2x^2y+z^3$ and $B=3xy^4+z$, then  $f(A+B)=f(2x^2y+3xy^4+z^3+z)=5xy+2z.$ On the other hand, $f(A)=2xy+z$ and $f(B)=3xy+z$, so $f(f(A)+f(B))=f(2xy+z +3xy+z)=5xy+2z$. Similarly, dropping exponents is compatible with multiplication up to applying $f$ after the product: $f(A\cdot B)=f(f(A)\cdot f(B)).$ Thus, $f(A\cdot B)=f(f(A)\cdot f(B))= f(f(A')\cdot f(B'))=f(A'\cdot B').$ For example, if $A=x^2y$ and $B=xy^3z$, then $f(A\cdot B)=f(x^3y^4z)=xyz$. On the other hand, $f(A)=xy$ and $f(B)=xyz$, so $f(f(A)\cdot f(B)) = f(xy\cdot xyz) = f(x^2y^2z) = xyz$. Therefore, \(A\cdot B \equiv_{\mathrm{Trio}}A'\cdot B'\). It follows that \(\equiv_{\mathrm{Trio}}\) is a congruence on \(\mathbb{N}[X]\).} The semiring \(\mathrm{Trio}(X)\) is naturally ordered. Indeed, every element can be represented as a polynomial whose monomials are square-free: $A\leq_{\mathrm{Trio}(X)} B$ iff every square-free monomial occurs in $B$ with coefficient at least as large as its coefficient in $A$.  Notice that neither addition nor multiplication is idempotent in Trio$(X)$.

    \item \(\mathrm{Why}(X)=(\mathcal{P}_{\text{fin}}(\mathcal{P}_{\text{fin}}(X)),\cup,\Cup,\emptyset,\{\emptyset\})\) is the why-provenance semiring. Its elements are finite sets of finite sets of indeterminates, addition is union, while multiplication is pairwise union: $A \Cup B=\{a\cup b \mid a\in A,\ b\in B\}.$ \(\mathrm{Why}(X)\) is idempotent hence naturally ordered: since its additive operation is union, its natural order is simply the inclusion: $A \leq_{\mathrm{Why}(X)} B$ iff $A \subseteq B.$ Notice that \(\mathrm{Why}(X)\) is additively idempotent, since addition is union. However, multiplication is not idempotent in general. Equivalently, \(\mathrm{Why}(X)\) is the quotient of \(\mathrm{Trio}(X)\) obtained by dropping coefficients. \vt{explain what this means}  Thus, if elements of \(\mathrm{Trio}(X)\) are represented as polynomials with square-free monomials and coefficients in \(\mathbb N\), then two such polynomials are identified in \(\mathrm{Why}(X)\) exactly when they have the same monomials with non-zero coefficient.

    \item \(\mathrm{Sorp}(X)=(\mathrm{Sorp}(X),+,\cdot,0,1)\) is the absorptive semiring, defined as the quotient of \(\mathbb{N}[X]\) by the smallest congruence satisfying the absorption law $a+ab=a.$ \(\mathrm{Sorp}(X)\) is naturally ordered. Since it is obtained as a quotient of \(\mathbb{N}[X]\) by the absorption law, each element can be represented by a finite sum of monomials with coefficient \(1\) where no monomial divides another.  In this representation, addition is polynomial addition followed by the elimination of all absorbed monomials. Thus, if $A,B \in \mathrm{Sorp}(X)$, the natural order is given by $A\leq_{\mathrm{Sorp}(X)} B$ iff for each monomial $a$ occurring in $A$ there exists a monomial $b$ occurring in $B$ such that $b$ divides $a$. Equivalently, $A\leq_{\mathrm{Sorp}(X)}B$ iff $A+B=B$.

    \item \(\mathrm{Which}(X)=(\mathcal{P}_{\text{fin}}(X)\cup\{\bot\},+,\cdot, \bot,\emptyset)\) is the original ``lineage'' semiring~\cite{CuiWidom}
    \footnote{This semiring was denoted $\mathrm{Lin}(X)$
    in~\cite{Green}. However, ``lineage'' was also used for $\mathrm{Trio}(X)$~\cite{BenjellounEtAlULDB2006} and even for $\mathrm{PosBool}(X)$~\cite{BookOnProbDBs}. Hence, $\mathrm{Lin}(X)$ was renamed $\mathrm{Which}(X)$
    in~\cite{Tannen2013BunemanFestschrift,GT2}.
    }
    For \(S,T\neq \bot\), we have $S+T=S\cdot T=S\cup T,$ while $\bot+S=S+\bot=S, \bot\cdot S=S\cdot \bot=\bot.$ $\mathrm{Which}(X)$ is naturally ordered and its natural order is given by: $A\leq_{\mathrm{Which}(X)} B$ iff either $A=\bot$, or $A,B\in\mathcal P_{\mathrm{fin}}(X)$ and $A\subseteq B$.

   \item \(\mathrm{PosBool}(X)=(\mathrm{PosBool}(X),\vee,\wedge,\bot,\top)\) is the semiring of positive Boolean formulas over \(X\), modulo Boolean equivalence. Equivalently, it is the free distributive lattice generated by \(X\). \(\mathrm{PosBool}(X)\) is naturally ordered: since its additive operation is disjunction, its natural order is the logical implication: $A\leq_{\mathrm{PosBool}(X)}B$ iff $A$ logically entails (in Boolean logic) $B$. 
\end{itemize}

\begin{Rmk} The provenance semirings above were introduced (not as semirings but as paradigms) in different papers in order to capture data provenance at different levels of abstraction.  As explained in~\cite{TannenGreen17} these semirings can be compared in terms of how detailed is the provenance they capture. In the Hasse diagram below,
$\mathbb{S}_1(X)$ is \emph{strictly more discerning/informative}
than $\mathbb{S}_2(X)$ when there is an edge (or several) from $\mathbb{S}_1(X)$ \emph{above} to $\mathbb{S}_2(X)$ \emph{below}:

\begin{center}
\begin{tikzpicture}[x=1.5cm,y=1.05cm,
                    every node/.style={inner sep=3pt}]
  \node (N)       at ( 0, 0) {$\mathbb{N}[X]$};
  \node (B)       at (-1,-1) {$\mathbb{B}[X]$};
  \node (Trio)    at ( 1,-1) {$\mathrm{Trio}(X)$};
  \node (Sorp)    at (-2,-2) {$\mathrm{Sorp}(X)$};
  \node (Why)     at ( 0,-2) {$\mathrm{Why}(X)$};
  \node (PosBool) at (-1,-3) {$\mathrm{PosBool}(X)$};
  \node (Which)   at ( 1,-3) {$\mathrm{Which}(X)$};

  \draw (N) -- (B);
  \draw (N) -- (Trio);
  \draw (B) -- (Sorp);
  \draw (B) -- (Why);
  \draw (Trio) -- (Why);
  \draw (Sorp) -- (PosBool);
  \draw (Why) -- (PosBool);
  \draw (Why) -- (Which);
\end{tikzpicture}
\end{center}
\end{Rmk}

Although the present paper focuses on the semirings listed above, other semirings of interest in annotated database theory include the tropical and Viterbi semirings. We do not present containment or equivalence results for databases annotated over these semirings here: the reason is that, although they are natural and important examples, their containment study requires different techniques, and we leave it for a different paper.

\begin{itemize}
    \item $\mathbb{T}=([0,\infty],\min,+,\infty,0)$ is the tropical semiring, where addition is given by the minimum and multiplication by the usual addition, extended in the expected way to \(\infty\). Its natural order is the reverse of the usual order: for \(a,b\in[0,\infty]\), $a\leq_{\mathbb{T}} b$ iff $a\geq b$ in the usual order. The semiring \(\mathbb{T}\) is additively idempotent, naturally ordered, and positive.

    \item $\mathbb{V}=([0,1],\max,\cdot,0,1)$ is the Viterbi semiring, where addition is given by the maximum and multiplication by the usual product. Its natural order coincides with the usual order on \([0,1]\): for \(p,q\in[0,1]\), $p\leq_{\mathbb{V}} q$ iff $p\leq q$ in the usual order. As for $\mathbb{T}$, the semiring \(\mathbb{V}\) is additively idempotent, naturally ordered and positive.
\end{itemize}

The two semirings are isomorphic via the map $\varphi\colon \mathbb{V}\longrightarrow\mathbb{T},$ $\varphi(p)=-\log(p),$ with the convention \(\varphi(0)=\infty\). Thus, working with the Viterbi presentation is sometimes more convenient, since its natural order agrees with the usual order on \([0,1]\), whereas the tropical presentation reverses the usual order.

\subsection{Semiring-annotated databases}
\label{sec:semiring-databases}

\vt{perhaps nerge; Semirings and semiring-annotated databases?}
In the next two definitions, we assume that $D$ is a fixed denumerable set, which  will be the \emph{domain} of elements occurring in databases. For  $n\geq 1$,  we write $D^n$ to denote the set of all $n$-tuples from $D$.  We also write  $D^0$ to denote the singleton  $\{ \langle \rangle\}$, where $\langle \rangle$ is the empty tuple (we assume that $\langle \rangle \ \notin D$).

\begin{Def}
Let  $\mathbb{S}=(S, +, \cdot,  0,1)$ be a semiring and let $n\geq 0$ be a non-negative integer. 

1. An $n$-ary $\mathbb{S}$-relation $R$ is a function $R \colon D^n\longrightarrow S$ of finite support, that is to say, the set $\supp(R) := \{(d_1,\ldots,d_n)\in D^n \mid  R(d_1,\ldots,d_n) \neq 0\}$ is finite.  We call $n$ the \emph{arity} of $R$.

2. An \emph{$\mathbb{S}$-relation} is an $n$-ary $\mathbb{S}$-relation $R$, for some $n\geq 0$.
\end{Def}

 Note that there are  $|S|$-many $0$-ary $\mathbb{S}$-relations. Indeed,  for every $a\in S$, we have the $0$-ary relation $R^0_a$ such that $R^0_a(\langle \rangle)= a$. If $a\neq 0$, then $\supp (R^0_a) = \{\langle \rangle\}$, while $\supp (R^0_0) = \emptyset$.

 As usual, a \emph{relational schema} or, simply, a \emph{schema} is a finite sequence $\tau= (R_1, \ldots, R_p)$ of relation symbols, where each relation symbol $R_i$ has a designated positive integer $r_i$ as its \emph{arity}.
 \begin{Def}
 Let  $\tau= (R_1, \ldots,R_p)$ be a schema. An  \emph{$\mathbb{S}$-database (or $\mathbb{S}$-instance) over $\tau$} is a finite sequence $\mathbf{I}=(R_1^\mathbf{I},\ldots,R_p^\mathbf{I})$ of\/ $\mathbb{S}$-relations such that the arity of each $\mathbb{S}$-relation $R_i^\mathbf{I}$ matches the arity of the  relation symbol $R_i$, $1\leq i\leq p$.
\end{Def}

Observe that a schema typically does not contain any 0-ary relations, so the  
$\mathbb{S}$-relations appearing in a database are always of arity $\geq 1$.

\subsection{(U)CQ\texorpdfstring{$^\neq$}{≠}s and their semiring semantics}

Let us start by recalling the definition of a conjunctive query:

\begin{Def}[(Unions of) Conjunctive Queries]\label{UCQdef}
Let $\tau$ be a database schema (i.e.\ a logical vocabulary formed by relation symbols). A \emph{conjunctive query} (in short, CQ) is a first-order $\tau$-formula of the form
$$P(\vec{u}) := \exists \vec{z} (R_1(\vec{u}_1) \wedge \ldots \wedge R_n(\vec{u}_n))$$
where $R_1, \ldots, R_n$ are relation symbols from $\tau$ (with possible repetitions), the sequences of variables $\vec{u}_1, \ldots, \vec{u}_n$ match the arity of the corresponding relations in the atom they appear in, and such that  for all $i\leq n$, the sequence $\vec{u}_i$ contains only variables in $\vec{u} \cup \vec{z}$. We say that the sequence of variables $\vec{u}$  is the \emph{head} of $P$ while the formula $R_1(\vec{u}_1) \wedge \ldots \wedge R_n(\vec{u}_n)$ is the \emph{body}. A \emph{Boolean CQ} is simply one with empty head. Finally, a \emph{union of CQs} (in short, UCQ), $\bigcup_{1\leq i \leq n} P_i$, is simply a disjunction where the $P_i$'s are CQs with the same head $\vec{u}$.  Sometimes, we abbreviate a $\bigcup_{1\leq i \leq n} P_i$ by $\overline{P}$. Importantly, our definition allows for repetitions of atoms in the body of a given  CQ. 
\end{Def}

There is a standard way to define semantics over semiring-annotated databases for these queries in the literature (e.g.~\cite{Green, Ioannidis}) as follows.

\begin{Def}[Semiring semantics of (U)CQs]
An \emph{assignment} is a unary function $v$ from the set of variables of a query $P$ into $D$. Assignments operate component-wise on tuples in the expected way. 
Let $P$ be a CQ in a schema $\tau$
$$
P(\vec{u}):= \exists \vec{z} (R_1\left(\vec{u}_1\right) \wedge \ldots \wedge R_n\left(\vec{u}_n\right))
$$
where $\vec{u}$ is an $n$-tuple of variables and let $\mathbf I$ be an $\mathbb{S}$-instance of the  schema $\tau$. The \emph{semantics of $P$ on $\mathbf I$}  is the $n$-ary $\mathbb{S}$-relation  defined, for all $n$-tuples $\vec{t}$ of elements from $D$, 
as $$
\llbracket P \rrbracket^\mathbf{I}(\vec{t}) \stackrel{\text {def}}{=} \sum_{v \ \text{such that} \ v(\vec{u})=\vec{t}} \prod_{i=1}^n R_i^\mathbf{I}\left(v\left(\vec{u}_i\right)\right)
$$
and the sums and products are in $\mathbb{S}$.\footnote{Obviously, an inductive version of this definition would have that $
\llbracket R \rrbracket^\mathbf{I}(\vec{t}) \stackrel{\text {def}}{=}  R_i^\mathbf{I}(v(\bar{t}_i))
$ and that $\exists$ corresponds to a sum and $\wedge$ to a product.} Similarly, for a UCQ of the form $\bigcup_{1\leq i \leq m} P_i$, its semantics is the $\mathbb{S}$-relation  defined, for all tuples $\vec{t}$ of elements from $D$,
as $$  \sum_{1\leq i\leq m}  \llbracket P_i \rrbracket^\mathbf{I}(\vec{t}).$$
Naturally, when dealing with Boolean (U)CQs, the corresponding $\mathbb{S}$-relation will have 0-arity.
\end{Def}

Notice that when the semiring is not multiplicatively idempotent, it makes a difference semantically whether we repeat atoms within the body. Next, we provide an example of how this works  in a concrete $\mathbb{S}$-instance.

\begin{Exm} \vt{Seems OK but I need to revisit still}
Let $\mathbb{L}$ be a semiring, let $a\in D$ be an element of the domain and let \(A\) be the \(\mathbb{L}\)-instance over the schema with unary predicates \(R,S,T\) such that $R^A,S^A,T^A$ all have support $\{a\}$ and 
such that $R^A(a)=x_1,S^A(a)=x_2, T^A(a)=x_3,$  where $x_1,x_2,x_3$
are distinct elements of $X$. 

Consider the unary UCQ \(q(y)\) with disjuncts $q_1(y):= R(y)\wedge S(y), q_2(y) := R(y)\wedge R(y)\wedge T(y), q_3(y) := R(y)\wedge S(y)\wedge T(y), q_4(y) := R(y)\wedge S(y)\wedge T(y), q_5(y) := S(y)\wedge T(y).$ The output of $q$ has only one element, $a$ and we compute its annotation in some of the provenance semirings built from $X$ and listed above.

If $\mathbb{L}$ is \(\mathbb{N}[X]\), the only non-zero valuation maps \(y\) to \(a\), and therefore the annotation of $q$ is $x_1x_2+x_1^2x_3+2x_1x_2x_3+x_2x_3.$ If we evaluate the same query over \(\mathbb{B}[X]\), coefficients are dropped, and the annotation of \(a\) becomes $x_1x_2+x_1^2x_3+x_1x_2x_3+x_2x_3.$ Over \(\mathrm{Trio}(X)\), exponents are dropped while coefficients are kept, so we obtain $x_1x_2+x_1x_3+2x_1x_2x_3+x_2x_3.$ Over \(\mathrm{Sorp}(X)\), the absorption law eliminates the monomial \(x_1x_2x_3\), since it is absorbed by both \(x_1x_2\) and \(x_2x_3\): the annotation is $x_1x_2+x_1^2x_3+x_2x_3.$ Over \(\mathrm{Why}(X)\), the annotation is the set of alternative witnesses $\big\{\{x_1,x_2\},\{x_1,x_3\},\{x_1,x_2,x_3\},\{x_2,x_3\}\big\}.$ Over \(\mathrm{PosBool}(X)\), the annotation is considered modulo Boolean equivalence, and becomes $(x_1\wedge x_2)\vee (x_1\wedge x_3)\vee (x_2\wedge x_3).$ Finally, over \(\mathrm{Which}(X)\), each of $x_1,x_2$  and $x_3$ occur in some derivation, so the annotation is $\{x_1,x_2,x_3\}$.
\end{Exm}

Next, we define queries that arise from adding expressions that describe identity relations between objects in a conjunctive query.

\begin{Def}[(Unions of) Conjunctive queries with disequations]
A \emph{conjunctive query with disequations} (in short, CQ$^\neq$) is simply a CQ where one is allowed literals of the form $x=y$\footnote{Our inclusion of expressions of the form $x=y$ is not strictly necessary, but rather merely a convenience, as obviously they can be eliminated by considering appropriate variable substitutions within a query (thus any safe query involving equations is semantically equivalent to one without them).} and $x\neq y$ in the body of the query. Furthermore,  we focus on \emph{safe} queries where the only variables allowed are those appearing in some relational atom of  the query. The notions of head and body are defined as before except that the body now also includes the new literals involving equality. A \emph{union of conjunctive queries with disequations} (in short, UCQ$^\neq$) is, as before, just a disjunction of CQ$^\neq$s with the same head.
\end{Def}

\begin{Def}[Semiring semantics of (U)CQ$^\neq$s] The only difference with the case of (U)CQs is what to do for the expressions $x=y$ and $x\neq y$. We simply assume that the relation $=$ has a fixed meaning where for any instance $\mathbf I$ and assignment $v$,
\[
\llbracket v(x)=v(y)\rrbracket^\mathbf{I} =
\begin{cases}
1 & \text{if } v(x)=v(y),\\
0 & \text{otherwise,}
\end{cases}
\qquad
\llbracket v(x)\neq v(y)\rrbracket^\mathbf{I} =
\begin{cases}
1 & \text{if } v(x)\neq v(y),\\
0 & \text{otherwise.}
\end{cases}
\]

\end{Def}

\begin{Conv} 
 We will restrict our attention to only \emph{consistent} CQ$^\neq$s, i.e.\ queries such that $x\neq x$ does not follow logically from the equations and disequations in the body of the query for any variable $x$. The reason is that the containment problem we will discuss below is otherwise  trivial  because any inconsistent query takes the value $0$, which is the minimum in any naturally ordered semiring. 
 \end{Conv}

\begin{Def} Let $P$ be a CQ$^\neq$.

1. A \emph{completion} $P'$ of  $P$ is any CQ$^\neq$ that comes from adding to $P$, for all variables $x,y$ that appear in a relational atom of $P$, either $x=y$ or $x\neq y$ (as long as the resulting CQ$^\neq$ $P'$  is also consistent). 

2. Given a completion $P'$, an \emph{identification} $P^{\mathrm{id}}$ of  $P'$ is a CQ$^\neq$ that comes from eliminating all equations by applying a \emph{canonical substitution} uniformly throughout $P'$: considering the equivalence relation on the variables of $P'$ given by $x\equiv y$ iff $x=y$ is a logical consequence of the equations and disequations in $P'$, a canonical substitution maps all elements in an equivalence class to one term of that class.\footnote{
Take $P$ to be a CQ$^\neq$. If we apply a canonical substitution to $P$ directly (not to a completion), we will obtain an equivalent query that contains only relational atoms and disequations.}     
\end{Def}

There might  be clearly more than one identification of a given completion but they are all isomorphic up to variable renaming. If there are no equations and only disequations in a completion, its only identification is itself. Given a query, its identifications are all the identifications of its completions.\footnote{The idea of identifications is roughly from~\cite{Nutt}, although the language there is richer and includes comparison relations so they talk about `linearizations' instead.} Observe that any identification $P^{\mathrm{id}}$ of  a completion $P'$ is logically equivalent to $
P'$ itself. Furthermore, where $\mathbb{S}$ is an ordered semiring, observe that for any $\mathbb{S}$-database instance $\mathbf I$, $\llbracket P' \rrbracket^\mathbf{I} =\llbracket P^{\mathrm{id}}\rrbracket^\mathbf{I}$ when $P'$ is a completion of a CQ$^\neq$ $P$ and $P^{\mathrm{id}}$ is an identification of $P'$. 

\begin{Exm}
Consider the queries $Q:= \exists x, y (R(x,y) \wedge R(y,x))$ and 
$P:= \exists x, y (R(x,y) \wedge x=y).$ Observe that $Q$ has the following two  possible identifications:
\begin{itemize}
\item[1.] $\exists x, y (R(x,y) \wedge R(y,x) \wedge x \neq y)$
\item[2.] $\exists x (R(x,x) \wedge R(x,x))$,
\end{itemize}
where $\exists x (R(x,x) \wedge R(x,x))$ comes from the completion $\exists x, y (R(x,y) \wedge R(y,x) \wedge x = y)$ and the canonical substitution that sends $y$ to $x$. Similarly, $P$ (which is already a completion of itself) has only the following identification:
\begin{itemize}
\item[1.] 
 $\exists x R(x,x)$.
\end{itemize}
\end{Exm}

\subsection{(U)CQNs and their semiring semantics}\label{section2.4}

Our next step is to introduce queries where negations appear in from of relational atoms instead of equations.

\begin{Def}[(Unions of) Conjunctive Queries with Negated Atoms]
A \emph{conjunctive query with negated atoms} (in short, CQN) is simply a CQ where one further allows literals of the form $\neg R(\vec{x})$  in the body of the query. Once more,  we focus on \emph{safe} queries where the only variables allowed are those appearing in some relational (unnegated) atom of  the query. The notions of head and body are defined as before except that the body now also includes the new literals. A \emph{union of conjunctive queries with negated atoms} (in short, UCQN) is, as before, just a disjunction of CQNs with the same head.
\end{Def}

In order to give semantics for (U)CQNs, we need to find a way to deal with negation in front of atoms. There is no obvious  answer in this case, as opposed to the semantics of disequations.  The approach we choose to follow is the  framework of Gr\"adel and Tannen~\cite{GT}, where first-order formulas are evaluated not simply as true or false, but as elements of a commutative semiring by means of so-called ``semiring interpretations''. However, other options are left as lines of future investigation~\cite{Ba, Suciu}.

\begin{Def}[Semiring semantics of (U)CQNs]
Let \(\tau\) be a finite relational vocabulary and let \(A\) be a finite set of elements. We denote by \(\mathrm{Facts}_A(\tau)\) the set of all ground relational atoms over \(\tau\) and \(A\) (i.e.~expressions of the form $R(\vec{a})$ where $R\in \tau$ and $\vec{a}$ is a tuple of elements from $A$), and by \(\mathrm{NegFacts}_A(\tau)\) the set of their negations. We write $\mathrm{Lit}_A(\tau)=\mathrm{Facts}_A(\tau)\cup \mathrm{NegFacts}_A(\tau)$ for the set of ground literals. An interpretation over a commutative semiring \(\mathbb S=(S,+,\cdot,0,1)\) (an $\mathbb S$-interpretation) is a map $\pi\colon \mathrm{Lit}_A(\tau)\longrightarrow S.$  Let $P$ be a CQN in a schema $\tau$ \vt{The ARBITRARY finite set $A$ is part of $\pi$? When we define containment we require that $A$ contains the active domain of an instance (compatibility) but and annotates the same way as the instance. \vt{THis is crucial and it is not well explained. Let's discuss.}}
$$
P(\vec{u}):= \exists \vec{z} (l_1\left(\vec{u}_1\right) \wedge \ldots \wedge l_n\left(\vec{u}_n\right)),
$$
where each $l_i(\vec{u}_i)$ is a literal. An assignment in this case is a mapping $v$ from the variables of $P$ to $A$. We extend $\pi$ to $P$ by letting, for each tuple $\vec{t}$ from $A$,
\[
\pi\llbracket P(\vec{t})\rrbracket
   =\sum_{{v \ \text{such that} \ v(\vec{u})=\vec{t}}} \prod_{i=1}^n \pi\llbracket l_i(v(\vec{u}_i))\rrbracket.
\]
\end{Def}

Finally, we may consider queries that involve the expressivity of both (U)CQNs and (U)CQ$^\neq$s  in the following manner.

\begin{Def}[Semiring semantics of (U)CQ$^\neq$Ns] We can define (U)CQ$^\neq$Ns as (U)CQs obtained by permitting equations, disequations and negated atoms to simultaneously appear in the body of the queries. Safety can be defined exactly as before.
One can then enrich the interpretation-based semantics above by allowing atomic formulas involving equality and setting
\[
\pi\llbracket a=b\rrbracket =
\begin{cases}
1 & \text{if } a=b,\\
0 & \text{otherwise,}
\end{cases}
\qquad
\pi\llbracket a\neq b\rrbracket =
\begin{cases}
1 & \text{if } a\neq b,\\
0 & \text{otherwise.}
\end{cases}
\]
\end{Def}
The framework of~\cite{GT} is, in fact,  defined for arbitrary first-order formulas, and hence allows negation to occur freely. In that setting, to compute the interpretation of a negated formula, the formula is first transformed into negation normal form, so that negation is pushed down to the atomic level. In this paper, we will only consider query languages in which negation is already restricted to atomic formulas. This allows us to work directly with positive and negative relational literals, without repeatedly referring to the transformation into negation normal form.

\subsection{Dual indeterminate polynomial semirings}\label{Dual indeterminate polynomial semirings}

We now move from the positive semirings introduced above to some special quotients (useful in handling with negation) following the construction of dual-indeterminate provenance semirings in~\cite{GT}. The semirings $\mathbb{N}[X]$, $\mathbb{B}[X]$, $\mathrm{Trio}(X)$, $\mathrm{Sorp}(X)$, $\mathrm{Why}(X)$ and $\mathrm{PosBool}(X)$ track provenance for positive database facts: to deal with negated relational atoms, we introduce a second copy of the provenance tokens used for negative facts.
\begin{Def}[\cite{GT}]
Let $X$ and $\overline{X}$ be two disjoint sets equipped with a fixed one-to-one correspondence. For every $x \in X$, we denote by $\overline{x} \in \overline{X}$ the corresponding element, and we call $x$ and $\overline{x}$ \emph{complementary tokens}. The elements of $X$ will be used as positive provenance tokens, while the elements of $\overline{X}$ will be used as negative provenance tokens. 
\end{Def}
\begin{Conv}
Given a ground atom $R(\vec{a})$, if $x \in X$ is used to annotate
$R(\vec{a})$, then the corresponding token $\overline{x}$ is reserved (but may not be used) for the negated literal $\neg R(\vec{a})$.
\end{Conv}

The first dual semiring is obtained by starting from the polynomial semiring $\mathbb{N}(X \cup \overline{X})$ and by identifying with $0$ all products that contain a pair of complementary tokens.
\begin{Def}[\cite{GT}]
We define $\mathbb{N}[X,\overline{X}]$ as the quotient of
$\mathbb{N}[X\cup \overline{X}]$ by the congruence generated by the equations $x\cdot \overline{x}=0$ for every $x\in X$. The elements of $\mathbb{N}[X,\overline{X}]$ are called \emph{dual-indeterminate provenance polynomials}.
\end{Def}
Notice that $\mathbb{N}[X,\overline{X}]$ is no longer positive in the sense of having no zero divisors. In fact, for every $x\in X$ we have $x\cdot \overline{x}=0$, although both $x$ and $\overline{x}$ are non-zero. We now define the corresponding dual versions of the other provenance semirings considered above.

\begin{itemize}
\item Following the same quotient construction used for $\mathbb{N}[X,\overline{X}]$,
we define the dual versions of \(\mathbb{B}[X,\overline{X}]\), \(\mathrm{Trio}(X,\overline{X})\), and \(\mathrm{Sorp}(X,\overline{X})\). More precisely, each of them is obtained by quotienting, respectively, \(\mathbb{B}[X\cup\overline{X}]\), \(\mathrm{Trio}(X\cup\overline{X})\), and \(\mathrm{Sorp}(X\cup\overline{X})\), by the congruence generated by \(x\cdot\overline{x}=0\), for every \(x\in X\). Hence, \(\mathbb{B}[X,\overline{X}]\) consists of Boolean provenance polynomials whose monomials contain no complementary pair; \(\mathrm{Trio}(X,\overline{X})\) consists of polynomials with coefficients in \(\mathbb{N}\), whose monomials are square-free and contain no complementary pair; and \(\mathrm{Sorp}(X,\overline{X})\) behaves as \(\mathrm{Sorp}(X\cup\overline{X})\), with the absorption law \(p+pq=p\), but with all monomials containing complementary variables removed.
\item We define \(\mathrm{Why}(X,\overline{X})\) by applying an analogous quotient construction to the why-provenance semiring. We start from $\mathrm{Why}(X\cup\overline{X}) = \bigl(\mathcal P_{\mathrm{fin}}(\mathcal P_{\mathrm{fin}}(X\cup\overline{X})),\cup, \Cup, \emptyset, \{\emptyset\}\bigr),$ where addition is union and multiplication is pairwise union. We then quotient \(\mathrm{Why}(X\cup\overline{X})\) by the congruence generated by the equations $\{\{x,\overline x\}\}=\emptyset$ for every $x\in X.$ This quotient removes all witnesses containing a complementary pair. Equivalently, if one views \(\mathrm{Why}\)-provenance in polynomial terms, \(\mathrm{Why}(X)\) is obtained from \(\mathrm{Trio}(X)\) by dropping coefficients. From this viewpoint, \(\mathrm{Why}(X,\overline{X})\) is obtained by the same quotient construction used for the other dual semirings: starting from polynomial representatives over \(X\cup\overline{X}\), one imposes $x\cdot\overline x=0$ for every $x\in X$, and then drops coefficients. 
 \item  Let \(\mathrm{PosForm}(X\cup\overline{X})\) be the set of positive propositional formulas over \(X\cup\overline{X}\), built from \(\vee,\wedge,\bot,\top\). We define an equivalence relation \(\approx\) on \(\mathrm{PosForm}(X\cup\overline{X})\) by setting $\varphi\approx\psi$ iff $\nu(\varphi)=\nu(\psi)$ for every valuation \(\nu\colon X\cup\overline{X}\longrightarrow\{0,1\}\) such that $\nu(x)\cdot\nu(\overline{x})=0$ for every $x\in X$. The relation \(\approx\) is a congruence with respect to \(\vee\) and \(\wedge\). We define $\mathrm{PosBool}(X,\overline{X}) =\big(\mathrm{PosForm}(X\cup\overline{X})/{\approx}, \vee,\wedge,\bot,\top \big).$ Thus, \(\mathrm{PosBool}(X,\overline{X})\) is the semiring of positive Boolean formulas over \(X\cup\overline{X}\), modulo equivalence under valuations in which no complementary pair \(x,\overline{x}\) is simultaneously true. Since quotients of distributive lattices by lattice congruences are distributive lattices, \(\mathrm{PosBool}(X,\overline{X})\) is again a distributive lattice. Note that \(\mathrm{PosBool}(X,\overline{X})\) is not isomorphic to the Boolean algebra over \(X\): although the quotient imposes \(x\wedge\overline{x}=\bot\), it does not impose \(x\vee\overline{x}=\top\).
\end{itemize}
We will not introduce a dual indeterminate version of $\mathrm{Which}(X)$ 
since, unlike the other provenance semirings considered below, \(\mathrm{Which}(X)\) does not fit neatly into the same dual-token construction (see the discussion before Theorem~\ref{containmentwhich modeldef} in \S \ref{The problem of containmentneq}). \vt{why?} Its treatment in the presence of negated relational atoms will therefore be discussed separately in the sections devoted to query containment. 

We include a proof of the following fact in Appendix~\ref{app:proofs1}.
\begin{restatable}{Pro}{firstprop}   
The semirings \(\mathbb N[X,\overline{X}]\), \(\mathbb B[X,\overline{X}]\), \(\mathrm{Trio}(X,\overline{X})\), \(\mathrm{Sorp}(X,\overline{X})\), \(\mathrm{Why}(X,\overline{X})\), and \(\mathrm{PosBool}(X,\overline{X})\) are naturally ordered. 
\end{restatable}

It is worth distinguishing \(\mathrm{PosBool}(X,\overline{X})\) from $\mathrm{BoolExp}(X)$, the Boolean algebra of propositional formulas over \(X\). In the latter, the symbol \(\neg x\) is the Boolean complement of \(x\), so both \(x\wedge \neg x=\bot\) and \(x\vee \neg x=\top\) hold. By contrast, in \(\mathrm{PosBool}(X,\overline{X})\), the quotient imposes the incompatibility \(x\wedge\overline x=\bot\), but not the law \(x\vee\overline x=\top\). Thus, \(\overline x\) should be understood as a negative provenance token, not as the Boolean complement of \(x\). The Boolean algebra of propositional formulas over \(X\), seen as the semiring $(\mathrm{BoolExp}(X),\vee,\wedge,\bot,\top),$ 
is nevertheless a lattice semiring, since every Boolean algebra is a bounded distributive lattice.

\subsection{Containment and equivalence}\label{section2.6}

We are finally able to define the two central notions  that we are concerned with in this paper, namely containment and equivalence between queries.

\begin{Def}[Containment and Equivalence over  Instances]
Let $\mathbb{S}$ be a naturally ordered semiring while $P$ and $Q$ are (U)CQ$^\neq$s in the schema $\tau$. Then, we write $P \sqsubseteq_{\mathbb{S}} Q$ (i.e.\ $P$ is \emph{$\mathbb{S}$-contained} in $Q$) if for each $\mathbb{S}$-instance $\mathbf I$ of the schema $\tau$  and tuple $\vec{t}$, we have that $\llbracket P(\vec{t})\rrbracket^\mathbf{I} \leq_\mathbb{S} \llbracket Q(\vec{t})\rrbracket^\mathbf{I}$. We write $P \equiv_\mathbb{S} Q$ (i.e.\ $P$ and $Q$ are \emph{$\mathbb{S}$-equivalent}) if $P \sqsubseteq_{\mathbb{S}} Q$  and $Q \sqsubseteq_{\mathbb{S}} P$. 
\end{Def}

This first notion of containment (the same one used in~\cite{Green}) works well for queries where there is no negation beyond disequations since in this case the semantics of all the logical operators and connectives is fixed. For the case where we allow negation in front of relational atoms, we have to be more careful.

Given an $\mathbb{S}$-instance $\mathbf I$, the \emph{active domain}  of\/ $\mathbf{I}$, denoted by $\mathrm{adom}(\mathbf{I})$, is the set of all elements of $D$ that occur in the support of  
a relation $R_i$ of\/ $\mathbf{I}$. An interpretation $\pi$ will be said to be \emph{compatible}\footnote{Roughly in the sense of \cite{Kolaitis3} except that we do not require $\pi$ to be model-defining.} with an 
 $\mathbb{S}$-instance $\mathbf I$ if 
 $\mathrm{adom}(\mathbf{I}) = A$ and for every relation symbol $R$ and tuple $\vec{t}$ of the appropriate arity from $A$, $\pi\llbracket R(\vec{t})\rrbracket=R^\mathbf{I}(\vec{t})$. 
In the sense of~\cite[Def. 3]{GT}, we can consider \emph{model-defining interpretations} (where exactly one of an atom or its negation takes the value $0$). The class of model-defining interpretations will be denoted by \emph{md} below.   If an interpretation $\pi$ is such that for any pair $R(\vec{a}), \neg R(\vec{a})$, exactly one takes  value 1, we call it \emph{dually model-defining}. This class of interpretations will be denoted by \emph{dmd} below. \emph{Provenance-tracking interpretations}~\cite[Def. 5]{GT} for dual indeterminate polynomial semirings  are mappings where positive literals are annotated by tokens in $X \cup \{0,1\}$ and negative ones by tokens in $\overline{X} \cup \{0,1\}$. The class of such interpretations will be denoted by \emph{pt}  below.

\begin{Def}[Containment and Equivalence  over Interpretations]
Let $\mathbb{S}$ be a naturally ordered semiring, $\theta$ a class of interpretations while $P$ and $Q$ are (U)CQNs. Then, we write $P \sqsubseteq_{\mathbb{S}}^\theta Q$ (i.e.\ $P$ is \emph{$\mathbb{S}$-contained over the class $\theta$} in $Q$) if for each $\mathbb{S}$-instance $\mathbf I$,  interpretation $\pi \in \theta$ compatible with $\mathbf I$, and tuple $\vec{t}$ from $A$ of the arity of the common head, we have that $\pi\llbracket P(\vec{t})\rrbracket \leq_\mathbb{S} \pi \llbracket Q(\vec{t})\rrbracket$. We write $P \equiv^\theta_\mathbb{S} Q$ (i.e.\ $P$ and $Q$ are \emph{$\mathbb{S}$-equivalent over the class $\theta$}) if $P \sqsubseteq^\theta_{\mathbb{S}} Q$  and $Q \sqsubseteq^\theta_{\mathbb{S}} P$.
\end{Def}
Notice that requiring $A=\mathrm{adom}(\mathbf I)$, rather than $\mathrm{adom}(\mathbf I)\subseteq A$, does not affect containment or equivalence. Indeed, consider any interpretation $\pi$ over a domain $A$, and define an $\mathbb{S}$-instance $\mathbf I$ by setting $R^\mathbf I(\vec{a})=\pi\llbracket R(\vec{a})\rrbracket$ for every positive ground atom over $A$, and $0$ otherwise. Let $B=\mathrm{adom}(\mathbf I)\subseteq A$. By construction, every positive relational atom involving an element outside $B$ has value $0$. Since we only consider safe queries, any assignment using an element outside $B$ contributes $0$ to the evaluation. Hence, restricting $\pi$ to $\mathrm{Lit}_B(\tau)$ preserves the value of every (U)CQ$^\neq$N on tuples over $B$, while all output tuples containing elements outside $B$ have value $0$. Moreover, the restricted interpretation is compatible with $\mathbf I$, since its domain is exactly $\mathrm{adom}(\mathbf I)$. Therefore, for any class $\theta$ closed under restriction, any counterexample to containment over a larger domain can also be obtained with $A=\mathrm{adom}(\mathbf I)$. This applies to $\mathrm{md}$, $\mathrm{dmd}$, $\mathrm{pt}$ and their intersections. We also allow $A=\varnothing$ when the active domain is empty.

One might observe that this second notion of containment  generalizes the first one if we consider (U)CQ$^\neq$s. For example, in such case, $P \sqsubseteq_{\mathbb{S}} Q$ iff $P \sqsubseteq^\text{\emph{md}}_{\mathbb{S}} Q$. This is because the semiring value of the queries for particular tuples is not affected by anything other than what happens in the database instance. \vt{This is quite subtle and we should discuss}

\subsection{Canonical databases and interpretations}\label{Canonical databases and interpretations}

We start with a familiar definition from the Boolean setting:

\begin{Def}[Canonical $\mathbb{B}$-databases]\label{candataB}
Given an identification  $P^{\mathrm{id}}$ of some CQ$^\neq$ $P$, one can build its \emph{canonical $\mathbb{B}$-database}, denoted by $\mathrm{can}_{\mathbb{B}}(P^{\mathrm{id}})$, in the usual manner. For any relation $R$ of the schema of $P^{\mathrm{id}}$
we let $R^{\mathrm{can}_{\mathbb{B}}(P^{\mathrm{id}})} \langle x_1, \ldots, x_n\rangle = \top$ if $R(x_1, \ldots, x_n)$ is an atom in  $P^{\mathrm{id}}$, and $R^{\mathrm{can}_{\mathbb{B}}(P^{\mathrm{id}})} \langle x_1, \ldots, x_n\rangle = \bot$ otherwise. 
\end{Def}

By construction, if the identification $P^{\mathrm{id}}$ is a formula $\phi(u)$, then  $u$ is an answer to the query $\phi$  in the database $\mathrm{can}_{\mathbb{B}}(P^{\mathrm{id}})$. Notice that the elements in the active domain of $\mathrm{can}_{\mathbb{B}}(P^{\mathrm{id}})$ are the variables of $P^{\mathrm{id}}$ and that, trivially, an equation $x=y$ or disequation $x\neq y$ holds between these objects iff it is a logical consequence of the equations and disequations in $P^{\mathrm{id}}$. In this paper, we use the extension to provenance semirings, introduced in~\cite{Green}, of the method of canonical databases involving \emph{abstractly tagged relations}. Roughly speaking, the \emph{abstractly tagged version of an $\mathbb{S}$-relation $R$} (unique up to renaming of tags) comes  from replacing each non-zero semiring annotation of a tuple in $R$ with its own tuple id. A canonical database is then built by replacing all relations by their abstractly tagged versions.

    \begin{Def}[Canonical abstractly tagged  databases] Let $\mathbb{S}_1 \in \{\mathbb{B}[X],\allowbreak \mathbb{N}[X],\allowbreak \mathrm{Trio}(X),\allowbreak \mathrm{Sorp}(X),\allowbreak \mathrm{Why}(X),\allowbreak \mathrm{Which}(X),\allowbreak \mathrm{PosBool}(X)\}$ and $\mathbb{S}_2$ be any semiring. Consider an $\mathbb{S}_2$-relation $R\colon D^n\longrightarrow S_2$.  The \emph{abstractly tagged $\mathbb{S}_1$-version of $R$} is the $\mathbb{S}_1$-relation $\text{ab}_{\mathbb{S}_1(R)}$ that
\begin{itemize}
\item[(1)] if $\mathbb{S}_1\in \{\mathbb{B}[X], \mathbb{N}[X], \mathrm{Trio}(X), \mathrm{Sorp}(X), \mathrm{PosBool}(X)\}$, then for any tuple $\vec{t}$ in the support of $R$,  $\text{ab}_{\mathbb{S}_1}(R)(\vec{t})= p_{\vec{t}}^R$ where $p_{\vec{t}}^R$ is a fresh variable from $X$,

\item[(2)] if $\mathbb{S}_1 =\mathrm{Which}(X)$, then for any tuple $\vec{t}$ in the support of $R$,  $\text{ab}_{\mathbb{S}_1}(R)(\vec{t})= \{p_{\vec{t}}^R\}$ where $p_{\vec{t}}^R$ is a fresh variable from $X$, and 

\item[(3)] if $\mathbb{S}_1=\mathrm{Why}(X)$, then for any tuple $\vec{t}$ in the support of $R$,  $\text{ab}_{\mathbb{S}_1}(R)(\vec{t})= \{\{p_{\vec{t}}^R\}\}$ where $p_{\vec{t}}^R$ is a fresh variable from $X$. 
\end{itemize}
Given an $\mathbb{S}_2$-database  $\mathbf{I}=(R_1^\mathbf{I},\ldots,R_p^\mathbf{I})$, its abstractly tagged  $\mathbb{S}_1$-version is $\text{ab}_{\mathbb{S}_1}(\mathbf{I})=(\text{ab}_{\mathbb{S}_1}(R_1^\mathbf{I}),\ldots,\text{ab}_{\mathbb{S}_1}(R_p^\mathbf{I}))$ where we make sure that no tags $p_{\vec{t}}^R$ are used repeatedly across the various relations. Finally, when  $P^{\mathrm{id}}$ is an identification of some CQ$^\neq$ $P$, then the \emph{canonical $\mathbb{S}_1$-database for $P^{\mathrm{id}}$} (in symbols, $\mathrm{can}_{\mathbb{S}_1}(P^{\mathrm{id}})$)  will be simply $\text{ab}_{\mathbb{S}_1}(\mathrm{can}_{\mathbb{B}}(P^{\mathrm{id}}))$. 
\end{Def}

For the case of $\mathrm{Why}(X)$, we will also use the following alternative version of the canonical database.

\begin{Def}[Occurrence-sensitive canonical $\mathrm{Why}(X)$-database]
Let $P^{\mathrm{id}}$ be an identification of a CQ$^\neq$ $P$, and let $A_1(\vec{z}_1),\ldots,A_n(\vec{z}_n)$ be the relational atoms occurring in the body of $P^{\mathrm{id}}$, where repeated occurrences of the same atom are indexed separately. Let $x_1,\ldots,x_n\in X$ be pairwise distinct fresh tags. We define the \emph{occurrence-sensitive canonical $\mathrm{Why}(X)$-database}\footnote{An occurrence-sensitive canonical construction of this kind already appears in~\cite{Kostylev}, where different occurrences of the same atom are assigned distinct annotation variables, which are combined in the annotation of the corresponding canonical tuple. This refines the usual canonical-database construction used in~\cite{Green}, where repeated identical atoms give rise to a single canonical tuple. In what follows, we use the occurrence-sensitive construction whenever the semantics under consideration can distinguish such multiplicities, while we retain the usual canonical database in settings, such as provenance-tracking $\mathrm{Why}$-interpretations, where repeated occurrences of the same relational literal are semantically indistinguishable.} of $P^{\mathrm{id}}$, denoted by $\mathrm{can}^{\text{occ}}_{\mathrm{Why}(X)}(P^{\mathrm{id}})$, as the $\mathrm{Why}(X)$-database having the same active domain as $\mathrm{can}_{\mathbb B}(P^{\mathrm{id}})$ and such that, for every relation symbol $R$ and tuple $\vec{t}$, $R^{\mathrm{can}^{\text{occ}}_{\mathrm{Why}(X)}(P^{\mathrm{id}})} (\vec{t}) = \bigl\{ \{x_i\} \mid A_i(\vec{z}_i)=R(\vec{t}),\ 1\leq i\leq n \bigr\},$ whenever $R(\vec{t})$ occurs in $P^{\mathrm{id}}$, and it is $\emptyset$ otherwise.

Thus, if the same relational atom $R(\vec{t})$ occurs at positions $i_1,\ldots,i_k$ in the body of $P^{\mathrm{id}}$, then $R^{\mathrm{can}^{\text{occ}}_{\mathrm{Why}(X)}(P^{\mathrm{id}})} (\vec{t}) = \bigl\{\{x_{i_1}\},\ldots,\{x_{i_k}\}\bigr\}.$
\end{Def}

For the case of $\mathrm{Trio}(X)$, ordinary abstract tags do not always retain enough information, since $\mathrm{Trio}(X)$ identifies monomials that differ only in their positive exponents. Following~\cite{Green}, we therefore also use $k$-abstract tags, which allow computations involving at most $k$ relational factors to be recovered.

We start by ordering the variables of $X=\left\{x_1, x_2, \ldots\right\}$. For every $k\geq 1$, we define the set $S_k \subseteq \mathrm{Trio}(X)$ of sums of $k$ fresh variables from $X$:
$$
S_k \stackrel{\text { def }}{=}\left\{\left(x_1+\ldots+x_k\right),\left(x_{k+1}+\ldots+x_{2 k}\right), \ldots\right\}.
$$
Thus, distinct elements of $S_k$ involve pairwise disjoint sets of variables. The \emph{$k$-abstractly tagged version} $\mathrm{ab}_k(\mathbf I)$ of a $\mathrm{Trio}(X)$-instance $\mathbf I$ is the $\mathrm{Trio}(X)$-instance obtained by replacing the annotation of every tuple in the support of $\mathbf I$ with a distinct fresh element of $S_k$.

\begin{Def}
In general, given a commutative semiring $\mathbb{K}=(K,+,\cdot,0,1)$ and a valuation $v\colon X\longrightarrow K$, we denote by $\operatorname{Eval}_v\colon \mathbb{N}[X]\longrightarrow K$ the evaluation map obtained by replacing every indeterminate $x\in X$ with $v(x)$ and evaluating the resulting expression in $\mathbb{K}$. More precisely, $\operatorname{Eval}_v$ is defined by $\operatorname{Eval}_v(x)=v(x), \operatorname{Eval}_v(0)=0, \operatorname{Eval}_v(1)=1,$ and, for every $F,G\in\mathbb{N}[X]$, $\operatorname{Eval}_v(F+G)
= \operatorname{Eval}_v(F)+\operatorname{Eval}_v(G)
$ and $ \operatorname{Eval}_v(F\cdot G) = \operatorname{Eval}_v(F)\cdot\operatorname{Eval}_v(G).$ Therefore, $\operatorname{Eval}_v$ is a semiring homomorphism. Moreover, by the
universal property of $\mathbb{N}[X]$, it is the unique semiring homomorphism
extending $v$.

\end{Def}

We define $\mathrm{Trio}_k(X) \subseteq \mathrm{Trio}(X)$ to be the set of annotations from $\mathrm{Trio}(X)$ obtained via calculations of degree at most $k$ involving elements of $S_k$:
$$
\mathrm{Trio}_k(X) \stackrel{\text { def }}{=}\left\{\operatorname{Eval}_v(a)\mid a \in \mathbb{N}[X],\ v\colon X \longrightarrow S_k \text{ s.t. } a \text{ has degree } \leq k \text{ and } v \text{ is injective }\right\}.
$$

We will also use the $k$-abstractly tagged construction in connection with canonical databases.

\begin{Def}[$k$-abstract canonical $\mathrm{Trio}(X)$-databases]\label{cankTrio}
Let $P^{\mathrm{id}}$ be an identification of a CQ$^\neq$ $P$ and let $k\geq 1$. We define the \emph{$k$-abstract canonical $\mathrm{Trio}(X)$-database} of $P^{\mathrm{id}}$, denoted by $\text{can}^{k}_{\mathrm{Trio}(X)}(P^{\mathrm{id}})$, as
$$
\text{can}^{k}_{\mathrm{Trio}(X)}(P^{\mathrm{id}})
\stackrel{\text{def}}{=}
\mathrm{ab}_k\left(\text{can}_{\mathrm{Trio}(X)}(P^{\mathrm{id}})\right).
$$
Thus, $\text{can}^{k}_{\mathrm{Trio}(X)}(P^{\mathrm{id}})$ has the same active domain and support as $\text{can}_{\mathrm{Trio}(X)}(P^{\mathrm{id}})$, but every tuple in its support is annotated with a distinct element of $S_k$. In particular, if $R(\vec{t})$ is a relational atom of $P^{\mathrm{id}}$, we may write its annotation as $x_{R(\vec{t}),1}+\ldots+x_{R(\vec{t}),k},
$ where all the variables used for distinct canonical tuples are pairwise different.
\end{Def}

The following property, introduced in~\cite[Prop. 7.24]{Green}, ensures that $k$-abstract tags retain polynomial computations of degree at most $k$. We include a proof in Appendix~\ref{app:proofs1}.

\begin{restatable}{Pro}{propseventwetyfour}\label{prop7.24} 
Suppose a valuation $v\colon X \longrightarrow S_k$ is injective. Then, for any $a, b \in \mathbb{N}[X]$ of degree $\leq k$, $a=b$ iff $\operatorname{Eval}_v(a)=\operatorname{Eval}_v(b)$. As a consequence, any element $c \in \mathrm{Trio}_k(X)$ decomposes uniquely into a sum of products of elements of $S_k$.
\end{restatable}

We will use the following consequence for CQ$^\neq$s. We include a proof in Appendix~\ref{app:proofs1}.

\begin{restatable}
    {Lem}{kabstractCQ}
\label{kabstractCQ}
Suppose $P$ and $Q$ are CQ$^\neq$s having at most $k$ relational atom occurrences, and suppose $\mathbf I$ is a $\mathrm{Trio}(X)$-instance such that $\llbracket P \rrbracket^\mathbf I \not\leq_{\mathrm{Trio}(X)} \llbracket Q \rrbracket^\mathbf I.$ Then $\llbracket P \rrbracket^{\mathrm{ab}_k(\mathbf I)}
\not\leq_{\mathrm{Trio}(X)} \llbracket Q \rrbracket^{\mathrm{ab}_k(\mathbf I)}.$
\end{restatable}

Finally, we need to adapt the technique of canonical databases for the setting of semiring interpretations in order to deal with negations on atoms in CQNs. 

\begin{Def}[Canonical interpretations]\label{CQNscan}
Let $\mathbb{S} \in \{\mathbb{B}[X], \mathbb{N}[X], \mathrm{Trio}(X), \mathrm{Sorp}(X), \mathrm{Why}(X),\mathrm{PosBool}(X) \}$, $P$ be a CQN  and $P^+$ is the CQ resulting from deleting all negated atoms from $P$. Then, we take $\pi^{\mathrm{can}_{\mathbb{S}}(P^+)}$ to be the following $\mathbb{S}[X, \overline{X}]$-interpretation compatible\footnote{Observe that, in general, not all $\mathbb{S}$-databases have a compatible provenance-tracking interpretation. It suffices to consider that a tuple can get annotated by a more complicated polynomial than a single tag (then there is no provenance-tracking interpretation that can take that polynomial as value for the literal).}  with  the $\mathbb{S}$-instance $\mathrm{can}_{\mathbb{S}}(P^+)$: 
\begin{itemize}

\item[(1)] if $\mathbb{S}\in \{\mathbb{B}[X], \mathbb{N}[X], \mathrm{Trio}(X), \mathrm{Sorp}(X), \mathrm{PosBool}(X)\}$, then 
\begin{itemize}
\item $\pi^{\mathrm{can}_{\mathbb{S}}(P^+)} \llbracket R(\vec{a}) \rrbracket = R^{\mathrm{can}_{\mathbb{S}}(P^+)}(\vec{a})$,

\item $\pi^{\mathrm{can}_{\mathbb{S}}(P^+)} \llbracket \neg R(\vec{a}) \rrbracket = \overline{r}^R_{\bar{a}}$, when $\neg R(\vec{a})$ appears in the body of $P$ and $\overline{r}^R_{\bar{a}}$ is a new tag from $\overline{X}$ that has not had its complementary token $r$ used as an annotation in   $\mathrm{can}_{\mathbb{S}}(P^+)$,

\item $\pi^{\mathrm{can}_{\mathbb{S}}(P^+)}$ gives $0$ in all other cases. 
\end{itemize}

\item[(2)] if $\mathbb{S} =\mathrm{Why}(X)$, then 
\begin{itemize}
\item $\pi^{\mathrm{can}_{\mathbb{S}}(P^+)} \llbracket R(\vec{a}) \rrbracket = R^{\mathrm{can}_{\mathbb{S}}(P^+)}(\vec{a})$,

\item $\pi^{\mathrm{can}_{\mathbb{S}}(P^+)} \llbracket \neg R(\vec{a}) \rrbracket = \{ \{ \overline{r}^R_{\bar{a}} \} \}$, when $\neg R(\vec{a})$ appears in the body of $P$ and $\overline{r}^R_{\bar{a}}$ is a new tag from $\overline{X}$ that has not had its complementary token $r^R_{\bar{a}}$ otherwise used in an annotation $\{ \{ r^R_{\bar{a}} \} \}$ in $\mathrm{can}_{\mathbb{S}}(P^+)$,

\item $\pi^{\mathrm{can}_{\mathbb{S}}(P^+)}$ gives $0$ in all other cases. 
\end{itemize}

\end{itemize}
Notice that $\pi^{\mathrm{can}_{\mathbb{S}}(P^+)}$ is provenance-tracking, but need not be model-defining. When $P$ contains no atom and its negation, it becomes model-defining if we modify the annotations of negated atoms not appearing in $P$ as follows, leaving all other annotations unchanged:
\begin{itemize}
\item For all negated atoms $\neg R(\vec{a})$ not appearing in the body of $P$ and such that $R(\vec{a})$ does not appear either,   $\pi^{\mathrm{can}_{\mathbb{S}}(P^+)}\llbracket \neg R(\vec{a}) \rrbracket$ is a distinguished  $\overline{p}^R_{\bar{a}}$ (or $\{\{\overline{p}^R_{\bar{a}}\}\}$ in the case of $\mathrm{Why}(X)$), where $\overline{p}^R_{\bar{a}}$ is a fresh tag whose complementary token is not used elsewhere. 
\item For all negated atoms $\neg R(\vec{a})$ not appearing in the body of $P$ but such that $R(\vec{a})$  appears  in the body of $P$,   $\pi^{\mathrm{can}_{\mathbb{S}}(P^+)}\llbracket \neg R(\vec{a}) \rrbracket=0$. 
\end{itemize}

\end{Def}

\subsection{Homomorphisms for containment} \label{Homomorphisms for containment}
Next, we will consider suitable  mappings between queries that will let us establish characterization theorems for containment over different semirings of interest. Even though these mappings are strictly speaking between the variables of the queries they easily induce mappings between the atoms (or literals) of the queries and we will use these ideas ambiguously. 

\begin{Def}[$\neq$-homomorphisms]
Let   $P$ and $Q$ be two CQ$^\neq$s.

1. A $\neq$-homomorphism  $h$ from $P$ to $Q$ is a function from the variables of $P$ to $Q$ that preserves the head, i.e. $h(\vec{u})=\vec{v}$, and preserves literals in the following sense (cf.~\cite{Nutt}):
\begin{itemize}
\item if the atom $l(\vec{x})$ appears in  $P$, the atom $l(h(\vec{x}))$ appears in  $Q$,
\item if the equation or disequation  $l(\vec{x})$ appears in  $P$,
then $l(h(\vec{x}))$ is a \emph{logical consequence} of the equations and disequations in $Q$.
\end{itemize}

2. A $\neq$-homomorphism  $h$ from $P$ to $Q$  is \emph{one-to-one} or \emph{injective} for relational atoms if the multiset of images of atoms (the image of an atom $A(\vec{x})$ is $A(h(\vec{x}))$) of $P$ is bag-contained in the multiset of relational atoms of $Q$.

3. A $\neq$-homomorphism  $h$ from $P$ to $Q$ is \emph{bag-onto} or   \emph{surjective} for relational atoms if the multiset of relational atoms of $Q$ is bag-contained in the multiset of images of atoms  of $P$.\footnote{Any such mapping that is surjective on relational atoms must also be surjective as a mapping on variables since every variable of $Q$ appears in some of the relational atoms.} 

4. A $\neq$-homomorphism  $h$ from $P$ to $Q$ is \emph{exact for relational atoms} if the mapping is both injective and surjective in the above sense.

\end{Def}

\begin{Def}
Given a CQ$^\neq$ $P$, we denote by $\text{RelOcc}(P)$ the set of relational atom occurrences in the body of $P$, where repeated occurrences of the same relational atom are indexed separately. We denote by $\text{RelSet}(P)$ the set of syntactically distinct relational atoms occurring in the body of $P$.\footnote{Thus, repeated occurrences of the same relational atom occur only once in $\text{RelSet}(P)$.}
Every element of $\text{RelOcc}(P)$ has an underlying relational atom in $\text{RelSet}(P)$. Moreover, if $h\colon P\longrightarrow Q$ is a $\neq$-homomorphism and $B\in\text{RelOcc}(P)$, we write $h(B)$ for the relational atom obtained by applying $h$ to the variables occurring in $B$. By definition of $\neq$-homomorphism, $h(B)\in\text{RelSet}(Q)$.
\end{Def}

\begin{Exm}
Consider the CQ$^\neq$ $P(x):=\exists y\,\bigl(R(x,y)\wedge R(x,y)\wedge R(y,x)\wedge S(y)\wedge x\neq y\bigr).$ Then $\text{RelOcc}(P)$ contains four relational atom occurrences, namely the two occurrences of $R(x,y)$, the occurrence of $R(y,x)$, and the occurrence of $S(y)$, while $\text{RelSet}(P)=\{R(x,y),\,R(y,x),\,S(y)\}.$ Thus, $|\text{RelOcc}(P)|=4$ and $|\text{RelSet}(P)|=3$. Notice that the disequation $x\neq y$ belongs to neither $\text{RelOcc}(P)$ nor $\text{RelSet}(P)$.
\end{Exm}

The following is a reformulation of  results in \cite{Klug, Kolaitis1, Nutt}\footnote{The existence of a single $\neq$-homomorphism is not enough to characterize containment between CQ$^\neq$s in the Boolean semiring. To see this, take the Boolean queries $P: \exists x,y,z(R(x,y) \wedge R(y,z) \wedge z\neq x)$ and $Q: \exists x,y (y\neq x \wedge R(x,y))$. Then, $P \sqsubseteq_\mathbb{B} Q$ (assume $P$, there are two cases to consider: (1) $y\neq x$, in which case we are done as then $Q$ holds, or (2) $y= x$, in which case $R(x,z)$ holds since $R(y,z)$ does and since $z\neq x$ we are done). However, it is easy to see that there is no $\neq$-homomorphism  from $Q$ to $P$.} that we wish to recall here :

\begin{Thm}[\cite{Klug, Kolaitis1, Nutt}]\label{classical} For all CQ$^\neq$s $P$ and $Q$ with the same tuple of free variables
,  the following are equivalent:
\begin{itemize}
\item[1.] $P \sqsubseteq_{\mathbb{B}} Q$.


\item[2.] For every identification $P^{\mathrm{id}}$ of $P$, there is a  $\neq$-homomorphism 
$h_{P^{\mathrm{id}}} \colon Q \longrightarrow P^{\mathrm{id}}$.

\end{itemize}
Moreover, the corresponding decision problem (i.e.\ $P \sqsubseteq_{\mathbb{B}} Q$?) is $\Pi^p_2$-complete.
\end{Thm}

\begin{Rmk}
    Notice that considering just completions instead of identifications in Theorem~\ref{classical} is not enough. For example $\exists x, y (x=y \wedge R(x,y))$ is clearly contained in $\exists x R(x,x)$, and the former is already a completion of itself but there is no canonical mapping from $\exists x R(x,x)$ to $\exists x, y (x=y \wedge R(x,y))$. By contrast, $\exists x R(x,x)$ is clearly an identification of  $\exists x, y (x=y \wedge R(x,y))$.
\end{Rmk}

What about the situation for UCQ$^\neq$s? Reformulating \cite[Thm. 3]{Klug}, we get  the following result:

\begin{Thm}[~\cite{Klug}] For UCQ$^\neq$s, $\bigcup_{1\leq i \leq n} P_i, \bigcup_{1\leq i \leq m} Q_i$ with the same tuple of free variables
,  the following are equivalent:
\begin{itemize}
\item[1.] $\bigcup_{1\leq i \leq n} P_i \sqsubseteq_{\mathbb{B}} \bigcup_{1\leq i \leq m} Q_i$.


\item[2.] For each $1\leq i \leq n$ and   every identification $P_i'$ of $P_i$, there is a $j \leq m$ and a  $\neq$-homomorphism 
$h_{P_i'} \colon Q_j \longrightarrow P_i'$.
\end{itemize}
Moreover, the corresponding decision problem (i.e.\ $\bigcup_{1\leq i \leq n} P_i \sqsubseteq_{\mathbb{B}} \bigcup_{1\leq i \leq m} Q_i$?) is $\Pi^p_2$-complete.
\end{Thm}

\begin{Rmk}\label{complexityremark}
The above containment problem is clearly  $\Pi^p_2$-hard as containment of CQ$^\neq$s over the Boolean semiring is $\Pi^p_2$-hard\cite{VdM}. For the upper bound, the complement  of the containment problem for UCQ$^\neq$s is in $\text{NP}^{\text{coNP}} = \text{NP}^{\text{NP}}=\Sigma^p_2$, and thus our actual problems are in $\text{coNP}^{\text{NP}}=\Pi^p_2$. Given an instance of a problem $(\bigcup_{1\leq i \leq n} P_i, \bigcup_{1\leq i \leq m} Q_i)$ consisting of two queries, we wish to check whether for some $1\leq i \leq n$ and   some identification $P_i'$ of $P_i$, there is no $1 \leq j \leq m$ with a $\neq$-homomorphism 
$h_{P_i'} \colon Q_j \longrightarrow P_i'$. We
can  guess a $P_i$ and an identification $P_i'$ (an identification is polynomial in size because it adds at most $n^2-n$ literals to $P_i$ where $n$ is the number of variables of $P_i$). Next, we can  check with the $\text{coNP}$ oracle that  there is no $j \leq m$ with a   $\neq$-homomorphism 
    $h_{P_i'} \colon Q_j \longrightarrow P_i'$. (The problem  of checking that  for some $1 \leq j \leq m$ there is a   $\neq$-homomorphism 
    $h_{P_i'} \colon Q_j \longrightarrow P_i'$ is in $\text{NP}$ and uses the fact that logical entailments between equations/disequations is polynomial time checkable \cite[Section 14.2]{Ullman}.)

\end{Rmk}

\begin{Def}[negation-homomorphisms]
Let $P(\vec{u})$ and $Q(\vec{v})$ be two CQNs. 
\begin{enumerate}
\item A negation-homomorphism $h$ from $P$ to $Q$ is a function from the variables of $P$ to the variables of $Q$ that preserves literals, i.e.
if the literal $l(\vec{x})$ appears in  $P$, the literal $l(h(\vec{x}))$ appears in  $Q$, and preserves the head, i.e. $h(\vec{u})=\vec{v}$.
    \item A negation-homomorphism  $h$ from $P$ to $Q$  is \emph{one-to-one} or \emph{injective} for relational literals if the multiset of images of literals (the image of a literal $l(\vec{x})$ is $l(h(\vec{x}))$) of $P$ is bag-contained in the multiset of relational literals of $Q$.

\item A negation-homomorphism  $h$ from $P$ to $Q$ is \emph{bag-onto} or \emph{surjective} for relational literals if the multiset of relational literals of $Q$ is bag-contained in the multiset of images of literals  of $P$.

\item A negation-homomorphism  $h$ from $P$ to $Q$ is \emph{set-onto} for relational literals if every syntactically distinct relational literal occurring in the body of $Q$ is the image under $h$ of some relational literal occurring in the body of $P$, disregarding repeated occurrences of the same literal.

\item A negation-homomorphism  $h$ from $P$ to $Q$ is \emph{exact for relational literals} if the mapping is both injective and surjective in the above sense.
\end{enumerate}

\end{Def}

A $\neq$-negation-homomorphism between CQ$^\neq$Ns preserves the head and relational literals as above, and equations and disequations in the sense of $\neq$-homomorphisms. The injective, bag-onto, set-onto and exact requirements concern only relational literals.

\begin{Def}
Given a CQN $P$, we denote by $\text{LitSet}(P)$ the set of syntactically distinct relational literals occurring in the body of $P$. Thus, repeated occurrences of the same relational literal occur only once in $\text{LitSet}(P)$. Positive and negative literals are regarded as distinct, so $R(\vec{x})$ and $\neg R(\vec{x})$ are distinct elements of $\text{LitSet}(P)$.

Equivalently, a negation-homomorphism $h\colon P\longrightarrow Q$ is set-onto for relational literals if $\text{LitSet}(Q)\subseteq h(\text{LitSet}(P)).$ Since $h$ is a negation-homomorphism, we always have $h(\text{LitSet}(P))\subseteq \text{LitSet}(Q)$, and hence $h$ is set-onto for relational literals if and only if $h(\text{LitSet}(P))=\text{LitSet}(Q).$
\end{Def}

\begin{Def}
Given a CQN $P$, we denote by $\text{red}(P)$ its \emph{duplicate-free reduction}, obtained by retaining exactly one
occurrence of each relational literal in $\text{LitSet}(P)$. The head, variables,  relation symbols, and polarities (negative or positive) of the literals are left unchanged.
\end{Def}

\begin{Exm}
Consider the CQN $P(x) := \exists y\, \bigl( R(x,y)\wedge R(x,y)\wedge R(y,x) \wedge \neg S(y)\wedge \neg S(y)\wedge \neg R(x,y) \bigr).$ Then $\text{LitSet}(P) = \{R(x,y),\,R(y,x),\,\neg S(y),\,\neg R(x,y)\}$ and its duplicate-free reduction is $\text{red}(P)(x) := \exists y\, \bigl( R(x,y)\wedge R(y,x) \wedge \neg S(y)\wedge \neg R(x,y) \bigr).$ Thus, $\text{red}(P)$ removes only repeated occurrences of syntactically identical relational literals. In particular, $R(x,y)$ and $\neg R(x,y)$ are both retained, since they are distinct elements of $\text{LitSet}(P)$.
\end{Exm}

\begin{Def}
Given a CQN $P$ we let a \emph{completion} $P^c$ of $P$ be any CQN (in the same free variables) that adds to $P$, for every relation symbol $R$ of the common schema $\tau$, and tuple $\vec{x}$  of variables of $P$ with length matching the arity of $R$, exactly one of the literals  $R(\vec{x})$ or $\neg R(\vec{x})$, making sure the resulting query does not contain any contradiction. 
\end{Def}

For example, the query $\exists x (Ryx \wedge \neg Sx)$ has completions $\exists x (Ryx \wedge \neg Sx \wedge \neg Sy \wedge Rxx \wedge Ryy \wedge Rxy)$, $\exists x (Ryx \wedge \neg Sx \wedge Sy \wedge \neg Rxx \wedge Ryy \wedge Rxy)$, $\exists x (Ryx \wedge \neg Sx \wedge Sy \wedge Rxx \wedge Ryy \wedge Rxy)$, $\exists x (Ryx \wedge \neg Sx \wedge Sy \wedge Rxx \wedge Ryy \wedge \neg Rxy)$ etc.  The following characterization is known in the literature (e.g.~\cite[Prop. 2.5]{Chein}):

\begin{Thm} \label{containnegbool}
For CQNs $P$ and $Q$ with the same tuple of free variables
, the following are equivalent\footnote{Under model-defining interpretations, a query containing an atom and its negation has value $0$ over every semiring. In this case containment is trivial, and the completion condition is vacuous because $P$ has no contradiction-free completion. This zero-value observation does not apply to arbitrary provenance-tracking or dually model-defining interpretations.}:
\begin{itemize}
\item[1.] $P \sqsubseteq_{\mathbb{B}}^\text{md} Q$,

\item[2.] For every completion $P^c$ of $P$, there is a negation-homomorphism $h_{P^c} \colon Q \longrightarrow P^c$.

\end{itemize}
In fact, the corresponding decision problem (i.e.\ $P \sqsubseteq^\text{md}_{\mathbb{B}} Q$?) is 
\begin{itemize}
\item[(i)] $\Pi^p_2$-complete for queries over schemas with arity bound $k$, for every fixed $k\geq 2$~\cite[Thm. 12.3]{Chein}, and
\item[(ii)] coNEXP-complete in general~\cite[Thm. 2]{Geck}.  
\end{itemize}
\end{Thm}

\begin{Exm}\label{complneed}
Consider queries $Q = \exists x, y (R(x,x) \wedge R(x,y) \wedge \neg R(y,y))$ and
$P = \exists x,y, z (R(x,x) \wedge  R(x,y) \wedge R(y,z) \wedge \neg R(z,z))$. We have that $P \sqsubseteq_{\mathbb{B}}^{\text{md}} Q$. Assume $P$ is satisfied, so there is a variable assignment $v$ for which $R(v(x),v(x)) \wedge  R(v(x),v(y)) \wedge R(v(y), v(z)) \wedge \neg R(v(z), v(z))$ holds. Thus, there are several cases to consider: (1) $R(v(x),v(z))$ holds, in which case we are done as then $Q$ holds, or (2) $\neg R(v(x),v(z))$ holds,  and $\neg R(v(y),v(y))$ (and we are done again), or (3) $\neg R(v(x),v(z))$ holds,  and $R(v(y),v(y))$ but then we are done again. However, there is no negation-homomorphism $h\colon Q \longrightarrow P$. This shows the need for considering completions in the characterization of containment.
\end{Exm}

\begin{Rmk}\label{complexityremarknegation}
For the upper bound complexity calculation  in Theorem~\ref{containnegbool}, the complement  of the containment problem for CQNs is in $\text{NP}^{\text{coNP}} = \text{NP}^{\text{NP}}=\Sigma^p_2$ if there is a fixed arity bound on the relations used while it is in NEXP otherwise. Given an instance of the problem $(P, Q)$ consisting of two queries, we wish to check whether for some completion $P'$ of $P$, there is no   negation-homomorphism 
$h_{P'} \colon Q \longrightarrow P'$. We
can  guess a  completion $P'$. The size of $P'$ is generally exponentially bounded in the maximum arity $k$ of the relations in the common schema $\tau$ since $P'$ adds at most $|\tau|(1+n+\ldots+n^k)$ new literals where $n$ is the number of variables in $P$ but  if we fix $k$,  we get a polynomial bound in $n$. Next, we can  check with the $\text{coNP}$ oracle that  there is no  negation-homomorphism 
    $h_{P'} \colon Q \longrightarrow P'$. When there is no fixed $k$, after guessing $P'$ (now of exponential size), we can iterate the process of checking that a mapping is not a negation-homomorphism over all $n^m$ mappings from $Q$ to $P'$ where $m$ is the number of variables in $Q$. The latter is doable in exponential time, and thus the whole process is in NEXP. The calculation for the lower bounds requires more sophisticated encoding arguments as usual~\cite{Chein, Geck}.
\end{Rmk}

Let $P$ be a CQ$^\neq$N. An identification now comes from first completing $P$ in the sense of  CQ$^\neq$s, then taking an identification in the old sense, and subsequently completing in the sense of CQNs.
The canonical database of an identification in this sense is simply the canonical database of the query resulting from deleting all negated atoms. For the Boolean results below, we consider only \emph{consistent identifications} in the sense that the resulting query does not contain both a relational atom \(R(\vec{x})\) and its negation \(\neg R(\vec{x})\).\footnote{This follows the  case of CQNs, where only completions yielding consistent formulas are considered in the characterization of containment. Notice that an identification may make a previously consistent query inconsistent by identifying the variables occurring in a positive and a negative relational literal.} The Boolean case of the complexity of containment for CQ$^\neq$Ns  is not difficult to settle  once we have this notion in hand, as we can see in the next theorem (a proof is in Appendix~\ref{app:proofs1}).

\begin{restatable}{Thm}{contnegdisbool} \label{contnegdisbool} For each CQ$^\neq$Ns $P$ and $Q$ with the same tuple of free variables
, the following are equivalent:
\begin{itemize}
\item[1.] $P \sqsubseteq^{\text{md}}_{\mathbb{B}}  Q$.
\item[2.] For every consistent identification\footnote{The consistency restriction is necessary. For instance, consider $P:=R(x)\wedge\neg R(y)\wedge S(x,y)$ and $Q:=S(u,v)\wedge u\neq v.$ We have $P\sqsubseteq^{\text{md}}_{\mathbb B}Q$, since every assignment satisfying \(P\) must map \(x\) and \(y\) to distinct elements. However, the completion of \(P\) containing \(x=y\) yields the identification $P^{\mathrm{id}}=R(x)\wedge\neg R(x)\wedge S(x,x),$ which is inconsistent. There is no \(\neq\)-negation-homomorphism from \(Q\) to this \(P^{\mathrm{id}}\): mapping \(S(u,v)\) to \(S(x,x)\) forces \(h(u)=h(v)=x\), and hence does not preserve \(u\neq v\). Thus, quantifying over all identifications would make Item~2 fail even though Item~1 holds.} $P^{\mathrm{id}}$ of $P$, there is a  $\neq$-negation-homomorphism 
$h_{P^{\mathrm{id}}} \colon Q \longrightarrow P^{\mathrm{id}}$.

\end{itemize}
In fact, the corresponding decision problem (i.e.\ $P \sqsubseteq^\text{md}_{\mathbb{B}} Q$?) is 
\begin{itemize}
\item[(i)] $\Pi^p_2$-complete for queries over schemas with arity bound $k$, for every $k$~\cite[Prop. 4.6]{Geck}, and
\item[(ii)] coNEXP-complete in general~\cite[Thm. 2]{Geck}. 
\end{itemize}
\end{restatable}

Similarly, we can show the following result for UCQ$^\neq$Ns (the proof is left to the reader as a simple exercise; not included in the Appendix).

\begin{Thm} \label{contundisbool}
For UCQ$^\neq$Ns, $\bigcup_{1\leq i \leq n} P_i, \bigcup_{1\leq i \leq m} Q_i$ with the same tuple of free variables
,  the following are equivalent:
\begin{itemize}

\item[1.] $\bigcup_{1\leq i \leq n} P_i \sqsubseteq^\text{md}_{\mathbb{B}}  \bigcup_{1\leq i \leq m} Q_i$.

\item[2.] For each $1\leq i \leq n$ and   every consistent identification $P_i'$ of $P_i$, there is a $j \leq m$ and a  $\neq$-negation-homomorphism 
$h_{P_i'} \colon Q_j \longrightarrow P_i'$.

\end{itemize}
In fact, the corresponding decision problem (i.e.\ $P \sqsubseteq^\text{md}_{\mathbb{B}} Q$?) is 
\begin{itemize}
\item[(i)] $\Pi^p_2$-complete for queries over schemas with arity bound $k$, for every $k$~\cite[Prop. 4.6]{Geck}, and
\item[(ii)] coNEXP-complete in general~\cite[Thm. 2]{Geck}. 
\end{itemize}
\end{Thm}

\section{
Conjunctive queries with disequations (CQ\texorpdfstring{$^\neq$}{≠}s)
}\label{CQneqsection}

In this section, we establish the decidability of the containment and equivalence problems for CQ$^\neq$s over various provenance semirings. We provide upper complexity bounds for both problems,as well as lower bounds for containment, except over $\mathbb{N}[X]$ and $\mathrm{Trio}(X)$, and for equivalence only over bounded distributive lattices.

\subsection{
Containment of CQ\texorpdfstring{$^\neq$}{≠}s
} 
\label{CQneqsection1}
The next  result summarizes our work on containment for the convenience of the reader. We still include in this section a few examples of the (auxiliary) theorems pointed to therein to give the reader a flavor for the ideas involved but most of them are relegated in their entirety to Appendix~\ref{app:proofs}.

\begin{Thm} \label{cqneqconatinmentsummary}
For CQ$^\neq$s $P$ and $Q$ with the same tuple of free variables
, the following statements hold.
\begin{enumerate}
\item ``$P \sqsubseteq_{\mathbb{L}} Q$?" (where $\mathbb{L}$ is a bounded distributive lattice, including $\mathbb{B}$ and $\mathrm{PosBool}(X)$) is $\Pi^p_2$-complete (Theorem \ref{contdisdistlatt}). 

\item ``$P \sqsubseteq_{\mathbb{N}[X]} Q$?" is in $\text{coNP}^\text{PP}$ (Theorem \ref{N[X]}). 

\item ``$P\sqsubseteq_{\mathrm{Sorp}(X)}Q$?" is $\Pi^p_2$-complete (Theorem \ref{sorpchneq}). 

\item ``$P \sqsubseteq_{\mathbb{B}[X]} Q$?" is $\Pi^p_2$-complete (Theorem \ref{diseqB[X]}). 

\item ``$P \sqsubseteq_{\mathrm{Why}(X)} Q$?" is $\Pi^p_2$-complete (Theorem \ref{whychneq}).  

\item ``$P\sqsubseteq_{\mathrm{Trio}(X)} Q$?" is in $\text{coNP}^\text{PP}$ (Theorem \ref{triocqneqcontainment}).

\item ``$P \sqsubseteq_{\mathrm{Which}(X)} Q$?" is $\Pi^p_2$-complete (Theorem \ref{linhard}).

\end{enumerate}

\end{Thm}

  We start with two examples of the auxiliary theorems involved in our summary result. These cases use exact homomorphisms in their characterizations. An interesting difference with the CQ case~\cite{Green} is that $\mathbb{B}[X]$ and  $\mathbb{N}[X]$ containment now have different characterizations and, in fact, there are CQ$^\neq$s $P$ and $Q$ such that $P \sqsubseteq_{\mathbb{B}[X]} Q$ but  $P \not \sqsubseteq_{\mathbb{N}[X]} Q$\footnote{For example, consider the Boolean queries $P = \exists x,y,z( R(x)\wedge R(y)\wedge S(z)\wedge x\neq y)$ and $Q= \exists x,y,x (R(x)\wedge R(y)\wedge S(z)\wedge x\neq y\wedge x\neq z)$. They are equivalent over $\mathbb{B}[X]$, but $P\not\sqsubseteq_{\mathbb{N}[X]}Q$.}. We include full proofs  because the case of $\mathbb{B}[X]$ shows how we adapt the techniques of~\cite{Green} while the $\mathbb{N}[X]$ case shows how the analogy with the CQ case breaks down.  

\begin{Thm}\label{diseqB[X]} For CQ$^\neq$s $P$ and $Q$ with the same tuple of free variables, the following are equivalent:
\begin{itemize}
\item[1.] $P \sqsubseteq_{\mathbb{B}[X]} Q$,
\item[2.] For every identification $P^{\mathrm{id}}$ of $P$, we have that $\llbracket P^{\mathrm{id}} \rrbracket^{\mathrm{can}(P^{\mathrm{id}})} \leq \llbracket Q \rrbracket^{\mathrm{can}(P^{\mathrm{id}})}$.
\item[3.] For every identification $P^{\mathrm{id}}$ of $P$, there is a $\neq$-homomorphism $h_{P^{\mathrm{id}}} \colon Q \longrightarrow P^{\mathrm{id}}$ exact for relational atoms.

\end{itemize}
Moreover, the corresponding decision problem (i.e.\ $P \sqsubseteq_{\mathbb{B}[X]} Q$?) is  $\Pi^p_2$-complete.
\end{Thm}
\begin{proof}
$(1) \implies (2):$ Suppose that $P \sqsubseteq_{\mathbb{B}[X]} Q$ and consider an identification $P^{\mathrm{id}}$ of a completion $P^c$ of $P$. We have that  $P^{\mathrm{id}}\sqsubseteq_{\mathbb{B}[X]}  P^c \sqsubseteq_{\mathbb{B}[X]} P$ since on any $\mathbb{B}[X]$-database instance $\mathbf I$, the polynomial corresponding to $P^c$ contains only monomials appearing in the one corresponding to $P$ (some of the monomials appearing in $P$ might have turned into $0$ in $P^c$). Thus, $\llbracket P^{\mathrm{id}} \rrbracket^{\mathrm{can}_{\mathbb{B}[X]}(P^{\mathrm{id}})} \leq \llbracket P^c \rrbracket^{\mathrm{can}_{\mathbb{B}[X]}(P^{\mathrm{id}})} \leq  \llbracket P \rrbracket^{\mathrm{can}_{\mathbb{B}[X]}(P^{\mathrm{id}})} \leq  \llbracket Q \rrbracket^{\mathrm{can}_{\mathbb{B}[X]}(P^{\mathrm{id}})}$.

$(2) \implies (3):$  Assume that (2) holds, so
for every identification $P^{\mathrm{id}}$ of any completion $P^c$ of $P$, we have that 
$$\llbracket P^{\mathrm{id}} \rrbracket^{\mathrm{can}_{\mathbb{B}[X]}(P^{\mathrm{id}})}(\vec{u}) \leq \llbracket Q \rrbracket^{\mathrm{can}_{\mathbb{B}[X]}(P^{\mathrm{id}})}(\vec{u}),$$
where $\vec{u}$ is the tuple of free variables of $P^{\mathrm{id}}$. The polynomial $\llbracket P^{\mathrm{id}} \rrbracket^{\mathrm{can}_{\mathbb{B}[X]}(P^{\mathrm{id}})}(\vec{u})$ has as a term a monomial $x_1 \cdot \ldots \cdot x_n $ containing every tag $p$ appearing in $\mathrm{can}_{\mathbb{B}[X]}(P^{\mathrm{id}})$ possibly with repetitions (such monomial is obtained using the identity assignment for the variables in $ P^{\mathrm{id}}$; noticing that the disequations in $P^{\mathrm{id}}$ all take value $1$ since $P$ and $P^c$ are consistent).  Then, by the definition of $\leq $ for $\mathbb{B}[X]$, we must have that the monomial $x_1 \cdot \ldots \cdot x_n $ appears in $\llbracket Q \rrbracket^{\mathrm{can}_{\mathbb{B}[X]}(P^{\mathrm{id}})}(\vec{u})$. The rest of the reasoning is as the corresponding part of Theorem~\ref{N[X]}.
 
$(3) \implies (1):$ Assume that for every identification $P^{\mathrm{id}}$ of any completion $P^c$ of $P$, there is a $\neq$-homomorphism $h_{P^{\mathrm{id}}}\colon Q \longrightarrow P^{\mathrm{id}}$ exact for relational atoms. In order to establish $P \sqsubseteq_{\mathbb{B}[X]} Q$ consider an arbitrary $\mathbb{B}[X]$-database instance $\mathbf I$ and a tuple $\vec{t}$. We need to show that any monomial in $\llbracket P \rrbracket^\mathbf{I}(\vec{t})$ appears  in $\llbracket Q \rrbracket^\mathbf{I}(\vec{t})$. Take an assignment $v$ of elements to the variables of $P$ justifying a monomial $x_1 \cdot \ldots \cdot x_n $ and consider the completion $ P^c$ of $P$ obtained by adding $x=y$ or $x\neq y$ depending on whether  $v(x) = v(y)$ or $v(x) \neq v(y)$. Take an identification $P^{\mathrm{id}}$ of $P^c$ by a canonical substitution $\theta$. We may observe that $\theta(x) = \theta(y)$ iff $v(x) = v(y)$. Then, $x_1 \cdot \ldots \cdot x_n $ is also a monomial in $\llbracket P^{\mathrm{id}} \rrbracket^\mathbf{I}(\vec{t})$ justified by the assignment $v$ itself since the new disequations added can only take value $1$ by construction according to the assignment $v$.  By assumption, there is a $\neq$-homomorphism $h_{P^{\mathrm{id}}}\colon Q \longrightarrow P^c$ exact for relational atoms. Then, $v \circ h_{P^{\mathrm{id}}}$ is an assignment of elements to the variables of $Q$ justifying the monomial $x_1 \cdot \ldots \cdot x_n $. If all the $x_i$s are values of relational atoms of $P$ according to $v$, then     if $x_i = \llbracket B(\theta(\vec{z})) \rrbracket^\mathbf{I}(v(\theta(\vec{z})))$ where  $B(\vec{z})$ is a relational atom of  $P$ (and thus $B(\theta(\vec{z}))$ is one of $P^{\mathrm{id}}$), then $x_i = \llbracket A(\vec{x}) \rrbracket^\mathbf{I}(v(h_{P^{\mathrm{id}}}(\vec{x})))$  for some relational atom $A(\vec{x})$ of $Q$ such that $A(h_{P^{\mathrm{id}}}(\vec{x})) = B(\theta(\vec{z}))$, which exists by surjectivity.   Furthermore, the multiplicity  with which  $x_i$ appears using the assignment $v \circ h_{P^{\mathrm{id}}} $ is exactly the multiplicity with which it appears in  $x_1 \cdot \ldots \cdot x_m $ given that $h_{P^{\mathrm{id}}}$ is both onto and injective on relational atoms.
Since equations and disequations can only take values $0$ or $1$, they can be disregarded in showing that we can justify a monomial.

For the complexity calculation, observe that the complement of  the above problem  is in $\text{NP}^{\text{coNP}} = \text{NP}^{\text{NP}}=\Sigma^p_2$, and thus our actual problem is in $\text{coNP}^{\text{NP}}=\Pi^p_2$. Given an instance of the problem $(P, Q)$ consisting of two queries, we wish to check whether there is an identification $P^{\mathrm{id}}$ of $P$ such that there is no $\neq$-homomorphism $h_{P^{\mathrm{id}}}\colon Q \longrightarrow P^{\mathrm{id}}$ exact for relational atoms. We can run our Turing machine to guess identifications of $P$ (checking that something is an identification involves checking consistency, but this can be done in polynomial time \cite[Section 14.2]{Ullman}) and for each check with the $\text{coNP}$ oracle that there are no  $\neq$-homomorphisms $h_{P^{\mathrm{id}}} \colon Q \longrightarrow P^{\mathrm{id}}$ exact for relational atoms.

Now we focus on the lower bound. We modify an argument from~\cite{VdM} spelled out in~\cite{ Kolaitis1}. Given a $\Pi_2$ quantified Boolean formula $\Phi$, we construct two Boolean CQ$^\neq$s $P_\Phi$ and $Q_\Phi$ such that $\Phi$ is satisfiable if and only if for every identification $P_\Phi'$ of $P_\Phi$ there is a $\neq$-homomorphism  $h_{P_\Phi'}\colon Q_\Phi \longrightarrow P_\Phi' $ exact for relational atoms. 
Assume that $\Phi$ is the quantified Boolean formula
$$
\forall p_1 \ldots \forall p_n \exists q_1 \ldots \exists q_m \alpha\left(p_1, \ldots, p_n, q_1, \ldots q_m\right)
$$
where $\alpha$ is a quantifier-free Boolean formula. For any connective $\delta$, we denote by $G^\delta_\alpha$ the multiset of subformulas of the formula $\alpha$ with $\delta$ as main connective in its syntax tree.
The queries are defined over the following the vocabulary (respecting the multiplicities of the subindices) $$ \text{ISTRUE}, (\text{OR}_{\theta_1\vee \theta_2})_{\theta_1\vee \theta_2\in  G^\vee_\alpha}, (\text{AND}_{\theta_1\wedge\theta_2})_{\theta_1\wedge \theta_2\in G^\wedge_\alpha}, (\text{NOT}_{\neg \theta})_{\neg \theta \in G^\neg_\alpha},  (P_i)_{ 0<i\leq n}$$ The predicates AND$_{\theta_1\wedge\theta_2}$ (ternary), OR$_{\theta_1\vee\theta_2}$ (ternary), NOT$_{\neg \theta}$ (binary) will be used to define the operators of Boolean algebra. Importantly, we take  each such predicate to denote a gate in the circuit representation of the formula $\alpha$. The predicate ISTRUE is unary and defines the truth value TRUE. Each $P_i$ is binary.



The query $P_\Phi$ is defined by the following existential formula:
$$
\begin{aligned}
& \exists f, t ((f \neq t) \wedge  \operatorname{ISTRUE}(t) \wedge
\bigwedge_{\theta_1\vee \theta_2\in G^\vee_\alpha}(\text { OR }_{\theta_1\vee \theta_2}(t, t, t) \wedge \text { OR }_{\theta_1\vee \theta_2}(t, f, t) \wedge \text { OR }_{\theta_1\vee \theta_2}(f, t, t) \wedge \text { OR }_{\theta_1\vee \theta_2}(f, f, f)) \wedge \\
& \bigwedge_{\theta_1\wedge \theta_2\in G^\wedge_\alpha}(\text { AND }_{\theta_1\wedge \theta_2}(t, t, t) \wedge \text { AND }_{\theta_1\wedge \theta_2}(t, f, f) \wedge \text { AND }_{\theta_1\wedge \theta_2}(f, t, f) \wedge \text { AND }_{\theta_1\wedge \theta_2}(f, f, f)) \wedge
\\
& \bigwedge_{\neg \theta\in G^\neg_\alpha} (\text { NOT }_{\neg \theta}(t, f) \wedge \text { NOT }_{\neg \theta}(f, t)) \wedge \\
& \bigwedge_{i=1}^{n} \exists u_i, v_i, w_i  (P_i\left(u_i, t\right) \wedge P_i\left(v_i, f\right) \wedge u_i \neq v_i \wedge P_i\left(w_i, t\right) \wedge P_i\left(w_i, f\right))).
\end{aligned}
$$

The query $Q_\Phi$ is defined as the existential closure of the conjunction of the following formulas (we consider a variable $x_\theta$ for each occurrence of a compound subformula $\theta$ in $\alpha$, and the variables $z_{p_i}$ and $z_{q_j}$ corresponding to the atoms $p_i$ and $q_j$ without caring for repeated occurrences):
$$
\operatorname{ISTRUE}(x_\alpha),$$
$$
\bigwedge_{i=1}^n ( P_i\left(g_i, z_{p_i}\right) \wedge P_i\left(h_i, z_{p_i}\right) \wedge g_i \neq h_i \wedge P_i\left(a_i, b_{i}\right) \wedge P_i\left(c_i, d_{i}\right)), $$
 where $a_i, b_{i}, c_i, d_{i}$ are  new variables in the formula (the last two atoms are fillers to match the body of $P_\Phi$),
$$\bigwedge_{\theta_1\wedge \theta_2 \in G^\wedge_\alpha}(\text { AND }_{\theta_1\wedge \theta_2}(x_{\theta_1}, x_{\theta_2}, x_{\theta_1\wedge \theta_2}) \wedge  \bigwedge_{i=1}^3 (\text { AND }_{\theta_1\wedge \theta_2}(a_{\theta_1\wedge \theta_2, i}, b_{\theta_1\wedge \theta_2, i}, c_{\theta_1\wedge \theta_2, i})),$$
where the atoms in the last big conjunction are all fillers and the variables appearing in them are all new,
$$\bigwedge_{\theta_1\vee \theta_2 \in G^\vee_\alpha}(\text { OR }_{\theta_1\vee \theta_2}(x_{\theta_1}, x_{\theta_2}, x_{\theta_1\vee \theta_2}) \wedge  \bigwedge_{i=1}^3 (\text { OR }_{\theta_1\vee \theta_2}(a_{\theta_1\vee \theta_2, i}, b_{\theta_1\vee \theta_2, i}, c_{\theta_1\vee \theta_2, i})),$$
where the atoms in the last big conjunction are all fillers and the variables appearing in them are all new,
$$\bigwedge_{\neg \theta \in G^\neg_\alpha}(\text { NOT}_{\neg \theta}(x_{\theta},  x_{\neg \theta}) \wedge  \text { NOT}_{\neg \theta}(a_{\theta}, b_{\theta}),$$
where the last atom is a filler and the variables in the filler atoms  are fresh and appear in no other atom.

It is immediate that the bodies of $P_\Phi, Q_\Phi$ contain the same multiplicities of atoms involving any given relational symbol.  We denote by $\mathcal{T}, \mathcal{F}$ the variables $t, f$ of $P_\Phi$ in a  given identification. We also use $\mathcal{U}_i, \mathcal{V}_i, \mathcal{W}_i$ to denote the variables $u_i, v_i, w_i$ of $P_\Phi$, respectively, in the identification. We now have to show that $\Phi$ is satisfiable if and only if for every identification $P_\Phi'$ of $P_\Phi$ there is a $\neq$-homomorphism  $h_{P_\Phi'}\colon Q_\Phi \longrightarrow P_\Phi' $ exact for relational atoms.

Assume that $\Phi$ is satisfiable. Let $P_\Phi'$  be an identification of  a completion $P_\Phi^c$  of  $P_\Phi$. Therefore, $P_\Phi'$ contains the predicates AND$_{\theta_1\wedge\theta_2}$, OR$_{\theta_1\vee\theta_2}$, NOT$_{\neg \theta}$, which specify the rules of the Boolean algebra on the variables $\mathcal{T}$ and $\mathcal{F}$ for all the occurrences of a subformulas $\theta_1\vee\theta_2$, $\theta_1\wedge \theta_2$ and $\neg \theta$ in $\alpha$. $P_\Phi'$ also contains, for each $i, 1 \ldots n$, relation symbols $P_i$ that appear in  four indexed relational occurrences each. Note that $\mathcal{U}_i \neq \mathcal{V}_i$ appears in $P_\Phi' $ (as required by $P_\Phi$), but the literals $\mathcal{W}_i = \mathcal{U}_i$, $\mathcal{W}_i = \mathcal{V}_i$, $\mathcal{W}_i \neq  \mathcal{U}_i$, $\mathcal{W}_i \neq \mathcal{V}_i$ will not all appear in $P_\Phi^c$ (but some combination will by definition of completions). 

We now show that there is there is a $\neq$-homomorphism  $h_{P_\Phi'}\colon Q_\Phi \longrightarrow P_\Phi' $ exact for relational atoms. We let $h_{P_\Phi'}$ send the variable $x_\alpha$ to $\mathcal{T}$ (so the atom $\operatorname{ISTRUE}(x_\alpha)$ has $\operatorname{ISTRUE}(t)$ as its image, and both atoms appear with exactly the same multiplicity in both queries). For each $i, 1 \ldots n$, we have now the following possibilities when  $P_\Phi'$ comes from the canonical substitution  $\pi$:

\noindent
{\sc Case 1.} $\mathcal{U}_i=\mathcal{W}_i$ appears in $P_\Phi^c$ and $\pi$ sends     $\mathcal{U}_i$ to $\mathcal{W}_i$. We let $h_{P_\Phi'}$ have  $g_i \mapsto \mathcal{V}_i, h_i\mapsto\mathcal{W}_i$, $z_{p_i} \mapsto\mathcal{F}$, $a_i, c_i\mapsto\mathcal{W}_i$ and $b_i, d_i\mapsto\mathcal{T}$. Following this mapping on atoms we get that  $P_i\left(g_i, z_{p_i}\right) \mapsto P_i\left(\mathcal{V}_i, \mathcal{F}\right)$, 
$ P_i\left(h_i, z_{p_i}\right) \mapsto P_i\left(\mathcal{W}_i, \mathcal{F}\right)$,
$ P_i\left(a_i, b_{i}\right) \mapsto P_i\left(\mathcal{W}_i, \mathcal{T}\right)$, and $ P_i\left(c_i, d_{i}\right) \mapsto P_i\left(\mathcal{W}_i, \mathcal{T}\right)$ (notice that the multiplicity of the atoms and images are matched in the body of both queries are required for exactness). 

\noindent
{\sc Case 2.} 
$\mathcal{U}_i=\mathcal{W}_i$ appears in $P_\Phi^c$ and $\pi$  sends   $\mathcal{W}_i$ to $\mathcal{U}_i$. We do as in {\sc Case 1} but let $h_i\mapsto\mathcal{U}_i$,  and $a_i, c_i\mapsto\mathcal{U}_i$.  We get that  $P_i\left(g_i, z_{p_i}\right) \mapsto P_i\left(\mathcal{V}_i, \mathcal{F}\right)$, 
$ P_i\left(h_i, z_{p_i}\right) \mapsto P_i\left(\mathcal{U}_i, \mathcal{F}\right)$,
$ P_i\left(a_i, b_{i}\right) \mapsto P_i\left(\mathcal{U}_i, \mathcal{T}\right)$, and $ P_i\left(c_i, d_{i}\right) \mapsto P_i\left(\mathcal{U}_i, \mathcal{T}\right)$ (notice again that the multiplicity of the atoms and images are matched in the body of both queries are required for exactness). 

\noindent
{\sc Case 3.} $\mathcal{U}_i \neq \mathcal{W}_i$ appears in $P_\Phi'$. We let $h_{P_\Phi'}$ have $g_i \mapsto \mathcal{U}_i, h_i\mapsto \mathcal{W}_i$, $z_{p_i} \mapsto \mathcal{T}$, $a_i\mapsto\mathcal{V}_i$, $ c_i\mapsto\mathcal{W}_i$, and $b_i, d_i\mapsto\mathcal{F}$.  We get that  $P_i\left(g_i, z_{p_i}\right) \mapsto P_i\left(\mathcal{U}_i, \mathcal{T}\right)$, 
$ P_i\left(h_i, z_{p_i}\right) \mapsto P_i\left(\mathcal{W}_i, \mathcal{T}\right)$,
$ P_i\left(a_i, b_{i}\right) \mapsto P_i\left(\mathcal{V}_i, \mathcal{F}\right)$, and $ P_i\left(c_i, d_{i}\right) \mapsto P_i\left(\mathcal{W}_i, \mathcal{F}\right)$ (once more the multiplicities of atoms on both sides match as required for the exactness of the mapping).

Note that since $\Phi$ is satisfiable, then regardless of what assignment of the variables $\mathcal{T}$ and $ \mathcal{F}$ in $P_\Phi'$ is made to the variables $z_{p_i}, i=1, \ldots, n$ (as long as it respects the equations that involve these variables in $Q_\Phi$ which our proposal in the previous paragraph does), there is an assignment to the remaining variables $x_\theta$ in $Q_\Phi$ sending them to $\mathcal{T}$ or $\mathcal{F}$
that preserves all the atoms and equations in $Q_\Phi$ when $x_\alpha$ is assigned to $\mathcal{T}$. Simply chase down the truth values assigned to the subformulas $\theta$ in the satisfying assignment making $\alpha$ true when the truth-values of the $p_1, \ldots, p_n$ match the assignments made to the $z_{p_1}, \ldots, z_{p_n}$. The variables in the filler atoms then can be assigned to $\mathcal{T}$ or $\mathcal{F}$ as needed to complete the remaining rows in the truth table of the main connective of each $\theta$. The way $Q_\Phi$ and $P_\Phi$ have been constructed then guaranties that the map we will get is exact in preserving multiplicities of atoms.

Next assume that for every identification $P_\Phi'$ of $P_\Phi$ there is a $\neq$-homomorphism  $h_{P_\Phi'}\colon Q_\Phi \longrightarrow P_\Phi' $ exact for relational atoms. Let us construct $2^n$ completions of $P_\Phi$, denoted as  $P_\Phi^1, P_\Phi^2, \ldots P_\Phi^{2^n}$, containing equations that make  the variables $\mathcal{U}_i, \mathcal{V}_i, \mathcal{W}_i$  identical to some of the variables in the set  $\{\mathcal{T}, \mathcal{F}\}$; and such that each completion describes one of the possible truth assignments to the $n$-many universally quantified variables of $\Phi$ (in the sense that if the assignment says that $p_i$ takes value true, we complete $P_\Phi$ for the variables $\mathcal{U}_i, \mathcal{V}_i, \mathcal{W}_i$ by adding $\mathcal{V}_i = \mathcal{W}_i$ and $\mathcal{U}_i \neq  \mathcal{W}_i$, while if the assignment says that $p_i$ takes value false, we add $\mathcal{U}_i= \mathcal{W}_i$ and $\mathcal{V}_i \neq  \mathcal{W}_i$). We put  $\mathcal{U}_i= \mathcal{T}$ and   $\mathcal{V}_i= \mathcal{F}$ in every such completion, and $\mathcal{W}_i= \mathcal{T}$ or $\mathcal{W}_i= \mathcal{F}$, depending on whether we have put $\mathcal{U}_i= \mathcal{W}_i$  or $\mathcal{V}_i = \mathcal{W}_i$.  The relation symbols $P_i$ in each of these completions appear in the atoms $P_i\left(\mathcal{U}_i, t\right), P_i\left(\mathcal{V}_i, f\right), P_i\left(\mathcal{W}_i, t\right), P_i\left(\mathcal{W}_i, f\right)$, for $i= 1, \ldots, n$. Notice that  these completions are all identical, except that for some we have the equation $\mathcal{W}_i=\mathcal{U}_i$ while for others we have  the equation $\mathcal{W}_i=\mathcal{V}_i$. By our hypothesis, for each identification $P_\Phi^{l'}, 1\leq l \leq 2^n$ of these completions, there is a $\neq$-homomorphism $h_{P_\Phi^{l'}}\colon Q_\Phi \longrightarrow P_\Phi^{l'}$ exact for relational atoms. For the identification of a completion in which $\mathcal{U}_i=\mathcal{W}_i$ appears, the variable $z_{p_i}$ from $Q_\Phi$ must be mapped to $\mathcal{F}$, and in the completion in which $\mathcal{V}_i=\mathcal{W}_i$ appears, the variable $z_{p_i}$ can only be mapped to $\mathcal{T}$. Other mappings will fail to  preserve the literals of $Q_\Phi$. For example, since the literals $P_i\left(g_i, z_{p_i}\right), P_i\left(h_i, z_{p_i}\right), g_i \neq h_i$ appear in $Q_\Phi$, if in $P_\Phi^{l'}$ we have $\mathcal{U}_i=\mathcal{W}_i$ and thus by construction also $\mathcal{V}_i\neq \mathcal{W}_i, P_i\left(\mathcal{U}_i, \mathcal{T}\right), P_i\left(\mathcal{V}_i, \mathcal{F}\right), P_i\left(\mathcal{W}_i, \mathcal{T}\right), P_i\left(\mathcal{W}_i, \mathcal{F}\right), \mathcal{U}_i\neq \mathcal{V}_i$, then if  we map $z_{p_i}$ to $\mathcal{T}$ we have that $P_i\left(\mathcal{U}_i, \mathcal{T}\right)$ and $P_i\left(\mathcal{W}_i, \mathcal{T}\right)$ are the only possible images of $P_i\left(g_i, z_{p_i}\right), P_i\left(h_i, z_{p_i}\right)$, but  $g_i \neq h_i$ fails to be preserved under this mapping. Then, no matter how we map  $z_{p_i}$, $i=1, \ldots, n$ to elements in the set $\{\mathcal{T}, \mathcal{F}\}$, we will be able to extend this mapping to a $\neq$-homomorphism  from $Q_\Phi$ to a suitable identification of $P_\Phi$.   Given the way $Q_\Phi$ is defined, this implies that $\exists q_1 \ldots \exists q_m \alpha\left(p_1, \ldots, p_n, q_1, \ldots, q_m\right)$ is satisfiable for all possible truth assignments of the universally quantified variables of $\Phi$. Thus, $\Phi$ is satisfiable.
\end{proof}

The $\mathbb{N}[X]$ case cannot be characterized by the existence of a homomorphism exact for relational atoms anymore as the following simple example shows. In fact, the example shows that no characterization involving the existence of mapping would work simply because exact mappings are some of the most demanding ones we can think of. Let
\[
 P=\exists x y z\,(R(x)\wedge R(y)\wedge R(z)\wedge x\ne y)
\]
and
\[
 Q=\exists  u v w\,(R(u)\wedge R(v)\wedge R(w)\wedge u\ne v\wedge u\ne w).
\]
Notice that $P\equiv_{\mathbb{B}[X]}Q$. Every identification of $P$ admits at least one exact map from $Q$. $P$ has the following identifications:
\begin{itemize}
\item[1.]  $\exists x y \,(R(x)\wedge R(y)\wedge R(x)\wedge x\ne y)$ (an exact mapping from $Q$ is $u \mapsto y$ and $v,w \mapsto x$)
\item[2.] $ \exists z y \,(R(z)\wedge R(y)\wedge R(z)\wedge z\ne y)$ (an exact mapping from $Q$ is $u \mapsto y$ and $v,w \mapsto z$)
\item[3.] $ \exists x z\,(R(x)\wedge R(z)\wedge R(z)\wedge x\ne z)$ (an exact mapping from $Q$ is $u \mapsto x$ and $v,w \mapsto z$)
\item[4.] $\exists x y \,(R(x)\wedge R(y)\wedge R(y)\wedge x\ne y)$ (an exact mapping from $Q$ is $u \mapsto x$ and $v,w \mapsto y$)

\item[5.] $\exists x y z\,(R(x)\wedge R(y)\wedge R(z)\wedge x\ne y \wedge x\ne z \wedge z\ne y )$ (an exact mapping from $Q$ is $u \mapsto x$, $v \mapsto y$ and $w \mapsto z$)

\end{itemize}
Nevertheless, on the two-element  instance  $\bf I$ with domain $\{a,b\}$  and where $R(a)=p$, $R(b)=q$, we have that
\[
 \llbracket P\rrbracket^{\bf I}=2p^2q+2pq^2,
 \qquad
 \llbracket Q\rrbracket^{\bf I}=p^2q+pq^2,
\]
and thus $P \not \sqsubseteq_{\mathbb{N}[X]} Q$.
The two distinct monomials in $\llbracket P\rrbracket^{\bf I}$ each have coefficient $2$, corresponding to two exact homomorphisms to the respective identification.
For  identification 1 containing two occurrences of $R(x)$ and one of $R(y)$ with $x\ne y$, there are two exact homomorphisms from $P$ to $P'$ (they send $(x,y,z)$ to $(x,y,x)$ and $(y,x,x)$, respectively) but only one exact term map from $Q$ to 1 (described above) since the image of $v,w$ must go to $x$ since $R(x)$ is the only atom appearing twice in the body of 1.   The idea of counting mappings in the characterization below for $\mathbb{N}[X]$ is similar to some techniques used in~\cite{Kostylev2} to handle the complexity of containment for UCQs.

\begin{Thm}\label{N[X]}Given any CQ$^\neq$s $R$ and $S$, we  let $|\mathrm{ExHom}(R,S)|$ denote the number of $\neq$-homomorphisms from $R$ to $S$ that are exact for relational atoms. If $P, Q$ have the same tuple of free variables,  the following are equivalent:
\begin{itemize}
\item[1.] $P \sqsubseteq_{\mathbb{N}[X]} Q$,

\item[2.] For every identification $P^{\mathrm{id}}$ of $P$,
$|\mathrm{ExHom}(P,P^{\mathrm{id}})|\leq |\mathrm{ExHom}(Q,P^{\mathrm{id}})|$.

\end{itemize}
Moreover, the corresponding decision problem (i.e.\ $P \sqsubseteq_{\mathbb{N}[X]} Q$?) is in $\text{coNP}^\text{PP}$.
\end{Thm}

\begin{proof}
$(1) \implies (2):$ 
Suppose that $P \sqsubseteq_{\mathbb{N}[X]} Q$ and consider an identification $P^{\mathrm{id}}$ of $P$. By assumption, $\llbracket P \rrbracket^{\mathrm{can}_{\mathbb{N}[X]}(P^{\mathrm{id}})} \leq  \llbracket Q \rrbracket^{\mathrm{can}_{\mathbb{N}[X]}(P^{\mathrm{id}})}$. Moreover, we have that $P^{\mathrm{id}}\sqsubseteq_{\mathbb{N}[X]}  P^c \sqsubseteq_{\mathbb{N}[X]} P$ since on any $\mathbb{N}[X]$-database instance $\mathbf I$, the polynomial corresponding to $P^c$ contains only monomials appearing in the one corresponding to $P$ (some of the monomials appearing in $P$ might have turned into $0$ in $P^c$). The polynomial $\llbracket P^{\mathrm{id}} \rrbracket^{\mathrm{can}_{\mathbb{N}[X]}(P^{\mathrm{id}})}(\vec{u})$ (and thus $\llbracket P \rrbracket^{\mathrm{can}_{\mathbb{N}[X]}(P^{\mathrm{id}})}(\vec{u})$) has as a term a monomial $x_1 \cdot \ldots \cdot x_n $ containing every tag $p$ appearing in $\mathrm{can}_{\mathbb{N}[X]}(P^{\mathrm{id}})$ possibly with repetitions (such monomial is obtained using the identity assignment for the variables in $ P^{\mathrm{id}}$; noticing that the disequations in $P^{\mathrm{id}}$ all take value $1$ since $P$ and $P^c$ are consistent). Take an assignment $v$ to the variables of $P$ justifying the monomial $x_1 \cdot \ldots \cdot x_n $ in 
$\llbracket P \rrbracket^{\mathrm{can}_{\mathbb{N}[X]}(P^{\mathrm{id}})}(\vec{u})$. We can see that $v$ is a $\neq$-homomorphism from $P$ to $P^{\mathrm{id}}$ exact for relational atoms. To show injectivity, notice that any tag $p$ that is the value of a relational atom $A(\vec{z})$ of $P$ according to $v$, appears in $x_1 \cdot \ldots \cdot x_n $ with at most the same multiplicity as the atom canonically annotated by $p$ (which is the image of $A(\vec{z})$ by $v$) does in  the body of $P^{\mathrm{id}}$ (since this monomial corresponds to the body of $P^{\mathrm{id}}$), which entails that the multiset of images by $v$ of relational atoms of $P$ is bag-contained in the multiset of relational atoms from $P^{\mathrm{id}}$. Surjectivity follows from the fact that $x_1 \cdot \ldots \cdot x_n $ corresponds to the multiset of  the images by $v$ of the multiset of relational atoms from $P$, so the former is bag-contained in the latter. Moreover, clearly  every $\neq$-homomorphism from $P$ to $P^{\mathrm{id}}$ exact for relational atoms must produce the monomial $x_1 \cdot \ldots \cdot x_n $ when seen as an assignment to the variables of $P$  on the instance $\mathrm{can}_{\mathbb{N}[X]}(P^{\mathrm{id}})$. Thus there is a bijection between justifications of the monomial $x_1 \cdot \ldots \cdot x_n $ in the polynomial $\llbracket P \rrbracket^{\mathrm{can}_{\mathbb{N}[X]}(P^{\mathrm{id}})}(\vec{u})$ and the mappings in $\mathrm{ExHom}(P,P^{\mathrm{id}})$. By a similar reasoning, there is a bijection between justifications of the monomial $x_1 \cdot \ldots \cdot x_n $ in the polynomial $\llbracket Q \rrbracket^{\mathrm{can}_{\mathbb{N}[X]}(P^{\mathrm{id}})}(\vec{u})$ and the mappings in $\mathrm{ExHom}(Q,P^{\mathrm{id}})$. Thus the coefficients of the monomial $x_1 \cdot \ldots \cdot x_n $ in $\llbracket P \rrbracket^{\mathrm{can}_{\mathbb{N}[X]}(P^{\mathrm{id}})}(\vec{u})$ and $\llbracket Q \rrbracket^{\mathrm{can}_{\mathbb{N}[X]}(P^{\mathrm{id}})}(\vec{u})$ correspond to $|\mathrm{ExHom}(P,P^{\mathrm{id}})|$ and $|\mathrm{ExHom}(Q,P^{\mathrm{id}})|$, respectively. By our hypothesis then $|\mathrm{ExHom}(P,P^{\mathrm{id}})|\leq |\mathrm{ExHom}(Q,P^{\mathrm{id}})|$.

$(2) \implies (1):$ We prove the contrapositive.  Assume that $P \not\sqsubseteq_{\mathbb{N}[X]} Q$, so for some $\mathbb{N}[X]$-database instance $\mathbf I$ and a tuple $\vec{t}$, $\llbracket P \rrbracket^\mathbf{I}(\vec{t}) \not\leq_{\mathbb{N}[X]} \llbracket Q\rrbracket^\mathbf{I}(\vec{t})$.  We wish to show that for some identification $P^{\mathrm{id}}$ of $P$,
$|\mathrm{ExHom}(Q,P^{\mathrm{id}})| < |\mathrm{ExHom}(P,P^{\mathrm{id}})| $.
By Lemma~\ref{lemma7.15bis} (this is essentially~\cite[Lemma 7.15]{Green}), $\llbracket P \rrbracket^\mathbf{J} (\vec{t})\not\leq_ {\mathbb{N}[X]} \llbracket Q \rrbracket^\mathbf{J}(\vec{t})$, where $\mathbf{J} = \text{ab}_{\mathbb{N}[X]}(\mathbf{I}) $. Thus  
some monomial $\mu=x_1 \cdot \ldots \cdot x_n $ in $\llbracket P \rrbracket^\mathbf{J} (\vec{t})$ appears with greater  coefficient than it does in $\llbracket Q \rrbracket^\mathbf{J}(\vec{t})$. The monomial $\mu$ came from a justifying variable assignment $v$ to the variables of $P$.

 Let $P^c_v$ be the completion of $P$ that  for every pair of variables $x,y$, it contains $x=y$ if $v(x)=v(y)$ and $x\neq y$ otherwise. Let $\theta$ be a canonical substitution for this completion, and let $ P^{\mathrm{id}}_v $ be the resulting identification. Define 
 $v_*(\theta_v(x)):=v(x)$. The map $v_*$ is well-defined and injective, since $ \theta(x)=\theta(y) $ iff $ v(x)=v(y)$.  Moreover, $v_*$ is a bijection between the variables of $P^{\mathrm{id}}_v$ and the domain elements occurring in the tuples whose tags occur in $\mu$. Indeed, every such domain element is the value under $v$ of a variable appearing in  a relational occurrence of $P$. Conversely, since $P$ is safe, every variable of $P$ occurs in a relational atom, so every value assumed by $v$ occurs in a tuple whose tag occurs in $\mu$. 
 
 We claim that, for every safe $\mathrm{CQ}^{\neq}$ $R$ with the same  head as $P^{\mathrm{id}}_v$, the coefficient of $\mu$ in the polynomial $\llbracket R\rrbracket^\mathbf{J}(\vec{t})$  is exactly  $\left| \mathrm{ExHom}(R,P^{\mathrm{id}}_v) \right|$.  Let $w$ be an assignment to the variables of $R$ which produces $\vec{t}$ for the head and justifies $\mu$. Since $\mathbf J$ is abstractly tagged, every value used by $w$ occurs in a tuple whose tag occurs in $\mu$, and therefore belongs to the range of $v_*$. Define  $h_w(x):=v_*^{-1}(w(x))$. The map $h_w$ preserves the  head and all equations and disequations. Furthermore, for every tuple $\vec{a}$, there is an  exponent in $\mu$ corresponding exactly  to the number of indexed relational occurrences of $R$ mapped by $w$ to $\vec{a}$. Since $v$ justifies $\mu$, this is exactly the number of indexed relational occurrences of $P^{\mathrm{id}}_v$ whose image under $v_*$ is $\vec{a}$. Consequently, the mapping is exact for relational atoms. Hence $h_w\in\mathrm{ExHom}(R,P^{\mathrm{id}}_v)$.  Conversely, if $ h\in\mathrm{ExHom}(R,P^{\mathrm{id}}_v)$,  then $ w_h:=v_*\circ h$ is an assignment to the variables of $R$ producing $\vec{t}$. It satisfies every equation and disequation of $R$, and exactness ensures that it uses, with the same multiplicities, precisely the ground relational tuples used by $v$. Therefore, $w_h$ contributes exactly $\mu$. The assignments $w$ and the exact homomorphisms $h$ correspond bijectively under $w\longmapsto v_*^{-1}\circ w,  h\longmapsto v_*\circ h$.   Thus $ \left|\mathrm{ExHom}(P,P^{\mathrm{id}}_v)\right| > \left|\mathrm{ExHom}(Q,P^{\mathrm{id}}_v)\right|$ which means that ultimately item~2 fails for the identification $P^{\mathrm{id}}_v$.

Finally, for the upper bound complexity calculation, we look at the complement of the problem stated in item 2 and we show it is in $\text{NP}^\text{PP}$ (and thus the actual problem we are interested in is in $\text{coNP}^\text{PP}$ as $\text{PP}$ is closed under complement). Given an input $(P, Q)$ for our problem, we start by guessing an identification $P^{\mathrm{id}}$ (checking that something is an identification involves checking consistency, but this can be done in polynomial time \cite[Section 14.2]{Ullman})  of $P$, and then using our $\text{PP}$  oracle to decide whether $ |\mathrm{ExHom}(Q,P^{\mathrm{id}})| < |\mathrm{ExHom}(P,P^{\mathrm{id}})|$. Observe that computing $|\mathrm{ExHom}(Q,P^{\mathrm{id}})|$ and $|\mathrm{ExHom}(P,P^{\mathrm{id}})|$ are clearly $\#P$ problems (we are counting the number of accepting instances of a decision problem that we saw to be in NP in the proof of Theorem~\ref{diseqB[X]}), and  thus computing $|\mathrm{ExHom}(P,P^{\mathrm{id}})| -|\mathrm{ExHom}(Q,P^{\mathrm{id}})|$ is in $\text{GapP}$ by the characterization in~\cite[Prop. 3.5]{Fenner} and, consequently, deciding $|\mathrm{ExHom}(P,P^{\mathrm{id}})| -|\mathrm{ExHom}(Q,P^{\mathrm{id}})| > 0$ (and thus $ |\mathrm{ExHom}(Q,P^{\mathrm{id}})| < |\mathrm{ExHom}(P,P^{\mathrm{id}})|$) is in $\text{PP}$ by~\cite[Prop. 4.2]{Fenner}.
\end{proof}

The final case we will deal with is $\mathrm{Trio}(X)$. We start with an observation on containment that displays a disanalogy with the  positive case~\cite{Green}.

\begin{Rmk}
For positive CQs, $\mathrm{Trio}(X)$-containment coincides with
$\mathrm{Why}(X)$-containment \cite{Green}. This is no longer the case when disequations are allowed. Indeed, consider the Boolean CQ$^\neq$s $P:=\exists x,y,z\,\bigl(R(x)\wedge R(y)\wedge R(z)\wedge x\neq y\bigr)$ and $Q:=\exists u,v,w\,\bigl(R(u)\wedge R(v)\wedge R(w)\wedge u\neq v\wedge u\neq w\bigr).$ Up to variable renaming, the identifications of $P$ correspond to the three possible equality patterns for $z$: either $z=x$, $z=y$, or $z$ is distinct from both $x$ and $y$. In these three cases, respectively, bag-onto $\neq$-homomorphisms from $Q$ to the corresponding identification of $P$ are given by $u\mapsto y,\ v\mapsto x,\ w\mapsto x,$ or $u\mapsto x,\ v\mapsto y,\ w\mapsto y,$ or $u\mapsto x,\ v\mapsto y,\ w\mapsto z.$ Hence, by Theorem~\ref{whychneq}, $P\sqsubseteq_{\mathrm{Why}(X)}Q.$ On the other hand, consider the $\mathrm{Trio}(X)$-database $\mathbf I$ with domain $\{a,b\}$ and $R^{\mathbf I}(a)=p,$ $R^{\mathbf I}(b)=q,$ where $p,q\in X$ are distinct. Then $\llbracket P\rrbracket^{\mathbf I}=4pq,$ $\llbracket Q\rrbracket^{\mathbf I}=2pq.$ Since the natural order of $\mathrm{Trio}(X)$ is coefficientwise, $4pq\not\leq_{\mathrm{Trio}(X)}2pq$, and therefore $P\not\sqsubseteq_{\mathrm{Trio}(X)}Q.$ Thus, the coincidence between $\mathrm{Why}(X)$ and $\mathrm{Trio}(X)$-containment for positive CQs does not extend to CQ$^\neq$s.
\end{Rmk}

Thus, the case of $\mathrm{Trio}(X)$ requires a separate treatment. The main issue is that, unlike $\mathrm{Why}(X)$, $\mathrm{Trio}(X)$ retains coefficients, and hence the multiplicity of derivations, while still identifying monomials that differ only in their positive exponents. Consequently, the existence of suitable $\neq$-homomorphisms alone is not sufficient to characterize containment. To retain the missing multiplicity information, we combine $\neq$-homomorphisms with the $k$-abstract tags introduced above, recording for each relational atom occurrence which of the $k$ tags associated with its image is selected.

\begin{Def}[$k$-indexed $\neq$-homomorphisms and $k$-tag signature]
Let $P$ and $Q$ be two CQ$^\neq$s and let $k\geq 1$. We write $[k]=\{1,\ldots,k\}$.
\begin{enumerate}
    \item  A \emph{$k$-indexed $\neq$-homomorphism} from $P$ to $Q$ is a pair $(h,\iota)$, where $h\colon P\longrightarrow Q$ is a $\neq$-homomorphism and $\iota\colon \text{RelOcc}(P)\longrightarrow [k].$ Thus, $\iota$ assigns an index to every relational atom occurrence of $P$.
    \item A \emph{$k$-tag signature} for $Q$ is a function $\sigma\colon \text{RelSet}(Q)\longrightarrow \mathcal{P}([k]).$ We say that a $k$-indexed $\neq$-homomorphism $(h,\iota)$ from $P$ to $Q$ \emph{realizes} $\sigma$ if, for every $A\in\text{RelSet}(Q)$, $\sigma(A)=\{\iota(B)\mid B\in\text{RelOcc}(P),\ h(B)=A\}.$ We write $h_\iota$ for the $k$-indexed $\neq$-homomorphism determined by $(h,\iota)$.
\end{enumerate}
Given $P,Q,k$ and a $k$-tag signature $\sigma$ for $Q$, we denote by $\text{IHom}_k(P,Q,\sigma)$ the set of all $k$-indexed $\neq$-homomorphisms from $P$ to $Q$ that realize $\sigma$.
\end{Def}

Notice that distinct relational atom occurrences of $P$ mapped by $h$ to the same relational atom of $Q$ are indexed independently and may therefore be assigned different indices by $\iota$. We now associate each $k$-tag signature with a square-free monomial. This will allow us to relate the indexed homomorphisms introduced above with the coefficients appearing in the evaluation of a query over a $k$-abstract canonical database. Let $P^{\mathrm{id}}$ be an identification of a CQ$^\neq$ $P$. Recall that, in $\text{can}^{k}_{\mathrm{Trio}(X)}(P^{\mathrm{id}})$, the canonical tuple corresponding to every $A\in\text{RelSet}(P^{\mathrm{id}})$ is annotated by a distinct element of $S_k$ $x_{A,1}+\ldots+x_{A,k}$. 

\begin{Def}[Square-free monomial associated to $\sigma$]  Given a $k$-tag signature $\sigma\colon\text{RelSet}(P^{\mathrm{id}})\longrightarrow\mathcal{P}([k]),$ we let $M_\sigma$ be the square-free monomial $$
M_\sigma\stackrel{\text{def}}{=}
\prod_{A\in\text{RelSet}(P^{\mathrm{id}})}
\prod_{i\in\sigma(A)}x_{A,i}.
$$   
\end{Def}
Thus, for every relational atom $A$, the monomial $M_\sigma$ contains exactly the tags whose indices belong to $\sigma(A)$. Since distinct canonical tuples are annotated using pairwise disjoint sets of variables, different $k$-tag signatures give rise to different square-free monomials. Conversely, every square-free monomial involving only variables occurring in the annotations of $\text{can}^{k}_{\mathrm{Trio}(X)}(P^{\mathrm{id}})$ uniquely determines a $k$-tag signature by recording for every $A\in\text{RelSet}(P^{\mathrm{id}})$ the indices of the variables $x_{A,i}$ occurring in the monomial.

The following observation explains the connection between these monomials and $k$-indexed $\neq$-homomorphisms.

\begin{Lem}\label{indexedTrio}
Let $P$ and $Q$ be CQ$^\neq$s, let $P^{\mathrm{id}}$ be an identification of $P$, and let $\vec{u}$ denote the tuple of free variables of $P^{\mathrm{id}}$. Suppose that $Q$ has the same head arity as $P^{\mathrm{id}}$. Let $k\geq 1$, and let $\sigma$ be a $k$-tag signature for $P^{\mathrm{id}}$. Then the coefficient of $M_\sigma$ in
$\llbracket Q \rrbracket^{\text{can}^{k}_{\mathrm{Trio}(X)}(P^{\mathrm{id}})}(\vec{u})$ is exactly $\left|\text{IHom}_k(Q,P^{\mathrm{id}},\sigma)\right|.$
\end{Lem}

\begin{proof}

Consider an assignment $v$ to the variables of $Q$ contributing a non-zero value to $\llbracket Q \rrbracket^{\text{can}^{k}_{\mathrm{Trio}(X)}(P^{\mathrm{id}})}\vec{u}).$ Since the active domain of $\text{can}^{k}_{\mathrm{Trio}(X)}(P^{\mathrm{id}})$ consists of all the variables of $P^{\mathrm{id}}$, we can see $v$ as a mapping from the variables of $Q$ to those of $P^{\mathrm{id}}$. By construction of the canonical database, every relational atom occurrence of $Q$ is mapped to a relational atom of $P^{\mathrm{id}}$. Moreover, since $v$ satisfies all equations and disequations of $Q$, their images are logical consequences of the body of $P^{\mathrm{id}}$. Since $v$ produces the tuple $\vec{u}$, it also preserves the head. Hence, by definition, $v$ induces a $\neq$-homomorphism $h\colon Q\longrightarrow P^{\mathrm{id}}$. 

Conversely every such $\neq$-homomorphism determines an assignment contributing a non-zero value to the evaluation above: fix one such $\neq$-homomorphism $h$, if $B\in\text{RelOcc}(Q)$ and $h(B)=A$, then the factor contributed by the occurrence $B$ is $x_{A,1}+\ldots+x_{A,k}.$ Therefore, when expanding the product corresponding to the contribution of $h$, one independently chooses, for every relational atom occurrence $B\in\text{RelOcc}(Q)$, an index $\iota(B)\in[k]$. Hence every term in this expansion corresponds precisely to a $k$-indexed $\neq$-homomorphism $h_\iota$.

Before dropping exponents, the term selected by $h_\iota$ is $\prod_{B\in\text{RelOcc}(Q)}
x_{h(B),\iota(B)}.$ In $\mathrm{Trio}(X)$ positive exponents are dropped. Consequently, the corresponding square-free monomial is $\prod_{A\in\text{RelSet}(P^{\mathrm{id}})}
\prod_{i\in
\{\iota(B)\mid B\in\text{RelOcc}(Q),\ h(B)=A\}}
x_{A,i}.$
By definition, this monomial is $M_\sigma$ exactly when $h_\iota$ realizes $\sigma$.

Finally, distinct $k$-indexed $\neq$-homomorphisms $h_\iota$ correspond to distinct choices in the evaluation, even when they give rise to the same square-free monomial. Since $\mathrm{Trio}(X)$ drops exponents but retains natural-number coefficients, all such choices contribute separately to the coefficient of that monomial. Therefore, the coefficient of $M_\sigma$ in $\llbracket Q \rrbracket^{\text{can}^{k}_{\mathrm{Trio}(X)}(P^{\mathrm{id}})}(\vec{u})$ is precisely the number of $k$-indexed $\neq$-homomorphisms from $Q$ to $P^{\mathrm{id}}$ realizing $\sigma$, that is, $\left|\text{IHom}_k(Q,P^{\mathrm{id}},\sigma)\right|.$
\end{proof}

Notice that Lemma~\ref{indexedTrio} gives a direct interpretation of the coefficients obtained on $k$-abstract canonical databases. In particular, since the natural order of $\mathrm{Trio}(X)$ compares the coefficients of square-free monomials, inequalities between query evaluations on such databases can equivalently be expressed as cardinality inequalities between the corresponding sets of indexed homomorphisms. The proof is similar to the $\mathbb{N}[X]$ case so we give it in Appendix~\ref{app:proofs}.

\begin{restatable}{Thm}{triocqneqcontainment}\label{triocqneqcontainment}
Let $P$ and $Q$ be CQ$^\neq$s with the same tuple of free variables, and let $k=\max\{|\text{RelOcc}(P)|,|\text{RelOcc}(Q)|,1\}.$ Then, the following are equivalent:
\begin{itemize}
\item[1.] $P\sqsubseteq_{\mathrm{Trio}(X)} Q$.

\item[2.] For every identification $P^{\mathrm{id}}$ of $P$, we have
$$
\llbracket P\rrbracket^{\text{can}^{k}_{\mathrm{Trio}(X)}(P^{\mathrm{id}})}
\leq_{\mathrm{Trio}(X)}
\llbracket Q\rrbracket^{\text{can}^{k}_{\mathrm{Trio}(X)}(P^{\mathrm{id}})}.
$$

\item[3.] For every identification $P^{\mathrm{id}}$ of $P$ and every $k$-tag signature $\sigma\colon\text{RelSet}(P^{\mathrm{id}})\longrightarrow\mathcal{P}([k]),
$ we have $\left|\text{IHom}_k(P,P^{\mathrm{id}},\sigma)\right|
\leq \left|\text{IHom}_k(Q,P^{\mathrm{id}},\sigma)\right|.
$
\end{itemize}

Moreover, the corresponding decision problem (i.e.\ $P\sqsubseteq_{\mathrm{Trio}(X)}Q$?) is in $\text{coNP}^{\text{PP}}$.
\end{restatable}

\subsection{ Equivalence of CQ\texorpdfstring{$^\neq$}{≠}s}
We now consider the complexity of equivalence of CQ$^\neq$s. The upper bounds follow from our previous containment results (since the problem of equivalence is, by definition, the same as two containment problems). We also establish matching lower bounds for all the semirings considered above, except for $\mathbb{N}[X]$ and $\mathrm{Trio}(X)$. For bounded distributive lattices, we use the following simple fact.

\begin{restatable}{Pro}{equivlatdis} \label{equivlatdis}
For CQ$^\neq$s $P$ and $Q$ with the same tuple of free variables and a bounded distributive lattice $\mathbb{L}$, the following holds:
\begin{itemize}
\item[(1)] $P \equiv_{\mathbb{L}} Q$ iff $P \sqsubseteq_{\mathbb{L}} Q$ and $Q \sqsubseteq_{\mathbb{L}} P$,

\item[(2)] $P \sqsubseteq_{\mathbb{L}} Q$ iff $P \equiv_{\mathbb{L}} P \wedge Q$.

\end{itemize}
\end{restatable}

\begin{proof}
(1) is immediate by definition (in fact, it holds for any semiring, not just lattices). For (2) assume that $P \equiv_{\mathbb{L}} P \wedge Q$, so for each $\mathbb{L}$-instance $\mathbf I$, and tuple $\vec{t}$, we have that $\llbracket P(\vec{t})\rrbracket^\mathbf{I} = \llbracket P(\vec{t})\rrbracket^\mathbf{I}\cdot \llbracket Q(\vec{t})\rrbracket^\mathbf{I}$, which in a lattice means that $\llbracket P(\vec{t})\rrbracket^\mathbf{I} \leq_\mathbb{L}\llbracket Q(\vec{t})\rrbracket^\mathbf{I}$. Thus, $P \sqsubseteq_{\mathbb{L}} Q$  as desired. The converse is entirely symmetric.
\end{proof}

We next establish a lower bound for equivalence over commutative idempotent semirings.
\begin{Thm}\label{cqneqequivalHard}
Let $\mathbb{S}$ be a commutative idempotent semiring. The equivalence problem for $\mathrm{CQ}^{\neq}$s over $\mathbb{S}$ is $\Pi_2^p$-hard.
\end{Thm} 
\ali{The proof can be placed in the appendix}
\begin{proof}
We give a polynomial-time reduction from the $3$-choosability problem for bipartite graphs~\cite[Lemma 6.2]{GutnerTarsiChoosability}. A \emph{list assignment} for a graph $G=(V,E)$ associates with each vertex $i\in V$ a set $L(i)$ of exactly three distinct colors. The lists associated with different vertices may have colors in common and don't need to be equal. A \emph{proper coloring} from these lists is a mapping $c$ that assigns to each vertex $i$ a color $c(i)\in L(i)$, with $c(i)\neq c(j)$ whenever $\{i,j\}\in E$. The graph is $3$\emph{-choosable} if every such list assignment admits a proper coloring. In this decision problem, the input is the graph: the property must hold for every choice of the lists (i.e. we ask whether, for every assignment of a three-element list of colors to each vertex, there exists a proper coloring that maps each vertex to a color from its list). Since only finitely many colors occur in the lists, we can represent each color by a distinct element of the domain $D$.

Given a bipartite graph $G=(\{1,\ldots,n\},E)$, we construct two Boolean $\mathrm{CQ}^{\neq}$s $P_G$ and $Q_G$. We may assume that $n\geq 1$, since the case of a graph with no vertices is trivial. We use a schema consisting of three relation symbols: a unary relation symbol $R$ and two binary relation symbols $N$ and $C$. We write $C$ for \emph{choice}, since the relation associates a list identifier with a color that can be chosen from that list. For every vertex $i$, introduce variables $s_i,x_{i,1},x_{i,2},x_{i,3}$ (all these variable names are distinct). The variable $s_i$ is used as an identifier for the list associated with the vertex $i$, while $x_{i,1},x_{i,2},x_{i,3}$ are used for its three colors. The atoms $$C(s_i,x_{i,1})\land C(s_i,x_{i,2})\land C(s_i,x_{i,3})$$ describe the three colors chosen for vertex $i$ from the list identified by $s_i$. We require these three colors to be distinct by adding $$D_i:=(x_{i,1}\neq x_{i,2})\land (x_{i,1}\neq x_{i,3})\land (x_{i,2}\neq x_{i,3}).$$

The unary symbol $R$ stands for \emph{root} and is used to specify the first identifier: in the instance constructed below, its support contains only the identifier associated with vertex $1$. The binary symbol $N$ stands for \emph{next}, and the atoms $N(s_i,s_{i+1})$ describe a directed path connecting the list identifiers in the order $1,\ldots,n$. The edges of $G$ will instead be expressed by disequations between colors. All three symbols are relation symbols, interpreted by $\mathbb{S}$-relations.

Let $$A_n:=R(s_1)\land \bigwedge_{i=1}^{n-1}N(s_i,s_{i+1}) \land\bigwedge_{i=1}^{n}\bigwedge_{j=1}^{3} C(s_i,x_{i,j}).$$ 
$A_n$ describes the lists through their identifiers: $R(s_1)$ specifies the first identifier, the atoms $N(s_i,s_{i+1})$ connect consecutive identifiers, and the atoms $C(s_i,x_{i,j})$ associate each identifier $s_i$ with the three colors $x_{i,1},x_{i,2},x_{i,3}$. The disequations $D_i$ will ensure that these three colors are distinct.

Let $\vec{s}=(s_1,\ldots,s_n)$ and let $\vec{x}$ contain all the variables $x_{i,j}$, we define
$$
P_G:=\exists\vec{s}\,\exists\vec{x}
\left(A_n\land\bigwedge_{i=1}^{n}D_i\right)
$$
and
$$
Q_G:=\exists\vec{s}\,\exists\vec{x}
\left(A_n\land\bigwedge_{i=1}^{n}D_i
\land\bigwedge_{\{i,j\}\in E}(x_{i,1}\neq x_{j,1})\right).
$$
Thus, $P_G$ selects three distinct colors for each vertex. In $Q_G$, the variable $x_{i,1}$ represents the color chosen for vertex $i$, and the additional disequations require adjacent vertices to receive different colors. In particular, $P_G$ does not require the selected lists to admit a proper coloring of $G$, while $Q_G$ requires such a coloring, represented by the values assigned to the variables $x_{i,1}$. The other two colors remain in the query, so that both queries use the same relational atoms with the same multiplicities. Every variable occurs in a relational atom, and hence both queries are safe. We show that $G$ is $3$-choosable if and only if $P_G\equiv_{\mathbb{S}}Q_G$.

($\Rightarrow$) Assume first that $G$ is $3$-choosable, and consider an arbitrary $\mathbb{S}$-instance $\mathbf{I}$. Take an assignment $v$ under which the body of $P_G$ has a non-zero value. All the disequations in $P_G$ must take value $1$ under $v$, and hence, for every $i$, the set $L_v(i):=\{v(x_{i,1}),v(x_{i,2}),v(x_{i,3})\}$ has exactly three distinct elements. These sets $L_v(i)$ form a list assignment for $G$. By the hypothesis of $3$-choosability, there is a proper coloring $c$ from these lists.

We now define another assignment $w$ by leaving the values of all the $s_i$ unchanged and reordering the three colors at each vertex so that the color chosen by the coloring $c$ appears first, as $x_{i,1}$. More precisely, let $w(s_i)=v(s_i)$ and $w(x_{i,1})=c(i)$, and assign the other two elements of $L_v(i)$ to $x_{i,2}$ and $x_{i,3}$ in either order. Every $D_i$ is still satisfied, since the same three distinct colors are used. Also, $w(x_{i,1})\neq w(x_{j,1})$ for every $\{i,j\}\in E$, because $c$ is a proper coloring. Thus, all the disequations in $Q_G$ take value $1$ under $w$.

For every $i$, changing the order of the three colors means multiplying the same three annotations in different order: by commutativity of multiplication,
$$ \prod_{j=1}^{3} C^{\mathbf{I}}(w(s_i),w(x_{i,j})) = \prod_{j=1}^{3} C^{\mathbf{I}}(v(s_i),v(x_{i,j})).$$
The values of the $R$ and $N$ atoms are unchanged, since $v$ and $w$ agree on all the variables $s_i$. Therefore, the value of the body of $Q_G$ under $w$ is exactly the value of the body of $P_G$ under $v$. This also works when different vertices use the same list identifier: we still multiply the same annotations the same number of times, possibly in a different order.

Conversely, any assignment $v$ under which the body of $Q_G$ has a non-zero value gives the same value to the body of $P_G$, since $Q_G$ only adds disequations to $P_G$.

Notice that the value of the body under a single assignment is an element of $\mathbb{S}$, obtained by multiplying the annotations of the relational atoms and the values of the disequations. When the annotations are polynomials, this value may itself be a polynomial with several monomials. The construction above preserves this entire value, since it preserves all the annotations being multiplied, with the same multiplicities.

We have shown that, whenever an assignment gives the body of $P_G$ a non-zero value $a \in \mathbb{S}$, there is an assignment giving the body of $Q_G$ exactly the same value $a$. Conversely, every non-zero value obtained from the body of $Q_G$ is also obtained from the body of $P_G$. Therefore, the same non-zero values occur when we evaluate the bodies of the two queries over all assignments.

To obtain the value of each query, we sum the values of its body over all assignments. Only finitely many assignments give a non-zero value, since the queries are safe and $\mathbf{I}$ has finite support. There may be several assignments giving the same value $a$, and their number may differ between the two queries, however, since addition in $\mathbb{S}$ is idempotent by hypothesis, we have $a+a=a$ for every $a\in\mathbb{S}$. This also applies when $a$ is an entire polynomial. 

Consequently, each query evaluation is equal to the sum in which every distinct non-zero value obtained from its body is included only once. Since these values are the same for both queries, we conclude that
$$\llbracket P_G\rrbracket^{\mathbf{I}}=\llbracket Q_G\rrbracket^{\mathbf{I}}.$$ Since $\mathbf{I}$ was arbitrary, we obtain $P_G\equiv_{\mathbb{S}}Q_G$.

($\Leftarrow$) We prove the contrapositive. Assume that $G$ is not $3$-choosable, we construct an instance on which $P_G$ and $Q_G$ have different values.

Choose lists $L(1),\ldots,L(n)$, each containing exactly three distinct colors, for which no proper coloring exists. We represent these lists in an $\mathbb{S}$-instance $\mathbf{I}$. Choose pairwise distinct elements $d_1,\ldots,d_n$ of $D$ outside $\bigcup_{i=1}^{n}L(i)$, and use $d_i$ as the identifier of the list $L(i)$.

The support of $R^{\mathbf{I}}$ is $\{d_1\}$, and the support of $N^{\mathbf{I}}$ consists precisely of the tuples $(d_i,d_{i+1})$ for $1\leq i<n$. The support of $C^{\mathbf{I}}$ consists precisely of the pairs $(d_i,a)$ with $a\in L(i)$, so that $C$ records the three colors belonging to each list. All these tuples are annotated by $1$, and every other tuple is annotated by $0$. In particular,
\[
C^{\mathbf{I}}(d_i,a)=
\begin{cases}
1 & \text{if }a\in L(i),\\
0 & \text{otherwise}.
\end{cases}
\]

The purpose of $R$ and $N$ is to ensure that each vertex $i$ uses its prescribed list $L(i)$. To see this, consider an assignment under which the body of $P_G$ has a non-zero value. Every relational atom must then have a non-zero value. The atom $R(s_1)$ forces $s_1$ to be assigned $d_1$, since this is the only element in the support of $R^{\mathbf{I}}$. If $s_i$ is assigned $d_i$, then $N(s_i,s_{i+1})$ forces $s_{i+1}$ to be assigned $d_{i+1}$, since $(d_i,d_{i+1})$ is the only tuple in the support of $N^{\mathbf{I}}$ whose first coordinate is $d_i$. Thus, every $s_i$ must be assigned $d_i$. The atoms $C(s_i,x_{i,j})$ then require the three values assigned to $x_{i,1},x_{i,2},x_{i,3}$ to belong to $L(i)$, and $D_i$ requires them to be distinct. Hence, these three values must be exactly the three elements of $L(i)$, in some order.

For $P_G$, we can therefore set $s_i=d_i$ and assign the three elements of each $L(i)$ to $x_{i,1},x_{i,2},x_{i,3}$ in any order. Every relational atom and every disequation then has value $1$, so the body of $P_G$ has value $1$ under this assignment. This is possible even though the lists admit no proper coloring, since $P_G$ imposes no disequations between colors associated with different vertices. Moreover, every assignment gives the body value $0$ or $1$, since all tuple annotations and all disequation values belong to $\{0,1\}$. Thus, the sum defining the value of $P_G$ contains at least one occurrence of $1$, and all its non-zero terms are $1$. By idempotence of addition, we obtain
\[
\llbracket P_G\rrbracket^{\mathbf{I}}=1.
\]

On the other hand, suppose that the body of $Q_G$ had a non-zero value under an assignment $w$. Since $Q_G$ contains the same $R$ and $N$ atoms as $P_G$, the same argument gives $w(s_i)=d_i$ for every $i$. The additional disequations in $Q_G$ also require $w(x_{i,1})\neq w(x_{j,1})$ whenever $\{i,j\}\in E$. Therefore, assigning the color $w(x_{i,1})$ to each vertex $i$ would give a proper coloring from the chosen lists. This contradicts the choice of $L(1),\ldots,L(n)$. Hence, the body of $Q_G$ has value $0$ under every assignment, and
\[
\llbracket Q_G\rrbracket^{\mathbf{I}}=0.
\]
Consequently, $P_G\not\equiv_{\mathbb{S}}Q_G$.

We have shown that $G$ is $3$-choosable if and only if $P_G\equiv_{\mathbb{S}}Q_G$. The queries have $4n$ variables and $4n$ relational atom occurrences, while $P_G$ has $3n$ disequations and $Q_G$ has $3n+|E|$. Thus, the construction is polynomial in the size of $G$. Since $3$-choosability for bipartite graphs is $\Pi_2^p$-complete~\cite[Lemma~6.2]{GutnerTarsiChoosability}, the equivalence problem for $\mathrm{CQ}^{\neq}$s over $\mathbb{S}$ is $\Pi_2^p$-hard.
\end{proof}

 Now we are ready to state the following result.

\begin{Thm} \label{cqneqequivalence}
For CQ$^\neq$s $P$ and $Q$ with the same tuple of free variables, the following statements hold.
\begin{enumerate}
\item ``$P \equiv_{\mathbb{L}} Q$?" (where $\mathbb{L}$ is a bounded distributive lattice, including $\mathbb{B}$ and $\mathrm{PosBool}(X)$) is $\Pi^p_2$-complete. 

\item ``$P \equiv_{\mathbb{N}[X]} Q$?" is in $\text{coNP}^\text{PP}$.

\item ``$P \equiv_{\mathrm{Sorp}(X)}Q$?" is $\Pi^p_2$-complete. 

\item ``$P \equiv_{\mathbb{B}[X]} Q$?"  is $\Pi^p_2$-complete. 

\item ``$P \equiv_{\mathrm{Why}(X)} Q$?" is  $\Pi^p_2$-complete. 

\item ``$P \equiv_{\mathrm{Trio}(X)} Q$?" is in $\text{coNP}^\text{PP}$.

\item ``$P \equiv_{\mathrm{Which}(X)} Q$?" is $\Pi^p_2$-complete.

\end{enumerate}
\end{Thm}

\begin{proof}
Notice that  item 1 of Proposition~\ref{equivlatdis} holds for all semirings, not just bounded distributive lattices. Thus, the upper bounds follows from Theorem~\ref{cqneqconatinmentsummary}. For bounded distributive lattices, the matching lower bound follows from item 2 of Proposition~\ref{equivlatdis}. For the semirings in items 3, 4, 5, and 7, which are commutative and have idempotent addition, Theorem~\ref{cqneqequivalHard} gives a matching $\Pi_2^p$ lower bound. Hence, equivalence is $\Pi_2^p$-complete in these cases as well.

\end{proof}

\section{Unions of conjunctive queries with disequations (UCQ\texorpdfstring{$^\neq$}{≠}s)}

In this section, we focus on the decidability of the containment and equivalence problems for UCQ$^\neq$s over the same  provenance semirings discussed in \S~\ref{CQneqsection}. We provide exact complexity calculations for both problems over bounded distributive lattices, $\mathbb{B}[X]$, $\mathrm{Why}(X)$, $\mathrm{Sorp}(X)$ and $\mathrm{Which}(X)$, while for $\mathbb{N}[X]$ and $\mathrm{Trio}(X)$ we provide upper bounds only. 

\subsection{Containment of UCQ\texorpdfstring{$^\neq$}{≠}s} \label{UCQneqsection1}
Once more, we provide  a result that summarizes our work on containment for UCQ$^\neq$s. The majority of the supporting theorems will be relegated in their entirety to Appendix~\ref{app:proofs2}.

\begin{Thm} \label{ucqneqconatinmentsummary}
For UCQ$^\neq$s $\overline{P}$ and $\overline{Q}$ with the same tuple of free variables, the following statements hold.
\begin{enumerate}
\item ``$\overline{P} \sqsubseteq_{\mathbb{L}} \overline{Q}$?" (where $\mathbb{L}$ is a bounded distributive lattice, including $\mathbb{B}$ and $\mathrm{PosBool}(X)$) is $\Pi^p_2$-complete (Theorem \ref{ucqdiseqconta}). 

\item ``$\overline{P} \sqsubseteq_{\mathbb{N}[X]} \overline{Q}$?" is in $\text{coNP}^\text{PP}$ (Theorem \ref{nxucqdiscont}). 

\item ``$\overline{P} \sqsubseteq_{\mathrm{Sorp}(X)}\overline{Q}$?" is $\Pi^p_2$-complete (Theorem \ref{usorphard}). 

\item ``$\overline{P} \sqsubseteq_{\mathbb{B}[X]} \overline{Q}$?" is $\Pi^p_2$-complete (Theorem \ref{ucqbxcont}). 

\item ``$\overline{P} \sqsubseteq_{\mathrm{Why}(X)} \overline{Q}$?" is $\Pi^p_2$-complete (Theorem \ref{ucqcontwhy}).  

\item ``$\overline{P} \sqsubseteq_{\mathrm{Trio}(X)} \overline{Q}$?" is in $\text{coNP}^\text{PP}$ (Theorem \ref{ucqdistrio}).

\item ``$\overline{P} \sqsubseteq_{\mathrm{Which}(X)} \overline{Q}$?" is $\Pi^p_2$-complete (Theorem \ref{ulinhard}).

\end{enumerate}

\end{Thm}

To give the reader a taste for the form our characterizations take in the case of $\mathrm{UCQ}^\neq$s,  we leave here the statement of the result for $\mathbb{B}[X]$.  We include a proof in Appendix~\ref{app:proofs2}.

\begin{restatable}{Thm}{ucqbxcont}\label{ucqbxcont} For UCQ$^\neq$s, $\bigcup_{1\leq i \leq n} P_i, \bigcup_{1\leq i \leq m} Q_i$ with the same tuple of free variables,  the following are equivalent:
\begin{itemize}
\item[1.] $\bigcup_{1\leq i \leq n} P_i  \sqsubseteq_{\mathbb{B}[X]} \bigcup_{1\leq i \leq m} Q_i$.
\item[2.]   For each $1\leq i \leq n$,   every identification $P_i^{id}$ of $P_i$, we have that  $\llbracket P_i^{id} \rrbracket^{\mathrm{can}_{\mathbb{B}[X]}(P_i^{id})} \leq \llbracket \bigcup_{1\leq i \leq m} Q_i \rrbracket^{\mathrm{can}_{\mathbb{B}[X]}(P_i^{id})}$.

\item[3.] For each $1\leq i \leq n$,  every identification $P^{\mathrm{id}}_i$ of $P_i$, there is a $j \leq m$ and a  $\neq$-homomorphism 
$h_{P^{\mathrm{id}}_i} \colon Q_j \longrightarrow P^{\mathrm{id}}_i$  exact for relational atoms.

\end{itemize}
Moreover, the corresponding decision problem (i.e.\ $\bigcup_{1\leq i \leq n} P_i \sqsubseteq_{\mathbb{B}[X]} \bigcup_{1\leq i \leq m} Q_i$?) is  $\Pi^p_2$-complete.
\end{restatable}

Just as in~\cite{Green}, the cases of  $\mathbb{N}[X]$ and $\mathrm{Trio}[X]$, deserve their own treatment, and we cannot directly apply the same methods as before. We will include the discussion on the   $\mathbb{N}[X]$ case  below and leave the rest to Appendix~\ref{app:proofs2}.

\begin{Rmk}
The following example shows that homomorphisms are not enough for the case of UCQs on $\mathbb{N}[X]$. Consider $P(x) := R(x)$ and $Q(x) := R(x) \vee R(x). $ For every query in the UCQ $Q$, there is a mapping from $P$ to the query, but $Q$ is not contained in $P$, since $p+p \not\leq p.$ \end{Rmk}

Nevertheless, we can still show that the problem has a $\text{coNP}^{\text{PP}}$ upper bound.
 The idea is to follow the same argument as~\cite{Green} for containment of regular UCQs and establish a `small counterexample' property. In particular, we show that if $\overline{P} \not \sqsubseteq_{\mathbb{N}[X]} \overline{Q}$, then $\llbracket \overline{P} \rrbracket^\mathbf{I} \not \leq_{\mathbb{N}[X]} \llbracket \overline{Q} \rrbracket^\mathbf{I}$ for some $\mathbb{N}[X]$-instance $\mathbf I$ whose size is bounded by the size of $\overline{P}$ (recall that by Def.~\ref{UCQdef} we use $\overline{P}, \overline{Q}$ to abbreviate unions of queries). We begin by bounding the sizes of the annotations of the required counterexample, by observing that if a counterexample exists, then its abstractly-tagged version is also a counterexample.

With a function $h\colon K \longrightarrow K'$, we can transform a $K$-relation $R$ into a $K'$-relation $h(R)$ by applying $h$ to each tuple annotation in $R$. Performing this transformation component-wise on the $K$-relations of a $K$-instance $\mathbf I$ transforms it into a $K'$-instance $h(I)$. We include a proof in Appendix~\ref{app:proofs2} (essentially like~\cite[Prop. 6.1]{Green}).

\begin{restatable}{Pro}{propsixonebis}\label{prop6.1bis}
Let $h\colon K_1 \longrightarrow K_2$, where $K_1$ and $K_2$ are commutative semirings. If $h$ is a semiring homomorphism, then $\llbracket \overline{Q} \rrbracket^{h(I)} = h(\llbracket \overline{Q} \rrbracket^\mathbf{I})$ for all $\overline{Q}$ a UCQ$^\neq$ and $K_1$-instances $\mathbf I$.
\end{restatable}



The following is essentially~\cite[Prop. 3.3]{Green}.  The simple proof is in Appendix~\ref{app:proofs2}. 

\begin{restatable}{Pro}{propthrethreebis}\label{prop3.3bis}
Let $K_1, K_2$ be naturally-ordered commutative semirings. If $h\colon K_1 \longrightarrow K_2$ is a semiring homomorphism, then for all $a,b \in K_1$
$$
a \leq_{K_1} b \Longrightarrow h(a) \leq_{K_2} h(b).
$$
\end{restatable}


The next useful lemma  follows~\cite[Lem. 7.15]{Green}.  The proof is in Appendix~\ref{app:proofs2}. 

\begin{restatable}{Lem}{lemmasevenfifteenbis}\label{lemma7.15bis}
If $\overline{P}, \overline{Q}$ are UCQ$^\neq$s and $\llbracket \overline{P} \rrbracket^\mathbf{I} \not\leq_{\mathbb{N}[X]} \llbracket \overline{Q} \rrbracket^\mathbf{I}$ for some $\mathbb{N}[X]$-instance $\mathbf I$, then $\llbracket \overline{P} \rrbracket^J \not\leq_{\mathbb{N}[X]} \llbracket \overline{Q} \rrbracket^J,$ where $J=\text{ab}_{\mathbb{N}[X]}(I)$.
\end{restatable}



Subsequently, we show that the number of tuples annotated with non-zero semiring values in a counterexample can also be bounded. This is very much as the small model theorem in~\cite[Thm. 7.16]{Green}. The proof is in Appendix~\ref{app:proofs2}.

\begin{restatable}{Thm}{thmsevensixteenbis}\label{thm7.16bis}
Let $\overline{P},\overline{Q}$ be UCQ$^\neq$s, and let $k$ be the maximum between $1$ and the maximum number of relational atom occurrences in the body of a CQ$^\neq$ in $\overline{P}$.\footnote{The intuition here is that this latter maximum is the degree of the multi-variable polynomial corresponding to the annotation of $\overline{P}$ in an abstractly-tagged database.} Then $\overline{P} \not\sqsubseteq_{\mathbb{N}[X]} \overline{Q}$ if and only if $\llbracket \overline{P} \rrbracket^\mathbf{I}
\not\leq_{\mathbb{N}[X]}
\llbracket \overline{Q} \rrbracket^\mathbf{I}$ for some abstractly-tagged $\mathbb{N}[X]$-instance $\mathbf I$ containing at most $k$ tuples annotated with non-zero values.
\end{restatable}

Now we are finally able to establish the result we are after for $\mathbb{N}[X]$.

\begin{Thm}\label{nxpspace} \label{nxucqdiscont}
$\mathbb{N}[X]$-containment of UCQ$^\neq$s (i.e.\ $\bigcup_{1\leq i \leq n} P_i \sqsubseteq_{\mathbb{N}[X]} \bigcup_{1\leq i \leq m} Q_i$?) is in $\text{coNP}^{\text{PP}}$.
\end{Thm}

\begin{proof} For the upper bound complexity calculation, we use the same counting argument as for CQ$^\neq$s over $\mathbb{N}[X]$. We look at the complement of the containment problem and show that it is in $\text{NP}^{\text{PP}}$.

Write $\overline P=\bigcup_{1\leq i\leq n}P_i$ and $\overline Q=\bigcup_{1\leq j\leq m}Q_j$. Suppose first that $\overline P\not\sqsubseteq_{\mathbb{N}[X]}\overline Q.$ By Lemma~\ref{lemma7.15bis}, there is an abstractly-tagged $\mathbb{N}[X]$-instance $\mathbf J$ and an output tuple $\vec t$ such that $\llbracket\overline P\rrbracket^\mathbf J(\vec t)
\not\leq_{\mathbb{N}[X]}
\llbracket\overline Q\rrbracket^\mathbf J(\vec t).$ Hence, there is a monomial $\mu$ which has a greater coefficient in $\llbracket\overline P\rrbracket^\mathbf J(\vec t)$ than in $\llbracket\overline Q\rrbracket^\mathbf J(\vec t)$. Since its coefficient on the left-hand side is positive, $\mu$ is justified by an assignment $v$ to the variables of some disjunct $P_i$ of $\overline P$.

Let $P^c_{i,v}$ be the completion of $P_i$ which, for every pair of variables $x,y$, contains $x=y$ if $v(x)=v(y)$ and $x\neq y$ otherwise. Let $\theta$ be a canonical substitution for this completion and let $P^{\mathrm{id}}_{i,v}$ be the resulting identification. Define $ v_*(\theta(x))=v(x)$. As in the proof for CQ$^\neq$s over $\mathbb{N}[X]$, $v_*$ is well-defined and injective. Moreover, since our queries are safe, it is a bijection between the variables of $P^{\mathrm{id}}_{i,v}$ and the domain elements occurring in the ground tuples whose tags occur in $\mu$.

We claim that, for every CQ$^\neq$ $R$ with the same  head, the coefficient of $\mu$ in $\llbracket R\rrbracket^\mathbf J(\vec t)$ is exactly $|\mathrm{ExHom}(R,P^{\mathrm{id}}_{i,v})|.$ Indeed, let $w$ be an assignment to the variables of $R$ which produces $\vec t$ and contributes $\mu$. Since $\mathbf J$ is abstractly tagged, every ground tuple used by $w$ has a tag occurring in $\mu$. By safety, every value used by $w$ therefore belongs to the range of $v_*$, and we can define $h_w=v_*^{-1}\circ w.$ The map $h_w$ preserves the ordered head and all equations and disequations. Furthermore, for every ground tuple, its tag occurs in $\mu$ with precisely the number of relational atom occurrences of $R$ mapped by $w$ to that tuple. Since $v$ also contributes $\mu$, this is precisely the multiplicity of the corresponding relational atom in $P^{\mathrm{id}}_{i,v}$. Hence $h_w$ is exact for relational atoms, and therefore $h_w\in\mathrm{ExHom}(R,P^{\mathrm{id}}_{i,v}).$ Conversely, if $h\in\mathrm{ExHom}(R,P^{\mathrm{id}}_{i,v})$, then $v_*\circ h$ is an assignment to the variables of $R$ which produces $\vec t$, satisfies all its equations and disequations, and, by exactness for relational atoms,  justifies exactly $\mu$. Thus assignments justifying $\mu$ and mappings in $\mathrm{ExHom}(R,P^{\mathrm{id}}_{i,v})$ are in bijection. Since union adds the contributions of its disjuncts, it follows that the coefficient of $\mu$ in $\llbracket\overline P\rrbracket^\mathbf J(\vec t)$ is $\sum_{1\leq \ell\leq n}
|\mathrm{ExHom}(P_\ell,P^{\mathrm{id}}_{i,v})|,$ while its coefficient in $\llbracket\overline Q\rrbracket^\mathbf J(\vec t)$ is $\sum_{1\leq j\leq m}
|\mathrm{ExHom}(Q_j,P^{\mathrm{id}}_{i,v})|.$ Therefore, $\sum_{1\leq \ell\leq n}
|\mathrm{ExHom}(P_\ell,P^{\mathrm{id}}_{i,v})| > \sum_{1\leq j\leq m}
|\mathrm{ExHom}(Q_j,P^{\mathrm{id}}_{i,v})|.$

Conversely, suppose that for some disjunct $P_i$ of $\overline P$ and some identification $P_i^{\mathrm{id}}$ of $P_i$ we have\[\sum_{1\leq \ell\leq n} |\mathrm{ExHom}(P_\ell,P_i^{\mathrm{id}})| > \sum_{1\leq j\leq m}
|\mathrm{ExHom}(Q_j,P_i^{\mathrm{id}})|.\] Consider the canonical $\mathbb{N}[X]$-database $\mathrm{can}_{\mathbb{N}[X]}(P_i^{\mathrm{id}})$ and the monomial obtained by multiplying the canonical tags corresponding to the relational atom occurrences of $P_i^{\mathrm{id}}$, with repetitions. By the same correspondence between assignments producing this monomial and exact $\neq$-homomorphisms used above, its coefficient in the evaluation of $\overline P$ is the left-hand side of the inequality, while its coefficient in the evaluation of $\overline Q$ is the right-hand side. Hence the coefficientwise order fails, and therefore $\overline P\not\sqsubseteq_{\mathbb{N}[X]}\overline Q.$ 

For a fixed identification $P_i^{\mathrm{id}}$, computing the quantity $\sum_{1\leq \ell\leq n}
|\mathrm{ExHom}(P_\ell,P_i^{\mathrm{id}})|$ is a $\#\text{P}$ function. Indeed, a nondeterministic polynomial-time machine guesses a disjunct $P_\ell$ and a mapping from its variables to the variables of $P_i^{\mathrm{id}}$, and accepts exactly when the mapping is a $\neq$-homomorphism exact for relational atoms. Thus the accepting computations count exactly all the mappings occurring in the sum above. By the same argument, $\sum_{1\leq j\leq m}
|\mathrm{ExHom}(Q_j,P_i^{\mathrm{id}})|$ is also a $\#\text{P}$ function. Consequently, the difference between these two quantities is in $\text{GapP}$ by~\cite[Prop. 3.5]{Fenner}, and deciding whether this difference is positive is in $\text{PP}$ by~\cite[Prop. 4.2]{Fenner}. A nondeterministic polynomial-time machine can therefore guess a disjunct $P_i$ and one of its identifications $P_i^{\mathrm{id}}$, and use a $\text{PP}$ oracle to check whether the above strict inequality holds. Hence non-containment is in $\text{NP}^{\text{PP}}$, and therefore $\mathbb{N}[X]$-containment of UCQ$^\neq$s is in $\text{coNP}^{\text{PP}}$.

\end{proof}

Using a similar method as for $\mathbb{N}[X]$, along the lines of~\cite{Green}, we can establish the same upper bound for $\mathrm{Trio}(X)$. The proof is in Appendix~\ref{app:proofs2}.

\subsection{Equivalence of UCQ\texorpdfstring{$^\neq$}{≠}s}

The following result (essentially from~\cite{Green}, although stated only for regular UCQs there) is key in being able to use the previous theorems to understand the complexity of equivalence between UCQ$^\neq$s:

\begin{Pro}[essentially Prop. 7.2 from~\cite{Green}] \label{propogreen}For UCQ$^\neq$s $\bar{P}, \bar{Q}$ and idempotent (in the sense that the operation $+$ is idempotent) semiring $\mathbb{S}$ (which includes $\mathbb{B}[X]$, $\mathrm{Why}(X)$, $\mathrm{Sorp}(X)$, bounded distributive lattices and $\mathrm{Which}(X)$, but excludes $\mathbb{N}[X]$ and $\mathrm{Trio}(X)$)  we have
\begin{itemize}
\item[(1)] $\bar{P}\sqsubseteq_\mathbb{S}\bar{Q}$ iff $\bar{P} \cup \bar{Q} \equiv_\mathbb{S}\bar{Q}$.

\item[(2)] $\bar{P}\equiv_\mathbb{S}\bar{Q}$ iff $\bar{P}\sqsubseteq_\mathbb{S}\bar{Q}$ and $\bar{Q}\sqsubseteq_\mathbb{S}\bar{P}$.

\end{itemize}
\end{Pro}

Combining Proposition~\ref{propogreen} with our characterizations of containment across different semirings leads directly to the following complexity results for query equivalence:

\begin{Thm}    \label{ucqdisequivalence}
 For UCQ$^\neq$s $\overline{P}$ and $\overline{Q}$ with the same tuple of free variables, the following statements hold.
\begin{enumerate}
\item ``$\overline{P} \equiv_{\mathbb{L}} \overline{Q}$?" (where $\mathbb{L}$ is a bounded distributive lattice, including $\mathbb{B}$ and $\mathrm{PosBool}(X)$) is $\Pi^p_2$-complete. 

\item ``$\overline{P} \equiv_{\mathbb{N}[X]} \overline{Q}$?" is in $\text{coNP}^\text{PP}$.

\item ``$\overline{P} \equiv_{\mathrm{Sorp}(X)}\overline{Q}$?" is $\Pi^p_2$-complete. 
\item ``$\overline{P} \equiv_{\mathbb{B}[X]} \overline{Q}$?" is $\Pi^p_2$-complete. 

\item ``$\overline{P} \equiv_{\mathrm{Why}(X)} \overline{Q}$?" is $\Pi^p_2$-complete. 

\item ``$\overline{P} \equiv_{\mathrm{Trio}(X)} \overline{Q}$?" is in $\text{coNP}^\text{PP}$.

\item ``$\overline{P} \equiv_{\mathrm{Which}(X)} \overline{Q}$?" is $\Pi^p_2$-complete.

\end{enumerate}
\end{Thm}
\begin{proof}
For bounded distributive lattices, $\mathbb{B}[X]$, $\mathrm{Why}(X)$, $\mathrm{Sorp}(X)$ and $\mathrm{Which}(X)$, the upper bounds follow from item 2 of Proposition~\ref{propogreen} and the corresponding containment results, namely Theorems~\ref{ucqdiseqconta}, \ref{ucqbxcont}, \ref{ucqcontwhy}, \ref{usorphard}, and~\ref{ulinhard}. The matching lower bounds follow from item 1 of Proposition~\ref{propogreen}, since containment is $\Pi^p_2$-complete for each of these semirings. For $\mathbb{N}[X]$ and $\mathrm{Trio}(X)$, the upper bounds follow from item 2 of Proposition~\ref{propogreen}, together with Theorems~\ref{nxucqdiscont} and~\ref{ucqdistrio}.
\end{proof}

\section{Conjunctive queries with negated atoms (CQNs)}

In this section, we focus our attention on CQs with negated atoms rather than disequations. Since we are working in the Gr\"adel--Tannen framework, the choice of classes of  interpretations under consideration ends up making an interesting difference for the complexity calculation, as we can see with bounded distributive lattices.

\subsection{Containment of CQNs}\label{The problem of containmentneq}

As before, we can summarize the results of this section into a single theorem (the proof is immediate from the citations of the theorems used).

\begin{Thm} For CQNs $P$ and $Q$ with the same tuple of free variables, the following statements hold. In the dually model-defining case of item (8)\footnote{For $\mathrm{Which}(X)$, we do not need to assume that $P$ is consistent.  Indeed, under model-defining interpretations, any time we have a query $P$ that contains an atom and its negation, its value is always $0$, so containment holds trivially.}, assume that $P$ contains no atom and its negation.
\begin{enumerate}
\item ``$P \sqsubseteq_{\mathbb{L}}^{\text{md}} Q$?" (where $\mathbb{L}$ is a bounded distributive lattice wit more than two elements) is NP-complete (Theorem~\ref{dsitrlatticeneg}). 
\item ``$P \sqsubseteq_{\mathbb{N}[X, \overline{X}]}^{\text{pt}} Q$?" is NP-complete (Theorem~\ref{N[X, X]}).   
\item ``$P \sqsubseteq_{\mathrm{Sorp}(X, \overline{X})}^{\text{pt}} Q$?" is NP-complete (Theorem~\ref{sorp(X,X)}). 
\item ``$P \sqsubseteq_{\mathbb{B}[X, \overline{X}]}^{\text{pt}} Q$?" is NP-complete (Theorem~\ref{bxcompletept}). 

\item ``$P \sqsubseteq_{\mathrm{Why}(X, \overline{X})}^{\text{pt}} Q$?" is NP-complete (Theorem~\ref{whyneg}). 

\item ``$P \sqsubseteq_{\mathrm{Trio}(X, \overline{X})}^{\text{pt}} Q$?" is NP-complete (Theorem~\ref{triocqncontainment}).

\item ``$P \sqsubseteq_{\mathrm{PosBool}(X, \overline{X})}^{\text{pt}} Q$?" is NP-complete (Theorem~\ref{posboolneg}).

\item   ``$P \sqsubseteq^\text{dmd}_{\mathbb{L}} Q$?"  (Theorem~\ref{dualmdinterdistrlattice}) and ``$P \sqsubseteq_{\mathrm{Which}(X)}^{\text{md}} Q$?" (Theorem~\ref{containmentwhich modeldef}) are
\begin{itemize}
\item[(i)] $\Pi^p_2$-complete for queries over schemas with arity bound $k$, for every fixed $k\geq 2$, and
\item[(ii)] coNEXP-complete in general.  
\end{itemize} 
\end{enumerate}
\end{Thm}

We start with some observations for bounded distributive lattices. Consider the queries in Example~\ref{complneed} and the following model-defining interpretation (we may assume that the domain are the variables of the query $P$ itself from Example~\ref{complneed}) on the four element Boolean lattice ${\bf 4} =\{0, a, b, 1\}$:

\[ \begin{array}{c|ccc} \pi\llbracket R(u,v)\rrbracket & v:=x & v:=y & v:=z \\ \hline u:=x & 1 & 1 & 0 \\ u:=y & 1 & a & 1 \\ u:=z & 1 & 1 & 0 \end{array} \qquad \begin{array}{c|ccc} \pi\llbracket \neg R(u,v)\rrbracket & v:=x & v:=y & v:=z \\ \hline u:=x & 0 & 0 & 1 \\ u:=y & 0 & 0 & 0 \\ u:=z & 0 & 0 & 1 \end{array} \]

We can observe that  

\[ \begin{aligned} \pi\llbracket R(x,x) \wedge R(x,y) \wedge \neg R(y,y)\rrbracket &= 0, \\ \pi \llbracket R(x,x) \wedge R(x,z) \wedge \neg R(z,z)\rrbracket &= 0, \\ \pi \llbracket R(y,y) \wedge R(y,x) \wedge \neg R(x,x)\rrbracket &= 0, \\ \pi \llbracket R(y,y) \wedge R(y,z) \wedge \neg R(z,z)\rrbracket &= a, \\ \pi \llbracket R(z,z) \wedge R(z,x) \wedge \neg R(x,x)\rrbracket &= 0, \\ \pi\llbracket R(z,z) \wedge R(z,y) \wedge \neg R(y,y)\rrbracket &= 0, \end{aligned} \]
and thus $\pi\llbracket Q \rrbracket = a$. By contrast, $\pi \llbracket R(x,x) \wedge  R(x,y) \wedge R(y,z) \wedge \neg R(z,z)\rrbracket = 1$ and thus $\pi\llbracket P \rrbracket = 1$. This means that despite the fact that for every completion $P^c$ of $P$, there is a negation-homomorphism $h_{P^c} \colon Q \longrightarrow P^c$ (applying the result for the Boolean case), we have 
$P \not \sqsubseteq_{\bf 4}^{\text{md}} Q$, and this condition is not sufficient for containment in that case. It turns out that for bounded distributive lattices different from the Boolean semiring and model-defining interpretations, containment has a similar characterization as for regular CQs.

\begin{restatable}{Thm}{cqnContainmentA} \label{dsitrlatticeneg}
For CQNs $P$ and $Q$ with the same tuple of free variables, and a bounded distributive lattice $\mathbb{L}$ with an intermediate element $\epsilon \neq \top, \bot$, where $P$ does not contain any atom and its negation, the following are equivalent:
\begin{itemize}
\item[1.] $P \sqsubseteq_{\mathbb{L}}^{\text{md}} Q$,

\item[2.] There is a negation-homomorphism $h \colon Q \longrightarrow P$.

\end{itemize}
Moreover, the corresponding decision problem (i.e.\ $P \sqsubseteq_{\mathbb{L}}^{\text{md}} Q$?) is NP-complete. 
\end{restatable}

The proof can be found in Appendix~\ref{app:cqn-containment-proofs}.

Since the characterization for containment in the Boolean case fails to generalize to model-defining interpretations in the bounded distributive lattice case, it is natural to ask for which interpretations it  actually holds. It turns out that dually model-defining interpretations do the trick. For these interpretations (which on the Boolean semiring coincide with model-defining ones) we immediately get the following result. (The counterexample above is model-defining but not dually model-defining as $\pi\llbracket R(y,y)\rrbracket = \frac{1}{2}$ and $\pi\llbracket \neg R(y,y)\rrbracket = 0$.) The proof of the next result can be found in Appendix~\ref{app:cqn-containment-proofs}.

\begin{restatable}{Thm}{cqnContainmentB} \label{dualmdinterdistrlattice}
For CQNs $P$ and $Q$ with the same tuple of free variables, and a bounded distributive lattice $\mathbb{L}$, where $P$ does not contain any atom and its negation, the following are equivalent:
\begin{itemize}
\item[1.] $P \sqsubseteq_{\mathbb{L}}^{\text{dmd}} Q$,

\item[2.] For every completion $P^c$ of $P$, there is a negation-homomorphism $h_{P^c} \colon Q \longrightarrow P^c$.

\end{itemize}
Moreover, the corresponding decision problem (i.e.\ $P \sqsubseteq^\text{dmd}_{\mathbb{L}} Q$?) is 
\begin{itemize}
\item[(i)] $\Pi^p_2$-complete for queries over schemas with arity bound $k$, for every fixed $k\geq 2$, and
\item[(ii)] coNEXP-complete in general.  
\end{itemize} 
\end{restatable}

The result below is an example of how we deal with containment over provenance-tracking interpretations. When $P$ contains no atom and its negation, we have that $P\sqsubseteq^{\text{mdpt}}_{\mathbb{N}[X,\overline{X}]}Q$ iff $P \sqsubseteq^{\text{pt}}_{\mathbb{N}[X, \overline{X}]} Q$.  To see the left-to-right direction, we can extend the canonical interpretation as in Definition~\ref{CQNscan}, and then apply the semiring homomorphism that sends the added tags to $0$ and fixes the original tags to recover the canonical inequality in item 2 in Theorem~\ref{N[X, X]}. The converse follows because every mdpt interpretation is pt. If $P$ contains an atom and its negation, its value under mdpt interpretations is always $0$, so mdpt containment holds trivially but this does not apply to pt interpretations. These observations are used in Theorem~\ref{equivalcqnsbags}. We relegate the remaining cases of this kind to Appendix~\ref{app:cqn-containment-proofs}.

\begin{restatable}{Thm}{cqnContainmentC}\label{N[X, X]} For CQNs $P$ and $Q$ with the same tuple of free variables, the following are equivalent:
\begin{itemize}
\item[1.] $P \sqsubseteq^{\text{pt}}_{\mathbb{N}[X, \overline{X}]} Q$,
\item[2.]  $ \pi^{\mathrm{can}_{\mathbb{N}[X]}(P^+)}\llbracket P\rrbracket  \leq_{\mathbb{N}[X, \overline{X}]} \pi^{\mathrm{can}_{\mathbb{N}[X]}(P^+)}\llbracket Q \rrbracket$,
\item[3.] There is a  negation-homomorphism $h\colon Q \longrightarrow P$ exact for relational literals.

\end{itemize}
Moreover, the corresponding decision problem (i.e.\ $P \sqsubseteq_{\mathbb{N}[X, \overline{X}]}^{\text{pt}} Q$?) is NP-complete. 
\end{restatable}

\begin{proof}
$(1) \implies (2):$ Obvious.

$(2) \implies (3):$ Assume that (2) holds, so
$$ \pi^{\mathrm{can}_{\mathbb{N}[X]}(P^+)}\llbracket P(\vec{u})\rrbracket  \leq_{\mathbb{N}[X, \overline{X}]} \pi^{\mathrm{can}_{\mathbb{N}[X]}(P^+)}\llbracket Q(\vec{u})\rrbracket,$$
where $\vec{u}$ is the tuple of free variables of $P$. The polynomial $\pi^{\mathrm{can}_{\mathbb{N}[X]}(P^+)}\llbracket P(\vec{u})\rrbracket$ has as a term a monomial $x_1 \cdot \ldots \cdot x_n $ containing every tag from $X \cup \overline{X}$ used to annotate literals by $\pi^{\mathrm{can}_{\mathbb{N}[X]}(P^+)}$,  possibly with repetitions  (such monomial is obtained using the identity assignment for the variables in $P$).  Then, by the definition of $\leq $ for $\mathbb{N}[X, \overline{X}]$, we must have that the monomial $x_1 \cdot \ldots \cdot x_n $ appears in $\pi^{\mathrm{can}_{\mathbb{N}[X]}(P^+)}\llbracket Q(\vec{u})\rrbracket$ with an equal or greater coefficient. Since $Q$ is a CQN, we must have that there is an assignment $v$ to the bound variables in $Q$ such that it produces the monomial $x_1 \cdot \ldots \cdot x_n $ as the value for the body of $Q$.
 Then, we get the desired 
negation-homomorphism $h\colon Q \longrightarrow P$ exact for literals by letting $h(x)$ be the variable $v(x)$ (recall that the elements of the active domain of $\mathrm{can}_{\mathbb{N}[X]}(P^+)$ are the variables of $P$). To show injectivity, notice that any tag $p$ or $\overline{p}$ that is the value of a literal of $Q$ under assignment  $v$, appears in $x_1 \cdot \ldots \cdot x_n $ with at most the same multiplicity as in  the body of $P$ (since this monomial corresponds to the body of $P$), which entails that the multiset of images by $h$ of literals of $Q$ is bag-contained in the multiset of literals from $P$. 
Surjectivity follows from the fact that $x_1 \cdot \ldots \cdot x_n $ corresponds to the multiset of literals from $P$ but also to the images by $h$ of the multiset of literals from $Q$, so the former is bag-contained in the latter.

$(3) \implies (1):$ Assume that there is a  negation-homomorphism $h\colon Q \longrightarrow P$ exact for literals. In order to establish $P \sqsubseteq^{\text{pt}}_{\mathbb{N}[X, \overline{X}]} Q$, consider an arbitrary $\mathbb{N}[X]$-database instance $\mathbf I$, provenance-tracking interpretation $\pi$ compatible with it, and a tuple $\vec{t}$. We need to show that any monomial in $\pi\llbracket P(\vec{t})\rrbracket$ appears with greater or equal coefficient in $\pi\llbracket Q(\vec{t})\rrbracket$. Now consider two assignments $v, v'$ of elements to the variables of $P$ justifying  a monomial $x_1 \cdot \ldots \cdot x_m $.  The assignments $v \circ h$ and $ v' \circ h$ justify the same monomial in $\pi\llbracket Q(\vec{t})\rrbracket$. For example, if $x_i = \pi\llbracket L(v(\vec{z}))\rrbracket$ where  $L(\vec{z})$ is a literal of  $P$, then $x_i =  \pi\llbracket L^*(v(h(\vec{x}))\rrbracket$  for some literal $L^*(\vec{x})$ of $Q$ such that $L^*(h(\vec{x})) = L(\vec{z})$, which exists by surjectivity.  Furthermore, the multiplicity  with which  $x_i$ appears using the assignment $v \circ h $ is exactly the multiplicity with which it appears in  $x_1 \cdot \ldots \cdot x_m $ given that $h$ is both onto and injective on relational atoms. 
Given that $h$ is surjective for literals, it must also be surjective as a mapping on variables, and thus if $v \neq v'$ then $v \circ h \neq v' \circ h$. Hence every justification of a monomial in $\pi\llbracket P(\vec{t})\rrbracket$ corresponds to a unique justification of a monomial in  $\pi\llbracket Q(\vec{t})\rrbracket$. 
We need this condition since coefficients correspond to the number of justifications: without uniqueness, it may happen that a monomial in $\pi\llbracket P(\vec{t})\rrbracket$   has a higher coefficient than the one in $\pi\llbracket Q(\vec{t})\rrbracket$. Since addition in $\mathbb{N}$ is monotone, the required inequality for the term coefficients holds.

The NP upper bound  is given by item 3 of the characterization in the current theorem. The NP lower bound is a result of the NP hardness of the same problem as in item 3  when there are no negations~\cite{Green}.
\end{proof}

Before presenting the containment result for $\mathrm{Which}(X)$, let us explain why we treat this semiring differently from the polynomial semirings. A natural first attempt would be to introduce a set of complementary tokens $\overline{X}$ and define a semiring $\mathrm{Which}(X,\overline{X})$ in which tokens from $X$ annotate positive relational literals, while tokens from $\overline{X}$ annotate negated relational literals. However, this approach does not seem to interact well with the algebraic structure of $\mathrm{Which}(X)$. Indeed, nonzero elements of $\mathrm{Which}(X)$ are flat sets of tokens, and both addition and multiplication are given by union. Hence, once tokens $x$ and $\overline{x}$ are allowed simultaneously, products or sums may create inconsistent sets containing both of them. Removing, collapsing, or quotienting such inconsistent sets in a way that preserves a semiring structure with the intended lineage behaviour appears problematic.\footnote{In particular, the usual construction with dual indeterminates, obtained by quotienting $\mathrm{Which}(X\cup\overline{X})$ by the congruence generated by $x\overline{x}=\bot$, cannot yield a nontrivial semiring. Indeed, since $x\overline{x}=x+\overline{x}$ and $\emptyset+x\overline{x}=x\overline{x}$ in $\mathrm{Which}(X\cup\overline{X})$, imposing $x\overline{x}=\bot$ forces $\emptyset=\emptyset+\bot=\emptyset+x\overline{x}=x\overline{x}=\bot$ in the quotient. Hence, all elements become identified, resulting in the trivial semiring.} For this reason, in the case of $\mathrm{Which}(X)$ we do not use provenance-tracking interpretations but we consider model-defining interpretations directly over $\mathrm{Which}(X)$. This is also faithful to the intended meaning of lineage. If a tuple belongs to the output of a query, then its lineage is the set of input tuples contributing to that output, while if the tuple does not belong to the output, then its lineage is $\bot$. Thus, for a ground atom $R(\vec{a})$, exactly one between $R(\vec{a})$ and $\neg R(\vec{a})$ should have non-bottom lineage in a classical model. This is precisely what model-defining interpretations enforce. \\

Let $P$ be a CQN and let $P^c$ be a negation completion of $P$. We now define a model-defining interpretation associated with $P^c$, denoted by $\pi^{P^c}_{\mathrm{Which}(X)}$. For each literal $L$ appearing in the body of $P$, let $x_L$ be a fresh variable in $X$, and assume that $x_L\neq x_{L'}$ whenever $L\neq L'$. Then, $\pi^{P^c}_{\mathrm{Which}(X)}$ is defined as follows:
\begin{itemize}
\item if $L$ is a literal appearing in the body of $P$, then $\pi^{P^c}_{\mathrm{Which}(X)}\llbracket L \rrbracket=\{x_L\};$
\item if $L$ is a literal appearing in the body of $P^c$ but not in the body of $P$, then $\pi^{P^c}_{\mathrm{Which}(X)}\llbracket L\rrbracket=\emptyset;$
\item if $L$ is a literal such that the complementary literal appears in the body of $P^c$, then $\pi^{P^c}_{\mathrm{Which}(X)}\llbracket L\rrbracket=\bot.$
\end{itemize}

Notice that $\pi^{P^c}_{\mathrm{Which}(X)}$ is model-defining. 
Moreover, since $\emptyset$ is the neutral element for multiplication, the literals we add from $P$ to $P^c$ do not contribute any new token to the value of the query. Hence, for the identity assignment on the variables of $P$, the value $\pi^{P^c}_{\mathrm{Which}(X)} \llbracket P\rrbracket$ is
\[
\prod_{L\in \text{Lit}(P)} \pi^{P^c}_{\mathrm{Which}(X)}\llbracket L\rrbracket
=
\bigcup_{L\in \text{Lit}(P)} \{x_L\}
=
\{x_L \mid \text{L is a literal in P}\}.
\]
The same value is obtained for the body of $P^c$, since every literal in $P^c$ but not in $P$ has value $\emptyset$.

\begin{restatable}{Thm}{cqnContainmentI}\label{containmentwhich modeldef}
For CQNs $P$ and $Q$ with the same tuple of free variables, the following are equivalent:
\begin{itemize}
\item[1.] $P \sqsubseteq_{\mathrm{Which}(X)}^{\text{md}} Q$,

\item[2.] For every completion $P^c$ of $P$,
\[
\pi^{P^c}_{\mathrm{Which}(X)}\llbracket P\rrbracket \leq_{\mathrm{Which}(X)}
\pi^{P^c}_{\mathrm{Which}(X)}\llbracket Q\rrbracket.
\]

\item[3.] For every completion $P^c$ of $P$ and every literal $L$ in the body of $P$, there is a negation-homomorphism
\[
h_{P^c,L} \colon Q \longrightarrow P^c
\]
such that $L$ is in the range of $h_{P^c,L}$.
\end{itemize}
Moreover, the corresponding decision problem (i.e.\ $P \sqsubseteq_{\mathrm{Which}(X)}^{\text{md}} Q$?) is 
\begin{itemize}
\item[(i)] $\Pi^p_2$-complete for queries over schemas with arity bound $k$, for every fixed $k\geq 2$, and
\item[(ii)] coNEXP-complete in general.  
\end{itemize} 
\end{restatable}


\begin{proof}
If $P$ contains an atom and its negation, item (1) holds because its value is always $\bot$, while items (2) and (3) are vacuous. Indeed, by definition, a completion cannot contain an atom and its negation. Thus, in this case, $P$ has no completions, and both conditions are trivially satisfied. Otherwise, let $Q(\vec{x})=\exists\vec{z}\,(A_1(\vec{x_1})\wedge\ldots\wedge A_n(\vec{x_n}))$ and $P(\vec{z})=\exists\vec{u}\,(B_1(\vec{z_1})\wedge\ldots\wedge B_m(\vec{z_m}))$.

$(1) \implies (2):$ trivial.

$(2) \implies (3):$ Assume that (2) holds. We want to show that for every completion $P^c$ of $P$ and every literal $L$ in the body of $P$, there is a negation-homomorphism $h_{P^c,L} \colon Q \longrightarrow P^c$ such that $L$ belongs to the image of $h_{P^c,L}$. Let $P^c$ be an arbitrary completion of $P$, and let $L$ be an arbitrary literal in the body of $P$. By definition of the interpretation $\pi^{P^c}_{\mathrm{Which}(X)}$, we have that $\pi^{P^c}_{\mathrm{Which}(X)}\llbracket L\rrbracket=\{x_L\}$, where $x_L$ is the fresh variable of $X$ associated with the literal $L$. Moreover, considering the identity assignment on the variables of $P$, we obtain that the value of the body of $P$ contains all the variables associated with the literals of $P$: $x_L \in \pi^{P^c}_{\mathrm{Which}(X)}\llbracket P(\vec{z})\rrbracket$ (simply the set-theoretic union). By hypothesis, $\pi^{P^c}_{\mathrm{Which}(X)}\llbracket P(\vec{z})\rrbracket \leq_{\mathrm{Which}(X)} \pi^{P^c}_{\mathrm{Which}(X)}\llbracket Q(\vec{x})\rrbracket$. Since the natural order on Which$(X)$ is the inclusion on $\wp(X)$ with the exception that $\bot \leq A$ for any $A \subseteq \wp(X)$, it follows that $x_L \in \pi^{P^c}_{\mathrm{Which}(X)}\llbracket Q(\vec{x})\rrbracket$. 
Since $\pi^{P^c}_{\mathrm{Which}(X)}\llbracket Q(\vec{x})\rrbracket$ is obtained as the sum, in Which$(X)$, of the values of the body of $Q$ over all assignments sending the free variables $\overline{x}$ to $\vec{z}$, and since this sum is given by the union, there must be an assignment $v$ from the variables of $Q$ to the variables of $P$ (that are the same of $P^c$) such that $v(\vec{x})=\vec{z}$ and $x_L$ belongs to $\pi^{P^c}_{\mathrm{Which}(X)}\llbracket A_1(v(\vec{x_1}))\rrbracket \cdot \ldots \cdot \pi^{P^c}_{\mathrm{Which}(X)}\llbracket A_n(v(\vec{x_n}))\rrbracket$. Multiplication in Which$(X)$ gives $\bot$ whenever one of the factors is $\bot$, so it follows that, for every $1 \leq i \leq n$, we have $\pi^{P^c}_{\mathrm{Which}(X)}\llbracket A_i(v(\vec{x_i}))\rrbracket \neq \bot$. But by construction of $\pi^{P^c}_{\mathrm{Which}(X)}$, a literal has value different from $\bot$ only if it belongs to the body of $P^c$, so, for every $1 \leq i \leq n$, the literal $A_i(v(\vec{x_i}))$ belongs to the body of $P^c$. Moreover, the variable $x_L$ can occur only in the value of the literal $L$, therefore, there is some $1 \leq i \leq n$ such that $A_i(v(\vec{x_i}))=L$. We now define $h_{P^c,L}$ by letting $h_{P^c,L}(x)=v(x)$ for every variable $x$ of $Q$. Since the active domain of $P^c$ consists of the variables of $P$, this is a mapping from the variables of $Q$ to the variables of $P$. Also, since the two queries have the same tuple of free variables and we are evaluating $Q$ on the tuple $\vec{z}$, the mapping sends the free variables of $Q$ to the corresponding free variables of $P$. Also, for every literal $A_i(\vec{x_i})$ of $Q$, we have shown that $A_i(h_{P^c,L}(\vec{x_i}))=A_i(v(\vec{x_i}))$ belongs to the body of $P^c$. Thus, $h_{P^c,L}$ is a negation-homomorphism from $Q$ to $P^c$. Since for some $1 \leq i \leq n$ we have $A_i(h_{P^c,L}(\vec{x_i}))=L$, the literal $L$ belongs to the image of $h_{P^c,L}$, as required.

$(3) \implies (1):$ Assume that for each completion $P^c$ of $P$ and each literal $L$ in the body of $P$, there is a negation-homomorphism $h_{P^c,L} \colon Q \longrightarrow P^c$ such that $L$ belongs to the image of $h_{P^c,L}$. We want to show that $P \sqsubseteq_{\mathrm{Which}(X)}^{\text{md}} Q$. Consider an arbitrary Which$(X)$-database instance $\mathbf I$, a model-defining interpretation $\pi$ compatible with $\mathbf I$, and an arbitrary tuple $\vec{t}$. We need to show that $\pi\llbracket P(\vec{t})\rrbracket \leq_{\mathrm{Which}(X)} \pi\llbracket Q(\vec{t})\rrbracket$. If $\pi\llbracket P(\vec{t})\rrbracket=\bot$, then there is nothing to prove. So assume that $\pi\llbracket P(\vec{t})\rrbracket \neq \bot$. First choose an assignment $v_0$ with a non-bottom contribution to $P(\vec t)$, and complete $P$ by adding, for each atom on its variables, the unique literal having non-bottom value under $v_0$. This gives a completion $P^c_{v_0}$. Choose any literal $L_0$ in the body of $P$. By (3), there is a negation-homomorphism $h_{P^c_{v_0},L_0}\colon Q\longrightarrow P^c_{v_0}$. The assignment $v_0\circ h_{P^c_{v_0},L_0}$ gives a non-bottom contribution to $Q(\vec t)$, so $\pi\llbracket Q(\vec t)\rrbracket\neq\bot$. It remains to prove inclusion between these two subsets of $X$. Let $x \in \pi\llbracket P(\vec{t})\rrbracket$. Since $\pi\llbracket P(\vec{t})\rrbracket$ is obtained as the sum of the values of the body of $P$ over all assignments sending the free variables $\vec{z}$ to $\vec{t}$, there is an assignment $v$ of elements of the domain of $\mathbf I$ to the bound variables of $P$ such that $v(\vec{z})=\vec{t}$ and $x$ belongs to $\pi\llbracket B_1(v(\vec{z_1})) \wedge \ldots \wedge B_m(v(\vec{z_m}))\rrbracket$. Since multiplication in Which$(X)$ gives us $\bot$ any time one of the factors is $\bot$, it follows that $\pi\llbracket B_j(v(\vec{z_j}))\rrbracket \neq \bot$ for every $1 \leq j \leq m$. Moreover, since multiplication between elements different from $\bot$ is the union, there is a literal $L$ in the body of $P$ such that $x \in \pi\llbracket L(v)\rrbracket$. We now build a completion $P^c_v$ of $P$ depending on the assignment $v$ and the interpretation $\pi$: for every relation symbol $R$ of the common schema $\tau$ and every tuple $\vec{w}$ of variables of $P$ with length that match the arity of $R$, we add to $P$ the literal $R(\vec{w})$ if $\pi\llbracket R(v(\vec{w}))\rrbracket\neq \bot$, and we add the literal $\neg R(\vec{w})$ if $\pi\llbracket \neg R(v(\vec{w}))\rrbracket\neq \bot$. Since $\pi$ is model-defining, exactly one of these two values is different from $\bot$, so only one of the two literals is added to $P^c$ (if a literal already belonged to the body of $P$, then we already said that its value under $v$ is different from $\bot$ so all the literals of $P$ are also in $P^c$). $P^c_v$ is a completion of $P$. By hypothesis, there is a negation-homomorphism $h_{P^c_v,L} \colon Q \longrightarrow P^c_v$ such that $L$ belongs to the image of $h_{P^c_v,L}$. So, for every literal $A_i(\vec{x_i})$ of $Q$, the literal $A_i(h_{P^c_v,L}(\vec{x_i}))$ in in the body of $P^c_v$, and for some $1 \leq i \leq n$ we have $A_i(h_{P^c_v,L}(\vec{x_i}))=L$.
Consider now the assignment $v \circ h_{P^c_v,L}$. Since $h_{P^c_v,L}$ is a negation-homomorphism, it sends the free variables of $Q$ to corresponding free variables of $P^c$ (and $P$), and so $v \circ h_{P^c_v,L}$ sends the tuple of free variables $\vec{x}$ to $\vec{t}$. Moreover, for every $1 \leq i \leq n$, the literal $A_i(h_{P^c_v,L}(\vec{x_i}))$ is in the body of $P^c_v$. By construction of $P^c_v$, this means that $\pi\llbracket A_i(v \circ h_{P^c_v,L}(\vec{x_i}))\rrbracket \neq \bot$ for every $1 \leq i \leq n$. Therefore, the value of the conjunction $A_1(v \circ h_{P^c_v,L}(\vec{x_1})) \wedge \ldots \wedge A_n(v \circ h_{P^c_v,L}(\vec{x_n}))$ is different from $\bot$.
Finally, since $L$ is in the image of $h_{P^c_v,L}$, there is some $1 \leq i \leq n$ such that $A_i(h_{P^c_v,L}(\vec{x_i}))=L$. Applying $v$, we have $A_i(v \circ h_{P^c_v,L}(\vec{x_i}))=L(v)$. Since $x \in \pi\llbracket L(v)\rrbracket$, it follows that $x$ belongs to the value of $Q$ under this assignment. Hence $x \in \pi\llbracket Q(\vec{t})\rrbracket$. We have shown that every $x \in \pi\llbracket P(\vec{t})\rrbracket$ also belongs to $\pi\llbracket Q(\vec{t})\rrbracket$. Therefore, $\pi\llbracket P(\vec{t})\rrbracket \subseteq \pi\llbracket Q(\vec{t})\rrbracket$, and thus $\pi\llbracket P(\vec{t})\rrbracket \leq_{\mathrm{Which}(X)} \pi\llbracket Q(\vec{t})\rrbracket$. But $\mathbf I$, $\pi$, and $\vec{t}$ were arbitrary so we can conclude that $P \sqsubseteq_{\mathrm{Which}(X)}^{\text{md}} Q$.

For the complexity calculation, membership in $\Pi^p_2$ or coNEXP follows by item 3 similarly to the Boolean case.  For hardness, we  may assume that the bound variables in two arbitrary CQNs $P$ and $Q$ are disjoint. We define $Q^*$ as the  conjunction of $P$ and $Q$ (still a CQN).
First, observe that 
\begin{itemize}
\item[(1)] for every completion $P'$ of $P$, and every literal $l(\vec{z})$ of $P$  there is a negation-homomorphism $g_{P'} \colon Q^* \longrightarrow P'$ with $l(\vec{z
})$ in the image of $g_{P'}$, 
\end{itemize}
entails that 
\begin{itemize}
\item[(2)] for every completion $P'$ of $P$,  there is a negation-homomorphism  $h_{P'}\colon Q \longrightarrow P' $,
\end{itemize}
as $h_{P'}$ can be obtained as the restriction of $g_{P'}$ to the variables of $Q$.
Second, we also have that (2) entails (1). For any mapping $h_{P'}\colon Q \longrightarrow P' $ can be extended to a mapping $g_{P'} \colon Q^* \longrightarrow P'$  with the property that every literal  $l(\vec{z})$  of $P$ is in the image of $g_{P'}$ by letting the map be the identity on the variables of $P$. Thus, $P  \sqsubseteq_{\mathrm{Which}(X)}^{\text{md}} Q^*$ iff $P  \sqsubseteq_{\mathbb{B}}^{\text{md}} Q$. Using Theorem~\ref{containnegbool}, then we get that the containment problem for CQNs over $\mathrm{Which}(X)$ and model-defining interpretations is $\Pi_2^p$-hard or coNEXP-hard.
\end{proof}

\begin{Exm}\label{complneedwhich}
Consider the following CQNs without free variables: $P=\exists x,y,z\, \bigl( R(x,x)\wedge R(x,y)\wedge R(y,z)\wedge \neg R(z,z)
\bigr)$ and $Q=\exists u,v,s,t\, \bigl( R(u,v)\wedge \neg R(v,v)\wedge R(s,t)\wedge R(t,t) \bigr).$ We have that $P \sqsubseteq_{\mathrm{Which}(X)}^{\text{md}} Q.$ However, the characterization would fail if completions were omitted. Indeed,
there is no negation-homomorphism $h\colon Q\longrightarrow P$ whose image contains the literal $R(x,y)$.\footnote{ To cover $R(x,y)$, either $R(u,v)$ or $R(s,t)$ must be mapped to $R(x,y)$ (literals $\neg R(v,v)$ and $R(t,t)$ cannot be mapped to $R(x,y)$ since they are diagonal). In the first case, we must have $h(u)=x,h(v)=y,$ but then $\neg R(v,v)$ is mapped to $\neg R(y,y)$, which is not in the body of $P$. In the second case, we must have $h(s)=x, h(t)=y,$ but then $R(t,t)$ is mapped to $R(y,y)$, which is not in the body of $P$. Therefore, no such mapping $Q\longrightarrow P$ exists. On the other hand, if we consider completions, this problem no longer occurs. For the literal $R(x,y)$, if a completion $P^c$ contains $\neg R(y,y)$, then the mapping $u\mapsto x, v\mapsto y, s\mapsto x, t\mapsto x$ maps $Q$ to $P^c$ and covers $R(x,y)$. If instead $P^c$ contains $R(y,y)$, then the mapping $u\mapsto y, v\mapsto z, s\mapsto x, t\mapsto y$ maps $Q$ to $P^c$ and covers $R(x,y)$. The remaining literals of $P$ are all covered by the same mapping $u\mapsto y, v\mapsto z, s\mapsto x, t\mapsto x$. This shows that completions are needed in the characterization of containment over $\mathrm{Which}(X)$.
To show that $P \sqsubseteq_{\mathrm{Which}(X)}^{\text{md}} Q$, let $\pi$ be an arbitrary model-defining interpretation over $\mathrm{Which}(X)$. We prove that $\pi\llbracket P \rrbracket\leq_{\mathrm{Which}(X)}\pi\llbracket Q \rrbracket$. In $\mathrm{Which}(X)$ the order is given by inclusion on non-bottom elements, and $\bot$ is the least element. Consider an arbitrary assignment $x\mapsto a,y\mapsto b, z\mapsto c$ for the variables of $P$. If this assignment gives bottom contribution to $P$, there is nothing to prove. Assume then that it gives a non-bottom contribution. So, $\pi\llbracket R(a,a)\rrbracket\cdot \pi\llbracket R(a,b)\rrbracket\cdot\pi\llbracket R(b,c)\rrbracket\cdot \pi\llbracket \neg R(c,c)\rrbracket\neq \bot.$ Since $\bot$ is absorbing for multiplication in $\mathrm{Which}(X)$, each factor must be non-bottom: $\pi\llbracket R(a,a)\rrbracket,\pi\llbracket R(a,b)\rrbracket,\pi\llbracket R(b,c)\rrbracket, \pi\llbracket \neg R(c,c)\rrbracket\neq \bot.$ Multiplication of non-bottom elements in $\mathrm{Which}(X)$ is union, so the contribution of this assignment to $P$ is $\pi\llbracket R(a,a)\rrbracket\cup \pi\llbracket R(a,b)\rrbracket\cup \pi\llbracket R(b,c)\rrbracket\cup \pi\llbracket \neg R(c,c)\rrbracket.$ We need to prove that this union is an element of $\pi\llbracket Q\rrbracket$.  Consider now the assignment for the variables of $Q$ given by $u\mapsto b,v\mapsto c, s\mapsto a, t\mapsto a.$ Under this assignment, the body of $Q$ becomes $R(b,c)\wedge \neg R(c,c)\wedge R(a,a)\wedge R(a,a).$ All these literals have non-bottom value by the assumption on the assignment for $P$. Therefore, this assignment gives a non-bottom contribution to $Q$, and this contribution contains all tags occurring in $\pi\llbracket R(a,a)\rrbracket\cup \pi\llbracket R(b,c) \rrbracket\cup \pi\llbracket \neg R(c,c)\rrbracket$. This means that $\pi\llbracket R(a,a)\rrbracket\cup \pi\llbracket R(b,c) \rrbracket\cup \pi\llbracket \neg R(c,c)\rrbracket  \leq_{\mathrm{Which}(X)} \pi\llbracket Q \rrbracket .$ It just remains to show that the tags occurring in $\pi\llbracket  R(a,b)\rrbracket $ are also contained in $\pi\llbracket Q\rrbracket $. Since $\pi$ is model-defining, exactly one between $R(b,b)$ and $\neg R(b,b)$ has non-bottom value. If $\pi\llbracket \neg R(b,b)\rrbracket \neq \bot,$ we consider the assignment for $Q$ given by $u\mapsto a,v\mapsto b, s\mapsto a, t\mapsto a.$ The body of $Q$ becomes $R(a,b)\wedge \neg R(b,b)\wedge R(a,a)\wedge R(a,a)$. Again all these literals have non-bottom value ($R(a,b)$ and $R(a,a)$ have non-bottom value because the assignment $x\mapsto a$, $y\mapsto b$, $z\mapsto c$ gives a non-bottom contribution to $P$, while $\neg R(b,b)$ has non-bottom value by assumption). This assignment gives a non-bottom contribution to $Q$, and this contribution contains all tags occurring in $\pi \llbracket R(a,b)\rrbracket$. If $\pi\llbracket  R(b,b)\rrbracket \neq \bot$, consider the assignment for $Q$ given by $u\mapsto b, v\mapsto c, s\mapsto a, t\mapsto b.$ Then, the body of $Q$ becomes $R(b,c)\wedge \neg R(c,c)\wedge R(a,b)\wedge R(b,b).$ All these literals have non-bottom value ($R(b,c)$, $\neg R(c,c)$ and $R(a,b)$ have non-bottom value because the assignment $x\mapsto a$, $y\mapsto b$, $z\mapsto c$ gives a non-bottom contribution to $P$, while $R(b,b)$ has non-bottom value by assumption). So this assignment gives a non-bottom contribution to $Q$, and this contribution again contains all tags occurring in $\pi\llbracket  R(a,b)\rrbracket $. Therefore, $\pi\llbracket  R(a,b)\rrbracket \leq_{\mathrm{Which}(X)}\pi\llbracket  Q\rrbracket$ too. We obtain $\pi\llbracket  R(a,a)\rrbracket \cup \pi\llbracket R(a,b)\rrbracket \cup \pi\llbracket R(b,c)\rrbracket \cup \pi\llbracket  \neg R(c,c)\rrbracket \leq_{\mathrm{Which}(X)}\pi\llbracket Q\rrbracket$. But the assignment for $P$ was arbitrary, so every non-bottom contribution to $P$ is contained in $\pi\llbracket Q\rrbracket$. Finally, $\pi\llbracket P\rrbracket $ is the sum, so the union, of all contributions of assignments for $P$. Therefore, $\pi\llbracket P\rrbracket \leq_{\mathrm{Which}(X)}\pi\llbracket Q\rrbracket$, and since $\pi$ was arbitrary, we can conclude that $P \sqsubseteq_{\mathrm{Which}(X)}^{\text{md}} Q.$}
\end{Exm}


\subsection{Equivalence of CQNs}\label{The problem of equivalence}
 Next we focus on the complexity of equivalence for CQNs over the semirings considered above and we establish exact complexity bounds for all of them.

 We will summarize the work in this section through the following theorem:

 \begin{Thm} For CQNs $P$ and $Q$ with the same tuple of free variables, the following statements hold. In item (10), assume that neither $P$ nor $Q$ contains any atom and its negation.
\begin{enumerate}
\item ``$P \equiv_{\mathbb{L}}^{\text{md}} Q$?" (where $\mathbb{L}$ is a bounded distributive lattice wit more than two elements) is NP-complete (Theorem~\ref{cqnslattice}). 
\item ``$P \equiv_{\mathbb{N}[X, \overline{X}]}^{\text{pt}} Q$?" is GI-complete (Theorem~\ref{equivcqnsvarious}). 
\item ``$P \equiv_{\mathbb{N}}^{\text{md}} Q$?" is GI-complete (Theorem~\ref{equivalcqnsbags}). 
\item ``$P \equiv_{\mathrm{Sorp}(X, \overline{X})}^{\text{pt}} Q$?" is GI-complete (Theorem~\ref{equivcqnsvarious}). 
\item ``$P \equiv_{\mathbb{B}[X, \overline{X}]}^{\text{pt}} Q$?" is GI-complete (Theorem~\ref{equivcqnsvarious}). 

\item ``$P \equiv_{\mathrm{Why}(X, \overline{X})}^{\text{pt}} Q$?" is GI-complete (Theorem~\ref{whycqnequivalence}). 

\item ``$P \equiv_{\mathrm{Trio}(X, \overline{X})}^{\text{pt}} Q$?" is GI-complete (Theorem~\ref{triocqnequivalencecomplexity}).

\item ``$P \equiv_{\mathrm{PosBool}(X, \overline{X})}^{\text{pt}} Q$?" is NP-complete (Theorem~\ref{equivcqnsvarious}).

\item  ``$P \equiv_{\mathrm{Which}(X)}^{\text{md}} Q$?" (Theorem~\ref{equivalencewhichmdcqn}) is 
\begin{itemize}
\item[(i)] $\Pi^p_2$-c for queries over schemas with arity bound $k$, for every $k$, and
\item[(ii)] coNEXP-c in general.  
\end{itemize} 

\item ``$P \equiv^\text{dmd}_{\mathbb{L}} Q$?"  (Theorem~\ref{cqnslattice})  is
\begin{itemize}
\item[(i)] $\Pi^p_2$-complete for queries over schemas with arity bound $k$, for every fixed $k\geq 2$, and
\item[(ii)] coNEXP-complete in general.  
\end{itemize}

\end{enumerate}

\end{Thm}

 Below we will provide one of the necessary results in  this section as an example but leave the rest to Appendix~\ref{app:cqn-equivalence-proofs}. 
Call two CQNs  \emph{isomorphic} (following  Chaudhuri \&  Vardi~\cite{Chad}) iff there are (both ways) \emph{negation} homomorphisms \emph{onto for atoms} and \emph{one-to-one on variables}. 
When two CQNs $P$ and $Q$ are isomorphic, we will write $P\simeq^\neg Q$. For the usual condition where the homomorphisms need not preserve negations, we write as $P\simeq Q$ (which is exactly the notion of isomorphism in~\cite{Chad}).

\begin{restatable}{Thm}{cqnEquivalenceC}\label{equivalcqns} For CQNs $P$ and $Q$ with the same tuple of free variables, we have that
\begin{itemize}

\item[(1)] $P \equiv^{\text{pt}}_{\mathbb{N}[X, \overline{X}]} Q$ iff $P\simeq^\neg Q$.

\item[(2)] $P \equiv^{\text{pt}}_{\mathbb{B}[X, \overline{X}]} Q$ iff $P\simeq^\neg Q$.
\item[(3)] $P \equiv^{\text{pt}}_{\mathrm{Sorp}(X, \overline{X})} Q$ iff $P\simeq^\neg Q$.

\end{itemize}

\end{restatable}

The proof can be found in Appendix~\ref{app:cqn-equivalence-proofs}.

%
%
%
%
%
%

\begin{Rmk}
Observe that it follows from Theorem~\ref{equivalcqns} that two CQs are equivalent over arbitrary $\mathbb{N}[X]$-instances if and only if they are equivalent over $\mathbb{N}[X]$-instances that use annotations only from $X \cup \{0,1\}$, and the same for $\mathbb{B}[X]$. This is because when we remove negations from the queries in question altogether, we get the usual isomorphisms \`a la  Chaudhuri \& Vardi~\cite{Chad}, which, using the characterizations in~\cite{Green}, gives us equivalence over arbitrary $\mathbb{N}[X]$-instances.
\end{Rmk}

Next, we construct  a function $\mathtt{tr}$  which converts any CQNs into a CQ in an expanded database schema (or logical vocabulary)  such that $P\simeq^\neg Q$ iff $\mathtt{tr}(P)\simeq \mathtt{tr}(Q)$. Take a CQN $P$ over a schema $\sigma$, and then consider a schema $\sigma^*$ that has a positive $R^+$ and negative $R^-$ relation symbol for each relation symbol $R$ in $\sigma$ (we can thus speak of the polarity of the symbols $R^+$ and $R^-$ as positive and negative, respectively). Then, let $\mathtt{tr}$ be defined inductively with  literals as the base case: $\mathtt{tr}(R(\vec{x})) = R^+(\vec{x})$ while $\mathtt{tr}(\neg R(\vec{x})) = R^-(\vec{x})$. Observe that the mapping $\mathtt{tr}$ is such that if $L(\vec{x})$ and $L'(\vec{z})$ are different literals then  $\mathtt{tr}(L(\vec{x}))\neq \mathtt{tr}(L'(\vec{z}))$. We obtain then the following fact:

\begin{restatable}{Lem}{cqnEquivalenceD}\label{transuse} Let $P$ and $Q$ be CQNs. A mapping $h\colon P \longrightarrow Q$ is a negation-homomorphism iff it is a homomorphism $h\colon \mathtt{tr}(P) \longrightarrow \mathtt{tr}(Q)$. 

\end{restatable}

The proof can be found in Appendix~\ref{app:cqn-equivalence-proofs}.

We are now ready to establish the following result:

\begin{restatable}{Thm}{cqnEquivalenceE}\label{equivcqnsvarious} Equivalence of  CQNs for provenance-tracking interpretations over $ \mathbb{B}[X, \overline{X}]$, $\mathbb{N}[X, \overline{X}]$ and $\mathrm{Sorp}(X, \overline{X})$ is GI complete exactly as for  CQs. Moreover, equivalence for provenance-tracking interpretations over 
$\mathrm{PosBool}(X, \overline{X})$ is  NP-complete.
\end{restatable}


\begin{proof}
The first part of the theorem is immediate from Theorem~\ref{equivalcqns} and Lemma~\ref{transuse}.  Evidently, the problem of determining $P\simeq^\neg Q$ that we get from Theorem~\ref{equivalcqns} has a GI lower bound, since isomorphism of CQs is GI-hard~\cite[Thm. 7.9 and Cor. 7.10]{Green} and, for positive queries, $P\simeq^\neg Q$ iff $P\simeq Q$. For the upper bound, we can use Lemma~\ref{transuse} to show that $P\simeq^\neg Q$ iff $\mathtt{tr}(P)\simeq \mathtt{tr}(Q)$ and thus use the  GI upper bounds established in~\cite{Green}. Now, for the last part of the theorem, we start by observing that Lemma~\ref{equivlat} works also for provenance-tracking interpretations, and thus we can apply it to $\mathrm{PosBool}(X, \overline{X})$ (which is a bounded distributive lattice).  Then, exactly the same reasoning as in the proof of Corollary~\ref{cqnslattice}, gives us the NP-completeness we are after. Finally, for $\mathrm{Which}(X)$ under model-defining interpretations, the upper bounds follow immediately from Theorem~\ref{containmentwhich modeldef}, and equivalence belongs to $\Pi^p_2$ for schemas with bounded arity and to coNEXP in the general case.
\end{proof}

Finally, we turn to equivalence over $\mathrm{Which}(X)$ with model-defining interpretations and we show that the upper bounds obtained from containment are exact.

\begin{restatable}{Thm}{cqnEquivalenceWhichMD} \label{equivalencewhichmdcqn}
Equivalence of CQNs over $\mathrm{Which}(X)$ with model-defining interpretations is
\begin{itemize}
\item[(i)] $\Pi^p_2$-complete for queries over schemas with a fixed arity bound $k\geq 2$, and
\item[(ii)] coNEXP-complete for unbounded arity.
\end{itemize}
\end{restatable}
The proof can be found in Appendix~\ref{app:cqn-equivalence-proofs}.

\section{Unions of conjunctive queries with negated atoms UCQNs}


In this section, we extend our characterizations of containment and equivalence from single conjunctive queries with negation (CQNs) to unions of conjunctive queries with negation (UCQNs).  Below, we systematically analyze how the structural conditions developed for CQNs generalize to UCQNs across model-defining ($\mathrm{md}$), dually model-defining ($\mathrm{dmd}$), and provenance-tracking ($\mathrm{pt}$) interpretations. We start with the first result that stands in contrast to the Boolean case.

\subsection{Containment of UCQNs}\label{The problem of containmentUCQN}

We begin yet again with a summarizing theorem:

\begin{Thm} For UCQNs $\overline{P}$ and $\overline{Q}$ with the same tuple of free variables, the following statements hold. In the dually model-defining case of item (8), assume that no disjunct of $\overline{P}$ contains any atom and its negation.
\begin{enumerate}
\item ``$\overline{P} \sqsubseteq_{\mathbb{L}}^{\text{md}} \overline{Q}$?" (where $\mathbb{L}$ has more than two elements) is NP-complete (Theorem~\ref{dsitrlatticenegucq}). 
\item ``$\overline{P} \sqsubseteq_{\mathbb{N}[X, \overline{X}]}^{\text{pt}} \overline{Q}$?" is in $\text{coNP}^\text{PP}$ (Theorem~\ref{nxpspace2}).   
\item ``$\overline{P} \sqsubseteq_{\mathrm{Sorp}(X, \overline{X})}^{\text{pt}} \overline{Q}$?" is NP-complete (Theorem~\ref{sorpchnegucq}). 
\item ``$\overline{P} \sqsubseteq_{\mathbb{B}[X, \overline{X}]}^{\text{pt}} \overline{Q}$?" is NP-complete (Theorem~\ref{bxucqnegcont}). 

\item ``$\overline{P} \sqsubseteq_{\mathrm{Why}(X, \overline{X})}^{\text{pt}} \overline{Q}$?" is NP-complete (Theorem~\ref{ucqequinegwhy}). 

\item ``$\overline{P} \sqsubseteq_{\mathrm{Trio}(X, \overline{X})}^{\text{pt}} \overline{Q}$?" is in $\text{coNP}^\text{PP}$  (Theorem~\ref{triopspace2}).

\item ``$\overline{P} \sqsubseteq_{\mathrm{PosBool}(X, \overline{X})}^{\text{pt}} \overline{Q}$?" is NP-complete (Theorem~\ref{posboolucqnneg}).

\item The containment decision problems  ``$\overline{P} \sqsubseteq^\text{dmd}_{\mathbb{L}} \overline{Q}$?"  (Theorem~\ref{dsitrlatticenegucqdual}) and ``$\overline{P} \sqsubseteq_{\mathrm{Which}(X)}^{\text{md}} \overline{Q}$?" (Theorem~\ref{ucqnegwhichmd}) is
\begin{itemize}
\item[(i)] $\Pi^p_2$-complete for queries over schemas with arity bound $k$, for every fixed $k\geq 2$, and
\item[(ii)] coNEXP-complete in general.  
\end{itemize}

\end{enumerate}

\end{Thm}

We now consider UCQN containment under the Boolean polynomial provenance semiring, $\mathbb{B}[X, \overline{X}]$ as our sample result in this section (everything else is relegated to Appendix~\ref{app:ucqn-containment-proofs}). As in the single-query case, tracking literal multiplicities forces the negation-homomorphism to be exact rather than set-onto, preserving the $\mathrm{NP}$-completeness bound.

\begin{restatable}{Thm}{ucqnContainmentE}\label{bxucqnegcont} For UCQNs, $\bigcup_{1\leq i \leq n} P_i, \bigcup_{1\leq i \leq m} Q_i$ with the same tuple of free variables, the following are equivalent:
\begin{itemize}
\item[1.] $\bigcup_{1\leq i \leq n} P_i  \sqsubseteq^{\text{pt}}_{\mathbb{B}[X, \overline{X}]} \bigcup_{1\leq i \leq m} Q_i$.
\item[2.]   For each $1\leq i \leq n$, we have that  $\pi^{\mathrm{can}_{\mathbb{B}[X]} (P_i)}\llbracket P_i \rrbracket \leq_{\mathbb{B}[X, \overline{X}]} \pi^{\mathrm{can}_{\mathbb{B}[X]}(P_i)} \llbracket \bigcup_{1\leq i \leq m} Q_i \rrbracket $.

\item[3.] For each $1\leq i \leq n$, there is a $j \leq m$ and a negation-homomorphism 
$h \colon Q_j \longrightarrow P_i$ exact for literals.

\end{itemize}
Moreover, the corresponding decision problem (i.e.\ $\bigcup_{1\leq i \leq n} P_i \sqsubseteq_{\mathbb{B}[X, \overline{X}]}^{\text{pt}} \bigcup_{1\leq i \leq m} Q_i $?)  is NP-complete. 
\end{restatable}

\begin{proof}
$(1) \implies (2):$ Obvious.

$(2) \implies (3):$  Assume that (2) holds, for each $1\leq i \leq n$, we have that 

$$\pi^{\mathrm{can}_{\mathbb{B}[X]}(P_i)}\llbracket P_i(\vec{u})\rrbracket \leq_{\mathbb{B}[X, \overline{X}]} \pi^{\mathrm{can}_{\mathbb{B}[X]}(P_i)} \llbracket \bigcup_{1\leq i \leq m} Q_i(\vec{u})\rrbracket,$$
where $\vec{u}$ is the tuple of free variables of $P_i$.

The polynomial $\pi^{\mathrm{can}_{\mathbb{B}[X]}(P_i)}\llbracket P_i(\vec{u})\rrbracket$ has as a term a monomial $x_1 \cdot \ldots \cdot x_n $ containing every tag from $X \cup \overline{X}$ used to annotate literals by $\pi^{\mathrm{can}_{\mathbb{B}[X]}\llbracket P_i\rrbracket}$, possibly with repetitions (such monomial is obtained using the identity assignment for the variables in $ P_i$).  Then, by the definition of $\leq $ for $\mathbb{B}[X, \overline{X}]$, we must have that the monomial $x_1 \cdot \ldots \cdot x_n $ appears in $\pi^{\mathrm{can}_{\mathbb{B}[X]}(P_i)} \llbracket \bigcup_{1\leq i \leq m} Q_i(\vec{u})\rrbracket$.

Since $\bigcup_{1\leq i \leq m} Q_i$ is a UCQN, we must have that there is a $j\leq m$, and an assignment $v$ to the bound variables in $Q_j$ 
such that it produces the monomial $x_1 \cdot \ldots \cdot x_n $ as the value for the body of $Q_j$.
Then, we get the desired negation-homomorphism $h \colon Q_j \longrightarrow P_i$ exact for relational atoms by letting $h(x)$ be the variable $v(x)$ (recall that the elements of the active domain of $\mathrm{can}_{\mathbb{B}[X]}(P_i)$ are the variables of $P_i$). To show injectivity, notice that any tag $p$ or $\overline{p}$ that is the value of a literal of $Q_j$ according to $v$, appears in $x_1 \cdot \ldots \cdot x_n $ with at most the same multiplicity as in  the body of $P_i$ (since this monomial corresponds to the body of $P_i$), which entails that the multiset of images by $h$ of literals of $Q_j$ is bag-contained in the multiset of literals from $P_i$. Surjectivity follows from the fact that $x_1 \cdot \ldots \cdot x_n $ corresponds to the multiset of literals from $P_i$ but also to the images by $h$ of the multiset of literals from $Q_j$, so the former is bag-contained in the latter.

$(3) \implies (1):$ Assume that for each $1\leq i \leq n$, there is a $j \leq m$ and a negation-homomorphism 
$h \colon Q_j \longrightarrow P_i$ exact for literals. In order to establish 
$\bigcup_{1\leq i \leq n} P_i  \sqsubseteq^{\text{pt}}_{\mathbb{B}[X, \overline{X}]} \bigcup_{1\leq i \leq m} Q_i$, 
consider an arbitrary $\mathbb{B}[X]$-database instance $\mathbf I$, an interpretation compatible with it and a tuple $\vec{t}$. We need to show that any monomial in 
$\pi\llbracket( \bigcup_{1\leq i \leq n} P_i)(\vec{t})\rrbracket$ appears in $\pi\llbracket(\bigcup_{1\leq i \leq m} Q_i)(\vec{t})\rrbracket$.
Take an assignment $v$ of elements to the variables of $P_i$ justifying a given monomial $x_1 \cdot \ldots \cdot x_n $. By assumption, there is a negation-homomorphism $h \colon Q_j \longrightarrow P_i$ exact for literals. Then, $v \circ h$ is an assignment of elements to the variables of $Q$ justifying the monomial $x_1 \cdot \ldots \cdot x_n $. If all the $x_i$s are values of literals of $P_i$ according to $v$, then if $x_i = \pi\llbracket B(v(\vec{z}))\rrbracket$ where  $B(\vec{z})$ is a literal of  $P_i$, then $x_i = \pi\llbracket B(v(h(\vec{z})))\rrbracket$  for some literal $A(\vec{x})$ of $Q$ such that $A(h(\vec{x})) = B(\vec{z})$, which exists by surjectivity.   
Furthermore, the multiplicity with which  $x_i$ appears using the assignment $v \circ h$ is exactly the multiplicity with which it appears in  $x_1 \cdot \ldots \cdot x_m $ given that $h$ is both onto and injective on literals.

The complexity calculation follows by reasoning as in Theorem~\ref{dsitrlatticenegucq}. For the lower bound, we use Theorem~\ref{bxcompletept}.
\end{proof}

\subsection{Equivalence of UCQNs}\label{The problem of equivalenceUCQN}

We now turn to the equivalence problem for UCQNs over the semirings considered above. We establish exact complexity bounds for bounded distributive lattices, $\mathrm{Which}(X)$, $\mathrm{PosBool}(X,\overline{X})$, $\mathrm{Sorp}(X,\overline{X})$, $\mathrm{Why}(X,\overline{X})$, and $\mathbb{B}[X,\overline{X}]$. For $\mathbb{N}[X,\overline{X}]$, $\mathrm{Trio}(X,\overline{X})$, and $\mathbb{N}$, we provide upper bounds only. Once more, the following result is key in being able to use the previous theorems to understand the complexity of equivalence between UCQNs:

\begin{restatable}{Pro}{ucqnEquivalenceA}\label{propogreen2} For UCQNs $\bar{P}, \bar{Q}$, additionally idempotent semiring $K$ (which includes 
$\mathrm{Sorp}(X, \overline{X})$, $\mathrm{Why}(X, \overline{X})$, bounded distributive lattices,  $\mathbb{B}[X, \overline{X}]$ 
 and $\mathrm{Which}(X)$, but excludes $\mathbb{N}[X,\overline{X}]$ and $\mathrm{Trio}(X, \overline{X})$) and any class of interpretations $\mathcal C$, we have
\begin{itemize}
\item[(1)] $\bar{P}\sqsubseteq_K^\mathcal{C}\bar{Q}$ iff $\bar{P} \cup \bar{Q} \equiv_K^{\mathcal C}\bar{Q}$.

\item[(2)] $\bar{P}\equiv_K^\mathcal{C}\bar{Q}$ iff $\bar{P}\sqsubseteq_K^\mathcal{C}\bar{Q}$ and $\bar{Q}\sqsubseteq_K^\mathcal{C}\bar{P}$. 

\end{itemize}

\end{restatable}

The proof can be found in Appendix~\ref{app:ucqn-equivalence-proofs}.  

We can next state the following result:
\begin{restatable}{Thm}{ucqnEquivalenceB} \label{ucqnegequivalence}

For UCQNs $\overline{P}$ and $\overline{Q}$ with the same tuple of free variables, the following statements hold. In the dually model-defining case of item (8), assume that no disjunct of either $\overline{P}$ or $\overline{Q}$ contains any atom and its negation.
\begin{enumerate}
\item ``$\overline{P} \equiv_{\mathbb{L}}^{\text{md}} \overline{Q}$?" (where $\mathbb{L}$ is a bounded distributive lattice wit more than two elements) is NP-complete. 
\item ``$\overline{P} \equiv_{\mathbb{N}[X, \overline{X}]}^{\text{pt}} \overline{Q}$?" and ``$\overline{P} \equiv_\mathbb{N}^{\text{md}} \overline{Q}$?" are in $\text{coNP}^\text{PP}$.   
\item ``$\overline{P} \equiv_{\mathrm{Sorp}(X, \overline{X})}^{\text{pt}} \overline{Q}$?" is NP-complete. 
\item ``$\overline{P} \equiv_{\mathbb{B}[X, \overline{X}]}^{\text{pt}} \overline{Q}$?" is NP-complete. 

\item ``$\overline{P} \equiv_{\mathrm{Why}(X, \overline{X})}^{\text{pt}} \overline{Q}$?" is NP-complete. 

\item ``$\overline{P} \equiv_{\mathrm{Trio}(X, \overline{X})}^{\text{pt}} \overline{Q}$?" is in $\text{coNP}^\text{PP}$.

\item ``$\overline{P} \equiv_{\mathrm{PosBool}(X, \overline{X})}^{\text{pt}} \overline{Q}$?" is NP-complete.

\item ``$\overline{P} \equiv^\text{dmd}_{\mathbb{L}} \overline{Q}$?"   and ``$\overline{P} \equiv_{\mathrm{Which}(X)}^{\text{md}} \overline{Q}$?"  are
\begin{itemize}
\item[(i)] $\Pi^p_2$-complete for queries over schemas with arity bound $k$, for every fixed $k\geq 2$, and
\item[(ii)] coNEXP-complete in general.  
\end{itemize}

\end{enumerate}

\end{restatable}

The proof can be found in Appendix~\ref{app:ucqn-equivalence-proofs}.
%

\section{Conjunctive queries with disequations and negated atoms (CQ\texorpdfstring{$^\neq$}{≠}Ns) }
In this section, an identification in the sense of CQ$^\neq$s is obtained by completing the equations and disequations of a query and applying a canonical substitution, while retaining its relational literals and their multiplicities. For provenance-tracking interpretations, we consider all such identifications, even those containing an atom and its negation. Instead, under model-defining interpretations, these identifications always have value $0$, so we only need to consider the consistent ones. In particular, for bounded distributive lattices we use consistent identifications directly, while for $\mathrm{Which}(X)$ we also consider their consistent completions in the sense of CQNs.

Note that, if $P$ has no consistent identification, its value is always $0$ under model-defining interpretations. Thus, containment holds trivially, and the corresponding conditions are vacuously satisfied. The case of mdpt interpretations is discussed in Remark~\ref{mdpt-identifications}.

\subsection{Containment of CQ\texorpdfstring{$^\neq$}{≠}Ns}\label{sec:cqneqn-containment}

We start this section with our next summarizing theorem.

\begin{Thm} For CQ$^\neq$Ns $P$ and $Q$ with the same tuple of free variables, the following statements hold.
\begin{enumerate}
\item ``$P \sqsubseteq_{\mathbb{L}}^{\text{md}} Q$?" (where $\mathbb{L}$ is a bounded distributive lattice wit more than two elements) is $\Pi^p_2$-complete (Theorem~\ref{cqneqnegmd}). 
\item ``$P \sqsubseteq_{\mathbb{N}[X, \overline{X}]}^{\text{pt}} Q$?" is in coNP$^\text{PP}$ (Theorem~\ref{N_CQNs}).

\item ``$P \sqsubseteq_{\mathrm{Sorp}(X, \overline{X})}^{\text{pt}} Q$?" is $\Pi^p_2$-complete  (Theorem~\ref{sorpcqneqnegcont}). 
\item ``$P \sqsubseteq_{\mathbb{B}[X, \overline{X}]}^{\text{pt}} Q$?" is $\Pi^p_2$-complete  (Theorem~\ref{b[X]cqneqnegcont}). 

\item ``$P \sqsubseteq_{\mathrm{Why}(X, \overline{X})}^{\text{pt}} Q$?" is  $\Pi^p_2$-complete  (Theorem~\ref{whycqneqnegcont}). 

\item ``$P \sqsubseteq_{\mathrm{Trio}(X, \overline{X})}^{\text{pt}} Q$?" is in coNP$^\text{PP}$ (Theorem~\ref{triocqneqn}).

\item ``$P \sqsubseteq_{\mathrm{PosBool}(X, \overline{X})}^{\text{pt}} Q$?" is  $\Pi^p_2$-complete   (Theorem~\ref{posboolcqneqncont}).

\item ``$P \sqsubseteq_{\mathrm{Which}(X)}^{\text{md}} Q$?" (Theorem~\ref{whichcqneqncont}) is
\begin{itemize}
\item[(i)] $\Pi^p_2$-complete for queries over schemas with arity bound $k$, for every $k$, and
\item[(ii)] coNEXP-complete in general.  
\end{itemize}

\end{enumerate}

\end{Thm}

Our first example in this section will pertain to the  $\mathrm{Sorp}(X, \overline{X})$ semiring. In this case,  containment for $\mathrm{CQ}^\neq\mathrm{Ns}$ can be characterized by literal-injective $\neq$-negation-homomorphisms across all query identifications, driving the computational complexity to $\Pi^p_2$-completeness.

\begin{restatable}{Thm}{cqneqnContainmentE} \label{sorpcqneqnegcont}For CQ$^\neq$Ns $P$ and $Q$ with the same tuple of free variables, the following are equivalent:
\begin{itemize}
\item[1.] $P \sqsubseteq_{\mathrm{Sorp}(X, \overline{X})}^{\text{pt}} Q$,
\item[2.] For every identification (in the sense of CQ$^\neq$s) $P^{\mathrm{id}}$ of $P$, we have that $ \pi^{\mathrm{can}_{\mathrm{Sorp}(X)}(P^{\mathrm{id}})}\llbracket P^{\mathrm{id}} \rrbracket  \leq_{\mathrm{Sorp}(X, \overline{X})} \pi^{\mathrm{can}_{\mathrm{Sorp}(X)}(P^{\mathrm{id}})}\llbracket Q \rrbracket$.
\item[3.] For every identification (in the sense of CQ$^\neq$s) $P^{\mathrm{id}}$ of $P$, there is an  $\neq$-negation-homomorphism $h_{P^{\mathrm{id}}}\colon Q \longrightarrow P^{\mathrm{id}}$ injective for relational literals.

\end{itemize}
Moreover, the corresponding decision problem (i.e.\ $P \sqsubseteq^{\text{pt}}_{\mathrm{Sorp}(X, \overline{X})} Q$?) is $\Pi^p_2$-complete.
\end{restatable}



\begin{proof}
$(1) \implies (2):$ Suppose that $P \sqsubseteq_{\mathrm{Sorp}(X, \overline{X})}^{\text{pt}} Q$ and consider an identification $P^{\mathrm{id}}$ of a completion $P^c$ of $P$. We have that  $P^{\mathrm{id}}\sqsubseteq_{\mathrm{Sorp}(X, \overline{X})}^{\text{pt}} P^c \sqsubseteq_{\mathrm{Sorp}(X, \overline{X})}^{\text{pt}} P$ since identification preserves query values, completion only removes assignment contributions, and addition is monotone in the natural order. Thus, $ \pi^{\mathrm{can}_{\mathrm{Sorp}(X)}(P^{\mathrm{id}})}\llbracket P^{\mathrm{id}} \rrbracket  \leq_{\mathrm{Sorp}(X, \overline{X})} \pi^{\mathrm{can}_{\mathrm{Sorp}(X)}(P^{\mathrm{id}})}\llbracket P^c \rrbracket     \leq_{\mathrm{Sorp}(X, \overline{X})} 
\pi^{\mathrm{can}_{\mathrm{Sorp}(X)}(P^{\mathrm{id}})}\llbracket P \rrbracket \leq_{\mathrm{Sorp}(X, \overline{X})} \pi^{\mathrm{can}_{\mathrm{Sorp}(X)}(P^{\mathrm{id}})}\llbracket Q \rrbracket$.

$(2) \implies (3):$  Assume that (2) holds, so
for every identification $P^{\mathrm{id}}$ of any completion $P^c$ of $P$, we have that 
$$
\pi^{\mathrm{can}_{\mathrm{Sorp}(X)}(P^{\mathrm{id}})}\llbracket P^{\mathrm{id}}(\vec{u}) \rrbracket  \leq_{\mathrm{Sorp}(X, \overline{X})} \pi^{\mathrm{can}_{\mathrm{Sorp}(X)}(P^{\mathrm{id}})}\llbracket Q(\vec{u}) \rrbracket,
$$
where $\vec{u}$ is the tuple of free variables of $P^{\mathrm{id}}$. The polynomial $\pi^{\mathrm{can}_{\mathrm{Sorp}(X)}(P^{\mathrm{id}})}\llbracket P^{\mathrm{id}}(\vec{u}) \rrbracket$ has as a term a monomial $x_1 \cdot \ldots \cdot x_n $ containing every tag 
(with some exponent, accounting for duplicate literals) 
from $X \cup \overline{X}$ used to annotate literals by $\pi^{\mathrm{can}_{\mathrm{Sorp}(X)}(P^{\mathrm{id}})}$ (such monomial is obtained using the identity assignment for the variables in $ P^{\mathrm{id}}$; noticing that the disequations in $P^{\mathrm{id}}$ all take value $1$ by construction of the comparison completion; the canonical relational tags are pairwise noncomplementary).

Then, by the definition of $\leq $ for $\mathrm{Sorp}(X, \overline{X})$, we must have that there is a  monomial $y_1 \cdot \ldots \cdot y_m $  in $\pi^{\mathrm{can}_{\mathrm{Sorp}(X)}(P^{\mathrm{id}})}\llbracket Q(\vec{u}) \rrbracket$ where only tags from $x_1 \cdot \ldots \cdot x_n $ appear, with equal or smaller exponents as in $x_1 \cdot \ldots \cdot x_n $. Since $Q$ is a  CQ$^\neq$N, we must have that there is an assignment $v$ to the bound variables in $Q$ over $\mathrm{can}_{\mathrm{Sorp}(X)}(P^{\mathrm{id}})$ such that the body (i.e.\ the quantifier-free matrix) of $Q$ evaluates to the monomial $y_1 \cdot \ldots \cdot y_m$.  Furthermore, the equations and disequations in $Q$ must all take value $1$ under the assignment $v$ in order to obtain the monomial  $y_1 \cdot \ldots \cdot y_m$ (instead of just $0$), and thus they must be  equations holding in $\mathrm{can}_{\mathrm{Sorp}(X)}(P^{\mathrm{id}})$ (and thus logically entailed by the body of $P^{\mathrm{id}}$) and disequations appearing in the body of $P^{\mathrm{id}}$ (since the latter contains a complete description of disequations in  $\mathrm{can}_{\mathrm{Sorp}(X)}(P^{\mathrm{id}})$).  Then, we get the desired 
 $\neq$- negation-homomorphism $h_{P^{\mathrm{id}}}\colon Q \longrightarrow P^{\mathrm{id}}$ injective for relational literals by letting $h_{P^{\mathrm{id}}}(x)$ be the variable $v(x)$ (recall that the elements of the active domain of $\mathrm{can}_{\mathrm{Sorp}(X)}(P^{\mathrm{id}})$ are the variables of $P^{\mathrm{id}}$). To see the injectivity, notice that any tag $p$ or $\overline{p}$ that is the value of a relational literal of $Q$ according to $v$, appears in $y_1 \cdot \ldots \cdot y_m$ with at most the same multiplicity as in  $x_1 \cdot \ldots \cdot x_n $ (since this monomial corresponds to the body of $P^{\mathrm{id}}$), and thus the multiset of images by $h_{P^{\mathrm{id}}}$ of relational literals of $Q$ is bag-contained in the multiset of relational literals from $P^{\mathrm{id}}$.

$(3) \implies (1):$ Assume that for every identification $P^{\mathrm{id}}$ of a completion $P^c$ of $P$, there is an  $\neq$-negation-homomorphism $h_{P^{\mathrm{id}}}\colon Q \longrightarrow P^{\mathrm{id}}$ injective for relational literals. In order to establish $P \sqsubseteq_{\mathrm{Sorp}(X, \overline{X})}^{\text{pt}} Q$, consider an arbitrary $\mathrm{Sorp}(X)$-database instance $\mathbf I$, a compatible interpretation $\pi$ and a tuple $\vec{t}$. We need to show that for any monomial in $\pi\llbracket P(\vec{t}) \rrbracket$ there is a monomial  with  equal or smaller exponents  in $\pi\llbracket Q(\vec{t}) \rrbracket$. Take an assignment $v$ of elements to the variables of $P$ justifying a monomial $x_1  \cdot \ldots \cdot x_n $ and consider the $\{=, \neq \}$-completion $ P^c$ of $P$ obtained by adding $x=y$ or $x\neq y$ depending on whether  $v(x) = v(y)$ or $v(x) \neq v(y)$. Take an identification $P^{\mathrm{id}}$ of $P^c$ by a canonical substitution $\theta$. We may observe that $\theta(x) = \theta(y)$ iff $v(x) = v(y)$. Then, $x_1 \cdot \ldots \cdot  x_n $ is also a monomial in $\pi\llbracket P^{\mathrm{id}}(\vec{t}) \rrbracket$ justified by the assignment $v$ itself since the new  disequations added can only take value $1$ by construction according to the assignment $v$. 
The assignment  $v \circ h_{P^{\mathrm{id}}} $  justifies a   monomial with smaller or equal exponents in $\pi\llbracket Q(\vec{t}) \rrbracket$. This is because for any literal $A(\vec{x})$ of $Q$ such that $A(h_{P^{\mathrm{id}}}(\vec{x})) = B(v(\vec{z}))$, where  $B(\vec{z})$ is a literal of $P$, if 
$x_i = \pi\llbracket B(v(\theta(\vec{z}))) \rrbracket$, 
then $x_i = \pi\llbracket A(h_{P^{\mathrm{id}}}(\theta(\vec{z}))) \rrbracket$. 
By the injectivity of $h_{P^{\mathrm{id}}}$, the images of the literals in the body of $Q$ (considered as a multiset) appear with at most the same multiplicity as in the body of $P^{\mathrm{id}}$ considered as multiset, which entails that $x_i$ appears as the value of a literal of $Q$ with at most the same multiplicity as it appears in the monomial $x_1  \cdot \ldots \cdot x_n $ (since the latter corresponds to the body of $P^{\mathrm{id}}$). Since equations and disequations can only take values $0$ or $1$, they can be disregarded in showing that we can justify a monomial.

The lower bound for the complexity calculation follows immediately from the hardness of the problem for  CQ$^\neq$s  (in particular, the hardness of deciding whether item 3 holds). Naturally, $\neq$-negation-homomorphisms injective  for relational literals are simply  $\neq$-homomorphisms  injective for relational literals for the case of CQ$^\neq$s. The upper bound follows by the same kind of reasoning as for CQ$^\neq$s.
\end{proof}

\begin{Rmk}
The use of provenance-tracking interpretations makes this case of\/ $\mathrm{Trio}$ different from the case of CQ$^\neq$s considered in Section~3. There, arbitrary $\mathrm{Trio}(X)$-annotations are sums of several tags. Consequently, when evaluating a query, each relational atom occurrence may independently select one of the tags occurring in the annotation of its image. This is the reason why we introduced $k$-indexed $\neq$-homomorphisms, which record not only the underlying $\neq$-homomorphism but also the tag selected by each relational atom occurrence.
In the present setting, instead, we consider provenance-tracking interpretations. Thus, every non-zero relational literal is annotated by a single provenance token or by $1$. Therefore, once an assignment is fixed, there is no additional choice of a tag inside the annotation of a relational literal, and no analogue of the $k$-indexed $\neq$-homomorphism is needed. Moreover, since $\mathrm{Trio}(X,\overline{X})$ identifies monomials that differ only in their positive exponents, repeated uses of the same relational literal do not affect the resulting square-free monomial: the relevant information is instead the number of different assignments producing it. This motivates counting $\neq$-negation-homomorphisms that are set-onto for relational literals.
\end{Rmk}


In order to relate the coefficients of square-free monomials with the cardinalities introduced above, it's useful to work only with provenance-tracking interpretations in which different non-zero ground relational literals are annotated by different tags. The following lemma shows that this can be assumed without loss of generality when looking for counterexamples to containment.

\begin{restatable}{Lem}{cqneqnContainmentH}\label{freshTrioPT}
Let $\mathbf I$ be a $\mathrm{Trio}(X)$-database instance and let $\pi$ be a provenance-tracking interpretation compatible with $\mathbf I$. Then there are a $\mathrm{Trio}(X)$-database instance $\mathbf I^*$, a provenance-tracking interpretation $\pi^*$ compatible with $\mathbf I^*$, and a semiring homomorphism $E\colon \mathrm{Trio} (X,\overline{X})\longrightarrow \mathrm{Trio}(X,\overline{X})$ such that distinct non-zero ground relational literals receive pairwise distinct tags under $\pi^*$, no two of the tags used by $\pi^*$ are complementary, and, for every CQ$^\neq$N $P$ and every tuple $\vec{t}$, $E\bigl(\pi^*\llbracket P(\vec{t})\rrbracket\bigr) = \pi\llbracket P(\vec{t})\rrbracket.$ Consequently, if for two CQ$^\neq$Ns $P$ and $Q$ and some tuple $\vec{t}$ we have
$$
\pi\llbracket P(\vec{t})\rrbracket
\not\leq_{\mathrm{Trio}(X,\overline{X})}
\pi\llbracket Q(\vec{t})\rrbracket, \quad \text{then} \quad \pi^*\llbracket P(\vec{t})\rrbracket
\not\leq_{\mathrm{Trio}(X,\overline{X})}
\pi^*\llbracket Q(\vec{t})\rrbracket.
$$
\end{restatable}

The proof can be found in Appendix~\ref{app:cqneqn-containment-proofs}.

The fresh-token construction above has the important consequence that a non-zero square-free monomial uniquely determines the set of distinct ground relational literals used by an assignment. We now use this observation to relate the coefficients of such monomials directly to the number of set-onto $\neq$-negation-homomorphisms.

\begin{restatable}{Lem}{cqneqnContainmentI}\label{countingTrioPT}
Let $\mathbf I$ be a $\mathrm{Trio}(X)$-database instance and let $\pi$ be a provenance-tracking interpretation compatible with $\mathbf I$ such that distinct non-zero ground relational literals are annotated by pairwise distinct tags and no two tags used by $\pi$ are complementary. Let $P$ be a CQ$^\neq$N, let $\vec{t}$ be a tuple, and let $v$ be an assignment to the variables of $P$ producing $\vec{t}$ and contributing a non-zero square-free monomial $\mu$ to $\pi\llbracket P(\vec{t})\rrbracket$. Let $P^c_v$ be the completion of $P$ determined by $v$, and let $P^{\mathrm{id}}_v$ be the corresponding identification. Then, for every CQ$^\neq$N $Q$ with the same head as $P$, the coefficient of $\mu$ in $\pi\llbracket Q(\vec{t})\rrbracket$ is exactly $\left|\mathrm{SOHom}(Q,P^{\mathrm{id}}_v)\right|$, where for any queries $P, Q$,  $\mathrm{SOHom}(P,Q)$  denotes the set of all $\neq$-negation-homomorphisms from $P$ to $Q$ that are set-onto for relational literals. 
\end{restatable}

The proof can be found in Appendix~\ref{app:cqneqn-containment-proofs}.

Using Lemma~\ref{countingTrioPT}, we obtain the following counting characterization and $\text{coNP}^{\text{PP}}$ upper bound for containment over $\mathrm{Trio}(X, \overline{X})$.

\begin{restatable}{Thm}{cqneqnContainmentJ}\label{triocqneqn}
For CQ$^\neq$Ns $P$ and $Q$ with the same tuple of free variables, the following are equivalent:
\begin{itemize}

\item[1.] $P\sqsubseteq^{\text{pt}}_{\mathrm{Trio}(X,\overline X)}Q$.

\item[2.] For every identification $P^{\mathrm{id}}$ of $P$ (in the sense of CQ$^\neq$s),
$$
|\mathrm{SOHom}(P,P^{\mathrm{id}})|
\leq
|\mathrm{SOHom}(Q,P^{\mathrm{id}})|.
$$

\end{itemize}
Moreover, the corresponding decision problem (i.e.\ $P\sqsubseteq^{\text{pt}}_{\mathrm{Trio}(X,\overline X)}Q$?) is in $\text{coNP}^{\text{PP}}$.
\end{restatable}

\begin{proof}
$(1)\implies(2):$
Suppose that $P\sqsubseteq^{\text{pt}}_{\mathrm{Trio}(X,\overline X)}Q$ and consider an identification $P^{\mathrm{id}}$ of a $\{=,\neq\}$-completion $P^c$ of $P$. Let $\theta$ be the canonical substitution used to obtain $P^{\mathrm{id}}$ from $P^c$, and let $\text{LitSet}(P^{\mathrm{id}})=\{L_1,\ldots,L_r\}.$ Consider a provenance-tracking interpretation $\pi$ over the domain formed by the variables of $P^{\mathrm{id}}$ which assigns to every $L_i$ a different fresh tag $x_i\in X\cup\overline X$ of the corresponding polarity, so that no two complementary tags are used, and let every other ground relational literal take value $0$. Let $\mathbf I$ be the $\mathrm{Trio}(X)$-database instance induced by the values of $\pi$ on positive ground relational atoms. Then $\pi$ is a provenance-tracking interpretation compatible with $\mathbf I$.

Consider the assignment $v_\theta$ to the variables of $P$ defined by $v_\theta(x):=\theta(x).$ By construction of the completion $P^c$, we have that $\theta(x)=\theta(y)$ if and only if $v_\theta(x)=v_\theta(y)$, and therefore the identification of $P$ induced by $v_\theta$ is precisely $P^{\mathrm{id}}$. Moreover, $v_\theta$ satisfies all equations and disequations of $P$ and contributes the non-zero square-free monomial $\mu_{P^{\mathrm{id}}}:=x_1\cdot\ldots\cdot x_r$ to $\pi\llbracket P(\theta(\vec{u}))\rrbracket$. Indeed, repeated occurrences of the same relational literal may only produce positive powers of the corresponding tag, which are dropped in $\mathrm{Trio}(X,\overline X)$.

By Lemma~\ref{countingTrioPT}, applied to $P$ and to $Q$, we have that the coefficient of  $\mu_{P^{\mathrm{id}}}$ in $\pi\llbracket P(\theta(\vec{u}))\rrbracket $ is $|\mathrm{SOHom}(P,P^{\mathrm{id}})|$ and the coefficient of $\mu_{P^{\mathrm{id}}}$ in $\pi\llbracket Q(\theta(\vec{u}))\rrbracket$ is $|\mathrm{SOHom}(Q,P^{\mathrm{id}})|$. By the containment hypothesis, $\pi\llbracket P(\theta(\vec{u}))\rrbracket
\leq_{\mathrm{Trio}(X,\overline X)}
\pi\llbracket Q(\theta(\vec{u}))\rrbracket.$ Since the natural order of $\mathrm{Trio}(X,\overline X)$ is the coefficientwise order on consistent square-free monomials, it follows that $|\mathrm{SOHom}(P,P^{\mathrm{id}})| \leq |\mathrm{SOHom}(Q,P^{\mathrm{id}})|.$ Since $P^{\mathrm{id}}$ was arbitrary, item 2 follows.

$(2)\implies(1):$
We prove the contrapositive. Suppose that $P\not\sqsubseteq^{\text{pt}}_{\mathrm{Trio}(X,\overline X)}Q.$ Then there are a $\mathrm{Trio}(X)$-database instance $\mathbf I$, a provenance-tracking interpretation $\pi$ compatible with $\mathbf I$, and a tuple $\vec{t}$ such that $\pi\llbracket P(\vec{t})\rrbracket
\not\leq_{\mathrm{Trio}(X,\overline X)}
\pi\llbracket Q(\vec{t})\rrbracket.$ By Lemma~\ref{freshTrioPT}, there are a $\mathrm{Trio}(X)$-database instance $\mathbf I^*$ and a provenance-tracking interpretation $\pi^*$ compatible with $\mathbf I^*$, in which distinct non-zero ground relational literals receive pairwise distinct tags and no two tags used by $\pi^*$ are complementary, such that $\pi^*\llbracket P(\vec{t})\rrbracket
\not\leq_{\mathrm{Trio}(X,\overline X)}
\pi^*\llbracket Q(\vec{t})\rrbracket.$

Since the natural order of $\mathrm{Trio}(X,\overline X)$ is coefficientwise, there is a consistent square-free monomial $\mu$ such that the coefficient of  $\mu$
in $\pi^*\llbracket P(\vec{t})\rrbracket$ is strictly bigger than the coefficient of $\mu$ in $
\pi^*\llbracket Q(\vec{t})\rrbracket.$ In particular, the coefficient of $\mu$ in $\pi^*\llbracket P(\vec{t})\rrbracket$ is non-zero, and hence there is an assignment $v$ to the variables of $P$, producing $\vec{t}$, whose contribution is $\mu$. Let $P^c_v$ be the $\{=,\neq\}$-completion of $P$ determined by $v$, obtained by adding $x=y$ if $v(x)=v(y)$ and $x\neq y$ otherwise, and let $P^{\mathrm{id}}_v$ be the corresponding identification.

By Lemma~\ref{countingTrioPT}, applied to this assignment, we have that the coefficient of $\mu$ in $\pi^*\llbracket P(\vec{t})\rrbracket = |\mathrm{SOHom}(P,P^{\mathrm{id}}_v)|$ and the coefficient of 
$\mu$
in $\pi^*\llbracket Q(\vec{t})\rrbracket$ is $|\mathrm{SOHom}(Q,P^{\mathrm{id}}_v)|$. Therefore, $|\mathrm{SOHom}(P,P^{\mathrm{id}}_v)| > |\mathrm{SOHom}(Q,P^{\mathrm{id}}_v)|$. Thus item 2 fails for the identification $P^{\mathrm{id}}_v$ of $P$. This proves the contrapositive and hence $(2)\implies(1)$.

Finally, for the upper bound complexity calculation, we look at the complement of the problem in item 2. We can guess an identification $P^{\mathrm{id}}$ of $P$ and use a PP oracle to decide whether $|\mathrm{SOHom}(P,P^{\mathrm{id}})| > |\mathrm{SOHom}(Q,P^{\mathrm{id}})|.
$ For a fixed identification $P^{\mathrm{id}}$, computing $|\mathrm{SOHom}(R,P^{\mathrm{id}})|$ is a $\#P$ problem: we guess a mapping from the variables of $R$ to those of $P^{\mathrm{id}}$ and check in polynomial time that it is a $\neq$-negation-homomorphism and that it is set-onto for relational literals. Hence $|\mathrm{SOHom}(P,P^{\mathrm{id}})| -|\mathrm{SOHom}(Q,P^{\mathrm{id}})|$ is in $\text{GapP}$, and deciding whether this difference is positive is in PP. Therefore, non-containment is in $\text{NP}^{\text{PP}}$, and the containment problem is in $\text{coNP}^{\text{PP}}$.
\end{proof}

\begin{Rmk}
Theorem~\ref{triocqneqn} makes very explicit the difference between $\mathrm{Why}(X,\overline X)$ and $\mathrm{Trio}(X,\overline X)$ in the presence of disequations. For $\mathrm{Why}(X,\overline X)$, containment only requires that, for every identification $P^{\mathrm{id}}$ of  $P$, $|\mathrm{SOHom}(Q,P^{\mathrm{id}})|>0,
$ since $\mathrm{Why}(X,\overline X)$ does not retain the number of derivations producing the same witness. In $\mathrm{Trio}(X,\overline X)$, instead, positive exponents are dropped but coefficients are retained. Thus, the coefficients of the relevant square-free monomials are captured by the cardinalities of the sets of such mappings. This is why Theorem~\ref{triocqneqn} requires $|\mathrm{SOHom}(P,P^{\mathrm{id}})| \leq |\mathrm{SOHom}(Q,P^{\mathrm{id}})|. $ In particular, since $\mathrm{SOHom}(P,P^{\mathrm{id}})\neq\emptyset$ for every identification $P^{\mathrm{id}}$, Trio-containment implies Why-containment, while the converse need not hold.
\end{Rmk}

\begin{Rmk}\label{mdpt-identifications}
For each of $\mathbb N[X,\overline X]$, $\mathbb B[X,\overline X]$, $\mathrm{Sorp}(X,\overline X)$, $\mathrm{Why}(X,\overline X)$, $\mathrm{Trio}(X,\overline X)$, and $\mathrm{PosBool}(X,\overline X)$, the containment characterization remains valid with $\sqsubseteq^{\mathrm{pt}}$ replaced by $\sqsubseteq^{\mathrm{mdpt}}$ after restricting every universal quantification over identifications to consistent identifications in the sense of CQ$^\neq$s. The mapping type and the exact or set-onto homomorphism counts remain unchanged.

For necessity, consider a consistent identification $P^{\mathrm{id}}$ of $P$ and its canonical interpretation $\pi$. Since $\pi$ may assign $0$ to both an atom and its negation, it is not necessarily model-defining. We can extend it as in Definition~\ref{CQNscan}, by assigning a fresh negative tag to every negated ground atom whose value, together with that of its complementary positive ground atom, is $0$, leaving all other annotations unchanged (and thus eliminating the problematic case described before). Since $P^{\mathrm{id}}$ is consistent, the resulting interpretation $\pi^*$ is both model-defining and provenance-tracking. By hypothesis, containment holds under $\pi^*$. Now consider the semiring homomorphism that leaves the original tags unchanged and maps every new tag, together with its complementary token, to $0$. This homomorphism is well-defined for all six semirings and, by construction, maps $\pi^*$ back to $\pi$. Since semiring homomorphisms preserve the natural order and commute with query evaluation, containment also holds under the original canonical interpretation $\pi$. Hence, the required homomorphisms or counting conditions follow as in the corresponding proofs for provenance-tracking interpretations.

For sufficiency, consider an arbitrary interpretation that is both model-defining and provenance-tracking. Every assignment giving a non-zero contribution to $P$ induces a consistent identification, since an atom and its negation cannot both have non-zero values. Therefore, the arguments used for provenance-tracking interpretations still apply, considering only consistent identifications. In the two counting cases, the fresh-token constructions preserve the zero annotations, and hence also the model-defining property. Finally, if $P$ has no consistent identification, its value is always $0$, and the conditions are vacuously satisfied.
\end{Rmk}

\subsection{Equivalence of CQ\texorpdfstring{$^\neq$}{≠}Ns}\label{sec:cqneqn-equivalence}

We summarize the results of this subsection in the following theorem.

\begin{Thm}
For CQ$^\neq$Ns $P$ and $Q$ with the same tuple of free variables, the following statements hold.
\begin{enumerate}
\item ``$P \equiv_{\mathbb{L}}^{\text{md}} Q$?'' (where $\mathbb{L}$ is a bounded distributive lattice with more than two elements) is $\Pi^p_2$-complete (Corollary~\ref{equivcqnn}).
\item ``$P \equiv_{\mathbb{N}[X,\overline{X}]}^{\text{pt}} Q$?'' is in coNP$^{\text{PP}}$ (Theorem~\ref{equivcqneqnegpt}).
\item ``$P \equiv_{\mathrm{Sorp}(X,\overline{X})}^{\text{pt}} Q$?'' is $\Pi^p_2$-c (Theorem~\ref{equivcqneqnegpt}).
\item ``$P \equiv_{\mathbb{B}[X,\overline{X}]}^{\text{pt}} Q$?'' is $\Pi^p_2$-c (Theorem~\ref{equivcqneqnegpt}).
\item ``$P \equiv_{\mathrm{Why}(X,\overline{X})}^{\text{pt}} Q$?'' is $\Pi^p_2$-c (Theorem~\ref{equivcqneqnegpt}).
\item ``$P \equiv_{\mathrm{Trio}(X,\overline{X})}^{\text{pt}} Q$?'' is in coNP$^{\text{PP}}$ (Theorem~\ref{equivcqneqnegpt}).
\item ``$P \equiv_{\mathrm{PosBool}(X,\overline{X})}^{\text{pt}} Q$?'' is $\Pi^p_2$-complete (Theorem~\ref{equivcqneqnegpt}).
\item ``$P \equiv_{\mathrm{Which}(X)}^{\text{md}} Q$?'' (Theorem~\ref{equivcqneqnegwhich}) is
\begin{itemize}
\item[(i)] $\Pi^p_2$-c for queries over schemas with a fixed arity bound, and
\item[(ii)] coNEXP-c in general.
\end{itemize}
\item ``$P \equiv_{\mathbb{B}}^{\text{md}} Q$?'' (Corollary~\ref{equivcqnn}) is
\begin{itemize}
\item[(i)] $\Pi^p_2$-complete for queries over schemas with a fixed arity bound, and
\item[(ii)] coNEXP-complete in general.
\end{itemize}
\item ``$P \equiv_{\mathbb{N}}^{\text{md}} Q$?'' is in coNP$^{\text{PP}}$ (Remark~\ref{bag-equivalence-bridge}).
\end{enumerate}
\end{Thm}

 We conclude this section by considering the complexity of equivalence for CQ$^\neq$Ns. As before, all the upper bounds follow immediately from the corresponding containment results, since equivalence is, by definition, the same as containment in both directions. For bounded distributive lattices, we obtain matching lower bounds by using the following simple observation. This is the analogue, in the present setting, of Proposition~\ref{equivlatdis} for CQ$^\neq$s and Lemma~\ref{equivlat} for CQNs.

\begin{restatable}{Lem}{cqneqnEquivalenceA}\label{equivlatneqneg}
For CQ$^\neq$Ns $P$ and $Q$ with the same tuple of free variables, bounded distributive lattice $\mathbb{L}$, and any class of interpretations $\mathcal{C}$, the following holds:
\begin{itemize}
    \item[(1)] $P \equiv^{\mathcal{C}}_\mathbb{L} Q$ iff $P \sqsubseteq^{\mathcal{C}}_\mathbb{L} Q$ and $Q \sqsubseteq^{\mathcal{C}}_\mathbb{L} P$;
    \item[(2)] $P \sqsubseteq^{\mathcal{C}}_\mathbb{L} Q$ iff $P \equiv^{\mathcal{C}}_\mathbb{L} P\wedge Q$.
\end{itemize}
\end{restatable}

The proof can be found in Appendix~\ref{app:cqneqn-equivalence-proofs}.


As in Lemma~\ref{equivlat}, item (1) of Lemma~\ref{equivlatneqneg} trivially holds for any naturally ordered semiring, not just lattices.

\begin{restatable}{Cor}{cqneqnEquivalenceB} \label{equivcqnn}
Equivalence of CQ$^\neq$Ns over bounded distributive lattices with more than two elements and model-defining interpretations is $\Pi^p_2$-complete. Over $\mathbb{B}$ and model-defining interpretations, it is $\Pi^p_2$-complete if there is a fixed arity bound on the relations allowed in the queries, and coNEXP-complete otherwise.
\end{restatable}

The proof can be found in Appendix~\ref{app:cqneqn-equivalence-proofs}.


We can now use the previous lemma, together with the containment results established above and Theorem~\ref{cqneqequivalHard}, to obtain the corresponding complexity bounds for provenance-tracking interpretations.

\begin{restatable}{Thm}{cqneqnEquivalenceC}\label{equivcqneqnegpt}
Equivalence of CQ$^\neq$Ns for provenance-tracking interpretations over $\mathrm{PosBool}(X,\overline{X})$, $\mathrm{Sorp}(X,\overline{X})$, $\mathrm{Why}(X,\overline{X})$, and $\mathbb{B}[X,\overline{X}]$ is $\Pi^p_2$-complete, while for $\mathbb{N}[X,\overline{X}]$ and $\mathrm{Trio}(X,\overline{X})$ equivalence is in $\text{coNP}^{\text{PP}}$.
\end{restatable}

The proof can be found in Appendix~\ref{app:cqneqn-equivalence-proofs}.


Finally, we obtain exact complexity bounds also for $\mathrm{Which}(X)$ under model-defining interpretations.

\begin{restatable}{Thm}{cqneqnEquivalenceD}\label{equivcqneqnegwhich}
Equivalence of CQ$^\neq$Ns over $\mathrm{Which}(X)$ with model-defining interpretations is $\Pi^p_2$-complete for queries over schemas with arity bound $k\geq 2$, and coNEXP-complete in general.
\end{restatable}

The proof can be found in Appendix~\ref{app:cqneqn-equivalence-proofs}.


\section{Unions of conjunctive queries with disequations and negated atoms (UCQ\texorpdfstring{$^\neq$}{≠}Ns)}

In this section, we leverage the results and techniques for CQ$^\neq$Ns in order to understand the complexity of the containment and equivalence problems for their unions.

\begin{Rmk}\label{mdpt-union-identifications}
For the six semirings in Remark~\ref{mdpt-identifications}, the containment characterizations in this section likewise hold with $\sqsubseteq^{\mathrm{pt}}$ replaced by $\sqsubseteq^{\mathrm{mdpt}}$ when identifications of each  disjunct of the query on the left
are restricted to consistent identifications in the sense of CQ$^\neq$s. Mapping conditions and counting inequalities remain unchanged. The fresh model-defining extension and the homomorphims replacing tokens with $0$ works in the case of unions too because semiring homomorphisms preserve finite sums. Conversely, each non-zero source contribution induces a consistent identification, so the same witness and counting arguments prove sufficiency while disjuncts with no consistent identification contribute $0$.
\end{Rmk}

\subsection{Containment of UCQ\texorpdfstring{$^\neq$}{≠}Ns}\label{sec:ucqneqn-containment}

We start this section with our next summarizing theorem.

\begin{Thm}
For UCQ$^\neq$Ns $\overline{P}=\bigcup_{1\leq i\leq n}P_i$ and $\overline{Q}=\bigcup_{1\leq j\leq m}Q_j$ with the same tuple of free variables, the following statements hold.
\begin{enumerate}
\item ``$\overline{P} \sqsubseteq_{\mathbb{L}}^{\text{md}} \overline{Q}$?'' (where $\mathbb{L}$ is a bounded distributive lattice with more than two elements) is $\Pi^p_2$-complete (Theorem~\ref{ucqneqnmdcont}).
\item ``$\overline{P} \sqsubseteq_{\mathbb{N}[X,\overline{X}]}^{\text{pt}} \overline{Q}$?'' is in coNP$^{\text{PP}}$ (Theorem~\ref{N[X]ucqneqncontainment}).
\item ``$\overline{P} \sqsubseteq_{\mathrm{Sorp}(X,\overline{X})}^{\text{pt}} \overline{Q}$?'' is $\Pi^p_2$-complete (Theorem~\ref{sorpucqneqncon}).
\item ``$\overline{P} \sqsubseteq_{\mathbb{B}[X,\overline{X}]}^{\text{pt}} \overline{Q}$?'' is $\Pi^p_2$-complete (Theorem~\ref{Bucqneqncont}).
\item ``$\overline{P} \sqsubseteq_{\mathrm{Why}(X,\overline{X})}^{\text{pt}} \overline{Q}$?'' is $\Pi^p_2$-complete (Theorem~\ref{whyucqneqncont}).
\item ``$\overline{P} \sqsubseteq_{\mathrm{Trio}(X,\overline{X})}^{\text{pt}} \overline{Q}$?'' is in coNP$^{\text{PP}}$ (Theorem~\ref{Trioucqneqncontainment}).
\item ``$\overline{P} \sqsubseteq_{\mathrm{PosBool}(X,\overline{X})}^{\text{pt}} \overline{Q}$?'' is $\Pi^p_2$-complete (Theorem~\ref{posboolucqneqncont}).
\item ``$\overline{P} \sqsubseteq_{\mathrm{Which}(X)}^{\text{md}} \overline{Q}$?'' (Theorem~\ref{whichucqneqncont}) is
\begin{itemize}
\item[(i)] $\Pi^p_2$-complete for queries over schemas with a fixed arity bound, and
\item[(ii)] coNEXP-complete in general.
\end{itemize}
\end{enumerate}
\end{Thm}

\begin{restatable}{Thm}{ucqneqnContainmentA}\label{ucqneqnmdcont}
Let $\overline P=\bigcup_{1\leq i\leq n}P_i$ and $\overline Q=\bigcup_{1\leq j\leq m}Q_j$ be UCQ$^\neq$Ns with the same tuple of free variables, and a bounded distributive lattice $\mathbb{L}$ with an intermediate element $\epsilon\neq\top,\bot$. Then, the following are equivalent:
\begin{itemize}

\item[1.] $\overline P\sqsubseteq_{\mathbb{L}}^{\mathrm{md}}\overline Q$.

\item[2.] For every $1\leq i\leq n$ and every consistent identification $P_i^{\mathrm{id}}$ (only in the sense of CQ$^\neq$s) of $P_i$, there is a $j\leq m$ and a $\neq$-negation-homomorphism
$$
h_{P_i^{\mathrm{id}}}\colon Q_j\longrightarrow P_i^{\mathrm{id}}.
$$

\end{itemize}
Moreover, the corresponding decision problem (i.e.\ $\overline P\sqsubseteq_{\mathbb{L}}^{\mathrm{md}}\overline Q$?) is $\Pi^p_2$-complete.
\end{restatable}

The proof can be found in Appendix~\ref{app:ucqneqn-containment-proofs}.

Next, we consider provenance-tracking containment for UCQ$^\neq$Ns over $\mathrm{Sorp}(X, \overline{X})$, where containment requires literal-injective $\neq$-negation-homomorphisms across all identifications $P_i^{\mathrm{id}}$.

\begin{restatable}{Thm}{ucqneqnContainmentB} \label{sorpucqneqncon}
For UCQ$^\neq$Ns, $\bigcup_{1\leq i \leq n} P_i, \bigcup_{1\leq i \leq m} Q_i$ with the same tuple of free variables,  the following are equivalent:
\begin{itemize}
\item[1.] $\bigcup_{1\leq i \leq n} P_i  \sqsubseteq_{\mathrm{Sorp}(X, \overline{X})}^{\text{pt}} \bigcup_{1\leq i \leq m} Q_i$,
\item[2.]
For each $1\leq i \leq n$, every identification (in the sense of CQ$^\neq$s) $P^{\mathrm{id}}_i$ of $P_i$, we have that  $$\pi^{\mathrm{can}_{\mathrm{Sorp}(X)}(P_i^{id})}\llbracket P_i^{\mathrm{id}} \rrbracket \leq_{\mathrm{Sorp}(X, \overline{X})} \pi^{\mathrm{can}_{\mathrm{Sorp}(X)}(P_i^{id})}\llbracket  \bigcup_{1\leq i \leq m} Q_i \rrbracket.$$
\item[3.]  For each $1\leq i \leq n$,   every identification $P^{\mathrm{id}}_i$ of $P_i$,  there is a $j \leq m$ and a $\neq$-negation-homomorphism 
$h_{P^{\mathrm{id}}_i} \colon Q_j \longrightarrow P^{\mathrm{id}}_i$  injective for literals.

\end{itemize}
Thus, the containment problem for UCQ$^\neq$Ns over $\mathrm{Sorp}(X, \overline{X})$  is $\Pi_2^p$-complete.
\end{restatable}

The proof can be found in Appendix~\ref{app:ucqneqn-containment-proofs}.

For $\mathrm{Why}(X, \overline{X})$, containment of UCQ$^\neq$Ns requires a set-onto $\neq$-negation-homomorphism across all identifications $P_i^{\mathrm{id}}$, retaining $\Pi_2^p$-completeness.

\begin{restatable}{Thm}{ucqneqnContainmentC} \label{whyucqneqncont}
For UCQ$^\neq$Ns $\bigcup_{1\leq i\leq n}P_i$, $\bigcup_{1\leq j\leq m}Q_j$ with the same tuple of free variables, the following are equivalent:
\begin{itemize}

\item[1.] $\bigcup_{1\leq i\leq n}P_i \sqsubseteq^{\text{pt}}_{\text{Why}(X,\overline X)}
\bigcup_{1\leq j\leq m}Q_j.$

\item[2.] For every $1\leq i\leq n$ and every identification $P_i^{id}$ of $P_i$ in the sense of CQ$^\neq$s,
$$
\pi^{\text{can}_{\text{Why}(X)}((P_i^{id})^+)}\llbracket P_i^{id} \rrbracket
\leq_{\text{Why}(X,\overline X)}
\pi^{\text{can}_{\text{Why}(X)}((P_i^{id})^+)}
\llbracket \bigcup_{1\leq j\leq m}Q_j \rrbracket.
$$

\item[3.] For every $1\leq i\leq n$ and every identification $P_i^{id}$ of $P_i$ in the sense of CQ$^\neq$s, there are some $1\leq j\leq m$ and a $\neq$-negation-homomorphism $h_{P_i^{id}}\colon Q_j\longrightarrow P_i^{id}$ set-onto for relational literals, i.e. $h_{P_i^{id}}\bigl(\text{LitSet}(Q_j)\bigr)= \text{LitSet}(P_i^{id}).$

\end{itemize}
Moreover, the corresponding containment problem is $\Pi_2^p$-complete.
\end{restatable}

The proof can be found in Appendix~\ref{app:ucqneqn-containment-proofs}.

For the Boolean polynomial semiring $\mathbb{B}[X, \overline{X}]$, UCQ$^\neq$N containment requires an exact $\neq$-negation-homomorphism across all identifications $P_i^{\mathrm{id}}$, retaining $\Pi^p_2$-completeness.

\begin{restatable}{Thm}{ucqneqnContainmentD} \label{Bucqneqncont}For UCQ$^\neq$Ns, $\bigcup_{1\leq i \leq n} P_i, \bigcup_{1\leq i \leq m} Q_i$ with the same tuple of free variables, the following are equivalent:
\begin{itemize}
\item[1.] $\bigcup_{1\leq i \leq n} P_i  \sqsubseteq_{\mathbb{B}[X, \overline{X}]}^{\text{pt}} \bigcup_{1\leq i \leq m} Q_i$.
\item[2.] For each $1\leq i \leq n$ and every identification $P_i^{id}$ of $P_i$, we have $\pi^{\mathrm{can}_{\mathbb{B}[X]}(P_i^{id})} \llbracket P_i^{id} \rrbracket  \leq_{\mathbb{B}[X, \overline{X}]} \pi^{\mathrm{can}_{\mathbb{B}[X]}(P_i^{id})} \llbracket \bigcup_{1\leq i \leq m} Q_i \rrbracket$.

\item[3.] For each $1\leq i \leq n$, every identification $P^{\mathrm{id}}_i$ of $P_i$,  there is a $j \leq m$ and a  $\neq$-negation-homomorphism 
$h_{P^{\mathrm{id}}_i} \colon Q_j \longrightarrow P^{\mathrm{id}}_i$  exact for literals.

\end{itemize}
Thus, the containment problem for UCQ$^\neq$Ns over $\mathbb{B}[X, \overline{X}]$  is $\Pi^p_2$-complete.
\end{restatable}

The proof can be found in Appendix~\ref{app:ucqneqn-containment-proofs}.

For $\mathrm{PosBool}(X, \overline{X})$, $\mathrm{UCQ}^\neq\mathrm{N}$ containment reduces to finding an unrestricted $\neq$-negation-homomorphism for each identification $P_i^{\mathrm{id}}$.

\begin{restatable}{Thm}{ucqneqnContainmentE} \label{posboolucqneqncont}
For UCQ$^\neq$Ns, $\bigcup_{1\leq i \leq n} P_i, \bigcup_{1\leq j \leq m} Q_j$ with the same tuple of free variables, the following are equivalent:
\begin{itemize}

\item[1.] 
$\bigcup_{1\leq i \leq n} P_i
\sqsubseteq^{\text{pt}}_{\mathrm{PosBool}(X, \overline{X})}
\bigcup_{1\leq j \leq m} Q_j$.

\item[2.]
For each $1\leq i\leq n$ and every identification (in the sense of
CQ$^\neq$s) $P_i^{id}$ of $P_i$, we have that
$$
\pi^{\mathrm{can}_{\mathrm{PosBool}(X)}(P_i^{id})} \llbracket P_i^{id} \rrbracket \leq_{\mathrm{PosBool}(X, \overline{X})} \pi^{\mathrm{can}_{\mathrm{PosBool}(X)}(P_i^{id})} \llbracket \bigcup_{1\leq j \leq m}Q_j
 \rrbracket.
$$

\item[3.]
For each $1\leq i\leq n$ and every identification (in the sense of
CQ$^\neq$s) $P_i^{id}$ of $P_i$, there is $j\leq m$ and a
$\neq$-negation-homomorphism $h_{P_i^{id}}\colon Q_j\longrightarrow P_i^{id}.$

\end{itemize}
Moreover, the corresponding decision problem is $\Pi^p_2$-complete.
\end{restatable}

The proof can be found in Appendix~\ref{app:ucqneqn-containment-proofs}.

We now extend our characterization of $\mathrm{UCQ}^\neq\mathrm{N}$ containment to model-defining interpretations over the $\mathrm{Which}(X)$ provenance semiring. As established in the following theorem, containment reduces to verifying that every inherited relational literal in each consistent identification $P_i^{\mathrm{id}}$ is covered by a $\neq$-negation-homomorphism from some target $Q_j$.

\begin{restatable}{Thm}{ucqneqnContainmentF} \label{whichucqneqncont}
For UCQ$^\neq$Ns, $\bigcup_{1\leq i \leq n} P_i, \bigcup_{1\leq j \leq m} Q_j$ with the same tuple of free variables, the following are equivalent. Here, an identification is formed by first taking an identification in the sense of CQ$^\neq$s and then a consistent relational completion over the common schema:
\begin{itemize}

\item[1.] 
$\bigcup_{1\leq i \leq n} P_i
\sqsubseteq^{\text{md}}_{\mathrm{Which}(X)}
\bigcup_{1\leq j \leq m} Q_j$.

\item[2.]
For each $1\leq i \leq n$ and every consistent identification $P_i^{id}$ of $P_i$, we have that
$$
\pi^{P_i^{id}}_{\mathrm{Which}(X)}(P_i)
\leq_{\mathrm{Which}(X)}
\pi^{P_i^{id}}_{\mathrm{Which}(X)}
\left(\bigcup_{1\leq j \leq m} Q_j\right).
$$

\item[3.]
For each $1\leq i \leq n$, every consistent identification $P_i^{id}$ of $P_i$, and every relational literal $L$ of $P_i^{id}$ inherited from $P_i$, there is a $j\leq m$ and a $\neq$-negation-homomorphism $ h_{P_i^{id},L}\colon Q_j \longrightarrow P_i^{id} $ such that $L$ belongs to the image of $h_{P_i^{id},L}$.

\end{itemize}
Moreover, the corresponding decision problem is
\begin{itemize}
\item[(i)] $\Pi^p_2$-complete for queries over schemas with arity bound $k$, for every $k$, and
\item[(ii)] coNEXP-complete in general.
\end{itemize}
\end{restatable}

The proof can be found in Appendix~\ref{app:ucqneqn-containment-proofs}.

For $\mathbb{N}[X, \overline{X}]$, the additive structure requires comparing the total number of exact $\neq$-negation-homomorphisms contributed by all disjuncts across each identification $P_i^{\mathrm{id}}$, placing $\mathrm{UCQ}^\neq\mathrm{N}$ containment in $\text{coNP}^{\text{PP}}$.

\begin{restatable}{Thm}{ucqneqnContainmentG}\label{N[X]ucqneqncontainment}
Let $\overline P=\bigcup_{1\leq i\leq n}P_i$ and $\overline Q=\bigcup_{1\leq j\leq m}Q_j$ be UCQ$^\neq$Ns with the same tuple of free variables. The following are equivalent:
\begin{itemize}

\item[1.] $\overline P\sqsubseteq^{\mathrm{pt}}_{\mathbb{N}[X,\overline X]}\overline Q$.

\item[2.] For every $1\leq i\leq n$ and every identification $P_i^{\mathrm{id}}$ of $P_i$ in the sense of CQ$^\neq$s,
$$
\sum_{1\leq \ell\leq n}
\left|\mathrm{ExNegHom}(P_\ell,P_i^{\mathrm{id}})\right|
\leq
\sum_{1\leq j\leq m}
\left|\mathrm{ExNegHom}(Q_j,P_i^{\mathrm{id}})\right|.
$$

\end{itemize}
Moreover, the corresponding containment problem is in $\text{coNP}^{\text{PP}}$.
\end{restatable}

The proof can be found in Appendix~\ref{app:ucqneqn-containment-proofs}.

For $\mathrm{Trio}(X, \overline{X})$, dropping positive exponents modifies the disjunct-wise mapping requirement to set-onto $\neq$-negation-homomorphisms, while additive aggregation yields an analogous sum-comparison condition in $\text{coNP}^{\text{PP}}$.

\begin{restatable}{Thm}{ucqneqnContainmentH}\label{Trioucqneqncontainment}
Let $\overline P=\bigcup_{1\leq i\leq n}P_i$ and $\overline Q=\bigcup_{1\leq j\leq m}Q_j$ be UCQ$^\neq$Ns with the same tuple of free variables. The following are equivalent:
\begin{itemize}

\item[1.] $\overline P\sqsubseteq^{\mathrm{pt}}_{\mathrm{Trio}(X,\overline X)}\overline Q$.

\item[2.] For every $1\leq i\leq n$ and every identification $P_i^{\mathrm{id}}$ of $P_i$ in the sense of CQ$^\neq$s,
$$
\sum_{1\leq \ell\leq n}
\left|\mathrm{SOHom}(P_\ell,P_i^{\mathrm{id}})\right|
\leq
\sum_{1\leq j\leq m}
\left|\mathrm{SOHom}(Q_j,P_i^{\mathrm{id}})\right|.
$$

\end{itemize}
Moreover, the corresponding containment problem is in $\text{coNP}^{\text{PP}}$.
\end{restatable}

The proof can be found in Appendix~\ref{app:ucqneqn-containment-proofs}.

\subsection{Equivalence of UCQ\texorpdfstring{$^\neq$}{≠}Ns}\label{sec:ucqneqn-equivalence}
  We now turn to the equivalence problem for UCQ$^\neq$Ns over the semirings considered above. We establish exact complexity bounds for bounded distributive lattices, $\mathrm{Which}(X)$, $\mathrm{PosBool}(X,\overline{X})$, $\mathrm{Sorp}(X,\overline{X})$, $\mathrm{Why}(X,\overline{X})$, and $\mathbb{B}[X,\overline{X}]$. For $\mathbb{N}[X,\overline{X}]$ and $\mathrm{Trio}(X,\overline{X})$, we provide upper bounds only. As with the case of UCQ$^\neq$s, we can state the following useful result (the proof is exactly as Proposition~\ref{propogreen2}):

\begin{Pro}\label{propogreen21}For UCQ$^\neq$Ns $\bar{P}, \bar{Q}$, idempotent semiring $\mathbb{S}$ (which includes $\mathbb{B}[X, \overline{X}]$, $\mathrm{Why}(X, \overline{X})$, $\mathrm{Sorp}(X, \overline{X})$, bounded distributive lattices, $\mathrm{PosBool}(X, \overline{X})$ and $\mathrm{Which}(X)$, but excludes $\mathbb{N}[X, \overline{X}]$ and $\mathrm{Trio}(X, \overline{X})$) and class of interpretations $\mathcal{C}$, we have
\begin{itemize}
\item[(1)] $\bar{P}\sqsubseteq^\mathcal{C}_\mathbb{S}\bar{Q}$ iff $\bar{P} \cup \bar{Q} \equiv^\mathcal{C}_\mathbb{S}\bar{Q}$.

\item[(2)] $\bar{P}\equiv^\mathcal{C}_\mathbb{S}\bar{Q}$ iff $\bar{P}\sqsubseteq^\mathcal{C}_\mathbb{S}\bar{Q}$ and $\bar{Q}\sqsubseteq^\mathcal{C}_\mathbb{S}\bar{P}$.

\end{itemize}

\end{Pro}
Using this proposition, we  immediately obtain complexity bounds for the equivalence problems we are interested in.

\begin{restatable}{Thm}{ucqneqnEquivalenceA} \label{ucqneqnequivalence}
For UCQ$^\neq$Ns $\overline{P}$ and $\overline{Q}$ with the same tuple of free variables, the following statements hold.
\begin{enumerate}
\item ``$\overline{P} \equiv_{\mathbb{L}}^{\text{md}} \overline{Q}$?'' (where $\mathbb{L}$ is a bounded distributive lattice with more than two elements) is $\Pi^p_2$-complete.
\item ``$\overline{P} \equiv_{\mathbb{N}[X,\overline{X}]}^{\text{pt}} \overline{Q}$?'' is in coNP$^{\text{PP}}$.
\item ``$\overline{P} \equiv_{\mathrm{Sorp}(X,\overline{X})}^{\text{pt}} \overline{Q}$?'' is $\Pi^p_2$-complete.
\item ``$\overline{P} \equiv_{\mathbb{B}[X,\overline{X}]}^{\text{pt}} \overline{Q}$?'' is $\Pi^p_2$-complete.
\item ``$\overline{P} \equiv_{\mathrm{Why}(X,\overline{X})}^{\text{pt}} \overline{Q}$?'' is $\Pi^p_2$-complete.
\item ``$\overline{P} \equiv_{\mathrm{Trio}(X,\overline{X})}^{\text{pt}} \overline{Q}$?'' is in coNP$^{\text{PP}}$.
\item ``$\overline{P} \equiv_{\mathrm{PosBool}(X,\overline{X})}^{\text{pt}} \overline{Q}$?'' is $\Pi^p_2$-complete.
\item ``$\overline{P} \equiv_{\mathrm{Which}(X)}^{\text{md}} \overline{Q}$?'' is
\begin{itemize}
\item[(i)] $\Pi^p_2$-complete for queries over schemas with a fixed arity bound, and
\item[(ii)] coNEXP-complete in general.
\end{itemize}
\item ``$\overline{P} \equiv_{\mathbb{N}}^{\text{md}} \overline{Q}$?'' is in coNP$^{\text{PP}}$ (Remark~\ref{bag-equivalence-bridge}).
\end{enumerate}
\end{restatable}

The proof can be found in Appendix~\ref{app:ucqneqn-equivalence-proofs}.

\section{Concluding remarks}

In addition to the work presented here, we already have  results on containment for CQNs and CQ$^\neq$s  for  the tropical, Viterbi, the arctic, and \L ukasiewicz semirings, which we will report on in a separate paper. We have exact complexity characterizations of all these problems. For CQNs over model-defining interpretations, the complexity stays identical to the case of CQs (NP-complete~\cite{Kostylev, Carb}) while for CQ$^\neq$s it becomes $\Pi^p_2$-complete.

However, there are still  various directions and open questions for future research:
\begin{itemize}

\item[1.] A number of alternative notions of negation exist in the literature~\cite{Ba,GT, Suciu}. It is natural to study containment of CQNs for those other notions. What is the complexity of the containment problem in those cases?

\item[2.] In some cases, our calculations only give upper bounds for the complexity. Can we produce exact calculations?

\item[3.] Another issue is containment for other classes of interpretations different from the ones considered here (provenance-tracking and  model-defining interpretations). Can the complexity of the problem increase for some natural classes?

\item[4.] Finally, can we produce a deeper algebraic analysis of the containment problem for queries with negative information along the lines of~\cite{Kostylev}?

\end{itemize}

\section*{Acknowledgments} GB and CN wish to thank Phokion Kolaitis for very insightful conversations regarding~\cite{Green}.

\appendix

\section{Omitted  Proofs from \S~\ref{Dual indeterminate polynomial semirings},  \ref{Canonical databases and interpretations} and \ref{Homomorphisms for containment}}
\label{app:proofs1}

In this Appendix we include full statements and proofs of all results that have only been stated  in the body of the paper.

\firstprop*

\begin{proof}
The natural orders on the dual semirings are inherited from the natural orders of the corresponding positive semirings over \(X\cup\overline{X}\), restricted to consistent representatives. We now check, case by case, that these induced preorders coincide with the natural preorders and are partial orders, hence all these semirings are naturally ordered.
\begin{itemize}
    \item For \(\mathbb{N}[X,\overline{X}]\), every equivalence class \([P]\) has a unique consistent representative \(P^c\), obtained by deleting all monomials containing a complementary pair \(x,\overline{x}\). We define $ [P]\leq_{\mathbb{N}[X,\overline{X}]}[Q]$ iff $P^c\leq_{\mathbb{N}[X\cup\overline{X}]}Q^c$, where the natural order on \(\mathbb{N}[X\cup\overline{X}]\) is the usual coefficientwise natural order. We show that this is the natural order on \(\mathbb{N}[X,\overline{X}]\). Assume first that \([P]\leq_{\mathbb{N}[X,\overline{X}]}[Q]\), then \(P^c\leq_{\mathbb{N}[X\cup\overline{X}]}Q^c\), and hence there exists an \(R\in\mathbb{N}[X\cup\overline{X}]\) such that $P^c+R=Q^c$; since \(P^c\) and \(Q^c\) contain only consistent monomials, \(R\) can be chosen as the coefficientwise difference \(Q^c-P^c\), and hence it also contains only consistent monomials; therefore \([R]\in\mathbb{N}[X,\overline{X}]\) and $[P]+[R]=[Q].$ Conversely, if there exists \([R]\in\mathbb{N}[X,\overline{X}]\) such that $[P]+[R]=[Q],$ then, by taking consistent representatives, we have $P^c+R^c=Q^c$, thus \(P^c\leq_{\mathbb{N}[X\cup\overline{X}]}Q^c\), and hence \([P]\leq_{\mathbb{N}[X,\overline{X}]}[Q]\). It remains to check that this relation is a partial order. This follows from the fact that \(\leq_{\mathbb{N}[X\cup\overline{X}]}\) is a partial order on consistent representatives. Indeed, reflexivity and transitivity are immediate; as for antisymmetry, if \([P]\leq[Q]\) and \([Q]\leq[P]\), then \(P^c\leq Q^c\) and \(Q^c\leq P^c\); hence \(P^c=Q^c\), and therefore \([P]=[Q]\). Thus, \(\mathbb{N}[X,\overline{X}]\) is naturally ordered.
\item For \(\mathbb{B}[X,\overline{X}]\), every equivalence class \([P]\) has a unique consistent representative \(P^c\), obtained by deleting all monomials containing a complementary pair \(x,\overline{x}\). We define $[P]\leq_{\mathbb{B}[X,\overline{X}]}[Q] $ iff $P^c\leq_{\mathbb{B}[X\cup\overline{X}]}Q^c,$ where the order on \(\mathbb{B}[X\cup\overline{X}]\) is the natural order (inclusion of monomials). To show that this is the natural order on \(\mathbb{B}[X,\overline{X}]\), assume first that \([P]\leq_{\mathbb{B}[X,\overline{X}]}[Q]\). Then \(P^c\leq_{\mathbb{B}[X\cup\overline{X}]}Q^c\); hence, by the natural order of \(\mathbb{B}[X\cup\overline{X}]\), there exists \(R\in\mathbb{B}[X\cup\overline{X}]\) such that $P^c+R=Q^c$; since \(P^c\) and \(Q^c\) contain only consistent monomials, \(R\) can be chosen as the set of monomials occurring in \(Q^c\) and not in \(P^c\); hence \(R\) also contains only consistent monomials, so \([R]\in\mathbb{B}[X,\overline{X}]\) and $[P]+[R]=[Q]$. Conversely, suppose that there exists \([R]\in\mathbb{B}[X,\overline{X}]\) such that $[P]+[R]=[Q]$; taking consistent representatives, this gives $P^c+R^c=Q^c$, thus \(P^c\leq_{\mathbb{B}[X\cup\overline{X}]}Q^c\), and therefore $[P]\leq_{\mathbb{B}[X,\overline{X}]}[Q]$. Hence the order defined above coincides with the natural preorder of \(\mathbb{B}[X,\overline{X}]\). It remains to check that it is a partial order. Reflexivity and transitivity are immediate, for antisymmetry, if \([P]\leq[Q]\) and \([Q]\leq[P]\), then \(P^c\leq Q^c\) and \(Q^c\leq P^c\); hence \(P^c=Q^c\), and therefore \([P]=[Q]\). Thus, \(\mathbb{B}[X,\overline{X}]\) is naturally ordered.
\item For \(\mathrm{Trio}(X,\overline{X})\), every equivalence class \([P]\) has a unique representative \(P^c\) whose monomials are square-free and contain no complementary pair \(x,\overline{x}\). We call such representatives consistent. We define $[P]\leq_{\mathrm{Trio}(X,\overline{X})}[Q]$ iff $P^c\leq_{\mathrm{Trio}(X\cup\overline{X})}Q^c,$ where the natural order on \(\mathrm{Trio}(X\cup\overline{X})\) is the coefficientwise order on square-free monomials. We show that this is the natural order on \(\mathrm{Trio}(X,\overline{X})\). Assume first that $[P]\leq_{\mathrm{Trio}(X,\overline{X})}[Q]$, then $P^c\leq_{\mathrm{Trio}(X\cup\overline{X})}Q^c$; since \(\mathrm{Trio}(X\cup\overline{X})\) is naturally ordered, there exists \(R\in\mathrm{Trio}(X\cup\overline{X})\) such that $P^c+R=Q^c$; moreover, \(R\) can be chosen as the coefficientwise difference \(Q^c-P^c\); since both \(P^c\) and \(Q^c\) contain only square-free monomials with no complementary pair, the same holds for \(R\), hence \([R]\) is an element of \(\mathrm{Trio}(X,\overline{X})\), and $[P]+[R]=[Q]$. Conversely, suppose that there exists \([R]\in\mathrm{Trio}(X,\overline{X})\) such that $[P]+[R]=[Q]$; taking consistent representatives, we get $P^c+R^c=Q^c$; thus $P^c\leq_{\mathrm{Trio}(X\cup\overline{X})}Q^c$, and therefore $[P]\leq_{\mathrm{Trio}(X,\overline{X})}[Q]$. Hence the order defined above coincides with the natural preorder of \(\mathrm{Trio}(X,\overline{X})\). This relation is a partial order: reflexivity and transitivity follow from the natural order on \(\mathrm{Trio}(X\cup \overline{X})\). For antisymmetry, assume that $[P]\leq_{\mathrm{Trio}(X,\overline{X})}[Q]$ and $[Q]\leq_{\mathrm{Trio}(X,\overline{X})}[P]$, then $P^c\leq_{\mathrm{Trio}(X\cup\overline{X})}Q^c$ and $Q^c\leq_{\mathrm{Trio}(X\cup\overline{X})}P^c$; since the coefficientwise order is antisymmetric, we have \(P^c=Q^c\), and therefore \([P]=[Q]\). Thus, \(\mathrm{Trio}(X,\overline{X})\) is naturally ordered.
\item For \(\mathrm{Sorp}(X,\overline{X})\), every equivalence class \([P]\) has a unique consistent representative \(P^c\), obtained by removing all monomials containing a complementary pair \(x,\overline{x}\), and then applying the absorption law \(p+pq=p\), so that no monomial properly divides another. We define $[P]\leq_{\mathrm{Sorp}(X,\overline{X})}[Q]$ iff $P^c\leq_{\mathrm{Sorp}(X\cup\overline{X})}Q^c,$ where the natural order on \(\mathrm{Sorp}(X\cup\overline{X})\) is the absorption order.  Every monomial of \(P^c\) is absorbed by some monomial of \(Q^c\), that is, for every monomial \(n\) of \(P^c\), there exists a monomial \(m\) of \(Q^c\) such that \(m\) divides \(n\). We show that this is the natural order on \(\mathrm{Sorp}(X,\overline{X})\). Assume first that $[P]\leq_{\mathrm{Sorp}(X,\overline{X})}[Q]$, then $P^c\leq_{\mathrm{Sorp}(X\cup\overline{X})}Q^c$; since \(\mathrm{Sorp}(X\cup\overline{X})\) is naturally ordered, there exists \(R\in\mathrm{Sorp}(X\cup\overline{X})\) such that $P^c+R=Q^c$. In fact, since addition in \(\mathrm{Sorp}(X\cup\overline{X})\) is idempotent, we can just take \(R=Q^c\); so \([R]\) is an element of \(\mathrm{Sorp}(X,\overline{X})\), and we get $[P]+[R]=[Q]$. Conversely, suppose that there exists \([R]\in\mathrm{Sorp}(X,\overline{X})\) such that $[P]+[R]=[Q]$. Taking consistent representatives, this means that the representative of \(P^c+R^c\) is \(Q^c\); hence, in \(\mathrm{Sorp}(X\cup\overline{X})\), there exists \(R^c\) such that $P^c+R^c=Q^c$; therefore $P^c\leq_{\mathrm{Sorp}(X\cup\overline{X})}Q^c$, and so $[P]\leq_{\mathrm{Sorp}(X,\overline{X})}[Q]$. Hence the order defined above coincides with the natural preorder of \(\mathrm{Sorp}(X,\overline{X})\). This relation is also a partial order. Reflexivity and transitivity follow from the natural order on \(\mathrm{Sorp}(X\cup\overline{X})\); for antisymmetry, assume that $[P]\leq_{\mathrm{Sorp}(X,\overline{X})}[Q]$ and $[Q]\leq_{\mathrm{Sorp}(X,\overline{X})}[P]$, then $P^c\leq_{\mathrm{Sorp}(X\cup\overline{X})}Q^c$ and $Q^c\leq_{\mathrm{Sorp}(X\cup\overline{X})}P^c$; since the natural order on \(\mathrm{Sorp}(X\cup\overline{X})\) is antisymmetric, we have \(P^c=Q^c\), and therefore \([P]=[Q]\). Thus, \(\mathrm{Sorp}(X,\overline{X})\) is naturally ordered.
\item For \(\mathrm{Why}(X,\overline{X})\), we define the order by using the natural order of \(\mathrm{Why}(X\cup\overline{X})\) restricted to consistent witnesses. For \(A,B\in\mathcal{P}_{\mathrm{fin}}(\mathcal{C}(X,\overline{X}))\), we set $A\leq_{\mathrm{Why}(X,\overline{X})}B$ iff $A\subseteq B$. We show that this is the natural order on \(\mathrm{Why}(X,\overline{X})\). Assume first that $A\leq_{\mathrm{Why}(X,\overline{X})}B$, then \(A\subseteq B\); taking \(C=B\), we have $A\cup C=A\cup B=B$; hence there exists \(C\in\mathrm{Why}(X,\overline{X})\) such that \(A+C=B\). Conversely, suppose that there exists \(C\in\mathrm{Why}(X,\overline{X})\) such that $A+C=B$; since addition in \(\mathrm{Why}(X,\overline{X})\) is union, this means $A\cup C=B$; therefore \(A\subseteq B\), and so $A\leq_{\mathrm{Why}(X,\overline{X})}B$; thus, the order defined above coincides with the natural preorder. It remains to check that it is a partial order. This follows immediately from the fact that \(\subseteq\) is a partial order.
\item For \(\mathrm{PosBool}(X,\overline{X})\), we write \([\varphi]\) for the \(\approx\)-equivalence class of a positive formula \(\varphi\). We define the order by restricting the natural order of \(\mathrm{PosBool}(X\cup\overline{X})\) to compatible valuations. Namely, for \([\varphi],[\psi]\in\mathrm{PosBool}(X,\overline{X})\), we set $[\varphi]\leq_{\mathrm{PosBool}(X,\overline{X})}[\psi]$ iff, for every valuation \(\nu\colon X\cup\overline{X}\longrightarrow\{0,1\}\) such that \(\nu(x)\cdot\nu(\overline{x})=0\) for every \(x\in X\), we have $\nu(\varphi)=1$ implies $\nu(\psi)=1.$ We show that this is the natural order on \(\mathrm{PosBool}(X,\overline{X})\). Assume first that $[\varphi]\leq_{\mathrm{PosBool}(X,\overline{X})}[\psi]$, then \(\varphi\) implies \(\psi\) under all compatible valuations, so $\varphi\vee\psi\approx\psi$; therefore, taking \([\theta]=[\psi]\), we obtain $[\varphi]\vee[\theta]=[\varphi]\vee[\psi]=[\psi]$; thus, there exists \([\theta]\in\mathrm{PosBool}(X,\overline{X})\) such that $[\varphi]+[\theta]=[\psi]$. Conversely, suppose that there exists \([\theta]\in\mathrm{PosBool}(X,\overline{X})\) such that $[\varphi]+[\theta]=[\psi]$; since addition is \(\vee\), this means $[\varphi\vee\theta]=[\psi]$; let \(\nu\) be any compatible valuation, if \(\nu(\varphi)=1\), then $\nu(\varphi\vee\theta)=1$; since \(\varphi\vee\theta\approx\psi\), we get \(\nu(\psi)=1\); hence $[\varphi]\leq_{\mathrm{PosBool}(X,\overline{X})}[\psi].$ So the order defined above coincides with the natural preorder. It remains to check that this relation is a partial order. Reflexivity and transitivity are immediate from logical implication over compatible valuations. For antisymmetry, suppose that $[\varphi]\leq[\psi]$ and $[\psi]\leq[\varphi]$, then \(\varphi\) and \(\psi\) have the same truth value under every compatible valuation, hence \(\varphi\approx\psi\), and therefore $[\varphi]=[\psi].$ Thus, \(\mathrm{PosBool}(X,\overline{X})\) is naturally ordered.\qedhere
\end{itemize}
\end{proof}

\propseventwetyfour*

\begin{proof} Obviously, $a=b$ implies $\operatorname{Eval}_v(a)=\operatorname{Eval}_v(b)$. For the converse, suppose that $\operatorname{Eval}_v(a)=\operatorname{Eval}_v(b)$. Since $v$ is injective, for every variable $x_i$ occurring in $a$ or $b$, we may write \[ v(x_i)=x_{r_i,1}+\ldots+x_{r_i,k}, \] where the variables $x_{r_i,1},\ldots,x_{r_i,k}$ are the members of the corresponding element of $S_k$, and where the sets \[ \{x_{r_i,1},\ldots,x_{r_i,k}\} \] corresponding to distinct variables $x_i$ are pairwise disjoint. We show, by descending induction on the degree, that every monomial has the same coefficient in $a$ and in $b$. Let \[ \mu=x_{i_1}^{e_1}\cdot \ldots \cdot x_{i_m}^{e_m}, \] where $i_1<\ldots<i_m$, be a monomial of degree $d=e_1+\ldots+e_m\leq k$, and consider the square-free monomial \[ \tau_\mu= \prod_{j=1}^m\prod_{\ell=1}^{e_j}x_{r_{i_j},\ell}. \] The monomial $\mu$ contributes to $\tau_\mu$ in $\operatorname{Eval}_v(\mu)$ with coefficient $e_1!\cdot \ldots \cdot e_m!>0$. No different monomial of degree $d$ can contribute to $\tau_\mu$: in order to produce the $e_j$ distinct variables selected from the block associated with $x_{i_j}$, it must contain $x_{i_j}$ with exponent at least $e_j$, and equality of the total degrees then forces all these exponents to be exactly $e_j$. Moreover, a monomial of degree smaller than $d$ cannot contribute to $\tau_\mu$. For $d=k$, it follows directly from $\operatorname{Eval}_v(a)=\operatorname{Eval}_v(b)$ that the coefficient of $\mu$ is the same in $a$ and in $b$. Suppose now that the claim has already been proved for all monomials of degree strictly greater than $d$. Monomials of higher degree may also contribute to $\tau_\mu$, since exponents are dropped in $\mathrm{Trio}(X)$, but by the induction hypothesis their total contribution is the same on the two sides. After removing these equal contributions, the remaining contribution to $\tau_\mu$ comes only from $\mu$. Hence, if $\alpha_\mu$ and $\beta_\mu$ are the coefficients of $\mu$ in $a$ and $b$, respectively, we obtain \[ \alpha_\mu e_1!\cdot \ldots \cdot e_m! = \beta_\mu e_1!\cdot \ldots \cdot e_m!, \] and therefore $\alpha_\mu=\beta_\mu$. The degree-zero case follows in the same way from the coefficient of $1$. Thus $a=b$. The uniqueness of the decomposition of every element of $\mathrm{Trio}_k(X)$ follows immediately. \end{proof}

\kabstractCQ*

\begin{proof}
This follows by the same argument used for $\mathrm{Trio}(X)$ in~\cite{Green}. Let $\mathbf I^{\prime}=\mathrm{ab}_k(\mathbf I)$, and let $v\colon S_k\longrightarrow \mathrm{Trio}(X)$ record how the annotations of $\mathbf I$ that have been replaced by elements of $S_k$, so that applying $v$ to $\mathbf I^{\prime}$ gives back $\mathbf I$. By the same argument as in~\cite[Lemma 7.25]{Green}, the map $v$ extends uniquely to the computations in $\mathrm{Trio}_k(X)$ and behaves as a semiring homomorphism as long as these computations remain of degree at most $k$; moreover, this extension is compatible with the natural order and commutes with query evaluation. Since $P$ and $Q$ have at most $k$ relational atom occurrences, every contribution to their evaluation is a product of at most $k$ tuple annotations. Equations and disequations contribute only $0$ or $1$, and hence do not increase the degree of these computations. Therefore, all the relevant calculations over $\mathbf I^{\prime}$ belong to $\mathrm{Trio}_k(X)$. Consequently, $\llbracket P \rrbracket^{\mathbf I^{\prime}} \leq_{\mathrm{Trio}(X)} \llbracket Q \rrbracket^{\mathbf I^{\prime}}$ would imply $\llbracket P \rrbracket^\mathbf I
\leq_{\mathrm{Trio}(X)} \llbracket Q \rrbracket^\mathbf I,$ contrary to the assumption.
\end{proof}

\contnegdisbool*

\begin{proof}$(1) \implies (2)$: Assume that $P \sqsubseteq^{\text{md}}_{\mathbb{B}}  Q$. Given a consistent identification $P^{\mathrm{id}}$ of a completion $P'$ of $P$, consider the canonical database $\mathrm{can}_{\mathbb{B}}(P^{id+})$. If $P^{\mathrm{id}}$ is the formula $\phi(u)$, $u$ is an answer to the query $P^{\mathrm{id}}$ in  $\mathrm{can}_{\mathbb{B}}(P^{id+})$. Since clearly $P^{\mathrm{id}} \sqsubseteq^{\text{md}}_{\mathbb{B}}  P' \sqsubseteq^{\text{md}}_{\mathbb{B}}  P$, $u$ is an answer to $P$ in $\mathrm{can}_{\mathbb{B}}(P^{id+})$ and by the containment assumption it must be an answer to $Q$ in $\mathrm{can}_{\mathbb{B}}(P^{id+})$. This means that there is a $\neq$-negation-homomorphism $h$ from $Q$  to $P^{\mathrm{id}}$ by sending each variable $x$ of  $Q$ to a variable $y$ of $P^{\mathrm{id}}$  such that $y$ is a witness of $x$ in $\mathrm{can}_{\mathbb{B}}(P^{id+})$. For if $R(x_1, \ldots, x_n)$ is an atom in $Q$, we have that $(y_1, \ldots, y_n) \in R^{\mathrm{can}_{\mathbb{B}}(P^{id+})}$, and thus $R(y_1, \ldots, y_n)$ is an atom of $P^{\mathrm{id}}$, while if $\neg R(x_1, \ldots, x_n)$ is a negated atom in $Q$, we have that $(y_1, \ldots, y_n) \notin R^{\mathrm{can}_{\mathbb{B}}(P^{id+})}$, and thus $\neg R(y_1, \ldots, y_n)$ is an atom of $P^{\mathrm{id}}$. Moreover, if the equation (disequation)  $l(x_1,x_2)$ appears in  $Q$, $l(y_1,y_2)$ holds for the elements $y_1,y_2$ of $\mathrm{can}_{\mathbb{B}}(P^{id+})$  that act as witnesses for $x_1,x_2$,  and thus 
the atom $l(h(x), h(y))$ is a logical consequence of the body of  $P^{\mathrm{id}}$.

$(2) \implies (1)$: Assume that (2) holds, that is, for every consistent identification $P^{\mathrm{id}}$ of $P$, there is a  $\neq$-negation-homomorphism 
$h_{P^{\mathrm{id}}} \colon Q \longrightarrow P^{\mathrm{id}}$. Suppose for a contradiction now that $P \not\sqsubseteq^{\text{md}}_{\mathbb{B}} Q$. Thus, there is a database instance $\mathbf I$ such that $\vec{a} \in P^{\bf I}$ but $\vec{a} \notin Q^{\bf I}$ for some tuple $\vec{a}$. Consider the  CQ$^\neq$N,  denoted by $O(\vec{a})$, obtained by building the conjunction of all literals of the form $R(\overline{b})$ (where $\overline{b}$ is a tuple in $R^\mathbf{I}$), $\neg R(\overline{b})$ (where $\overline{b}$ is not a tuple in $R^\mathbf{I}$ and all the elements of $\overline{b}$ are in the active domain of $\mathbf I$),  and $b\neq c$  (for every two elements $b,c$ in the active domain of $\mathbf I$), and quantifying existentially every variable different from those in the tuple $\vec{a}$. Since $\vec{a} \in P^\mathbf{I}$, there is clearly a $\neq$-homomorphism 
$h_{P}\colon P \longrightarrow O$ by sending the variables in $P$ to the elements that witness them in $\mathbf I$. Similarly, since $\vec{a} \notin Q^\mathbf{I}$, there is no  $\neq$-negation-homomorphism 
$h_{Q}\colon Q \longrightarrow O$. Let $P^{\mathrm{id}}$ be a consistent identification obtained by completing $P$  by adding $x\neq y$ if  $h_{P}(x)\neq h_{P}(y)$ is in $O$,  or $x = y$ if  $h_{P}(x)= h_{P}(y)$ is in $O$, then identifying variables according to a canonical substitution $\pi$, and, finally, adding the relational  literal $l(\overline {x})$ every time $l(h_{P}(\overline {x}))$ is in $O$, for all variables $x,y, \overline{x}$ of $P$. Now, for any such identification $P^{\mathrm{id}}$,  there is a $\neq$-negation-homomorphism 
$h_{P^{\mathrm{id}}}\colon P^{\mathrm{id}} \longrightarrow O$ obtained by sending each variable $\pi(y)$ in $P^{\mathrm{id}}$ to $h_{P}(y)$ (this is well-defined as when $\pi(y) = \pi(z)$, we must have $y=z$ as a logical consequence of the body of $P^{\mathrm{id}}$, and then  we have that $h_{P}(y) = h_{P}(z)$). To see this, suppose that $R(x_1, \ldots, x_n)$ is an atom in $P^{\mathrm{id}}$ (where $x_1, \ldots, x_n$ are not necessarily distinct). Then, $R(y_1, \ldots, y_n)$  (where again $y_1, \ldots, y_n$ are not necessarily distinct) must be an atom of $P^{\mathrm{id}}$ where $\pi(y_i)= x_i$, and thus $R(h_{P}(y_1), \ldots, h_{P}(y_n))$ is an atom of $O$ (similarly for  $\neg R(x_1, \ldots, x_n)$ and $\neg R(h_{P}(y_1), \ldots, h_{P}(y_n))$). Moreover, if $\pi(y) \neq  \pi(z)$ appears in $P^{\mathrm{id}}$, this means that we must have $y\neq z$ as a logical consequence of the body of $P^{\mathrm{id}}$, and $h_{P}(y)\neq h_{P}(z)$ appears $O$.
By hypothesis,  there is a  $\neq$-negation-homomorphism 
$g_{P^{\mathrm{id}}}\colon Q \longrightarrow P^{\mathrm{id}}$. But  then  $g_{P^{\mathrm{id}}}\circ h_{P^{\mathrm{id}}}$ is a $\neq$-negation-homomorphism 
from $Q$ to $O$, which is a contradiction.

The lower bound for the complexity calculation follows immediately from the hardness of the problem for  CQNs.  The upper bound follows by the same kind of reasoning as for CQNs.
\end{proof}

\section{Omitted  Proofs from \S~\ref{CQneqsection1}} \label{app:proofs}

 We start by analyzing the simplest case of our problem, namely, when the characterization of containment coincides with that of the Boolean case.

\begin{restatable}{Thm}{contdisdistlatt} \label{contdisdistlatt}
Let $P$ and $Q$ be CQ$^\neq$s 
with the same tuple of free variables, and\/ $\mathbb{L}$ a bounded distributive lattice.  Then, the following are equivalent:
\begin{itemize}
\item[1.]  $P \sqsubseteq_\mathbb{L} Q$.

\item[2.] For every identification $P^{\mathrm{id}}$ of $P$, there is a  $\neq$-homomorphism 
$h_{P^{\mathrm{id}}} \colon Q \longrightarrow P^{\mathrm{id}}$.

\end{itemize}
Moreover, the corresponding decision problem (i.e.\ $P \sqsubseteq_{\mathbb{L}} Q$?) is $\Pi^p_2$-complete.
\end{restatable}

\begin{proof}

Let $Q(\vec{x})$ be  $\exists \vec{z} (A_1(\vec{x_1}) \wedge \ldots \wedge A_n(\vec{ x_n}))$ and $P(\vec{z})$ be $\exists \vec{u} (B_1(\vec{ y_1}) \wedge \ldots \wedge B_m(\vec{ y_m}))$, where $\overline{x}, \vec{z} $ and $\vec{z}, \vec{u}$ form partitions of $\vec{x_1}, \ldots , \vec{ x_n}$ and $\vec{z_1}, \ldots , \overline{ y_m}$, respectively.

 $(1) \implies (2)$: Suppose that $P \sqsubseteq_\mathbb{L} Q$. Consider an identification $P^{\mathrm{id}}$ of a completion $P'$ of $P$. Now build a canonical database $\mathrm{can}_{\mathbb{B}}(P^{\mathrm{id}})$ as in the Boolean case (Definition~\ref{candataB}). We have that $P^{\mathrm{id}} \sqsubseteq_\mathbb{L} P' \sqsubseteq_\mathbb{L} P$ since the possible values of the new literals of $P'$ are $\top$ and $\bot$, which means that the semiring value in any database instance of $P^{\mathrm{id}}$  can only be at most that of $P$. By construction, $  \llbracket P^{\mathrm{id}} \rrbracket^{\mathrm{can}_{\mathbb{B}}(P^{\mathrm{id}})}(\vec{z}) = \top$, and thus, $\llbracket P \rrbracket^{\mathrm{can}_{\mathbb{B}}(P^{\mathrm{id}})}(\vec{z}) = \top$ and from the hypothesis, $\llbracket Q \rrbracket^{\mathrm{can}_{\mathbb{B}}(P^{\mathrm{id}})}(\overline{x}) = \top$ since the lattice is bounded. Consequently,  for some assignment $v$, it must be that $\llbracket A_1(v(\vec{ x_1})) \wedge \ldots \wedge A_n(v(\vec{ x_n})) \rrbracket^{\mathrm{can}_{\mathbb{B}}(P^{\mathrm{id}})}=\top$ and thus $\llbracket A_i(v(\vec{ x_i}))\rrbracket^{\mathrm{can}_{\mathbb{B}}(P^{\mathrm{id}})} =\top$ for $i\leq n$ since we are in a lattice.
But then, the mapping $h_{P^{\mathrm{id}}}$ that sends each variable $x$ to  the $y$ such that $v(x) = y$, is a homomorphism from $Q$ to $P^{\mathrm{id}}$ since by construction of $\mathrm{can}_{\mathbb{B}}(P^{\mathrm{id}})$, we have that $\llbracket A_i(v(\vec{ x_i}))\rrbracket^{\mathrm{can}_{\mathbb{B}}(P^{\mathrm{id}})} =\top$ only if there is a  $B_{j_i}(\overline{ y_{j_i}})$ in $P^{\mathrm{id}}$ such that $A_i(v(\vec{ x_i})) = B_{j_i}(\overline{ y_{j_i}})$.

  $(2) \implies (1)$: Assume that for every identification $P^{\mathrm{id}}$ of $P$, there is a  $\neq$-homomorphism 
$h_{P^{\mathrm{id}}}\colon Q \longrightarrow P^{\mathrm{id}}$. We wish to show that $P \sqsubseteq_\mathbb{L} Q$. Consider an arbitrary $\mathbb{L}$-database instance $\mathbf I$ and  an arbitrary assignment $v$ of elements to the bound variables of $P$. In this case,   $\llbracket B_1(v(\overline{ y_1})) \wedge \ldots \wedge B_m(v(\overline{ y_m})) \rrbracket^\mathbf{I}$ is the infimum of the values $\llbracket B_i(v(\overline{ y_i}))  \rrbracket^\mathbf{I}$. Let $P'$ be the completion of $P$ obtained by adding $x\neq y$ if  $v(x)\neq v(y)$, or $x = y$ if  $v(x)= v(y)$, for all variables $x,y$ of $P$, and $P^{\mathrm{id}}$ an identification of $P'$ by a canonical substitution $\theta$.  We may observe that $\theta(x) = \theta(y)$ iff $v(x) = v(y)$.  Moreover, for each literal of $A_i(\vec{ x_i})$ of $Q$ there is a literal $B'_{j_i}(\theta(\overline{ y_{j_i}}))$ of $P^{\mathrm{id}}$ such that $A_i(h_{P^{\mathrm{id}}}(\vec{ x_i})) = B'_{j_i}(\theta(\overline{ y_{j_i}}))$. Thus, $v \circ h_{P^{\mathrm{id}}}$ is an assignment of elements to the bound variables of $Q$ where $\llbracket A_i(v \circ h_{P^{\mathrm{id}}}(\vec{ x_i}))\rrbracket^\mathbf{I} = \llbracket B'_{j_i}(v(\theta(\overline{ y_{j_i}})))\rrbracket^\mathbf{I}$, and, since  all the newly added literals to $P$ in order to form $P'$ must take value $\top$ in $\mathbf{I}$ under $v$ (and are hence inessential in calculations of infima), we must have that $\llbracket  B_1(v (\theta(\overline{ y_1}))) \wedge \ldots \wedge B_m(v (\theta(\overline{ y_m})))\rrbracket^\mathbf{I}$ is at most $\llbracket A_1(v \circ h_{P^{\mathrm{id}}}(\vec{ x_1})) \wedge \ldots \wedge A_n(v \circ h_{P^{\mathrm{id}}}(\vec{ x_n}))\rrbracket^\mathbf{I}$. Observe that $\theta(x)=\theta(y)$ iff $v(x) = v(y)$, so $\llbracket B'_{j_i}(v(\theta(\overline{ y_{j_i}}))\rrbracket^\mathbf{I}= \llbracket  B'_{j_i}(v(\overline{ y_{j_i}}))\rrbracket^\mathbf{I}$. This means that for any tuple of elements $\vec{t}$, we have that  $\llbracket P(\vec{z})\rrbracket^\mathbf{I}(\vec{t})$ (the supremum of all values $\llbracket  B_1(e (\overline{ y_1})) \wedge \ldots \wedge B_m(e (\overline{ y_m}))\rrbracket^\mathbf{I}$ for all assignments $e$ to the variables where $e(\vec{z}) = \vec{t}$) is at most 
$\llbracket Q(\vec{x}) \rrbracket^\mathbf{I}(\vec{t})$  (the supremum of all values $\llbracket  A_1(e (\vec{ x_1})) \wedge \ldots \wedge A_n(e (\vec{ x_n}))\rrbracket^\mathbf{I}$ for all assignments $e$ to the variables where $e(\overline{x}) = \vec{t}$).

Finally, the complexity calculation follows immediately from Theorem~\ref{classical}.
\end{proof}

 The result that we present below for the semiring $\mathrm{Which}(X)$ uses families of \emph{containment coverings} in the sense of a set of homomorphisms such that for each atom of a query there is  one mapping with that atom in its image.

\begin{restatable}{Thm}{linhard}\label{linhard} For CQ$^\neq$s $P$ and $Q$ with the same tuple of free variables,  the following are equivalent:
\begin{itemize}
\item[1.] $P \sqsubseteq_{\mathrm{Which}(X)} Q$,
\item[2.] For every identification $P^{\mathrm{id}}$ of $P$, we have that $\llbracket P^{\mathrm{id}} \rrbracket^{\mathrm{can}_{\mathrm{Which}(X)}(P^{\mathrm{id}})} \leq \llbracket Q \rrbracket^{\mathrm{can}_{\mathrm{Which}(X)}(P^{\mathrm{id}})}$.
\item[3.] For every identification $P^{\mathrm{id}}$ of $P$, and every relational atom $R(\vec{z})$ of $P^{\mathrm{id}}$ there is a  $\neq$-homomorphism $h_{P^{\mathrm{id}}}\colon Q \longrightarrow P^{\mathrm{id}}$ with $R(\vec{z
})$ in the image of $h_{P^{\mathrm{id}}}$.

\end{itemize}
Moreover, the corresponding decision problem (i.e.\ $P \sqsubseteq_{\mathrm{Which}(X)} Q$?) is $\Pi^p_2$-complete.
\end{restatable}

\begin{proof}
$(1) \implies (2):$ Suppose that $P \sqsubseteq_{\mathrm{Which}(X)} Q$ and consider an identification $P^{\mathrm{id}}$ of a completion $P^c$ of $P$. We can observe  that  $P^{\mathrm{id}}\sqsubseteq_{\mathrm{Which}(X)}  P^c \sqsubseteq_{\mathrm{Which}(X)} P$. To see this,  take  any $\mathrm{Which}(X)$-database instance $\mathbf I$. On the one hand, if $\llbracket P^c \rrbracket^\mathbf{I}(\vec{t})\subseteq X$, then $\llbracket P^c \rrbracket^\mathbf{I}(\vec{t})$ contains only elements of $X$ appearing already in $\llbracket P \rrbracket^\mathbf{I}(\vec{t}) \subseteq X$. This is because the possible values of the new equations and disequations added to $P$ to form $P^c$ are $\bot$ and $\emptyset$. On the other hand,
if $\llbracket P^c \rrbracket^\mathbf{I}(\vec{t})= \bot$, obviously we have that $\llbracket P^c \rrbracket^\mathbf{I}(\vec{t})\leq \llbracket P \rrbracket^\mathbf{I}(\vec{t})$.
Thus, $\llbracket P^{\mathrm{id}} \rrbracket^{\mathrm{can}_{\mathrm{Which}(X)}(P^{\mathrm{id}})} \leq  \llbracket P \rrbracket^{\mathrm{can}_{\mathrm{Which}(X)}(P^{\mathrm{id}})} \leq  \llbracket Q \rrbracket^{\mathrm{can}_{\mathrm{Which}(X)}(P^{\mathrm{id}})}$.

$(2) \implies (3):$ Assume that (2) holds, so
for every identification $P^{\mathrm{id}}$ of a completion $P^c$ of $P$, we have that 
$$\llbracket P^{\mathrm{id}} \rrbracket^{\mathrm{can}_{\mathrm{Which}(X)}(P^{\mathrm{id}})} \leq  \llbracket Q \rrbracket^{\mathrm{can}_{\mathrm{Which}(X)}(P^{\mathrm{id}})},$$
where $\vec{u}$ is the tuple of free variables of $P^{\mathrm{id}}$. Let us abbreviate $\mathrm{can}_{\mathrm{Which}(X)}(P^{\mathrm{id}})$ as $\mathbf I$ in what follows. We observe that if $\{x_1\}, \ldots, \{x_n\}$ are all the abstract tags of tuples in $\mathbf I$, the set $\{x_1, \ldots, x_n\}=\llbracket P^{\mathrm{id}} \rrbracket^\mathbf{I}(\vec{u})$. To see this, consider the assignment to the bound variables of $P^{\mathrm{id}}$ that sends every variable to itself in $\mathbf I$. Under this assignment, the semiring values of the relational atoms of $P^{\mathrm{id}}$ are exactly $\{x_1\}, \ldots, \{x_n\}$ while the value of all other  disequations is $\emptyset$. Thus, we have that $\llbracket P^{\mathrm{id}} \rrbracket^\mathbf{I}(\vec{u})$ is the union of $\{x_1\} \cup \ldots \cup \{x_n\} \cup \emptyset \ldots \cup \emptyset = \{x_1\} \cup \ldots \cup \{x_n\}$ and potentially  some further sets which are subsets of $\{x_1, \ldots, x_n\}$ (since the latter is the top element of the subsemiring of $\mathrm{Which}(X)$ generated by $\{x_1\}  \ldots  \{x_n\}$), but then $\{x_1, \ldots, x_n\}= \llbracket P^{\mathrm{id}} \rrbracket^\mathbf{I}(\vec{u})$, as desired. By our hypothesis, we must have that $\{x_1, \ldots, x_n\} =\llbracket Q \rrbracket^\mathbf{I}(\vec{u})$ as well since $\{x_1, \ldots, x_n\}$ is the top element of the subsemiring of possible values of queries in $\mathbf I$. The only way for this to happen in $\mathbf I$ is for there to be, for each $x_i$, an assignment $v_i$ of objects in  $\mathbf I$  to the variables of $Q$ such that the abstract tag $\{x_i\}$ corresponding to an atom $R(\vec{z})$ is taken by some relational atom $R(\vec{z})$ of $Q$ and that all equations and disequations in $Q$  take value $\emptyset$ (and thus must correspond to equations and disequations appearing in $P^{\mathrm{id}}$ by construction of $\mathbf{I}$). From $v$ we can produce the desired $\neq$-homomorphism $h_{P^{\mathrm{id}}} \colon Q \longrightarrow P^{\mathrm{id}}$  by letting  $h_{P^{\mathrm{id}}}(x)$ be the variable $v(x)$.

$(3) \implies (1)$:  Let $Q(\vec{x})$ be  $\exists \vec{z} (A_1(\vec{x_1}) \wedge \ldots \wedge A_n(\vec{ x_n}))$ and $P(\vec{z})$ be $\exists \vec{u} (B_1(\vec{ y_1}) \wedge \ldots \wedge B_m(\vec{ y_m}))$, where $\overline{x}, \vec{z} $ and $\vec{z}, \vec{u}$ form partitions of $\overline{x_1}, \ldots , \overline{ x_n}$ and $\vec{z_1}, \ldots , \overline{ y_m}$, respectively, where the literals in these queries can be either relational atoms, equations or disequations. Assume that (3) holds and we want to show that $P \sqsubseteq_{\mathrm{Which}(X)} Q$. Take an arbitrary $\mathrm{Which}(X)$-database instance $\mathbf I$ and tuple $\vec{t}$. If $\llbracket P\rrbracket^{\mathbf I}(\vec t)=\bot$, the inequality is immediate. Otherwise, choose an assignment $w$ producing $\vec t$ with non-bottom contribution to $P$, and let $F$ be its equality identification. Applying (3) to any relational atom of $F$ gives a $\neq$-homomorphism $h\colon Q\to F$. Composing $h$ with the assignment induced by $w$ gives a non-bottom contribution to $Q$ at $\vec t$. It remains to show that for any $x\in X$, if  $x\in \llbracket P \rrbracket^\mathbf{I}(\vec{t})$ then $x\in \llbracket Q \rrbracket^\mathbf{I}(\vec{t})$. But if $x\in \llbracket P \rrbracket^\mathbf{I}(\vec{t})$, we have that there must be an assignment $v$ of elements to the bound variables of $P$ with non-bottom contribution such that $x\in \llbracket B_i(\overline{ y_i}) \rrbracket^\mathbf{I}(v(\overline{ y_i}))$ for some $B_i$.
Consider next the completion $ P^c$ of $P$ obtained by adding $x=y$ or $x\neq y$ depending on whether  $v(x) = v(y)$ or $v(x) \neq v(y)$. Take an identification $P^{\mathrm{id}}$ of $P^c$ by a canonical substitution $\theta$. We may observe that $\theta(x) = \theta(y)$ iff $v(x) = v(y)$.  Since the new expressions in $ P^{\mathrm{id}}$ all take value 
$\emptyset$ under $v$ and the canonical substitution $\theta$ matches the variables identified under $v$, we must have that $ \llbracket B_i(\overline{ y_i}) \rrbracket^\mathbf{I}(v(\overline{ y_i})) \subseteq \llbracket P^c \rrbracket^\mathbf{I}(\vec{t})$ and thus $x\in \llbracket P^c \rrbracket^\mathbf{I}(\vec{t})$. By assumption we have a $\neq$-homomorphism $h_{P^{\mathrm{id}}}\colon Q \longrightarrow P^{\mathrm{id}}$ with $\theta(B_i(\overline{ y_i}))$ in its image. 
 But then $v \circ   h_{P^{\mathrm{id}}}$ is an assignment of elements to the bound variables of $Q$ with non-bottom contribution such that  $\llbracket A_j(\vec{ x_j}) \rrbracket^\mathbf{I}(v\circ h_{P^{\mathrm{id}}}(\vec{ x_j})) =\llbracket B_i(\overline{ y_i}) \rrbracket^\mathbf{I}(v(\overline{ y_i}))$ for some $A_j$. Thus, $x\in \llbracket Q \rrbracket^\mathbf{I}(\vec{t})$ as desired.

For the upper bound, non-containment can be checked in $\mathrm{NP}^{\mathrm{coNP}}$: guess an identification $P^{\mathrm{id}}$ and a relational atom in it, and use a coNP oracle to check that no $\neq$-homomorphism $Q\longrightarrow P^{\mathrm{id}}$ covers that atom. A proposed covering map can be verified in polynomial time, so containment is in $\Pi^p_2$. For hardness, from the above, containment is equivalent to the statement that for every identification $P^c$ of $P$, and every relational atom $R(\vec{z})$ of $P^{\mathrm{id}}$ there is a $\neq$-homomorphism $h_{P^{\mathrm{id}}} \colon Q \longrightarrow P^{\mathrm{id}}$ with $R(\vec{z
})$ in the image of $h_{P^{\mathrm{id}}}$. Given an instance formed by  two  CQ$^\neq$s, $P$ and $Q$, of the  containment problem over the Boolean semiring, we may assume that the bound variables in $P$ and $Q$ are disjoint. We define then $Q^*$ as the  conjunction of $P$ and $Q$.
First, observe that 
\begin{itemize}
\item[(1)] for every identification $P'$ of $P$, and every relational atom $R(\vec{z})$ of $P'$  there is a $\neq$-homomorphism $g_{P'} \colon Q^* \longrightarrow P'$ with $R(\vec{z
})$ in the image of $g_{P'}$, 
\end{itemize}
entails that 
\begin{itemize}
\item[(2)] for every identification $P'$ of $P$,  there is a $\neq$-homomorphism  $h_{P'}\colon Q \longrightarrow P' $,
\end{itemize}
as $h_{P'}$ can be obtained as the restriction of $g_{P'}$ to the variables of $Q$.
Second, we also have that (2) entails (1). For any mapping $h_{P'}\colon Q \longrightarrow P' $ can be extended to a mapping $g_{P'} \colon Q^* \longrightarrow P'$  with the property that every relational atom $R(\vec{z})$  of $P'$ is in the image of $g_{P'}$ by letting the map be the identity on the variables of $P$. Thus, $P  \sqsubseteq_{\mathrm{Which}(X)} Q^*$ iff $P  \sqsubseteq_{\mathbb{B}} Q$. Using Theorem~\ref{classical},  we then get that the containment problem for CQ$^\neq$s over $\mathrm{Which}(X)$  is $\Pi_2^p$-hard.
\end{proof}

  As we will see below,  the semiring of absorptive polynomials over $X$ requires a family of injective  mappings in order to obtain a characterization for containment.  

\begin{restatable}{Thm}{sorpchneq}\label{sorpchneq} For CQ$^\neq$s $P$ and $Q$ with the same tuple of free variables,  the following are equivalent:
\begin{itemize}
\item[1.] $P \sqsubseteq_{\mathrm{Sorp}(X)} Q$,
\item[2.] For every identification $P^{\mathrm{id}}$ of $P$, we have that $\llbracket P^{\mathrm{id}} \rrbracket^{\mathrm{can}_{\mathrm{Sorp}(X)}(P^{\mathrm{id}})} \leq \llbracket Q \rrbracket^{\mathrm{can}_{\mathrm{Sorp}(X)}(P^{\mathrm{id}})}$.
\item[3.] For every identification $P^{\mathrm{id}}$ of $P$, there is a $\neq$-homomorphism $h_{P^{\mathrm{id}}}\colon Q \longrightarrow P^{\mathrm{id}}$ injective for relational atoms.

\end{itemize}
Moreover, the corresponding decision problem (i.e.\ $P \sqsubseteq_{\mathrm{Sorp}(X)} Q$?) is $\Pi^p_2$-complete.
\end{restatable}

\begin{proof}
$(1) \implies (2):$ Suppose that $P \sqsubseteq_{\mathrm{Sorp}(X)} Q$ and consider an identification $P^{\mathrm{id}}$ of a completion $P^c$ of $P$. We have that  $P^{\mathrm{id}}\sqsubseteq_{\mathrm{Sorp}(X)}  P^c \sqsubseteq_{\mathrm{Sorp}(X)} P$ since identification preserves query values, completion only removes assignment contributions, and addition is monotone in the natural order. Thus, $\llbracket P^{\mathrm{id}} \rrbracket^{\mathrm{can}_{\mathrm{Sorp}(X)}(P^{\mathrm{id}})} \leq \llbracket P^c \rrbracket^{\mathrm{can}_{\mathrm{Sorp}(X)}(P^{\mathrm{id}})} \leq  \llbracket P \rrbracket^{\mathrm{can}_{\mathrm{Sorp}(X)}(P^{\mathrm{id}})} \leq  \llbracket Q \rrbracket^{\mathrm{can}_{\mathrm{Sorp}(X)}(P^{\mathrm{id}})}$.

$(2) \implies (3):$  Assume that (2) holds, so
for every identification $P^{\mathrm{id}}$ of any completion $P^c$ of $P$, we have that 
$$\llbracket P^{\mathrm{id}} \rrbracket^{\mathrm{can}_{\mathrm{Sorp}(X)}(P^{\mathrm{id}})}(\vec{u}) \leq \llbracket Q \rrbracket^{\mathrm{can}_{\mathrm{Sorp}(X)}(P^{\mathrm{id}})}(\vec{u}),$$
where $\vec{u}$ is the tuple of free variables of $P^{\mathrm{id}}$. The polynomial $\llbracket P^{\mathrm{id}} \rrbracket^{\mathrm{can}_{\mathrm{Sorp}(X)}(P^{\mathrm{id}})}(\vec{u})$ has as a term a monomial $x_1 \cdot \ldots \cdot x_n $ containing every tag $p$ (with some exponent, accounting for duplicate atoms) appearing in $\mathrm{can}_{\mathrm{Sorp}(X)}(P^{\mathrm{id}})$ (such monomial is obtained using the identity assignment for the variables in $ P^{\mathrm{id}}$; noticing that the disequations in $P^{\mathrm{id}}$ all take value $1$ since $P$ and $P^c$ are consistent).  Then, by the definition of $\leq $ for $\mathrm{Sorp}(X)$, we must have that there is a  monomial $y_1 \cdot \ldots \cdot y_m $  in $\llbracket Q \rrbracket^{\mathrm{can}_{\mathrm{Sorp}(X)}(P^{\mathrm{id}})}(\vec{u})$ where  only tags $p$ from $x_1 \cdot \ldots \cdot x_n $ appear, with equal or smaller exponents as in $x_1 \cdot \ldots \cdot x_n $. Since $Q$ is a  CQ$^\neq$, we must have that there is an assignment $v$ to the bound variables in $Q$ over $\mathrm{can}_{\mathrm{Sorp}(X)}(P^{\mathrm{id}})$ such that the body (i.e.\ the quantifier-free matrix) of $Q$ evaluates to the monomial $y_1 \cdot \ldots \cdot y_m$.  Furthermore, the equations and disequations in $Q$ must all take value $1$ under the assignment $v$ in order to obtain the monomial  $y_1 \cdot \ldots \cdot y_m$ (instead of just $0$), and thus they must be  equations holding in $\mathrm{can}_{\mathbb{N}[X]}(P^{\mathrm{id}})$ (and thus logically entailed by the body of $P^{\mathrm{id}}$) and disequations appearing in the body of $P^{\mathrm{id}}$ (since the latter contains a complete description of disequations in  $\mathrm{can}_{\mathrm{Sorp}(X)}(P^{\mathrm{id}})$).  Then, we get the desired 
 $\neq$-homomorphism $h_{P^{\mathrm{id}}}\colon Q \longrightarrow P^{\mathrm{id}}$ injective for relational atoms by letting $h_{P^{\mathrm{id}}}(x)$ be the variable $v(x)$ (recall that the elements of the active domain of $\mathrm{can}_{\mathbb{N}[X]}(P^{\mathrm{id}})$ are the variables of $P^{\mathrm{id}}$). To see the injectivity, notice that any tag $p$ that is the value of a relational atom of $Q$ according to $v$, appears in $y_1 \cdot \ldots \cdot y_m$ with at most the same multiplicity as in  $x_1 \cdot \ldots \cdot x_n $ (since this monomial corresponds to the body of $P^{\mathrm{id}}$), and thus the multiset of images by $h_{P^{\mathrm{id}}}$ of relational atoms of $Q$ is bag-contained in the multiset of relational atoms from $P^{\mathrm{id}}$.
 
$(3) \implies (1):$ Assume that for every identification $P^{\mathrm{id}}$ of a completion $P^c$ of $P$, there is a $\neq$-homomorphism $h_{P^{\mathrm{id}}}\colon Q \longrightarrow P^{\mathrm{id}}$ injective for relational atoms. In order to establish $P \sqsubseteq_{\mathrm{Sorp}(X)} Q$, consider an arbitrary $\mathrm{Sorp}(X)$-database instance $\mathbf I$ and a tuple $\vec{t}$. We need to show that for any monomial in $\llbracket P \rrbracket^\mathbf{I}(\vec{t})$ there is a monomial  with  equal or smaller exponents  in $\llbracket Q \rrbracket^\mathbf{I}(\vec{t})$. Take an assignment $v$ of elements to the variables of $P$ justifying a monomial $x_1  \cdot \ldots \cdot x_n $ and consider the completion $ P^c$ of $P$ obtained by adding $x=y$ or $x\neq y$ depending on whether  $v(x) = v(y)$ or $v(x) \neq v(y)$. Take an identification $P^{\mathrm{id}}$ of $P^c$ by a canonical substitution $\theta$. We may observe that $\theta(x) = \theta(y)$ iff $v(x) = v(y)$. Then, $x_1 \cdot \ldots \cdot  x_n $ is also a monomial in $\llbracket P^{\mathrm{id}} \rrbracket^\mathbf{I}(\vec{t})$ justified by the assignment $v$ itself since the new  disequations added can only take value $1$ by construction according to the assignment $v$. 
  The assignment  $v \circ h_{P^{\mathrm{id}}} $  justifies a   monomial with smaller or equal exponents in $\llbracket Q \rrbracket^\mathbf{I}(\vec{t})$. This is because for any relational atom $A(\overline{x})$ of $Q$ such that $A(h_{P^{\mathrm{id}}}(\overline{x})) = B(\theta(\vec{z}))$, where  $B(\vec{z})$ is a relational atom of  $P$, if $x_i = \llbracket B(\theta(\vec{z})) \rrbracket^\mathbf{I}(v(\theta(\vec{z})))$, then $x_i = \llbracket A(\overline{x}) \rrbracket^\mathbf{I}(v(h_{P^{\mathrm{id}}}(\overline{x})))$. By the injectivity of $h_{P^{\mathrm{id}}}$, the images of the relational atoms in the body of $Q$ (considered as a multiset) appear with at most the same multiplicity as in the body of $P^{\mathrm{id}}$ considered as multiset, which entails that $x_i$ appears as the value of an atom of $Q$ with at most the same multiplicity as it appears in the monomial $x_1  \cdot \ldots \cdot x_n $ (since the latter corresponds to the body of $P^{\mathrm{id}}$). Since equations and disequations can only take values $0$ or $1$, they can be disregarded in showing that we can justify a monomial.

Finally, the upper bound in the complexity calculation follows exactly as for Theorem~\ref{diseqB[X]}. For hardness, the proof strategy is similar  to that of Theorem~\ref{linhard}. We simply use instead the characterization of containment obtained above. We need to modify the queries in the proof of Theorem~\ref{linhard} a bit. This time we leave $Q$ alone, and modify $P$ instead, to obtain $P^*$, by adding, for all the relational symbols in $P$, as many copies of the atoms involving this symbol in $P$ as times the symbol appears in an atom in $Q$. This now makes the mapping $h_{P'}$ clearly injective for relational atoms, considered as a mapping from $Q$ into $P^{*'}$ (an identification of $P^*$).
\end{proof}

For the next semiring, we first observe that the usual onto condition on variables is not sufficient for containment over $\text{Why}(X)$ (cf.~\cite{Green}), since repeated occurrences of the same relational atom may produce additional provenance annotations.
\begin{Exm}\label{whyexm}
For example, consider $P(x):= R(x)\land R(x),$ $Q(y) := R(y)$, and a $\mathrm{Why}(X)$-instance $\mathbf I$ such that $R^\mathbf I(a)=\{\{p\},\{q\}\}$. There is an onto homomorphism from $Q$ to $P$. However, $\llbracket P\rrbracket^\mathbf I(a)=\{\{p\},\{q\},\{p,q\}\},$ while $\llbracket Q\rrbracket^\mathbf I(a)=\{\{p\},\{q\}\}.$ Thus, $P$ is not contained in $Q$.
\end{Exm}

This motivates the use of bag-onto homomorphisms following~\cite{Green}, which take into account the occurrences of relational atoms, together with occurrence-sensitive canonical databases. 

\begin{restatable}{Thm}{whychneq}\label{whychneq}
For CQ$^\neq$s $P$ and $Q$ with the same tuple of free variables, the following are equivalent:
\begin{itemize}
\item[1.]
$P\sqsubseteq_{\mathrm{Why}(X)} Q$.

\item[2.]
For every identification $P^{\mathrm{id}}$ of $P$, we have $\llbracket P^{\mathrm{id}}\rrbracket^{\mathrm{can}^{\text{occ}}_{\mathrm{Why}(X)}(P^{\mathrm{id}})} \leq_{\mathrm{Why}(X)} \llbracket Q\rrbracket^{ \mathrm{can}^{\text{occ}}_{\mathrm{Why}(X)}(P^{\mathrm{id}})}$.

\item[3.]
For every identification $P^{\mathrm{id}}$ of $P$, there is a $\neq$-homomorphism $h_{P^{\mathrm{id}}}\colon Q\longrightarrow P^{\mathrm{id}}$ bag-onto for relational atoms.
\end{itemize}

Moreover, the corresponding decision problem (i.e.\ $P\sqsubseteq_{\mathrm{Why}(X)}Q$?) is $\Pi^p_2$-complete.
\end{restatable}

\begin{proof}
$(1)\Rightarrow(2):$
Suppose that $P\sqsubseteq_{\mathrm{Why}(X)}Q$ and consider an identification $P^{\mathrm{id}}$ of a completion $P^c$ of $P$. We have $P^{\mathrm{id}}\sqsubseteq_{\mathrm{Why}(X)}P^c \sqsubseteq_{\mathrm{Why}(X)}P$. Indeed, $P^{\mathrm{id}}$ is semantically equivalent to $P^c$, while the equations and disequations added to $P$ in order to obtain $P^c$ can only eliminate assignments from the evaluation of $P$. Hence, $P^{\mathrm{id}}\sqsubseteq_{\mathrm{Why}(X)}P \sqsubseteq_{\mathrm{Why}(X)}Q.$ Applying this containment to $\mathrm{can}^{\text{occ}}_{\mathrm{Why}(X)}(P^{\mathrm{id}})$ gives $\llbracket P^{\mathrm{id}}\rrbracket^{ \mathrm{can}^{\text{occ}}_{\mathrm{Why}(X)}(P^{\mathrm{id}})} \leq_{\mathrm{Why}(X)} \llbracket Q\rrbracket^{ \mathrm{can}^{\text{occ}}_{\mathrm{Why}(X)}(P^{\mathrm{id}})}.$

$(2)\Rightarrow(3):$
Assume that (2) holds and fix an identification $P^{\mathrm{id}}$ of a completion $P^c$ of $P$. Let $A_1(\vec{z}_1),\ldots,A_n(\vec{z}_n)$ be the relational atom occurrences of $P^{\mathrm{id}}$ (where the atoms are not necessarily syntactically distinct) and let $x_1,\ldots,x_n$ be their corresponding tags in $\mathrm{can}^{\text{occ}}_{\mathrm{Why}(X)}(P^{\mathrm{id}})$. Let us abbreviate this database as $\mathbf I$.

Consider the identity assignment to the variables of $P^{\mathrm{id}}$. For each occurrence $A_i(\vec{z}_i)$, we can choose the witness $\{x_i\}$ from its annotation in $\mathbf I$. Moreover, all disequations in $P^{\mathrm{id}}$ take value $1$ under this assignment. Hence the pairwise-union product contains the witness $W_{P^{\mathrm{id}}}:=\{x_1,\ldots,x_n\}$ and therefore $W_{P^{\mathrm{id}}}\in\llbracket P^{\mathrm{id}}(\vec{u})\rrbracket^{\mathbf I}.$ Since the natural order on $\mathrm{Why}(X)$ is set inclusion, by item 2 we also have $W_{P^{\mathrm{id}}}\in \llbracket Q(\vec{u})\rrbracket^{\mathbf I}.$

Let $B_1(\overline{x}_1),\ldots,B_m(\overline{x}_m)$ be the relational atom occurrences of $Q$. By the semantics of $\mathrm{Why}(X)$, there is an assignment $v$ to the variables of $Q$, producing $\vec{u}$ and satisfying all equations and disequations of $Q$, and, for every $1\leq j\leq m$, there is a witness $V_j\in B_j^{\mathbf I}(v(\overline{x}_j))$ such that $W_{P^{\mathrm{id}}}=V_1\cup\ldots\cup V_m.$

By construction of $\mathbf I$, every witness occurring in a non-zero tuple annotation is one of the singleton sets $\{x_1\},\ldots,\{x_n\}$. Hence, for every $j$, there is a unique $i$ such that $V_j=\{x_i\}.$

Define $\alpha\colon\{1,\ldots,m\}\longrightarrow\{1,\ldots,n\}$ by setting $\alpha(j)=i$ if and only if $V_j=\{x_i\}$. Then $\{x_1,\ldots,x_n\} = \{x_{\alpha(1)}\}\cup\ldots\cup \{x_{\alpha(m)}\}.$ Hence, for every $i\in\{1,\ldots,n\}$, the tag $x_i$ must occur among $x_{\alpha(1)},\ldots,x_{\alpha(m)}$. Since the tags $x_1,\ldots,x_n$ are pairwise distinct, for every $i$ there is some $j$ such that $\alpha(j)=i$. Therefore, $\alpha$ is surjective.

Moreover, $\{x_{\alpha(j)}\} = V_j \in B_j^{\mathbf I}(v(\overline{x}_j)).$ By the definition of $\mathbf I$, the tag $x_{\alpha(j)}$ occurs only in the annotation of the canonical tuple corresponding to the occurrence $A_{\alpha(j)}(\vec{z}_{\alpha(j)})$. Hence $B_j(v(\overline{x}_j)) = A_{\alpha(j)}(\vec{z}_{\alpha(j)}).$ We can define $h_{P^{\mathrm{id}}}(x):=v(x)$ for every variable $x$ of $Q$. The previous equality shows that $h_{P^{\mathrm{id}}} \bigl(B_j(\overline{x}_j)\bigr) = A_{\alpha(j)} (\vec{z}_{\alpha(j)}) $ for every relational atom occurrence of $Q$ and the head is preserved since $v$ produces the tuple $\vec{u}$. Moreover, all equations and disequations of $Q$ take value $1$ under $v$, and hence they are preserved by $h_{P^{\mathrm{id}}}$ by construction of the identification $P^{\mathrm{id}}$. Therefore, $h_{P^{\mathrm{id}}}\colon Q\longrightarrow P^{\mathrm{id}}$ is a $\neq$-homomorphism.

It remains to show that $h_{P^{\mathrm{id}}}$ is bag-onto for relational atoms. Since $\alpha$ is surjective, for every occurrence $A_i(\vec{z}_i)$ of $P^{\mathrm{id}}$ there is an occurrence $B_j(\overline{x}_j)$ of $Q$ such that $\alpha(j)=i$ and hence $h_{P^{\mathrm{id}}}\bigl(B_j(\overline{x}_j)\bigr)=A_i(\vec{z}_i).$ Moreover, the same occurrence of $Q$ cannot be associated by $\alpha$ with two different occurrences of $P^{\mathrm{id}}$. Therefore, for the different occurrences of $P^{\mathrm{id}}$ we can choose different preimage occurrences in $Q$. Hence the multiset of relational atoms of $P^{\mathrm{id}}$ is bag-contained in the multiset of images under $h_{P^{\mathrm{id}}}$ of the relational atoms of $Q$, and $h_{P^{\mathrm{id}}}$ is bag-onto for relational atoms.

$(3)\Rightarrow(1):$
Assume that for every identification $P^{\mathrm{id}}$ of $P$ there is a $\neq$-homomorphism $h_{P^{\mathrm{id}}}\colon Q\longrightarrow P^{\mathrm{id}}$ bag-onto for relational atoms. Consider an arbitrary $\mathrm{Why}(X)$-database instance $\mathbf I$ and a tuple $\vec{t}$. We show that every witness in $\llbracket P(\vec{t})\rrbracket^{\mathbf I}$ also belongs to $\llbracket Q(\vec{t})\rrbracket^{\mathbf I}$.

Let $W\in\llbracket P(\vec{t})\rrbracket^{\mathbf I}.$ Write the relational atom occurrences of $P$ as $C_1(\vec{z}_1),\ldots,C_r(\vec{z}_r)$. By the semantics of $\mathrm{Why}(X)$, there is an assignment $v$ to the variables of $P$, producing $\vec{t}$ on the free variables and satisfying all equations and disequations of $P$, and there are witnesses $W_i\in C_i^{\mathbf I}(v(\vec{z}_i)),$ $ 1\leq i\leq r,$ such that $W=W_1\cup\ldots\cup W_r.$

Consider now the completion $P^c$ of $P$ obtained by adding $x=y$ or $x\neq y$ depending on whether $v(x)=v(y)$ or $v(x)\neq v(y)$, and let $P^{\mathrm{id}}$ be an identification of $P^c$ obtained by a canonical substitution $\theta$. We have $\theta(x)=\theta(y)$ if and only if $v(x)=v(y).$ Define now an assignment $v^{id}$ to the variables of $P^{\mathrm{id}}$ by $v^{id}(\theta(x)):=v(x).$ If $C_i(\vec{z}_i)$ is the $i$-th relational occurrence of $P$, then $C_i(\theta(\vec{z}_i))$ is the corresponding occurrence of $P^{\mathrm{id}}$; in particular, repeated occurrences are not removed by the identification. We have $v^{id}(\theta(\vec{z}_i))=v(\vec{z}_i)$, and therefore $W_i\in C_i^{\mathbf I}\bigl(v^{id}(\theta(\vec{z}_i))\bigr)$ for every $1\leq i\leq r$. All the disequations added in forming $P^c$ take value $1$ under $v^{id}$ by construction, hence the same witnesses $W_1,\ldots,W_r$ yield the witness $W$ for $P^{\mathrm{id}}$.

By assumption, there is a $\neq$-homomorphism $h_{P^{\mathrm{id}}}\colon Q\longrightarrow P^{\mathrm{id}}$ bag-onto for relational atoms. Let $B_1(\overline{x}_1),\ldots,B_m(\overline{x}_m)$ be the relational atom occurrences of $Q$. Since $h_{P^{\mathrm{id}}}$ is bag-onto for relational atoms, then there is a surjective map $\beta\colon\{1,\ldots,m\}\longrightarrow\{1,\ldots,r\}$ between the relational atom occurrences of $Q$ and the ones of $P^{\mathrm{id}}$such that $h_{P^{\mathrm{id}}} \bigl(B_j(\overline{x}_j)\bigr) = C_{\beta(j)} \bigl(\theta(\vec{z}_{\beta(j)})\bigr)$ for every $1\leq j\leq m$. Such a map exists because the multiset of relational atoms of $P^{\mathrm{id}}$ is bag-contained in the multiset of images under $h_{P^{\mathrm{id}}}$ of the relational atoms of $Q$.

Consider now the assignment $v^{id}\circ h_{P^{\mathrm{id}}}$ to the variables of $Q$. Since $h_{P^{\mathrm{id}}}$ preserves the head, this assignment produces $\vec{t}$. Moreover, since $h_{P^{\mathrm{id}}}$ is a $\neq$-homomorphism, all equations and disequations of $Q$ take value $1$ under this assignment. For every relational atom occurrence $B_j(\overline{x}_j)$ of $Q$, we have $B_j^{\mathbf I} \bigl((v^{id}\circ h_{P^{\mathrm{id}}})(\overline{x}_j)\bigr) = C_{\beta(j)}^{\mathbf I} \bigl(v^{id} (\theta(\vec{z}_{\beta(j)}))\bigr).$ Thus, for this occurrence of $Q$, we can choose the witness
$W_{\beta(j)}$. The union of all witnesses selected in this way is $\bigcup_{1\leq j\leq m}W_{\beta(j)}= \bigcup_{1\leq i\leq r}W_i = W$, where the first equality follows from the surjectivity of $\beta$: every $W_i$ occurs at least once, and additional repetitions do not affect a set union. Hence $W\in\llbracket Q(\vec{t})\rrbracket^{\mathbf I}$ too. Since $\mathbf I$, $\vec{t}$ and $W\in\llbracket  P(\vec{t})\rrbracket^{\mathbf I}$ were arbitrary, we obtain $P\sqsubseteq_{\mathrm{Why}(X)}Q.$

For the upper bound, non-containment can be checked in $\mathrm{NP}^{\mathrm{coNP}}$: guess an identification $P^{\mathrm{id}}$ and use a coNP oracle to check that no bag-onto $\neq$-homomorphism $Q\longrightarrow P^{\mathrm{id}}$ exists. Such a homomorphism can be guessed and verified in polynomial time, so containment is in $\Pi^p_2$. For hardness, the proof strategy is the same as for Theorem~\ref{linhard}, using the characterization in item~3. Given an instance $P,Q$ of the containment problem for CQ$^\neq$s over the Boolean semiring, we may assume that the bound variables of $P$ and $Q$ are disjoint, and define $Q^\ast$ as the conjunction of $P$ and $Q$. For every identification $P^{\mathrm{id}}$ of $P$, any $\neq$-homomorphism $h_{P^{\mathrm{id}}}\colon Q\longrightarrow P^{\mathrm{id}}$ can be extended to a
$\neq$-homomorphism $g_{P^{\mathrm{id}}}\colon Q^\ast\longrightarrow P^{\mathrm{id}}$ bag-onto for relational atoms, by letting $g_{P^{\mathrm{id}}}$ coincide on the $P$-component with the canonical substitution used to obtain
$P^{\mathrm{id}}$. Conversely, the restriction to the $Q$-component of any such bag-onto $\neq$-homomorphism is a $\neq$-homomorphism $Q\longrightarrow P^{\mathrm{id}}$. Thus, $P\sqsubseteq_{\mathbb B} Q$ if and only if $ P\sqsubseteq_{\mathrm{Why}(X)} Q^\ast.$ By Theorem~\ref{classical}, the containment problem is therefore
$\Pi^p_2$-hard, and hence $\Pi^p_2$-complete.
\end{proof}

\triocqneqcontainment*

\begin{proof}
$(1)\implies(2)$: Suppose that $P\sqsubseteq_{\mathrm{Trio}(X)}Q$ and consider an identification $P^{\mathrm{id}}$ of $P$. Since $\text{can}^{k}_{\mathrm{Trio}(X)}(P^{\mathrm{id}})$ is a $\mathrm{Trio}(X)$-database, by the containment hypothesis we immediately have $\llbracket P\rrbracket^{\text{can}^{k}_{\mathrm{Trio}(X)}(P^{\mathrm{id}})}
(\vec{u})
\leq_{\mathrm{Trio}(X)} 
\llbracket Q\rrbracket^{\text{can}^{k}_{\mathrm{Trio}(X)}(P^{\mathrm{id}})}
(\vec{u}).$

$(2)\implies(1)$: We prove the contrapositive. Suppose that $P\not\sqsubseteq_{\mathrm{Trio}(X)}Q$, then, there are a $\mathrm{Trio}(X)$-database $\mathbf I$ and a tuple $\vec{a}$ such that $\llbracket P\rrbracket^{\mathbf I}(\vec{a})
\not\leq_{\mathrm{Trio}(X)}
\llbracket Q\rrbracket^{\mathbf I}(\vec{a}).$ By Lemma~\ref{kabstractCQ}, we may consider the $k$-abstractly tagged database $\mathbf I'=\mathrm{ab}_k(\mathbf I)$ and still have $\llbracket P\rrbracket^{\mathbf I'}(\vec{a})
\not\leq_{\mathrm{Trio}(X)}
\llbracket Q\rrbracket^{\mathbf I'}(\vec{a}).$ Since the natural order of $\mathrm{Trio}(X)$ is coefficientwise, there is a square-free monomial $\mu$ whose coefficient in $\llbracket P\rrbracket^{\mathbf I'}(\vec{a})$ is strictly larger than its coefficient in $\llbracket Q\rrbracket^{\mathbf I'}(\vec{a})$. In particular, the coefficient of $\mu$ in $\llbracket P\rrbracket^{\mathbf I'}(\vec{a})$ is non-zero. Thus, we can fix an assignment $v$ to the variables of $P$ producing $\vec{a}$, together with one choice of a tag from the annotation of each relational atom occurrence, which contributes the monomial $\mu$. Let $S$ be the set of distinct ground relational atoms used by $v$, and let $\mathbf J$ be the restriction of $\mathbf I'$ to the tuples in $S$. We claim that the coefficient of $\mu$ is unchanged on both sides after this restriction. Indeed, every tuple in $S$ contributes at least one tag occurring in $\mu$. Since $\mathbf I'$ is $k$-abstractly tagged, the sets of tags occurring in the annotations of distinct tuples are pairwise disjoint. Hence, any assignment of either $P$ or $Q$ contributing $\mu$ cannot use a tuple outside $S$, since this would introduce in the resulting square-free monomial a tag which does not occur in $\mu$. Conversely, deleting tuples outside $S$ cannot remove any assignment contributing $\mu$. Therefore, $\llbracket P\rrbracket^{\mathbf J}(\vec{a})
\not\leq_{\mathrm{Trio}(X)}
\llbracket Q\rrbracket^{\mathbf J}(\vec{a}).$ We now construct an identification of $P$ from the assignment $v$. Consider the completion $P^c$ obtained by adding, for every pair of variables $x,y$ of $P$, the equation $x=y$ if $v(x)=v(y)$, and the disequation $x\neq y$ otherwise. This completion is consistent since it is satisfied by $v$. Let $P^{\mathrm{id}}$ be an identification of $P^c$, by construction, two variables of $P$ are identified in $P^{\mathrm{id}}$ exactly when they have the same value under $v$.

Define a mapping $\rho\colon\text{vars}(P^{\mathrm{id}})\longrightarrow\mathrm{adom}(\mathbf J)$ by sending each variable of $P^{\mathrm{id}}$ to the common value under $v$ of the variables of $P$ that are identified with it. The mapping is well-defined and injective by the construction of $P^{\mathrm{id}}$. Moreover, it is surjective. In fact, every element of $\text{adom}(\mathbf J)$ occurs in a tuple used by $v$, while, since the queries are safe, the value assigned by $v$ to every variable of $P$ occurs in one of these tuples. Hence $\rho$ is a bijection.

Under this bijection, the syntactically distinct relational atoms of $P^{\mathrm{id}}$ correspond exactly to the distinct ground relational atoms in $S$. Indeed, every relational atom of $P^{\mathrm{id}}$ is inherited from a relational atom of $P$, and its image under $\rho$ is precisely the corresponding tuple used by $v$. Conversely, every tuple in $S$ is obtained in this way. Moreover, two occurrences that become the same relational atom in $P^{\mathrm{id}}$ are mapped by $v$ to the same ground tuple, as required.

Both $\mathbf J$ and $\text{can}^{k}_{\mathrm{Trio}(X)}(P^{\mathrm{id}})$ annotate each of these distinct tuples with a distinct element of $S_k$. Therefore, up to a renaming of domain elements and tags, $\mathbf J=\text{can}^{k}_{\mathrm{Trio}(X)}(P^{\mathrm{id}}).$ Under this renaming, the tuple $\vec{a}=v(\vec{u})$ corresponds to the tuple of free variables $\vec{u}$ of $P^{\mathrm{id}}$. Consequently, $\llbracket P\rrbracket^{\text{can}^{k}_{\mathrm{Trio}(X)}(P^{\mathrm{id}})}
(\vec{u})
\not\leq_{\mathrm{Trio}(X)}
\llbracket Q\rrbracket^{\text{can}^{k}_{\mathrm{Trio}(X)}(P^{\mathrm{id}})}
(\vec{u}),
$ contradicting item 2. Thus, $P\sqsubseteq_{\mathrm{Trio}(X)}Q$.

$(2)\Longleftrightarrow(3)$: Fix an identification $P^{\mathrm{id}}$ of $P$. As observed above, every square-free monomial involving variables occurring in the annotations of $\text{can}^{k}_{\mathrm{Trio}(X)}(P^{\mathrm{id}})$ is uniquely of the form $M_\sigma$ for some $k$-tag signature $\sigma\colon\text{RelSet}(P^{\mathrm{id}})\longrightarrow\mathcal{P}([k]).$ Since the natural order of $\mathrm{Trio}(X)$ is coefficientwise, we have $\llbracket P\rrbracket^{\text{can}^{k}_{\mathrm{Trio}(X)}(P^{\mathrm{id}})}
(\vec{u})
\leq_{\mathrm{Trio}(X)}
\llbracket Q\rrbracket^{\text{can}^{k}_{\mathrm{Trio}(X)}(P^{\mathrm{id}})}
(\vec{u})$ if and only if, for every $k$-tag signature $\sigma$, the coefficient of $M_\sigma$ on the $P$-side is at most its coefficient on the $Q$-side. By Lemma~\ref{indexedTrio}, these two coefficients are respectively $\left|\text{IHom}_k(P,P^{\mathrm{id}},\sigma)\right|$ and $\left|\text{IHom}_k(Q,P^{\mathrm{id}},\sigma)\right|.$ Thus, the above inequality holds if and only if $\left|\text{IHom}_k(P,P^{\mathrm{id}},\sigma)\right| \leq \left|\text{IHom}_k(Q,P^{\mathrm{id}},\sigma)\right|$ for every $k$-tag signature $\sigma$. Since $P^{\mathrm{id}}$ was arbitrary, items 2 and 3 are equivalent.

Finally, for the upper bound complexity calculation, we look at the complement of the problem stated in item 3 and show that it is in $\text{NP}^{\text{PP}}$. Given an input $(P,Q)$ for our problem, we start by guessing an identification $P^{\mathrm{id}}$ of $P$ and a $k$-tag signature $\sigma\colon\text{RelSet}(P^{\mathrm{id}})\longrightarrow\mathcal{P}([k]).$ Both objects have polynomial-size encodings: an identification is polynomial in the size of $P$, and $k$ is bounded by the number of relational atom occurrences in the input queries. We then use a $\text{PP}$ oracle to decide whether $\left|\text{IHom}_k(P,P^{\mathrm{id}},\sigma)\right| > \left|\text{IHom}_k(Q,P^{\mathrm{id}},\sigma)\right|.$ For fixed $R,P^{\mathrm{id}}$, and $\sigma$, computing $\left|\text{IHom}_k(R,P^{\mathrm{id}},\sigma)\right|$ is a $\#\text{P}$ problem. Indeed, we can nondeterministically guess a mapping $h$ from the variables of $R$ to the variables of $P^{\mathrm{id}}$ and, for every relational atom occurrence $B\in\text{RelOcc}(R)$, an index $\iota(B)\in[k]$. We can then check in polynomial time that $h$ is a $\neq$-homomorphism, that it preserves the head, and that, for every $A\in\text{RelSet}(P^{\mathrm{id}})$, $\sigma(A)=\{\iota(B)\mid B\in\text{RelOcc}(R),\ h(B)=A\}.$ Thus, the accepting computations are precisely the elements of $\text{IHom}_k(R,P^{\mathrm{id}},\sigma)$. It follows that both $\left|\text{IHom}_k(P,P^{\mathrm{id}},\sigma)\right| $ and $\left|\text{IHom}_k(Q,P^{\mathrm{id}},\sigma)\right|$ are $\#\text{P}$ functions, and therefore their difference is in $\text{GapP}$. Consequently, deciding whether $\left|\text{IHom}_k(P,P^{\mathrm{id}},\sigma)\right| - \left|\text{IHom}_k(Q,P^{\mathrm{id}},\sigma)\right|>0$ is in $\text{PP}$. Hence, non-containment is in $\text{NP}^{\text{PP}}$, and the containment problem is in $\text{coNP}^{\text{PP}}$.
\end{proof}

\section{Omitted  Proofs from \S~\ref{UCQneqsection1}} \label{app:proofs2}

As it turns out, for UCQ$^\neq$s the  characterization from the Boolean case again extends to the case of bounded distributive lattices (and thus the complexity remains identical).

\begin{restatable}{Thm}{ucqdiseqconta}  \label{ucqdiseqconta}

For UCQ$^\neq$s, $\bigcup_{1\leq i \leq n} P_i, \bigcup_{1\leq i \leq m} Q_i$ with the same tuple of free variables, and $\mathbb{L}$ a bounded distributive lattice, the following are equivalent:
\begin{itemize}

\item[1.] $\bigcup_{1\leq i \leq n} P_i \sqsubseteq_\mathbb{L}  \bigcup_{1\leq i \leq m} Q_i$.

\item[2.] For each $1\leq i \leq n$ and   every identification $P_i'$ of $P_i$, there is a $j \leq m$ and a  $\neq$-homomorphism 
$h_{P_i'} \colon Q_j \longrightarrow P_i'$.

\end{itemize}
Moreover, the corresponding decision problem (i.e.\ $\bigcup_{1\leq i \leq n} P_i \sqsubseteq_{\mathbb{L}} \bigcup_{1\leq i \leq m} Q_i$?) is $\Pi^p_2$-complete.
\end{restatable}

\begin{proof}
 $(1) \implies (2)$: Suppose that $\bigcup_{1\leq i \leq n} P_i \sqsubseteq_\mathbb{L}  \bigcup_{1\leq i \leq m} Q_i$.
Consider $P_i$ and an identification $P_i'$ of a completion  $P_i^*$ of $P_i$.  Now build a canonical database $\mathrm{can}_{\mathbb{B}}(P_i')$. Now, we have that  $P_i'\sqsubseteq_\mathbb{L} P_i^*\sqsubseteq_\mathbb{L} P_i$ since the possible values of the new literals of $P_i'$ are $\top$ and $\bot$, which means that the semiring value in any database instance of $P_i'$  can only be at most that of $P_i$. By construction, $\llbracket P_i'(\vec{z})\rrbracket^{\mathrm{can}_{\mathbb{B}}(P_i')} = \top$, and thus, $\llbracket P_i(\vec{z})\rrbracket^{\mathrm{can}_{\mathbb{B}}(P_i')} = \top = \llbracket\bigcup_{1\leq i \leq n} P_i(\vec{z}) \rrbracket^{\mathrm{can}_{\mathbb{B}}(P_i')}$ and from the hypothesis, $\llbracket\bigcup_{1\leq i \leq m} Q_i(\overline{x})\rrbracket^{\mathrm{can}_{\mathbb{B}}(P_i')} = \top$ since the lattice is bounded, which entails that $\llbracket  Q_j(\overline{x}) \rrbracket^{\mathrm{can}_{\mathbb{B}}(P_i')} = \top$ for some $1 \leq j\leq m$, by construction of $\mathrm{can}_{\mathbb{B}}(P_i')$ since $\top$ or $\bot$ are the only possibilities of values for  $\llbracket Q_j)(\overline{x}) \rrbracket^{\mathrm{can}_{\mathbb{B}}(P_i')}$ for all $j\leq m$. Let $Q_j(\overline{x})$ be  $\exists \vec{z} (A_1(\vec{x_1}) \wedge \ldots \wedge A_n(\vec{ x_n}))$ and $P_i(\vec{z})$ be $\exists \vec{u} (B_1(\vec{ y_1}) \wedge \ldots \wedge B_m(\vec{ y_m}))$, where $\overline{x}, \vec{z} $ and $\vec{z}, \vec{u}$ form partitions of $\overline{x_1}, \ldots , \overline{ x_n}$ and $\vec{z_1}, \ldots , \overline{ y_n}$, respectively. Consequently,  for some assignment $v$, it must be that $\llbracket A_1(v(\vec{ x_1})) \wedge \ldots \wedge A_n(v(\vec{ x_n}))\rrbracket^{\mathrm{can}_{\mathbb{B}}(P_i')}=\top$ and thus $\llbracket A_i(v(\vec{ x_i})\rrbracket^{\mathrm{can}_{\mathbb{B}}(P_i')} =\top$ for $i\leq n$ since we are in a lattice.
But then, the mapping $h_{P_i'}$ that sends each variable $x$ to  the $y$ such that $v(x) = y$, is a homomorphism from $Q_j$ to $P_i'$ since by construction of $\mathrm{can}_{\mathbb{B}}(P_i')$, we have that $\llbracket A_i(v(\vec{ x_i})\rrbracket^{\mathrm{can}_{\mathbb{B}}(P_i')} =\top$ only if there is a  $B_{j_i}(\overline{ y_{j_i}})$ such that $A_i(v(\vec{ x_i})) = B_{j_i}(\overline{ y_{j_i}})$.

  $(2) \implies (1)$: Assume that for each $1\leq i \leq n$ and   every  identification $P_i'$ of $P_i$, there is a $j \leq m$ and a  $\neq$-homomorphism 
$h_{P_i'} \colon Q_j \longrightarrow P_i'$. We wish to show that $\bigcup_{1\leq i \leq n} P_i \sqsubseteq_\mathbb{L}  \bigcup_{1\leq i \leq m} Q_i$. Consider an arbitrary $\mathbb{L}$-database instance $\mathbf I$ and  an arbitrary assignment $v$ of elements to the bound variables of $\bigcup_{1\leq i \leq n} P_i$. Let  $P_i(\vec{z})$ be $\exists \vec{u} (B_1(\vec{ y_1}) \wedge \ldots \wedge B_m(\vec{ y_m}))$. Take  $P_i^*$ to be the completion of $P_i$ obtained by adding $x\neq y$ if  $v(x)\neq v(y)$, or $x = y$ if  $v(x)= v(y)$, for all variables $x,y$ of $P_i$, and $P_i'$ an identification of $P_i'$ by a canonical substitution $\theta$.   Let $Q_j(\overline{x})$ be  $\exists \vec{z} (A_1(\vec{x_1}) \wedge \ldots \wedge A_n(\vec{ x_n}))$.  Moreover, for each literal of $A_i(\vec{ x_i})$ of $Q_j$ there is a literal $B'_{j_i}(\theta(\overline{ y_{j_i}}))$ of $P_i'$ such that $A_i(h_{P_i'}(\vec{ x_i})) = B'_{j_i}(\theta(\overline{ y_{j_i}}))$. Thus, $v \circ h_{P_i'}$ is an assignment of elements to the bound variables of $Q_j$ where $A_i(v \circ h_{P'}(\vec{ x_i})) = B'_{j_i}(v(\theta(\overline{ y_{j_i}})))$, and, since  all the newly added literals to $P_i$ in order to form $P_i^*$ (and from it $P_i'$) must take value $\top$ in $\mathbf I$ under $v$ (and are hence inessential in calculations of infima), we must have that $\llbracket B_1(v (\theta(\overline{ y_1}))) \wedge \ldots \wedge B_m(v (\theta(\overline{ y_m})))\rrbracket^\mathbf{I}$ is at most $\llbracket A_1(v \circ h_{P_i'}(\vec{ x_1})) \wedge \ldots \wedge A_n(v \circ h_{P_i'}(\vec{ x_n}))\rrbracket^\mathbf{I}$. Observe that $\theta(x)=\theta(y)$ iff $v(x) = v(y)$, so $\llbracket B'_{j_i}(v(\theta(\overline{ y_{j_i}})))\rrbracket^\mathbf{I}= \llbracket B'_{j_i}(v(\overline{ y_{j_i}}))\rrbracket^\mathbf{I}$. This means that for any tuple of elements $\vec{t}$, we have that  $\llbracket P_i(\vec{z})\rrbracket^\mathbf{I}(\vec{t})$ (the supremum of all values $\llbracket  B_1(e (\overline{ y_1})) \wedge \ldots \wedge B_m(e (\overline{ y_m}))\rrbracket^\mathbf{I}$ for all assignments $e$ to the variables where $e(\vec{z}) = \vec{t}$) is at most 
$\llbracket \bigcup_{1\leq i \leq m} Q_i (\overline{x})\rrbracket^\mathbf{I}(\vec{t})$  (the supremum of all values $\llbracket A_1(e (\vec{ x_1})) \wedge \ldots \wedge A_n(e (\vec{ x_n}))\rrbracket^\mathbf{I}$ for all assignments $e$ to the variables where $e(\overline{x}) = \vec{t}$).

The complexity calculation follows the same argument as Remark~\ref{complexityremark}.
\end{proof}

The following case has again exactly the same complexity
as the Boolean case. Here the lower bound is immediate from the result for CQ$^\neq$s and thus we only need to work out the upper bound. 

\begin{restatable}{Thm}{ulinhard}\label{ulinhard} For UCQ$^\neq$s, $\bigcup_{1\leq i \leq n} P_i, \bigcup_{1\leq i \leq m} Q_i$ with the same tuple of free variables,  the following are equivalent:
\begin{itemize}
\item[1.] $\bigcup_{1\leq i \leq n} P_i \sqsubseteq_{\mathrm{Which}(X)}  \bigcup_{1\leq i \leq m} Q_i$,
\item[2.]  For each $1\leq i \leq n$,   every identification $P_i^{id}$ of $P_i$, we have that\[\llbracket P_i^{id} \rrbracket^{\mathrm{can}_{\mathrm{Which}(X)}(P_i^{id})} \leq \llbracket \bigcup_{1\leq i \leq m} Q_i \rrbracket^{\mathrm{can}_{\mathrm{Which}(X)}(P_i^{id})},\]
\item[3.]   For each $1\leq i \leq n$,   every identification $P_i^{id} $ of $P_i$, and every relational atom $R(\vec{z})$ of $P_i$, there is a $j \leq m$ and a $\neq$-homomorphism 
$h_{P_i^{id} } \colon Q_j \longrightarrow P_i^{id}$ with $R(\vec{z
})$ in the image of $h_{P_i^{id} }$.
\end{itemize}
Moreover, the corresponding decision problem (i.e.\ $\bigcup_{1\leq i \leq n} P_i \sqsubseteq_{\mathrm{Which}(X)} \bigcup_{1\leq i \leq m} Q_i$?) is $\Pi^p_2$-complete.
\end{restatable}

\begin{proof}
$(1) \implies (2):$ Suppose that $\bigcup_{1\leq i \leq n} P_i \sqsubseteq_{\mathrm{Which}(X)}  \bigcup_{1\leq i \leq m} Q_i$.  Consider an identification $P^{\mathrm{id}}_i$ of a completion $P^c_i$ of  $P_i$ for $1\leq i \leq n$. We can observe  that  $P^{\mathrm{id}}_i\sqsubseteq_{\mathrm{Which}(X)}  P^c_i \sqsubseteq_{\mathrm{Which}(X)} P_i$ and $P_i \sqsubseteq_{\mathrm{Which}(X)} \bigcup_{1\leq i \leq n} P_i$.  To see this,  take  any $\mathrm{Which}(X)$-database instance $\mathbf I$. On the one hand, if $\llbracket P^c_i \rrbracket^\mathbf{I}(\vec{t})\subseteq X$, then $\llbracket P^c_i \rrbracket^\mathbf{I}(\vec{t})$ contains only elements of $X$ appearing already in $\llbracket P_i \rrbracket^\mathbf{I}(\vec{t}) \subseteq X$. This is because the possible values of the new equations and disequations added to $P$ to form $P^c$ are $\bot$ and $\emptyset$. On the other hand,
if $\llbracket P^c_i \rrbracket^\mathbf{I}(\vec{t})= \bot$, obviously we have that $\llbracket P^c_i \rrbracket^\mathbf{I}(\vec{t})\leq \llbracket P_i \rrbracket^\mathbf{I}(\vec{t})$. Thus, $\llbracket P_i^{id}  \rrbracket^{\mathrm{can}_{\mathrm{Which}(X)}(P_i^{id} )} \leq \llbracket P_i \rrbracket^{\mathrm{can}_{\mathrm{Which}(X)}(P_i^{id} )} \leq \llbracket \bigcup_{1\leq i \leq n} P_i \rrbracket^{\mathrm{can}_{\mathrm{Which}(X)}(P_i^{id} )} \leq \llbracket \bigcup_{1\leq i \leq m} Q_i\rrbracket^{\mathrm{can}_{\mathrm{Which}(X)}(P_i^{id} )} $.

$(2) \implies (3):$ Assume that (2) holds, so
 for each $1\leq i \leq n$,   every completion $P_i'$ of $P_i$, we have that  $$\llbracket P_i' \rrbracket^{\mathrm{can}_{\mathrm{Which}(X)}(P_i')} (\vec{u}) \leq \llbracket \bigcup_{1\leq i \leq m} Q_i \rrbracket^{\mathrm{can}_{\mathrm{Which}(X)}(P_i')}(\vec{u}),$$
where $\vec{u}$ is the tuple of free variables of $P_i'$. Let us abbreviate $\mathrm{can}_{\mathrm{Which}(X)}(P_i')$ as $\mathbf I$ in what follows. We observe that if $\{x_1\}, \ldots, \{x_n\}$ are all the abstract tags of tuples in $\mathbf I$, then we have that $\{x_1, \ldots, x_n\}=\llbracket P_i' \rrbracket^\mathbf{I}(\vec{u})$.   By our hypothesis, we must have $\{x_1, \ldots, x_n\}= \llbracket \bigcup_{1\leq i \leq m} Q_i  \rrbracket^\mathbf{I}(\vec{u})$ (since $\{x_1, \ldots, x_n\}$ is the top element of the subsemiring of $\mathrm{Which}(X)$ generated by $\{x_1\}  \ldots  \{x_n\}$). The only way for this to happen in $\mathbf I$ is for there to be, for each $x_i$, a $Q_j$ with $j\leq m$, and an assignment $v_i$ of objects in  $\mathbf I$  to the variables of $Q_j$ such that the abstract tag $\{x_i\}$ is taken by some relational atom $R(\vec{z})$ of $Q_j$ and that all equations and disequations in $Q_j$  take value $\emptyset$ (and thus must correspond to equations and disequations appearing in $P^c$ by construction of $\mathbf I$). From $v_i$ we can produce the desired $\neq$-homomorphism $h_{P_i'}\colon Q_j \longrightarrow P_i'$ with $R(\vec{z})$ in its image by letting  $h_{P^c}(x)$ be the variable $v(x)$ is an equivalence class representative of.

$(3) \implies (1)$:  Assume that (3) holds and we want to show that $\bigcup_{1\leq i \leq n} P_i \sqsubseteq_{\mathrm{Which}(X)}  \bigcup_{1\leq i \leq m} Q_i$. Take an arbitrary $\mathrm{Which}(X)$-database instance $\mathbf I$ and tuple $\vec{t}$. 

If $\llbracket\bigcup_{1\leq i\leq n}P_i\rrbracket^{\mathbf I}(\vec t)=\bot$, the inequality is immediate. Otherwise, choose a disjunct $P_i$ and an assignment $w$ producing $\vec t$ with non-bottom contribution, and let $F$ be its equality identification. Applying (3) to any relational atom of $F$ gives some $j$ and a $\neq$-homomorphism $h\colon Q_j\to F$. Composing $h$ with the assignment induced by $w$ gives a non-bottom contribution to $Q_j$ at $\vec t$, so the right-hand union has non-bottom value.

It remains to show that for any $x\in X$, if  $x\in \llbracket \bigcup_{1\leq i \leq n} P_i \rrbracket^\mathbf{I}(\vec{t})$ then $x\in \llbracket \bigcup_{1\leq i \leq m} Q_i \rrbracket^\mathbf{I}(\vec{t})$. But if $x\in \llbracket  \bigcup_{1\leq i \leq n} P_i  \rrbracket^\mathbf{I}(\vec{t})$,  then there is an $1\leq i\leq n$ such that  $x\in \llbracket  P_i \rrbracket^\mathbf{I}(\vec{t})$ (let $P_i(\vec{z})$ be $\exists \vec{u} (B_1(\vec{ y_1}) \wedge \ldots \wedge B_m(\vec{ y_m}))$) and there must be an assignment $v$ of elements to the bound variables of $P_i$ with non-bottom contribution such that $x\in \llbracket B_s(\overline{ y_s}) \rrbracket^\mathbf{I}(v(\overline{ y_s}))$ for some $B_s$.
Consider next the completion $P^c_i$ of $P_i$ obtained by adding $x=y$ or $x\neq y$ depending on whether  $v(x) = v(y)$ or $v(x) \neq v(y)$. 

Take an identification $P^{\mathrm{id}}_i$ of $P^c_i$ by a canonical substitution $\theta$. We may observe that $\theta(x) = \theta(y)$ iff $v(x) = v(y)$. Since the new expressions in $ P^{\mathrm{id}}_i$ all take value $\emptyset$ under $v$, we must have that $ \llbracket B_s(\overline{ y_s}) \rrbracket^\mathbf{I}(v(\overline{ y_s})) \subseteq \llbracket P^{\mathrm{id}}_i \rrbracket^\mathbf{I}(\vec{t})$ and thus $x\in \llbracket P^{\mathrm{id}}_i \rrbracket^\mathbf{I}(\vec{t})$. By assumption we have a $Q_j$ (let $Q_j(\overline{x})$ be  $\exists \vec{z} (A_1(\vec{x_1}) \wedge \ldots \wedge A_n(\vec{ x_n}))$) and a  $\neq$-homomorphism $h_{P^{\mathrm{id}}_i}\colon Q_j \longrightarrow P^{\mathrm{id}}_i$ with $\theta(B_s(\overline{ y_s}))$ in its image.  But then $v \circ   h_{P^{\mathrm{id}}_i}$ is an assignment of elements to the bound variables of $Q_j$ with non-bottom contribution such that  $\llbracket A_r(\vec{ x_r}) \rrbracket^\mathbf{I}(v\circ h_{P^{\mathrm{id}}_i}(\vec{ x_r})) =\llbracket B_s(\overline{ y_s}) \rrbracket^\mathbf{I}(v(\overline{ y_s}))$ for some $A_r$. Thus, $x\in \llbracket Q_j \rrbracket^\mathbf{I}(\vec{t})$ as desired.

Finally, the hardness of the problem follows from the hardness of containment for  CQ$^\neq$s on $\mathrm{Which}(X)$ from Theorem~\ref{linhard}, while the upper bound follows by a similar argument as in Remark~\ref{complexityremark}. 
\end{proof}

 Once more, we find the same complexity as in the Boolean case due to the result for CQ$^\neq$s but by a different argument for the upper bound.  

\begin{restatable}{Thm}{usorphard} \label{usorphard} For UCQ$^\neq$s, $\bigcup_{1\leq i \leq n} P_i, \bigcup_{1\leq i \leq m} Q_i$ with the same tuple of free variables,  the following are equivalent:
\begin{itemize}
\item[1.] $\bigcup_{1\leq i \leq n} P_i  \sqsubseteq_{\mathrm{Sorp}(X)} \bigcup_{1\leq i \leq m} Q_i$,
\item[2.]
For each $1\leq i \leq n$,   every identification $P^{\mathrm{id}}_i$ of $P_i$, we have that  $\llbracket P_i^{id} \rrbracket^{\mathrm{can}_{\mathrm{Sorp}(X)}(P_i^{id})} \leq \llbracket \bigcup_{1\leq i \leq m} Q_i \rrbracket^{\mathrm{can}_{\mathrm{Sorp}(X)}(P_i^{id})}$.
\item[3.]  For each $1\leq i \leq n$,   every identification $P^{\mathrm{id}}_i$ of $P_i$,  there is a $j \leq m$ and a  $\neq$-homomorphism 
$h_{P^{\mathrm{id}}_i} \colon Q_j \longrightarrow P^{\mathrm{id}}_i$  injective for relational atoms.

\end{itemize}
Moreover, the corresponding decision problem (i.e.\ $\bigcup_{1\leq i \leq n} P_i \sqsubseteq_{\mathrm{Sorp}(X)} \bigcup_{1\leq i \leq m} Q_i$?) is $\Pi^p_2$-complete.
\end{restatable}

\begin{proof}
$(1) \implies (2):$ 
Suppose that $\bigcup_{1\leq i \leq n} P_i \sqsubseteq_{\mathrm{Sorp}(X)} \bigcup_{1\leq i \leq m} Q_i$ and consider an identification $P^{\mathrm{id}}_i$ of a completion $P^c_i$ of $P_i$ for $1\leq i \leq n$. We have that  $P^{\mathrm{id}}_i\sqsubseteq_{\mathrm{Sorp}(X)}  P^c_i \sqsubseteq_{\mathrm{Sorp}(X)} \bigcup_{1\leq i \leq n} P_i $ since identification preserves query values, completion only removes assignment contributions, and addition is monotone in the natural order. Thus, $\llbracket P^{\mathrm{id}}_i \rrbracket^{\mathrm{can}_{\mathrm{Sorp}(X)}(P^{\mathrm{id}}_i)} \leq \llbracket P^c_i \rrbracket^{\mathrm{can}_{\mathrm{Sorp}(X)}(P^{\mathrm{id}}_i)} \leq  \llbracket P_i \rrbracket^{\mathrm{can}_{\mathrm{Sorp}(X)}(P^{\mathrm{id}}_i)} \leq  \llbracket \bigcup_{1\leq i \leq n} P_i \rrbracket^{\mathrm{can}_{\mathrm{Sorp}(X)}(P^{\mathrm{id}}_i)} \leq  \llbracket \bigcup_{1\leq i \leq m} Q_i \rrbracket^{\mathrm{can}_{\mathrm{Sorp}(X)}(P^{\mathrm{id}}_i)}$.

$(2) \implies (3):$ Assume that (2) holds, so for each $1\leq i \leq n$,   every identification $P^{\mathrm{id}}$ of $P$, we have that  $$\llbracket P_i^{id} \rrbracket^{\mathrm{can}_{\mathrm{Sorp}(X)}(P_i^{id})}(\vec{u}) \leq \llbracket \bigcup_{1\leq i \leq m} Q_i \rrbracket^{\mathrm{can}_{\mathrm{Sorp}(X)}(P_i^{id})}(\vec{u}),$$
where $\vec{u}$ is the tuple of free variables of $P^{\mathrm{id}}_i$. The polynomial $\llbracket P^{\mathrm{id}}_i \rrbracket^{\mathrm{can}_{\mathrm{Sorp}(X)}(P^{\mathrm{id}}_i)}(\vec{u})$ has as a term a monomial $x_1 \cdot \ldots \cdot x_n $ containing every tag $p$ (with some exponent, accounting for duplicate atoms) appearing in $\mathrm{can}_{\mathrm{Sorp}(X)}(P^{\mathrm{id}}_i)$ (such monomial is obtained using the identity assignment for the variables in $ P^{\mathrm{id}}_i$; noticing that the disequations in $P^{\mathrm{id}}_i$ all take value $1$ since $P_i$ and $P^c_i$ are consistent).  Then, by the definition of $\leq $ for $\mathrm{Sorp}(X)$, we must have that there is a  monomial $y_1 \cdot \ldots \cdot y_m $  in $\llbracket \bigcup_{1\leq i \leq m} Q_i \rrbracket^{\mathrm{can}_{\mathrm{Sorp}(X)}(P^{\mathrm{id}}_i)}(\vec{u})$ where  only tags $p$ from $x_1 \cdot \ldots \cdot x_n $ appear, with equal or smaller exponents as in $x_1 \cdot \ldots \cdot x_n $. Since $\bigcup_{1\leq i \leq m} Q_i$ is a UCQ$^\neq$, we must have that there is a $Q_j$ with $j\leq m$ and an assignment $v$ to the bound variables in $Q_j$ over $\mathrm{can}_{\mathrm{Sorp}(X)}(P^{\mathrm{id}}_i)$ such that the body (i.e.\ the quantifier-free matrix) of $Q_j$ evaluates to the monomial $y_1 \cdot \ldots \cdot y_m$. 
 Furthermore, the equations and disequations in $Q_j$ must all take value $1$ under the assignment $v$ in order to obtain the monomial  $y_1 \cdot \ldots \cdot y_m$ (instead of just $0$), and thus they must be  equations holding in $\mathrm{can}_{\mathbb{N}[X]}(P^{\mathrm{id}}_i)$ (and thus logically entailed by the body of $P^{\mathrm{id}}_i$) and disequations appearing in the body of $P^{\mathrm{id}}_i$ (since the latter contains a complete description of disequations in  $\mathrm{can}_{\mathrm{Sorp}(X)}(P^{\mathrm{id}}_i)$).
 Then, we get the desired 
 $\neq$-homomorphism $h_{P^{\mathrm{id}}_i}\colon Q_j \longrightarrow P^{\mathrm{id}}_i$ injective for relational atoms by letting $h_{P^{\mathrm{id}}_i}(x)$ be the variable $v(x)$ (recall that the elements of the active domain of $\mathrm{can}_{\mathbb{N}[X]}(P^{\mathrm{id}}_i)$ are the variables of $P^{\mathrm{id}}_i$).
 To see the injectivity, notice that any tag $p$ that is the value of a relational atom of $Q_j$ according to $v$, appears in $y_1 \cdot \ldots \cdot y_m$ with at most the same multiplicity as in  $x_1 \cdot \ldots \cdot x_n $ (since this monomial corresponds to the body of $P^{\mathrm{id}}_i$), and thus the multiset of images by $h_{P^{\mathrm{id}}_i}$ of relational atoms of $Q_j$ is bag-contained in the multiset of relational atoms from $P^{\mathrm{id}}_i$.
 
$(3) \implies (1):$    Assume that (3) holds and we want to show that $\bigcup_{1\leq i \leq n} P_i \sqsubseteq_{\mathrm{Sorp}(X)}  \bigcup_{1\leq i \leq m} Q_i$. Thus, for each $1\leq i \leq n$, every identification $P^{\mathrm{id}}_i$ of $P_i$,  there is a $j \leq m$ and a $\neq$-homomorphism 
$h_{P^{\mathrm{id}}_i} \colon Q_j \longrightarrow P^{\mathrm{id}}_i$ with $R(\vec{z
})$ injective for relational atoms. In order to establish $\bigcup_{1\leq i \leq n} P_i \sqsubseteq_{\mathrm{Sorp}(X)}  \bigcup_{1\leq i \leq m} Q_i$, consider an arbitrary $\mathrm{Sorp}(X)$-database instance $\mathbf I$ and a tuple $\vec{t}$. We need to show that for any monomial in $\llbracket \bigcup_{1\leq i \leq n} P_i  \rrbracket^\mathbf{I}(\vec{t})$ there is a monomial  with  equal or smaller exponents  in $\llbracket \bigcup_{1\leq i \leq m} Q_i \rrbracket^\mathbf{I}(\vec{t})$. Take an assignment $v$ of elements to the variables of $P_i$ (where $i\leq n$) justifying a monomial $x_1  \cdot \ldots \cdot x_n $ and consider the completion $ P^c_i$ of $P_i$ obtained by adding $x=y$ or $x\neq y$ depending on whether  $v(x) = v(y)$ or $v(x) \neq v(y)$. Take an identification $P^{\mathrm{id}}_i$ of $P^c_i$ by a canonical substitution $\theta$. We may observe that $\theta(x) = \theta(y)$ iff $v(x) = v(y)$. Then, $x_1 \cdot \ldots \cdot  x_n $ is also a monomial in $\llbracket P^{\mathrm{id}}_i \rrbracket^\mathbf{I}(\vec{t})$ justified by the assignment $v$ itself since the new  disequations added can only take value $1$ by construction according to the assignment $v$.
The assignment  $v \circ h_{P^{\mathrm{id}}} $  justifies a   monomial with smaller or equal exponents in $\llbracket Q_j \rrbracket^\mathbf{I}(\vec{t})$. This is because for any relational atom $A(\overline{x})$ of $Q_j$ such that $A(h_{P^{\mathrm{id}}}(\overline{x})) = B(\theta(\vec{z}))$, where  $B(\vec{z})$ is a relational atom of  $P_i$, if $x_i = \llbracket B(\theta(\vec{z})) \rrbracket^\mathbf{I}(v(\theta(\vec{z})))$, then $x_i = \llbracket A(\overline{x}) \rrbracket^\mathbf{I}(v(h_{P^{\mathrm{id}}_i}(\overline{x})))$. By the injectivity of $h_{P^{\mathrm{id}}_i}$, the images of the relational atoms in the body of $Q_j$ (considered as a multiset) appear with the same multiplicity as in the body of $P^{\mathrm{id}}_i$ considered as multiset, which entails that $x_i$ appears as the value of an atom of $Q_j$ with at most the same multiplicity as it appears in the monomial $x_1  \cdot \ldots \cdot x_n $ (since the latter corresponds to the body of $P^{\mathrm{id}}_i$). Since equations and disequations can only take values $0$ or $1$, they can be disregarded in showing that we can justify a monomial.

Finally, the hardness of the problem follows from the hardness of containment for CQ$^\neq$s on $\mathrm{Sorp}(X)$ from Theorem~\ref{sorpchneq}, while the upper bound follows by a similar argument as in Remark~\ref{complexityremark}.
\end{proof}

We now extend our characterization of containment under Why-provenance to unions of conjunctive queries with disequations. As shown below, $\mathrm{Why}(X)$-containment reduces to testing the existence of a relational-atom bag-onto $\neq$-homomorphism from a target query to every identification of each source query, placing the decision problem at the second level of the polynomial hierarchy. 

\begin{restatable}{Thm}{ucqcontwhy}\label{ucqcontwhy}
For UCQ$^\neq$s, $\bigcup_{1\leq i\leq n}P_i,\bigcup_{1\leq j\leq m}Q_j$ with the same tuple of free variables, the following are equivalent:
\begin{itemize}

\item[1.] $ \bigcup_{1\leq i\leq n}P_i\sqsubseteq_{\mathrm{Why}(X)}\bigcup_{1\leq j\leq m}Q_j$.

\item[2.] For each $1\leq i\leq n$ and every identification $P_i^{id}$ of $P_i$, we have that
$$
\llbracket P_i^{id}\rrbracket^{\text{can}^{\text{occ}}_{\mathrm{Why}(X)}(P_i^{id})}
\leq_{\mathrm{Why}(X)}
\llbracket \bigcup_{1\leq j\leq m}Q_j\rrbracket^{\text{can}^{\text{occ}}_{\mathrm{Why}(X)}(P_i^{id})}.
$$

\item[3.] For each $1\leq i\leq n$ and every identification $P_i^{id}$ of $P_i$, there is a $j\leq m$ and a $\neq$-homomorphism
$h_{P_i^{id}}\colon Q_j\longrightarrow P_i^{id}$ bag-onto for relational atoms.

\end{itemize}
Moreover, the corresponding decision problem (i.e.\ $\bigcup_{1\leq i\leq n}P_i\sqsubseteq_{\mathrm{Why}(X)}\bigcup_{1\leq j\leq m}Q_j$?) is $\Pi^p_2$-complete.
\end{restatable}

\begin{proof}
$(1)\Rightarrow(2):$ Suppose that $\bigcup_{1\leq i\leq n}P_i\sqsubseteq_{\mathrm{Why}(X)}\bigcup_{1\leq j\leq m}Q_j$ and consider an identification $P_i^{id}$ of a completion $P_i^c$ of $P_i$. We have $P_i^{id}\sqsubseteq_{\mathrm{Why}(X)}P_i^c\sqsubseteq_{\mathrm{Why}(X)}P_i$. Indeed, $P_i^{id}$ is semantically equivalent to $P_i^c$, while the equations and disequations added to $P_i$ in order to obtain $P_i^c$ can only eliminate assignments from the evaluation of $P_i$. Moreover, since addition in $\mathrm{Why}(X)$ is set-theoretic union, we clearly have $P_i\sqsubseteq_{\mathrm{Why}(X)}\bigcup_{1\leq k\leq n}P_k$. Hence $P_i^{id}\sqsubseteq_{\mathrm{Why}(X)}\bigcup_{1\leq k\leq n}P_k\sqsubseteq_{\mathrm{Why}(X)}\bigcup_{1\leq j\leq m}Q_j.$ Applying this containment to $\text{can}^{\text{occ}}_{\mathrm{Why}(X)}(P_i^{id})$ gives $\llbracket P_i^{id}\rrbracket^{\text{can}^{\text{occ}}_{\mathrm{Why}(X)}(P_i^{id})}\leq_{\mathrm{Why}(X)}\llbracket \bigcup_{1\leq j\leq m}Q_j\rrbracket^{\text{can}^{\text{occ}}_{\mathrm{Why}(X)}(P_i^{id})},$ as required.

$(2)\Rightarrow(3):$ Assume that item 2 holds and fix $1\leq i\leq n$ and an identification $P_i^{id}$ of $P_i$. Let $A_1(\vec{z}_1),\ldots,A_r(\vec{z}_r)$ be the relational atom occurrences of $P_i^{id}$, where the atoms are not necessarily syntactically distinct, and let $x_1,\ldots,x_r$ be their corresponding tags in $\text{can}^{\text{occ}}_{\mathrm{Why}(X)}(P_i^{id})$. Let us abbreviate this database as $\mathbf I$. Consider the identity assignment to the variables of $P_i^{id}$. For each occurrence $A_k(\vec{z}_k)$, we can choose the witness $\{x_k\}$ from its annotation in $\mathbf I$, while all equations and disequations of $P_i^{id}$ take value $1$ under this assignment. Hence the pairwise-union product contains the witness $W_{P_i^{id}}:=\{x_1,\ldots,x_r\},$ and therefore $W_{P_i^{id}}\in\llbracket P_i^{id}(\vec{u})\rrbracket^{\mathbf I}$. By item 2 and the fact that the natural order on $\mathrm{Why}(X)$ is set inclusion, we have $W_{P_i^{id}}\in\llbracket \bigcup_{1\leq j\leq m}Q_j(\vec{u})\rrbracket^{\mathbf I}.$ Since addition in $\mathrm{Why}(X)$ is set-theoretic union, $\llbracket \bigcup_{1\leq j\leq m}Q_j(\vec{u})\rrbracket^{\mathbf I}=\bigcup_{1\leq j\leq m}\llbracket Q_j(\vec{u})\rrbracket^{\mathbf I}.$ Thus there is some $1 \leq j\leq m$ such that $W_{P_i^{id}}\in\llbracket Q_j(\vec{u})\rrbracket^{\mathbf I}$.

Let $B_1(\overline{x}_1),\ldots,B_s(\overline{x}_s)$ be the relational atom occurrences of $Q_j$. By the semantics of $\mathrm{Why}(X)$, there is an assignment $v$ to the variables of $Q_j$, producing $\vec{u}$ and satisfying all equations and disequations of $Q_j$, and, for every $1\leq \ell\leq s$, there is a witness $V_\ell\in B_\ell^{\mathbf I}(v(\overline{x}_\ell))$ such that $W_{P_i^{id}}=V_1\cup\ldots\cup V_s.$ By construction of $\mathbf I$, every witness occurring in a non-zero tuple annotation is one of the singleton sets $\{x_1\},\ldots,\{x_r\}$. Hence, for every $\ell$, there is a unique $k$ such that $V_\ell=\{x_k\}$. Define $\alpha\colon\{1,\ldots,s\}\longrightarrow\{1,\ldots,r\}$ by setting $\alpha(\ell)=k$ if and only if $V_\ell=\{x_k\}$. Then $\{x_1,\ldots,x_r\}=\{x_{\alpha(1)}\}\cup\ldots\cup\{x_{\alpha(s)}\}.$ Hence, for every $k\in\{1,\ldots,r\}$, there is some $\ell$ such that $\alpha(\ell)=k$, and therefore $\alpha$ is surjective.

Moreover, $\{x_{\alpha(\ell)}\}=V_\ell\in B_\ell^{\mathbf I}(v(\overline{x}_\ell))$. By the definition of $\mathbf I$, the tag $x_{\alpha(\ell)}$ occurs only in the annotation of the canonical tuple determined by the occurrence $A_{\alpha(\ell)}(\vec{z}_{\alpha(\ell)})$. Hence $B_\ell(v(\overline{x}_\ell))=A_{\alpha(\ell)}(\vec{z}_{\alpha(\ell)}).$ Since every variable of $Q_j$ occurs in a relational atom, the value assigned by $v$ to each variable belongs to the active domain of $\mathbf I$, and hence corresponds to a variable occurring in $P_i^{id}$. We can thus define $h_{P_i^{id}}(x):=v(x)$ for every variable $x$ of $Q_j$. The previous equality shows that $h_{P_i^{id}}$ preserves relational atoms and the head is preserved since $v$ produces the tuple $\vec{u}$. Moreover, all equations and disequations of $Q_j$ take value $1$ under $v$, and, since $P_i^{id}$ comes from a completion of $P_i$, they are preserved by $h_{P_i^{id}}$. Therefore, $h_{P_i^{id}}\colon Q_j\longrightarrow P_i^{id}$ is a $\neq$-homomorphism.

It remains to show that $h_{P_i^{id}}$ is bag-onto for relational atoms. Since $\alpha$ is surjective, for every occurrence $A_k(\vec{z}_k)$ of $P_i^{id}$ there is an occurrence $B_\ell(\vec{x}_\ell)$ of $Q_j$ such that $\alpha(\ell)=k$ and hence $h_{P_i^{id}}\bigl(B_\ell(\vec{x}_\ell)\bigr)=A_k(\vec{z}_k).$ Moreover, the same occurrence of $Q_j$ cannot be associated by $\alpha$ with two different occurrences of $P_i^{id}$. Therefore, for the different occurrences of $P_i^{id}$ we can choose different preimage occurrences in $Q_j$. Hence the multiset of relational atoms of $P_i^{id}$ is bag-contained in the multiset of images under $h_{P_i^{id}}$ of the relational atoms of $Q_j$, and $h_{P_i^{id}}$ is bag-onto for relational atoms.

$(3)\Rightarrow(1):$ Assume that item 3 holds and consider an arbitrary $\mathrm{Why}(X)$-database instance $\mathbf I$ and a tuple $\vec{t}$. We show that every witness in $\llbracket\bigcup_{1\leq i\leq n}P_i(\vec{t})\rrbracket^{\mathbf I}$ also belongs to $\llbracket\bigcup_{1\leq j\leq m}Q_j(\vec{t})\rrbracket^{\mathbf I}$. Let $W\in\llbracket\bigcup_{1\leq i\leq n}P_i(\vec{t})\rrbracket^{\mathbf I}.$ Since addition in $\mathrm{Why}(X)$ is set-theoretic union, there is some $1 \leq i\leq n$ such that $W\in\llbracket P_i(\vec{t})\rrbracket^{\mathbf I}$. Write the relational atom occurrences of $P_i$ as $C_1(\vec{z}_1),\ldots,C_r(\vec{z}_r)$. By the semantics of $\mathrm{Why}(X)$, there is an assignment $v$ to the variables of $P_i$, producing $\vec{t}$ on the free variables and satisfying all equations and disequations of $P_i$, and there are witnesses $W_k\in C_k^{\mathbf I}(v(\vec{z}_k)),$ with $ 1\leq k\leq r,$ such that $W=W_1\cup\ldots\cup W_r$.

Consider now the completion $P_i^c$ of $P_i$ determined by $v$, obtained by adding $x=y$ or $x\neq y$ depending on whether $v(x)=v(y)$ or $v(x)\neq v(y)$, and let $P_i^{id}$ be an identification of $P_i^c$. As usual, we use the same symbol $v$ for the assignment induced on the variables of $P_i^{id}$. The relational atom occurrences of $P_i$ give corresponding relational atom occurrences in $P_i^{id}$ and, in particular, repeated occurrences are not removed by the identification. Hence the same witnesses $W_1,\ldots,W_r$ can be chosen for the corresponding relational atom occurrences of $P_i^{id}$, and all the disequations added in forming $P_i^c$ take value $1$ under $v$. Thus the same witnesses yield $W$ for $P_i^{id}$.

By item 3, there is a $j\leq m$ and a $\neq$-homomorphism $h_{P_i^{id}}\colon Q_j\longrightarrow P_i^{id}$ bag-onto for relational atoms. Let $B_1(\overline{x}_1),\ldots,B_s(\overline{x}_s)$ be the relational atom occurrences of $Q_j$. Since $h_{P_i^{id}}$ is bag-onto for relational atoms, there is a surjective map $\beta\colon\{1,\ldots,s\}\longrightarrow\{1,\ldots,r\}$ between the relational atom occurrences of $Q_j$ and the corresponding occurrences of $P_i^{id}$ such that $h_{P_i^{id}}\bigl(B_\ell(\vec{x}_\ell)\bigr)=C_{\beta(\ell)}$ for every $1\leq\ell\leq s$, where on the right-hand side we identify each occurrence of $P_i$ with its corresponding occurrence in $P_i^{id}$.

Consider now the assignment $v\circ h_{P_i^{id}}$ to the variables of $Q_j$. Since $h_{P_i^{id}}$ preserves the head, this assignment produces $\vec{t}$. Moreover, since $h_{P_i^{id}}$ is a $\neq$-homomorphism, all equations and disequations of $Q_j$ take value $1$ under this assignment. For every relational atom occurrence $B_\ell(\vec{x}_\ell)$ of $Q_j$, we have $B_\ell^{\mathbf I}\bigl((v\circ h_{P_i^{id}})(\overline{x}_\ell)\bigr) = C_{\beta(\ell)}^{\mathbf I}\bigl(v(\vec{z}_{\beta(\ell)})\bigr).$ Thus, for this occurrence of $Q_j$, we can choose the witness $W_{\beta(\ell)}$. The union of all witnesses selected in this way is $\bigcup_{1\leq\ell\leq s}W_{\beta(\ell)} = \bigcup_{1\leq k\leq r}W_k = W, $ where the first equality follows from the surjectivity of $\beta$: every $W_k$ occurs at least once, and additional repetitions do not affect a set union. Hence $W\in\llbracket Q_j(\vec{t})\rrbracket^{\mathbf I}$, and therefore $W\in\llbracket\bigcup_{1\leq j\leq m}Q_j(\vec{t})\rrbracket^{\mathbf I}. $ Since $\mathbf I$, $\vec{t}$, and $W\in\llbracket\bigcup_{1\leq i\leq n}P_i(\vec{t})\rrbracket^{\mathbf I}$ were arbitrary, we obtain $\bigcup_{1\leq i\leq n}P_i\sqsubseteq_{\mathrm{Why}(X)}\bigcup_{1\leq j\leq m}Q_j.$

Finally, the lower bound for the complexity calculation follows immediately from the $\Pi^p_2$-hardness of containment for CQ$^\neq$s over $\mathrm{Why}(X)$, since CQ$^\neq$s are a particular case of UCQ$^\neq$s. For the upper bound, by item 3, the complement of the containment problem is in $\Sigma^p_2$. Indeed, we can guess an $1 \leq i\leq n$ and an identification $P_i^{id}$ of $P_i$, and then use a coNP oracle to check that there is no $j\leq m$ and no $\neq$-homomorphism $h\colon Q_j\longrightarrow P_i^{id}$ bag-onto for relational atoms. The existence of such a $j$ and such a mapping can be checked in NP. Hence the containment problem is in $\Pi^p_2$, and therefore it is $\Pi^p_2$-complete.
\end{proof}

\ucqbxcont*

\begin{proof}
$(1) \implies (2):$ Suppose that $\bigcup_{1\leq i \leq n} P_i  \sqsubseteq_{\mathbb{B}[X]} \bigcup_{1\leq i \leq m} Q_i$ and consider an identification $P_i^{id}$ of a completion $P_i^c$ of $P_i$. We have that  $P_i^{id}\sqsubseteq_{\mathbb{B}[X]}  P_i^c \sqsubseteq_{\mathbb{B}[X]} P_i$ since on any $\mathbb{B}[X]$-database instance $\mathbf I$, the polynomial corresponding to $P_i^c$ contains only monomials appearing in the one corresponding to $P_i$ (some of the monomials appearing in $P_i$ might have turned into $0$ in $P_i^c$). Thus, $\llbracket P_i^{id} \rrbracket^{\mathrm{can}_{\mathbb{B}[X]}(P_i^{id})} \leq \llbracket P_i^c \rrbracket^{\mathrm{can}_{\mathbb{B}[X]}(P_i^{id})} \leq  \llbracket P_i \rrbracket^{\mathrm{can}_{\mathbb{B}[X]}(P_i^{id})} \leq  \llbracket \bigcup_{1\leq i \leq n} P_i\rrbracket^{\mathrm{can}_{\mathbb{B}[X]}(P_i^{id})}  \leq \llbracket \bigcup_{1\leq i \leq m} Q_i \rrbracket^{\mathrm{can}_{\mathbb{B}[X]}(P_i^{id})}$.

$(2) \implies (3):$  Assume that (2) holds, for each $1\leq i \leq n$,   every identification $P_i^{id}$ of $P_i$, we have that  $$\llbracket P_i^{id} \rrbracket^{\mathrm{can}_{\mathbb{B}[X]}(P_i^{id})}(\vec{u}) \leq \llbracket \bigcup_{1\leq i \leq m} Q_i \rrbracket^{\mathrm{can}_{\mathbb{B}[X]}(P_i^{id})}(\vec{u}),$$
where $\vec{u}$ is the tuple of free variables of $P_i^{id}$. The polynomial $\llbracket P_i^{id} \rrbracket^{\mathrm{can}_{\mathbb{B}[X]}(P_i^{id})}(\vec{u})$ has as a term a monomial $x_1 \cdot \ldots \cdot x_n $ containing every tag $p$ appearing in $\mathrm{can}_{\mathbb{B}[X]}(P_i^{id})$ possibly with repetitions (such monomial is obtained using the identity assignment for the variables in $ P_i^{id}$; noticing that the disequations in $P_i^{id}$ all take value $1$ since $P_i$ and $P_i^c$ are consistent).  Then, by the definition of $\leq $ for $\mathbb{B}[X]$, we must have that the monomial $x_1 \cdot \ldots \cdot x_n $ appears in $\llbracket \bigcup_{1\leq i \leq m} Q_i\rrbracket^{\mathrm{can}_{\mathbb{B}[X]}(P_i^{id})}(\vec{u})$. Since $\bigcup_{1\leq i \leq m} Q_i$ is a  UCQ$^\neq$, we must have that there is a $j\leq m$, and an assignment $v$ to the bound variables in $Q_j$ over $\mathrm{can}_{\mathbb{N}[X]}(P_i^{id})$ such that it produces the monomial $x_1 \cdot \ldots \cdot x_n $ as the value for the body of $Q_j$.
Furthermore, the disequations in $Q_j$ must all take value $1$ under the assignment $v$ in order to obtain the monomial  $x_1 \cdot \ldots \cdot x_n $ (and not $0$), and thus they must be  equations holding in $\mathrm{can}_{\mathbb{N}[X]}(P_i^{id})$ (and thus logically entailed by the body of $P_i^{id}$) and disequations appearing in the body of $P_i^{id}$ (since the latter contains a complete description of disequations in  $\mathrm{can}_{\mathbb{N}[X]}(P_i^{id})$).  Then, we get the desired 
 $\neq$-homomorphism $h_{P_i^{id}}\colon Q_j \longrightarrow P_i^{id}$ exact for relational atoms by letting $h_{P_i^{id}}(x)$ be the variable $v(x)$ (recall that the elements of the active domain of $\mathrm{can}_{\mathbb{N}[X]}(P_i^{id})$ are the variables of $P_i^{id}$). To show injectivity, notice that any tag $p$ that is the value of a relational atom of $Q_j$ according to $v$, appears in $x_1 \cdot \ldots \cdot x_n $ with at most the same multiplicity as in  the body of $P_i^{id}$ (since this monomial corresponds to the body of $P^{\mathrm{id}}$), which entails that the multiset of images by $h_{P_i^{id}}$ of relational atoms of $Q_j$ is bag-contained in the multiset of relational atoms from $P_i^{id}$. Surjectivity follows from the fact that $x_1 \cdot \ldots \cdot x_n $ corresponds to the multiset of relational atoms from $P_i^{id}$ but also to the images by $h_{P_i^{id}}$ of the multiset of relational atoms from $Q_j$, so the former is bag-contained in the latter.

$(3) \implies (1):$ Assume that for each $1\leq i \leq n$,   every identification $P^{\mathrm{id}}_i$ of $P_i$,  there is a $j \leq m$ and a $\neq$-homomorphism 
$h_{P^{\mathrm{id}}_i} \colon Q_j \longrightarrow P^{\mathrm{id}}_i$  exact for relational atoms. In order to establish $\bigcup_{1\leq i \leq n} P_i  \sqsubseteq_{\mathbb{B}[X]} \bigcup_{1\leq i \leq m} Q_i$, consider an arbitrary $\mathbb{B}[X]$-database instance $\mathbf I$ and a tuple $\vec{t}$. We need to show that any monomial in $\llbracket \bigcup_{1\leq i \leq n} P_i  \rrbracket^\mathbf{I}(\vec{t})$ appears  in $\llbracket \bigcup_{1\leq i \leq m} Q_i \rrbracket^\mathbf{I}(\vec{t})$. Take an assignment $v$ of elements to the variables of $P_i$ justifying a given monomial $x_1 , \ldots , x_n $ and consider the completion $ P_i^c$ of $P_i$ obtained by adding $x=y$ or $x\neq y$ depending on whether  $v(x) = v(y)$ or $v(x) \neq v(y)$. Take an identification $P_i^{id}$ of $P_i^c$ by a canonical substitution $\theta$. We may observe that $\theta(x) = \theta(y)$ iff $v(x) = v(y)$. Then, $x_1 , \ldots , x_n $ is also a monomial in $\llbracket P_i^{id} \rrbracket^\mathbf{I}(\vec{t})$ justified by the assignment $v$ itself since the new disequations added can only take value $1$ by construction according to the assignment $v$.  By assumption, there is a $\neq$-homomorphism $h_{P_i^{id}}\colon Q_j \longrightarrow P_i^c$ exact for relational atoms. Then, $v \circ h_{P_i^{id}}$ is an assignment of elements to the variables of $Q$ justifying the monomial $x_1 , \ldots , x_n $. If all the $x_i$s are values of relational atoms of $P_i$ according to $v$, then     if $x_i = \llbracket B(\theta(\vec{z})) \rrbracket^\mathbf{I}(v(\theta(\vec{z})))$ where  $B(\vec{z})$ is a relational atom of  $P_i$ (and thus $B(\theta(\vec{z}))$ is one of $P_i^{id}$), then $x_i = \llbracket A(\overline{x}) \rrbracket^\mathbf{I}(v(h_{P_i^{id}}(\overline{x})))$  for some relational atom $A(\overline{x})$ of $Q$ such that $A(h_{P_i^{id}}(\overline{x})) = B(\theta(\vec{z}))$, which exists by surjectivity.   Furthermore, the multiplicity  with which  $x_i$ appears using the assignment $v \circ h_{P_i^{id}} $ is exactly the multiplicity with which it appears in  $x_1 \ldots x_m $ given that $h_{P_i^{id}}$ is both onto and injective on relational atoms.
Since equations and disequations can only take values $0$ or $1$, they can be disregarded in showing that we can justify a monomial.

The upper bound complexity calculation follows by a similar argument as in Remark~\ref{complexityremark}. The lower bound is immediate from the case of CQ$^\neq$s.
\end{proof}

\propsixonebis*

\begin{proof}
If the query consists of one relation with only one atom, the statement holds by construction of $h$. If $\overline{Q}$ consists only of an equation or disequation, the property holds since equations and disequations take values in $\{0,1\}$, and $h(0_{K_1})=0_{K_2}$ and $h(1_{K_1})=1_{K_2}$.

Recall that, for a CQ$^\neq$, $\llbracket Q \rrbracket^\mathbf{I}(t) \stackrel{\text{def}}{=} \sum_{v \text{ s.t. } v(\vec{u})=t} \prod_{i=1}^n R_i^\mathbf{I}\left(v\left(\vec{u}_i\right)\right),$ where equations and disequations contribute either $0$ or $1$. Then, for a UCQ$^\neq$ $\overline Q=\bigcup_{1\leq j\leq m}Q_j$, $\llbracket \overline{Q} \rrbracket^\mathbf{I}(t) = \sum_{j=1}^{m} \llbracket Q_j \rrbracket^\mathbf{I}(t).$ Since $h$ preserves $0$, $1$, finite sums and finite products, applying $h$ to the evaluation of each assignment, and then to the sum over all assignments and all disjuncts, gives $h(\llbracket \overline{Q} \rrbracket^\mathbf{I}(t)) = \llbracket \overline{Q} \rrbracket^{h(I)}(t). $
\end{proof}

\propthrethreebis*

\begin{proof}
If $a \leq_{K_1} b$, there exists a $c \in K_1$ such that $b=a+c$. Applying the homomorphism, $h(b)=h(a+c)=h(a)+h(c),$ so $h(a)\leq_{K_2}h(b)$.
\end{proof}

\lemmasevenfifteenbis*

\begin{proof}
Suppose $\llbracket \overline{P} \rrbracket^\mathbf{I} \not\leq_{\mathbb{N}[X]} \llbracket \overline{Q} \rrbracket^\mathbf{I}$ for some $\mathbb{N}[X]$-instance $\mathbf I$. Let $\mathbb{N}[X]'$ denote the subsemiring of $\mathbb{N}[X]$ generated by the annotations actually occurring in $\mathbf I$. Let $v\colon X \longrightarrow \mathbb{N}[X]'$ be the valuation which maps each variable used as an annotation in $J$ to the corresponding annotation used in $\mathbf I$, and maps all the remaining variables to $0$.

By the universality of $\mathbb{N}[X]$, $v$ can be extended uniquely to a semiring homomorphism $\text{Eval}_v\colon \mathbb{N}[X] \longrightarrow \mathbb{N}[X]'.$ Observe that $\text{Eval}_v(J)=I$. It follows, using Proposition~\ref{prop6.1bis}, that $\text{Eval}_v(\llbracket \overline{P} \rrbracket^J) = \llbracket \overline{P} \rrbracket^\mathbf{I}$ and $\text{Eval}_v(\llbracket \overline{Q} \rrbracket^J) = \llbracket \overline{Q} \rrbracket^\mathbf{I}.$ If $\llbracket \overline{P} \rrbracket^J \leq_{\mathbb{N}[X]} \llbracket \overline{Q} \rrbracket^J$, Proposition~\ref{prop3.3bis} would imply $\llbracket \overline{P} \rrbracket^\mathbf{I}
\leq_{\mathbb{N}[X]}
\llbracket \overline{Q} \rrbracket^\mathbf{I},$
a contradiction. Hence $\llbracket \overline{P} \rrbracket^J \not\leq_{\mathbb{N}[X]} \llbracket \overline{Q} \rrbracket^J$, as required.
\end{proof}

\thmsevensixteenbis*

\begin{proof}
$(\Leftarrow)$ is trivial.

For $(\Rightarrow)$, suppose $\overline{P} \not\sqsubseteq_{\mathbb{N}[X]} \overline{Q}$. Then, for some $\mathbb{N}[X]$-instance $\mathbf I$, we have $\llbracket \overline{P} \rrbracket^\mathbf{I}
\not\leq_{\mathbb{N}[X]}
\llbracket \overline{Q} \rrbracket^\mathbf{I}.$ By Lemma~\ref{lemma7.15bis}, we may assume that $\mathbf I$ is abstractly-tagged. Choose some tuple $\vec{t}$ such that $\llbracket \overline{P} \rrbracket^\mathbf{I}(\vec{t})
\not\leq_{\mathbb{N}[X]}
\llbracket \overline{Q} \rrbracket^\mathbf{I}(\vec{t}).$ There must be some monomial $\mu$ in the polynomial $\llbracket \overline{P} \rrbracket^\mathbf{I}(\vec{t})$ with coefficient $a$ such that the same monomial $\mu$ in the polynomial $\llbracket \overline{Q} \rrbracket^\mathbf{I}(\vec{t})$ has coefficient $b$ and $a>b$.

Since $a>0$, there is some disjunct $P_i$ of $\overline P$ and an assignment $v$ to the variables of $P_i$, producing $\vec t$, whose contribution is $\mu$. Hence all equations and disequations of $P_i$ take value $1$ under $v$. Moreover, every distinct tag occurring in $\mu$ annotates a ground tuple used by $v$ for some relational atom occurrence of $P_i$. Since $P_i$ contains at most $k$ relational atom occurrences, at most $k$ distinct tuple annotations occur in $\mu$.

Now let $\mathbf I'$ be the $\mathbb{N}[X]$-instance obtained from $\mathbf I$ by discarding, i.e.\ setting their annotations to $0$, all tuples whose abstract tags do not occur in $\mu$. Thus $\mathbf I'$ contains at most $k$ tuples annotated with non-zero values.

We show that the coefficient of $\mu$ is unchanged in the evaluation of either $\overline P$ or $\overline Q$. Consider an assignment to a disjunct of either query whose contribution under $\mathbf I$ is $\mu$. If this assignment used a tuple discarded in $\mathbf I'$, its contribution would contain the fresh tag annotating that tuple. Since $\mathbf I$ is abstractly-tagged and that tag does not occur in $\mu$, the contribution could not be equal to $\mu$. Hence every assignment contributing $\mu$ under $\mathbf I$ uses only tuples retained in $\mathbf I'$ and contributes the same monomial under $\mathbf I'$. Conversely, every assignment contributing $\mu$ under $\mathbf I'$ also contributes $\mu$ under $\mathbf I$, since $\mathbf I'$ agrees with $\mathbf I$ on all retained tuples. Notice that repeated occurrences of the same tuple cause no problem, they simply produce the corresponding positive power of its tag in $\mu$.

Consequently, $\mu$ still has coefficient $a$ in $\llbracket \overline{P} \rrbracket^{\mathbf I'}(\vec t)$ and coefficient $b$ in $\llbracket \overline{Q} \rrbracket^{\mathbf I'}(\vec t)$. Since $a>b$, $\llbracket \overline{P} \rrbracket^{\mathbf I'}(\vec t)
\not\leq_{\mathbb{N}[X]}
\llbracket \overline{Q} \rrbracket^{\mathbf I'}(\vec t),$ and therefore $\llbracket \overline{P} \rrbracket^{\mathbf I'}
\not\leq_{\mathbb{N}[X]}
\llbracket \overline{Q} \rrbracket^{\mathbf I'}.$
\end{proof}

Recall that, for $\mathrm{Trio}(X)$, replacing every non-zero tuple annotation by a single fresh tag is not sufficient. For example, consider $P(x):=R(x)\wedge R(x) $ and $Q(x):=R(x),$ and a $\mathrm{Trio}(X)$-instance $\mathbf I$ such that $R^\mathbf I(a)=2$. Then $\llbracket P\rrbracket^\mathbf I(a)=4$ and $\llbracket Q\rrbracket^\mathbf I(a)=2,$ and hence $\llbracket P\rrbracket^\mathbf I(a)
\not\leq_{\mathrm{Trio}(X)}
\llbracket Q\rrbracket^\mathbf I(a).$ If the annotation of $R(a)$ is replaced by a single fresh tag $r$, both evaluations become $r$, since positive exponents are dropped in $\mathrm{Trio}(X)$. On the other hand, replacing it by the $2$-abstract tag $r+s$ gives $(r+s)^2=r+s+2rs$ for $P$, while the evaluation of $Q$ is $r+s$. Thus, using sufficiently large abstract tags allows us to retain the information that can be lost by dropping positive exponents.

We first make explicit the bounded-degree order property that will be used below. Let $Y=\{y_1,\ldots,y_s\}$ be a finite set of variables and, for every $i$, let $B_i=x_{i,1}+\ldots+x_{i,k},$ where all the variables occurring in different $B_i$'s are pairwise distinct. 

\begin{restatable}{Lem}{trioordertransferUCQneq}\label{trioordertransferUCQneq}
Let $F,G\in\mathbb{N}[Y]$ be polynomials of degree at most $k$, and let $a_1,\ldots,a_s\in\mathrm{Trio}(X)$. If $F(B_1,\ldots,B_s)
\leq_{\mathrm{Trio}(X)}
G(B_1,\ldots,B_s),$ then $F(a_1,\ldots,a_s)
\leq_{\mathrm{Trio}(X)}
G(a_1,\ldots,a_s).$
\end{restatable}

\begin{proof}
For every $i$, write $a_i$ as a finite sum of square-free monomials, where a monomial occurring with coefficient $c$ is considered as occurring $c$ times in the sum. Fix a square-free monomial $\mu$.

For integers $q_1,\ldots,q_s$ between $0$ and $k$, consider the square-free monomial $\tau=\prod_{i=1}^{s}\prod_{j=1}^{q_i}x_{i,j},$ where the corresponding product is empty when $q_i=0$. Consider now a monomial $m=y_1^{e_1}\ldots y_s^{e_s}$ of degree at most $k$. In the expansion of $m(B_1,\ldots,B_s)$, the number of terms that give $\tau$ after dropping positive exponents is the number of ways of assigning the $e_i$ occurrences of $y_i$ to the $q_i$ variables $x_{i,1},\ldots,x_{i,q_i}$ in such a way that all these $q_i$ variables are used, for every $i$.

The same number occurs when evaluating $m(a_1,\ldots,a_s)$. Indeed, fix, for every $i$, $q_i$ distinct summands of $a_i$. The number of ways in which the $e_i$ occurrences of $y_i$ can choose these summands so that all of them are used is exactly the same as above. Therefore, for fixed $q_1,\ldots,q_s$, the contributions to $\mu$ under the substitution $y_i\mapsto a_i$ are obtained from the coefficient of the corresponding monomial $\tau$ under the substitution $y_i\mapsto B_i$ by multiplying it by the number of choices of $q_i$ distinct summands of each $a_i$ whose product, after dropping positive exponents, is $\mu$.

Suppose that $\tau$ occurs in $F(B_1,\ldots,B_s)$ with coefficient $c$ and in $G(B_1,\ldots,B_s)$ with coefficient $d$. By hypothesis, $c\leq d$ for every possible choice of $q_1,\ldots,q_s$. Moreover, the numbers by which these coefficients are multiplied in order to obtain the contributions to $\mu$ are non-negative and are the same for $F$ and $G$. Summing over all possible choices of $q_1,\ldots,q_s$, it follows that the coefficient of $\mu$ in $F(a_1,\ldots,a_s)$ is at most its coefficient in $G(a_1,\ldots,a_s)$. Since $\mu$ was arbitrary, $F(a_1,\ldots,a_s) \leq_{\mathrm{Trio}(X)} G(a_1,\ldots,a_s),$ as required.
\end{proof}

We can now show that a counterexample can always be replaced by a sufficiently abstractly tagged one as in the case of $\mathbb{N}[X]$. 

\begin{restatable}{Lem}{trioabstractUCQneq}\label{trioabstractUCQneq}
Let $\overline P,\overline Q$ be UCQ$^\neq$s such that every CQ$^\neq$ occurring in either union has at most $k$ relational atom occurrences. If $\llbracket\overline P\rrbracket^\mathbf I
\not\leq_{\mathrm{Trio}(X)}
\llbracket\overline Q\rrbracket^\mathbf I$ for some $\mathrm{Trio}(X)$-instance $\mathbf I$, then $\llbracket\overline P\rrbracket^{\mathrm{ab}_k(\mathbf I)}
\not\leq_{\mathrm{Trio}(X)}
\llbracket\overline Q\rrbracket^{\mathrm{ab}_k(\mathbf I)}.$
\end{restatable}

\begin{proof}
Choose an output tuple $\vec t$ such that $\llbracket\overline P\rrbracket^\mathbf I(\vec t)
\not\leq_{\mathrm{Trio}(X)}
\llbracket\overline Q\rrbracket^\mathbf I(\vec t).$ For every tuple $e$ annotated with a non-zero value in $\mathbf I$, choose a fresh variable $y_e$. Consider the instance obtained by replacing the annotation of every such tuple by $y_e$. Evaluating $\overline P$ and $\overline Q$ on $\vec t$ over this instance gives two polynomials $F,G\in\mathbb{N}[Y]$, where $Y$ is the set of the variables $y_e$.

Every monomial occurring in $F$ or $G$ is contributed by an assignment to some CQ$^\neq$ in one of the two unions. Equations and disequations contribute only $0$ or $1$, while every relational atom occurrence contributes one tuple annotation. Hence $F$ and $G$ have degree at most $k$.

For every supported tuple $e$, let $B_e$ be its $k$-abstract tag in $\mathrm{ab}_k(\mathbf I)$. Then $F((B_e)_e) = \llbracket\overline P\rrbracket^{\mathrm{ab}_k(\mathbf I)}(\vec t)$ and $G((B_e)_e) = \llbracket\overline Q\rrbracket^{\mathrm{ab}_k(\mathbf I)}(\vec t).$ On the other hand, substituting the original annotation $\mathbf I(e)$ for every $y_e$ gives $F((\mathbf I(e))_e) = \llbracket\overline P\rrbracket^\mathbf I(\vec t)$ and $G((\mathbf I(e))_e)=\llbracket\overline Q\rrbracket^\mathbf I(\vec t).$ 
If $\llbracket\overline P\rrbracket^{\mathrm{ab}_k(\mathbf I)}(\vec t)
\leq_{\mathrm{Trio}(X)}
\llbracket\overline Q\rrbracket^{\mathrm{ab}_k(\mathbf I)}(\vec t),$ then Lemma~\ref{trioordertransferUCQneq} would imply $\llbracket\overline P\rrbracket^\mathbf I(\vec t)
\leq_{\mathrm{Trio}(X)}
\llbracket\overline Q\rrbracket^\mathbf I(\vec t),$ a contradiction. Hence the $k$-abstractly tagged instance is also a counterexample.
\end{proof}

Next, we show that the number of tuples annotated with non-zero values in a counterexample can also be bounded by a small model theorem. 

\begin{restatable}{Thm}{trioucqneqsmall}\label{trioucqneqsmall}
Let $\overline P,\overline Q$ be UCQ$^\neq$s, and let $k$ be the maximum between $1$ and the maximum number of relational atom occurrences in the body of a CQ$^\neq$ occurring in $\overline P$ or $\overline Q$. Then $\overline P\not\sqsubseteq_{\mathrm{Trio}(X)}\overline Q$ if and only if $\llbracket\overline P\rrbracket^\mathbf I
\not\leq_{\mathrm{Trio}(X)}
\llbracket\overline Q\rrbracket^\mathbf I$ for some $k$-abstractly tagged $\mathrm{Trio}(X)$-instance $\mathbf I$ containing at most $k$ tuples annotated with non-zero values.
\end{restatable}

\begin{proof}
$(\Leftarrow)$ is trivial.

For $(\Rightarrow)$, suppose $\overline P\not\sqsubseteq_{\mathrm{Trio}(X)}\overline Q.$ Then, for some $\mathrm{Trio}(X)$-instance $\mathbf J$, $\llbracket\overline P\rrbracket^\mathbf J
\not\leq_{\mathrm{Trio}(X)}
\llbracket\overline Q\rrbracket^\mathbf J.$ By Lemma~\ref{trioabstractUCQneq}, we may consider the $k$-abstractly tagged instance $\mathbf I=\mathrm{ab}_k(\mathbf J)$. Choose some tuple $\vec t$ such that $\llbracket\overline P\rrbracket^\mathbf I(\vec t)
\not\leq_{\mathrm{Trio}(X)}
\llbracket\overline Q\rrbracket^\mathbf I(\vec t).$ Since the natural order of $\mathrm{Trio}(X)$ compares the coefficients of square-free monomials, there must be some square-free monomial $\mu$ in $\llbracket\overline P\rrbracket^\mathbf I(\vec t)$ with coefficient $a$ such that the same monomial $\mu$ in $\llbracket\overline Q\rrbracket^\mathbf I(\vec t)$ has coefficient $b$ and $a>b$.

Since $a>0$, there is some disjunct $P_i$ of $\overline P$, an assignment $v$ to the variables of $P_i$ producing $\vec t$, and a choice of one variable from the block annotating the image of each relational atom occurrence of $P_i$, whose product, after dropping positive exponents, is $\mu$. In particular, all equations and disequations of $P_i$ are satisfied by $v$.

Let $\mathbf I'$ be the instance obtained from $\mathbf I$ by retaining precisely those tuples whose block annotations contain at least one variable occurring in $\mu$, and by assigning $0$ to every other tuple. Every block containing a variable occurring in $\mu$ annotates a tuple used by the contribution described above. Since $P_i$ contains at most $k$ relational atom occurrences, $\mathbf I'$ contains at most $k$ tuples annotated with non-zero values.

We show that the coefficient of $\mu$ is unchanged in the evaluation of either query. Consider an assignment to a disjunct of $\overline P$ or $\overline Q$, together with a choice of one variable from the block corresponding to each relational atom occurrence, whose contribution under $\mathbf I$ is $\mu$. If this contribution used a tuple discarded in $\mathbf I'$, at least one variable from the block annotating that tuple would occur in the resulting product. Since the blocks used for distinct tuples are pairwise disjoint, this variable cannot come from any other tuple. Moreover, although $\mathrm{Trio}(X)$ drops positive exponents, it never removes the last occurrence of a variable. Hence that variable would occur in $\mu$, contradicting the definition of $\mathbf I'$. Therefore, every contribution to $\mu$ under $\mathbf I$ uses only tuples retained in $\mathbf I'$ and remains unchanged.

Conversely, since $\mathbf I'$ agrees with $\mathbf I$ on every retained tuple, every contribution to $\mu$ under $\mathbf I'$ was already a contribution to $\mu$ under $\mathbf I$. Consequently, $\mu$ still has coefficient $a$ in $\llbracket\overline P\rrbracket^{\mathbf I'}(\vec t)$ and coefficient $b$ in $\llbracket\overline Q\rrbracket^{\mathbf I'}(\vec t)$. Since $a>b$, $\llbracket\overline P\rrbracket^{\mathbf I'}(\vec t) \not\leq_{\mathrm{Trio}(X)}
\llbracket\overline Q\rrbracket^{\mathbf I'}(\vec t),$ and therefore $\mathbf I'$ is the required counterexample.
\end{proof}

We are now ready to prove the upper bound result for $\mathrm{Trio}(X)$ by an analogous reasoning to the $\mathbb{N}[X]$. 

\begin{restatable}{Thm}{ucqdistrio}\label{ucqdistrio}\label{triopspace}
$\mathrm{Trio}(X)$-containment of UCQ$^\neq$s is in $\text{coNP}^{\text{PP}}$.
\end{restatable}

\begin{proof}
For the upper bound complexity calculation, we use the same counting argument as for CQ$^\neq$s over $\mathrm{Trio}(X)$. We look at the complement of the containment problem and show that it is in $\text{NP}^{\text{PP}}$.

Write $\overline P=\bigcup_{1\leq i\leq n}P_i$ and $\overline Q=\bigcup_{1\leq j\leq m}Q_j$. Suppose first that $\overline  P\not\sqsubseteq_{\mathrm{Trio}(X)}\overline Q.$
By the construction in the proof of Theorem~\ref{trioucqneqsmall}, we may consider a $k$-abstractly tagged counterexample $\mathbf I$, an output tuple $\vec t$, and a square-free monomial $\mu$ whose coefficient in $\llbracket\overline P\rrbracket^\mathbf I(\vec t)$ is strictly greater than its coefficient in $\llbracket\overline Q\rrbracket^\mathbf I(\vec t)$. Moreover, since the coefficient of $\mu$ on the left-hand side is positive, we can fix a disjunct $P_i$, an assignment $v$ to its variables producing $\vec t$, and, for every relational atom occurrence of $P_i$, one choice of a variable from the block annotating its image tuple, such that the resulting square-free monomial is $\mu$.

Let $P^c_{i,v}$ be the completion of $P_i$ which, for every pair of variables $x,y$, contains $x=y$ if $v(x)=v(y)$ and $x\neq y$ otherwise. Let $P^{\mathrm{id}}_{i,v}$ be the corresponding identification. As in the proof of Theorem~\ref{triocqneqcontainment}, the assignment $v$ induces a bijection between the variables of $P^{\mathrm{id}}_{i,v}$ and the domain elements occurring in the tuples used by the contribution to $\mu$. Under this bijection, the syntactically distinct relational atoms of $P^{\mathrm{id}}_{i,v}$ correspond exactly to the distinct ground relational tuples involved in the counterexample. Since these tuples are annotated in $\mathbf I$ by pairwise disjoint elements of $S_k$, up to a renaming of domain elements and tags we may identify $\mathbf I$ with $\text{can}^{k}_{\mathrm{Trio}(X)}(P^{\mathrm{id}}_{i,v})$.

Under this renaming, the monomial $\mu$ determines a $k$-tag signature $\sigma\colon\text{RelSet}(P^{\mathrm{id}}_{i,v})\longrightarrow\mathcal{P}([k])$ by letting $\sigma(A)$ contain exactly those indices of the variables from the $k$-abstract tag of $A$ that occur in $\mu$. Thus $\mu$ corresponds to the monomial $M_\sigma$ associated with $\sigma$.

By Lemma~\ref{indexedTrio}, for every CQ$^\neq$ $R$ with the same head arity, the coefficient of $M_\sigma$ in $\llbracket R\rrbracket^{\text{can}^{k}_{\mathrm{Trio}(X)}(P^{\mathrm{id}}_{i,v})}$ is exactly $\left|\text{IHom}_k(R,P^{\mathrm{id}}_{i,v},\sigma)\right|.$ Since union adds the contributions of its disjuncts, the coefficient of $M_\sigma$ in the evaluation of $\overline P$ is therefore $\sum_{1\leq \ell\leq n}
\left|\text{IHom}_k(P_\ell,P^{\mathrm{id}}_{i,v},\sigma)\right|,$ while its coefficient in the evaluation of $\overline Q$ is $\sum_{1\leq j\leq m}
\left|\text{IHom}_k(Q_j,P^{\mathrm{id}}_{i,v},\sigma)\right|.$ Since $\mu$ had a strictly greater coefficient on the left, we obtain $\sum_{1\leq \ell\leq n}
\left|\text{IHom}_k(P_\ell,P^{\mathrm{id}}_{i,v},\sigma)\right|  > \sum_{1\leq j\leq m}
\left|\text{IHom}_k(Q_j,P^{\mathrm{id}}_{i,v},\sigma)\right|.$

Conversely, suppose that for some disjunct $P_i$ of $\overline P$, some identification $P_i^{\mathrm{id}}$ of $P_i$, and some $k$-tag signature $\sigma\colon\text{RelSet}(P_i^{\mathrm{id}})\longrightarrow\mathcal{P}([k]),$ we have $\sum_{1\leq \ell\leq n}
\left|\text{IHom}_k(P_\ell,P_i^{\mathrm{id}},\sigma)\right| > \sum_{1\leq j\leq m} \left|\text{IHom}_k(Q_j,P_i^{\mathrm{id}},\sigma)\right|.$ Consider the $k$-abstract canonical database $\text{can}^{k}_{\mathrm{Trio}(X)}(P_i^{\mathrm{id}})$. By Lemma~\ref{indexedTrio} and the fact that union adds the evaluations of its disjuncts, the two sides of the above inequality are exactly the coefficients of $M_\sigma$ in $\llbracket\overline P\rrbracket^{\text{can}^{k}_{\mathrm{Trio}(X)}(P_i^{\mathrm{id}})}$ and $\llbracket\overline Q\rrbracket^{\text{can}^{k}_{\mathrm{Trio}(X)}(P_i^{\mathrm{id}})},$ respectively. Hence the coefficientwise order fails, and therefore $\overline P\not\sqsubseteq_{\mathrm{Trio}(X)}\overline Q.$

It remains to calculate the complexity of checking the existence of such an identification and signature. An identification $P_i^{\mathrm{id}}$ is obtained from a completion of $P_i$, which fixes, for every pair of variables $x,y$, whether $x=y$ or $x\neq y$. Thus a completion, and hence the corresponding identification, has polynomial size and can be guessed nondeterministically. Moreover, $k$ is bounded by the number of relational atom occurrences in the input queries, and a $k$-tag signature for $P_i^{\mathrm{id}}$ can be represented by specifying, for every atom in $\text{RelSet}(P_i^{\mathrm{id}})$, a subset of $[k]$. Hence the signature also has polynomial size.

For fixed $P_i^{\mathrm{id}}$ and $\sigma$, the quantity $\sum_{1\leq \ell\leq n}
\left|\text{IHom}_k(P_\ell,P_i^{\mathrm{id}},\sigma)\right|$ is a $\#\text{P}$ function. Indeed, a nondeterministic polynomial-time machine guesses a disjunct $P_\ell$, a mapping $h$ from its variables to the variables of $P_i^{\mathrm{id}}$, and, for every relational atom occurrence $B\in\text{RelOcc}(P_\ell)$, an index $\iota(B)\in[k]$. It accepts exactly when $h$ is a $\neq$-homomorphism preserving the head and the pair $(h,\iota)$ realizes $\sigma$. Thus the accepting computations are precisely the pairs consisting of a disjunct $P_\ell$ and an element of $\text{IHom}_k(P_\ell,P_i^{\mathrm{id}},\sigma)$, and their number is exactly the sum above. By the same argument, $\sum_{1\leq j\leq m}
\left|\text{IHom}_k(Q_j,P_i^{\mathrm{id}},\sigma)\right|$ is also a $\#\text{P}$ function.

Consequently, the difference between these two quantities is in $\text{GapP}$ by~\cite[Prop. 3.5]{Fenner}, and deciding whether this difference is positive is in $\text{PP}$ by~\cite[Prop. 4.2]{Fenner}. A nondeterministic polynomial-time machine can therefore guess a disjunct $P_i$, one of its identifications $P_i^{\mathrm{id}}$, and a $k$-tag signature $\sigma$, and use a $\text{PP}$ oracle to check whether the above strict inequality holds. Hence non-containment is in $\text{NP}^{\text{PP}}$, and therefore $\mathrm{Trio}(X)$-containment of UCQ$^\neq$s is in $\text{coNP}^{\text{PP}}$.
\end{proof}

\section{Omitted  Proofs from \S~\ref{The problem of containmentneq}} \label{app:cqn-containment-proofs}

\cqnContainmentA*

\begin{proof}
Let $Q(\vec{x})$ be  $\exists \vec{z} (A_1(\vec{x_1}) \wedge \ldots \wedge A_n(\vec{ x_n}))$ where each $A_i(\vec{ x_i})$ is a literal, and $P(\vec{z})$ be $\exists \vec{u} (B_1(\vec{ y_1}) \wedge \ldots \wedge B_m(\vec{ y_m}))$,   where each $B_j(\vec{ x_j})$ is a literal and where $\overline{x}, \vec{z} $ and $\vec{z}, \vec{u}$ form partitions of $\overline{x_1}, \ldots , \overline{ x_n}$ and $\vec{z_1}, \ldots , \overline{ y_n}$, respectively.

 $(1) \implies (2)$: Suppose that $P \sqsubseteq_{\mathbb{L}}^{\text{md}} Q$, so for each $\mathbb{L}$-instance $\mathbf I$ and model-defining interpretation $\pi$ compatible with it, and tuple $\vec{t}$, we have that $\pi\llbracket P(\vec{t})\rrbracket \leq_\mathbb{L} \pi\llbracket Q(\vec{t})\rrbracket$.  Now build an $\mathbb{L}$-database denoted by $\mathrm{can}_{\mathbb{L}}(P)$ where the elements are variables as in the Boolean case, and for any relation symbol $R$, 
we let $R^{\mathrm{can}_{\mathbb{L}}(P)}(x_1, \ldots, x_n) = \top$ if $R(x_1, \ldots, x_n)$ is an atom in  $P$,  $R^{\mathrm{can}_{\mathbb{L}}(P)}(x_1, \ldots, x_n) = \bot$  if  $\neg R(x_1, \ldots, x_n)$ is a literal in  $P$, and $R^{\mathrm{can}_{\mathbb{L}}(P)}(x_1, \ldots, x_n) = \epsilon$ otherwise. Take next an    interpretation $\pi^*$ compatible with $\mathrm{can}_{\mathbb{L}}(P)$ defined as follows: $\pi^*\llbracket R(\vec{a})\rrbracket= R^{\mathrm{can}_{\mathbb{L}}(P)}(\vec{a})$, $\pi^*\llbracket\neg R(\vec{a})\rrbracket=\top$ if $\pi^*\llbracket R(\vec{a})\rrbracket=\bot$, and  $\pi^*\llbracket\neg R(\vec{a})\rrbracket=\bot$ if $\pi^*\llbracket R(\vec{a})\rrbracket = \epsilon$ or $\pi^*\llbracket R(\vec{a})\rrbracket = \top$.   By construction, $\pi^*\llbracket P(\vec{z})\rrbracket = \top$, and thus, from the hypothesis, $\pi^*\llbracket Q(\vec{x})\rrbracket = \top$ since the lattice is bounded. Consequently,  for some assignment $v$, it must be that $\pi^*\llbracket A_1(v(\vec{ x_1})) \wedge \ldots \wedge A_n(v(\vec{ x_n}))\rrbracket=\top$ and thus $\pi^*\llbracket A_i(v(\vec{ x_i})\rrbracket =\top$ for $i\leq n$ since we are in a lattice.
But then, the mapping $h$ that sends each variable $x$ to  the $y$ such that $v(x) = y$, is a homomorphism from $Q$ to $P$ since by construction of $\pi^*$, we have that $\pi^*\llbracket A_i(v(\vec{ x_i})\rrbracket =\top$ only if there is a  $B_{j_i}(\overline{ y_{j_i}})$ in $P$ such that $A_i(v(\vec{ x_i})) = B_{j_i}(\overline{ y_{j_i}})$.
 
  $(2) \implies (1)$: Assume that  there is a negation-homomorphism $h \colon Q \longrightarrow P$. We wish to show that $P \sqsubseteq_{\mathbb{L}}^{\text{md}} Q$. Consider an arbitrary $\mathbb{L}$-database instance $\mathbf I$,  an interpretation $\pi$ compatible with $\mathbf I$,   and  an arbitrary assignment $v$ of elements to the bound variables of $P$. In this case,   $\pi\llbracket B_1(v(\overline{ y_1})) \wedge \ldots \wedge B_m(v(\overline{ y_m}))\rrbracket$ is the infimum of the values $\pi\llbracket B_i(v(\overline{ y_i})) \rrbracket$.    For each literal of $A_i(\vec{ x_i})$ of $Q$ there is a literal $B'_{j_i}(\overline{ y_{j_i}})$ of $P$ such that $A_i(h(\vec{ x_i})) = B'_{j_i}(\overline{ y_{j_i}})$. Thus, $v \circ h$ is an assignment of elements to the bound variables of $Q$ where $\pi\llbracket A_i(v \circ h(\vec{ x_i}))\rrbracket = \pi\llbracket B'_{j_i}(v(\overline{ y_{j_i}}))\rrbracket$, and we must have that $\pi\llbracket B_1(v (\overline{ y_1})) \wedge \ldots \wedge B_m(v (\overline{ y_m}))\rrbracket$ is at most $\pi\llbracket A_1(v \circ h(\vec{ x_1})) \wedge \ldots \wedge A_n(v \circ h(\vec{ x_n}))\rrbracket$.  This means that for any tuple of elements $\vec{t}$, we have that  $\pi\llbracket P(\vec{t})\rrbracket$ (the supremum of all values $\pi\llbracket B_1(e (\overline{ y_1})) \wedge \ldots \wedge B_m(e (\overline{ y_m}))\rrbracket$ for all assignments $e$ to the variables) is at most 
$\pi\llbracket Q(\vec{t})\rrbracket$  (the supremum of all values $\pi\llbracket A_1(e (\vec{ x_1})) \wedge \ldots \wedge A_n(e (\vec{ x_n}))\rrbracket$ for all assignments $e$ to the variables).

Finally, the NP upper bound on the complexity is immediate given item 2 of the characterization in the current theorem. The hardness of the problem is given by the fact that the problem without negation is NP-hard~\cite{Green}.
\end{proof}

\cqnContainmentB*

\begin{proof}
Let $Q(\vec{x})$ be  $\exists \vec{z} (A_1(\vec{x_1}) \wedge \ldots \wedge A_n(\vec{ x_n}))$ where each $A_i(\vec{ x_i})$ is a literal, and $P(\vec{z})$ be $\exists \vec{u} (B_1(\vec{ y_1}) \wedge \ldots \wedge B_m(\vec{ y_m}))$,   where each $B_j(\vec{ x_j})$ is a literal and where $\overline{x}, \vec{z} $ and $\vec{z}, \vec{u}$ form partitions of $\overline{x_1}, \ldots , \overline{ x_n}$ and $\vec{z_1}, \ldots , \overline{ y_n}$, respectively.

 $(1) \implies (2)$: Suppose that $P \sqsubseteq_{\mathbb{L}}^{\text{dmd}} Q$, so for each $\mathbb{L}$-instance $\mathbf I$ and dually model-defining interpretation $\pi$ compatible with it, and tuple $\vec{t}$, we have that $\pi\llbracket P(\vec{t})\rrbracket \leq_\mathbb{L} \pi\llbracket Q(\vec{t})\rrbracket$. Consider a completion $P^c$ of $P$. Now  consider the database $\mathrm{can}_\mathbb{B}(P^c)$ once more. Take next the canonical interpretation $\pi^*$ compatible with $\mathrm{can}_\mathbb{B}(P^c)$. We have then that $\pi^*\llbracket R(\vec{a})\rrbracket= R^{\mathrm{can}_\mathbb{B}(P^c)}(\vec{a})$ while $\pi^*\llbracket\neg R(\vec{a})\rrbracket=\top$ if $\pi^*\llbracket R(\vec{a})\rrbracket=\bot$, and $\pi^*\llbracket\neg R(\vec{a})\rrbracket=\bot$ if $\pi^*\llbracket R(\vec{a})\rrbracket = \top$.  Now, we have that  $ \pi^*\llbracket P^{c}(\bar{t})\rrbracket \leq_\mathbb{L}  \pi^*\llbracket P(\bar{t})\rrbracket$ for any tuple $\bar{t}$ since the possible values of the new literals of $P^c$ are $\top$ and $\bot$, which means that the semiring value  of $P^c(\bar{t})$  can only be at most that of $P(\bar{t})$. By construction, $\pi^*\llbracket P^c(\vec{z})\rrbracket = \top$, and thus, $\pi^*\llbracket P(\vec{z})\rrbracket = \top$ and from the hypothesis, $\pi^*\llbracket Q(\vec{x})\rrbracket = \top$ since the lattice is bounded. Consequently,  for some assignment $v$, it must be that $\pi^*\llbracket A_1(v(\vec{ x_1})) \wedge \ldots \wedge A_n(v(\vec{ x_n}))\rrbracket=\top$ and thus $\pi^*\llbracket A_i(v(\vec{ x_i})\rrbracket =\top$ for $i\leq n$ since we are in a lattice.
But then, the mapping $h_{P^c}$ that sends each variable $x$ to  the $y$ such that $v(x) = y$, is a homomorphism from $Q$ to $P^{c}$ since by construction of $\pi^*$, we have that $\pi^*\llbracket A_i(v(\vec{ x_i})\rrbracket =\top$ only if there is a  $B_{j_i}(\overline{ y_{j_i}})$ in $P^c$ such that $A_i(v(\vec{ x_i})) = B_{j_i}(\overline{ y_{j_i}})$.

  $(2) \implies (1)$: Assume that for every completion $P^c$ of $P$, there is a negation-homomorphism $h_{P^c} \colon Q \longrightarrow P^c$. We wish to show that $P \sqsubseteq_{\mathbb{L}}^{\text{dmd}} Q$. Consider an arbitrary $\mathbb{L}$-database instance $\mathbf I$,  a dually model-defining interpretation $\pi$ compatible with $\mathbf I$,   and  an arbitrary assignment $v$ of elements to the bound variables of $P$. In this case,   $\pi\llbracket B_1(v(\overline{ y_1})) \wedge \ldots \wedge B_m(v(\overline{ y_m}))\rrbracket$ is the infimum of the values $\pi\llbracket B_i(v(\overline{ y_i})) \rrbracket$. Let $P^c$ retain all literals of $P$. For every relation symbol $R$ of the common schema and every tuple $\overline{x}$ of variables of $P$ of the appropriate arity such that neither $R(\vec{x})$ nor $\neg R(\vec{x})$ occurs in $P$, add $R(\vec{x})$ if $\pi\llbracket R(v(\overline{x}))\rrbracket=\top$, and add $\neg R(\vec{x})$ otherwise. Since $P$ contains no atom and its negation and $\pi$ is dually model-defining, this yields a completion of $P$ in which every newly added literal has value $\top$ under $v$.   For each literal of $A_i(\vec{ x_i})$ of $Q$ there is a literal $B'_{j_i}(\overline{ y_{j_i}})$ of $P^c$ such that $A_i(h_{P^c}(\vec{ x_i})) = B'_{j_i}(\overline{ y_{j_i}})$. Thus, $v \circ h_{P^c}$ is an assignment of elements to the bound variables of $Q$ where $\pi\llbracket A_i(v \circ h_{P^c}(\vec{ x_i}))\rrbracket = \pi\llbracket B'_{j_i}(v(\overline{ y_{j_i}}))\rrbracket$, and, since  all the newly added literals to $P$ in order to form $P^c$  will  be inessential in calculations of infima (as they will take value $\top$), we must have that $\pi\llbracket B_1(v (\overline{ y_1})) \wedge \ldots \wedge B_m(v (\overline{ y_m}))\rrbracket$ is at most $\pi\llbracket A_1(v \circ h_{P^c}(\vec{ x_1})) \wedge \ldots \wedge A_n(v \circ h_{P^c}(\vec{ x_n}))\rrbracket$.  This means that for any tuple of elements $\vec{t}$, we have that  $\pi\llbracket P(\vec{t})\rrbracket$ (the supremum of all values $\pi\llbracket B_1(e (\overline{ y_1})) \wedge \ldots \wedge B_m(e (\overline{ y_m}))\rrbracket$ for all assignments $e$ to the variables) is at most 
$\pi\llbracket Q(\vec{t})\rrbracket$  (the supremum of all values $\pi\llbracket A_1(e (\vec{ x_1})) \wedge \ldots \wedge A_n(e (\vec{ x_n}))\rrbracket$ for all assignments $e$ to the variables).

For the complexity calculations, by item 2 of the present theorem,  $P \sqsubseteq_{\mathbb{L}}^{\text{dmd}} Q$ iff $P \sqsubseteq_{\mathbb{B}}^{\text{md}} Q$, and thus we get the same complexity as the Boolean case.
\end{proof}

When moving from the semiring of polynomials over the natural numbers $\mathbb{N}[X, \overline{X}]$ to that of absorptive polynomials $\mathrm{Sorp}(X, \overline{X})$, the strict multiplicity requirement on literal matches is relaxed. Specifically, while containment over $\mathbb{N}[X, \overline{X}]$ requires a negation-homomorphism that is \emph{exact} for literals, over $\mathrm{Sorp}(X, \overline{X})$ it suffices for the homomorphism to be merely \emph{injective}.

\begin{restatable}{Thm}{cqnContainmentD}\label{sorp(X,X)}  For CQNs $P$ and $Q$ with the same tuple of free variables, the following are equivalent:
\begin{itemize}
\item[1.] $P \sqsubseteq^{\text{pt}}_{\mathrm{Sorp}(X, \overline{X})} Q$,
\item[2.]  $ \pi^{\mathrm{can}_{\mathrm{Sorp}(X)}(P^+)}\llbracket P\rrbracket  \leq_{\mathrm{Sorp}(X, \overline{X})} \pi^{\mathrm{can}_{\mathrm{Sorp}(X)}(P^+)}\llbracket Q\rrbracket$.
\item[3.] There is a  negation-homomorphism $h\colon Q \longrightarrow P$ injective for relational literals.

\end{itemize}
Moreover, the corresponding decision problem (i.e.\ $P \sqsubseteq_{\mathrm{Sorp}(X, \overline{X})}^{\text{pt}} Q$?) is NP-complete. 
\end{restatable}

\begin{proof}
$(1) \implies (2):$ Obvious.

$(2) \implies (3):$ Assume that (2) holds, so
$$ \pi^{\mathrm{can}_{\mathrm{Sorp}(X)}(P^+)}\llbracket P(\vec{u})\rrbracket  \leq_{\mathrm{Sorp}(X, \overline{X})} \pi^{\mathrm{can}_{\mathrm{Sorp}(X)}(P^+)}\llbracket Q(\vec{u})\rrbracket,$$
where $\vec{u}$ is the tuple of free variables of $P$. The polynomial $\pi^{\mathrm{can}_{\mathrm{Sorp}(X)}(P^+)}\llbracket P(\vec{u})\rrbracket$ has as a term a monomial $x_1 \cdot \ldots \cdot x_n $ containing every tag from $X \cup \overline{X}$ used to annotate literals by $\pi^{\mathrm{can}_{\mathrm{Sorp}(X)}(P^+)}$,  with some exponent, accounting for duplicate atoms (such monomial is obtained using the identity assignment for the variables in $P$).  Then, by the definition of $\leq $ for $\mathrm{Sorp}(X, \overline{X})$, we must have that there is a monomial $y_1 \cdot \ldots \cdot y_m $ in $\pi^{\mathrm{can}_{\mathrm{Sorp}(X)}(P^+)}\llbracket Q(\vec{u})\rrbracket$ where  only tags from $x_1 \cdot \ldots \cdot x_n $ appear but with equal or smaller exponents as in $x_1 \cdot \ldots \cdot x_n $. Since $Q$ is a  CQN, we must have that there is an assignment $v$ to the bound variables in $Q$ such that it produces the monomial $y_1 \cdot \ldots \cdot y_m$ as the value for the body of $Q$.
  Then, we get the desired 
negation-homomorphism $h\colon Q \longrightarrow P$ injective for literals by letting $h(x)$ be the variable $v(x)$ (recall that the elements of the active domain of $\mathrm{can}_{\mathrm{Sorp}(X)}(P^+)$ are the variables of $P$). To show injectivity, notice that any tag $p$ or $\overline{p}$ that is the value of a literal of $Q$ under assignment  $v$, appears in $y_1 \cdot \ldots \cdot y_m$  with at most the same multiplicity as in  $x_1 \cdot \ldots \cdot x_n $ which in turn corresponds to the body of $P$, which entails that the multiset of images by $h$ of literals of $Q$ is bag-contained in the multiset of literals from $P$.

$(3) \implies (1):$ 
Assume that there is a  negation-homomorphism $h\colon Q \longrightarrow P$ injective for literals. In order to establish $P \sqsubseteq^{\text{pt}}_{\mathrm{Sorp}(X, \overline{X})} Q$, consider an arbitrary $\mathrm{Sorp}(X)$-database instance $\mathbf I$, provenance-tracking interpretation $\pi$ compatible with it, and a tuple $\vec{t}$. We need to show that for any monomial $x_1 \cdot \ldots \cdot x_n$ in $\pi\llbracket P(\vec{t})\rrbracket$, there is a monomial  $y_1 \cdot \ldots \cdot y_m$ in $\pi\llbracket Q(\vec{t})\rrbracket$ where  only tags from $x_1 \cdot \ldots \cdot x_n $ appear but with equal or smaller exponents. Now consider an assignment $v$ of elements to the variables of $P$ justifying  a monomial $x_1 \cdot \ldots \cdot x_m$.  The assignment $v \circ h$   justifies a   monomial with smaller or equal exponents  in $\pi\llbracket Q(\vec{t})\rrbracket$. For example, for any literal $L(\vec{z})$ of $Q$, $ \pi\llbracket L(v\circ h (\vec{z}))\rrbracket = \pi\llbracket L(v  (\overline{x}))\rrbracket$ where $h(\vec{z}) = \overline{x} $ and $L(\vec{x})$ is a literal of $P$. Furthermore, the multiplicity  with which a tag $x_i$ appears as a value of a literal using the assignment $v \circ h $ is at most the multiplicity with which it appears in  $x_1 \ldots x_m $ given that $h$ is  injective on literals. 

The NP upper bound  is given by item 3 of the characterization in the current theorem. The NP lower bound is a result of the NP hardness of the same problem as in item 3  when there are no negations~\cite[Thm. 4.9]{Kostylev}.
\end{proof}

Turning to the Boolean polynomial semiring $\mathbb{B}[X, \overline{X}]$, we observe a return to the stricter condition required under $\mathbb{N}[X, \overline{X}]$. Specifically, provenance-tracking containment over $\mathbb{B}[X, \overline{X}]$ demands that the negation-homomorphism be \emph{exact} for literals, rather than merely \emph{injective} as in $\mathrm{Sorp}(X, \overline{X})$.

\begin{restatable}{Thm}{cqnContainmentE}\label{bxcompletept} For CQNs $P$ and $Q$ with the same tuple of free variables, the following are equivalent:
\begin{itemize}
\item[1.] $P \sqsubseteq_{\mathbb{B}[X, \overline{X}]}^{\text{pt}} Q$,
\item[2.] $ \pi^{\mathrm{can}_{\mathbb{B}[X]}(P^+)}\llbracket P\rrbracket  \leq_{\mathbb{B}[X, \overline{X}]} \pi^{\mathrm{can}_{\mathbb{B}[X]}(P^+)}\llbracket Q\rrbracket$,
\item[3.] There is a negation-homomorphism $h \colon Q \longrightarrow P$ exact for relational literals.

\end{itemize}
Moreover, the corresponding decision problem (i.e.\ $P \sqsubseteq_{\mathbb{B}[X, \overline{X}]}^{\text{pt}} Q$?) is NP-complete. 
\end{restatable}

\begin{proof}
$(1) \implies (2):$ Obvious.

$(2) \implies (3):$  Assume that (2) holds, so 

$$ 
\pi^{\mathrm{can}_{\mathbb{B}[X]}(P^+)}\llbracket P(\vec{u})\rrbracket  \leq_{\mathbb{B}[X, \overline{X}]} \pi^{\mathrm{can}_{\mathbb{B}[X]}(P^+)}\llbracket Q(\vec{u})\rrbracket,
$$
where $\vec{u}$ is the tuple of free variables of $P$. The polynomial $\pi^{\mathrm{can}_{\mathbb{B}[X]}(P^+)}\llbracket P(\vec{u})\rrbracket$ has as a term a monomial $x_1 \cdot \ldots \cdot x_n $ containing every tag from $X \cup \overline{X}$ used to annotate literals by $\pi^{\mathrm{can}_{\mathbb{B}[X]}(P^+)}$, possibly with repetitions (such monomial is obtained using the identity assignment for the variables in $P$). Then, by the definition of $\leq $ for $\mathbb{B}[X, \overline{X}]$, we must have that the monomial $x_1 \cdot \ldots \cdot x_n $ appears in $\pi^{\mathrm{can}_{\mathbb{B}[X]}(P^+)}\llbracket Q(\vec{u})\rrbracket$. The rest of the reasoning is as the corresponding part of Theorem~\ref{N[X, X]}.

$(3) \implies (1):$ Assume that there is a negation-homomorphism $h\colon Q \longrightarrow P$ exact for literals. In order to establish $P \sqsubseteq_{\mathbb{B}[X, \overline{X}]}^{\text{pt}} Q$, consider an arbitrary $\mathbb{N}[X]$-database instance $\mathbf I$, a provenance-tracking interpretation $\pi$ compatible with it and a tuple $\vec{t}$. We need to show that any monomial in $\pi\llbracket P(\vec{t})\rrbracket$ appears  in $\pi\llbracket Q(\vec{t})\rrbracket$. 
Take an assignment $v$ of elements to the variables of $P$ justifying a monomial $x_1 \cdot \ldots \cdot x_n $.  Then, $v \circ h$ is an assignment of elements to the variables of $Q$ justifying the monomial $x_1 \cdot \ldots \cdot x_n $. 
If all the $x_i$s are values of literals of $P$ according to $v$, then if $x_i = \pi\llbracket L(v(\vec{z}))\rrbracket$ where  $L(\vec{z})$ is a literal of $P$, then $x_i = \pi\llbracket L^*(v(h(\overline{x})))\rrbracket$ for some literal $L^*(\overline{x})$ of $Q$ such that $L^*(h(\overline{x})) = L(\vec{z})$, which exists by surjectivity.   Furthermore, the multiplicity with which $x_i$ appears using the assignment $v \circ h $ is exactly the multiplicity with which it appears in  $x_1 \ldots x_m $ given that $h$ is both onto and injective on literals.

The NP upper bound  is given by item 3 of the characterization in the current theorem. The NP lower bound is a result of the NP hardness of the same problem as in item 3  when there are no negations~\cite{Green}.
\end{proof}

Now we examine provenance-tracking containment under the extended Why-provenance semiring, $\mathrm{Why}(X, \overline{X})$. Because both addition and multiplication in $\mathrm{Why}(X, \overline{X})$ are set-theoretic operations (union and pairwise union), duplicate occurrences of literals do not generate distinct provenance tokens (and this are the only kind of annotations we are concerned with in provenance-tracking interpretations). As a result, the structural requirement on the negation-homomorphism relaxes from bag-surjectivity  from the CQ case to being merely \emph{set-onto} for relational literals. This stands in contrast to the need for bag-onto mappings in Theorem~\ref{whychneq}. Clearly, the non-containment between CQs from Example~\ref{whyexm} counts as a proper containment on provenance-tracking interpretations using Theorem~\ref{whyneg} below.

\begin{restatable}{Thm}{cqnContainmentF}\label{whyneg}
For CQNs $P$ and $Q$ with the same tuple of free variables, the following are equivalent:
\begin{itemize}
\item[1.]
$P\sqsubseteq^{\text{pt}}_{\mathrm{Why}(X,\overline{X})}Q$.

\item[2.]$\pi^{\text{can}_{\mathrm{Why}(X)}(P^+)}\llbracket P \rrbracket\leq_{\mathrm{Why}(X,\overline{X})}\pi^{\text{can}_{\mathrm{Why}(X)}(P^+)}\llbracket Q \rrbracket.$

\item[3.]
There is a negation-homomorphism $h\colon Q\longrightarrow P$ set-onto for relational literals, i.e. $h(\text{LitSet}(Q))=\text{LitSet}(P)$.
\end{itemize}

Moreover, the corresponding decision problem (i.e.\ $P\sqsubseteq^{\text{pt}}_{\mathrm{Why}(X,\overline{X})}Q$?) is NP-complete.
\end{restatable}

\begin{proof}
$(1)\implies(2):$
Suppose that $P\sqsubseteq^{\text{pt}}_{\mathrm{Why}(X,\overline{X})}Q$. The interpretation $\pi^{\text{can}_{\mathrm{Why}(X)}(P^+)}$ is provenance-tracking, and hence the inequality in item 1 holds, in particular, on this interpretation. Therefore, $\pi^{\text{can}_{\mathrm{Why}(X)}(P^+)}\llbracket P \rrbracket \leq_{\mathrm{Why}(X,\overline{X})} \pi^{\text{can}_{\mathrm{Why}(X)}(P^+)}\llbracket Q \rrbracket.$

$(2)\implies(3):$
Assume that (2) holds, so $\pi^{\text{can}_{\mathrm{Why}(X)}(P^+)} \llbracket P(\vec{u}) \rrbracket \leq_{\mathrm{Why}(X,\overline{X})} \pi^{\text{can}_{\mathrm{Why}(X)}(P^+)} \llbracket Q(\vec{u}) \rrbracket.$ Let us abbreviate $\text{can}_{\mathrm{Why}(X)}(P^+)$ as $\mathbf I$ and $\pi^{\text{can}_{\mathrm{Why}(X)}(P^+)}$ as $\pi^{\mathbf I}$. Let $\text{LitSet}(P)=\{L_1,\ldots,L_r\}.$ By construction of $\pi^{\mathbf I}$, each distinct relational literal $L_i$ is annotated by a different tag $x_i\in X\cup\overline X$, that is, $\pi^{\mathbf I}\llbracket L_i \rrbracket=\{\{x_i\}\}.$ Repeated occurrences of the same literal receive the same annotation, since they correspond to the same ground relational literal in the canonical interpretation. Moreover, by the definition of the canonical interpretation, no pair of complementary tags occurs among $x_1,\ldots,x_r$: the tags used for negative literals are chosen so that their complementary positive tags are not used in the canonical database. Hence the witness $W_P:=\{x_1,\ldots,x_r\}$ contains no complementary pair and is therefore not reduced to $\emptyset$ in the quotient defining $\mathrm{Why}(X,\overline X)$.

Consider now the assignment to the bound variables of $P$ that sends every variable to itself in $\mathbf I$. Under this assignment, every occurrence of a literal $L_i$ has value $\{\{x_i\}\}$. Since multiplication in $\mathrm{Why}(X,\overline X)$ is pairwise union and $W_P$ contains no complementary pair, their product is $\{W_P\}$ (indeed, multiplication is pairwise union, so, for instance, $\{\{x\}\}\Cup\{\{y\}\}=\{\{x,y\}\}$ and $\{\{x,y\}\}\Cup\{\{z\}\}=\{\{x,y,z\}\}$, and the same reasoning
iterated over all the factors yields $\{\{x_1,\ldots,x_r\}\}
=\{W_P\}$). Notice that repeated occurrences of the same literal do not change this value, since $\{\{x_i\}\}\Cup\{\{x_i\}\} = \{\{x_i\}\}$. Therefore, $W_P\in \pi^{\mathbf I}\llbracket P(\vec{u}) \rrbracket.$ 

The natural order on $\mathrm{Why}(X,\overline X)$ is set inclusion. Thus, by the assumed inequality, $W_P\in \pi^{\mathbf I}\llbracket Q(\vec{u}) \rrbracket.$ By the semantics of CQs, there is an assignment $v$ of elements of $\mathbf I$ to the bound variables of $Q$, extending the identity assignment on $\vec{u}$, whose contribution to $\pi^{\mathbf I}\llbracket Q(\vec{u}) \rrbracket$ contains $W_P$.

Every relational literal of $Q$ must take a non-zero value under $v$, since otherwise the product corresponding to this assignment would be $0$ (recall that $0=\emptyset$ in $\mathrm{Why}(X,\overline X)$ and that, for every $A\in\mathrm{Why}(X,\overline X)$, $\emptyset\Cup A=A\Cup\emptyset=\emptyset$). By construction of $\pi^{\mathbf I}$, a ground relational literal has non-zero value only if it is one of the copies of the relational literals occurring in $P$. Hence, by letting $h(x):=v(x)$ for every variable $x$ of $Q$, we obtain a negation-homomorphism $h\colon Q\longrightarrow P.$ Indeed, $h$ preserves the free variables and, for every relational literal $L(\vec{x})$ occurring in $Q$, $L(h(\overline{x}))$ occurs in the body of $P$.

It remains to show that $h$ is set-onto. Since every non-zero annotation in $\pi^{\mathbf I}$ is a singleton, the product associated with the assignment $v$ is obtained by taking pairwise unions of singleton outer sets. Hence, before passing to the quotient, it has the form $\{W_v\}$, where $W_v$ is the union of all the tags contributed by the relational literals of $Q$ under $v$. If $W_v$ contains a complementary pair, this unique witness is removed by the quotient defining $\mathrm{Why}(X,\overline X)$ and the contribution of $v$ is therefore $\emptyset$. Otherwise, the contribution of $v$ is precisely $\{W_v\}$. We already know that $W_P$ belongs to the contribution of $v$. Hence this contribution cannot be $\emptyset$, and must therefore be $\{W_v\}$. Since $\{W_v\}$ contains only one witness and $W_P\in\{W_v\}$, it follows that $W_v=W_P$. Thus the union of the tags contributed by the relational literals of $Q$ under $v$ is exactly $W_P$. 

Therefore, for every $1\leq i\leq r$, the tag $x_i$ must occur as the annotation of at least one relational literal of $Q$ under $v$. Since $x_i$ occurs in the annotation of no other canonical relational literal, for each $L_i\in\text{LitSet}(P)$ there is a relational literal $A$ of $Q$ such that $h(A)=L_i.$ Hence $\text{LitSet}(P) \subseteq h(\text{LitSet}(Q)). $ The reverse inclusion follows from the fact that $h$ is a negation-homomorphism. Thus $h(\text{LitSet}(Q)) =\text{LitSet}(P),$ and $h$ is set-onto for relational literals.

$(3)\implies(1):$
Assume that there is a negation-homomorphism $h\colon Q\longrightarrow P$ such that $h(\text{LitSet}(Q)) = \text{LitSet}(P).$ Write $ P(\vec{z}) := \exists\vec{u}\, \bigl( B_1(\vec{z}_1)\wedge\ldots\wedge B_m(\vec{z}_m) \bigr)$ and $Q(\vec{x}) := \exists\vec{z}\, \bigl( A_1(\vec{x}_1)\wedge\ldots\wedge A_n(\overline{x}_n) \bigr).$ Consider an arbitrary $\mathrm{Why}(X)$-database instance $\mathbf I$, a provenance-tracking interpretation $\pi$ compatible with $\mathbf I$, and a tuple $\vec{t}$. We show that $\pi\llbracket P(\vec{t}) \rrbracket \subseteq \pi\llbracket Q(\vec{t}) \rrbracket.$ Let $W\in\pi\llbracket P(\vec{t}) \rrbracket,$ then there is an assignment $v$ to the variables of $P$, producing $\vec{t}$ on the free variables, such that $W$ belongs to the contribution of $v$ to $\pi\llbracket P(\vec{t}) \rrbracket$. Since this contribution is non-zero, every relational literal of $P$ takes a non-zero value under $v$. Since $\pi$ is provenance-tracking, for every ground relational literal $G$ such that $\pi\llbracket G \rrbracket\neq 0$, there is a unique finite set $S_G\subseteq X\cup\overline X$ such that $\pi\llbracket G \rrbracket=\{S_G\},$ where $S_G=\emptyset$ when $\pi\llbracket G \rrbracket=1$, and $S_G$ is a singleton token set otherwise. Let us write $s_\pi\llbracket G \rrbracket:=S_G$.

Thus, for every $1\leq i\leq m$, $\pi\llbracket B_i(v(\vec{z}_i)) \rrbracket =\{s_\pi\llbracket B_i(v(\vec{z}_i)) \rrbracket\}.$ By the definition of multiplication in $\mathrm{Why}(X,\overline X)$, before passing to the quotient the product of these singleton sets is the singleton whose unique witness is their union $\left\{ \bigcup_{1\leq i\leq m} s_\pi\llbracket B_i(v(\vec{z}_i)) \rrbracket \right\}.$ If this union contains a pair of complementary tokens, its unique witness is removed by the quotient defining
$\mathrm{Why}(X,\overline X)$, and the product is therefore $\emptyset$. Otherwise, the witness is preserved. Since the contribution of $v$ contains $W$, it is non-zero, and hence the latter case must occur. Therefore, $W= \bigcup_{1\leq i\leq m} s_\pi\llbracket B_i(v(\vec{z}_i)) \rrbracket.$

Consider now the assignment $v\circ h$ to the variables of $Q$. Since $h$ preserves the tuple of free variables, $v\circ h$ also produces $\vec{t}$. Furthermore, for every relational literal $A_j(\overline{x}_j)$ of $Q$, $(v\circ h)(A_j(\overline{x}_j)) = v(h(A_j(\overline{x}_j))),$ and $h(A_j(\overline{x}_j))$ is a relational literal occurring in $P$, so all these literals have non-zero value under $\pi$.

Since $h(\text{LitSet}(Q)) = \text{LitSet}(P),$ applying the assignment $v$ to both sides gives $\bigl\{ v(h(A)) :  A\in\text{LitSet}(Q) \bigr\} = \bigl\{v(B) : B\in\text{LitSet}(P) \bigr\}.$ Notice that distinct literals may become the same ground literal under $v$, but this does not affect the equality, since we are considering ordinary sets and not multi-sets. 

Applying $s_\pi$ to the ground literals we obtain $\bigl\{ s_\pi\llbracket v(h(A)) \rrbracket : A\in\text{LitSet}(Q) \bigr\} = \bigl\{ s_\pi\llbracket v(B) \rrbracket : B\in\text{LitSet}(P) \bigr\}. $ Taking the union of the members of these sets, and observing that repeated occurrences of a relational literal do not change a set union, yields $\bigcup_{1\leq j\leq n} s_\pi\llbracket A_j(v\circ h(\overline{x}_j)) \rrbracket = \bigcup_{1\leq i\leq m} s_\pi\llbracket B_i(v(\vec{z}_i)) \rrbracket.$ Since we have already shown that $W= \bigcup_{1\leq i\leq m} s_\pi\llbracket B_i(v(\vec{z}_i)) \rrbracket$, the right side is exactly $W$. Hence $\bigcup_{1\leq j\leq n} s_\pi\llbracket A_j(v\circ  h(\overline{x}_j)) \rrbracket = W.$ 

Since $W$ is the unique witness of the non-zero contribution of $v$ to $P$, it is not removed by the quotient defining $\mathrm{Why}(X,\overline X)$, and therefore it contains no pair of complementary tokens. Hence the union $\bigcup_{1\leq j\leq n} s_\pi\llbracket A_j(v\circ h(\overline{x}_j)) \rrbracket=W$ also contains no complementary pair. Its corresponding singleton witness is therefore preserved in the quotient, so the product of the values of the relational literals of $Q$ under $v\circ h$ is exactly $\{W\}$. Consequently, $W\in\pi\llbracket Q(\vec{t}) \rrbracket.$ Since $\mathbf I$, $\pi$, $\vec{t}$, and $W\in\pi\llbracket P(\vec{t}) \rrbracket$ were arbitrary, we obtain $P\sqsubseteq^{\text{pt}}_{\mathrm{Why}(X,\overline X)}Q.$

Finally, for the complexity calculation the NP upper bound follows immediately from item 3, since we can guess a mapping $h\colon Q\longrightarrow P$ and check in polynomial time that it is a negation-homomorphism and set-onto for relational literals. For hardness, we adapt the argument of~\cite[Corollary~7.7]{Green}. Recall that a graph $G$ is $3$-colorable if and only if there is a
graph homomorphism from $G$ to the complete graph $K_3$. Given an undirected graph $G$, let $Q_G$ be the Boolean CQ obtained by replacing every edge $\{u,v\}$ of $G$ by the two atoms $E(u,v)$ and $E(v,u)$, and let $Q_{K_3}$ be the Boolean CQ encoding the loop-free complete directed graph on three vertices. We may assume that the variables of $Q_G$ and $Q_{K_3}$ are disjoint, and let $Q'_G$ be the Boolean CQ whose body is the conjunction of the bodies of $Q_G$ and $Q_{K_3}$. If $G$ is $3$-colorable, the corresponding graph homomorphism gives a homomorphism $Q_G\longrightarrow Q_{K_3}$. Extending it by the identity on the $Q_{K_3}$-component of $Q'_G$ yields a homomorphism $h\colon Q'_G\longrightarrow Q_{K_3}$ which is set-onto for relational literals, since every relational literal of $Q_{K_3}$ is in the image of its corresponding copy in $Q'_G$. Conversely, if there is a set-onto homomorphism $h\colon Q'_G\longrightarrow Q_{K_3}$, its restriction to the $Q_G$-component is a graph homomorphism from $G$ to $K_3$, and hence $G$ is $3$-colorable. Notice that all the queries used in the reduction are positive and contain no repeated relational literals, so negation-homomorphisms coincide with ordinary homomorphisms in this case. Therefore, $G\text{ is $3$-colorable}$ if and only if $Q'_G\longrightarrow Q_{K_3}$ by a set-onto homomorphism. By item 3, this is equivalent to $Q_{K_3} \sqsubseteq^{\mathrm{pt}}_{\mathrm{Why}(X,\overline X)} Q'_G.$ Thus the containment problem is NP-hard, and therefore NP-complete.
\end{proof}

Remarkably, the set-onto characterization obtained for $\mathrm{Why}(X, \overline{X})$ extends directly to the square-free provenance semiring $\mathrm{Trio}(X, \overline{X})$. Despite $\mathrm{Trio}(X, \overline{X})$ tracking coefficients while $\mathrm{Why}(X, \overline{X})$ collapses them set-theoretically, provenance-tracking interpretations ensure that containment in both semirings is governed by the exact same syntactic condition.

\begin{restatable}{Thm}{cqnContainmentG}\label{triocqncontainment}
For CQNs $P$ and $Q$ with the same tuple of free variables, $P\sqsubseteq^{\mathrm{pt}}_{\mathrm{Why}(X,\overline{X})}Q$ if and only if $P\sqsubseteq^{\mathrm{pt}}_{\mathrm{Trio}(X,\overline{X})}Q.$ 

Thus, the corresponding decision problem for containment in the case of $\mathrm{Trio}(X,\overline{X})$ (i.e. $P\sqsubseteq^{\mathrm{pt}}_{\mathrm{Trio}(X,\overline{X})}Q$?) is NP-complete.
\end{restatable}

\begin{proof}
By the characterization obtained above for $\mathrm{Why}(X,\overline{X})$, we know that $P\sqsubseteq^{\mathrm{pt}}_{\mathrm{Why}(X,\overline{X})}Q$ if and only if there is a negation-homomorphism $h\colon Q\longrightarrow P$ set-onto for relational literals, i.e. $h(\text{LitSet}(Q))=\text{LitSet}(P).$ We show that the same condition characterizes containment over $\mathrm{Trio}(X,\overline{X})$.

Assume first that there is a negation-homomorphism $h\colon Q\longrightarrow P$ such that $h(\text{LitSet}(Q))=\text{LitSet}(P)$. Consider an arbitrary $\mathrm{Trio}(X)$-database instance $I$, a provenance-tracking interpretation $\pi$ compatible with it, and a tuple $\vec{t}$. We show that $\pi\llbracket P(\vec{t}) \rrbracket\leq_{\mathrm{Trio}(X,\overline{X})}\pi\llbracket Q(\vec{t}) \rrbracket.$ Since the natural order of $\mathrm{Trio}(X,\overline{X})$ is the coefficientwise order on consistent square-free monomials, it is enough to show that every such monomial occurs in $\pi\llbracket Q(\vec{t}) \rrbracket$ with coefficient at least as large as its coefficient in $\pi\llbracket P(\vec{t}) \rrbracket$.

Let $\mu$ be a consistent square-free monomial occurring in $\pi\llbracket P(\vec{t}) \rrbracket$, and consider an assignment $v$ to the variables of $P$, producing $\vec{t}$ on the free variables, whose contribution to $\pi\llbracket P(\vec{t}) \rrbracket$ is $\mu$. Since $\pi$ is provenance-tracking, every non-zero ground literal is annotated by a single token or by $1$. Thus, before passing to the quotient defining $\mathrm{Trio}(X,\overline{X})$, the contribution of $v$ is a product of provenance tags, possibly with repetitions. Since this contribution is non-zero, the product contains no complementary pair, and after replacing every positive exponent by $1$ its square-free representative is precisely $\mu$.

Consider now the assignment $v\circ h$ to the variables of $Q$. Since $h$ preserves the tuple of free variables, $v\circ h$ also produces $\vec{t}$. Moreover, from $h(\text{LitSet}(Q))=\text{LitSet}(P)$ we obtain $\{v(h(A))\mid A\in\text{LitSet}(Q)\}=\{v(L)\mid L\in\text{LitSet}(P)\}.$ Notice that distinct relational literals may become the same ground relational literal under $v$, but this does not affect the equality, since we are considering ordinary sets and not multisets. Hence the set of distinct non-unit provenance tags occurring in the contribution of $v\circ h$ is exactly the same as the set of distinct non-unit provenance tags occurring in the contribution of $v$. The multiplicities with which these tags occur may be different, but they only affect their positive exponents, which are then dropped in $\mathrm{Trio}(X,\overline{X})$. Moreover, since the contribution of $v$ is non-zero, this common set of tags contains no complementary pair. Therefore, $v\circ h$ contributes exactly the same square-free monomial $\mu$ to $\pi\llbracket Q(\vec{t}) \rrbracket$.

It remains to check that the coefficients are the same. First observe that $h$ is surjective on variables. Indeed, let $y$ be a variable of $P$. Since $P$ is safe, $y$ occurs in some positive relational literal $L\in\text{LitSet}(P)$. Since $h$ is set-onto for relational literals, there is a relational literal $A\in\text{LitSet}(Q)$ such that $h(A)=L$. Therefore, at the position of $L$ occupied by $y$, the corresponding position of $A$ is a variable $x$ such that $h(x)=y$. Hence every variable of $P$ belongs to the image of $h$.

Suppose now that $v$ and $v'$ are two distinct assignments to the variables of $P$. Then there is a variable $y$ of $P$ such that $v(y)\neq v'(y)$. Since $h$ is surjective on variables, there is a variable $x$ of $Q$ such that $h(x)=y$. Hence $(v\circ h)(x)=v(y)\neq v'(y)=(v'\circ h)(x),$ and therefore $v\circ h\neq v'\circ h$. Thus, the map that sends every assignment $v$ of $P$ to the assignment $v\circ h$ of $Q$ is injective.

Now fix a consistent square-free monomial $\mu$. Under a provenance-tracking interpretation, every assignment contributes either $0$ or a single square-free monomial with coefficient $1$. Hence the coefficient of $\mu$ in $\pi\llbracket P(\vec{t}) \rrbracket$ is exactly the number of assignments to the variables of $P$ producing $\vec{t}$ and contributing $\mu$. As shown above, every such assignment $v$ determines an assignment $v\circ h$ to the variables of $Q$ producing $\vec{t}$ and contributing the same monomial $\mu$, and distinct assignments of $P$ determine distinct assignments of $Q$. Therefore, the coefficient of $\mu$ in $\pi\llbracket P(\vec{t}) \rrbracket$ is at most its coefficient in $\pi\llbracket Q(\vec{t}) \rrbracket$. Since this holds for every consistent square-free monomial $\mu$, we obtain $\pi\llbracket P(\vec{t}) \rrbracket\leq_{\mathrm{Trio}(X,\overline{X})}\pi\llbracket Q(\vec{t}) \rrbracket.$ Since $I$, $\pi$, and $\vec{t}$ were arbitrary, we conclude that $P\sqsubseteq^{\mathrm{pt}}_{\mathrm{Trio}(X,\overline{X})}Q.$

Conversely, assume that $P\sqsubseteq^{\mathrm{pt}}_{\mathrm{Trio}(X,\overline{X})}Q.$ Let $\mathbf{I}=\operatorname{can}_{\mathrm{Trio}(X)}(P^+)$ and abbreviate the corresponding canonical provenance-tracking interpretation as $\pi^{\mathbf I}$. Let $\text{LitSet}(P)=\{L_1,\ldots,L_r\}.$ By construction of $\pi^{\mathbf I}$, each distinct relational literal $L_i$ is annotated by a different tag $x_i\in X\cup\overline{X}$, while repeated occurrences of the same relational literal receive the same annotation. Moreover, by the definition of the canonical interpretation, no pair of complementary tags occurs among $x_1,\ldots,x_r$. Hence $M_P:=x_1\cdot\ldots\cdot x_r$ is a non-zero consistent square-free monomial in $\mathrm{Trio}(X,\overline{X})$.

Consider the assignment to the variables of $P$ that sends every variable to itself in $I$. If $L_i$ occurs $m_i\geq 1$ times in the body of $P$, then these occurrences contribute the factor $x_i^{m_i}$. Thus, before dropping exponents, the contribution of this assignment has the form $x_1^{m_1}\ldots x_r^{m_r},$ whose square-free representative in $\mathrm{Trio}(X,\overline{X})$ is precisely $M_P$. Therefore, the coefficient of $M_P$ in $\pi^{\mathbf I}\llbracket P(\vec{u}) \rrbracket$ is non-zero.

By the containment assumption we have $\pi^{\mathbf I}\llbracket P(\vec{u}) \rrbracket\leq_{\mathrm{Trio}(X,\overline{X})}\pi^{\mathbf I}\llbracket Q(\vec{u}) \rrbracket.$ Since the natural order of $\mathrm{Trio}(X,\overline{X})$ is coefficientwise on square-free monomials, the coefficient of $M_P$ in $\pi^{\mathbf I}\llbracket Q(\vec{u}) \rrbracket$ is also non-zero. Since $\pi^{\mathbf I}$ is provenance-tracking, every assignment to the variables of $Q$ contributes either $0$ or a single square-free monomial with coefficient $1$. Hence there is an assignment $v$ to the variables of $Q$, producing $\vec{u}$, whose contribution to $\pi^{\mathbf I}\llbracket Q(\vec{u}) \rrbracket$ is $M_P$.

Every relational literal of $Q$ must take a non-zero value under $v$, since otherwise the corresponding product would be $0$. By construction of $\pi^{\mathbf I}$, a ground literal has non-zero value only if it corresponds to a relational literal occurring in $P$. Moreover every variable of $Q$ occurs in some positive relational literal, and therefore the value assigned by $v$ to every variable belongs to the active domain of $I$ which consists of the variables of $P$. Thus, by letting $h(x):=v(x)$ for every variable $x$ of $Q$, we obtain a negation-homomorphism $h\colon Q\longrightarrow P.$ 

It remains to show that $h$ is set-onto for relational literals. Since the contribution of $v$ is exactly $M_P=x_1\cdot \ldots \cdot x_r,$ every tag $x_i$, $1\leq i\leq r$, must occur among the annotations of the literals of $Q$ under $v$. By construction of $\pi^{\mathbf I}$, the tag $x_i$ occurs only as the annotation associated with the literal $L_i$. Therefore, for every $L_i\in\text{LitSet}(P)$ there is a literal $A\in\text{LitSet}(Q)$ such that $h(A)=L_i$. Hence $\text{LitSet}(P)\subseteq h(\text{LitSet}(Q)).$ The reverse inclusion follows immediately from the fact that $h$ is a negation-homomorphism. Therefore, $h(\text{LitSet}(Q))=\text{LitSet}(P),$ and $h$ is set-onto for relational literals. By the characterization obtained above for $\mathrm{Why}(X,\overline{X})$, it follows that $P\sqsubseteq^{\mathrm{pt}}_{\mathrm{Why}(X,\overline{X})}Q.$

Finally, for the complexity calculation, the NP upper bound follows from the equivalent characterization in terms of set-onto negation-homomorphisms. For hardness, the result follows immediately from the NP-hardness of containment for CQNs over $\mathrm{Why}(X,\overline{X})$, since the two containment problems are equivalent by the result just proved.
\end{proof}

Finally, we consider containment under the positive Boolean provenance semiring, $\mathrm{PosBool}(X, \overline{X})$. Because logical conjunction satisfies the absorption property ($\varphi \wedge \psi \leq \varphi$), a target query $Q$ no longer needs to cover all literals of $P$ to establish containment. Consequently, the set-onto requirement drops entirely, and provenance-tracking containment over $\mathrm{PosBool}(X, \overline{X})$ reduces to the existence of an unrestricted negation-homomorphism.

\begin{restatable}{Thm}{cqnContainmentH}\label{posboolneg}
For CQNs $P$ and $Q$ with the same tuple of free variables, the following are equivalent:
\begin{itemize}
\item[1.] $P \sqsubseteq^{\text{pt}}_{\mathrm{PosBool}(X,\overline{X})} Q$,
\item[2.] $\pi^{\mathrm{can}_{\mathrm{PosBool}(X)}(P)}\llbracket P(\bar u)\rrbracket \leq_{\mathrm{PosBool}(X,\overline{X})} \pi^{\mathrm{can}_{\mathrm{PosBool}(X)}(P)}\llbracket Q(\bar u)\rrbracket$,
\item[3.] There is a negation-homomorphism $h \colon Q \longrightarrow P$.
\end{itemize}
Moreover, the corresponding decision problem (i.e.\ $P \sqsubseteq_{\mathrm{PosBool}(X,\overline{X})}^{\text{pt}} Q$?) is NP-complete. 
\end{restatable}

\begin{proof}
$(1)\Rightarrow(2)$: Obvious.

$(2)\Rightarrow(3)$: Assume that $(2)$ holds, so
\[
\pi^{\mathrm{can}_{\mathrm{PosBool}(X)}(P)}\llbracket P(\bar u)\rrbracket
\leq_{\mathrm{PosBool}(X,\overline{X})}
\pi^{\mathrm{can}_{\mathrm{PosBool}(X)}(P)}\llbracket Q(\bar u)\rrbracket.
\]
Let us abbreviate $\mathrm{can}_{\mathrm{PosBool}(X)}(P)$ by $\mathbf I$ and $\pi^{\mathrm{can}_{\mathrm{PosBool}(X)}(P)}$ by $\pi^\mathbf{I}$.\\
Consider $P(\bar u)=\exists \bar y\,(B_1(\bar t_1)\wedge \ldots \wedge B_m(\bar t_m)), 
Q(\bar u)=\exists \bar z\,(A_1(\bar x_1)\wedge \ldots \wedge A_n(\bar x_n))$. 
Consider the assignment on the variables of $P$ that sends every variable of $P$ to itself in the active domain of $\mathbf I$ (identity assigment). Since the value of a conjunctive query is the join of the values of its body under all assignments, the conjunction $\tau_P:=\pi^{\mathbf I}\llbracket B_1(\bar t_1)\rrbracket\wedge \ldots \wedge \pi^\mathbf{I}\llbracket B_m(\bar t_m)\rrbracket$ is one of the disjuncts occurring in $\pi^{\mathbf I}\llbracket P(\bar u)\rrbracket$.\\
By construction of $\pi^{\mathbf I}$, each conjunct in $\tau_P$ is either a tag from $X$ or a tag from $\overline{X}$, and no complementary pair $x,\overline x$ can occur simultaneously in $\tau_P$ (if $\overline r$ is used to annotate a negative literal of $P$, then by construction its complementary token $r$ is not used as an annotation of any positive literal in the canonical instance). So there is the following admissible (in the sense that $\nu(x) \cdot \nu(\overline{x}) = 0$) valuation $\nu_{\tau_P}\colon X\cup \overline{X} \longrightarrow \{0,1\}$ such that $\nu_{\tau_P}(t)=1$ exactly for the tags occurring in $\tau_P$, and $\nu_{\tau_P}(t)=0$ for all other tags.\\
By construction, $\nu_{\tau_P}(\tau_P)=1.$ Since $\tau_P$ is one of the disjuncts of $\pi^{\mathbf I}\llbracket P(\bar u)\rrbracket$, we get $\nu_{\tau_P}\bigl(\pi^{\mathbf I}\llbracket P(\bar u) \rrbracket\bigr)=1.$\\
Now, by the natural order on $\mathrm{PosBool}(X,\overline{X})$, from the hypothesis $\pi^{\mathbf I}\llbracket P(\bar u)\rrbracket
\leq_{\mathrm{PosBool}(X,\overline{X})}
\pi^{\mathbf I}\llbracket Q(\bar u)\rrbracket $ (recall that it means that for any valuation $v\colon X \cup \overline{X} \longrightarrow \{0,1\}$ such that $v(x) \cdot v(\overline{x})= 0$ for any $x \in X$, if $v(\pi^{\mathbf I}\llbracket P(\bar u)\rrbracket) = 1$ then $v(\pi^{\mathbf I}\llbracket Q(\bar u)\rrbracket) = 1$)  we obtain $\nu_{\tau_P}\bigl(\pi^{\mathbf I}\llbracket Q(\bar u)\rrbracket\bigr)=1.$\\
By the definition of the semantics of $Q$, the value $\pi^{\mathbf I}\llbracket Q(\bar u)\rrbracket$ is the join of the values $\pi^{\mathbf I}\llbracket A_1(f_i(\bar x_1))\rrbracket\wedge \ldots \wedge \pi^{\mathbf I}\llbracket A_n(f_i(\bar x_n))\rrbracket $ over all assignments $f_i$ of the bound variables of $Q$ into the active domain of $\mathbf I$ extending the identity on the free variables $\bar u$. Since $\nu_{\tau_P}\bigl(\pi^{\mathbf I}\llbracket Q(\bar u)\rrbracket \bigr)=1$, there must be such an assignment $f$ satisfying $\nu_{\tau_P}\Bigl(
\pi^{\mathbf I}\llbracket A_1(f(\bar x_1))\rrbracket\wedge \ldots \wedge \pi^{\mathbf I}\llbracket A_n(f(\bar x_n))\rrbracket
\Bigr)=1.$ So each conjunct must be evaluated $1$ under $\nu_{\tau_P}$: that means that for each literal $A_i$ with $i \leq n$ $\nu_{\tau_P}\bigl(\pi^{\mathbf I}\llbracket A_i(f(\bar x_i))\rrbracket\bigr)=1$. \\
Now, by construction of $\pi^{\mathbf I}$, a literal takes a value different from $\bot$ only if it is a literal that occurs in the body of $P$. But, since $\nu_{\tau_P}(\bot)=0$, it follows that for every $1 \leq i\leq n$, the literal $A_i(f(\bar x_i))$ must occurs in the body of $P$.\\
Define now a mapping $h$ from the variables of $Q$ to the terms of $P$ by $h(x):=f(x)$ for every variable $x$ of $Q$. Since $f$ extends the identity on the free variables $\bar u$, the map $h$ also preserves the head variables $\vec{u}$. Moreover, for every literal $A_i(\bar x_i)$ of $Q$, the literal $A_i(h(\bar x_i)) = A_i(f(\bar x_i))$ occurs in the body of $P$. Therefore, $h$ is a negation-homomorphism from $Q$ to $P$.

$(3)\Rightarrow(1)$: Assume that there is a negation-homomorphism $h\colon Q\longrightarrow P$ Take an arbitrary $\mathrm{PosBool}(X)$-instance $\mathbf I$, an arbitrary provenance-tracking interpretation $\pi$ compatible with $\mathbf I$, and an arbitrary tuple $\bar t$.
\[
P(\bar u)=\exists \bar y\,(B_1(\bar t_1)\wedge \ldots \wedge B_m(\bar t_m)),
\]
\[
Q(\bar u)=\exists \bar z\,(A_1(\bar x_1)\wedge \ldots \wedge A_n(\bar x_n)).
\]
We have that
\[
\pi\llbracket P(\bar t)\rrbracket
=
\bigvee_{e}
\Bigl(
\pi\llbracket B_1(e_i(\bar t_1))\rrbracket \wedge \ldots \wedge \pi\llbracket B_m(e_i(\bar t_m))\rrbracket
\Bigr),
\]
where $e_i$ ranges over all possible assignments of elements of $\mathbf I$ to the bound variables of $P$ extending $\bar u\to \bar t$. Now, fix one of these assignment $e$, and let $
\tau_e :=
\pi\llbracket B_1(e(\bar t_1))\rrbracket\wedge \ldots \wedge \pi\llbracket B_m(e(\bar t_m))\rrbracket.$ Then, $\tau_e$ is one of the disjuncts occurring in $\pi\llbracket P(\bar t)\rrbracket$.\\
Since $h$ is a negation-homomorphism from $Q$ to $P$, the composition $e\circ h
$ is an assignment of elements of $\mathbf I$ to the bound variables of $Q$ extending $\bar u\to \bar t$. Hence, the formula $\sigma_e :=
\pi\llbracket A_1((e\circ h)(\bar x_1))\rrbracket\wedge \ldots \wedge \pi\llbracket A_n((e\circ h)(\bar x_n))\rrbracket$ is one of the disjuncts occurring in $\pi\llbracket Q(\bar t)\rrbracket$.
Since $h$ is a negation-homomorphism, for every literal $A_i(\bar x_i)$ of $Q$, the literal $A_i(h(\bar x_i))$ occurs in the body of $P$. Therefore, every conjunct occurring in $\sigma_e$ is also one of the conjuncts occurring in $\tau_e$ (in $\tau_e$ there can be maybe more conjuncts). 
Since in $\mathrm{PosBool}(X,\overline{X})$ by the natural order we have $\varphi \land \psi \leq \varphi$, it follows that $\tau_e \leq_{\mathrm{PosBool}(X,\overline{X})} \sigma_e,$ and since $\sigma_e$ is one of the disjuncts of $\pi\llbracket Q(\bar t)\rrbracket$, we also have $\sigma_e \leq_{\mathrm{PosBool}(X,\overline{X})} \pi\llbracket Q(\bar t) \rrbracket.$
Then, by transitivity, $\tau_e \leq_{\mathrm{PosBool}(X,\overline{X})} \pi\llbracket Q(\bar t)\rrbracket.$
But this holds for every possible assignment, so taking the join over all such $e$ we obtain that for arbitrary $\mathbf I$, $\pi$, and $\bar t$: $\pi\llbracket P(\bar t)\rrbracket \leq_{\mathrm{PosBool}(X,\overline{X})} \pi\llbracket Q(\bar t)\rrbracket,$ so $P \sqsubseteq^{\text{pt}}_{\mathrm{PosBool}(X,\overline{X})} Q.$

Finally, the complexity of the problem can be calculated exactly as in Theorem~\ref{dsitrlatticeneg}.
\end{proof}

\cqnContainmentI*

\begin{proof}
If $P$ contains an atom and its negation, item (1) holds because its value is always $\bot$, while items (2) and (3) are vacuous. Otherwise, let $Q(\vec{x})=\exists\vec{z}\,(A_1(\vec{x_1})\wedge\ldots\wedge A_n(\overline{x_n}))$ and $P(\vec{z})=\exists\vec{u}\,(B_1(\vec{z_1})\wedge\ldots\wedge B_m(\vec{z_m}))$.

$(1) \implies (2):$ trivial.

$(2) \implies (3):$ Assume that (2) holds. We want to show that for every completion $P^c$ of $P$ and every literal $L$ in the body of $P$, there is a negation-homomorphism $h_{P^c,L} \colon Q \longrightarrow P^c$ such that $L$ belongs to the image of $h_{P^c,L}$. Let $P^c$ be an arbitrary completion of $P$, and let $L$ be an arbitrary literal in the body of $P$. By definition of the interpretation $\pi^{P^c}_{\mathrm{Which}(X)}$, we have that $\pi^{P^c}_{\mathrm{Which}(X)}\llbracket L\rrbracket=\{x_L\}$, where $x_L$ is the fresh variable of $X$ associated with the literal $L$. Moreover, considering the identity assignment on the variables of $P$, we obtain that the value of the body of $P$ contains all the variables associated with the literals of $P$: $x_L \in \pi^{P^c}_{\mathrm{Which}(X)}\llbracket P(\vec{z})\rrbracket$ (simply the union). By hypothesis, $\pi^{P^c}_{\mathrm{Which}(X)}\llbracket P(\vec{z})\rrbracket \leq_{\mathrm{Which}(X)} \pi^{P^c}_{\mathrm{Which}(X)}\llbracket Q(\vec{x})\rrbracket$. Since the natural order on Which$(X)$ is the inclusion on $\wp(X)$ with the exception that $\bot \leq A$ for any $A \subseteq \wp(X)$, it follows that $x_L \in \pi^{P^c}_{\mathrm{Which}(X)}\llbracket Q(\vec{x})\rrbracket$. 
Since $\pi^{P^c}_{\mathrm{Which}(X)}\llbracket Q(\vec{x})\rrbracket$ is obtained as the sum, in Which$(X)$, of the values of the body of $Q$ over all assignments sending the free variables $\overline{x}$ to $\vec{z}$, and since this sum is given by the union, there must be an assignment $v$ from the variables of $Q$ to the variables of $P$ (that are the same of $P^c$) such that $v(\overline{x})=\vec{z}$ and $x_L$ belongs to $\pi^{P^c}_{\mathrm{Which}(X)}\llbracket A_1(v(\overline{x_1}))\rrbracket \cdot \ldots \cdot \pi^{P^c}_{\mathrm{Which}(X)}\llbracket A_n(v(\overline{x_n}))\rrbracket$. Multiplication in Which$(X)$ gives $\bot$ whenever one of the factors is $\bot$, so it follows that, for every $1\leq i \leq n$, we have $\pi^{P^c}_{\mathrm{Which}(X)}\llbracket A_i(v(\overline{x_i}))\rrbracket \neq \bot$. But by construction of $\pi^{P^c}_{\mathrm{Which}(X)}$, a literal has value different from $\bot$ only if it belongs to the body of $P^c$, so, for every $1 \leq i \leq n$, the literal $A_i(v(\overline{x_i}))$ belongs to the body of $P^c$. Moreover, the variable $x_L$ can occur only in the value of the literal $L$, therefore, there is some $1 \leq i \leq n$ such that $A_i(v(\overline{x_i}))=L$. We now define $h_{P^c,L}$ by letting $h_{P^c,L}(x)=v(x)$ for every variable $x$ of $Q$. Since the active domain of $P^c$ consists of the variables of $P$, this is a mapping from the variables of $Q$ to the variables of $P$. Also, since the two queries have the same tuple of free variables and we are evaluating $Q$ on the tuple $\vec{z}$, the mapping sends the free variables of $Q$ to the corresponding free variables of $P$. Also, for every literal $A_i(\vec{x_i})$ of $Q$, we have shown that $A_i(h_{P^c,L}(\overline{x_i}))=A_i(v(\overline{x_i}))$ belongs to the body of $P^c$. Thus, $h_{P^c,L}$ is a negation-homomorphism from $Q$ to $P^c$. Since for some $1 \leq i \leq n$ we have $A_i(h_{P^c,L}(\overline{x_i}))=L$, the literal $L$ belongs to the image of $h_{P^c,L}$, as required.

$(3) \implies (1):$ Assume that for each completion $P^c$ of $P$ and each literal $L$ in the body of $P$, there is a negation-homomorphism $h_{P^c,L} \colon Q \longrightarrow P^c$ such that $L$ belongs to the image of $h_{P^c,L}$. We want to show that $P \sqsubseteq_{\mathrm{Which}(X)}^{\text{md}} Q$. Consider an arbitrary Which$(X)$-database instance $\mathbf I$, a model-defining interpretation $\pi$ compatible with $\mathbf I$, and an arbitrary tuple $\vec{t}$. We need to show that $\pi\llbracket P(\vec{t})\rrbracket \leq_{\mathrm{Which}(X)} \pi\llbracket Q(\vec{t})\rrbracket$. If $\pi\llbracket P(\vec{t})\rrbracket=\bot$, then there is nothing to prove. So assume that $\pi\llbracket P(\vec{t})\rrbracket \neq \bot$. First choose an assignment $v_0$ with a non-bottom contribution to $P(\vec t)$, and complete $P$ by adding, for each atom on its variables, the unique literal having non-bottom value under $v_0$. This gives a completion $P^c_{v_0}$. Choose any literal $L_0$ in the body of $P$. By (3), there is a negation-homomorphism $h_{P^c_{v_0},L_0}\colon Q\longrightarrow P^c_{v_0}$. The assignment $v_0\circ h_{P^c_{v_0},L_0}$ gives a non-bottom contribution to $Q(\vec t)$, so $\pi\llbracket Q(\vec t)\rrbracket\neq\bot$. It remains to prove inclusion between these two subsets of $X$. Let $x \in \pi\llbracket P(\vec{t})\rrbracket$. Since $\pi\llbracket P(\vec{t})\rrbracket$ is obtained as the sum of the values of the body of $P$ over all assignments sending the free variables $\vec{z}$ to $\vec{t}$, there is an assignment $v$ of elements of the domain of $\mathbf I$ to the bound variables of $P$ such that $v(\vec{z})=\vec{t}$ and $x$ belongs to $\pi\llbracket B_1(v(\vec{z_1})) \wedge \ldots \wedge B_m(v(\vec{z_m}))\rrbracket$. Since multiplication in Which$(X)$ gives us $\bot$ any time one of the factors is $\bot$, it follows that $\pi\llbracket B_j(v(\vec{z_j}))\rrbracket \neq \bot$ for every $1 \leq j \leq m$. Moreover, since multiplication between elements different from $\bot$ is the union, there is a literal $L$ in the body of $P$ such that $x \in \pi\llbracket L(v)\rrbracket$. We now build a completion $P^c_v$ of $P$ depending on the assignment $v$ and the interpretation $\pi$: for every relation symbol $R$ of the common schema $\tau$ and every tuple $\vec{w}$ of variables of $P$ with length that match the arity of $R$, we add to $P$ the literal $R(\vec{w})$ if $\pi\llbracket R(v(\vec{w}))\rrbracket\neq \bot$, and we add the literal $\neg R(\vec{w})$ if $\pi\llbracket \neg R(v(\vec{w}))\rrbracket\neq \bot$. Since $\pi$ is model-defining, exactly one of these two values is different from $\bot$, so only one of the two literals is added to $P^c$ (if a literal already belonged to the body of $P$, then we already said that its value under $v$ is different from $\bot$ so all the literals of $P$ are also in $P^c$). $P^c_v$ is a completion of $P$. By hypothesis, there is a negation-homomorphism $h_{P^c_v,L} \colon Q \longrightarrow P^c_v$ such that $L$ belongs to the image of $h_{P^c_v,L}$. So, for every literal $A_i(\vec{x_i})$ of $Q$, the literal $A_i(h_{P^c_v,L}(\vec{x_i}))$ in in the body of $P^c_v$, and for some $1 \leq i \leq n$ we have $A_i(h_{P^c_v,L}(\vec{x_i}))=L$.
 Consider now the assignment $v \circ h_{P^c_v,L}$. Since $h_{P^c_v,L}$ is a negation-homomorphism, it sends the free variables of $Q$ to corresponding free variables of $P^c$ (and $P$), and so $v \circ h_{P^c_v,L}$ sends the tuple of free variables $\overline{x}$ to $\vec{t}$. Moreover, for every $1 \leq i \leq n$, the literal $A_i(h_{P^c_v,L}(\vec{x_i}))$ is in the body of $P^c_v$. By construction of $P^c_v$, this means that $\pi\llbracket A_i(v \circ h_{P^c_v,L}(\vec{x_i}))\rrbracket \neq \bot$ for every $1 \leq i \leq n$. Therefore, the value of the conjunction $A_1(v \circ h_{P^c_v,L}(\vec{x_1})) \wedge \ldots \wedge A_n(v \circ h_{P^c_v,L}(\vec{x_n}))$ is different from $\bot$.
 Finally, since $L$ is in the image of $h_{P^c_v,L}$, there is some $1 \leq i \leq n$ such that $A_i(h_{P^c_v,L}(\vec{x_i}))=L$. Applying $v$, we have $A_i(v \circ h_{P^c_v,L}(\vec{x_i}))=L(v)$. Since $x \in \pi\llbracket L(v)\rrbracket$, it follows that $x$ belongs to the value of $Q$ under this assignment. Hence $x \in \pi\llbracket Q(\vec{t})\rrbracket$. We have shown that every $x \in \pi\llbracket P(\vec{t})\rrbracket$ also belongs to $\pi\llbracket Q(\vec{t})\rrbracket$. Therefore, $\pi\llbracket P(\vec{t})\rrbracket \subseteq \pi\llbracket Q(\vec{t})\rrbracket$, and thus $\pi\llbracket P(\vec{t})\rrbracket \leq_{\mathrm{Which}(X)} \pi\llbracket Q(\vec{t})\rrbracket$. But $\mathbf I$, $\pi$, and $\vec{t}$ were arbitrary so we can conclude that $P \sqsubseteq_{\mathrm{Which}(X)}^{\text{md}} Q$.

For the complexity calculation, membership in $\Pi^p_2$ or coNEXP follows by item 3 similarly to the Boolean case.  For hardness, we  may assume that the bound variables in two arbitrary CQNs $P$ and $Q$ are disjoint. We define $Q^*$ as the  conjunction of $P$ and $Q$ (still a CQN).
First, observe that 
\begin{itemize}
\item[(1)] for every completion $P'$ of $P$, and every literal $l(\vec{z})$ of $P$  there is a negation-homomorphism $g_{P'} \colon Q^* \longrightarrow P'$ with $l(\vec{z
})$ in the image of $g_{P'}$, 
\end{itemize}
entails that 
\begin{itemize}
\item[(2)] for every completion $P'$ of $P$,  there is a negation-homomorphism  $h_{P'}\colon Q \longrightarrow P' $,
\end{itemize}
as $h_{P'}$ can be obtained as the restriction of $g_{P'}$ to the variables of $Q$.
Second, we also have that (2) entails (1). For any mapping $h_{P'}\colon Q \longrightarrow P' $ can be extended to a mapping $g_{P'} \colon Q^* \longrightarrow P'$  with the property that every literal  $l(\vec{z})$  of $P$ is in the image of $g_{P'}$ by letting the map be the identity on the variables of $P$. Thus, $P  \sqsubseteq_{\mathrm{Which}(X)}^{\text{md}} Q^*$ iff $P  \sqsubseteq_{\mathbb{B}}^{\text{md}} Q$. Using Theorem~\ref{containnegbool}, then we get that the containment problem for CQNs over $\mathrm{Which}(X)$ and model-defining interpretations is $\Pi_2^p$-hard or coNEXP-hard.
\end{proof}

\section{Omitted  Proofs from \S~\ref{The problem of equivalence}} \label{app:cqn-equivalence-proofs}

 We start with a useful result for bounded distributive lattices.

\begin{restatable}{Lem}{cqnEquivalenceA} \label{equivlat}
For CQNs $P$ and $Q$ with the same tuple of free variables, bounded distributive lattice $\mathbb{L}$ and any class of interpretations $\mathcal{C}$, the following holds:
\begin{itemize}
\item[(1)] $P \equiv_{\mathbb{L}}^\mathcal{C} Q$ iff $P \sqsubseteq_{\mathbb{L}}^\mathcal{C} Q$ and $Q \sqsubseteq_{\mathbb{L}}^\mathcal{C} P$,

\item[(2)] $P \sqsubseteq_{\mathbb{L}}^\mathcal{C} Q$ iff $P \equiv_{\mathbb{L}}^\mathcal{C} P \wedge Q$.

\end{itemize}
\end{restatable}

\begin{proof}
(1) is immediate. For (2) assume that $P \equiv_{\mathbb{L}}^\mathcal{C} P \wedge Q$, so for each $\mathbb{L}$-instance $\mathbf I$ and interpretation $\pi$ in $\mathcal{C}$ compatible with $\mathbf I$, and tuple $\vec{t}$, we have that $\pi \llbracket P(\vec{t})\rrbracket = \pi\llbracket P(\vec{t})\rrbracket \cdot \pi\llbracket Q(\vec{t})\rrbracket $, which in a lattice means that $\pi\llbracket P(\vec{t})\rrbracket  \leq_\mathbb{L}\pi\llbracket Q(\vec{t})\rrbracket $. Thus, $P \sqsubseteq_{\mathbb{L}}^\mathcal{C} Q$  as desired. The converse is entirely symmetric.
\end{proof}

In fact, item (1) of Lemma~\ref{equivlat} trivially holds for any naturally ordered semiring, not just lattices.

\begin{restatable}{Thm}{cqnEquivalenceB}\label{cqnslattice}
Equivalence of CQNs over bounded distributive lattices with more than two elements and model-defining interpretations is NP-complete while, for pairs of CQNs neither of which contains any atom and its negation, over dual model-defining interpretations (regardless of the cardinality of the lattice) it is $\Pi^p_2$-complete if there is a fixed arity bound $k\geq 2$ on the relations allowed in the queries or coNEXP-complete for unbounded arity.
\end{restatable}

\begin{proof}
For model-defining interpretations, first handle contradictory queries: two such queries are equivalent, whereas exactly one contradictory query cannot be equivalent to a noncontradictory one, since the latter has a Boolean-valued model with value $1$. For the remaining model-defining case, and for the stated dually model-defining case, membership follows from item (1) of Lemma~\ref{equivlat} and the respective containment upper bounds. For hardness, rename bound variables apart and use item (2) of Lemma~\ref{equivlat}: $P\sqsubseteq Q$ if and only if $P\equiv P\wedge Q$. In the dually model-defining case the source $P$ is noncontradictory. If the CQN obtained by joining the bodies of $P$ and $Q$ contains an atom and its negation, the containment instance is negative: a Boolean-valued interpretation satisfying $P$ witnesses failure of containment. In that case output a fixed pair of non-equivalent, noncontradictory CQNs. Otherwise, both queries produced by the reduction are noncontradictory. Thus the reduction also proves hardness for the stated restricted dually model-defining problem.
\end{proof}

\cqnEquivalenceC*

\begin{proof}
(1):
Assume first that  $P \equiv^{\text{pt}}_{\mathbb{N}[X, \overline{X}]} Q$. Then, using Theorem~\ref{N[X, X]}, we have that

\begin{itemize}

\item[(i)] There is a negation-homomorphism $h \colon Q \longrightarrow P$ exact for literals.

\item[(ii)] There is a negation-homomorphism $g \colon P \longrightarrow Q$ exact for literals.

\end{itemize}
Both $h$ and $g$ are also surjective on variables. Since we have surjections from both the variables in $P$ to the ones in $Q$ and viceversa, both sets  of variables have the same finite size. Then, both  $h$ and $g$ must be injective on variables too.  For the converse, if $P\simeq^\neg Q$, then the result follows immediately using Theorem~\ref{N[X, X]}, since  $P \equiv^{\text{pt}}_{\mathbb{N}[X, \overline{X}]} Q$ is equivalent to the two statements $P \sqsubseteq^{\text{pt}}_{\mathbb{N}[X, \overline{X}]} Q$ and $Q \sqsubseteq^{\text{pt}}_{\mathbb{N}[X, \overline{X}]} P$. 

Items (2) follows similarly.

(3): Assume first that $P \equiv^{\text{pt}}_{\mathrm{Sorp}(X, \overline{X})} Q$. Then, using Theorem~\ref{sorp(X,X)}, we have that

\begin{itemize}

\item[(i)] There is a negation-homomorphism $h \colon Q \longrightarrow P$ injective for relational literals.

\item[(ii)] There is a negation-homomorphism $g \colon P \longrightarrow Q$ injective for relational literals.

\end{itemize}
Let $m$ and $n$ be the numbers of relational literal occurrences in $P$ and $Q$, respectively. Applying a mapping to the variables does not change the number of literal occurrences, since repeated images are counted with multiplicity. Since $h$ is injective for relational literals, we have that $n\leq m$, while $g$ gives $m\leq n$. Thus $m=n$. The multiset of images under each mapping is therefore bag-contained in the multiset of relational literals of the corresponding query and has the same finite size. Hence both $h$ and $g$ are exact for relational literals.

By the same reasoning as in item (1), both $h$ and $g$ are also surjective and injective on variables, and thus $P\simeq^\neg Q$.  For the converse, if $P\simeq^\neg Q$, then the result follows immediately using Theorem~\ref{sorp(X,X)}, since  $P \equiv^{\text{pt}}_{\mathrm{Sorp}(X, \overline{X})} Q$ is equivalent to the two statements $P \sqsubseteq^{\text{pt}}_{\mathrm{Sorp}(X, \overline{X})} Q$ and $Q \sqsubseteq^{\text{pt}}_{\mathrm{Sorp}(X, \overline{X})} P$.
\end{proof}

\cqnEquivalenceD*

\begin{proof}
Assume  $h\colon P \longrightarrow Q$ is a negation-homomorphism and we wish to show that it is a homomorphism from $\mathtt{tr}(P)$ to $\mathtt{tr}(Q)$.  Suppose that an atom $A(\overline{x})$ appears in  $\mathtt{tr}(P)$, to show that the atom $A(h(\overline{x}))$ appears in  $\mathtt{tr}(Q)$. By construction of $\mathtt{tr}(P)$, $A(\overline{x}) = \mathtt{tr}(L(\vec{x}) )$ for some literal $L(\vec{x})$ in $P$. But then, by hypothesis, $L(h(\overline{x}))$ appears in the body of $Q$. Thus, $\mathtt{tr}(L(h(\overline{x})))= A(h(\overline{x}))$ is in the body of  $\mathtt{tr}(Q)$. For the converse assume now that $h\colon \mathtt{tr}(P) \longrightarrow \mathtt{tr}(Q)$ is a homomorphism to show that it is a negation-homomorphism from $P$ to $Q$. Assume then that $L(\vec{x})$ is in the body of $P$, so $\mathtt{tr}(L(\vec{x}))$ is in the body of $\mathtt{tr}(P)$ and thus $\mathtt{tr}(L(h(\overline{x})))$ in the body of $\mathtt{tr}(Q)$ by hypothesis. By construction of $\mathtt{tr}(Q)$ this must mean that $L(h(\overline{x}))$ is actually in the body of $Q$ and we are done.
\end{proof}



The next case that we deal with is the bag semiring and model-defining interpretations. This will be essentially the same as for $\mathbb{N}[X, \overline{X}]$.

\begin{restatable}{Thm}{cqnEquivalenceF}\label{equivalcqnsbags} For CQNs $P$ and $Q$ with the same tuple of free variables, we have that
 $P \equiv^{\text{md}pt}_{\mathbb{N}[X, \overline{X}]} Q$ iff $P\equiv^{\text{md}}_{\mathbb{N}} Q$ (i.e.\ equivalence for model-defining and provenance-tracking interpretations on $\mathbb{N}[X, \overline{X}]$ is the same as equivalence for model-defining interpretations on $\mathbb{N}$). Thus, the equivalence of CQNs for model-defining interpretations and $\mathbb{N}$ is GI-complete.

\end{restatable}


\begin{proof}
Suppose first that $P\not\equiv^{\mathrm{mdpt}}_{\mathbb N[X,\overline X]}Q$, witnessed by a model-defining provenance-tracking interpretation $\pi$ and an output tuple $\vec t$. Replace every non-zero ground relational annotation, including $1$, by its own fresh token of the correct polarity, choosing all tokens from distinct complementary pairs, and leave zero annotations unchanged. The resulting interpretation $\pi^*$ is still model-defining and provenance-tracking. Sending each used token to its original annotation and every unused complementary token to $0$ defines a semiring homomorphism $E$ with $E\circ\pi^*=\pi$. Hence $\pi^*\llbracket P(\vec t)\rrbracket\neq\pi^*\llbracket Q(\vec t)\rrbracket$.

Since no complementary tokens are used by $\pi^*$, these two values are distinct ordinary polynomials in finitely many independent indeterminates. They can be separated by assigning positive integers to those indeterminates as a non-zero polynomial over $\mathbb Z$ cannot vanish on every tuple of positive integers~\cite[Lemma 2.1]{Alon}. Assign $1$ to any remaining used tokens and $0$ to all unused tokens. This assignment respects every relation $x\overline x=0$ and therefore extends to a semiring homomorphism $H\colon\mathbb N[X,\overline X]\longrightarrow\mathbb N$. Every non-zero ground annotation of $\pi^*$ has positive image under $H$, so $H\circ\pi^*$ is model-defining. It distinguishes $P$ and $Q$, proving $P\not\equiv^{\mathrm{md}}_{\mathbb N}Q$.

For the other direction of the result, we must assume that  $P \equiv^{\text{mdpt}}_{\mathbb{N}[X, \overline{X}]} Q$ to show that $P\equiv^{\text{md}}_{\mathbb{N}} Q$. It suffices to show that $P \sqsubseteq^{\text{mdpt}}_{\mathbb{N}[X, \overline{X}]} Q$ only if  $P\sqsubseteq^{\text{md}}_{\mathbb{N}} Q$. Hence, suppose $P \sqsubseteq^{\text{mdpt}}_{\mathbb{N}[X, \overline{X}]} Q$ and consider an arbitrary model-defining interpretation $\pi$ (compatible with an $\mathbb{N}$-instance $\mathbf I$). We need to show that for any tuple $\bar{t}$, $\pi\llbracket  P(\bar{t})\rrbracket  \leq_\mathbb{N} \pi\llbracket Q(\bar{t})\rrbracket $. Given $\pi$, we define  an $\mathbb{N}[X, \overline{X}]$-interpretation, denoted $\pi^*$, as:
\begin{itemize}
    \item $\pi^*\llbracket R(\vec{a})\rrbracket $ is a new tag $p \in X$ if $\pi\llbracket R(\vec{a})\rrbracket  \neq 0$,
    \item $\pi^*\llbracket \neg R(\vec{a})\rrbracket $ is a new tag $\overline{p} \in \overline{X} $ such that $p$ is not used elsewhere if $\pi\llbracket \neg R(\vec{a})\rrbracket  \neq 0$,
    \item $\pi^*$ associates $0$ to all the other literals (just as $\pi$ does)
\end{itemize} 
Since $\pi$ is model-defining, then so is $\pi^*$ (since the only change is that non-zero annotations have been replaced with annotations from $X \cup \overline{X}$) and, by construction, we have that it is also provenance-tracking. Let $v\colon X \cup X' \longrightarrow \mathbb{N}$ map the variables used by $\pi^*$ to the corresponding annotation used by $\pi$; the variables of $X \cup X'$ that are not used as annotations in $\pi^*$ are sent to 0, since we need $v(p) \cdot v(\overline{p})=0$. 
By the universality of $\mathbb{N}[X, \overline{X}]$ \cite[Prop. 10]{GT}, $v$ can be extended uniquely to a semiring homomorphism $\text{Eval}_v\colon \mathbb{N}[X, \overline{X}] \longrightarrow \mathbb{N}$. 
Observe that $\text{Eval}_v\circ \pi^* = \pi$ by construction. 
 Thus, $\text{Eval}_v (\pi^*\llbracket P(\bar{t})\rrbracket ) = \pi\llbracket P(\bar{t})\rrbracket $ and $\text{Eval}_v (\pi^*\llbracket Q(\bar{t})\rrbracket ) = \pi\llbracket Q(\bar{t})\rrbracket$ by~\cite[Prop. 2]{GT}. By the assumption $P \sqsubseteq^{\text{md}pt}_{\mathbb{N}[X, \overline{X}]} Q$, so $\pi^*\llbracket P(\bar{t})\rrbracket  \leq_{\mathbb{N}[X, \overline{X}]} \pi^*\llbracket Q(\bar{t})\rrbracket $ 
and by Proposition~\ref{prop3.3bis} (i.e.~\cite[Prop. 3.3]{Green}), $ \pi\llbracket P(\bar{t})\rrbracket  = \text{Eval}_v (\pi^*\llbracket P(\bar{t})\rrbracket ) \leq_\mathbb{N}\text{Eval}_v (\pi^*\llbracket Q(\bar{t})\rrbracket ) = \pi\llbracket Q(\bar{t})\rrbracket $. Since $\pi$ was arbitrary, we must have that  $P\sqsubseteq^{\text{md}}_{\mathbb{N}} Q$, as desired.

For the complexity claim, a CQN containing an atom and its negation is identically zero under model-defining interpretations. Two such queries are therefore equivalent, while a CQN without such a pair is not equivalent to zero, as its model-defining canonical interpretation witnesses. These cases are checked in polynomial time. For two remaining CQNs, the model-defining provenance-tracking version of Theorem~\ref{N[X, X]} gives negation-isomorphism exactly as in Theorem~\ref{equivalcqns}. The GI upper bound follows as in Theorem~\ref{equivcqnsvarious}, and GI-hardness already holds for CQs without negation.
\end{proof}

\begin{Rmk}\label{bag-equivalence-bridge}
The equivalence between $\mathbb N$ under model-defining interpretations and $\mathbb N[X,\overline X]$ under model-defining provenance-tracking interpretations also holds for UCQNs and for (U)CQ$^\neq$Ns. The proof above uses only preservation of sums and products while equations and disequations contribute $0$ or $1$ and are preserved as well. By Remarks~\ref{mdpt-identifications} and~\ref{mdpt-union-identifications}, containment under model-defining provenance-tracking interpretations is decided by the corresponding counting criteria restricted to consistent identifications. This restriction is checked in polynomial time and leaves the $\mathrm{coNP}^{\mathrm{PP}}$ upper bounds unchanged. Applying these bounds in both directions therefore places model-defining bag equivalence of (U)CQ$^\neq$Ns, and hence UCQNs, in $\mathrm{coNP}^{\mathrm{PP}}$.
\end{Rmk}

Finally, we deal with the case of  $\mathrm{Why}(X,\overline{X})$ (and thus $\mathrm{Trio}(X,\overline{X})$) which requires independent treatment from the previous cases.

\begin{restatable}{Lem}{cqnEquivalenceG}\label{whyduplicateelimination}
For every 
CQN $P$, $P \equiv^{\text{pt}}_{\mathrm{Why}(X,\overline{X})} \text{red}(P).$
\end{restatable}


\begin{proof}
Let $P(\vec{u}) := \exists \vec{z}\, \bigl(L_1\wedge\ldots\wedge L_m\bigr),$ where $L_1,\ldots,L_m$ are the relational-literal occurrences in the body of $P$, not necessarily distinct. Let $\text{LitSet}(P)=\{K_1,\ldots,K_r\}.$ Thus, $ \text{red}(P)(\vec{u}) := \exists \vec{z}\,\bigl(K_1\wedge\ldots\wedge K_r\bigr),$ where exactly one occurrence of each syntactically distinct relational
literal of $P$ is retained.

Fix an arbitrary provenance-tracking interpretation $\pi$, an arbitrary tuple $\vec{a}$, and an assignment $v$ of the variables of $P$ such that $v(\vec{u})=\vec{a}$. Since removing repeated occurrences of relational literals does not remove any variable from the query, $P$ and $\text{red}(P)$ have the same variables and the same head. Hence the same assignments are considered in the evaluation of the two queries.

We show that $v$ gives exactly the same contribution to $P(\vec{a})$ and to $\text{red}(P)(\vec{a})$. First suppose that, for some $K_j\in\text{LitSet}(P)$, $\pi\bigl(v(K_j)\bigr)=0.$ Since $K_j$ occurs in both $P$ and $\text{red}(P)$, the semiring product associated with $v$ contains a zero factor in both queries. Therefore, the contribution of $v$ is zero in both cases. Suppose now that $\pi\bigl(v(K_j)\bigr)\neq 0$ for every $1\leq j\leq r$. Since $\pi$ is provenance-tracking, for each $j$ there is a unique witness $S_j$ such that $\pi\bigl(v(K_j)\bigr)=\{S_j\}.$ For every occurrence $L_i$ of $P$, there is some $j(i)\in\{1,\ldots,r\}$ such that $L_i=K_{j(i)}.$ Hence $\pi\bigl(v(L_i)\bigr) = \pi\bigl(v(K_{j(i)})\bigr)= \{S_{j(i)}\}$. Using pairwise-union multiplication in $\mathrm{Why}(X,\overline{X})$, the
contribution of $v$ to $P(\vec{a})$ is therefore represented by $\prod_{i=1}^{m}\pi\bigl(v(L_i)\bigr) = \left\{\bigcup_{i=1}^{m} S_{j(i)}\right\}$. On the other hand, the contribution of the same assignment to $\text{red}(P)(\vec{a})$ is represented by $\prod_{j=1}^{r}\pi\bigl(v(K_j)\bigr) = \left\{ \bigcup_{j=1}^{r} S_j \right\}$. By definition of $\text{LitSet}(P)$, every $K_j$ occurs at least once among $L_1,\ldots,L_m$. Thus the map $i\longmapsto j(i)$ is surjective onto $\{1,\ldots,r\}$. Consequently, the family $S_{j(1)},\ldots,S_{j(m)}$ contains exactly the same distinct sets as $S_1,\ldots,S_r$, possibly with repetitions. Since set union is idempotent, $\bigcup_{i=1}^{m}S_{j(i)} =\bigcup_{j=1}^{r}S_j$, therefore $\prod_{i=1}^{m}\pi\bigl(v(L_i)\bigr) = \prod_{j=1}^{r}\pi\bigl(v(K_j)\bigr).$

Notice that this equality also covers the case in which the resulting witness contains a pair of complementary tokens: in $\mathrm{Why}(X,\overline{X})$, any witness containing such a pair has value $0$. Since the union of the witnesses is the same for $P$ and for $\text{red}(P)$, this happens for one query if and only if it happens for the other, hence the two contributions are still equal.

We have therefore shown that, for every assignment $v$ with $v(\vec{u})=\vec{a}$, its contribution to $P(\vec{a})$ is equal to its contribution to $\text{red}(P)(\vec{a})$. Since the two queries have the same assignments for the same output tuple, taking the sum over all such assignments yields $\pi\llbracket  P(\vec{a})\rrbracket =\pi\llbracket \text{red}(P)(\vec{a})\rrbracket.$ Since $\pi$ and $\vec{a}$ were arbitrary, we conclude that $P\equiv^{\text{pt}}_{\mathrm{Why}(X,\overline{X})} \text{red}(P).$
\end{proof}

\begin{restatable}{Lem}{cqnEquivalenceH}\label{whybackforth}
Let $P$ and $Q$ be finite
and duplicate-free CQNs. If there are set-onto negation-homomorphisms $h\colon Q\longrightarrow P$ and $g\colon P\longrightarrow Q,$ then $P\simeq_{\neg} Q.$
\end{restatable}


\begin{proof}
We first show that $h$ induces a bijection between the variables of $Q$ and those of $P$. Let $x$ be a variable of $P$, then $x$ occurs in some positive relational literal $L\in\text{LitSet}(P)$. Since $h$ is set-onto, there is a literal $K\in\text{LitSet}(Q)$ such that $h(K)=L$. Hence, at the argument position occupied by $x$ in $L$ the corresponding term of $K$ is mapped to $x$. Therefore, every variable of $P$ is the image under $h$ of some variable of $Q$, and thus the number of variables of $P$ is less or equal to the number of variables of Q. Applying the same argument to $g$ gives that the number of variables of $Q$ is less or equal to the number of variables of $P$. Hence, the two finite variable sets have the same cardinality. It follows that $h$ maps the variables of $Q$ bijectively to the variables of $P$. Consequently, if two distinct  literals $K_1,K_2$ of $Q$ had the same image under $h$, then preservation of the relation symbol, negations, and argument positions, together with injectivity on terms, would imply $K_1=K_2$, a contradiction. Thus $h$ is injective on $\text{LitSet}(Q)$. Since $h$ is also set-onto, $h(\text{LitSet}(Q))=\text{LitSet}(P),$ it induces a bijection between the distinct relational literals of $Q$ and those of $P$. As $P$ and $Q$ are duplicate-free, these are precisely their relational-literal occurrences, therefore $h$ is a bijection on variables and maps the relational literals of $Q$ bijectively onto those of $P$, preserving relation symbols, argument positions, and negations. Hence $h$ is a negation-isomorphism, and $P\simeq_{\neg}Q.$
\end{proof}

Combining duplicate elimination (Lemma~\ref{whyduplicateelimination}) with the back-and-forth property for duplicate-free queries (Lemma~\ref{whybackforth}), we obtain our next result: equivalence over $\mathrm{Why}(X,\overline{X})$ reduces precisely to syntactic isomorphism between the reduced queries.

\begin{restatable}{Thm}{cqnEquivalenceI}\label{whycqnredequivalence}
For CQNs $P$ and $Q$ with the same tuple of free variables, the following are equivalent:
\begin{itemize}
    \item[1.] 
    $P \equiv^{\text{pt}}_{\mathrm{Why}(X,\overline{X})} Q$.

    \item[2.]
    $\text{red}(P) \simeq_{\neg} \text{red}(Q)$.
\end{itemize}
\end{restatable}


\begin{proof}
$(1)\Rightarrow(2)$ Assume that $P \equiv^{\text{pt}}_{\mathrm{Why}(X,\overline{X})} Q.$ Then, by definition of equivalence, $P \sqsubseteq^{\text{pt}}_{\mathrm{Why}(X,\overline{X})} Q$ and $Q \sqsubseteq^{\text{pt}}_{\mathrm{Why}(X,\overline{X})} P.$ By the characterization of containment for CQNs over provenance-tracking interpretations, there are set-onto negation-homomorphisms $h\colon Q\longrightarrow P$ and $g\colon P\longrightarrow Q.$ In particular, $h(\text{LitSet}(Q))=\text{LitSet}(P)$ and $g(\text{LitSet}(P))=\text{LitSet}(Q).$ By definition of duplicate-free reduction, $\text{LitSet}(\text{red}(P))= \text{LitSet}(P)$ and $\text{LitSet}(\text{red}(Q)) =\text{LitSet}(Q).$ Moreover, the reduction does not change the terms, the head, relation symbols, or polarities of the literals. Hence the same mappings induce set-onto negation-homomorphisms $h\colon \text{red}(Q) \longrightarrow \text{red}(P)$ and $g\colon \text{red}(P) \longrightarrow \text{red}(Q).$ The queries $\text{red}(P)$ and $\text{red}(Q)$ are duplicate-free, therefore, by
Lemma~\ref{whybackforth}, $\text{red}(P) \simeq_{\neg} \text{red}(Q)$. 

$(2)\Rightarrow(1)$ Assume now that $\text{red}(P) \simeq_{\neg} \text{red}(Q).$ Let $h\colon \text{red}(Q) \longrightarrow \text{red}(P)$ be a negation-isomorphism. Then both $h$ and its inverse $h^{-1}\colon \text{red}(P) \longrightarrow \text{red}(Q)$ are set-onto negation-homomorphisms. Hence, by the characterization of
containment for CQNs over provenance-tracking interpretations, $\text{red}(P) \sqsubseteq^{\text{pt}}_{\mathrm{Why}(X,\overline{X})}\text{red}(Q)$ and $\text{red}(Q)\sqsubseteq^{\text{pt}}_{\mathrm{Why}(X,\overline{X})} \text{red}(P).$ Therefore, $\text{red}(P) \equiv^{\text{pt}}_{\mathrm{Why}(X,\overline{X})}\text{red}(Q).$ By Lemma~\ref{whyduplicateelimination}, $P \equiv^{\text{pt}}_{\mathrm{Why}(X,\overline{X})} \text{red}(P)$ and $Q\equiv^{\text{pt}}_{\mathrm{Why}(X,\overline{X})}\text{red}(Q).$ Combining these equivalences, we obtain $P \equiv^{\text{pt}}_{\mathrm{Why}(X,\overline{X})} Q$.
\end{proof}

Theorem~\ref{whycqnredequivalence} reduces semantic equivalence under $\mathrm{Why}(X, \overline{X})$ to checking negation-isomorphism between duplicate-free queries. Because testing query isomorphism inherently captures the structure of underlying relational graphs, this syntactic reduction allows us to settle the exact computational complexity of the equivalence problem.

\begin{restatable}{Thm}{cqnEquivalenceJ}\label{whycqnequivalence}
Equivalence of CQNs for provenance-tracking interpretations over $\mathrm{Why}(X, \overline{X})$ is GI-complete.
\end{restatable}


\begin{proof}
We first show that the problem is reducible in polynomial time to graph isomorphism. Let $P$ and $Q$ be two CQNs with the same tuple of free variables. By Theorem~\ref{whycqnredequivalence}, $P\equiv^{\text{pt}}_{\mathrm{Why}(X,\overline{X})}Q$ if and only if $\text{red} (P)\simeq_{\neg}\text{red}(Q).$ Moreover, $\text{red}(P)$ and $\text{red}(Q)$ can be computed in polynomial time. It is therefore enough to reduce negation-isomorphism of duplicate-free CQNs to graph isomorphism. Given a duplicate-free CQN $P$, we construct a vertex-colored graph $G_P$ as follows. We introduce one vertex for each variable of $P$, all having the same color. For every relational literal $L=\pm R(t_1,\ldots,t_r)$ occurring in the body of $P$, we introduce a literal vertex whose color records the relation symbol $R$, the polarity of $L$, and its arity $r$. For every such literal $L$ and every argument position $1\leq k\leq r$, we introduce a vertex $p_{L,k}$, whose color records the position $k$, and connect it both to the vertex corresponding to $L$ and to the variable vertex corresponding to $t_k$. Finally, for each position $k$ in the head of $P$, we introduce a head-position vertex with a color specific to $k$ and connect it to the variable occurring in the $k$-th position of the head. The size of $G_P$ is polynomial in the size of $P$. We claim that $\text{red}(P)\simeq_{\neg}\text{red}(Q)$ if and only if $G_{\text{red}(P)} \simeq_{\text{col}} G_{\text{red}(Q)}.$ Indeed, suppose first that there is a negation-isomorphism between $\text{red}(P)$ and $\text{red}(Q)$, it maps variable vertices according to the corresponding bijection on variables and literal vertices according to the induced bijection on relational literals. Each argument-position vertex $p_{L,k}$ is mapped to the vertex corresponding to the same argument position of the image of $L$, and each head-position vertex is mapped to the vertex corresponding to the same head position. Since a negation-isomorphism preserves relation symbols, polarities, arities, argument positions, and the head, this gives a color-preserving graph isomorphism. Conversely, suppose that there is a color-preserving isomorphism between $G_{\text{red}(P)}$ and $G_{\text{red}(Q)}$. The colors force variable vertices to be mapped to variable vertices, and literal vertices to literal vertices having the same relation symbol, polarity, and arity. Moreover, a vertex representing the $k$-th argument position can only be mapped to another vertex representing the $k$-th argument position. Since such a vertex is adjacent both to its literal vertex and to the variable occurring in that argument position, preservation of adjacency determines all the ordered arguments of every relational literal. Similarly, the head-position vertices ensure that the head is preserved. Hence the restriction of the graph isomorphism to the variable vertices induces a bijection on variables whose extension maps the relational literals of one query bijectively onto those of the other, preserving their relation symbols, polarities, and argument positions. Therefore, it induces a negation-isomorphism between $\text{red}(P)$ and $\text{red}(Q)$, proving that $\text{red}(P)\simeq_{\neg}\text{red}(Q)$ if and only if $G_{\text{red}(P)} \simeq_{\text{col}} G_{\text{red}(Q)}.$

It remains to remove the colors next. For this, consider two colored graphs $G$ and $H$ obtained as above, and let $N=\max\{|V(G)|,|V(H)|\}.$ Let the colors occurring in the two graphs be indexed by $1,\ldots,c$, using the same index for the same color in both graphs. For every vertex of color $i$, attach to it by a single edge a fresh cycle of length $N+i+3$. Let $U(G)$ and $U(H)$ denote the resulting uncolored graphs. The edge connecting an original vertex to its attached cycle is a bridge, since removing it disconnects the newly added cycle from the rest of the graph. In particular, this edge cannot belong to a simple cycle. Moreover, since each original graph has at most $N$ vertices, no cycle of length $N+i+3$ can occur entirely inside the original graph, thus the newly introduced cycles can be distinguished from all cycles already present in the original graphs, and their lengths encode the colors of the vertices to which they are attached. Consequently, every isomorphism from $U(G)$ to $U(H)$ must map an attached cycle of length $N+i+3$ to another attached cycle of the same length. Hence it maps the original vertex to which this cycle is attached to a vertex having the same original color. Its restriction to the original vertices gives a color-preserving isomorphism from $G$ to $H$. Conversely, every color-preserving isomorphism from $G$ to $H$ extends to an isomorphism from $U(G)$ to $U(H)$ by mapping, for each vertex, its attached cycle onto the cycle of the same length attached to its image. Hence $G\simeq_{\text{col}}H$ if and only if $U(G)\simeq U(H).$ This construction is polynomial since at most one cycle of length $N+c+3$ is added for each original vertex.

By $\text{red}(P)\simeq_{\neg}\text{red}(Q)$ if and only if $G_{\text{red}(P)} \simeq_{\text{col}} G_{\text{red}(Q)}$ and $G\simeq_{\text{col}}H$ if and only if $U(G)\simeq U(H)$, together with the polynomial-time computation of the duplicate-free reductions, we obtain a polynomial-time reduction from equivalence of CQNs over provenance-tracking interpretations to graph isomorphism.

We now prove GI-hardness. Let $G=(V,E)$ be a finite simple undirected graph. Let $\text{Vtx}$ be a unary relation symbol and $\text{Edge}$ a binary relation symbol. The former is used to represent the vertices of the graph, while the latter represents its edges. For every $v\in V$, introduce a variable $x_v$, and define the Boolean positive CQ $P_G := \bigwedge_{v\in V}\text{Vtx}(x_v) \ \wedge\ \bigwedge_{\{u,v\}\in E} \left( \text{Edge}(x_u,x_v) \wedge \text{Edge}(x_v,x_u) \right).$ The atoms $\text{Vtx}(x_v)$ ensure that every vertex of $G$ is represented in the query, including isolated vertices, which would otherwise occur in no $\text{Edge}$ atom. For every undirected edge $\{u,v\}\in E$, the two atoms $\text{Edge}(x_u,x_v)$ and $\text{Edge}(x_v,x_u)$ represent its two orientations, so that the construction does not depend on an arbitrary choice of orientation. Since $G$ is simple, it has neither loops nor multiple edges, so the construction introduces no repeated $\text{Edge}$ atoms, and there is exactly one atom $\text{Vtx}(x_v)$ for each vertex. Thus $P_G$ is duplicate-free. Moreover it is positive and Boolean. We claim that, for any finite simple undirected graphs $G$ and $H$, $G\simeq H$ if and only if $P_G\simeq P_H.$ Suppose first that $f\colon V(G)\longrightarrow V(H)$ is a graph isomorphism. The bijection $x_v\longmapsto x_{f(v)}$ preserves every $\text{Vtx}$ atom. Moreover, if $\{u,v\}\in E(G)$, then $\{f(u),f(v)\}\in E(H)$, and therefore it preserves both $\text{Edge}(x_u,x_v)$ and $\text{Edge}(x_v,x_u)$, hence it induces an isomorphism between $P_G$ and $P_H$. Conversely, suppose that $P_G\simeq P_H$. Since there is exactly one variable $x_v$ for each vertex $v$, and every such variable occurs in the atom $\text{Vtx}(x_v)$, a query isomorphism induces a bijection $x_v\longmapsto x_{f(v)}$ and therefore a bijection $f\colon V(G)\longrightarrow V(H).$ If $\{u,v\}\in E(G)$, then $\text{Edge}(x_u,x_v)$ occurs in $P_G$. Preservation of relational atoms implies that $\text{Edge}(x_{f(u)},x_{f(v)})$ occurs in $P_H$, and hence $\{f(u),f(v)\}\in E(H)$. Applying the same argument to the inverse query isomorphism gives the converse implication. Therefore, $\{u,v\}\in E(G)$ if and only if $\{f(u),f(v)\}\in E(H),$ and $f$ is a graph isomorphism. This proves that $G\simeq H$ if and only if $P_G\simeq P_H.$

Finally, $P_G$ and $P_H$ are positive and duplicate-free. Hence ordinary query isomorphism coincides with negation-isomorphism, and $\text{red}(P_G)=P_G,$ $\text{red}(P_H)=P_H.$ By Theorem~\ref{whycqnredequivalence} and the fact that $G\simeq H$ if and only if $P_G\simeq P_H$, $G\simeq H$ if and only if $P_G\simeq P_H$ if and only if $P_G\simeq_{\neg}P_H$ if and only if $P_G\equiv^{\text{pt}}_{\mathrm{Why}(X,\overline{X})}P_H.$ The construction of $P_G$ is polynomial in the size of $G$. Therefore, graph isomorphism reduces in polynomial time to the equivalence problem.

Combining the two polynomial-time reductions, we conclude that the equivalence problem for CQNs for provenance-tracking interpretations over $\mathrm{Why}(X, \overline{X})$ is GI-complete.
\end{proof}

As a direct consequence of the containment equivalence established earlier (Theorem~\ref{triocqncontainment}), query equivalence over the square-free provenance semiring $\mathrm{Trio}(X, \overline{X})$ coincides completely with equivalence over $\mathrm{Why}(X, \overline{X})$. This immediately transfers the GI-completeness result to $\mathrm{Trio}(X, \overline{X})$.

\begin{restatable}{Thm}{cqnEquivalenceK}\label{triocqnequivalence}
For CQNs $P$ and $Q$ with the same tuple of free variables,
$$
P\equiv^{\text{pt}}_{\mathrm{Trio}(X,\overline{X})}Q
\quad\text{if and only if}\quad
P\equiv^{\text{pt}}_{\mathrm{Why}(X,\overline{X})}Q.
$$
\end{restatable}


\begin{proof}
By definition, $P\equiv^{\text{pt}}_{\mathrm{Trio}(X,\overline{X})}Q$ if and only if $P\sqsubseteq^{\text{pt}}_{\mathrm{Trio}(X,\overline{X})}Q$ and $Q\sqsubseteq^{\text{pt}}_{\mathrm{Trio}(X,\overline{X})}P.$ By Theorem~\ref{triocqncontainment}, these two containments hold if and only if $P\sqsubseteq^{\text{pt}}_{\mathrm{Why}(X,\overline{X})}Q$ and $Q\sqsubseteq^{\text{pt}}_{\mathrm{Why}(X,\overline{X})}P,$ which, by definition, is equivalent to $P\equiv^{\text{pt}}_{\mathrm{Why}(X,\overline{X})}Q.$\end{proof}

\begin{restatable}{Thm}{cqnEquivalenceL}\label{triocqnequivalencecomplexity}
Equivalence of CQNs for provenance-tracking interpretations over $\mathrm{Trio}(X,\overline{X})$ is GI-complete.
\end{restatable}


\begin{proof} 
By Theorem~\ref{triocqnequivalence}, for every pair of CQNs $P$ and $Q$, $P\equiv^{\text{pt}}_{\mathrm{Trio}(X,\overline{X})}Q$ if and only if $P\equiv^{\text{pt}}_{\mathrm{Why}(X,\overline{X})}Q.$ Hence the two equivalence problems coincide. By Theorem~\ref{whycqnequivalence}, equivalence of CQNs for provenance-tracking interpretations over $\mathrm{Why}(X,\overline{X})$ is GI-complete. Therefore, the corresponding equivalence problem over $\mathrm{Trio}(X,\overline{X})$ is GI-complete as well.
\end{proof}

\cqnEquivalenceWhichMD*
\begin{proof}
The upper bounds follow from Theorem~\ref{containmentwhich modeldef}. Indeed, equivalence holds if and only if containment holds in both directions, and both $\Pi^p_2$ and coNEXP are closed under finite intersections.

For the lower bounds, we give a polynomial-time reduction from containment of CQNs over $\mathbb{B}$ with model-defining interpretations, using Theorem~\ref{containnegbool}. Let $P(\vec{u})$ and $Q(\vec{u})$ be two CQNs over a common schema $\tau$, we can assume that $\tau$ contains only the relation symbols occurring in these queries. Introduce two fresh unary relation symbols $T$ and $U$. For each $F\in\{P,Q\}$, let $F^*$ be the query obtained from $F$ by adding $T(y)\wedge U(y)$ to its body for every variable $y$ of $F$. The tuple of free variables is left unchanged.

We next define a Boolean CQN $H$ whose role is to contain all the tags that $P^*$ and $Q^*$ can contribute, whenever the value of $H$ is different from $\bot$. Let $$H_T:=\exists t\,T(t) \quad  \text{ and } \quad H_{TU}:=\exists s\,(T(s)\wedge U(s)).$$ For every relation symbol $R\in\tau$ of arity $r$, let $$
H_R^+ :=\exists z_1\cdots z_r\, \left( \bigwedge_{i=1}^{r}T(z_i) \wedge R(z_1,\ldots,z_r) \right) \quad \text{ and } \quad H_R^- :=\exists w_1\cdots w_r\, \left( \bigwedge_{i=1}^{r}T(w_i) \wedge\neg R(w_1,\ldots,w_r) \right).$$
Then we define $$H:=H_T\wedge H_{TU}\wedge \bigwedge_{R\in\tau}(H_R^+\wedge H_R^-).$$

All bound variables of $H$, $P^*$ and $Q^*$ are chosen pairwise distinct, while $P^*$ and $Q^*$ retain the common tuple of free variables $\vec{u}$. Define
$$
\widehat P(\vec{u}):=H\wedge P^*(\vec{u}),
\qquad
\widehat Q(\vec{u}):=H\wedge P^*(\vec{u})\wedge Q^*(\vec{u}).
$$
The atoms with $T$ ensure that all these queries are safe. We show that
$$
P\sqsubseteq^{\text{md}}_{\mathbb{B}}Q
\quad\text{if and only if}\quad
\widehat P\equiv^{\text{md}}_{\mathrm{Which}(X)}\widehat Q.
$$

$(\Rightarrow):$ Assume that $P\sqsubseteq^{\text{md}}_{\mathbb{B}}Q$. Consider an arbitrary $\mathrm{Which}(X)$-database instance $\mathbf I$, a compatible model-defining interpretation $\pi$, and an output tuple $\vec{t}$ over $A:=\mathrm{adom}(\mathbf I)$. If $\pi\llbracket H\rrbracket=\bot$, then both $\pi\llbracket\widehat P(\vec{t})\rrbracket$ and $\pi\llbracket\widehat Q(\vec{t})\rrbracket$ are $\bot$.  Otherwise, let $W:=\pi\llbracket H\rrbracket\neq\bot$. Since the queries forming $H$ have disjoint variables, their assignments can be chosen independently and combined into an assignment for $H$, so the value of $H$ is the product of their values. Since $\pi\llbracket H\rrbracket\neq\bot$, all these values are non-bottom, so their product is their union. In particular, the value of each query forming $H$ is a subset of $W$.

Let $F\in\{P,Q\}$ and let $v$ be an assignment giving a non-bottom contribution to $F^*$. Every literal used by $v$ must have non-bottom value. In particular, both $T(a)$ and $U(a)$ have non-bottom value for every value $a$ used by $v$. We show that each annotation used by $v$ is contained in the value of one of the queries forming $H$, and hence in $W$:
\begin{itemize}
\item For $T(a)$, the assignment $t\mapsto a$ gives a contribution to $H_T$ equal to $\pi\llbracket T(a)\rrbracket$. Hence
$\pi\llbracket T(a)\rrbracket\subseteq\pi\llbracket H_T\rrbracket\subseteq W$.

\item For $U(a)$, use the assignment $s\mapsto a$ in $H_{TU}=\exists s\,(T(s)\wedge U(s))$. Both atoms are already known to have non-bottom value, so this assignment contributes
$\pi\llbracket T(a)\rrbracket\cup\pi\llbracket U(a)\rrbracket$.
This contribution contains the annotation of $U(a)$, giving
$\pi\llbracket U(a)\rrbracket\subseteq\pi\llbracket H_{TU}\rrbracket\subseteq W$.

\item For a literal $L=R(a_1,\ldots,a_r)$ or $L=\neg R(a_1,\ldots,a_r)$, assign $z_i\mapsto a_i$ in $H_R^+$ or $w_i\mapsto a_i$ in $H_R^-$. Since $L$ and all the atoms $T(a_i)$ have non-bottom value, this gives a non-bottom contribution equal to the union of their annotations. It contains $\pi\llbracket L\rrbracket$ and is included in the value of the corresponding query, which is a subset of $W$, so $\pi\llbracket L\rrbracket\subseteq W$.
\end{itemize}

The contribution of $v$ is given by the union of these annotations, so it is a subset of $W$ too. Since $v$ is arbitrary, every non-bottom contribution to $F^*$ is a subset of $W$. Therefore,
$$
\pi\llbracket F^*(\vec{a})\rrbracket\neq\bot
\quad \text{implies}\quad
\pi\llbracket F^*(\vec{a})\rrbracket\subseteq W
$$
for $F\in\{P,Q\}$ and every output tuple $\vec{a}$.

We now use the assumption $P\sqsubseteq^{\text{md}}_{\mathbb{B}}Q$, and we construct from $\pi$ a Boolean interpretation that records which literals have non-bottom value. Let
$$
C:=\{a\in A\mid
\pi\llbracket T(a)\rrbracket\neq\bot
\text{ and }
\pi\llbracket U(a)\rrbracket\neq\bot\}.
$$
This set is nonempty because $\pi\llbracket H_{TU}\rrbracket\neq\bot$. We now define a new boolean interpretations $\pi_0$ with domain $C$. For every literal $L$, define
$$
\pi_0\llbracket L\rrbracket:=
\begin{cases}
\top,&\text{if }\pi\llbracket L\rrbracket\neq\bot,\\
\bot,&\text{otherwise.}
\end{cases}
$$
Since $\pi$ is model-defining, so is $\pi_0$.

An assignment gives a non-bottom contribution to $F^*$ exactly when its values belong to $C$ and all the literals of $F$ have non-bottom value under $\pi$, equivalently Boolean value $\top$ under $\pi_0$. Therefore, for $F\in\{P,Q\}$ and every output tuple $\vec{a}$ over $C$,
$$
\pi\llbracket F^*(\vec{a})\rrbracket\neq\bot
\quad \text{if and only if}\quad
\pi_0\llbracket F(\vec{a})\rrbracket=\top.
$$

Let $\mathbf J$ be the finite Boolean instance consisting of the positive ground atoms true under $\pi_0$. By safety, every assignment satisfying $P$ or $Q$ uses only elements of $\mathrm{adom}(\mathbf J)$, so if we restrict $\pi_0$ to this active domain it preserves all satisfying assignments, and we can apply the Boolean containment assumption also to $\pi_0$.

If $\pi\llbracket P^*(\vec{t})\rrbracket=\bot$, both $\widehat P$ and $\widehat Q$ have value $\bot$, since both contain $P^*$. Otherwise, the atoms we added in $T$ and $U$ ensure that $\vec{t}$ is over $C$. The equivalence above and the containment assumption give $\pi_0\llbracket P(\vec{t})\rrbracket=\top$, then $\pi_0\llbracket Q(\vec{t})\rrbracket=\top$, and $\pi\llbracket Q^*(\vec{t})\rrbracket\neq\bot$. Both query values are therefore non-bottom subsets of $W$. Since multiplication of non-bottom values is given by the union, we obtain
$$
\begin{aligned}
\pi\llbracket\widehat P(\vec{t})\rrbracket
&=W\cup\pi\llbracket P^*(\vec{t})\rrbracket=W,\\
\pi\llbracket\widehat Q(\vec{t})\rrbracket
&=W\cup\pi\llbracket P^*(\vec{t})\rrbracket
\cup\pi\llbracket Q^*(\vec{t})\rrbracket=W.
\end{aligned}
$$
Thus $\widehat P$ and $\widehat Q$ have the same value in both cases and since $\mathbf I$, $\pi$, and $\vec{t}$ were arbitrary, we can conclude that $\widehat P\equiv^{\text{md}}_{\mathrm{Which}(X)}\widehat Q$.

$(\Leftarrow):$ We prove the contrapositive. Suppose $P\not\sqsubseteq^{\text{md}}_{\mathbb{B}}Q$. Then there are a Boolean instance $\mathbf J$, a compatible model-defining interpretation $\pi_0$, and a tuple $\vec{t}$ such that $\pi_0\llbracket P(\vec{t})\rrbracket=\top$ and $\pi_0\llbracket Q(\vec{t})\rrbracket=\bot.$ Extend $\pi_0$ to a nonempty finite set $D\supseteq\mathrm{adom}(\mathbf J)$ by giving value $\bot$ to every positive atom involving a new element and $\top$ to its negation. By safety, any assignment using a new element gives value $\bot$ to the body of $P$ or $Q$, so the values of the two query values are unchanged.

To obtain different values for $\widehat P$ and $\widehat Q$, take two distinct fresh elements $d_1,d_2$ and let $A:=D\cup\{d_1,d_2\}$. For every $a\in A$, define $$\pi\llbracket T(a)\rrbracket:=\emptyset \text{ and } \pi\llbracket U(a)\rrbracket:=\begin{cases}\emptyset,&\text{if }a\in D,\\\bot,&\text{otherwise.}\end{cases}$$ 

For each $R\in\tau$ of arity $r$ and each $\vec{a}\in A^r$, define $$\pi\llbracket R(\vec{a})\rrbracket:=\begin{cases}\emptyset,&\text{if }\vec{a}\in D^r\text{ and }\pi_0\llbracket R(\vec{a})\rrbracket=\top,\\\emptyset,&\text{if }\vec{a}=(d_1,\ldots,d_1),\\\bot,&\text{otherwise.}\end{cases}$$

For every ground relational atom $L$ define $$\pi\llbracket\neg L\rrbracket:=\begin{cases}\emptyset,&\text{if }\pi\llbracket L\rrbracket=\bot,\\\bot,&\text{otherwise.}\end{cases}$$

This interpretation $\pi$ is model-defining and compatible with the finite $\mathrm{Which}(X)$-instance $\mathbf I$ given by its positive annotations. Moreover, $\mathrm{adom}(\mathbf I)=A$, since every $T(a)$ has value $\emptyset\neq\bot$.

Every query forming $H$ has an assignment with value $\emptyset$. For $H_T$ and $H_{TU}$, choose any element of the nonempty set $D$. For each $R\in\tau$, assign all variables of $H_R^+$ to $d_1$ and all variables of $H_R^-$ to $d_2$. The atoms in $T$ and the literals $R(d_1,\ldots,d_1)$ and $\neg R(d_2,\ldots,d_2)$ all have value $\emptyset$ by construction. Since all annotations are either $\bot$ or $\emptyset$, we obtain $\pi\llbracket H\rrbracket=\emptyset.$

The added atoms in $U$ force every assignment giving a non-bottom contribution to $P^*$ or $Q^*$ to use only elements of $D$. On $D$, the original literals have non-bottom value exactly when they have value $\top$ under $\pi_0$, and all atoms in $T$ and $U$ have value $\emptyset$. Hence $\pi\llbracket P^*(\vec{t})\rrbracket=\emptyset,$ $\pi\llbracket Q^*(\vec{t})\rrbracket=\bot.$ It follows that $\pi\llbracket\widehat P(\vec{t})\rrbracket=\emptyset\neq\bot=\pi\llbracket\widehat Q(\vec{t})\rrbracket.$ Thus $\widehat P$ and $\widehat Q$ are not equivalent.

Finally, the construction is polynomial in the size of $P$ and $Q$. We add two unary atoms for each variable and, for each occurring relation symbol of arity $r$, two Boolean queries with $r$ unary atoms and one relational literal. Moreover, the only new relation symbols are unary, so every fixed arity bound $k\geq 2$ is preserved. Hence the reduction gives both $\Pi^p_2$-hardness for bounded arity and coNEXP-hardness for unbounded arity, by Theorem~\ref{containnegbool}. Together with the upper bounds, this proves the result.

\end{proof}

\section{Omitted  Proofs from \S~\ref{The problem of containmentUCQN}} \label{app:ucqn-containment-proofs}

\begin{restatable}{Thm}{ucqnContainmentA} \label{dsitrlatticenegucq}
For UCQNs, $\bigcup_{1\leq i \leq n} P_i, \bigcup_{1\leq i \leq m} Q_i$ with the same tuple of free variables, and a bounded distributive lattice $\mathbb{L}$ with an intermediate element $\epsilon \neq \top, \bot$, where no disjunct $P_i$ contains any atom and its negation, the following are equivalent:
\begin{itemize}
\item[1.] $\bigcup_{1\leq i \leq n} P_i  \sqsubseteq_{\mathbb{L}}^{\text{md}} \bigcup_{1\leq i \leq m} Q_i$,

\item[2.]   For each $1\leq i \leq n$, there is a $j \leq m$ and a negation-homomorphism 
$h \colon Q_j \longrightarrow P_i$.

\end{itemize}
Moreover, the corresponding decision problem (i.e.\ $\bigcup_{1\leq i \leq n} P_i \sqsubseteq_{\mathbb{L}}^{\text{md}} \bigcup_{1\leq i \leq m} Q_i $?)  is NP-complete. 
\end{restatable}


\begin{proof}
The argument is similar to the proof of Theorem~\ref{dsitrlatticeneg}. For the complexity calculation, the NP lower bound follows from the hardness of the problem for CQNs.  Furthermore,  the containment problem for UCQNs is in $\text{NP}$. Given an instance of a problem $(\bigcup_{1\leq i \leq n} P_i, \bigcup_{1\leq i \leq m} Q_i)$ consisting of two queries, we wish to check whether for all $1\leq i \leq n$, there is a $j \leq m$ with a negation-homomorphism 
$h_{P_i} \colon Q_j \longrightarrow P_i$. Fixing $\mathbf I$,  we
can  guess a $Q_j$, a mapping $h_{P_i} \colon Q_j \longrightarrow P_i$ and  check in polynomial time whether $h_{P_i'}$  is a  negation-homomorphism. Since NP is closed under intersection, our problem is in NP.
\end{proof}

We next turn to dually model-defining ($\mathrm{dmd}$) interpretations over bounded distributive lattices. In contrast to model-defining interpretations, where checking each   disjunct  $P_i$ of the query on the left  directly is sufficient, dually model-defining semantics require considering every completion $P_i^c$ of each disjunct $P_i$. This additional requirement increases the computational complexity from $\mathrm{NP}$-completeness to $\Pi^p_2$-completeness under bounded arity, and coNEXP-completeness in the general case.

\begin{restatable}{Thm}{ucqnContainmentB} \label{dsitrlatticenegucqdual}
For UCQNs, $\bigcup_{1\leq i \leq n} P_i, \bigcup_{1\leq i \leq m} Q_i$ with the same tuple of free variables, and a bounded distributive lattice $\mathbb{L}$, where no disjunct $P_i$ contains any atom and its negation, the following are equivalent:
\begin{itemize}
\item[1.] $\bigcup_{1\leq i \leq n} P_i \sqsubseteq_{\mathbb{L}}^{\text{dmd}} \bigcup_{1\leq i \leq m} Q_i$,

\item[2.] For each $1\leq i \leq n$ and   every completion $P_i^c$ of $P_i$, there is a $j \leq m$ and a  negation-homomorphism 
$h_{P_i^c} \colon Q_j \longrightarrow P_i^c$.

\end{itemize}
Moreover, the corresponding decision problem (i.e.\ $\bigcup_{1\leq i \leq n} P_i \sqsubseteq_{\mathbb{L}}^{\text{dmd}} \bigcup_{1\leq i \leq m} Q_i $?) is 
\begin{itemize}
\item[(i)] $\Pi^p_2$-complete for queries over schemas with arity bound $k$, for every fixed $k\geq 2$, and
\item[(ii)] coNEXP-complete in general.  
\end{itemize}  
\end{restatable}


\begin{proof}
Similar to the proof of Theorem~\ref{dualmdinterdistrlattice}. For the complexity calculation, the lower bound follows from the hardness of the same problem for CQNs. Furthermore, for the upper bound,  the complement  of the containment problem for UCQNs is in $\text{NP}^{\text{coNP}} = \text{NP}^{\text{NP}}=\Sigma^p_2$ if there is a fixed arity bound on the relations used while it is in NEXP otherwise. Given an instance of the problem $(\bigcup_{1\leq i \leq n} P_i, \bigcup_{1\leq i \leq m} Q_i)$
consisting of two UCQNs, we wish to check
whether for some $1\leq i \leq n$ and   some completion $P_i^c$ of $P_i$, there is no $1 \leq j \leq m$ with a   negation-homomorphism 
$h_{P_i'} \colon Q_j \longrightarrow P_i'$.
We
can  guess a $P_i$ and a completion $P_i^c$. The size of $P_i^c$ is generally exponentially bounded in the maximum arity $k$ of the relations appearing in $P_i$ since $P_i^c$ adds at most $n^{k-(k-1)} + \ldots  +n^{k-1}+n^k$ new literals where $n$ is the number of variables in $P_i$ but  if we fix $k$,  we get a polynomial bound in $n$. Next, we can  check with the $\text{coNP}$ oracle that  there is  no $1\leq j \leq m$ with a   negation-homomorphism 
    $h_{P_i^c} \colon Q_j \longrightarrow P_i^c$. When there is no fixed $k$, after guessing $P_i^c$ (now of exponential size), we can iterate the process of checking that a mapping is not a negation-homomorphism over all $n^m$ mappings from $Q_j$ to $P_i^c$ where $m$ is the number of variables in $Q_j$. The latter is doable in exponential time, and thus the whole process is in NEXP. 
\end{proof}

Having analyzed lattice semantics under model-defining and dually model-defining interpretations, we now turn to provenance-tracking ($\text{pt}$) interpretations over polynomial semirings, starting with the semiring of absorptive polynomials $\mathrm{Sorp}(X, \overline{X})$. In contrast to dually model-defining interpretations, which need evaluating query completions, provenance-tracking semantics allow each  disjunct $P_i$ to be checked directly against target disjuncts $Q_j$. This brings the decision complexity back down to $\mathrm{NP}$-completeness via literal-injective negation-homomorphisms.

\begin{restatable}{Thm}{ucqnContainmentC}\label{sorpchnegucq} For UCQNs, $\bigcup_{1\leq i \leq n} P_i, \bigcup_{1\leq i \leq m} Q_i$ with the same tuple of free variables, the following are equivalent:
\begin{itemize}
\item[1.] $\bigcup_{1\leq i \leq n} P_i  \sqsubseteq^{\text{pt}}_{\mathrm{Sorp}(X, \overline{X})} \bigcup_{1\leq i \leq m} Q_i$,
\item[2.]
For each $1\leq i \leq n$, we have that  $\pi^{\mathrm{can}_{\mathrm{Sorp}(X)}(P_i)}\llbracket P_i \rrbracket \leq_{\mathrm{Sorp}(X, \overline{X})} \pi^{\mathrm{can}_{\mathrm{Sorp}(X)}(P_i)}\llbracket  \bigcup_{1\leq i \leq m} Q_i \rrbracket $,
\item[3.]  For each $1\leq i \leq n$, there is a $j \leq m$ and a negation-homomorphism 
$h \colon Q_j \longrightarrow P_i$  injective for literals.

\end{itemize}
Moreover, the corresponding decision problem (i.e.\ $\bigcup_{1\leq i \leq n} P_i \sqsubseteq_{\mathrm{Sorp}(X, \overline{X})}^{\text{pt}} \bigcup_{1\leq i \leq m} Q_i $?)  is NP-complete. 
\end{restatable}


\begin{proof}
$(1) \implies (2):$ Obvious. 

$(2) \implies (3):$ Assume that (2) holds, so for each $1\leq i \leq n$, we have that  
$$
\pi^{\mathrm{can}_{\mathrm{Sorp}(X)}(P_i)}\llbracket P_i(\vec{u}) \rrbracket \leq_{\mathrm{Sorp}(X, \overline{X})} \pi^{\mathrm{can}_{\mathrm{Sorp}(X)}(P_i)}\llbracket  (\bigcup_{1\leq i \leq m} Q_i) (\vec{u}) \rrbracket.
$$

The polynomial $\pi^{\mathrm{can}_{\mathrm{Sorp}(X)}(P_i)}\llbracket P_i(\vec{u}) \rrbracket$ has as a term a monomial $x_1 \cdot \ldots \cdot x_n $ containing every tag from $X \cup \overline{X}$ used to annotate literals by $\pi^{\mathrm{can}_{\mathrm{Sorp}(X)}(P_i)}$, with some exponent, accounting for duplicate atoms (such monomial is obtained using the identity assignment for the variables in $ P_i$. Then, by the definition of $\leq $ for $\mathrm{Sorp}(X, \overline{X})$, we must have that there is a monomial $y_1 \cdot \ldots \cdot y_m $  in $\pi^{\mathrm{can}_{\mathrm{Sorp}(X)}(P_i)}\llbracket  (\bigcup_{1\leq i \leq m} Q_i)(\vec{u}) \rrbracket$ where only tags from $x_1 \cdot \ldots \cdot x_n $ appear, with equal or smaller exponents as in $x_1 \cdot \ldots \cdot x_n $. Since $\bigcup_{1\leq i \leq m} Q_i$ is a  UCQN, we must have that there is a $Q_j$ with $j\leq m$ and an assignment $v$ to the bound variables in $Q_j$ 
such that the body (i.e.\ the quantifier-free matrix) of $Q_j$ evaluates to the monomial $y_1 \cdot \ldots \cdot y_m$. 
Then, we get the desired negation-homomorphism $h \colon Q_j \longrightarrow P_i$ injective for literals by letting $h(x)$ be the variable $v(x)$ (recall that the elements of the active domain of $\mathrm{can}_{\mathrm{Sorp}(X,\overline{X})}(P_i)$ are the variables of $P_i$).

To show the injectivity, notice that any tag $p$ or $\overline{p}$ that is the value of a literal of $Q_j$ according to $v$, appears in $y_1 \cdot \ldots \cdot y_m$ with at most the same multiplicity as in  $x_1 \cdot \ldots \cdot x_n $ (since this monomial corresponds to the body of $P_i$), and thus the multiset of images by $h$ of literals of $Q_j$ is bag-contained in the multiset of literals from $P_i$.
 
$(3) \implies (1):$    Assume that (3) holds and we want to show that $\bigcup_{1\leq i \leq n} P_i \sqsubseteq_{\mathrm{Sorp}(X, \overline{X})}  \bigcup_{1\leq i \leq m} Q_i$.   Thus, for each $1\leq i \leq n$, there is a $j \leq m$ and a negation-homomorphism $h \colon Q_j \longrightarrow P_i$ 
injective for literals. In order to establish $\bigcup_{1\leq i \leq n} P_i \sqsubseteq_{\mathrm{Sorp}(X, \overline{X})}  \bigcup_{1\leq i \leq m} Q_i$, consider an arbitrary $\mathrm{Sorp}(X, \overline{X})$-database instance $\mathbf I$, a provenance-tracking interpretation $\pi$ compatible with it and a tuple $\vec{t}$. We need to show that for any monomial in $\pi\llbracket ( \bigcup_{1\leq i \leq n} P_i) \rrbracket(\vec{t})$ there is a monomial  with  equal or smaller exponents in $\pi\llbracket (\bigcup_{1\leq i \leq m} Q_i)(\vec{t}) \rrbracket$. 
Take an assignment $v$ of elements to the variables of $P_i$ (where $i\leq n$) justifying a monomial $x_1  \cdot \ldots \cdot x_n $.
The assignment  $v \circ h $  justifies a monomial with smaller or equal exponents in 
$\pi\llbracket Q_j(\vec{t}) \rrbracket$. 
This is because for any literal $A(\overline{x})$ of $Q_j$ such that 
$A(h(\overline{x})) = B(\vec{z})$, 
where  $B(\vec{z})$ is a literal of  $P_i$, if 
$x_i = \pi\llbracket B(v(\vec{z})) \rrbracket$, then $x_i = \pi\llbracket A(v(h(\vec{z}))) \rrbracket $. 

By the injectivity of $h$, the images of the literals in the body of $Q_j$ (considered as a multiset) appear with the same multiplicity as in the body of $P_i$ considered as multiset, which entails that $x_i$ appears as the value of a literal of $Q_j$ with at most the same multiplicity as it appears in the monomial $x_1  \cdot \ldots \cdot x_n $ (since the latter corresponds to the body of $P_i$).

The complexity calculation follows by similar reasoning as Theorem~\ref{dsitrlatticenegucq}. For the lower bound we use Theorem~\ref{sorp(X,X)}.
\end{proof}

Moving from $\mathrm{Sorp}(X, \overline{X})$ to the extended Why-provenance semiring $\mathrm{Why}(X, \overline{X})$, containment for UCQNs is characterized by set-onto negation-homomorphisms over relational literals, while maintaining an $\mathrm{NP}$-completeness complexity bound.

\begin{restatable}{Thm}{ucqnContainmentD}\label{ucqequinegwhy}
For UCQNs, $\bigcup_{1\leq i\leq n}P_i,\bigcup_{1\leq j\leq m}Q_j$ with the same tuple of free variables, the following are equivalent:
\begin{itemize}

\item[1.] $ \bigcup_{1\leq i\leq n}P_i\sqsubseteq^{\text{pt}}_{\mathrm{Why}(X,\overline X)}\bigcup_{1\leq j\leq m}Q_j$.

\item[2.] For each $1\leq i\leq n$, we have
$$
\pi^{\text{can}_{\mathrm{Why}(X)}(P_i^+)}\llbracket P_i \rrbracket
\leq_{\mathrm{Why}(X,\overline X)}
\pi^{\text{can}_{\mathrm{Why}(X)}(P_i^+)}
\llbracket \bigcup_{1\leq j\leq m}Q_j \rrbracket.
$$

\item[3.] For each $1\leq i\leq n$, there is a $j\leq m$ and a negation-homomorphism $h_i\colon Q_j\longrightarrow P_i$ set-onto for relational literals, i.e. $h_i\bigl(\text{LitSet}(Q_j)\bigr)=\text{LitSet}(P_i).$

\end{itemize}
Moreover, the corresponding decision problem (i.e.\ $\bigcup_{1\leq i\leq n}P_i\sqsubseteq^{\text{pt}}_{\mathrm{Why}(X,\overline X)}\bigcup_{1\leq j\leq m}Q_j$?) is NP-complete.
\end{restatable}

\begin{proof}
$(1)\Rightarrow(2):$ Suppose that $\bigcup_{1\leq i\leq n}P_i\sqsubseteq^{\text{pt}}_{\mathrm{Why}(X,\overline X)}\bigcup_{1\leq j\leq m}Q_j$ and fix $1\leq i\leq n$. Since addition in $\mathrm{Why}(X,\overline X)$ is set-theoretic union, we clearly have $P_i\sqsubseteq^{\text{pt}}_{\mathrm{Why}(X,\overline X)}\bigcup_{1\leq k\leq n}P_k,$ and hence $P_i\sqsubseteq^{\text{pt}}_{\mathrm{Why}(X,\overline X)}\bigcup_{1\leq j\leq m}Q_j.$ The interpretation $\pi^{\text{can}_{\mathrm{Why}(X)}(P_i^+)}$ is provenance-tracking, and therefore the latter containment holds, in particular, on this interpretation. Thus $\pi^{\text{can}_{\mathrm{Why}(X)}(P_i^+)}\llbracket P_i \rrbracket
\leq_{\mathrm{Why}(X,\overline X)}
\pi^{\text{can}_{\mathrm{Why}(X)}(P_i^+)}
\llbracket \bigcup_{1\leq j\leq m}Q_j \rrbracket,$ as required.

$(2)\Rightarrow(3):$ Assume that item 2 holds and fix $1\leq i\leq n$. Let us abbreviate $\text{can}_{\mathrm{Why}(X)}(P_i^+)$ as $\mathbf I$ and $\pi^{\text{can}_{\mathrm{Why}(X)}(P_i^+)}$ as $\pi^{\mathbf I}$. Let $\text{LitSet}(P_i)=\{L_1,\ldots,L_r\}$. By construction of $\pi^{\mathbf I}$, each distinct relational literal $L_k$ is annotated by a different tag $x_k\in X\cup\overline X$, that is, $\pi^{\mathbf I}\llbracket L_k \rrbracket=\{\{x_k\}\}$. Repeated occurrences of the same relational literal receive the same annotation, since they correspond to the same ground relational literal in the canonical interpretation. Moreover, by the definition of the canonical interpretation, no pair of complementary tags occurs among $x_1,\ldots,x_r$: the tags used for negative literals are chosen so that their complementary positive tags are not used in the canonical database. Hence the witness $W_{P_i}:=\{x_1,\ldots,x_r\}$ contains no complementary pair and is therefore not reduced to $\emptyset$ in the quotient defining $\mathrm{Why}(X,\overline X)$.

Consider now the assignment to the bound variables of $P_i$ that sends every variable to itself in $\mathbf I$. Under this assignment, every occurrence of a literal $L_k$ has value $\{\{x_k\}\}$. Since multiplication in $\mathrm{Why}(X,\overline X)$ is pairwise union and $W_{P_i}$ contains no complementary pair, the contribution of this assignment contains the witness $W_{P_i}$. Notice that repeated occurrences of the same literal do not change this value, since they contribute the same tag. Therefore, $W_{P_i}\in\pi^{\mathbf I}\llbracket P_i(\vec{u}) \rrbracket.$ By item 2 and the fact that the natural order on $\mathrm{Why}(X,\overline X)$ is set inclusion, we have $W_{P_i}\in\pi^{\mathbf I}\llbracket \left(\bigcup_{1\leq j\leq m}Q_j\right)(\vec{u}) \rrbracket.$ Since addition in $\mathrm{Why}(X,\overline X)$ is set-theoretic union,$\pi^{\mathbf I}\llbracket \left(\bigcup_{1\leq j\leq m}Q_j\right)(\vec{u}) \rrbracket = \bigcup_{1\leq j\leq m}\pi^{\mathbf I}\llbracket Q_j(\vec{u}) \rrbracket.$ Hence there is some $1 \leq j\leq m$ such that $W_{P_i}\in\pi^{\mathbf I}\llbracket Q_j(\vec{u}) \rrbracket.$

By the semantics of CQNs, there is an assignment $v$ of elements of $\mathbf I$ to the bound variables of $Q_j$, extending the identity assignment on $\vec{u}$, whose contribution to $\pi^{\mathbf I}\llbracket Q_j(\vec{u}) \rrbracket$ contains $W_{P_i}$. Every relational literal of $Q_j$ must take a non-zero value under $v$, since otherwise the product corresponding to this assignment would be $0$. By construction of $\pi^{\mathbf I}$, a ground relational literal has non-zero value only if it is one of the relational literals occurring in $P_i$. Hence, by letting $h_i(x):=v(x)$ for every variable $x$ of $Q_j$, we obtain a negation-homomorphism $h_i\colon Q_j\longrightarrow P_i$. Indeed, $h_i$ preserves the free variables and, for every relational literal $L(\vec{x})$ occurring in $Q_j$, $L(h_i(\overline{x}))$ occurs in the body of $P_i$.

It remains to show that $h_i$ is set-onto for relational literals. Since every non-zero annotation in $\pi^{\mathbf I}$ is a singleton, the product associated with the assignment $v$ is obtained by taking pairwise unions of singleton outer sets. Hence, before passing to the quotient, it has the form $\{W_v\}$, where $W_v$ is the union of all the tags contributed by the relational literals of $Q_j$ under $v$. If $W_v$ contains a complementary pair, this unique witness is removed by the quotient defining $\mathrm{Why}(X,\overline X)$ and the contribution of $v$ is therefore $\emptyset$. Otherwise, the contribution of $v$ is precisely $\{W_v\}$. We already know that $W_{P_i}$ belongs to the contribution of $v$. Hence this contribution cannot be $\emptyset$, and must therefore be $\{W_v\}$. Since $\{W_v\}$ contains only one witness and $W_{P_i}\in\{W_v\}$, it follows that $W_v=W_{P_i}$. Thus the union of the tags contributed by the relational literals of $Q_j$ under $v$ is exactly $W_{P_i}$. Therefore, for every $1\leq k\leq r$, the tag $x_k$ must occur as the annotation of at least one relational literal of $Q_j$ under $v$. Since $x_k$ occurs in the annotation of no other canonical relational literal, for each $L_k\in\text{LitSet}(P_i)$ there is a relational literal $A$ of $Q_j$ such that $h_i(A)=L_k$. Hence $\text{LitSet}(P_i)\subseteq h_i(\text{LitSet}(Q_j)).$ The reverse inclusion follows from the fact that $h_i$ is a negation-homomorphism. Thus $h_i(\text{LitSet}(Q_j))=\text{LitSet}(P_i),$ and $h_i$ is set-onto for relational literals.

$(3)\Rightarrow(1):$ Assume that item 3 holds. Consider an arbitrary $\mathrm{Why}(X)$-database instance $\mathbf I$, a provenance-tracking interpretation $\pi$ compatible with $\mathbf I$, and a tuple $\vec{t}$. We show that $\pi\llbracket \left(\bigcup_{1\leq i\leq n}P_i\right)(\vec{t}) \rrbracket \subseteq \pi\llbracket \left(\bigcup_{1\leq j\leq m}Q_j\right)(\vec{t}) \rrbracket.$
Let $W\in\pi\llbracket \left(\bigcup_{1\leq i\leq n}P_i\right)(\vec{t}) \rrbracket.$ Since addition in $\mathrm{Why}(X,\overline X)$ is set-theoretic union, there is some $1\leq i\leq n$ such that $W\in\pi\llbracket P_i(\vec{t}) \rrbracket$. Write $P_i(\vec{z}):=\exists\vec{u}\,\bigl(B_1(\vec{z}_1)\wedge\ldots\wedge B_r(\vec{z}_r)\bigr).$ Then there is an assignment $v$ to the variables of $P_i$, producing $\vec{t}$ on the free variables, such that $W$ belongs to the contribution of $v$ to $\pi\llbracket P_i(\vec{t}) \rrbracket$. Since this contribution is non-zero, every relational literal of $P_i$ takes a non-zero value under $v$. Since $\pi$ is provenance-tracking, for every ground relational literal $G$ such that $\pi\llbracket G \rrbracket\neq 0$, there is a unique finite set $S_G\subseteq X\cup\overline X$ such that $\pi\llbracket G \rrbracket=\{S_G\}$, where $S_G=\emptyset$ when $\pi\llbracket G \rrbracket=1$, and $S_G$ is a singleton token set otherwise. Let us write $s_\pi\llbracket G \rrbracket:=S_G$.

By the definition of multiplication in $\mathrm{Why}(X,\overline X)$, before passing to the quotient the contribution of $v$ is the singleton whose unique witness is $\bigcup_{1\leq k\leq r}s_\pi\llbracket B_k(v(\vec{z}_k)) \rrbracket.$ Since the contribution of $v$ contains $W$, it is non-zero, and hence this union contains no pair of complementary tokens. Therefore,  $W=\bigcup_{1\leq k\leq r}s_\pi\llbracket B_k(v(\vec{z}_k)) \rrbracket.$ Repeated occurrences of the same relational literal do not change this union, and hence $W=\bigcup_{L\in\text{LitSet}(P_i)}s_\pi\llbracket v(L) \rrbracket.$

By item 3, there is a $j\leq m$ and a set-onto negation-homomorphism $h_i\colon Q_j\longrightarrow P_i$ such that $h_i(\text{LitSet}(Q_j))=\text{LitSet}(P_i).$ Write $Q_j(\overline{x}):=\exists\vec{z}\,\bigl(A_1(\vec{x}_1)\wedge\ldots\wedge A_s(\overline{x}_s)\bigr).$

Consider now the assignment $v\circ h_i$ to the variables of $Q_j$. Since $h_i$ preserves the tuple of free variables, $v\circ h_i$ also produces $\vec{t}$. Furthermore, for every relational literal $A$ of $Q_j$, $h_i(A)$ is a relational literal occurring in $P_i$, and therefore $\pi\llbracket v(h_i(A)) \rrbracket\neq 0$. Since $h_i(\text{LitSet}(Q_j))=\text{LitSet}(P_i)$, applying the assignment $v$ to both sides gives $\bigl\{v(h_i(A)):A\in\text{LitSet}(Q_j)\bigr\}=\bigl\{v(L):L\in\text{LitSet}(P_i)\bigr\}.$ Notice that distinct relational literals may become the same ground relational literal under $v$, but this does not affect the equality, since we are considering ordinary sets and not multisets. Applying $s_\pi$ to the ground relational literals and taking the union of the members of these sets yields $\bigcup_{A\in\text{LitSet}(Q_j)}s_\pi\llbracket v(h_i(A)) \rrbracket=\bigcup_{L\in\text{LitSet}(P_i)}s_\pi\llbracket v(L) \rrbracket=W.$ Since $W$ is the unique witness of the non-zero contribution of $v$ to $P_i$, it contains no pair of complementary tokens. Hence the same union obtained from the relational literals of $Q_j$ under $v\circ h_i$ contains no complementary pair and its corresponding witness is preserved in the quotient defining $\mathrm{Why}(X,\overline X)$. Therefore, the assignment $v\circ h_i$ contributes $W$ to $\pi\llbracket Q_j(\vec{t}) \rrbracket$, and consequently $W\in\pi\llbracket Q_j(\vec{t}) \rrbracket\subseteq
\pi\llbracket \left(\bigcup_{1\leq j\leq m}Q_j\right)(\vec{t}) \rrbracket.$ Since $\mathbf I$, $\pi$, $\vec{t}$, and $W\in\pi\llbracket (\bigcup_{1\leq i\leq n}P_i)(\vec{t}) \rrbracket$ were arbitrary, we conclude that $\bigcup_{1\leq i\leq n}P_i
\sqsubseteq^{\text{pt}}_{\mathrm{Why}(X,\overline X)}
\bigcup_{1\leq j\leq m}Q_j.$

Finally, for the complexity calculation, the NP lower bound follows from the hardness of the containment problem for CQNs over provenance-tracking interpretations from Theorem~\ref{whyneg}. For the upper bound, by item 3, for each $1\leq i\leq n$ we can guess a $j\leq m$ and a mapping $h_i\colon Q_j\longrightarrow P_i$, and check in polynomial time that $h_i$ is a negation-homomorphism and that $h_i(\text{LitSet}(Q_j))=\text{LitSet}(P_i)$. Since there are only polynomially many disjuncts $P_i$, all these mappings can be guessed and checked in polynomial time. Hence the containment problem is in NP, and therefore it is NP-complete.
\end{proof}

We continue by extending our UCQN characterization to $\mathrm{PosBool}(X, \overline{X})$, where lattice absorption eliminates structural restrictions on the negation-homomorphisms while maintaining $\mathrm{NP}$-completeness.

\begin{restatable}{Thm}{ucqnContainmentF}\label{posboolucqnneg}
For UCQNs $\bigcup_{1\leq i \leq n} P_i$ and 
$\bigcup_{1\leq j \leq m} Q_j$ with the same tuple of free variables 
$\vec{u}$, the following are equivalent:
\begin{itemize}
\item[1.] 
$\bigcup_{1\leq i \leq n} P_i
\sqsubseteq^{\text{pt}}_{\mathrm{PosBool}(X,\overline{X})}
\bigcup_{1\leq j \leq m} Q_j$,

\item[2.]
For each $1\leq i \leq n$, we have $\pi^{\mathrm{can}_{\mathrm{PosBool}(X)}(P_i)}\llbracket P_i(\vec{u}) \rrbracket
\leq_{\mathrm{PosBool}(X,\overline{X})}
\pi^{\mathrm{can}_{\mathrm{PosBool}(X)}(P_i)}
\llbracket 
\left(\bigcup_{1\leq j \leq m} Q_j\right)(\vec{u})
 \rrbracket.$

\item[3.]
For each $1\leq i \leq n$, there is a $j\leq m$ and a negation-homomorphism $h\colon Q_j \longrightarrow P_i.$
\end{itemize}
Moreover, the corresponding decision problem (i.e.\ $\bigcup_{1\leq i \leq n} P_i \sqsubseteq_{\mathrm{PosBool}(X,\overline{X})}^{\text{pt}} \bigcup_{1\leq i \leq m} Q_i $?)  is NP-complete. 
\end{restatable}


\begin{proof}
$(1)\Rightarrow(2)$: Obvious.

$(2)\Rightarrow(3)$: Assume that (2) holds. Then, for each $1\leq i\leq n$, we have that
\[
\pi^{\mathrm{can}_{\mathrm{PosBool}(X)}(P_i)}\llbracket P_i(\vec{u}) \rrbracket
\leq_{\mathrm{PosBool}(X,\overline{X})}
\pi^{\mathrm{can}_{\mathrm{PosBool}(X)}(P_i)}
\llbracket 
\left(\bigcup_{1\leq j \leq m} Q_j\right)(\vec{u})
 \rrbracket,
\]
where $\vec{u}$ is the tuple of free variables of $P_i$. We abbreviate $\mathrm{can}_{\mathrm{PosBool}(X)}(P_i)$ by $I_i$ in what follows. Consider $P_i(\vec{z})=
\exists \vec{u}\,
(B_1(\vec{z_1})\wedge \ldots \wedge B_r(\vec{z_r}))$ and the assignment on the variables of $P_i$ that sends every variable of $P_i$ to itself in $I_i$. Under this assignment, the body of $P_i$ takes the value $\tau_{P_i}:=
\pi^{I_i}\llbracket B_1(\vec{z_1}) \rrbracket\wedge \ldots \wedge
\pi^{I_i}\llbracket B_r(\vec{z_r}) \rrbracket.$ Thus, $\tau_{P_i}$ is one of the disjuncts 
occurring in $\pi^{I_i}\llbracket P_i(\vec{z}) \rrbracket$. By construction of $\pi^{I_i}$, each conjunct in $\tau_{P_i}$ is either a tag from $X$ or a tag from $\overline{X}$, and no complementary pair $x,\overline{x}$ can occur simultaneously in $\tau_{P_i}$. So we can define an admissible valuation $\nu_{\tau_{P_i}}\colon X\cup \overline{X} \longrightarrow \{0,1\}$ such that $\nu_{\tau_{P_i}}(t)=1$ only for the tags that occur in $\tau_{P_i}$, and $\nu_{\tau_{P_i}}(t)=0$ for all others. By construction $\nu_{\tau_{P_i}}(\tau_{P_i})=1$ and since $\tau_{P_i}$ is one of the disjuncts of $\pi^{I_i}\llbracket P_i(\vec{z}) \rrbracket$, we get $\nu_{\tau_{P_i}}\bigl(\pi^{I_i}\llbracket P_i(\vec{z}) \rrbracket\bigr)=1.$ 
Now, by the natural order on $\mathrm{PosBool}(X,\overline{X})$, from the hypothesis $\pi^{I_i}\llbracket P_i(\vec{z}) \rrbracket \leq_{\mathrm{PosBool} (X,\overline{X})} \pi^{I_i} \llbracket  \left(\bigcup_{1\leq j \leq m} Q_j\right)(\vec{z})  \rrbracket$ we obtain $\nu_{\tau_{P_i}}
\left(\pi^{I_i}\llbracket \left(\bigcup_{1\leq j \leq m} Q_j\right)(\vec{z}) \rrbracket\right)=1$ too. The value $\pi^{I_i}\llbracket \left(\bigcup_{1\leq j \leq m} Q_j\right)(\vec{z}) \rrbracket$ is the join of the values of the literals $Q_j$ under all assignments of their bound variables into $I_i$ extending the identity on the free variables
$\vec{z}$ so there exist $j\leq m$ and an assignment $f$ to the bound variables of $Q_j$ such that, if $Q_j(\overline{x})= \exists \vec{z}\, (A_1(\vec{x_1})\wedge \ldots \wedge A_s(\overline{x_s}))$ then $\nu_{\tau_{P_i}}\Bigl(\pi^{I_i}\llbracket A_1(f(\overline{x_1})) \rrbracket\wedge \ldots \wedge \pi^{I_i}\llbracket A_s(f(\overline{x_s})) \rrbracket\Bigr)=1.$ 
That means that for every $1\leq \ell \leq s$, we have $\nu_{\tau_{P_i}}\bigl(\pi^{I_i}\llbracket A_\ell(f(\overline{x_\ell})) \rrbracket\bigr)=1.$ But  by construction of $\pi^{I_i}$, a literal takes a value different from $[\bot]$ only if it occurs in the body of $P_i$: it follows that, for every $1\leq \ell \leq s$, the literal $A_\ell(f(\overline{x_\ell}))$ occurs in the body of $P_i$. We can now define a map $h$ from the variables of $Q_j$ to the ones of $P_i$ by letting $h(x):=f(x)$ for every variable $x$ of $Q_j$. Since $f$ extends the identity on the free variables $\overline{x}$, the map $h$ preserves the head variables. Moreover, for every literal $A_\ell(\vec{x_\ell})$ of $Q_j$, the literal $A_\ell(h(\overline{x_\ell}))=A_\ell(f(\overline{x_\ell}))$ occurs in the body of $P_i$. Therefore, $h$ is a negation-homomorphism $h\colon Q_j \longrightarrow P_i$. Since $\mathbf I$ was arbitrary, this proves (3).

$(3)\Rightarrow(1)$: Assume that (3) holds and we want to show that $\bigcup_{1\leq i \leq n} P_i \sqsubseteq^{\text{pt}}_{\mathrm{PosBool}(X,\overline{X})} \bigcup_{1\leq j \leq m} Q_j.$ Take an arbitrary $\mathrm{PosBool}(X)$-database instance $\mathbf I$, an arbitrary provenance-tracking interpretation $\pi$ compatible with it, and an arbitrary tuple $\vec{t}$. We need to show that $\pi\llbracket \left(\bigcup_{1\leq i \leq n} P_i\right)(\vec{t}) \rrbracket\leq_{\mathrm{PosBool}(X,\overline{X})}\pi\llbracket \left(\bigcup_{1\leq j \leq m} Q_j\right)(\vec{t}) \rrbracket.$ Fix $\mathbf I$ and take $P_i(\vec{z})=\exists \vec{u}\,(B_1(\vec{z_1})\wedge \ldots \wedge B_r(\vec{z_r})).$ Let $f$ be an assignment of elements of $\mathbf I$ to the bound variables of $P_i$ extending $\vec{z}\mapsto \vec{t}$, and consider the corresponding value $\pi\llbracket B_1(f(\vec{z_1})) \rrbracket\wedge \ldots \wedge
\pi\llbracket B_r(f(\vec{z_r})) \rrbracket.$ Then, this value is one of the disjuncts occurring in $\pi\llbracket P_i(\vec{t}) \rrbracket$. By assumption, there is a $j\leq m$ and a negation-homomorphism $h\colon Q_j \longrightarrow P_i.$ Let $ Q_j(\overline{x})=
\exists \vec{z}\, (A_1(\vec{x_1})\wedge \ldots \wedge A_s(\overline{x_s})).$ Then, $f\circ h$ is an assignment of elements of $\mathbf I$ to the bound variables of $Q_j$
extending $\overline{x}\mapsto \vec{t}$. So, $
\pi\llbracket A_1((f\circ h)(\overline{x_1})) \rrbracket\wedge \ldots \wedge
\pi\llbracket A_s((f\circ h)(\overline{x_s})) \rrbracket$ is one of the disjuncts occurring in $\pi\llbracket Q_j(\vec{t}) \rrbracket$. Since $h$ is a negation-homomorphism, for every literal $A_\ell(\vec{x_\ell})$ of $Q_j$, the literal $A_\ell(h(\overline{x_\ell}))$ occurs in the body of $P_i$. Therefore, every conjunct occurring in $
\pi\llbracket A_1((f\circ h)(\overline{x_1})) \rrbracket\wedge \ldots \wedge
\pi\llbracket A_s((f\circ h)(\overline{x_s})) \rrbracket$ also
occurs in a conjuct in $\pi\llbracket B_1(f(\vec{z_1})) \rrbracket\wedge \ldots \wedge \pi\llbracket B_r(f(\vec{z_r})) \rrbracket$. 
So we can write $
\pi\llbracket B_1(f(\vec{z_1})) \rrbracket\wedge \ldots \wedge \pi\llbracket B_r(f(\vec{z_r})) \rrbracket=\pi\llbracket A_1((f\circ h)(\overline{x_1})) \rrbracket\wedge \ldots \wedge
\pi\llbracket A_s((f\circ h)(\overline{x_s})) \rrbracket\wedge \rho$ for some conjunction $\rho$ of values of literals of $P_i$. Since $\mathrm{PosBool}(X,\overline{X})$ is a bounded distributive lattice, we have that $\pi\llbracket A_1((f\circ h)(\overline{x_1})) \rrbracket\wedge \ldots \wedge
\pi\llbracket A_s((f\circ h)(\overline{x_s})) \rrbracket\wedge \rho \leq_{\mathrm{PosBool}(X,\overline{X})} \pi\llbracket A_1((f\circ h)(\overline{x_1})) \rrbracket\wedge \ldots \wedge
\pi\llbracket A_s((f\circ h)(\overline{x_s})) \rrbracket.$ So $
\pi\llbracket B_1(f(\vec{z_1})) \rrbracket\wedge \ldots \wedge \pi\llbracket B_r(f(\vec{z_r})) \rrbracket \leq_{\mathrm{PosBool}(X,\overline{X})} \pi\llbracket A_1((f\circ h)(\overline{x_1})) \rrbracket\wedge \ldots \wedge
\pi\llbracket A_s((f\circ h)(\overline{x_s})) \rrbracket$ and since $\pi\llbracket A_1((f\circ h)(\overline{x_1})) \rrbracket\wedge \ldots \wedge
\pi\llbracket A_s((f\circ h)(\overline{x_s})) \rrbracket$ is one of the disjuncts occurring in $\pi\llbracket Q_j(\vec{t}) \rrbracket$,
and $Q_j$ is one of the disjuncts of $\bigcup_{1\leq j\leq m}Q_j$, we have $\pi\llbracket A_1((f\circ h)(\overline{x_1})) \rrbracket\wedge \ldots \wedge\pi\llbracket A_s((f\circ h)(\overline{x_s})) \rrbracket\leq_{\mathrm{PosBool}(X,\overline{X})}\pi\llbracket \left(\bigcup_{1\leq j \leq m} Q_j\right)(\vec{t}) \rrbracket.$ By transitivity, $\pi\llbracket B_1(f(\vec{z_1})) \rrbracket\wedge \ldots \wedge\pi\llbracket B_r(f(\vec{z_r})) \rrbracket\leq_{\mathrm{PosBool}(X,\overline{X})}\pi\llbracket (\bigcup_{1\leq j \leq m} Q_j)\vec{t} \rrbracket )$ This holds for every $1\leq i\leq n$ and for every assignment $f$ contributing to $\pi\llbracket P_i(\vec{t}) \rrbracket$, taking the join over all such disjuncts gives $\pi\llbracket \left(\bigcup_{1\leq i \leq n} P_i\right)(\vec{t})
 \rrbracket \leq_{\mathrm{PosBool}(X,\overline{X})}\pi\llbracket  \left(\bigcup_{1\leq j \leq m} Q_j\right)(\vec{t})  \rrbracket.$ Therefore, $\bigcup_{1\leq i \leq n} P_i \sqsubseteq^{\text{pt}}_{\mathrm{PosBool}(X,\overline{X})} \bigcup_{1\leq j \leq m} Q_j. $

The complexity calculation follows by similar reasoning as Theorem~\ref{dsitrlatticenegucq}. For the lower bound we use Theorem~\ref{posboolneg}.

\end{proof}

As in the case of CQNs, we consider UCQNs over $\mathrm{Which}(X)$ under model-defining interpretations rather than provenance-tracking interpretations, for the reasons discussed above. We obtain the following characterization of containment.

\begin{restatable}{Thm}{ucqnContainmentG}\label{ucqnegwhichmd}
For UCQNs, $\bigcup_{1\leq i \leq n} P_i, \bigcup_{1\leq j \leq m} Q_j$ with the same tuple of free variables, the following are equivalent:
\begin{itemize}
\item[1.] $\bigcup_{1\leq i \leq n} P_i  \sqsubseteq_{\mathrm{Which}(X)}^{\text{md}} \bigcup_{1\leq j \leq m} Q_j$,

\item[2.]
For each $1\leq i \leq n$ and every completion $P_i^c$ of $P_i$, we have that
\[
\pi^{P_i^c}_{\mathrm{Which}(X)}(P_i)
\leq_{\mathrm{Which}(X)}
\pi^{P_i^c}_{\mathrm{Which}(X)}\left(\bigcup_{1\leq j \leq m} Q_j\right).
\]

\item[3.]
For each $1\leq i \leq n$, every completion $P_i^c$ of $P_i$, and every literal $L$ in the body of $P_i$, there is a $j\leq m$ and a negation-homomorphism
\[
h_{P_i^c,L}\colon Q_j \longrightarrow P_i^c
\]
such that $L$ belongs to the range of $h_{P_i^c,L}$.
\end{itemize}
Moreover, the corresponding decision problem (i.e.\ $\bigcup_{1\leq i \leq n} P_i \sqsubseteq_{\mathrm{Which}(X)}^{\text{md}} \bigcup_{1\leq i \leq m} Q_i $?) is $\Pi^p_2$-complete for queries over schemas with a fixed arity bound $k\geq 2$, and coNEXP-complete in general.
\end{restatable}


\begin{proof}
$(1) \implies (2):$ Obvious.

$(2) \implies (3):$ Assume that (2) holds. We want to show that for each $1\leq i \leq n$, every completion $P_i^c$ of $P_i$, and every literal $L$ in the body of $P_i$, there is a $j\leq m$ and a negation-homomorphism $h_{P_i^c,L}\colon Q_j \longrightarrow P_i^c$ such that $L$ belongs to the image of $h_{P_i^c,L}$. Let $P_i^c$ be an arbitrary completion of $P_i$, and let $L$ be an arbitrary literal in the body of $P_i$. By definition of the interpretation $\pi^{P_i^c}_{\mathrm{Which}(X)}$, we have that $\pi^{P_i^c}_{\mathrm{Which}(X)}(L)=\{x_L\}$, where $x_L$ is the fresh variable of $X$ associated with the literal $L$. Moreover, considering the identity assignment on the variables of $P_i$, we obtain that the value of the body of $P_i$ contains all the variables associated with the literals of $P_i$. In particular, $x_L\in \pi^{P_i^c}_{\mathrm{Which}(X)}(P_i(\vec{u}))$. By hypothesis, we have that
\[
\pi^{P_i^c}_{\mathrm{Which}(X)}(P_i(\vec{u}))
\leq_{\mathrm{Which}(X)}
\pi^{P_i^c}_{\mathrm{Which}(X)}
\left(\left(\bigcup_{1\leq j \leq m} Q_j\right)(\vec{u})\right).
\]
Since the natural order on Which$(X)$ is the inclusion on $\wp(X)$, with the exception that $\bot \leq A$ for every $A\subseteq X$, it follows that $x_L\in \pi^{P_i^c}_{\mathrm{Which}(X)} \left(\left(\bigcup_{1\leq j \leq m} Q_j\right)(\vec{u})\right).$ Recall that the value of a union of queries is obtained by the sum of the values of its disjuncts, and that the sum in Which$(X)$ is given by union. Hence, there is some $1 \leq j\leq m$ such that $x_L\in \pi^{P_i^c}_{\mathrm{Which}(X)}(Q_j(\vec{u}))$. Let $Q_j(\overline{x})= \exists \vec{z}\, (A_1(\vec{x_1})\wedge \ldots \wedge A_R(\vec{x_r})).$ Then, since $\pi^{P_i^c}_{\mathrm{Which}(X)}(Q_j(\vec{u}))$ is obtained as the sum, in Which$(X)$, of the values of the body of $Q_j$ over all assignments sending the free variables $\overline{x}$ to $\vec{u}$, and since this sum is given by union, there must be an assignment $v$ from the variables of $Q_j$ to the variables of $P_i$ such that $v(\overline{x})=\vec{u}$ and $x_L\in \pi^{P_i^c}_{\mathrm{Which}(X)} A_1(v(\overline{x_1})))\cdot \ldots \cdot \pi^{P_i^c}_{\mathrm{Which}(X)}(A_r(v(\overline{x_r}))).$ Multiplication in Which$(X)$ gives $\bot$ whenever one of the factors is $\bot$, so it follows that, for every $s\leq r$, $\pi^{P_i^c}_{\mathrm{Which}(X)}(A_s(v(\overline{x_s})))\neq \bot.$ But by construction of $\pi^{P_i^c}_{\mathrm{Which}(X)}$, a literal has value different from $\bot$ only if it belongs to the body of $P_i^c$. Hence, for every $s\leq r$, the literal $A_s(v(\overline{x_s}))$ belongs to the body of $P_i^c$. Moreover, the variable $x_L$ can occur only in the value of the literal $L$. Therefore, there is some $s\leq r$ such that $A_s(v(\overline{x_s}))=L.$ We now define $h_{P_i^c,L}$ by letting $h_{P_i^c,L}(x)=v(x)$ for every variable $x$ of $Q_j$. Since the active domain of $P_i^c$ consists of the variables of $P_i$, this is a mapping from the variables of $Q_j$ to the variables of $P_i$. Also, since the two UCQNs have the same tuple of free variables and we are evaluating $Q_j$ on the tuple $\vec{u}$, the mapping sends the free variables of $Q_j$ to the corresponding free variables of $P_i$. Moreover, for every literal $A_s(\overline{x_s})$ of $Q_j$, we have shown that $A_s(h_{P_i^c,L}(\overline{x_s})) = A_s(v(\overline{x_s}))$ belongs to the body of $P_i^c$. Thus, $h_{P_i^c,L}$ is a negation-homomorphism from $Q_j$ to $P_i^c$. Since for some $s\leq r$ we have $A_s(h_{P_i^c,L}(\overline{x_s}))=L,$ the literal $L$ belongs to the image of $h_{P_i^c,L}$, as required.

$(3) \implies (1):$ Assume that (3) holds and we want to show that $\bigcup_{1\leq i \leq n} P_i \sqsubseteq_{\mathrm{Which}(X)}^{\text{md}} \bigcup_{1\leq j \leq m} Q_j.$ Take an arbitrary Which$(X)$-database instance $\mathbf I$, a model-defining interpretation $\pi$ compatible with $\mathbf I$, and a tuple $\vec{t}$. If the value of the left union is $\bot$, there is nothing to prove. Otherwise, choose a  disjunct $P_{i_0}$ and an assignment $v_0$ with a non-bottom contribution. Complete $P_{i_0}$ by choosing for each atom on its variables the unique literal with non-bottom value under $v_0$. By (3), applied to this completion and any literal of $P_{i_0}$, there are a target disjunct $Q_{j_0}$ and a negation-homomorphism into the completion. Composing with $v_0$ gives a non-bottom contribution to $Q_{j_0}(\vec t)$, so the value of the right union is non-bottom. It remains to show that, for every $x\in X$, if $x\in \pi\llbracket \left(\bigcup_{1\leq i \leq n} P_i\right)(\vec{t}) \rrbracket,$ then $x\in \pi\llbracket \left(\bigcup_{1\leq j \leq m} Q_j\right)(\vec{t}) \rrbracket.$ Let $P_i(\vec{z})= \exists \vec{u}\, (B_1(\vec{z_1})\wedge \ldots \wedge B_q(\vec{z_q}))$ be a disjunct of $P$. Since the value of a union is obtained as the sum of the values of its disjuncts, and since the sum in Which$(X)$ is given by union, if $x\in
\pi\llbracket \left(\bigcup_{1\leq i \leq n} P_i\right)(\vec{t}) \rrbracket,$ then there is some $i\leq n$ such that $x\in \pi\llbracket P_i(\vec{t}) \rrbracket.$ Thus, there must be an assignment $v$ of elements to the bound variables of $P_i$ such that $v(\vec{z})=\vec{t}$ and $x\in \pi\llbracket B_1(v(\vec{z_1})) \rrbracket\cdot \ldots \cdot \pi\llbracket B_q(v(\vec{z_q})) \rrbracket.$ Since multiplication in Which$(X)$ gives $\bot$ whenever one of the factors is $\bot$, we have that $\pi\llbracket B_s(v(\vec{z_s})) \rrbracket\neq \bot$ for every $s\leq q$. Moreover, since multiplication between elements different from $\bot$ is union, there is a literal $L$ in the body of $P_i$ such that $x\in \pi\llbracket L(v) \rrbracket.$ We now build a completion $P_i^c$ of $P_i$ depending on the assignment $v$ and the interpretation $\pi$. For every relation symbol $R$ of the common schema $\tau$ and every tuple $\vec{w}$ of variables of $P_i$ with length that matches the arity of $R$, we add to $P_i$ the literal $R(\vec{w})$ if $\pi\llbracket R(v(\vec{w})) \rrbracket\neq \bot$, and we add the literal $\neg R(\vec{w})$ if $\pi\llbracket \neg R(v(\vec{w})) \rrbracket\neq \bot.$ Since $\pi$ is model-defining, exactly one of these two values is different from $\bot$, so only one of the two literals is added to $P_i^c$. Moreover, if a literal already belonged to the body of $P_i$, then its value under $v$ is different from $\bot$, so all the literals of $P_i$ are also in $P_i^c$. Therefore, $P_i^c$ is a completion of $P_i$. By assumption, applied to $P_i^c$ and to the literal $L$, there is a $j\leq m$ and a negation-homomorphism $h_{P_i^c,L}\colon Q_j \longrightarrow P_i^c$ such that $L$ belongs to the image of $h_{P_i^c,L}$. Let $Q_j(\overline{x})= \exists \vec{z}\, (A_1(\vec{x_1})\wedge \ldots \wedge A_R(\vec{x_r})).$ But then $v\circ h_{P_i^c,L}$ is an assignment of elements to the bound variables of $Q_j$ such that $(v\circ h_{P_i^c,L})(\overline{x})=\vec{t}$. Moreover, for every $s\leq r$, the literal $A_s(h_{P_i^c,L}(\overline{x_s}))$ belongs to the body of $P_i^c$. By construction of $P_i^c$, this implies that $\pi\llbracket A_s(v(h_{P_i^c,L}(\overline{x_s}))) \rrbracket\neq \bot$ for every $s\leq r$. Finally, since $L$ belongs to the image of $h_{P_i^c,L}$, there is some $s\leq r$ such that $A_s(h_{P_i^c,L}(\overline{x_s}))=L.$ Applying $v$, we get $A_s(v(h_{P_i^c,L}(\overline{x_s})))=L(v).$ Since $x\in \pi\llbracket L(v) \rrbracket$, it follows that $x\in \pi\llbracket A_1(v(h_{P_i^c,L}(\overline{x_1}))) \rrbracket\cdot \ldots \cdot \pi\llbracket A_r(v(h_{P_i^c,L}(\overline{x_r}))) \rrbracket.$ Therefore, $x\in \pi\llbracket Q_j(\vec{t}) \rrbracket$. Since $Q_j$ is a disjunct of $\bigcup_{1\leq j \leq m} Q_j$, we obtain that $x\in
\pi\llbracket \left(\bigcup_{1\leq j \leq m} Q_j\right)(\vec{t}) \rrbracket$. Hence every $x\in \pi\llbracket (\bigcup_{1\leq i \leq n} P_i)(\vec{t}) \rrbracket$ also belongs to $\pi\llbracket \left(\bigcup_{1\leq j \leq m} Q_j\right)(\vec{t}) \rrbracket$. Thus, $\pi\llbracket \left(\bigcup_{1\leq i \leq n} P_i\right)(\vec{t}) \rrbracket \leq_{\mathrm{Which}(X)} \pi\llbracket \left(\bigcup_{1\leq j \leq m} Q_j\right)(\vec{t}) \rrbracket.$ Since $\mathbf I$, $\pi$, and $\vec{t}$ were arbitrary, we conclude that $\bigcup_{1\leq i \leq n} P_i
\sqsubseteq_{\mathrm{Which}(X)}^{\text{md}} \bigcup_{1\leq j \leq m} Q_j.$

For the complexity calculation, we might start with the upper bound by  observing that the complement  of the containment problem for UCQNs is in $\text{NP}^{\text{coNP}} = \text{NP}^{\text{NP}}=\Sigma^p_2$, and thus our actual problems are in $\text{coNP}^{\text{NP}}=\Pi^p_2$. Given an instance of a problem $(\bigcup_{1\leq i \leq n} P_i, \bigcup_{1\leq i \leq m} Q_i)$ consisting of two queries, we wish to check whether for some $1\leq i \leq n$,   some completion $P_i^c$ of $P_i$ and some literal $L$ of $P_i$, there is no $1 \leq j \leq m$ with a   negation-homomorphism 
$h_{P_i^c, L} \colon Q_j \longrightarrow P_i^c$ such that $L$ belongs to the range of $h_{P_i^c, L}$. We
can  guess a $P_i$, literal $L$ of $P_i$,  and a completion $P_i^c$, and for each check with the $\text{coNP}$ oracle that  there is no $1 \leq j \leq m$ with a   negation-homomorphism 
    $h_{P_i^c, L} \colon Q_j \longrightarrow P_i^c$. (The problem  of checking that  for some $1 \leq j \leq m$ there is a   negation-homomorphism 
    $h_{P_i'} \colon Q_j \longrightarrow P_i'$ with $L$ in its range is in $\text{NP}$.) For the $\Pi^p_2$-hardness, we can use Theorem~\ref{containmentwhich modeldef}. 
\end{proof}

Once more, we treat the $\mathbb{N}[X, \overline{X}]$ case separately. 

\begin{restatable}{Thm}{ucqnContainmentH}\label{nxpspace2}
$\mathbb{N}[X, \overline{X}]$-containment of UCQNs is decidable in coNP$^\text{PP}$.    
\end{restatable}


\begin{proof}
As we will show in Theorem~\ref{N[X]ucqneqncontainment}, $\mathbb{N}[X,\overline{X}]$-containment of UCQ$^\neq$Ns is in coNP$^\text{PP}$. Since UCQNs are a particular case of UCQ$^\neq$Ns, obtained by disallowing inequality atoms, the same upper bound immediately applies here.
\end{proof}

The $\mathrm{Trio}(X,\overline{X})$ case deserves its own treatment as well. 

\begin{restatable}{Thm}{ucqnContainmentI}\label{triopspace2}
$\mathrm{Trio}(X,\overline{X})$-containment of UCQNs is decidable in coNP$^\text{PP}$.
\end{restatable}


\begin{proof}
As in the case of $\mathbb{N}[X,\overline{X}]$, the complexity upper bound will follow from the more general result for UCQ$^\neq$Ns established later in \ref{Trioucqneqncontainment}. In particular, since UCQNs form a subclass of UCQ$^\neq$Ns, obtained by disallowing inequality atoms, the coNP$^\text{PP}$ upper bound for $\mathrm{Trio}(X,\overline{X})$-containment of UCQ$^\neq$Ns immediately yields the same upper bound for UCQNs.
\end{proof}

\section{Omitted  Proofs from \S~\ref{The problem of equivalenceUCQN}} \label{app:ucqn-equivalence-proofs}

\ucqnEquivalenceA*

\begin{proof} Item (2) follows by definition. For item (1), suppose that  $\bar{P}\sqsubseteq_K^\mathcal{C}\bar{Q}$, so, by monotonicity, $\bar{P} \cup \bar{Q}\sqsubseteq_K^\mathcal{C}\bar{Q} \cup \bar{Q}$ and by idempotency, $\bar{Q} \cup \bar{Q}\sqsubseteq_K^\mathcal{C} \bar{Q}$, so $\bar{P} \cup \bar{Q}\sqsubseteq_K^\mathcal{C} \bar{Q}$. Furthermore, $\bar{Q}\sqsubseteq_K^\mathcal{C} \bar{P} \cup \bar{Q}$ follows by definition of the natural order of the semiring.
\end{proof}

\ucqnEquivalenceB*

\begin{proof}
Under model-defining interpretations, first delete any disjunct containing an atom and its negation, since it is identically zero; an empty union is interpreted as zero. The containment characterizations apply to the remaining disjuncts.

The upper bounds for bounded distributive lattices under model-defining and dually model-defining interpretations, for $\mathrm{Which}(X)$ under model-defining interpretations, and for $\mathrm{PosBool}(X,\overline{X})$, $\mathrm{Sorp}(X,\overline{X})$, $\mathrm{Why}(X,\overline{X})$, and $\mathbb{B}[X,\overline{X}]$ under provenance-tracking interpretations follow from item (2) of Proposition~\ref{propogreen2} and the corresponding containment results, namely Theorems~\ref{dsitrlatticenegucq}, \ref{dsitrlatticenegucqdual}, \ref{ucqnegwhichmd}, \ref{posboolucqnneg}, \ref{sorpchnegucq}, \ref{ucqequinegwhy}, and~\ref{bxucqnegcont}. For these semirings, the lower bounds follow from item (1) of Proposition~\ref{propogreen2}. Finally, the upper bounds for $\mathbb{N}[X,\overline{X}]$ and $\mathrm{Trio}(X,\overline{X})$ follow from the fact that equivalence amounts to containment in both directions, together with Theorems~\ref{nxpspace2} and~\ref{triopspace2}.

For $\mathbb N$ under model-defining interpretations, Remark~\ref{bag-equivalence-bridge} identifies equivalence with $\mathbb N[X,\overline X]$-equivalence under model-defining provenance-tracking interpretations. Its $\mathrm{coNP}^{\mathrm{PP}}$ upper bound follows by checking the consistent-identification counting criterion in both directions, as explained in that remark.
\end{proof}

\section{Omitted  Proofs from \S~\ref{sec:cqneqn-containment}} \label{app:cqneqn-containment-proofs}

 We start with the simplest case where the complexity diverges from the Boolean case. Here the complexity does not change with the arity bound of the relational schema as in Theorem~\ref{contnegdisbool}.
\begin{restatable}{Thm}{cqneqnContainmentA}  \label{cqneqnegmd}
Let $P$ and $Q$ be CQ$^\neq$Ns 
with the same tuple of free variables,  and a bounded distributive lattice $\mathbb{L}$ with an intermediate element $\epsilon \neq \top, \bot$.  Then, the following are equivalent:
\begin{itemize}
\item[1.]  $P \sqsubseteq_{\mathbb{L}}^{\text{md}} Q$.

\item[2.] For every consistent identification $P^{\mathrm{id}}$ (only in the sense of CQ$^\neq$s) of $P$, there is a  $\neq$-negation-homomorphism 
$h_{P^{\mathrm{id}}} \colon Q \longrightarrow P^{\mathrm{id}}$.

\end{itemize}
Moreover, the corresponding decision problem (i.e.\ $P \sqsubseteq_{\mathbb{L}}^{\text{md}} Q$?) is $\Pi^p_2$-complete.
\end{restatable}


\begin{proof}

Let $Q(\vec{x})$ be $\exists \vec{z} (A_1(\vec{x_1}) \wedge \ldots \wedge A_n(\vec{ x_n}))$ and $P(\vec{z})$ be $\exists \vec{u} (B_1(\vec{ y_1}) \wedge \ldots \wedge B_m(\vec{ y_m}))$, where $\overline{x}, \vec{z} $ and $\vec{z}, \vec{u}$ form partitions of $\overline{x_1}, \ldots , \overline{ x_n}$ and $\vec{z_1}, \ldots , \overline{ y_n}$, respectively.

 $(1) \implies (2)$: Suppose that $P \sqsubseteq_{\mathbb{L}}^{\text{md}} Q$, so for each $\mathbb{L}$-instance $\mathbf I$ and model-defining interpretation $\pi$ compatible with it, and tuple $\vec{t}$, we have that $\pi\llbracket P(\vec{t}) \rrbracket \leq_\mathbb{L} \pi\llbracket Q(\vec{t}) \rrbracket$. Consider a consistent identification $P^{\mathrm{id}}$ of a $\neq$-completion $P'$ of $P$.  Now build a canonical database $\mathbf{I}$ where the elements are variables as in the Boolean case, and for any relation $R$ 
we let $R^\mathbf{I}(x_1, \ldots, x_n) = \top$ if $R(x_1, \ldots, x_n)$ is an atom in  $P^{\mathrm{id}}$,  $R^\mathbf{I}(x_1, \ldots, x_n) = \bot$  if  $\neg R(x_1, \ldots, x_n)$ is a literal in  $P^{\mathrm{id}}$, and $R^\mathbf{I}(x_1, \ldots, x_n) = \epsilon$ otherwise.    
 Take next an    interpretation $\pi^*$ compatible with $\mathbf{I}$ defined as follows: $\pi^*\llbracket R(\vec{a}) \rrbracket= R^\mathbf{I}(\vec{a})$, $\pi^*\llbracket \neg R(\vec{a}) \rrbracket=\top$ if $\pi^*\llbracket  R(\vec{a}) \rrbracket=\bot$, and  $\pi^*\llbracket \neg R(\vec{a}) \rrbracket=\bot$ if $\pi^*\llbracket  R(\vec{a}) \rrbracket = \epsilon$ or $\pi^*\llbracket  R(\vec{a}) \rrbracket = \top$. 
Next, we have that  $P^{\mathrm{id}} \sqsubseteq_\mathbb{L}  P' \sqsubseteq_\mathbb{L}  P$ since the possible values of the new literals of $P'$ are $\top$ and $\bot$, which means that the semiring value  of $P^{\mathrm{id}}$ in any interpretation can only be at most that of $P$. By construction, $\pi^*\llbracket P^{\mathrm{id}} \rrbracket(\vec{z}) = \top$, and thus, $\pi^*\llbracket P \rrbracket(\vec{z}) = \top$ and from the hypothesis, $\pi^*\llbracket Q \rrbracket(\overline{x}) = \top$ since the lattice is bounded. Consequently,  for some assignment $v$, it must be that $\pi^*\llbracket A_1(v(\vec{ x_1})) \wedge \ldots \wedge A_n(v(\vec{ x_n})) \rrbracket=\top$ and thus $\pi^*\llbracket A_i(v(\vec{ x_i})) \rrbracket =\top$ for $i\leq n$ since we are in a lattice.
But then, the mapping $h_{P^{\mathrm{id}}}$ that sends each variable $x$ to  the $y$ such that $v(x) = y$, is a homomorphism from $Q$ to $P^{\mathrm{id}}$ since by construction of $\pi^*$,
we have that $\pi^*\llbracket A_i(v(\vec{ x_i})) \rrbracket =\top$ only if there is a  $B_{j_i}(\overline{ y_{j_i}})$ in $P^{\mathrm{id}}$ such that $A_i(v(\vec{ x_i})) = B_{j_i}(\overline{ y_{j_i}})$.

  $(2) \implies (1)$: Assume that for every  consistent identification $P^{\mathrm{id}}$ of $P$, there is a  $\neq$-negation-homomorphism 
$h_{P^{\mathrm{id}}}\colon Q \longrightarrow P^{\mathrm{id}}$.  We wish to show that $P \sqsubseteq_{\mathbb{L}}^{\text{md}} Q$. Consider an arbitrary $\mathbb{L}$-database instance $\mathbf I$,  a model-defining interpretation $\pi$ compatible with $\mathbf I$, and an assignment $v$ to the variables of $P$ producing the tuple under consideration. A contribution equal to $\bot$ can be ignored; otherwise, every comparison of $P$ is satisfied and every relational literal has non-bottom value. In this case,   $\pi\llbracket B_1(v(\overline{ y_1})) \wedge \ldots \wedge B_m(v(\overline{ y_m})) \rrbracket$ is the infimum of the values $\pi\llbracket B_i(v(\overline{ y_i}))  \rrbracket$. Let $P'$ be the completion of $P$ obtained by adding $x\neq y$ if  $v(x)\neq v(y)$, or $x = y$ if  $v(x)= v(y)$, for all variables $x,y$ of $P$, and $P^{\mathrm{id}}$ an identification of $P'$ by a canonical substitution $\theta$.  We may observe that $\theta(x) = \theta(y)$ iff $v(x) = v(y)$. The identification is consistent, since opposite ground literals cannot both have non-bottom values under $\pi$.  Moreover, for each literal of $A_i(\vec{ x_i})$ of $Q$ there is a literal $B'_{j_i}(\theta(\overline{ y_{j_i}}))$ of $P^{\mathrm{id}}$ such that $A_i(h_{P^{\mathrm{id}}}(\vec{ x_i})) = B'_{j_i}(\theta(\overline{ y_{j_i}}))$. Thus, $v \circ h_{P^{\mathrm{id}}}$ is an assignment of elements to the bound variables of $Q$ where $\pi\llbracket A_i(v \circ h_{P^{\mathrm{id}}}(\vec{ x_i})) \rrbracket = \pi\llbracket B'_{j_i}(v(\theta(\overline{ y_{j_i}}))) \rrbracket$, and, since  all the newly added literals to $P$ in order to form $P'$ must take value $\top$ in $\pi$ under $v$ (and are hence inessential in calculations of infima), we must have that $\pi\llbracket B_1(v (\theta
(\overline{ y_1}))) \wedge \ldots \wedge B_m(v (\theta(\overline{ y_m}))) \rrbracket$ is at most $\pi\llbracket A_1(v \circ h_{P^{\mathrm{id}}}(\vec{ x_1})) \wedge \ldots \wedge A_n(v \circ h_{P^{\mathrm{id}}}(\vec{ x_n})) \rrbracket$. Observe that $\theta(x)=\theta(y)$ iff $v(x) = v(y)$, so $\pi\llbracket B'_{j_i}(v(\theta(\overline{ y_{j_i}}))) \rrbracket= \pi\llbracket B'_{j_i}(v(\overline{ y_{j_i}})) \rrbracket$. This means that for any tuple of elements $\vec{t}$, we have that  $\pi\llbracket P(\vec{z}) \rrbracket(\vec{t})$ (the supremum of all values $\pi\llbracket B_1(e (\overline{ y_1})) \wedge \ldots \wedge B_m(e (\overline{ y_m})) \rrbracket$ for all assignments $e$ to the variables) is at most 
$\pi\llbracket Q(\vec{x}) \rrbracket(\vec{t})$  (the supremum of all values $\pi\llbracket A_1(e (\vec{ x_1})) \wedge \ldots \wedge A_n(e (\vec{ x_n})) \rrbracket$ for all assignments $e$ to the variables).

The lower bound for the complexity calculation follows immediately from the hardness of the problem for  CQ$^\neq$s  (in particular, the hardness of deciding whether item 2 holds). Naturally, $\neq$-negation-homomorphisms are simply  $\neq$-homomorphisms  for the case of CQ$^\neq$s. The upper bound follows by the same kind of reasoning as for CQ$^\neq$s.
\end{proof}

We first show that, for the purpose of witnessing non-containment, we may assume that different non-zero ground relational literals are annotated by different fresh provenance tokens.

\begin{restatable}{Lem}{cqneqnContainmentB}\label{freshNPT}
Let $\mathbf I$ be an $\mathbb{N}[X]$-database instance and $\pi$ a provenance-tracking interpretation compatible with $\mathbf I$. Then there are an $\mathbb{N}[X]$-database instance $\mathbf I^*$, a provenance-tracking interpretation $\pi^*$ compatible with $\mathbf I^*$, and a semiring homomorphism $E\colon\mathbb{N}[X,\overline X]\longrightarrow\mathbb{N}[X,\overline X]$ such that distinct ground relational literals taking non-zero values under $\pi^*$ receive pairwise distinct tags, with the correct polarity such that no two tags used by $\pi^*$ are complementary, and, for every CQ$^\neq$N $R$ and tuple $\vec t$, $E\bigl(\pi^*\llbracket R(\vec t)\rrbracket\bigr) = \pi\llbracket R(\vec t)\rrbracket.$ Consequently, if $\pi\llbracket P(\vec t)\rrbracket
\not\leq_{\mathbb{N}[X,\overline X]}
\pi\llbracket Q(\vec t)\rrbracket,
$ then $\pi^*\llbracket P(\vec t)\rrbracket \not\leq_{\mathbb{N}[X,\overline X]} \pi^*\llbracket Q(\vec t)\rrbracket.$
\end{restatable}


\begin{proof}
For every positive ground relational literal $G$ such that $\pi\llbracket G \rrbracket\neq 0$, choose a fresh complementary pair $p_G,\overline{p_G}$, and set $\pi^*\llbracket G \rrbracket:=p_G.$ For every negative ground relational literal $H$ such that $\pi\llbracket H \rrbracket\neq 0$, choose a different fresh complementary pair $n_H,\overline{n_H}$, and set $\pi^*\llbracket H \rrbracket:=\overline{n_H}.$ All these complementary pairs are chosen pairwise distinct. Ground relational literals taking value $0$ under $\pi$ continue to take value $0$ under $\pi^*$. Notice that this construction also replaces an original annotation equal to $1$ by a fresh tag. 

Let $\mathbf I^*$ be the $\mathbb{N}[X]$-database instance induced by the values assigned by $\pi^*$ to positive ground relational atoms. Define a map on the generators by $E(p_G):=\pi\llbracket G \rrbracket,$ $E(\overline{p_G}):=0,$ and $E(\overline{n_H}):=\pi\llbracket H \rrbracket,$ $E(n_H):=0,$ and send every remaining generator to $0$. For every complementary pair $x,\overline x$, at least one of $E(x)$ and $E(\overline x)$ is $0$, and hence $E(x)E(\overline x)=0.$ The map extends to a semiring homomorphism $E\colon\mathbb{N}[X,\overline X]\longrightarrow\mathbb{N}[X,\overline X].$

By construction, for every ground relational literal $L$, $E(\pi^*\llbracket L \rrbracket)=\pi\llbracket L \rrbracket.$ Hence, for every assignment to a CQ$^\neq$N, $E$ maps its contribution under $\pi^*$ to its contribution under $\pi$. Equations and disequations contribute only $0$ or $1$, which are preserved by $E$. Taking the finite sum over all assignments gives $E\bigl(\pi^*\llbracket R(\vec t)\rrbracket\bigr) = \pi\llbracket R(\vec t)\rrbracket.$

Finally, semiring homomorphisms between naturally ordered semirings preserve the natural order. Therefore, $\pi^*\llbracket P(\vec t)\rrbracket \leq_{\mathbb{N}[X,\overline X]} \pi^*\llbracket Q(\vec t)\rrbracket $ would imply $ \pi\llbracket P(\vec t)\rrbracket \leq_{\mathbb{N}[X,\overline X]} \pi\llbracket Q(\vec t)\rrbracket.$
\end{proof}

The next lemma relates the coefficient of a monomial to the number of exact $\neq$-negation-homomorphisms.

\begin{restatable}{Lem}{cqneqnContainmentC}\label{countingNPT}
Let $P,R$ be CQ$^\neq$Ns with the same head, and let $\pi$ be a provenance-tracking interpretation such that distinct non-zero ground relational literals receive pairwise distinct tags, with correct polarity and no two tags used by $\pi$ are complementary. Suppose that an assignment $v$ to the variables of $P$ produces the tuple $\vec t$ and justifies a non-zero monomial $\mu$ to $\pi\llbracket P(\vec t)\rrbracket.$ Let $P^c_v$ be the completion (in the sense of CQ$^{\neq}$s) of $P$ obtained by adding $x=y$ whenever $v(x)=v(y)$ and $x\neq y$ whenever $v(x)\neq v(y)$, and let $P^{\mathrm{id}}_v$ be the corresponding identification. Then the coefficient of $\mu$ in $\pi\llbracket R(\vec t)\rrbracket$ is exactly $|\mathrm{ExNegHom}(R,P^{\mathrm{id}}_v)|$, where for any queries $P, Q$,  $\mathrm{ExNegHom}(P,Q)$  denotes the set of all $\neq$-negation-homomorphisms from $P$ to $Q$ that are exact for relational literals. 
\end{restatable}


\begin{proof}
Let $\theta_v$ be the canonical substitution used to obtain $P^{\mathrm{id}}_v$ from $P^c_v$. Define $v_*(\theta_v(x)):=v(x)$ for every variable $x$ of $P$. This map is well-defined and injective. 

We show that assignments to the variables of $R$ producing $\vec t$ and contributing $\mu$ are in bijection with the elements of $\mathrm{ExNegHom}(R,P^{\mathrm{id}}_v)$.

Let $w$ be an assignment to the variables of $R$ producing $\vec t$ and contributing $\mu$. We first show that every value used by $w$ belongs to the image of $v_*$. Let $x$ be a variable of $R$. Since $R$ is safe, $x$ occurs in some positive relational atom $A(\vec y)$. The value of $A(w(\vec y))$ under $\pi$ is a non-zero tag occurring in $\mu$. Since $v$ contributes the same monomial $\mu$ and different ground relational literals receive different tags, the same positive ground atom $A(w(\vec y))$ must also be used by $v$. Therefore, every component of $w(\vec y)$, and in particular $w(x)$, belongs to the image of $v_*$.

We can thus define $h_w(x):=v_*^{-1}(w(x)).$ Since $v$ and $w$ produce the same ordered tuple $\vec t$, the map $h_w$ preserves the head. Consider a relational literal $A(\vec x)$ of $R$. The tag of the signed ground literal $A(w(\vec x))$ occurs in $\mu$. Since $v$ contributes the same monomial, the same signed ground literal is used by $v$. Its inverse image through $v_*$ is therefore a relational literal of $P^{\mathrm{id}}_v$, and this literal is precisely $A(h_w(\vec x))$. Thus $h_w$ preserves relational literals, including their polarity.

Moreover, if $x=y$ occurs in $R$, then $w(x)=w(y)$ and hence $h_w(x)=h_w(y)$. If $x\neq y$ occurs in $R$, then $w(x)\neq w(y)$ and, by injectivity of $v_*$, $h_w(x)\neq h_w(y)$. Since $P^{\mathrm{id}}_v$ is obtained from the equality type determined by $v$, the corresponding equation or disequation occurs in $P^{\mathrm{id}}_v$. Hence $h_w$ is a $\neq$-negation-homomorphism $h_w\colon R\longrightarrow P^{\mathrm{id}}_v.$

It remains to show that $h_w$ is exact for relational literals. Fix $L\in\text{LitSet}(P^{\mathrm{id}}_v)$, and let $z_L$ be the tag assigned by $\pi$ to the signed ground relational literal $v_*(L)$. Since $v$ contributes $\mu$, the exponent with which $z_L$ occurs in $\mu$ is exactly the number of occurrences of $L$ in the body of $P^{\mathrm{id}}_v$. On the other hand, since $w$ contributes the same monomial, the exponent of $z_L$ in $\mu$ is exactly the number of relational literal occurrences of $R$ whose image under $h_w$ is $L$. Hence these two numbers are equal for every $L\in\text{LitSet}(P^{\mathrm{id}}_v)$. Therefore, $h_w$ is exact for relational literals, and $h_w\in\mathrm{ExNegHom}(R,P^{\mathrm{id}}_v).$

Conversely, let $h\in\mathrm{ExNegHom}(R,P^{\mathrm{id}}_v)$ and define $w_h:=v_*\circ h.$ Since $h$ preserves the head, $w_h$ produces $\vec t$. Since $h$ preserves equations and disequations and $v_*$ is injective, all comparisons of $R$ are satisfied by $w_h$. Moreover, every relational literal of $R$ is mapped by $h$ to a relational literal of $P^{\mathrm{id}}_v$, and hence its ground image under $w_h$ has non-zero value under $\pi$.

For every $L\in\text{LitSet}(P^{\mathrm{id}}_v)$, exactness for relational literals implies that the number of relational literal occurrences of $R$ mapped by $h$ to $L$ is exactly the number of occurrences of $L$ in $P^{\mathrm{id}}_v$. Therefore, the tag associated with $v_*(L)$ occurs in the contribution of $w_h$ with exactly the same exponent with which it occurs in the contribution of $v$. No other tag occurs. Hence $w_h$ contributes exactly $\mu$.

The two constructions are inverse. Indeed, for every assignment $w$ contributing $\mu$, $w_{h_w}=v_*\circ v_*^{-1}\circ w=w,$ while for every $h\in\mathrm{ExNegHom}(R,P^{\mathrm{id}}_v)$, $h_{w_h}=v_*^{-1}\circ v_*\circ h=h.$ Thus assignments producing $\vec t$ and contributing $\mu$ are in bijection with the elements of $\mathrm{ExNegHom}(R,P^{\mathrm{id}}_v)$.

Under a provenance-tracking interpretation, every assignment contributes either $0$ or one copy of a monomial. Hence the coefficient of $\mu$ in $\pi\llbracket R(\vec t)\rrbracket$ is exactly the number of assignments contributing $\mu$, and therefore it is $|\mathrm{ExNegHom}(R,P^{\mathrm{id}}_v)|,$ as required.
\end{proof}

We now characterize containment over $\mathbb{N}[X, \overline{X}]$ for $\mathrm{CQ}^\neq\mathrm{Ns}$ by comparing the counts of exact $\neq$-negation-homomorphisms for each identification $P^{\mathrm{id}}$, which yields a $\mathrm{coNP}^{\mathrm{PP}}$ upper bound.

\begin{restatable}{Thm}{cqneqnContainmentD}\label{N_CQNs}
For CQ$^\neq$Ns $P,Q$ with the same tuple of free variables, the following are equivalent:
\begin{itemize}

\item[1.] $P\sqsubseteq^{\text{pt}}_{\mathbb{N}[X,\overline X]}Q$.

\item[2.] For every identification $P^{\mathrm{id}}$ of $P$ in the sense of CQ$^\neq$s,
$$
|\mathrm{ExNegHom}(P,P^{\mathrm{id}})|
\leq
|\mathrm{ExNegHom}(Q,P^{\mathrm{id}})|.
$$

\end{itemize}
Moreover, the corresponding decision problem (i.e.\ $P\sqsubseteq^{\text{pt}}_{\mathbb{N}[X,\overline X]}Q$?) is in $\text{coNP}^{\text{PP}}$.
\end{restatable}


\begin{proof}
$(1)\implies(2):$
Suppose that $P\sqsubseteq^{\text{pt}}_{\mathbb{N}[X,\overline X]}Q$ and consider an identification $P^{\mathrm{id}}$ of a completion (in the sense of CQ$^\neq$s) $P^c$ of $P$. Let $\theta$ be the canonical substitution used to obtain $P^{\mathrm{id}}$ from $P^c$, and let $\text{LitSet}(P^{\mathrm{id}})=\{L_1,\ldots,L_r\}.$

Consider a provenance-tracking interpretation $\pi$ which assigns to every $L_i$ a different fresh tag $x_i\in X\cup\overline X$, with the correct polarity, chosen so that no two tags used by $\pi$ are complementary, and assigns value $0$ to every other ground relational literal. Let $\mathbf I$ be the $\mathbb{N}[X]$-database instance induced by the values of $\pi$ on positive ground relational atoms. Then $\pi$ is a provenance-tracking interpretation compatible with $\mathbf I$.

Consider the assignment $v_\theta$ to the variables of $P$ defined by $v_\theta(x):=\theta(x).$ By construction of the completion and identification, this assignment satisfies all equations and disequations of $P$ and produces the tuple $\theta(\overline u)$. Its contribution is the non-zero monomial $\mu$ obtained by taking, for every $L\in\text{LitSet}(P^{\mathrm{id}})$, the tag assigned to $L$, repeated as many times as $L$ occurs in the body of $P^{\mathrm{id}}$. Moreover, the identification of $P$ induced by $v_\theta$ is precisely $P^{\mathrm{id}}$.

By Lemma~\ref{countingNPT}, applied first to $P$ and then to $Q$, the coefficient of $\mu$ in $\pi\llbracket P(\theta(\overline u))\rrbracket$ is $|\mathrm{ExNegHom}(P,P^{\mathrm{id}})|,$ while its coefficient in $\pi\llbracket Q(\theta(\overline u))\rrbracket$ is $|\mathrm{ExNegHom}(Q,P^{\mathrm{id}})|.$ By the containment hypothesis, $\pi\llbracket P(\theta(\overline u))\rrbracket \leq_{\mathbb{N}[X,\overline X]} \pi\llbracket Q(\theta(\overline u))\rrbracket.$ Since the natural order of $\mathbb{N}[X,\overline X]$ is coefficientwise on consistent monomials, it follows that $|\mathrm{ExNegHom}(P,P^{\mathrm{id}})| \leq |\mathrm{ExNegHom}(Q,P^{\mathrm{id}})|.$ Since $P^{\mathrm{id}}$ was arbitrary, item 2 follows.

$(2)\implies(1):$
We prove the contrapositive. Suppose that $P\not\sqsubseteq^{\text{pt}}_{\mathbb{N}[X,\overline X]}Q. $ Then there are an $\mathbb{N}[X]$-database instance $\mathbf I$, a provenance-tracking interpretation $\pi$ compatible with $\mathbf I$, and a tuple $\vec t$ such that $\pi\llbracket P(\vec t)\rrbracket\not\leq_{\mathbb{N}[X,\overline X]}\pi\llbracket Q(\vec t)\rrbracket.$ By Lemma~\ref{freshNPT}, there are an $\mathbb{N}[X]$-database instance $\mathbf I^*$ and a provenance-tracking interpretation $\pi^*$ compatible with $\mathbf I^*$, in which distinct non-zero ground relational literals receive pairwise distinct tags with the correct polarity and no two tags used by $\pi^*$ are complementary, such that $\pi^*\llbracket P(\vec t)\rrbracket \not\leq_{\mathbb{N}[X,\overline X]} \pi^*\llbracket Q(\vec t)\rrbracket.$

Since the natural order of $\mathbb{N}[X,\overline X]$ is coefficientwise on consistent monomials, there is a consistent monomial $\mu$ whose coefficient in $\pi^*\llbracket P(\vec t)\rrbracket$ is strictly greater than its coefficient in $\pi^*\llbracket Q(\vec t)\rrbracket.$ In particular, the coefficient of $\mu$ on the left-hand side is non-zero. Hence there is an assignment $v$ to the variables of $P$, producing $\vec t$, whose contribution is exactly $\mu$.

Let $P^c_v$ be the completion (in the sense of CQ$^\neq$s) of $P$ determined by $v$, obtained by adding $x=y$ whenever $v(x)=v(y)$ and $x\neq y$ whenever $v(x)\neq v(y)$, and let $P^{\mathrm{id}}_v$ be the corresponding identification. 

By Lemma~\ref{countingNPT}, the coefficient of $\mu$ in $\pi^*\llbracket P(\vec t)\rrbracket$ is $|\mathrm{ExNegHom}(P,P^{\mathrm{id}}_v)|,$ while its coefficient in $\pi^*\llbracket Q(\vec t)\rrbracket$ is $|\mathrm{ExNegHom}(Q,P^{\mathrm{id}}_v)|. $ Therefore, $|\mathrm{ExNegHom}(P,P^{\mathrm{id}}_v)| > |\mathrm{ExNegHom}(Q,P^{\mathrm{id}}_v)|.$ Thus item 2 fails for the identification $P^{\mathrm{id}}_v$. This proves the contrapositive and hence $(2)\implies(1)$.

Finally, for the upper bound complexity calculation, we look at the complement of the problem in item 2. We can guess an identification $P^{\mathrm{id}}$ of $P$ and use a $\text{PP}$ oracle to decide whether $|\mathrm{ExNegHom}(P,P^{\mathrm{id}})| > |\mathrm{ExNegHom}(Q,P^{\mathrm{id}})|.$ An identification $P^{\mathrm{id}}$ is obtained only from a completion (in the sense of CQ$^\neq$s) of $P$: for every pair of variables the completion fixes whether they are equal or distinct, and the corresponding identification quotients equal variables. Thus its description has polynomial size. 

For fixed CQ$^\neq$Ns $R$ and $P^{\mathrm{id}}$, computing $|\mathrm{ExNegHom}(R,P^{\mathrm{id}})|$ is a $\#\text{P}$ problem. Indeed, a nondeterministic polynomial-time machine guesses a mapping from the variables of $R$ to the variables of $P^{\mathrm{id}}$ and checks that it preserves the ordered head, positive and negative relational literals, equations and disequations, and that it is exact for relational literals. The latter condition can be checked in polynomial time by verifying, for every $L\in\text{LitSet}(P^{\mathrm{id}})$, that the number of relational literal occurrences of $R$ mapped to $L$ is exactly the number of occurrences of $L$ in $P^{\mathrm{id}}$.

Hence $|\mathrm{ExNegHom}(P,P^{\mathrm{id}})|-|\mathrm{ExNegHom}(Q,P^{\mathrm{id}})|$ is in $\text{GapP}$ by~\cite[Prop. 3.5]{Fenner}, and deciding whether this difference is positive is in $\text{PP}$ by~\cite[Prop. 4.2]{Fenner}. Therefore, non-containment is in $\text{NP}^{\text{PP}}$, and consequently $\mathbb{N}[X,\overline X]$-containment of CQ$^\neq$Ns under provenance-tracking interpretations is in $\text{coNP}^{\text{PP}}$.
\end{proof}

\begin{restatable}{Thm}{cqneqnContainmentF} \label{b[X]cqneqnegcont}For CQ$^\neq$Ns $P$ and $Q$ with the same tuple of free variables,  the following are equivalent:
\begin{itemize}
\item[1.] $P \sqsubseteq_{\mathbb{B}[X, \overline{X}]}^{\text{pt}} Q$,
\item[2.] For every identification (in the sense of CQ$^\neq$s) $P^{\mathrm{id}}$ of $P$, we have that $ \pi^{\mathrm{can}_{\mathbb{B}[X]}(P^{\mathrm{id}})}\llbracket P^{\mathrm{id}} \rrbracket  \leq_{\mathbb{B}[X, \overline{X}]} \pi^{\mathrm{can}_{\mathbb{B}[X]}(P^{\mathrm{id}})}\llbracket Q \rrbracket$.
\item[3.] For every identification (in the sense of CQ$^\neq$s) $P^{\mathrm{id}}$ of $P$, there is a $\neq$-negation-homomorphism $h_{P^{\mathrm{id}}} \colon Q \longrightarrow P^{\mathrm{id}}$ exact for relational literals.

\end{itemize}
Moreover, the corresponding decision problem (i.e.\ $P \sqsubseteq^{\text{pt}}_{\mathbb{B}[X, \overline{X}]} Q$?) is $\Pi^p_2$-complete.
\end{restatable}


\begin{proof}
$(1) \implies (2):$ Suppose that $P \sqsubseteq_{\mathbb{B}[X, \overline{X}]}^{\text{pt}} Q$ and consider an identification $P^{\mathrm{id}}$ of a completion $P^c$ of $P$. We have that  $P^{\mathrm{id}}\sqsubseteq_{\mathbb{B}[X, \overline{X}]}^{\text{pt}}  P^c \sqsubseteq_{\mathbb{B}[X, \overline{X}]}^{\text{pt}} P$ since on any $\mathbb{B}[X]$-database instance $\mathbf I$ and compatible interpretation $\pi$, the polynomial $\pi\llbracket P^c \rrbracket$ contains only monomials appearing in the one corresponding to $P$ (some of the monomials appearing in $P$ might have turned into $0$ in $P^c$). Thus, $ \pi^{\mathrm{can}_{\mathbb{B}[X]}(P^{\mathrm{id}})}\llbracket P^{\mathrm{id}} \rrbracket  \leq_{\mathbb{B}[X, \overline{X}]} \pi^{\mathrm{can}_{\mathbb{B}[X]}(P^{\mathrm{id}})}\llbracket P^c \rrbracket  \leq_{\mathbb{B}[X, \overline{X}]} \pi^{\mathrm{can}_{\mathbb{B}[X]}(P^{\mathrm{id}})}\llbracket P \rrbracket \leq_{\mathbb{B}[X, \overline{X}]} \pi^{\mathrm{can}_{\mathbb{B}[X]}(P^{\mathrm{id}})}\llbracket Q \rrbracket$.

$(2) \implies (3):$  Assume that (2) holds, so
for every identification $P^{\mathrm{id}}$ of any completion $P^c$ of $P$, we have that 
$$\pi^{\mathrm{can}_{\mathbb{B}[X]}(P^{\mathrm{id}})}\llbracket P^{\mathrm{id}}(\vec{u}) \rrbracket  \leq_{\mathbb{B}[X, \overline{X}]} \pi^{\mathrm{can}_{\mathbb{B}[X]}(P^{\mathrm{id}})}\llbracket Q(\vec{u}) \rrbracket,$$
where $\vec{u}$ is the tuple of free variables of $P^{\mathrm{id}}$. The polynomial $\pi^{\mathrm{can}_{\mathbb{B}[X]}(P^{\mathrm{id}})}\llbracket P^{\mathrm{id}}(\vec{u}) \rrbracket$ has as a term a monomial $x_1 \cdot \ldots \cdot x_n $ containing every tag from $X \cup \overline{X}$ used to annotate literals by $\pi^{\mathrm{can}_{\mathbb{B}[X]}(P^{\mathrm{id}})}$ possibly with repetitions (such monomial is obtained using the identity assignment for the variables in $ P^{\mathrm{id}}$; noticing that the disequations in $P^{\mathrm{id}}$ all take value $1$ by construction of the comparison completion; the tags of the relational literals are chosen pairwise noncomplementary).  Then, by the definition of $\leq $ for $\mathbb{B}[X, \overline{X}]$, we must have that the monomial $x_1 \cdot \ldots \cdot x_n $ appears in $\pi^{\mathrm{can}_{\mathbb{B}[X]}(P^{\mathrm{id}})}\llbracket Q(\vec{u}) \rrbracket$. The rest of the reasoning is as the corresponding part of Theorem~\ref{N_CQNs}.

$(3) \implies (1):$ Assume that for every identification $P^{\mathrm{id}}$ of any completion $P^c$ of $P$, there is an  $\neq$-negation-homomorphism $h_{P^{\mathrm{id}}}\colon Q \longrightarrow P^{\mathrm{id}}$ exact for literals. In order to establish $P \sqsubseteq_{\mathbb{B}[X, \overline{X}]}^{\text{pt}} Q$ consider an arbitrary $\mathbb{B}[X]$-database instance $\mathbf I$, a compatible interpretation $\pi$ and a tuple $\vec{t}$. We need to show that any monomial in $\pi\llbracket P(\vec{t}) \rrbracket$ appears  in $\pi\llbracket Q(\vec{t}) \rrbracket$. Take an assignment $v$ of elements to the variables of $P$ justifying a monomial $x_1 \cdot \ldots \cdot x_n$ from $\mathbb{B}[X, \overline{X}]$ and consider the $\{ =, \neq\}$-completion $ P^c$ of $P$ obtained by adding $x=y$ or $x\neq y$ depending on whether  $v(x) = v(y)$ or $v(x) \neq v(y)$. Take an identification $P^{\mathrm{id}}$ of $P^c$ by a canonical substitution $\theta$. We may observe that $\theta(x) = \theta(y)$ iff $v(x) = v(y)$. Then, $x_1 \cdot \ldots \cdot x_n $ is also a monomial in $\pi\llbracket P^{\mathrm{id}}(\vec{t}) \rrbracket$ justified by the assignment $v$ itself since the new disequations added can only take value $1$ by construction according to the assignment $v$.  By assumption, there is an  $\neq$-negation-homomorphism $h_{P^{\mathrm{id}}}\colon Q \longrightarrow P^{\mathrm{id}}$ exact for literals. Then, $v \circ h_{P^{\mathrm{id}}}$ is an assignment of elements to the variables of $Q$ justifying the monomial $x_1 \cdot \ldots \cdot x_n $. If all the $x_i$s are values of relational literals of $P$ according to $v$, then if $x_i = \pi\llbracket B(v(\theta(\vec{z}))) \rrbracket$ where  $B(\vec{z})$ is a literal of $P$ (and thus $B(\theta(\vec{z}))$ is one of $P^{\mathrm{id}}$), then $x_i = \pi\llbracket A(v(h_{P^{\mathrm{id}}}(\overline{x}))) \rrbracket$ for some literal $A(\overline{x})$ of $Q$ such that $A(h_{P^{\mathrm{id}}}(\overline{x})) = B(\theta(\vec{z}))$, which exists by surjectivity. Furthermore, the multiplicity  with which $x_i$ appears using the assignment $v \circ h_{P^{\mathrm{id}}} $ is exactly the multiplicity with which it appears in  $x_1 \cdot \ldots \cdot x_m$ given that $h_{P^{\mathrm{id}}}$ is both onto and injective on relational literals.
Since equations and disequations can only take values $0$ or $1$, they can be disregarded in showing that we can justify a monomial.

Finally, the calculation of complexity for the upper bound follows the same kind of reasoning as for CQ$^\neq$s.
\end{proof}

For $\mathrm{Why}(X, \overline{X})$, containment of $\mathrm{CQ}^\neq\mathrm{Ns}$ requires a set-onto $\neq$-negation-homomorphism across all identifications $P^{\mathrm{id}}$, giving $\Pi^p_2$-completeness.

\begin{restatable}{Thm}{cqneqnContainmentG} \label{whycqneqnegcont}
For CQ$^\neq$Ns $P$ and $Q$ with the same tuple of free variables, the following are equivalent:
\begin{itemize}

\item[1.] $P \sqsubseteq^{\text{pt}}_{\mathrm{Why}(X,\overline X)} Q.$

\item[2.] For every identification $P^{id}$ of $P$ in the sense of CQ$^\neq$s, $\pi^{\mathrm{can}_{\mathrm{Why}(X)}((P^{id})^+)}    \llbracket P^{id} \rrbracket \leq_{\mathrm{Why}(X,\overline X)} \pi^{\mathrm{can}_{\mathrm{Why}(X)}((P^{id})^+)} \llbracket Q \rrbracket$.

\item[3.] For every identification $P^{id}$ of $P$ in the sense of CQ$^\neq$s, there is a $\neq$-negation-homomorphism $h_{P^{id}}\colon Q\longrightarrow P^{id}$ set-onto for relational literals i.e. $h_{P^{id}}\bigl(\text{LitSet}(Q)\bigr) = \text{LitSet}(P^{id}).$

\end{itemize}
Moreover, the corresponding decision problem (i.e.\ $P\sqsubseteq^{\text{pt}}_{\mathrm{Why}(X,\overline X)}Q$?) is $\Pi^p_2$-complete.
\end{restatable}


\begin{proof}
$(1)\Rightarrow(2)$:
Suppose that $P\sqsubseteq^{\text{pt}}_{\mathrm{Why}(X,\overline X)}Q$ and consider an identification $P^{id}$ of a $\{=,\neq\}$-completion $P^c$ of $P$. By construction, $P^{id}$ is semantically equivalent to $P^c$, while $P^c$ is obtained from $P$ only by adding equations and disequations. Since these can only eliminate assignments, and take value $1$ whenever they are satisfied, we have $P^{id} \sqsubseteq^{\text{pt}}_{\mathrm{Why}(X,\overline X)} P^c \sqsubseteq^{\text{pt}}_{\mathrm{Why}(X,\overline X)} P.$ Hence, by the containment assumption, $P^{id} \sqsubseteq^{\text{pt}}_{\mathrm{Why}(X,\overline X)} Q.$ The interpretation $\pi^{\mathrm{can}_{\mathrm{Why}(X)}((P^{id})^+)}$ is provenance-tracking, and therefore the latter containment holds, in particular, on this interpretation. Thus $\pi^{\mathrm{can}_{\mathrm{Why}(X)}((P^{id})^+)} \llbracket P^{id} \rrbracket \leq_{\mathrm{Why} (X,\overline X)} \pi^{\mathrm{can}_{\mathrm{Why}(X)}((P^{id})^+)} \llbracket Q \rrbracket$, as required.

$(2)\Rightarrow(3)$: Assume that item 2 holds and fix an identification $P^{id}$ of $P$. Let $I:=\mathrm{can}_{\mathrm{Why}(X)}((P^{id})^+)$ and abbreviate $\pi^{\mathrm{can}_{\mathrm{Why}(X)}((P^{id})^+)}$ as $\pi^{\mathbf I}$. Let $\text{LitSet}(P^{id})=\{L_1,\ldots,L_r\}$. By construction of $\pi^{\mathbf I}$, each distinct relational literal $L_i$ is associated with a different tag $x_i\in X\cup\overline X$, and $\pi^{\mathbf I}\llbracket L_i \rrbracket=\{\{x_i\}\}$. Repeated occurrences of the same relational literal receive the same annotation. Moreover, by construction of the canonical interpretation, no pair of complementary tags occurs among $x_1,\ldots,x_r$. In particular, if both $R(\overline a)$ and $\neg R(\overline a)$ occur in $P^{id}$, the tag used for the negative literal is chosen so that its complementary positive tag is not used as the annotation of $R(\overline a)$ in the canonical database. Hence the witness formed by the tags of the relational literals of $P^{id}$ is not removed by the quotient defining $\mathrm{Why}(X,\overline X)$.

Let $W_{P^{id}}:=\{x_1,\ldots,x_r\}$. Consider the assignment to the bound variables of $P^{id}$ that sends every variable to itself. Under this assignment all disequations of $P^{id}$ take value $\{\emptyset\}$, while every occurrence of a relational literal $L_i$ takes the value $\{\{x_i\}\}$. Since multiplication in $\mathrm{Why}(X,\overline X)$ is pairwise union, and repeated occurrences of the same literal do not change the resulting union, the corresponding contribution contains the witness $W_{P^{id}}$. Hence $W_{P^{id}}\in\pi^{\mathbf I}\llbracket P^{id}(\overline u) \rrbracket$. By item 2 and the fact that the natural order on $\mathrm{Why}(X,\overline X)$ is set inclusion, we have $W_{P^{id}}\in\pi^{\mathbf I}\llbracket Q(\overline u) \rrbracket$. Thus, there is an assignment $v$ to the variables of $Q$, producing the tuple $\overline u$, whose contribution contains $W_{P^{id}}$. In particular, all equations and disequations of $Q$ are satisfied by $v$, and every relational literal of $Q$ has non-zero value under $v$. If $A(\overline x)$ is a positive relational literal of $Q$, then $\pi^{\mathbf I}\llbracket A(v(\overline x)) \rrbracket\neq 0$, and, by construction of the canonical database $I$, $A(v(\overline x))$ corresponds to a positive relational literal occurring in $P^{id}$. Since $Q$ is safe, every variable of $Q$ occurs in some positive relational literal, and therefore the value assigned by $v$ to each variable is one of the variables occurring in $P^{id}$. We can thus define $h_{P^{id}}(x):=v(x)$ for every variable $x$ of $Q$. Similarly, if $A(\overline x)$ is a negative relational literal of $Q$, the fact that $\pi^{\mathbf I}\llbracket A(v(\overline x)) \rrbracket\neq 0$ and the definition of the canonical interpretation imply that $A(h_{P^{id}}(\overline x))$ is a negative relational literal of $P^{id}$. Hence $h_{P^{id}}$ preserves positive and negative relational literals, and the fact that $v$ produces the canonical tuple $\overline u$ gives preservation of the head. Moreover, if $x=y$ occurs in $Q$, then $v(x)=v(y)$ and therefore $h_{P^{id}}(x)=h_{P^{id}}(y)$. If $x\neq y$ occurs in $Q$, then $v(x)\neq v(y)$, so $h_{P^{id}}(x)$ and $h_{P^{id}}(y)$ are distinct representatives in $P^{id}$; since $P^{id}$ is obtained from a $\{=,\neq\}$-completion, the corresponding disequation is retained in $P^{id}$. Thus $h_{P^{id}}\colon Q\longrightarrow P^{id}$ is a $\neq$-negation-homomorphism. 

It remains to show that it is set-onto for relational literals. Since $\pi^{\mathbf I}$ is provenance-tracking, every non-zero relational literal under $v$ has a singleton annotation, and therefore the contribution of $v$ has a unique witness, obtained as the union of the tags contributed by the relational literals of $Q$. Since this contribution contains $W_{P^{id}}$, its unique witness must be exactly $W_{P^{id}}$. Consequently, for every $1\leq i\leq r$, the tag $x_i$ must occur in the annotation of some relational literal $A$ of $Q$ under $v$. By construction of $\pi^{\mathbf I}$, the tag $x_i$ occurs only in the annotation corresponding to the signed relational literal $L_i$. Hence $h_{P^{id}}(A)=L_i$. Therefore, $\text{LitSet}(P^{id})\subseteq h_{P^{id}}(\text{LitSet}(Q))$. The reverse inclusion follows immediately from the fact that $h_{P^{id}}$ is a $\neq$-negation-homomorphism. Hence $h_{P^{id}}(\text{LitSet}(Q))=\text{LitSet}(P^{id})$, and $h_{P^{id}}$ is set-onto for relational literals.

$(3)\Rightarrow(1)$: Assume item 3, we write $P(\vec y) := \exists\overline u\,\bigl(B_1(\vec y_1)\wedge\ldots\wedge B_m(\vec y_m)\wedge E_P\bigr)$ and $Q(\overline x) := \exists\overline z\,\bigl(A_1(\overline x_1)\wedge\ldots\wedge A_n(\overline x_n)\wedge E_Q\bigr)$, where the $B_i$'s and $A_j$'s are relational literals and $E_P,E_Q$ denote the conjunctions of equations and disequations occurring in the respective queries. Consider an arbitrary $\mathrm{Why}(X)$-database instance $\mathbf I$, a provenance-tracking interpretation $\pi$ compatible with $\mathbf I$, and a tuple $\vec t$. We show that $\pi^\mathbf{I}\llbracket P(\vec t) \rrbracket\subseteq\pi^\mathbf{I}\llbracket Q(\vec t) \rrbracket$. Let $W\in\pi^\mathbf{I}\llbracket P(\vec t) \rrbracket$. Then there is an assignment $v$ to the variables of $P$, producing $\vec t$ on the free variables, whose contribution contains $W$. Since this contribution is non-zero, all equations and disequations of $P$ are satisfied by $v$ and every relational literal $B_i$ has non-zero value under $v$. Since $\pi$ is provenance-tracking, for every ground relational literal $G$ with $\pi\llbracket G \rrbracket\neq 0$ there is a unique finite set $S_G\subseteq X\cup\overline X$ such that $\pi\llbracket G \rrbracket=\{S_G\}$. Let $s_\pi\llbracket G \rrbracket:=S_G$. Hence the contribution of $v$ has the unique witness $\bigcup_{1\leq i\leq m}s_\pi\llbracket B_i(v(\vec y_i)) \rrbracket$, and therefore $W=\bigcup_{1\leq i\leq m}s_\pi\llbracket B_i(v(\vec y_i)) \rrbracket$. Consider now the $\{=,\neq\}$-completion $P^c$ of $P$ determined by $v$, obtained by adding $x=y$ whenever $v(x)=v(y)$ and $x\neq y$ whenever $v(x)\neq v(y)$, and let $P^{id}$ be the corresponding identification obtained through a canonical substitution $\theta$. By construction, $\theta(x)=\theta(y)$ if and only if $v(x)=v(y)$. Define an assignment $v_*$ to the variables of $P^{id}$ by $v_*(\theta(x)):=v(x)$. This is well-defined, and for every relational literal $L$ of $P$ we have $v_*(\theta(L))=v(L)$. Notice that $P^{id}$ need not be consistent in the sense of containing no pair of opposite relational literals. This causes no problem here, since item 3 quantifies over all identifications of $P$ in the sense of CQ$^\neq$s. Thus item 3 applies to $P^{id}$. Several relational literals of $P$ may become the same literal in $P^{id}$, and repeated occurrences may already be present in $P$. However, all occurrences that become the same literal have the same ground value under $v_*$ and therefore the same set $s_\pi$. Since ordinary set union is idempotent, repetitions do not affect the resulting witness. Consequently, $W=\bigcup_{L\in\text{LitSet}(P^{id})}s_\pi\llbracket v_*(L) \rrbracket$. By item 3, there is a set-onto $\neq$-negation-homomorphism $h_{P^{id}}\colon Q\longrightarrow P^{id}$ such that $h_{P^{id}}(\text{LitSet}(Q))=\text{LitSet}(P^{id})$. Consider the assignment $w:=v_*\circ h_{P^{id}}$ to the variables of $Q$. Since $h_{P^{id}}$ preserves the head, $w$ produces the tuple $\vec t$. Moreover, $h_{P^{id}}$ preserves equations and disequations, and $v_*$ satisfies all the corresponding comparisons in $P^{id}$; hence all equations and disequations of $Q$ are satisfied by $w$ and contribute the multiplicative identity $\{\emptyset\}$. For every relational literal $A$ of $Q$, $h_{P^{id}}(A)$ belongs to $\text{LitSet}(P^{id})$, and therefore $\pi\llbracket v_*(h_{P^{id}}(A)) \rrbracket\neq 0$. Hence every relational literal of $Q$ has non-zero value under $w$. Since $\pi$ is provenance-tracking, the contribution of $w$ has the unique witness $\bigcup_{A\in\text{LitSet}(Q)}s_\pi\llbracket v_*(h_{P^{id}}(A)) \rrbracket$. By the set-onto property $h_{P^{id}}(\text{LitSet}(Q))=\text{LitSet}(P^{id})$, the sets occurring in this union are exactly those occurring in $\bigcup_{L\in\text{LitSet}(P^{id})}s_\pi\llbracket v_*(L) \rrbracket$, possibly with repetitions. Therefore, $\bigcup_{A\in\text{LitSet}(Q)}s_\pi\llbracket v_*(h_{P^{id}}(A)) \rrbracket=\bigcup_{L\in\text{LitSet}(P^{id})}s_\pi\llbracket v_*(L) \rrbracket=W$. Thus the assignment $w$ contributes $W$ to $\pi^\mathbf{I}\llbracket Q(\vec t) \rrbracket$, and consequently $W\in\pi^\mathbf{I}\llbracket Q(\vec t) \rrbracket$. Since $\mathbf I$, $\pi$, $\vec t$, and $W\in\pi^\mathbf{I}\llbracket P(\vec t) \rrbracket$ were arbitrary, we conclude that $P\sqsubseteq^{\mathrm{pt}}_{\mathrm{Why}(X,\overline X)}Q$.

Finally, the upper bound in the complexity calculation follows by the same kind of reasoning as for CQ$^\neq$s. For the lower bound, we use the same reduction as for CQ$^\neq$s, i.e. given an instance $P,Q$ of Boolean CQ$^\neq$ containment, we consider the query $Q^\ast$ obtained by conjoining $Q$ with a suitably copy of $P$. The latter component guarantees that every $\neq$-homomorphism $Q\to P^{id}$ extends to a $\neq$-homomorphism $Q^\ast\to P^{id}$ which is set-onto for relational literals, while the converse follows by restriction to the $Q$-component. In the absence of negated relational atoms, $\neq$-negation-homomorphisms coincide with $\neq$-homomorphisms. Hence the problem is $\Pi^p_2$-hard, and therefore $\Pi^p_2$-complete.
\end{proof}

\cqneqnContainmentH*

\begin{proof}
For every positive ground relational literal $G$ such that $\pi\llbracket G\rrbracket\neq 0$, choose a fresh token $p_G\in X$, and for every negative ground relational literal $H$ such that $\pi\llbracket H\rrbracket\neq 0$, choose a fresh token $\overline{p}_H\in\overline{X}$. We choose all these tokens from pairwise distinct complementary pairs. Define $\pi^*$ by letting $\pi^*\llbracket G\rrbracket=p_G$ for every positive $G$ such that $\pi\llbracket G\rrbracket\neq 0$, and $\pi^*\llbracket H\rrbracket=\overline{p}_H$ for every negative $H$ such that $\pi\llbracket H\rrbracket\neq 0$. If $\pi\llbracket L\rrbracket=0$, we let $\pi^*\llbracket L\rrbracket=0$. Thus, distinct non-zero ground relational literals receive distinct tags, positive and negative literals receive tags of the corresponding polarity, and no two tags used by $\pi^*$ are complementary.

Let $\mathbf I^*$ be the $\mathrm{Trio}(X)$-database instance obtained by setting, for every positive ground atom $R(\vec{a})$, $R^{\mathbf I^*} (\vec{a})=\pi^*\llbracket R(\vec{a})\rrbracket.$ Since $\pi$ is compatible with $\mathbf I$, a positive ground atom has non-zero value under $\pi^*$ exactly when it has non-zero value under $\pi$. Hence $\mathbf I$ and $\mathbf I^*$ have the same support and $\pi^*$ is a provenance-tracking interpretation compatible with $\mathbf I^*$.

We now define $E$ on the fresh complementary pairs used above. For every positive ground relational literal $G$ with $\pi\llbracket G\rrbracket\neq 0$, let $E(p_G)=\pi\llbracket G\rrbracket$ and $E(\overline{p}_G)=0.$ For every negative ground relational literal $H$ with $\pi\llbracket H\rrbracket\neq 0$, let $E(\overline{p}_H)=\pi\llbracket H\rrbracket $ and $E(p_H)=0.$ All remaining complementary pairs may be mapped to $0$. Since $\pi$ is provenance-tracking, every non-zero value appearing on the right-hand side above is either $1$ or a single provenance token and hence all these values are multiplicatively idempotent in $\mathrm{Trio}(X,\overline{X})$. Moreover, for every complementary pair $x,\overline{x}$, at least one between $E(x)$ and $E(\overline{x})$ is $0$, and therefore $E(x)\cdot E(\overline{x})=0.$ Thus the map on generators respects both the identification of positive exponents and the equations $x\cdot \overline{x}=0$ defining $\mathrm{Trio}(X,\overline{X})$, and consequently extends to a semiring homomorphism $E\colon \mathrm{Trio}(X,\overline{X})\longrightarrow \mathrm{Trio}(X,\overline{X}).$

By construction, for every ground relational literal $L$, $E\bigl(\pi^*\llbracket L\rrbracket\bigr) =\pi\llbracket L\rrbracket.$ Equations and disequations take only the values $0$ and $1$, which are preserved by $E$. Therefore, for every assignment $v$ to the variables of a CQ$^\neq$N $P$, applying $E$ to the contribution of $v$ under $\pi^*$ gives exactly the contribution of the same assignment under $\pi$. Since $E$ preserves finite sums and products, it follows that $E\bigl(\pi^*\llbracket P(\vec{t})\rrbracket\bigr) = \pi\llbracket P(\vec{t})\rrbracket$
for every tuple $\vec{t}$.

Finally, suppose that $\pi\llbracket P(\vec{t})\rrbracket
\not\leq_{\mathrm{Trio}(X,\overline{X})}
\pi\llbracket Q(\vec{t})\rrbracket.$ If $\pi^*\llbracket P(\vec{t})\rrbracket
\leq_{\mathrm{Trio}(X,\overline{X})}
\pi^*\llbracket Q(\vec{t})\rrbracket,$ then, since semiring homomorphisms preserve the natural order, applying $E$ and using the equality proved above would give $\pi\llbracket P(\vec{t})\rrbracket
\leq_{\mathrm{Trio}(X,\overline{X})}
\pi\llbracket Q(\vec{t})\rrbracket,$ a contradiction. Hence the required non-containment also holds under $\pi^*$.
\end{proof}

\cqneqnContainmentI*

\begin{proof}
Let $P^c_v$ be obtained by adding, for every pair of variables $x,y$ of $P$, $x=y$ if $v(x)=v(y)$ and $x\neq y$ otherwise. Let $\theta_v$ be a canonical substitution for $P^c_v$, so that $P^{\mathrm{id}}_v$ is the resulting identification. We keep denoting by $v$ the induced assignment to the variables of $P^{\mathrm{id}}_v$, defined by $v(\theta_v(x)):=v(x).$ This is well-defined and injective, since $\theta_v(x)=\theta_v(y)$ if and only if $v(x)=v(y)$.

Let $\mathcal G_v$ be the set of distinct ground relational literals used by $v$. Since distinct non-zero ground relational literals receive distinct tags under $\pi$, and no two of these tags are complementary, the square-free monomial $\mu$ contains exactly the tags corresponding to the literals in $\mathcal G_v$. Repeated occurrences of the same ground relational literal do not affect $\mu$, since positive exponents are dropped in $\mathrm{Trio}(X,\overline X)$.

Moreover, $v$ induces a bijection between $\text{LitSet}(P^{\mathrm{id}}_v)$ and $\mathcal G_v$. Indeed, every relational literal of $P^{\mathrm{id}}_v$ is the image under $\theta_v$ of a relational literal of $P$, and applying $v$ gives the corresponding ground relational literal used by the original assignment. Conversely, every ground relational literal used by the original assignment is obtained in this way. Finally, two distinct relational literals of $P^{\mathrm{id}}_v$ cannot have the same ground image under $v$, since $v$ is injective.

We now establish the claimed correspondence. Let $w$ be an assignment to the variables of $Q$ producing $\vec{t}$ and contributing $\mu$. We first observe that every value used by $w$ belongs to the image of $v$. Let $x$ be a variable of $Q$. Since $Q$ is safe, $x$ occurs in some positive relational literal $A(\vec{z})$ of $Q$. Since $w$ contributes the non-zero monomial $\mu$, the tag of the ground literal $A(w(\vec{z}))$ occurs in $\mu$. Hence $A(w(\vec{z}))$ belongs to $\mathcal G_v$, and therefore every component of $w(\vec{z})$, in particular $w(x)$, belongs to the image of $v$.

We can thus define $h_w(x):=v^{-1}(w(x))$ for every variable $x$ of $Q$. Since $v$ and $w$ produce the same tuple $\vec{t}$, the map $h_w$ preserves the head. If $A(\overline{x})$ is a relational literal of $Q$, then the tag of $A(w(\overline{x}))$ occurs in $\mu$. Hence $A(w(\overline{x}))$ belongs to $\mathcal G_v$, and its inverse image under $v$ is a relational literal of $P^{\mathrm{id}}_v$. This inverse image is precisely $A(h_w(\overline{x}))$. Thus $h_w$ preserves relational literals, including their polarity.

Furthermore, if $x=y$ occurs in $Q$, then $w(x)=w(y)$, and therefore $h_w(x)=h_w(y)$. If $x\neq y$ occurs in $Q$, then $w(x)\neq w(y)$, and thus $h_w(x)\neq h_w(y)$. By construction of the completion $P^c_v$ and of the corresponding identification $P^{\mathrm{id}}_v$, the required equations and disequations are therefore preserved. Hence $h_w$ is a $\neq$-negation-homomorphism from $Q$ to $P^{\mathrm{id}}_v$.

Finally, since $w$ contributes exactly $\mu$, the distinct tags occurring in its contribution are exactly those corresponding to $\mathcal G_v$. Since $v$ gives a bijection between $\text{LitSet}(P^{\mathrm{id}}_v)$ and $\mathcal G_v$, every relational literal of $P^{\mathrm{id}}_v$ is the image of some relational literal of $Q$. Thus $h_w\bigl(\text{LitSet}(Q)\bigr)=\text{LitSet}(P^{\mathrm{id}}_v),$ and hence $h_w\in\mathrm{SOHom}(Q,P^{\mathrm{id}}_v)$.

Conversely, let $h\in\mathrm{SOHom}(Q,P^{\mathrm{id}}_v)$ and define $w_h:=v\circ h.$ Since $h$ preserves the head, $w_h$ produces $\vec{t}$. Moreover, since $h$ preserves equations and disequations and $v$ is injective, all equations and disequations of $Q$ are satisfied by $w_h$. Every relational literal of $Q$ is mapped by $h$ to a relational literal of $P^{\mathrm{id}}_v$, and therefore its ground image under $w_h$ belongs to $\mathcal G_v$ and has non-zero value under $\pi$. Since $h$ is set-onto for relational literals, the set of distinct ground relational literals used by $w_h$ is exactly $\mathcal G_v$. Consequently, the distinct tags occurring in its contribution are exactly those occurring in $\mu$. Possible repetitions only change positive exponents, which are dropped in $\mathrm{Trio}(X,\overline X)$. Hence $w_h$ contributes exactly $\mu$.

The two constructions are inverse. Indeed, for every assignment $w$ contributing $\mu$, $w_{h_w}=v\circ v^{-1}\circ w=w,$ while for every $h\in\mathrm{SOHom}(Q,P^{\mathrm{id}}_v)$, $h_{w_h}=v^{-1}\circ v\circ h=h.$ Therefore, assignments to the variables of $Q$ producing $\vec{t}$ and contributing $\mu$ are in bijection with the elements of $\mathrm{SOHom}(Q,P^{\mathrm{id}}_v)$.

Under a provenance-tracking interpretation, every assignment contributes either $0$ or one copy of a square-free monomial. Hence the coefficient of $\mu$ in $\pi\llbracket Q(\vec{t})\rrbracket$ is exactly the number of assignments contributing $\mu$, and therefore it is $\left|\mathrm{SOHom}(Q,P^{\mathrm{id}}_v)\right|,$ as required.
\end{proof}

For $\mathrm{PosBool}(X, \overline{X})$, $\mathrm{CQ}^\neq\mathrm{N}$ containment is reduced to the existence of an unrestricted $\neq$-negation-homomorphism across all identifications $P^{\mathrm{id}}$.

\begin{restatable}{Thm}{cqneqnContainmentK} \label{posboolcqneqncont}
For CQ$^\neq$Ns $P$ and $Q$ with the same tuple of free variables, the following are equivalent:
\begin{itemize}
\item[1.] $P\sqsubseteq^{pt}_{\mathrm{PosBool}(X,\overline{X})}Q$.
\item[2.] For every identification (in the sense of CQ$^\neq$s) $P^{\mathrm{id}}$ of $P$, we have that $\pi^{\mathrm{can}_{\mathrm{PosBool}(X)}(P^{\mathrm{id}})}\llbracket P^{\mathrm{id}} \rrbracket
\leq_{\mathrm{PosBool}(X,\overline{X})}
\pi^{\mathrm{can}_{\mathrm{PosBool}(X)}(P^{\mathrm{id}})}\llbracket Q \rrbracket$.
\item[3.] For every identification (in the sense of CQ$^\neq$s) $P^{\mathrm{id}}$ of $P$, there is a $\neq$-negation-homomorphism $h_{P^{\mathrm{id}}}\colon Q\longrightarrow P^{\mathrm{id}}.$
\end{itemize}
Moreover, the corresponding decision problem (i.e.\ $P \sqsubseteq_{\mathrm{PosBool}(X,\overline{X})} Q$?) is $\Pi^p_2$-complete.
\end{restatable}


\begin{proof}
$(1)\implies(2):$ Suppose that 
$P\sqsubseteq^{pt}_{\mathrm{PosBool}(X,\overline{X})}Q$
and consider an identification $P^{\mathrm{id}}$ of a completion $P^c$ of $P$. We have that $P^{\mathrm{id}}\sqsubseteq^{pt}_{\mathrm{PosBool}(X,\overline{X})}P^c
\sqsubseteq^{pt}_{\mathrm{PosBool}(X,\overline{X})}P,$ since on any $\mathrm{PosBool}(X)$-database instance $\mathbf I$ and compatible interpretation $\pi$, the value $\pi\llbracket P^c \rrbracket$ is obtained from $\pi\llbracket P \rrbracket$ by possibly replacing some contributions with $\bot$, while the comparison atoms satisfied by an assignment take value $\top$ and do not affect the corresponding contribution. Thus, $\pi^{\mathrm{can}_{\mathrm{PosBool}(X)}(P^{\mathrm{id}})}\llbracket P^{\mathrm{id}} \rrbracket \leq_{\mathrm{PosBool}(X,\overline{X})}
\pi^{\mathrm{can}_{\mathrm{PosBool}(X)}(P^{\mathrm{id}})}\llbracket P^c \rrbracket\leq_{\mathrm{PosBool}(X,\overline{X})}
\pi^{\mathrm{can}_{\mathrm{PosBool}(X)}(P^{\mathrm{id}})}\llbracket P \rrbracket\leq_{\mathrm{PosBool}(X,\overline{X})}
\pi^{\mathrm{can}_{\mathrm{PosBool}(X)}(P^{\mathrm{id}})}\llbracket Q \rrbracket.$

$(2)\implies(3):$ Assume that (2) holds, so for every identification $P^{\mathrm{id}}$ of any completion $P^c$ of $P$, we have that
\[
\pi^{\mathrm{can}_{\mathrm{PosBool}(X)}(P^{\mathrm{id}})}
\llbracket P^{\mathrm{id}}(\vec{u}) \rrbracket
\leq_{\mathrm{PosBool}(X,\overline{X})}
\pi^{\mathrm{can}_{\mathrm{PosBool}(X)}(P^{\mathrm{id}})}
\llbracket Q(\vec{u}) \rrbracket,
\]
where $\vec{u}$ is the tuple of free variables of $P^{\mathrm{id}}$. Let $x_1 \wedge \ldots \wedge x_n$ be the conjunction of all tags from $X\cup\overline{X}$ used to annotate the relational literals of $P^{\mathrm{id}}$ by
$\pi^{\mathrm{can}_{\mathrm{PosBool}(X)}(P^{\mathrm{id}})}$. This conjunction is obtained using the identity assignment for the variables of $P^{\mathrm{id}}$, since all the disequations in $P^{\mathrm{id}}$ take value $\top$. By construction of the canonical interpretation, no pair of complementary tags occurs in $x_1 \wedge \ldots \wedge x_n$, even when $P^{\mathrm{id}}$ contains opposite relational literals. Consider the compatible Boolean valuation $\nu$ assigning value $1$ to all tags occurring in $x_1 \wedge \ldots \wedge x_n$ and value $0$ to all the remaining tags. Then, $\nu\left(
\pi^{\mathrm{can}_{\mathrm{PosBool}(X)}(P^{\mathrm{id}})}
\llbracket P^{\mathrm{id}}(\vec{u}) \rrbracket
\right)=1$. By the definition of the order in
$\mathrm{PosBool}(X,\overline{X})$, we also have $\nu\left(
\pi^{\mathrm{can}_{\mathrm{PosBool}(X)}(P^{\mathrm{id}})}
\llbracket Q(\vec{u}) \rrbracket
\right)=1.$ Since $Q$ is a CQ$^\neq$N, there is an assignment $v$ to the bound variables of $Q$ over $\mathrm{can}_{\mathrm{PosBool}(X)}(P^{\mathrm{id}})$ such that the body of $Q$ takes the value $1$ under $\nu$. Hence every relational literal of $Q$ is mapped by $v$ to a relational literal occurring in $P^{\mathrm{id}}$. Moreover, all equations and disequations of $Q$ take value $\top$ under $v$, and therefore they are preserved in $P^{\mathrm{id}}$. We obtain the desired $\neq$-negation-homomorphism $h_{P^{\mathrm{id}}}\colon Q\longrightarrow P^{\mathrm{id}}$ by letting $h_{P^{\mathrm{id}}}(x)$ be the variable $v(x)$, recalling that the elements of the active domain of $\mathrm{can}_{\mathrm{PosBool}(X)}(P^{\mathrm{id}})$ are the variables of $P^{\mathrm{id}}$.

$(3)\implies(1):$ Assume that for every identification $P^{\mathrm{id}}$ of any completion $P^c$ of $P$, there is a $\neq$-negation-homomorphism $h_{P^{\mathrm{id}}}\colon Q\longrightarrow P^{\mathrm{id}}.$ In order to establish $P\sqsubseteq^{\mathrm{pt}}_{\mathrm{PosBool}(X,\overline{X})}Q,$ consider an arbitrary $\mathrm{PosBool}(X)$-database instance $\mathbf I$, a compatible interpretation $\pi$, and a tuple $\vec{t}$. Contributions equal to $\bot$ need no further consideration. Take an assignment $v$ to the variables of $P$, producing $\vec t$ and contributing a non-zero value, and consider the $\{=,\neq\}$-completion $P^c$ of $P$ obtained by adding $x=y$ or $x\neq y$ depending on whether $v(x)=v(y)$ or $v(x)\neq v(y)$. Take an identification $P^{\mathrm{id}}$ of $P^c$ by a canonical substitution $\theta$. We may observe that $\theta(x)=\theta(y)$ if and only if $v(x)=v(y).$ The contribution of $v$ to $\pi\llbracket P(\vec{t}) \rrbracket$ is also a contribution to
$\pi\llbracket P^{\mathrm{id}}(\vec{t}) \rrbracket$, since the new disequations added take value $\top$ by construction according to the assignment $v$. By assumption, there is a $\neq$-negation-homomorphism $h_{P^{\mathrm{id}}}\colon Q\longrightarrow P^{\mathrm{id}}.$ Then, $v\circ h_{P^{\mathrm{id}}}$ is an assignment of elements to the variables of $Q$. Every relational literal of $Q$ is mapped to a relational literal of $P^{\mathrm{id}}$, and all equations and disequations of $Q$ take value $\top$ under this assignment. Let $\varphi_v$ be the conjunction of the values of the relational literals of $P^{\mathrm{id}}$ under $v$, and let $\psi_v$ be the conjunction of the values of the relational literals of $Q$ under $v\circ h_{P^{\mathrm{id}}}$. Since every factor occurring in $\psi_v$ also occurs in $\varphi_v$, possibly with repetitions, we have $\varphi_v\leq_{\mathrm{PosBool}(X,\overline{X})}\psi_v.$ Indeed, repetitions do not affect the value, since multiplication in
$\mathrm{PosBool}(X,\overline{X})$ is idempotent, and a conjunction is below each conjunction obtained by dropping some of its factors. Hence every non-zero contribution to $\pi\llbracket P(\vec{t}) \rrbracket$ is below a contribution to $\pi\llbracket Q(\vec{t}) \rrbracket$. Since addition is given by join, it follows that $\pi\llbracket P(\vec{t}) \rrbracket
\leq_{\mathrm{PosBool}(X,\overline{X})}
\pi\llbracket Q(\vec{t}) \rrbracket.$ Therefore, $P\sqsubseteq^{\mathrm{pt}}_{\mathrm{PosBool}(X,\overline{X})}Q.$

Finally, the complexity calculations here follow exactly as for the case of distributive lattices.
\end{proof}

We say that a relational literal of $P^{\mathrm{id}}$ is inherited from $P$ if it is of the form $\theta(L)$, where $L$ is a relational literal occurring in the body of $P$ and $\theta$ is the canonical substitution used in the construction of $P^{\mathrm{id}}$. 

We associate with $P^{\mathrm{id}}$ a model-defining $\mathrm{Which}(X)$-interpretation, denoted by $\pi_{\mathrm{Which}(X)}^{P^{\mathrm{id}}}$, over the domain formed by the variables of $P^{\mathrm{id}}$. For every distinct relational literal $L$ of $P^{\mathrm{id}}$ inherited from $P$, let $x_L$ be a fresh variable in $X$, with $x_L\neq x_{L'}$ whenever $L\neq L'$, and set $\pi_{\mathrm{Which}(X)}^{P^{\mathrm{id}}}(L)=\{x_L\}.$ If $L$ occurs in the body of $P^{\mathrm{id}}$ but is not inherited from $P$, we set $\pi_{\mathrm{Which}(X)}^{P^{\mathrm{id}}}(L)=\emptyset.$ Finally, if the complementary literal of $L$ occurs in the body of
$P^{\mathrm{id}}$, we set $\pi_{\mathrm{Which}(X)}^{P^{\mathrm{id}}}(L)=\bot.$ Since $P^{\mathrm{id}}$ is a consistent completion in the sense of CQNs, for every ground relational atom over its domain exactly one between the atom and its negation occurs in $P^{\mathrm{id}}$. Hence $\pi_{\mathrm{Which}(X)}^{P^{\mathrm{id}}}$ is model-defining.

\begin{restatable}{Thm}{cqneqnContainmentL} \label{whichcqneqncont}
For CQ$^\neq$Ns $P$ and $Q$ with the same tuple of free variables, the following are equivalent. Here, an identification is formed by first taking an identification in the sense of CQ$^\neq$s and then a consistent relational completion over the common schema:
\begin{itemize}
\item[1.]
$P\sqsubseteq_{\mathrm{Which}(X)}^{\text{md}}Q$.

\item[2.]
For every consistent 
identification $P^{\mathrm{id}}$ of $P$, we have that $\pi_{\mathrm{Which}(X)}^{P^{\mathrm{id}}}(P) \leq_{\mathrm{Which}(X)} \pi_{\mathrm{Which}(X)}^{P^{\mathrm{id}}}(Q).$

\item[3.]
For every consistent 
identification $P^{\mathrm{id}}$ of $P$ and every
relational literal $L$ of $P^{\mathrm{id}}$ inherited from $P$ (i.e.\ for every relational literal $L$ of $P$), there is a
$\neq$-negation-homomorphism
\[
h_{P^{\mathrm{id}},L}\colon Q\longrightarrow P^{\mathrm{id}}
\]
such that $L$ belongs to the image of $h_{P^{\mathrm{id}},L}$.
\end{itemize}
Moreover, the corresponding decision problem (i.e.\ $P \sqsubseteq^{\mathrm{md}}_{\mathrm{Which}(X)} Q$?) is
\begin{itemize}
    \item[(i)] $\Pi^p_2$-complete for queries over schemas with arity bound $k$, for every $k$, and
    \item[(ii)] coNEXP-complete in general.
\end{itemize}
\end{restatable}


\begin{proof}
Let $Q(\vec{x}) = \exists\vec{z}
\bigl(A_1(\vec{x}_1)\wedge\ldots\wedge
A_n(\overline{x}_n) \bigr)$ and $P(\vec{z}) = \exists\vec{w}
\bigl(B_1(\vec{z}_1)\wedge\ldots\wedge
B_m(\vec{z}_m)\bigr),$ where the $A_i$s and $B_i$s may be relational literals, equations, or
disequations.

$(1)\implies(2):$
Assume that $P\sqsubseteq_{\mathrm{Which}(X)}^{\text{md}}Q.$ For every consistent identification $P^{\mathrm{id}}$ of $P$, the interpretation $\pi_{\mathrm{Which}(X)}^{P^{\mathrm{id}}}$ is
model-defining. Hence, by the containment assumption, $\pi_{\mathrm{Which}(X)}^{P^{\mathrm{id}}}(P)
\leq_{\mathrm{Which}(X)} \pi_{\mathrm{Which}(X)}^{P^{\mathrm{id}}}(Q).$

$(2)\implies(3):$
Assume that (2) holds. Consider a consistent identification $P^{\mathrm{id}}$ of $P$. Let $P^c$ be the $\{=,\neq\}$-completion from which $P^{\mathrm{id}}$ is obtained, and let $\theta$ be the corresponding canonical substitution. Let $L$ be a relational literal of $P^{\mathrm{id}}$ inherited from $P$. Then, $L=\theta(L_0)$ for some relational literal $L_0$ occurring in the body of $P$. Consider the assignment $v_\theta$ from the variables of $P$ given by $v_\theta(x)=\theta(x).$ Since $P^c$ is a completion of $P$, all equations and disequations occurring in $P$ are satisfied by $v_\theta$. Moreover, for every relational literal $B$ of $P$, the literal $\theta(B)$ occurs in $P^{\mathrm{id}}$ and is inherited from $P$. Therefore, $\pi_{\mathrm{Which}(X)}^{P^{\mathrm{id}}} \bigl(\theta(B)\bigr) = \{x_{\theta(B)}\}.$ In particular, the contribution of $v_\theta$ to $\pi_{\mathrm{Which}(X)}^{P^{\mathrm{id}}} (P(\theta(\vec{u})))$ is non-bottom and contains $x_L$. Thus, $x_L\in \pi_{\mathrm{Which}(X)}^{P^{\mathrm{id}}} \bigl(P(\theta(\vec{u}))\bigr).$ By assumption, $\pi_{\mathrm{Which}(X)}^{P^{\mathrm{id}}} \bigl(P(\theta(\vec{u}))\bigr) \leq_{\mathrm{Which}(X)} \pi_{\mathrm{Which}(X)}^{P^{\mathrm{id}}} \bigl(Q(\theta(\vec{u}))\bigr),$ and hence $x_L\in \pi_{\mathrm{Which}(X)}^{P^{\mathrm{id}}} \bigl(Q(\theta(\vec{u}))\bigr).$ It follows that there is an assignment $v$ to the variables of $Q$ over the domain of $P^{\mathrm{id}}$, sending its free variables to $\theta(\vec{u})$, such that the value of the body of $Q$ under $v$ is non-bottom and contains $x_L$. Since this contribution is non-bottom, every relational literal $A_i(\vec{x}_i)$ occurring in $Q$ has non-bottom value under $v$. By the definition of $\pi_{\mathrm{Which}(X)}^{P^{\mathrm{id}}}$, this means that $A_i\bigl(v(\overline{x}_i)\bigr)$ occurs in the body of $P^{\mathrm{id}}$. Moreover, all equations and disequations of $Q$ are satisfied under $v$, since otherwise the contribution would be bottom. Therefore, by setting $h_{P^{\mathrm{id}},L}(x)=v(x),$ we obtain a $\neq$-negation-homomorphism $h_{P^{\mathrm{id}},L}\colon Q\longrightarrow P^{\mathrm{id}}.$ Finally, the token $x_L$ occurs only in the annotation of $L$: different inherited relational literals receive different singleton annotations, while relational literals added only in the final completion receive $\emptyset$. Since the contribution of $v$ contains $x_L$, there is a relational literal $A_i(\vec{x}_i)$ of $Q$ such that $A_i\bigl(h_{P^{\mathrm{id}},L}(\overline{x}_i)\bigr)=L.$ Thus, $L$ belongs to the image of $h_{P^{\mathrm{id}},L}$.

$(3)\implies(1):$
Assume that (3) holds. Consider an arbitrary $\mathrm{Which}(X)$-database instance $\mathbf I$, a compatible model-defining interpretation $\pi$, and a tuple $\vec{t}$. We show that $\pi\llbracket P(\vec{t}) \rrbracket \leq_{\mathrm{Which}(X)} \pi\llbracket Q(\vec{t}) \rrbracket.$ If $\pi\llbracket P(\vec{t}) \rrbracket=\bot$, then the inequality is immediate. Suppose $\pi\llbracket P(\vec{t}) \rrbracket\neq\bot$. First, choose any assignment giving a non-bottom contribution to $P(\vec t)$, form its consistent identification and relational completion as below, and choose any inherited relational literal. By item (3), a corresponding homomorphism from $Q$ exists. Composing it with the chosen assignment makes every relational literal of $Q$ non-bottom and satisfies every comparison, so $\pi\llbracket Q(\vec t)\rrbracket\neq\bot$. This also handles the case $\pi\llbracket P(\vec t)\rrbracket=\emptyset$. Now let $x\in\pi\llbracket P(\vec{t}) \rrbracket$.  Then, there is an assignment $v$ to the variables of $P$, sending $\vec{u}$ to $\vec{t}$, whose contribution to $\pi\llbracket P(\vec{t}) \rrbracket$ is non-bottom and contains $x$. Since the product of non-bottom elements in $\mathrm{Which}(X)$ is their union, there is a relational literal $L_0$ in the body of $P$ such that $x\in\pi\bigl(L_0(v)\bigr).$ Consider the $\{=,\neq\}$-completion $P^c_v$ of $P$ obtained by adding $y=z$ or $y\neq z$ according to whether $v(y)=v(z)$ or $v(y)\neq v(z)$. Let $\theta$ be the corresponding canonical substitution. We keep denoting by $v$ the induced assignment on the variables occurring after the substitution, defined by $v(\theta(y))=v(y).$ This is well-defined by the construction of $P^c_v$. We complete the resulting query in the sense of CQNs as follows. For every relation symbol $R$ in the common schema and every tuple $\vec{z}$ of its variables of the appropriate arity, we add $R(\vec{z})$ if $\pi\bigl(R(v(\vec{z}))\bigr)\neq\bot,$ and we add $\neg R(\vec{z})$ if $\pi\bigl(\neg  R(v(\vec{z}))\bigr)\neq\bot.$ Since $\pi$ is model-defining, exactly one of these two values is different from $\bot$. Hence this construction yields a consistent identification $P^{\mathrm{id}}_v$ of $P$. Moreover, every relational literal occurring in $P^{\mathrm{id}}_v$ has non-bottom value under $v$. Let $L=\theta(L_0),$ then $L$ is a relational literal of $P^{\mathrm{id}}_v$ inherited from $P$ and $x\in\pi\bigl(L(v)\bigr).$ By assumption, there is a $\neq$-negation-homomorphism $h_{P^{\mathrm{id}}_v,L}\colon Q\longrightarrow P^{\mathrm{id}}_v$ such that $L$ belongs to the image of $h_{P^{\mathrm{id}}_v,L}$. Consider the assignment $v\circ h_{P^{\mathrm{id}}_v,L}$ to the variables of $Q$. For every relational literal $A_i(\vec{x}_i)$ occurring in $Q$, the literal $A_i\bigl(h_{P^{\mathrm{id}}_v,L}(\overline{x}_i)\bigr)$ occurs in $P^{\mathrm{id}}_v$. By the construction of $P^{\mathrm{id}}_v$, we have $\pi\Bigl(A_i\bigl( v(h_{P^{\mathrm{id}}_v,L}(\overline{x}_i)) \bigr) \Bigr) \neq\bot.$ Furthermore, since $h_{P^{\mathrm{id}}_v,L}$ is a $\neq$-negation-homomorphism, all equations and disequations in $Q$ are satisfied by $v\circ h_{P^{\mathrm{id}}_v,L}$. Therefore, this assignment gives a non-bottom contribution to $\pi\llbracket Q(\vec{t}) \rrbracket$. Since $L$ belongs to the image of $h_{P^{\mathrm{id}}_v,L}$, there is a relational literal $A_i(\vec{x}_i)$ of $Q$ such that $A_i\bigl(h_{P^{\mathrm{id}}_v,L}(\overline{x}_i)\bigr)=L.$ Consequently, $A_i\bigl(v(h_{P^{\mathrm{id}}_v,L}(\overline{x}_i)) \bigr) = L(v).$ Since $x\in\pi\bigl(L(v)\bigr),$ and the product of the non-bottom values of the relational literals of $Q$ is their union, the contribution of $v\circ h_{P^{\mathrm{id}}_v,L}$ contains $x$. Hence $x\in\pi\llbracket Q(\vec{t}) \rrbracket.$ We have shown that every token belonging to $\pi\llbracket P(\vec{t}) \rrbracket$ also belongs to $\pi\llbracket Q(\vec{t}) \rrbracket$. Therefore, $\pi\llbracket P(\vec{t}) \rrbracket \leq_{\mathrm{Which}(X)} \pi\llbracket Q(\vec{t}) \rrbracket.$ Since $\mathbf I$, $\pi$, and $\vec{t}$ were arbitrary, we conclude that $P\sqsubseteq_{\mathrm{Which}(X)}^{\text{md}}Q.$

For the complexity calculation, the lower bounds follow from the hardness of the corresponding containment problem for CQNs, which are a special case of CQ$^\neq$Ns. For the upper bounds, by item 3, the complement of the containment problem is in $\mathrm{NP}^{\mathrm{coNP}}=\Sigma^p_2$ if there is a fixed arity bound on the relations used, while it is in $\mathrm{NEXP}$ otherwise. Indeed, we can guess a consistent identification $P^{\mathrm{id}}$ of $P$ and a relational literal $L$ of $P^{\mathrm{id}}$ inherited from $P$, and check that there is no $\neq$-negation-homomorphism
$h_{P^{\mathrm{id}},L}\colon Q\longrightarrow P^{\mathrm{id}}$ such that $L$ belongs to the image of $h_{P^{\mathrm{id}},L}$. The $\{=,\neq\}$-completion and the identification have polynomial size, while the subsequent completion in the sense of CQNs may add, for a relation of arity $k$ and at most $n$ variables, up to $n^k$ relational literals. Hence, if $k$ is fixed, $P^{\mathrm{id}}$ has polynomial size, whereas in the general case it may have exponential size. Thus, the containment problem is in $\Pi^p_2$ in the bounded-arity case and in $\mathrm{coNEXP}$ in general.
\end{proof}

\section{Omitted  Proofs from \S~\ref{sec:cqneqn-equivalence}} \label{app:cqneqn-equivalence-proofs}

\cqneqnEquivalenceA*

\begin{proof}
Item (1) is immediate by definition. For item (2), assume first that
$P \equiv^{\mathcal{C}}_\mathbb{L} P\wedge Q$. Then, for each $\mathbb{L}$-instance $\bf I$, interpretation $\pi$ in $\mathcal{C}$ compatible with $\bf I$, and tuple $\vec{t}$, we have $\pi\llbracket P(\vec{t})\rrbracket
= \pi\llbracket P(\vec{t})\rrbracket \cdot \pi\llbracket Q(\vec{t})\rrbracket.$ Since multiplication in a bounded distributive lattice is meet, this is equivalent to $\pi\llbracket P(\vec{t})\rrbracket \leq_\mathbb{L} \pi\llbracket  Q(\vec{t})\rrbracket.$ Thus, $P \sqsubseteq^{\mathcal{C}}_\mathbb{L} Q$. Conversely, if $P \sqsubseteq^{\mathcal{C}}_\mathbb{L} Q$, then for every such $\bf I$, $\pi$, and $\vec{t}$, $\pi\llbracket P(\vec{t})\rrbracket
= \pi\llbracket P(\vec{t})\rrbracket \cdot \pi\llbracket Q(\vec{t})\rrbracket,$ and hence $P \equiv^{\mathcal{C}}_\mathbb{L} P\wedge Q$.
\end{proof}

\cqneqnEquivalenceB*

\begin{proof}
Membership follows from Lemma~\ref{equivlatneqneg}, item (1), and the corresponding upper bounds for containment. For bounded distributive lattices with more than two elements under model-defining interpretations, the lower bound follows from Theorem~\ref{cqneqequivalence}, by restricting to CQ$^\neq$s without negated relational atoms. This gives $\Pi^p_2$-hardness. For $\mathbb{B}$ under model-defining interpretations, the lower bounds follow from Corollary~\ref{cqnslattice}, since model-defining and dually model-defining interpretations coincide over $\mathbb{B}$, and CQNs are a particular case of CQ$^\neq$Ns. This gives $\Pi^p_2$-hardness for bounded arity and coNEXP-hardness in the general case. 
\end{proof}

\cqneqnEquivalenceC*

\begin{proof}
The upper bounds follow immediately from item (1) of Lemma~\ref{equivlatneqneg}, which, as observed above, holds for any naturally ordered semiring, together with the corresponding containment results. For $\mathrm{PosBool}(X,\overline{X})$, we can obtain the matching lower bound. Indeed, $\mathrm{PosBool}(X,\overline{X})$ is a bounded distributive lattice and Lemma~\ref{equivlatneqneg} applies also to provenance-tracking interpretations. Thus, by item (2), containment reduces to equivalence, and the result follows from the $\Pi^p_2$-hardness of $\mathrm{PosBool}(X,\overline{X})$-containment established above.

For $\mathrm{Sorp}(X,\overline{X})$, $\mathrm{Why}(X,\overline{X})$, and $\mathbb{B}[X,\overline{X}]$, which are commutative and have idempotent addition, we use the same reduction from $3$-choosability for bipartite graphs as in the proof of Theorem~\ref{cqneqequivalHard}. The queries $P_G$ and $Q_G$ constructed there are CQ$^\neq$s, and hence also CQ$^\neq$Ns. We only need to check that the reduction works also for provenance-tracking interpretations. If $G$ is $3$-choosable, the two queries are equivalent over arbitrary instances, and hence also under provenance-tracking interpretations. If $G$ is not $3$-choosable, the counterexample instance constructed in that proof uses only annotations $0$ and $1$. We can then define a provenance-tracking interpretation compatible with this instance by assigning value $0$ to every negated ground relational atom. Since $P_G$ and $Q_G$ contain no negated relational atoms, their values remain $1$ and $0$, respectively. Thus, the same reduction gives $\Pi^p_2$-hardness in all three cases.
\end{proof}

\cqneqnEquivalenceD*

\begin{proof}
By item (1) of Lemma~\ref{equivlatneqneg}, which holds for any naturally ordered semiring, two CQ$^\neq$Ns $P$ and $Q$ are equivalent over Which$(X)$ if and only if $P \sqsubseteq^{\text{md}}_{\mathrm{Which}(X)} Q$ and $Q \sqsubseteq^{\text{md}}_{\mathrm{Which}(X)} P.$ The result then follows immediately from the corresponding upper bounds for containment established above, namely, $\Pi^p_2$ in the presence of a fixed arity bound and coNEXP in the general case.

For the lower bounds, every CQN is also a CQ$^\neq$N without disequations.  The two lower bounds therefore follow from Theorem~\ref{equivalencewhichmdcqn}.
\end{proof}

\section{Omitted  Proofs from \S~\ref{sec:ucqneqn-containment}} \label{app:ucqneqn-containment-proofs}

\ucqneqnContainmentA*

\begin{proof}
  $(1)\implies(2):$
Suppose that $\overline P\sqsubseteq_{\mathbb{L}}^{\mathrm{md}}\overline Q$. Fix $1\leq i\leq n$ and consider a consistent identification $P_i^{\mathrm{id}}$ of a completion (in the sense of CQ$^\neq$s) $P_i^c$ of $P_i$. Let $\theta$ be the canonical substitution used to obtain $P_i^{\mathrm{id}}$ from $P_i^c$. As in the proof of Theorem~\ref{cqneqnegmd}, build a canonical database $\mathbf I$ whose elements are the variables of $P_i^{\mathrm{id}}$, and, for every relation $R$, let $R^\mathbf{I}(x_1, \ldots, x_n) = \top$ if $R(x_1, \ldots, x_n)$ is an atom in  $P^{\mathrm{id}}_i$,  $R^\mathbf{I}(x_1, \ldots, x_n) = \bot$  if  $\neg R(x_1, \ldots, x_n)$ is a literal in  $P^{\mathrm{id}}_i$, and $R^\mathbf{I}(x_1, \ldots, x_n) = \epsilon$ otherwise. 
Consider next a model-defining interpretation $\pi^*$ compatible with $\mathbf I$ defined by $\pi^*\llbracket R(\overline a) \rrbracket=R^\mathbf I(\overline a)$, $\pi^*\llbracket \neg R(\overline a) \rrbracket=\top$ if $\pi^*\llbracket R(\overline a) \rrbracket=\bot$, and $\pi^*\llbracket \neg R(\overline a) \rrbracket=\bot$ otherwise. Consider the assignment $v_\theta$ to the variables of $P_i$ defined by $v_\theta(x)=\theta(x)$. By construction of the completion, all equations and disequations occurring in $P_i$ are satisfied by $v_\theta$. Moreover, every relational literal $L$ of $P_i$ is mapped to the relational literal $\theta(L)$ occurring in $P_i^{\mathrm{id}}$, and hence it takes value $\top$ under $\pi^*$. Therefore, the contribution of $v_\theta$ to $\pi^*\llbracket P_i(\theta(\overline u)) \rrbracket$ is $\top$, and thus $\pi^*\llbracket P_i(\theta(\overline u)) \rrbracket=\top.$ Since $P_i$ is one of the disjuncts of $\overline P$, we also have $\pi^*\llbracket \overline P(\theta(\overline u)) \rrbracket=\top.$ By the containment assumption, $\pi^*\llbracket \overline Q(\theta(\overline u)) \rrbracket=\top.$

Notice that, by construction of $\pi^*$, every relational literal takes value in the set $\{\bot,\epsilon,\top\}$, while equations and disequations take value either $\bot$ or $\top$. Hence, if the value of the union $\overline Q$ is $\top$, there must be some $1\leq j\leq m$ such that $\pi^*\llbracket Q_j(\theta(\overline u)) \rrbracket=\top,$ and, similarly, there must be an assignment $v$ to the variables of $Q_j$, producing $\theta(\overline u)$, whose contribution is $\top$.

It follows that every relational literal of $Q_j$ takes value $\top$ under $v$, and all equations and disequations occurring in $Q_j$ are satisfied. By construction of $\pi^*$, a relational literal takes value $\top$ only if it occurs in $P_i^{\mathrm{id}}$. Therefore, the mapping $h_{P_i^{\mathrm{id}}}$ defined by $h_{P_i^{\mathrm{id}}}(x):=v(x)$ is a $\neq$-negation-homomorphism from $Q_j$ to $P_i^{\mathrm{id}}$. Indeed, it preserves the head and all relational literals, while equations and disequations are preserved since $P_i^{\mathrm{id}}$ contains the complete description of the equality type of its variables. Thus item 2 follows.  

$(2)\implies(1):$
Assume that item 2 holds. Consider an arbitrary $\mathbb L$-database instance $\mathbf I$, a compatible model-defining interpretation $\pi$, and a tuple $\vec t$. We want to show that $\pi\llbracket \overline P(\vec t) \rrbracket \leq_{\mathbb L} \pi\llbracket \overline Q(\vec t) \rrbracket.$ It is enough to show that every contribution of an assignment to every disjunct $P_i$ is below $\pi\llbracket \overline Q(\vec t) \rrbracket$. Fix $1\leq i\leq n$ and an assignment $v$ to the variables of $P_i$ producing $\vec t$. If some equation or disequation of $P_i$ is not satisfied, its contribution is $\bot$, and there is nothing to prove. The same holds if one of the relational literals has value $\bot$. We may therefore suppose that the contribution of $v$ is different from $\bot$. Let $P_{i,v}^c$ be the completion of $P_i$ obtained by adding $x=y$ if $v(x)=v(y)$ and $x\neq y$ if $v(x)\neq v(y)$, and let $P_{i,v}^{\mathrm{id}}$ be the corresponding identification, obtained by a canonical substitution $\theta_v$.

We first observe that $P_{i,v}^{\mathrm{id}}$ is consistent: suppose that both $R(\vec y)$ and $\neg R(\vec y)$ occur in $P_{i,v}^{\mathrm{id}}$, then there are relational literals $R(\overline x)$ and $\neg R(\overline x')$ in $P_i$ such that $\theta_v(\overline x)=\theta_v(\overline x').$ By construction of the completion, this implies $v(\overline x)=v(\overline x').$ Since $\pi$ is model-defining, one of
$\pi\llbracket R(v(\overline x)) \rrbracket$ and $\pi\llbracket \neg R(v(\overline x)) \rrbracket$ must be $\bot$, and hence the contribution of $v$ would be $\bot$, a contradiction. Thus $P_{i,v}^{\mathrm{id}}$ is a consistent identification of $P_i$.

By item 2, there are some $1 \leq j\leq m$ and a $\neq$-negation-homomorphism $h_{P_{i,v}^{\mathrm{id}}}\colon Q_j\longrightarrow P_{i,v}^{\mathrm{id}}.$
Define an assignment $v_*$ on the variables of $P_{i,v}^{\mathrm{id}}$ by $v_*(\theta_v(x)):=v(x),$ this is well-defined by construction. Consider now the assignment $w:=v_*\circ h_{P_{i,v}^{\mathrm{id}}}$ to the variables of $Q_j$. Since $h_{P_{i,v}^{\mathrm{id}}}$ preserves the head, $w$ produces $\vec t$, moreover all equations and disequations of $Q_j$ are satisfied under $w$.

Let $B_1,\ldots,B_r$ be the relational literal occurrences in the body of $P_i$, and let $A_1,\ldots,A_q$ be those occurring in $Q_j$. For every $1\leq s\leq q$, the relational literal $h_{P_{i,v}^{\mathrm{id}}}(A_s)$ occurs in $P_{i,v}^{\mathrm{id}}$. Hence there is some relational literal occurrence $B_{\rho(s)}$ of $P_i$ such that
$h_{P_{i,v}^{\mathrm{id}}}(A_s)=\theta_v(B_{\rho(s)}).$ Therefore, $\pi\llbracket w(A_s) \rrbracket = \pi\llbracket v_*(h_{P_{i,v}^{\mathrm{id}}}(A_s)) \rrbracket=\pi\llbracket v(B_{\rho(s)}) \rrbracket.$

Since all equations and disequations are satisfied, they take value $\top$ and do not affect the infima. Let $c=
\bigwedge_{1\leq k\leq r}\pi\llbracket v(B_k) \rrbracket$ be the contribution of $v$ to $P_i$, and let $d=\bigwedge_{1\leq s\leq q}\pi\llbracket w(A_s) \rrbracket$ be the contribution of $w$ to $Q_j$. For every $s\leq q$, we have $c\leq_{\mathbb L}\pi\llbracket v(B_{\rho(s)}) \rrbracket=\pi\llbracket w(A_s) \rrbracket.$
Thus $c$ is a lower bound of all the values occurring in the infimum defining $d$, and hence $c\leq_{\mathbb L}d.$ Moreover, $d\leq_{\mathbb L}\pi\llbracket Q_j(\vec t) \rrbracket
\leq_{\mathbb L}\pi\llbracket \overline Q(\vec t) \rrbracket.$ Therefore, every contribution of every assignment to $P_i$ is below $\pi\llbracket \overline Q(\vec t) \rrbracket$. Taking the supremum over all assignments gives $\pi\llbracket P_i(\vec t) \rrbracket\leq_{\mathbb L}\pi\llbracket \overline Q(\vec t) \rrbracket.$
Since this holds for every $1\leq i\leq n$, taking the supremum over all the disjuncts gives $\pi\llbracket \overline P(\vec t) \rrbracket \leq_{\mathbb L} \pi\llbracket \overline Q(\vec t) \rrbracket.$
Since $\mathbf I$, $\pi$, and $\vec t$ were arbitrary, we conclude that $\overline P\sqsubseteq_{\mathbb L}^{\mathrm{md}}\overline Q.$

For the complexity calculation, the lower bound follows immediately from Theorem~\ref{cqneqnegmd}, since CQ$^\neq$Ns are a particular case of UCQ$^\neq$Ns. For the upper bound, by item 2, non-containment can be checked by nondeterministically guessing a disjunct $P_i$ and a consistent identification $P_i^{\mathrm{id}}$. Such an identification has polynomial size. For fixed $i$ and $P_i^{\mathrm{id}}$, deciding whether there are some $1 \leq j\leq m$ and a $\neq$-negation-homomorphism $h_{P_i^{\mathrm{id}}}\colon Q_j\longrightarrow P_i^{\mathrm{id}}$ is in NP. Hence the absence of such a $j$ and homomorphism can be checked in coNP. Therefore, non-containment is in $\mathrm{NP}^{\mathrm{coNP}}=\Sigma^p_2,$ and containment is in $\Pi^p_2$. Together with the lower bound, the problem is $\Pi^p_2$-complete. 
\end{proof}

\ucqneqnContainmentB*

\begin{proof}
$(1) \implies (2):$ 
Suppose that $\bigcup_{1\leq i \leq n} P_i \sqsubseteq_{\mathrm{Sorp}(X, \overline{X})}^{\text{pt}} \bigcup_{1\leq i \leq m} Q_i$ and consider an identification $P^{\mathrm{id}}_i$ of a completion $P^c_i$ of $P_i$ for $1\leq i \leq n$. We have that  $P^{\mathrm{id}}_i \sqsubseteq_{\mathrm{Sorp}(X, \overline{X})}^{\text{pt}} P^c_i \sqsubseteq_{\mathrm{Sorp}(X, \overline{X})}^{\text{pt}} \bigcup_{1\leq i \leq n} P_i $ since identification preserves query values, completion only removes assignment contributions, and addition is monotone in the natural order. Thus, 
$\pi^{\mathrm{can}_{\mathrm{Sorp}(X)}(P^{\mathrm{id}}_i)}\llbracket P^{\mathrm{id}}_i \rrbracket  \leq_{\mathrm{Sorp}(X, \overline{X})} \pi^{\mathrm{can}_{\mathrm{Sorp}(X)}(P^{\mathrm{id}}_i)}\llbracket P_i^c \rrbracket     \leq_{\mathrm{Sorp}(X, \overline{X})} 
\pi^{\mathrm{can}_{\mathrm{Sorp}(X)}(P^{\mathrm{id}}_i)}\llbracket P_i \rrbracket \leq_{\mathrm{Sorp}(X, \overline{X})} \pi^{\mathrm{can}_{\mathrm{Sorp}(X)}(P^{\mathrm{id}}_i)}\llbracket \bigcup_{1\leq i \leq m} Q_i \rrbracket$.

$(2) \implies (3):$ Assume that (2) holds, so
for each $1\leq i \leq n$, every identification $P^{\mathrm{id}}_i$ of $P_i$, we have that

$$\pi^{\mathrm{can}_{\mathrm{Sorp}(X)}(P_i^{id})}\llbracket P_i^{id}(\vec{u}) \rrbracket \leq_{\mathrm{Sorp}(X, \overline{X})} \pi^{\mathrm{can}_{\mathrm{Sorp}(X)}(P_i^{id})}\llbracket  (\bigcup_{1\leq i \leq m} Q_i) (\vec{u}) \rrbracket.$$

where $\vec{u}$ is the tuple of free variables of $P^{\mathrm{id}}_i$. The polynomial $\pi^{\mathrm{can}_{\mathrm{Sorp}(X)}(P_i^{id})}\llbracket P_i^{id}(\vec{u}) \rrbracket$ has as a term a monomial $x_1 \cdot \ldots \cdot x_n $ containing every tag (with some exponent, accounting for duplicate atoms) from $X \cup \overline{X}$ used to annotate literals by $\pi^{\mathrm{can}_{\mathrm{Sorp}(X)}(P_i^{id})}$ (such monomial is obtained using the identity assignment for the variables in $ P^{\mathrm{id}}_i$; noticing that the disequations in $P^{\mathrm{id}}_i$ all take value $1$ by construction of the comparison completion; the canonical relational tags are pairwise noncomplementary). Note that this monomial cannot be absorbed by another one, since every non-zero monomial produced by an assignment to $P^{\mathrm{id}}_i$ has the same total degree, while a proper divisor would have strictly smaller degree.  
Then, by the definition of $\leq $ for $\mathrm{Sorp}(X, \overline{X})$, we must have that there is a  monomial $y_1 \cdot \ldots \cdot y_m $  in $\pi^{\mathrm{can}_{\mathrm{Sorp}(X)}(P_i^{id})}\llbracket  (\bigcup_{1\leq i \leq m} Q_i) (\vec{u}) \rrbracket$ where only tags from $x_1 \cdot \ldots \cdot x_n $ appear, with equal or smaller exponents as in $x_1 \cdot \ldots \cdot x_n $. Since $\bigcup_{1\leq i \leq m} Q_i$ is a  UCQ$^\neq$N, we must have that there is a $Q_j$ with $j\leq m$ and an assignment $v$ to the bound variables in $Q_j$ over $\mathrm{can}_{\mathrm{Sorp}(X)}(P^{\mathrm{id}}_i)$ such that the body (i.e.\ the quantifier-free matrix) of $Q_j$ evaluates to the monomial $y_1 \cdot \ldots \cdot y_m$. Since every tag occurring in $y_1, \ldots, y_m$ also occurs in $x_1, \ldots, x_n$, and canonical tags uniquely identify relational literals, every relational literal of $Q_j$ is mapped by $v$ to a relational literal of $P^{\mathrm{id}}_i$.
Furthermore, the equations and disequations in $Q_j$ must all take value $1$ under the assignment $v$ in order to obtain the monomial  $y_1 \cdot \ldots \cdot y_m$ (instead of just $0$), and thus they must be  equations holding in $\mathrm{can}_{\mathrm{Sorp}(X, \overline{X})}(P^{\mathrm{id}}_i)$ (and thus logically entailed by the body of $P^{\mathrm{id}}_i$) and disequations appearing in the body of $P^{\mathrm{id}}_i$ (since the latter contains a complete description of disequations in  $\mathrm{can}_{\mathrm{Sorp}(X)}(P^{\mathrm{id}}_i)$).
Then, we get the desired 
$\neq$-negation-homomorphism $h_{P^{\mathrm{id}}_i}\colon Q_j \longrightarrow P^{\mathrm{id}}_i$ injective for relational literals by letting $h_{P^{\mathrm{id}}_i}(x)$ be the variable $v(x)$ (recall that the elements of the active domain of $\mathrm{can}_{\mathrm{Sorp}[X]}(P^{\mathrm{id}}_i)$ are the variables of $P^{\mathrm{id}}_i$).
To see the injectivity, notice that any tag $p$ that is the value of a literal of $Q_j$ according to $v$, appears in $y_1 \cdot \ldots \cdot y_m$ with at most the same multiplicity as in  $x_1 \cdot \ldots \cdot x_n $ (since this monomial corresponds to the body of $P^{\mathrm{id}}_i$), and thus the multiset of images by $h_{P^{\mathrm{id}}_i}$ of literals of $Q_j$ is bag-contained in the multiset of literals from $P^{\mathrm{id}}_i$.
 
$(3) \implies (1):$ Assume that (3) holds and we want to show that $\bigcup_{1\leq i \leq n} P_i \sqsubseteq_{\mathrm{Sorp}(X, \overline{X})}^{\text{pt}}  \bigcup_{1\leq i \leq m} Q_i$. Thus, for each $1\leq i \leq n$, every identification $P^{\mathrm{id}}_i$ of $P_i$, there is a $j \leq m$ and a $\neq$-negation-homomorphism 
$h_{P^{\mathrm{id}}_i} \colon Q_j \longrightarrow P^{\mathrm{id}}_i$ injective for relational literals. In order to establish $\bigcup_{1\leq i \leq n} P_i \sqsubseteq^{\text{pt}}_{\mathrm{Sorp}(X, \overline{X})}  \bigcup_{1\leq i \leq m} Q_i$, consider an arbitrary $\mathrm{Sorp}(X)$-database instance $\mathbf I$, a compatible interpretation $\pi$ and a tuple $\vec{t}$. We need to show that for any monomial in 
$\pi^\mathbf{I}\llbracket \bigcup_{1\leq i \leq n} P_i(\vec{t}) \rrbracket$ there is a monomial  with  equal or smaller exponents  in $\pi^\mathbf{I}\llbracket  \bigcup_{1\leq i \leq m} Q_i(\vec{t}) \rrbracket$. 
Take an assignment $v$ of elements to the variables of $P_i$ (where $i\leq n$) justifying a monomial $x_1  \cdot \ldots \cdot x_n $ and consider the completion $ P^c_i$ of $P_i$ obtained by adding $x=y$ or $x\neq y$ depending on whether  $v(x) = v(y)$ or $v(x) \neq v(y)$. Take an identification $P^{\mathrm{id}}_i$ of $P^c_i$ by a canonical substitution $\theta$. We may observe that $\theta(x) = \theta(y)$ iff $v(x) = v(y)$. This identification may contain opposite relational literals; unrestricted provenance-tracking interpretations do not force their annotations to be complementary. Then, $x_1 \cdot \ldots \cdot  x_n $ is also a monomial in $\pi^\mathbf{I}\llbracket P^{\mathrm{id}}_i(\vec{t}) \rrbracket$ justified by the assignment $v$ itself since the new  disequations added can only take value $1$ by construction according to the assignment $v$.
The assignment  $v \circ h_{P^{\mathrm{id}}_i} $  justifies a   monomial with smaller or equal exponents in $\pi^\mathbf{I}\llbracket Q_j(\vec{t}) \rrbracket$. This is because for any literal $A(\overline{x})$ of $Q_j$ such that $A(h_{P^{\mathrm{id}}_i}(\overline{x})) = B(\theta(\vec{z}))$, where  $B(\vec{z})$ is a literal of  $P_i$, if 
$x_i = \pi\llbracket B(v(\theta(\vec{z}))) \rrbracket$, then $x_i = \pi\llbracket A(h_{P^{\mathrm{id}}_i}(\theta(\vec{z}))) \rrbracket$. 
By the injectivity of $h_{P^{\mathrm{id}}_i}$, the images of the literals in the body of $Q_j$ (considered as a multiset) appear with with at most the same multiplicity as in the body of $P^{\mathrm{id}}_i$ considered as multiset, which entails that $x_i$ appears as the value of a literal of $Q_j$ with at most the same multiplicity as it appears in the monomial $x_1  \cdot \ldots \cdot x_n $ (since the latter corresponds to the body of $P^{\mathrm{id}}_i$). Since equations and disequations can only take values $0$ or $1$, they can be disregarded in showing that we can justify a monomial.

Finally, the lower bound for the complexity calculation follows immediately from the $\Pi^p_2$-hardness of containment for CQ$^\neq$Ns over $\mathrm{Sorp}(X,\overline{X})$, since CQs are a particular case of UCQs.
 For the upper bound, observe that the complement of the containment problem is in $\mathrm{NP}^{\mathrm{coNP}}=\mathrm{NP}^{\mathrm{NP}}=\Sigma^p_2$. Hence the containment problem is in $\Pi^p_2$, and therefore it is $\Pi^p_2$-complete.
\end{proof}

\ucqneqnContainmentC*

\begin{proof}
$(1)\Rightarrow(2):$
Suppose that $\bigcup_{1\leq i\leq n}P_i
\sqsubseteq^{\text{pt}}_{\text{Why}(X,\overline X)}
\bigcup_{1\leq j\leq m}Q_j$ and consider an identification $P_i^{id}$ of a completion (in the sense of CQ$^{\neq}$s) $P_i^c$ of some $P_i$. By construction, $P_i^{id}$ is semantically equivalent to $P_i^c$, while $P_i^c$ is obtained from $P_i$ only by adding equations and disequations. Since these can only eliminate assignments, and take value $1$ whenever they are satisfied, we have $P_i^{id}
\sqsubseteq^{\text{pt}}_{\text{Why}(X,\overline X)}
P_i^c \sqsubseteq^{\text{pt}}_{\text{Why}(X,\overline X)} P_i.$ Moreover, $P_i \sqsubseteq^{\text{pt}}_{\text{Why}(X,\overline X)} \bigcup_{1\leq i\leq n}P_i.$ Hence, by the containment assumption, $P_i^{id}\sqsubseteq^{\text{pt}}_{\text{Why}(X,\overline X)} \bigcup_{1\leq j\leq m}Q_j.$ The interpretation $\pi^{\text{can}_{\text{Why}(X)}((P_i^{id})^+)}$ is provenance-tracking, and therefore the latter containment holds, in particular, on this interpretation. Thus $\pi^{\text{can}_{\text{Why}(X)}((P_i^{id})^+)}\llbracket P_i^{id} \rrbracket \leq_{\text{Why}(X,\overline X)} \pi^{\text{can}_{\text{Why}(X)}((P_i^{id})^+)}
\llbracket \bigcup_{1\leq j\leq m}Q_j \rrbracket,$ as required.

$(2)\Rightarrow(3):$
Assume that item 2 holds and fix an identification $P_i^{id}$ of some $P_i$. Let $\mathbf I:=\text{can}_{\text{Why}(X)}((P_i^{id})^+)$ and abbreviate $\pi^{\text{can}_{\text{Why}(X)}((P_i^{id})^+)}$ as $\pi^{\mathbf I}$. Let $\text{LitSet}(P_i^{id}) = \{L_1,\ldots,L_r\}.$ By construction of the canonical interpretation $\pi^{\mathbf I}$, each distinct relational literal $L_k$ is annotated by a different tag $x_k\in X\cup\overline X$, that is, $\pi^{\mathbf I}\llbracket L_k \rrbracket=\{\{x_k\}\}.$ Repeated occurrences of the same relational literal receive the same annotation. Moreover, the tags for negative literals are chosen so that their complementary positive tags are not used, hence no complementary pair occurs among $x_1,\ldots,x_r$.

Let $W_{P_i^{id}} :=\{x_1,\ldots,x_r\}.$ Consider the assignment to the variables of $P_i^{id}$ that sends every variable to itself in $\mathbf I$. Under this assignment, all equations and disequations of $P_i^{id}$ are satisfied, while every occurrence of $L_k$ contributes $\{\{x_k\}\}$. Since multiplication in $\text{Why}(X,\overline X)$ is pairwise union, $W_{P_i^{id}}\in\pi^{\mathbf I}\llbracket P_i^{id}(\overline u) \rrbracket.$

By item 2, $W_{P_i^{id}}\in\pi^{\mathbf I}\llbracket \bigcup_{1\leq j\leq m}Q_j(\overline u) \rrbracket.$ Since the evaluation of a union is the union of the evaluations of its disjuncts, then there is some $1\leq j\leq m$ such that $W_{P_i^{id}} \in \pi^{\mathbf I}\llbracket Q_j(\overline u) \rrbracket.$ Hence there is an assignment $w$ to the variables of $Q_j$ producing $\overline u$ whose contribution contains $W_{P_i^{id}}$.

Under a provenance-tracking interpretation, a fixed non-zero contribution consists of a single witness. Thus the contribution of $w$ is exactly $\{W_{P_i^{id}}\}.$ Since the domain of the canonical interpretation consists of the variables of $P_i^{id}$, the assignment $w$ induces a map $h_{P_i^{id}}\colon Q_j\longrightarrow P_i^{id}.$ This map preserves the head, and every relational literal of $Q_j$ must be mapped to a relational literal having non-zero value under $\pi^{\mathbf I}$. By construction of the canonical interpretation, these are precisely the relational literals of $P_i^{id}$. Equations and disequations are also preserved, hence $h_{P_i^{id}}$ is a $\neq$-negation-homomorphism.

It remains to show that it is set-onto for relational literals. Let $L_k\in\text{LitSet}(P_i^{id}).$ Since $x_k\in W_{P_i^{id}},$ and the contribution of $w$ is exactly $\{W_{P_i^{id}}\}$, some relational literal of $Q_j$ must contribute the tag $x_k$. Since $x_k$ is assigned only to $L_k$, it follows that $L_k$ is in the image of $h_{P_i^{id}}$. Therefore, $\text{LitSet}(P_i^{id})\subseteq h_{P_i^{id}}\bigl(\text{LitSet}(Q_j)\bigr).$ The reverse inclusion follows from the fact that $h_{P_i^{id}}$ preserves relational literals. Hence $h_{P_i^{id}}\bigl(\text{LitSet}(Q_j)\bigr) = \text{LitSet}(P_i^{id}),$ as required.

$(3)\Rightarrow(1):$
Assume that item 3 holds. Consider an arbitrary $\text{Why}(X)$-database instance $\mathbf I$, a provenance-tracking interpretation $\pi$ compatible with $\mathbf I$, and a tuple $\vec t$. Let $W\in \pi^{\mathbf I} \llbracket  \bigcup_{1\leq i\leq n}P_i(\vec t) \rrbracket$, we show that $W\in \pi^{\mathbf I}\llbracket  \bigcup_{1\leq j\leq m}Q_j(\vec t) \rrbracket.$

Since the evaluation of a union is the union of the evaluations of its disjuncts, there are some $1\leq i\leq n$ and an assignment $v$ to the variables of $P_i$, producing $\vec t$, whose contribution contains $W$. A fixed non-zero contribution under a provenance-tracking interpretation consists of a single witness, hence the contribution of $v$ is exactly $\{W\}$.

Consider the completion (in the sense of CQ$^{\neq}$s) $P_i^c$ determined by $v$, obtained by adding $x=y$ whenever $v(x)=v(y)$ and $x\neq y$ whenever $v(x)\neq v(y)$, and let $P_i^{id}$ be the corresponding identification. Let $\theta$ be the canonical substitution defining $P_i^{id}$ and define $v_*(\theta(x)):=v(x).$ This map is well-defined and it is injective, since variables having the same value under $v$ are identified. 

The distinct ground literals used by $v$ are precisely $\{ v_*(L) \mid L\in\text{LitSet}(P_i^{id})\}. $ Indeed, every relational literal $K$ of $P_i$ becomes the literal $\theta(K)$ of $P_i^{id}$ and $v_*(\theta(K))=v(K).$ Conversely, every relational literal of $P_i^{id}$ is the image under $\theta$ of a relational literal of $P_i$, since the completion only adds equations and disequations.

By item 3, there are some $1\leq j\leq m$ and a $\neq$-negation-homomorphism $h_{P_i^{id}}\colon Q_j\longrightarrow P_i^{id}$ set-onto for relational literals. Hence $h_{P_i^{id}}\bigl(\text{LitSet}(Q_j)\bigr) = \text{LitSet}(P_i^{id}). $ Consider the assignment $w:=v_*\circ h_{P_i^{id}}$ to the variables of $Q_j$. Since $h_{P_i^{id}}$ preserves the head, $w$ produces $\vec t$, moreover, all equations and disequations of $Q_j$ are satisfied.

By set-ontoness, $\{ w(L)\mid L\in\text{LitSet}(Q_j)\} = \{ v_*(L)\mid L\in\text{LitSet}(P_i^{id}) \}.$ By the observation above, this is precisely the set of distinct ground relational literals used by $v$. Thus $v$ and $w$ use exactly the same distinct ground relational literals.

Under a provenance-tracking interpretation, every non-zero ground relational literal is annotated either by $1=\{\emptyset\}$ or by a singleton provenance token. Since multiplication in $\text{Why}(X,\overline X)$ is pairwise set union, repetitions of the same signed ground relational literal do not affect the resulting witness. Therefore, the contribution of an assignment depends only on the set of distinct ground relational literals that it uses. Since $v$ and $w$ use exactly the same such literals, and the contribution of $v$ is $\{W\}$, the contribution of $w$ is also $\{W\}$. Hence $W\in \pi^{\mathbf I}\llbracket Q_j(\vec t) \rrbracket \subseteq \pi^{\mathbf I} \llbracket  \bigcup_{1\leq j\leq m}Q_j(\vec t) \rrbracket.$ Since $\mathbf I$, $\pi$, $\vec t$, and $W$ were arbitrary, $\bigcup_{1\leq i\leq n}P_i \sqsubseteq^{\text{pt}}_{\text{Why}(X,\overline X)} \bigcup_{1\leq j\leq m}Q_j.$

It remains to calculate the complexity. By item 3, non-containment holds if and only if there are some $1\leq i\leq n$ and an identification $P_i^{id}$ of $P_i$ such that, for every $1\leq j\leq m$, there is no $\neq$-negation-homomorphism $h\colon Q_j\longrightarrow P_i^{id}$ set-onto for relational literals. An identification $P_i^{id}$ has polynomial size. Indeed, it is obtained from a $\{=,\neq\}$-completion of $P_i$, which fixes, for every pair of variables, whether they are equal or distinct, and then identifies variables required to be equal. Thus at most quadratically many comparisons are added, and no completion with respect to relational literals is required. For fixed $i$ and $P_i^{id}$, deciding whether there are some $j$ and a $\neq$-negation-homomorphism $h\colon Q_j\longrightarrow P_i^{id}$ set-onto for relational literals is in $\text{NP}$. Indeed, a nondeterministic polynomial-time machine guesses $j$ and a mapping from the variables of $Q_j$ to those of $P_i^{id}$, and checks that the mapping preserves the head, relational literals, equations and disequations, and that $h\bigl(\text{LitSet}(Q_j)\bigr)= \text{LitSet}(P_i^{id}).$ All these conditions can be checked in polynomial time.

Therefore, non-containment can be decided by nondeterministically guessing $i$ and an identification $P_i^{id}$, and using a $\text{coNP}$ test to verify that no such $j$ and $h$ exist. Hence non-containment is in $\text{NP}^{\text{coNP}} = \Sigma_2^p,$ and containment is in $\Pi_2^p$.

Finally, the lower bound follows immediately from the $\Pi_2^p$-hardness of containment for CQ$^\neq$Ns over $\text{Why}(X,\overline X)$, since CQ$^\neq$Ns are the particular case of UCQ$^\neq$Ns consisting of a single disjunct. Therefore, the containment problem is $\Pi_2^p$-complete.
\end{proof}

\ucqneqnContainmentD*

\begin{proof}
$(1) \implies (2):$ Suppose that $\bigcup_{1\leq i \leq n} P_i  \sqsubseteq_{\mathbb{B}[X, \overline{X}]}^{\text{pt}} \bigcup_{1\leq i \leq m} Q_i$ and consider an identification $P_i^{id}$ of a completion $P_i^c$ of $P_i$. We have that  $P_i^{id} \sqsubseteq_{\mathbb{B}[X, \overline{X}]}^{\text{pt}}  P_i^c \sqsubseteq_{\mathbb{B}[X, \overline{X}]}^{\text{pt}} P_i$ since on any $\mathbb{B}[X]$-database instance $\mathbf I$ and compatible interpretation $\pi$, the polynomial $\pi^\mathbf{I}\llbracket P_i^c \rrbracket$ contains only monomials appearing in $\pi^\mathbf{I}\llbracket P_i \rrbracket$ (some of the monomials appearing in $P_i$ might have turned into $0$ in $P_i^c$). Thus, 

\[\begin{aligned}\pi^{\mathrm{can}_{\mathbb{B}[X]}(P_i^{id})}\llbracket P_i^{id} \rrbracket
&\leq_{\mathbb{B}[X, \overline{X}]} \pi^{\mathrm{can}_{\mathbb{B}[X]}(P_i^{id})} \llbracket P_i^c \rrbracket \\ &\leq_{\mathbb{B}[X, \overline{X}]}  \pi^{\mathrm{can}_{\mathbb{B}[X]}(P_i^{id})} \llbracket P_i \rrbracket \\ &\leq_{\mathbb{B}[X, \overline{X}]} \pi^{\mathrm{can}_{\mathbb{B}[X]}(P_i^{id})} \llbracket \bigcup_{1\leq i \leq n} P_i \rrbracket \\ &\leq_{\mathbb{B}[X, \overline{X}]} \pi^{\mathrm{can}_{\mathbb{B}[X]}(P_i^{id})}\llbracket \bigcup_{1\leq i \leq m} Q_i \rrbracket.\end{aligned}\]

$(2) \implies (3):$  Assume that (2) holds, for each $1\leq i \leq n$, every identification $P_i^{id}$ of $P_i$, we have that 

$$\pi^{\mathrm{can}_{\mathbb{B}[X]}(P_i^{id})} \llbracket P_i^{id}(\vec{u}) \rrbracket  \leq_{\mathbb{B}[X, \overline{X}]} \pi^{\mathrm{can}_{\mathbb{B}[X]}(P_i^{id})} \llbracket (\bigcup_{1\leq i \leq m} Q_i)(\vec{u}) \rrbracket$$
where $\vec{u}$ is the tuple of free variables of $P_i^{id}$. The polynomial $\pi^{\mathrm{can}_{\mathbb{B}[X]}(P_i^{id})} \llbracket P_i^{id}(\vec{u}) \rrbracket$ has as a term a monomial $x_1 \cdot \ldots \cdot x_n $ containing every tag from $X \cup \overline{X}$ used to annotate literals by $\pi^{\mathrm{can}_{\mathbb{B}[X]}(P_i^{id})}$ possibly with repetitions (such monomial is obtained using the identity assignment for the variables in $ P_i^{id}$; noticing that the disequations in $P_i^{id}$ all take value $1$ by construction of the comparison completion; the tags of the relational literals are chosen pairwise noncomplementary). Then, by the definition of $\leq $ for $\mathbb{B}[X, \overline{X}]$, we must have that the monomial $x_1 \cdot \ldots \cdot x_n $ appears in $\pi^{\mathrm{can}_{\mathbb{B}[X]}(P_i^{id})} \llbracket (\bigcup_{1\leq i \leq m} Q_i)(\vec{u}) \rrbracket$. 
Since $\bigcup_{1\leq i \leq m} Q_i$ is a UCQ$^\neq$N, we must have that there is a $j\leq m$, and an assignment $v$ to the bound variables in $Q_j$ over $\mathrm{can}_{\mathbb{B}[X, \overline{X}]}(P_i^{id})$ such that it produces the monomial $x_1 \cdot \ldots \cdot x_n $ as the value for the body of $Q_j$.
Furthermore, the disequations in $Q_j$ must all take value $1$ under the assignment $v$ in order to obtain the monomial  $x_1\cdot \ldots \cdot x_n $ (and not $0$), and thus they must be  equations holding in $\mathrm{can}_{\mathbb{N}[X]}(P_i^{id})$ (and thus logically entailed by the body of $P_i^{id}$) and disequations appearing in the body of $P_i^{id}$ (since the latter contains a complete description of disequations in  $\mathrm{can}_{\mathbb{B}[X, \overline{X}]}(P_i^{id})$).  Then, we get the desired 
$\neq$-negation-homomorphism $h_{P_i^{id}}\colon Q_j \longrightarrow P_i^{id}$ exact for relational literals by letting $h_{P_i^{id}}(x)$ be the variable $v(x)$ (recall that the elements of the active domain of $\mathrm{can}_{\mathbb{B}[X, \overline{X}]}(P_i^{id})$ are the variables of $P_i^{id}$). To show injectivity, notice that any tag $p$ that is the value of a relational literal of $Q_j$ according to $v$, appears in $x_1 \cdot \ldots \cdot x_n $ with at most the same multiplicity as in  the body of $P_i^{id}$ (since this monomial corresponds to the body of $P^{\mathrm{id}}$), which entails that the multiset of images by $h_{P_i^{id}}$ of relational literals of $Q_j$ is bag-contained in the multiset of relational literals from $P_i^{id}$. Surjectivity follows from the fact that $x_1 \cdot \ldots \cdot x_n $ corresponds to the multiset of relational literals from $P_i^{id}$ but also to the images by $h_{P_i^{id}}$ of the multiset of relational literals from $Q_j$, so the former is bag-contained in the latter.
 
$(3) \implies (1):$ Assume that for each $1\leq i \leq n$,   every identification $P^{\mathrm{id}}_i$ of $P_i$,  there is a $j \leq m$ and a  $\neq$-negation-homomorphism 
$h_{P^{\mathrm{id}}_i} \colon Q_j \longrightarrow P^{\mathrm{id}}_i$  exact for relational literals. In order to establish $\bigcup_{1\leq i \leq n} P_i  \sqsubseteq_{\mathbb{B}[X, \overline{X}]}^{\text{pt}} \bigcup_{1\leq i \leq m} Q_i$, consider an arbitrary $\mathbb{B}[X]$-database instance $\mathbf I$, a compatible interpretation $\pi$ and a tuple $\vec{t}$. We need to show that any monomial in $\pi^\mathbf{I} \llbracket (\bigcup_{1\leq i \leq n} P_i)(\vec{t}) \rrbracket$ appears  in $\pi\llbracket ( \bigcup_{1\leq i \leq m} Q_i)(\vec{t}) \rrbracket$. Take an assignment $v$ of elements to the variables of $P_i$ justifying a given monomial $x_1 , \ldots , x_n $ and consider the completion $P_i^c$ of $P_i$ obtained by adding $x=y$ or $x\neq y$ depending on whether $v(x) = v(y)$ or $v(x) \neq v(y)$. Take an identification $P_i^{id}$ of $P_i^c$ by a canonical substitution $\theta$. We may observe that $\theta(x) = \theta(y)$ iff $v(x) = v(y)$. Then, $x_1 , \ldots , x_n $ is also a monomial in $\pi^\mathbf{I} \llbracket P_i^{id}(\vec{t}) \rrbracket$ justified by the assignment $v$ itself since the new disequations added can only take value $1$ by construction according to the assignment $v$. By assumption, there is a $\neq$-negation-homomorphism $h_{P_i^{id}}\colon Q_j \longrightarrow P_i^{id}$ exact for relational literals. Then, $v \circ h_{P_i^{id}}$ is an assignment of elements to the variables of $Q$ justifying the monomial $x_1 , \ldots , x_n $. If all the $x_i$s are values of relational literals of $P_i$ according to $v$, then if 
$x_i = \pi^\mathbf{I}\llbracket B(v(\theta(\vec{z}))) \rrbracket$ where  $B(\vec{z})$ is a relational literal of $P_i$ (and thus $B(\pi(\vec{z}))$ is one of $P_i^{id}$), then $x_i = \pi^\mathbf{I}\llbracket A(v(h_{P_i^{id}}(\overline{x}))) \rrbracket$ for some relational literal $A(\overline{x})$ of $Q$ such that $A(h_{P_i^{id}}(\overline{x})) = B(\theta(\vec{z}))$, which exists by surjectivity. 
Furthermore, the multiplicity with which $x_i$ appears using the assignment $v \circ h_{P_i^{id}}$ is exactly the multiplicity with which it appears in  $x_1 \ldots x_m$ given that $h_{P_i^{id}}$ is both onto and injective on relational literals.
Since equations and disequations can only take values $0$ or $1$, they can be disregarded in showing that we can justify a monomial.

Finally, the lower bound for the complexity calculation follows immediately from the $\Pi^p_2$-hardness of containment for CQ$^\neq$Ns over $\mathrm{Sorp}(X,\overline{X})$, since CQs are a particular case of UCQs.
 For the upper bound, observe that the complement of the containment problem is in $\mathrm{NP}^{\mathrm{coNP}}=\mathrm{NP}^{\mathrm{NP}}=\Sigma^p_2$. Hence the containment problem is in $\Pi^p_2$, and therefore it is $\Pi^p_2$-complete.
\end{proof}

\ucqneqnContainmentE*

\begin{proof}
$(1)\Rightarrow(2)$: Suppose that $\bigcup_{1\leq i \leq n} P_i
\sqsubseteq^{\text{pt}}_{\mathrm{PosBool}(X, \overline{X})}
\bigcup_{1\leq j \leq m} Q_j$ and consider an identification $P_i^{id}$ of a completion $P_i^c$ of $P_i$, for some $1\leq i\leq n$. We have that
$P_i^{id} \sqsubseteq^{\text{pt}}_{\mathrm{PosBool}(X, \overline{X})}
P_i^c \sqsubseteq^{\text{pt}}_{\mathrm{PosBool}(X, \overline{X})}
P_i \sqsubseteq^{\text{pt}}_{\mathrm{PosBool}(X, \overline{X})}
\bigcup_{1\leq k \leq n}P_k. $ Indeed, on any $\mathrm{PosBool}(X)$-database instance $\mathbf I$ and compatible
interpretation $\pi$, the value $\pi\llbracket P_i^c \rrbracket$ is obtained from
$\pi\llbracket P_i \rrbracket$ by possibly replacing some contributions with $\bot$, while
the comparison atoms satisfied by an assignment take value $\top$ and
do not affect the corresponding contribution. Moreover, $P_i^{id}$ is
obtained from $P_i^c$ by applying a canonical substitution, and
$P_i$ is one of the disjuncts of $\bigcup_{1\leq k \leq n}P_k$.
Thus, by the containment assumption,
\begin{align}
 \pi^{\mathrm{can}_{\mathrm{PosBool}(X)}(P_i^{id})}
\llbracket P_i^{id}(\vec{u}) \rrbracket &\leq_{\mathrm{PosBool}(X, \overline{X})}  \pi^{\mathrm{can}_{\mathrm{PosBool}(X)}(P_i^{id})}
\llbracket 
\left(\bigcup_{1\leq k \leq n}P_k\right)(\vec{u})
 \rrbracket \nonumber \\
   &\leq_{\mathrm{PosBool}(X, \overline{X})}  \pi^{\mathrm{can}_{\mathrm{PosBool}(X)}(P_i^{id})}
\llbracket 
\left(\bigcup_{1\leq j \leq m}Q_j\right)(\vec{u})
 \rrbracket \nonumber
\end{align}

$(2)\Rightarrow(3)$:
Assume that (2) holds. Fix $1\leq i\leq n$ and an identification
$P_i^{id}$ of $P_i$. Let $\pi^{\mathrm{can}} = \pi^{\mathrm{can}_{\mathrm{PosBool}(X)}(P_i^{id})}.$ Then $ \pi^{\mathrm{can}}\llbracket P_i^{id}(\vec{u}) \rrbracket \leq_{\mathrm{PosBool}(X, \overline{X})}
\pi^{\mathrm{can}} \llbracket  \left(\bigcup_{1\leq j \leq m}Q_j\right)(\vec{u})  \rrbracket.$ Let $ \tau=x_1\wedge\ldots\wedge x_r$ be the conjunction of all tags from $X\cup\overline X$ used by
$\pi^{\mathrm{can}}$ to annotate the relational literals of
$P_i^{id}$. This conjunction is obtained by using the identity
assignment for the variables of $P_i^{id}$, since all equations and
disequations satisfied by this assignment take value $\top$.
Consider the compatible Boolean valuation $\nu$ assigning value $1$
to all tags occurring in $\tau$ and value $0$ to all the remaining
tags. Then $\nu\bigl(\pi^{\mathrm{can}}\llbracket P_i^{id}(\vec{u}) \rrbracket\bigr)=1.$ By the definition of the natural order in $\mathrm{PosBool}(X,\overline X)$, it follows that $\nu\left( \pi^{\mathrm{can}} \llbracket  \left(\bigcup_{1\leq j \leq m}Q_j\right)(\vec{u})  \rrbracket \right)=1.$
Now, $\pi^{\mathrm{can}} \llbracket  \left(\bigcup_{1\leq j \leq m}Q_j\right)(\vec{u})  \rrbracket= \bigvee_{1\leq j \leq m} \pi^{\mathrm{can}}\llbracket Q_j(\vec{u}) \rrbracket, $ and each $\pi^{\mathrm{can}}\llbracket Q_j(\vec{u}) \rrbracket$ is itself the disjunction of the contributions produced by the assignments to the variables of $Q_j$. Hence there must be some $1 \leq j\leq m$ and an assignment $v$ to the variables of $Q_j$ such that the value of the body of $Q_j$ under $v$ evaluates to $1$ according to $\nu$. Therefore, every relational literal of $Q_j$ is mapped by $v$ to a relational literal occurring in $P_i^{id}$. Indeed, the tags made true by $\nu$ are precisely those used for relational literals of $P_i^{id}$, while every other literal has value $0$ under $\nu$ (either its annotation is $\bot$ or it uses a fresh tag outside this set). Moreover, every equation and disequation occurring in $Q_j$ must take value $\top$ under $v$; hence the corresponding equation or disequation is logically entailed by the body of $P_i^{id}$. Since $v$ contributes to the value at $\vec{u}$, it also preserves the tuple of free variables. Thus, identifying the elements of the canonical database with the variables of $P_i^{id}$, we obtain the desired
$\neq$-negation-homomorphism $h_{P_i^{id}}\colon Q_j\longrightarrow P_i^{id}$ by letting $h_{P_i^{id}}(x)$ be the variable corresponding to $v(x)$.

$(3)\Rightarrow(1)$:
Assume that (3) holds. We want to show that $\bigcup_{1\leq i \leq n} P_i
\sqsubseteq^{\text{pt}}_{\mathrm{PosBool}(X, \overline{X})}
\bigcup_{1\leq j \leq m} Q_j.$ Consider an arbitrary $\mathrm{PosBool}(X)$-database instance $\mathbf I$, a compatible provenance-tracking interpretation $\pi$, and a tuple $\vec{t}$. Fix $1\leq i\leq n$ and an assignment $v$ to the variables of $P_i$ such that $v(\vec{u})=\vec{t}$. If some equation or disequation of $P_i$ is not satisfied by $v$, its contribution is $\bot$ and there is nothing to prove. Thus, suppose that $v$ gives a non-zero contribution to $\pi\llbracket P_i(\vec{t}) \rrbracket$. Consider the $\{=,\neq\}$-completion $P_i^c$ of $P_i$ obtained by adding $x=y$ if $v(x)=v(y)$, and $x\neq y$ if $v(x)\neq v(y)$, for all variables $x,y$ of $P_i$. Let $P_i^{id}$ be an identification of $P_i^c$ obtained by a canonical substitution $\theta$. We have that $\theta(x)=\theta(y)\quad\text{iff}\quad v(x)=v(y).$ The assignment $v$ induces an assignment $v^{id}$ to the variables of
$P_i^{id}$ by setting $v^{id}(\theta(x))=v(x).$ This is well-defined by the previous observation. Moreover, all the new equations and disequations added in forming $P_i^c$ take value $\top$ under this assignment. Therefore, the contribution of $v$ to $\pi\llbracket P_i(\vec{t}) \rrbracket$ is also the contribution of $v^{id}$ to $\pi\llbracket P_i^{id}(\vec{t}) \rrbracket$. By assumption, there is some $1 \leq j\leq m$ and a $\neq$-negation-homomorphism $h_{P_i^{id}}\colon Q_j\longrightarrow P_i^{id}.$ Then $v^{id}\circ h_{P_i^{id}}$ is an assignment to the variables of $Q_j$ producing the same tuple $\vec{t}$. Every relational literal of $Q_j$ is mapped to a relational literal of $P_i^{id}$, while all equations and disequations of $Q_j$ take value $\top$ under this assignment. Let $\varphi_v$ be the conjunction of the values of the relational literals of $P_i^{id}$ under $v^{id}$, and let $\psi_v$ be the conjunction of the values of the relational literals of $Q_j$ under $v^{id}\circ h_{P_i^{id}}$. Since every factor occurring in $\psi_v$ also occurs in $\varphi_v$, possibly with repetitions, we have $\varphi_v \leq_{\mathrm{PosBool}(X, \overline{X})} \psi_v.$ Indeed, multiplication in $\mathrm{PosBool}(X,\overline X)$ is idempotent, so repetitions do not affect the value, and in a distributive lattice a conjunction is below every conjunction obtained by dropping some of its factors. Hence the contribution of $v$ to $\pi\llbracket P_i(\vec{t}) \rrbracket$ is below the contribution of $v^{id}\circ h_{P_i^{id}}$ to $\pi\llbracket Q_j(\vec{t}) \rrbracket$. Therefore, $\varphi_v
\leq_{\mathrm{PosBool}(X, \overline{X})} \pi\llbracket Q_j(\vec{t}) \rrbracket
\leq_{\mathrm{PosBool}(X, \overline{X})} \pi\llbracket 
\left(\bigcup_{1\leq j \leq m}Q_j\right)(\vec{t})
 \rrbracket.$ Since this holds for every assignment $v$ to every $P_i$, taking the join over all assignments gives $\pi\llbracket P_i(\vec{t}) \rrbracket
\leq_{\mathrm{PosBool}(X, \overline{X})}
\pi\llbracket 
\left(\bigcup_{1\leq j \leq m}Q_j\right)(\vec{t})
 \rrbracket$ for every $1\leq i\leq n$. Taking now the join over all disjuncts
$P_i$, we obtain $\pi\llbracket 
\left(\bigcup_{1\leq i \leq n}P_i\right)(\vec{t})
 \rrbracket \leq_{\mathrm{PosBool}(X, \overline{X})}
\pi\llbracket  \left(\bigcup_{1\leq j \leq m}Q_j\right)(\vec{t})
 \rrbracket.$ Since $\mathbf I$, $\pi$, and $\vec{t}$ were arbitrary, the desired containment follows.

Finally, the lower bound for the complexity calculation follows
immediately from the $\Pi^p_2$-hardness of containment for
CQ$^\neq$Ns over $\mathrm{PosBool}(X,\overline X)$, since CQs are a
particular case of UCQs. For the upper bound, by item 3 the complement
of the containment problem is in
$\text{NP}^{\text{coNP}}=\text{NP}^{\text{NP}}=\Sigma^p_2$. Hence the containment problem is in $\Pi^p_2$, and therefore it is
$\Pi^p_2$-complete.
\end{proof}

\ucqneqnContainmentF*

\begin{proof}
$(1) \implies (2):$
Assume that $\bigcup_{1\leq i \leq n} P_i \sqsubseteq^{\text{md}}_{\mathrm{Which}(X)} \bigcup_{1\leq j \leq m} Q_j$. For each $1\leq i \leq n$ and every identification $P_i^{id}$ of $P_i$, the interpretation $\pi^{P_i^{id}}_{\mathrm{Which}(X)}$ is model-defining. Since $P_i$ is one of the disjuncts of $\bigcup_{1\leq k \leq n}P_k$, we have $\pi^{P_i^{id}}_{\mathrm{Which}(X)}(P_i) \leq_{\mathrm{Which}(X)}
\pi^{P_i^{id}}_{\mathrm{Which}(X)} \left(\bigcup_{1\leq k \leq n}P_k\right).$ Hence, by the containment assumption, $\pi^{P_i^{id}}_{\mathrm{Which}(X)}(P_i) \leq_{\mathrm{Which}(X)}
\pi^{P_i^{id}}_{\mathrm{Which}(X)}\left(\bigcup_{1\leq j \leq m}Q_j\right).$

$(2) \implies (3):$
Assume that (2) holds. Fix $1\leq i \leq n$ and an identification $P_i^{id}$ of $P_i$. Let $P_i^c$ be the $\{=,\neq\}$-completion from which $P_i^{id}$ is obtained, and let $\theta$ be the corresponding canonical substitution. Let $L$ be a relational literal of $P_i^{id}$ inherited from $P_i$. Then, $L=\theta(L_0)$ for some relational literal $L_0$ occurring in the body of $P_i$. Consider the assignment $v_\theta$ from the variables of $P_i$ given by $v_\theta(x)=\theta(x)$. Since $P_i^c$ is a completion of $P_i$, all equations and disequations occurring in $P_i$ are satisfied by $v_\theta$. Moreover, for every relational literal $B$ of $P_i$, the literal $\theta(B)$ occurs in $P_i^{id}$ and is inherited from $P_i$. Therefore, $\pi^{P_i^{id}}_{\mathrm{Which}(X)}(\theta(B)) = \{x_{\theta(B)}\}.$ In particular, the contribution of $v_\theta$ to $\pi^{P_i^{id}}_{\mathrm{Which}(X)}(P_i(\theta(\vec{u})))$
is non-bottom and contains $x_L$. Thus, $x_L\in \pi^{P_i^{id}}_{\mathrm{Which}(X)}(P_i(\theta(\vec{u}))).$ By assumption, $ \pi^{P_i^{id}}_{\mathrm{Which}(X)} (P_i(\theta(\vec{u})))
\leq_{\mathrm{Which}(X)} \pi^{P_i^{id}}_{\mathrm{Which}(X)} \left(
\left(\bigcup_{1\leq j \leq m}Q_j\right) (\theta(\vec{u}))
\right),$ and hence $ x_L\in \pi^{P_i^{id}}_{\mathrm{Which}(X)}
\left( \left(\bigcup_{1\leq j \leq m}Q_j\right)
(\theta(\vec{u})) \right).$ Since the value of the UCQ is obtained as the sum of the values of its disjuncts, and addition in $\mathrm{Which}(X)$ is union, there must be some $1 \leq j\leq m$ and an assignment $v$ to the variables of $Q_j$ over the domain of $P_i^{id}$, sending the free variables to $\theta(\vec{u})$, such that the value of the body of $Q_j$ under $v$ is non-bottom and contains $x_L$. Since this contribution is non-bottom, every relational literal $A(\overline{x})$ occurring in $Q_j$ has non-bottom value under $v$. By the definition of $\pi^{P_i^{id}}_{\mathrm{Which}(X)}$, this means that $A(v(\overline{x}))$ occurs in the body of $P_i^{id}$. Moreover, all equations and disequations of $Q_j$ are satisfied under $v$, since otherwise the contribution would be bottom. Thus, the corresponding equations are logically entailed by the body of $P_i^{id}$, while the corresponding disequations appear in the body of $P_i^{id}$. Therefore, by setting $h_{P_i^{id},L}(x)=v(x),$ we obtain a $\neq$-negation-homomorphism $ h_{P_i^{id},L}\colon Q_j\longrightarrow P_i^{id}.
$ Finally, the token $x_L$ occurs only in the annotation of $L$ as different inherited relational literals receive different singleton annotations, while relational literals added only in the final completion receive $\emptyset$. Since the contribution of $v$ contains $x_L$, there is a relational literal $A(\overline{x})$ of $Q_j$ such that $A(h_{P_i^{id},L}(\overline{x}))=L.$ Thus, $L$ belongs to the image of $h_{P_i^{id},L}$.

$(3) \implies (1):$
Assume that (3) holds and consider an arbitrary $\mathrm{Which}(X)$-database instance $\mathbf I$, a compatible model-defining interpretation $\pi$, and a tuple $\vec{t}$. We want to show that $\pi\llbracket 
\left(\bigcup_{1\leq i \leq n}P_i\right)(\vec{t})  \rrbracket
\leq_{\mathrm{Which}(X)} \pi\llbracket 
\left(\bigcup_{1\leq j \leq m}Q_j\right)(\vec{t}) \rrbracket.$ If
$\pi\llbracket (\bigcup_{1\leq i \leq n}P_i)(\vec{t}) \rrbracket=\bot$, the inequality is immediate. Suppose otherwise. Fix a disjunct $P_i$ and an assignment $v$ to the variables of $P_i$, sending $\vec{u}$ to $\vec{t}$, whose contribution to $\pi\llbracket P_i(\vec{t}) \rrbracket$ is non-bottom. Consider the $\{=,\neq\}$-completion $P_i^c$ of $P_i$ obtained by adding $x=y$ or $x\neq y$ depending on whether $v(x)=v(y)$ or $v(x)\neq v(y)$, and let $\theta$ be the corresponding canonical substitution. We keep denoting by $v$ the induced assignment on the variables occurring after the substitution, defined by $ v(\theta(x))=v(x).$ This is well-defined by construction. We complete the resulting query in the sense of CQNs as follows. For every relation symbol $R$ in the common schema and every tuple $\vec{z}$ of its variables of the appropriate arity, we add $R(\vec{z})$ if $\pi\llbracket R(v(\vec{z})) \rrbracket\neq\bot,$ and we add $\neg R(\vec{z})$ if $\pi\llbracket \neg R(v(\vec{z})) \rrbracket\neq\bot.$ Since $\pi$ is model-defining, exactly one of these two values is different from $\bot$. Hence this construction yields an identification $P_{i,v}^{id}$ of $P_i$. Moreover, every relational literal occurring in $P_{i,v}^{id}$ has non-bottom value under $v$, and every relational literal $\theta(L_0)$ obtained from a relational literal $L_0$ of $P_i$ is inherited from $P_i$. We first observe that $\pi\llbracket  \left(\bigcup_{1\leq j \leq m}Q_j\right)(\vec{t}) \rrbracket\neq\bot.$ Indeed, choose any relational literal $L_0$ occurring in the body of $P_i$, and let $L=\theta(L_0)$. By assumption, there is some $1 \leq j\leq m$ and a $\neq$-negation-homomorphism $h_{P_{i,v}^{id},L}\colon Q_j\longrightarrow P_{i,v}^{id} $ such that $L$ belongs to the image of $h_{P_{i,v}^{id},L}$. Then $ v\circ h_{P_{i,v}^{id},L} $ is an assignment to the variables of $Q_j$ producing the tuple $\vec{t}$. For every relational literal $A(\overline{x})$ of $Q_j$, the literal $A(h_{P_{i,v}^{id},L}(\overline{x}))$ occurs in $P_{i,v}^{id}$ and hence has non-bottom value under $v$. Furthermore, since $h_{P_{i,v}^{id},L}$ is a $\neq$-negation-homomorphism, all equations and disequations of $Q_j$ are satisfied under $v\circ h_{P_{i,v}^{id},L}$. Therefore, this assignment gives a non-bottom contribution to $\pi\llbracket Q_j(\vec{t}) \rrbracket$, and hence $\pi\llbracket  \left(\bigcup_{1\leq j \leq m}Q_j\right)(\vec{t})  \rrbracket\neq\bot. $ Now let $x$ be any token occurring in the contribution of $v$ to $\pi\llbracket P_i(\vec{t}) \rrbracket$. Since multiplication between non-bottom elements in $\mathrm{Which}(X)$ is union, there is a relational literal $L_0$ occurring in the body of $P_i$ such that $x\in\pi\llbracket L_0(v) \rrbracket.$ Let $L=\theta(L_0)$. Then $L$ is a relational literal of $P_{i,v}^{id}$ inherited from $P_i$ and $x\in\pi\llbracket L(v) \rrbracket.$ By assumption, there is some $1 \leq j\leq m$ and a $\neq$-negation-homomorphism $
h_{P_{i,v}^{id},L}\colon Q_j\longrightarrow P_{i,v}^{id}$ such that $L$ belongs to the image of $h_{P_{i,v}^{id},L}$. Consider the assignment
$v\circ h_{P_{i,v}^{id},L}$. As above, every relational literal of $Q_j$ has non-bottom value under this assignment, and all equations and disequations are satisfied. Since $L$ belongs to the image of  $h_{P_{i,v}^{id},L}$, there is a relational literal $A(\overline{x})$ of $Q_j$ such that $ A(h_{P_{i,v}^{id},L}(\overline{x}))=L.$ Consequently, $A(v(h_{P_{i,v}^{id},L}(\overline{x})))=L(v). $ Since $x\in\pi\llbracket L(v) \rrbracket$, and the product of the non-bottom values of the relational literals of $Q_j$ is their union, the contribution of $v\circ h_{P_{i,v}^{id},L}$ contains $x$. Hence, $  x\in\pi\llbracket Q_j(\vec{t}) \rrbracket
\subseteq\pi\llbracket  \left(\bigcup_{1\leq j \leq m}Q_j\right)(\vec{t})
 \rrbracket. $ Thus, every token occurring in the contribution of $v$ to $\pi\llbracket P_i(\vec{t}) \rrbracket$ also occurs in $\pi\llbracket (\bigcup_{1\leq j \leq m}Q_j)(\vec{t}) \rrbracket$, and we have already shown that the latter value is non-bottom. Therefore, the contribution of $v$ is below $\pi\llbracket (\bigcup_{1\leq j \leq m}Q_j)(\vec{t}) \rrbracket$ in the natural order of $\mathrm{Which}(X)$. Since this holds for every non-bottom contribution of every disjunct $P_i$, taking their union gives $\pi\llbracket  \left(\bigcup_{1\leq i \leq n}P_i\right)(\vec{t})  \rrbracket \leq_{\mathrm{Which}(X)} \pi\llbracket 
\left(\bigcup_{1\leq j \leq m}Q_j\right)(\vec{t})  \rrbracket.$ Since $\mathbf I$, $\pi$, and $\vec{t}$ were arbitrary, we conclude that $\bigcup_{1\leq i \leq n}P_i\sqsubseteq^{\text{md}}_{\mathrm{Which}(X)}
\bigcup_{1\leq j \leq m}Q_j.$

For the complexity calculation, the lower bounds follow from the hardness of the corresponding containment problem for CQNs, which are a particular case of UCQ$^\neq$Ns. For the upper bounds, by item 3, the complement of the containment problem is in $\text{NP}^{\text{coNP}}=\Sigma^p_2$
if there is a fixed arity bound on the relations used, while it is in NEXP otherwise. Therefore, the containment problem is in $\Pi^p_2$ in the bounded-arity case and in coNEXP in general, and the corresponding lower bounds give the claimed completeness results.
\end{proof}

\ucqneqnContainmentG*

\begin{proof}
We first observe that Lemma~\ref{freshNPT} extends immediately to UCQ$^\neq$Ns. Indeed, if $\overline R=\bigcup_{1\leq r\leq s}R_r$, then $\pi\llbracket\overline R(\vec t)\rrbracket= \sum_{r=1}^{s}\pi\llbracket R_r(\vec t)\rrbracket,$ and, since the homomorphism $E$ of Lemma~\ref{freshNPT} preserves finite sums, $E\bigl(\pi^*\llbracket\overline R(\vec t)\rrbracket\bigr)=\pi\llbracket\overline R(\vec t)\rrbracket.$ Hence a failure of containment for two UCQ$^\neq$Ns is also preserved by the fresh-token construction.

$(1)\Rightarrow(2):$
Suppose that $\overline P\sqsubseteq^{\mathrm{pt}}_{\mathbb{N}[X,\overline X]}\overline Q.$ Fix $1\leq i\leq n$ and an identification $P_i^{\mathrm{id}}$ of a completion (in the sense of CQ$^{\neq}$s) $P_i^c$ of $P_i$. Let $\theta$ be the canonical substitution used to obtain $P_i^{\mathrm{id}}$ from $P_i^c$.

Consider a provenance-tracking interpretation $\pi$ over the domain formed by the variables of $P_i^{\mathrm{id}}$ which assigns to every $L\in\text{LitSet}(P_i^{\mathrm{id}})$ a different fresh tag, with correct polarity, chosen so that no two tags used by $\pi$ are complementary, and assigns value $0$ to every other ground relational literal. Let $\mathbf I$ be the $\mathbb{N}[X]$-database instance induced by the values of $\pi$ on positive ground relational atoms.

Consider the assignment $v_\theta$ to the variables of $P_i$ defined by $v_\theta(x):=\theta(x).$ By construction of the completion and identification, $v_\theta$ satisfies all equations and disequations of $P_i$ and produces the canonical output tuple $\theta(\overline u)$. Its contribution is a non-zero monomial $\mu$ containing, for every $L\in\text{LitSet}(P_i^{\mathrm{id}})$, the tag assigned to $L$, repeated as many times as $L$ occurs in the body of $P_i^{\mathrm{id}}$. Moreover, the identification induced by $v_\theta$ is precisely $P_i^{\mathrm{id}}$.

By Lemma~\ref{countingNPT}, for every $1\leq\ell\leq n$, the coefficient of $\mu$ in $\pi\llbracket P_\ell(\theta(\overline u))\rrbracket$ is exactly $\left|\mathrm{ExNegHom}(P_\ell,P_i^{\mathrm{id}})\right|.$ Since the evaluation of a union is the sum of the evaluations of its disjuncts, the coefficient of $\mu$ in $\pi\llbracket\overline P(\theta(\overline u))\rrbracket$ is therefore $\sum_{1\leq\ell\leq n}
\left|\mathrm{ExNegHom}(P_\ell,P_i^{\mathrm{id}})\right|.$ By the same argument, the coefficient of $\mu$ in $\pi\llbracket\overline Q(\theta(\overline u))\rrbracket$ is $\sum_{1\leq j\leq m}
\left|\mathrm{ExNegHom}(Q_j,P_i^{\mathrm{id}})\right|.$

By the containment hypothesis, $\pi\llbracket\overline P(\theta(\overline u))\rrbracket \leq_{\mathbb{N}[X,\overline X]} \pi\llbracket\overline Q(\theta(\overline u))\rrbracket.$ Since the natural order of $\mathbb{N}[X,\overline X]$ is coefficientwise on consistent monomials, it follows that $\sum_{1\leq\ell\leq n}
\left|\mathrm{ExNegHom}(P_\ell,P_i^{\mathrm{id}})\right|\leq \sum_{1\leq j\leq m}
\left|\mathrm{ExNegHom}(Q_j,P_i^{\mathrm{id}})\right|.$ Since $i$ and $P_i^{\mathrm{id}}$ were arbitrary, item 2 follows.

$(2)\Rightarrow(1):$
We prove the contrapositive. Suppose that $\overline P\not\sqsubseteq^{\mathrm{pt}}_{\mathbb{N}[X,\overline X]}\overline Q.$ Then there are an $\mathbb{N}[X]$-database instance $\mathbf I$, a provenance-tracking interpretation $\pi$ compatible with $\mathbf I$, and a tuple $\vec t$ such that $\pi\llbracket\overline P(\vec t)\rrbracket \not\leq_{\mathbb{N}[X,\overline X]} \pi\llbracket\overline Q(\vec t)\rrbracket.$ By the observation at the beginning of the proof and Lemma~\ref{freshNPT}, we may assume that distinct non-zero ground relational literals receive pairwise distinct tags with correct polarity and such that no two tags used by the interpretation are complementary.

Since the natural order of $\mathbb{N}[X,\overline X]$ is coefficientwise on consistent monomials, there is a consistent monomial $\mu$ whose coefficient in $\pi\llbracket\overline P(\vec t)\rrbracket$ is strictly greater than its coefficient in $\pi\llbracket\overline Q(\vec t)\rrbracket.$ In particular, the coefficient of $\mu$ on the left-hand side is non-zero. Hence there are some $1\leq i\leq n$ and an assignment $v$ to the variables of $P_i$ which produces $\vec t$ and contributes exactly $\mu$.

Let $P_{i,v}^c$ be the completion (in the sense of CQ$^{\neq}$s) of $P_i$ determined by $v$, obtained by adding $x=y$ whenever $v(x)=v(y)$ and $x\neq y$ whenever $v(x)\neq v(y)$, and let $P_{i,v}^{\mathrm{id}}$ be the corresponding identification.

Applying Lemma~\ref{countingNPT} to every disjunct $P_\ell$ of $\overline P$, we obtain that the coefficient of $\mu$ in $\pi\llbracket\overline P(\vec t)\rrbracket$ is $\sum_{1\leq\ell\leq n}
\left|\mathrm{ExNegHom}(P_\ell,P_{i,v}^{\mathrm{id}})\right|.$ Indeed, the coefficient contributed by each $P_\ell$ is $\left|\mathrm{ExNegHom}(P_\ell,P_{i,v}^{\mathrm{id}})\right|,$ and union adds the contributions of all indexed disjuncts. Notice in particular that other disjuncts of $\overline P$, besides $P_i$, may contribute to the same monomial $\mu$ and must therefore be included in the sum.

Likewise, applying Lemma~\ref{countingNPT} to every disjunct of $\overline Q$, the coefficient of $\mu$ in
$\pi\llbracket\overline Q(\vec t)\rrbracket$ is $\sum_{1\leq j\leq m}  \left|\mathrm{ExNegHom}(Q_j,P_{i,v}^{\mathrm{id}})\right|.$

By the choice of $\mu$, we therefore have $\sum_{1\leq\ell\leq n} \left|\mathrm{ExNegHom}(P_\ell,P_{i,v}^{\mathrm{id}})\right|>\sum_{1\leq j\leq m}\left|\mathrm{ExNegHom}(Q_j,P_{i,v}^{\mathrm{id}})\right|.$ Thus item 2 fails for the identification $P_{i,v}^{\mathrm{id}}$ of the left disjunct $P_i$. This proves the contrapositive and hence $(2)\Rightarrow(1)$.

Finally, for the upper bound complexity calculation, we look at the complement of the problem in item 2. A nondeterministic polynomial-time machine guesses a left disjunct $P_i$ and one of its identifications $P_i^{\mathrm{id}}$. Such an identification has polynomial size, since it is obtained only from a completion in the sense of CQ$^{\neq}$s, which fixes for every pair of variables whether they are equal or distinct, and then quotients variables that are required to be equal. In particular, no completion with respect to positive and negative relational literals is required.

For a fixed identification $P_i^{\mathrm{id}}$, consider the quantity $\sum_{1\leq\ell\leq n} \left|\mathrm{ExNegHom}(P_\ell,P_i^{\mathrm{id}})\right|.$ This is a $\#\text{P}$ function. Indeed, a nondeterministic polynomial-time machine guesses a disjunct index $\ell$ and a mapping from the variables of $P_\ell$ to the variables of $P_i^{\mathrm{id}}$, and accepts precisely when the guessed mapping belongs to $\mathrm{ExNegHom}(P_\ell,P_i^{\mathrm{id}}).$ Membership can be checked in polynomial time by verifying preservation of the ordered head, positive and negative relational literals, equations and disequations, and exactness for relational literals. Thus the accepting computations are in bijection with the pairs consisting of a disjunct index $\ell$ and an element of $\mathrm{ExNegHom}(P_\ell,P_i^{\mathrm{id}})$. Their number is exactly the sum above. 

By the same argument, $\sum_{1\leq j\leq m} \left|\mathrm{ExNegHom}(Q_j,P_i^{\mathrm{id}})\right|$ is a $\#\text{P}$ function. Consequently, the difference $\sum_{1\leq\ell\leq n} \left|\mathrm{ExNegHom}(P_\ell,P_i^{\mathrm{id}})\right| - \sum_{1\leq j\leq m} \left|\mathrm{ExNegHom}(Q_j,P_i^{\mathrm{id}})\right|$ is in $\text{GapP}$ by~\cite[Prop. 3.5]{Fenner}, and deciding whether this difference is positive is in $\text{PP}$ by~\cite[Prop. 4.2]{Fenner}. Hence non-containment is in $\text{NP}^{\text{PP}}$, and therefore $\mathbb{N}[X,\overline X]$-containment of UCQ$^\neq$Ns under provenance-tracking interpretations is in $\text{coNP}^{\text{PP}}$.
\end{proof}

\ucqneqnContainmentH*

\begin{proof}
We first observe that Lemma~\ref{freshTrioPT} extends immediately to UCQ$^\neq$Ns. Indeed, if $\overline R=\bigcup_{1\leq r\leq s}R_r$, then $\pi\llbracket\overline R(\vec t)\rrbracket = \sum_{r=1}^{s}\pi\llbracket R_r(\vec t)\rrbracket,$ and, since the homomorphism $E$ of Lemma~\ref{freshTrioPT} preserves finite sums, $E\bigl(\pi^*\llbracket\overline R(\vec t)\rrbracket\bigr)=\pi\llbracket\overline R(\vec t)\rrbracket.$ Hence a failure of containment for two UCQ$^\neq$Ns is also preserved by the construction.

$(1)\Rightarrow(2):$
Suppose that $\overline P\sqsubseteq^{\mathrm{pt}}_{\mathrm{Trio}(X,\overline X)}\overline Q$ and fix $1\leq i\leq n$ and an identification $P_i^{\mathrm{id}}$ of a completion (in the sense of CQ$^\neq$s) $P_i^c$ of $P_i$. Let $\theta$ be the canonical substitution used to obtain $P_i^{\mathrm{id}}$ from $P_i^c$, and let $\text{LitSet}(P_i^{\mathrm{id}}) = \{L_1,\ldots,L_r\}.$ Consider a provenance-tracking interpretation $\pi$ over the domain formed by the variables of $P_i^{\mathrm{id}}$ which assigns to every $L_k$ a different fresh tag $x_k\in X\cup\overline X$ of the corresponding polarity, so that no two complementary tags are used, and let every other ground relational literal take value $0$. Let $\mathbf I$ be the $\mathrm{Trio}(X)$-database instance induced by the values of $\pi$ on positive ground relational atoms. Then $\pi$ is a provenance-tracking interpretation compatible with $\mathbf I$.

Consider the assignment $v$ to the variables of $P_i$ defined by $v(x):=\theta(x).$ By construction of the completion and identification, $v$ satisfies all equations and disequations of $P_i$, produces the canonical output tuple $\theta(\overline u)$, and the identification of $P_i$ induced by $v$ is precisely $P_i^{\mathrm{id}}$. Moreover, $v$ contributes the non-zero square-free monomial $\mu:=x_1\cdot\ldots\cdot x_r$ to $\pi\llbracket P_i(\theta(\overline u))\rrbracket.$ Indeed, repeated occurrences of the same relational literal may only produce positive powers of the corresponding tag, which are dropped in $\mathrm{Trio}(X,\overline X)$.

By Lemma~\ref{countingTrioPT}, for every $1\leq\ell\leq n$, the coefficient of $\mu$ in $\pi\llbracket P_\ell(\theta(\overline u))\rrbracket$ is exactly $\left|\mathrm{SOHom}(P_\ell,P_i^{\mathrm{id}})\right|.$ Since the evaluation of a union is the sum of the evaluations of its disjuncts, the coefficient of $\mu$ in $\pi\llbracket\overline P(\theta(\overline u))\rrbracket$ is therefore $\sum_{1\leq\ell\leq n}
\left|\mathrm{SOHom}(P_\ell,P_i^{\mathrm{id}})\right|.$ By the same argument, the coefficient of $\mu$ in $\pi\llbracket\overline Q(\theta(\overline u))\rrbracket$ is $\sum_{1\leq j\leq m}
\left|\mathrm{SOHom}(Q_j,P_i^{\mathrm{id}})\right|.$ 

By the containment hypothesis, $\pi\llbracket\overline P(\theta(\overline u))\rrbracket \leq_{\mathrm{Trio}(X,\overline X)} \pi\llbracket\overline Q(\theta(\overline u))\rrbracket.$ Since the natural order of $\mathrm{Trio}(X,\overline X)$ is the coefficientwise order on consistent square-free monomials, it follows that $\sum_{1\leq\ell\leq n}
\left|\mathrm{SOHom}(P_\ell,P_i^{\mathrm{id}})\right|\leq \sum_{1\leq j\leq m}
\left|\mathrm{SOHom}(Q_j,P_i^{\mathrm{id}})\right|.$ Since $i$ and $P_i^{\mathrm{id}}$ were arbitrary, item 2 follows.

$(2)\Rightarrow(1):$
We prove the contrapositive. Suppose that $\overline P\not\sqsubseteq^{\mathrm{pt}}_{\mathrm{Trio}(X,\overline X)}\overline Q.$ Then there are a $\mathrm{Trio}(X)$-database instance $\mathbf I$, a provenance-tracking interpretation $\pi$ compatible with $\mathbf I$, and a tuple $\vec t$ such that $\pi\llbracket\overline P(\vec t)\rrbracket
\not\leq_{\mathrm{Trio}(X,\overline X)}
\pi\llbracket\overline Q(\vec t)\rrbracket.$ By the observation at the beginning of the proof and Lemma~\ref{freshTrioPT}, we may assume that distinct non-zero ground relational literals receive pairwise distinct tags and no two tags used by the interpretation are complementary.

Since the natural order of $\mathrm{Trio}(X,\overline X)$ is coefficientwise on consistent square-free monomials, there is a consistent square-free monomial $\mu$ whose coefficient in $\pi\llbracket\overline P(\vec t)\rrbracket$ is strictly greater than its coefficient in $\pi\llbracket\overline Q(\vec t)\rrbracket. $ In particular, the coefficient of $\mu$ on the left-hand side is non-zero. Hence there are some $1\leq i\leq n$ and an assignment $v$ to the variables of $P_i$ which produces $\vec t$ and contributes exactly $\mu$.

Let $P_{i,v}^c$ be the completion (in the sense of CQ$^\neq$s) of $P_i$ determined by $v$, obtained by adding $x=y$ whenever $v(x)=v(y)$ and $x\neq y$ whenever $v(x)\neq v(y)$, and let $P_{i,v}^{\mathrm{id}}$ be the corresponding identification.

Applying Lemma~\ref{countingTrioPT} to every disjunct $P_\ell$ of $\overline P$, we obtain that the coefficient of $\mu$ in $\pi\llbracket\overline P(\vec t)\rrbracket$ is\[\sum_{1\leq\ell\leq n}
\left|\mathrm{SOHom}(P_\ell,P_{i,v}^{\mathrm{id}})\right|.\] Indeed, the coefficient contributed by each $P_\ell$ is $\left|\mathrm{SOHom}(P_\ell,P_{i,v}^{\mathrm{id}})\right|,$ and union adds the contributions of all disjuncts. Notice in particular that other disjuncts of $\overline P$, besides $P_i$, may contribute to the same square-free monomial $\mu$ and must therefore be included in the sum.

Likewise, applying Lemma~\ref{countingTrioPT} to every disjunct of $\overline Q$, the coefficient of $\mu$ in $\pi\llbracket\overline Q(\vec t)\rrbracket$ is $\sum_{1\leq j\leq m}
\left|\mathrm{SOHom} (Q_j,P_{i,v}^{\mathrm{id}})\right|.$ 

By the choice of $\mu$, we therefore have $\sum_{1\leq\ell\leq n} \left|\mathrm{SOHom}(P_\ell,P_{i,v}^{\mathrm{id}})\right| > \sum_{1\leq j\leq m} \left|\mathrm{SOHom} (Q_j,P_{i,v}^{\mathrm{id}})\right|.$ Thus item 2 fails for the identification $P_{i,v}^{\mathrm{id}}$ of the left disjunct $P_i$. This proves the contrapositive and hence $(2)\Rightarrow(1)$.

Finally, for the upper bound complexity calculation, we look at the complement of the problem in item 2. A nondeterministic polynomial-time machine guesses a left disjunct $P_i$ and one of its identifications $P_i^{\mathrm{id}}$. Such an identification has polynomial size, since it is obtained only from a completion in the sense of CQ$^\neq$s, which fixes for every pair of variables whether they are equal or distinct, and then quotients variables that are required to be equal. In particular, no completion with respect to positive and negative relational literals is required.

For a fixed identification $P_i^{\mathrm{id}}$, consider the quantity $\sum_{1\leq\ell\leq n}
\left|\mathrm{SOHom}(P_\ell,P_i^{\mathrm{id}})\right|.$ This is a $\#\text{P}$ function. Indeed, a nondeterministic polynomial-time machine guesses a disjunct index $\ell$ and a mapping from the variables of $P_\ell$ to the variables of $P_i^{\mathrm{id}}$, and accepts precisely when the guessed mapping belongs to $\mathrm{SOHom}(P_\ell,P_i^{\mathrm{id}})$. Membership can be checked in polynomial time by verifying preservation of the ordered head, positive and negative relational literals, equations and disequations, and that the mapping is set-onto for relational literals. Thus the accepting computations are in bijection with the pairs consisting of a disjunct index $\ell$ and an element of $\mathrm{SOHom}(P_\ell,P_i^{\mathrm{id}})$. Their number is exactly the sum above.

By the same argument, $\sum_{1\leq j\leq m} \left|\mathrm{SOHom}(Q_j,P_i^{\mathrm{id}})\right|$ is a $\#\text{P}$ function. Consequently, the difference\[\sum_{1\leq\ell\leq n} \left|\mathrm{SOHom}(P_\ell,P_i^{\mathrm{id}})\right|- \sum_{1\leq j\leq m}\left|\mathrm{SOHom}(Q_j,P_i^{\mathrm{id}})\right|\]
is in $\text{GapP}$ by~\cite[Prop. 3.5]{Fenner}, and deciding whether this difference is positive is in $\text{PP}$ by~\cite[Prop. 4.2]{Fenner}. Hence non-containment is in $\text{NP}^{\text{PP}}$, and therefore $\mathrm{Trio}(X,\overline X)$-containment of UCQ$^\neq$Ns under provenance-tracking interpretations is in $\text{coNP}^{\text{PP}}$.
\end{proof}

\section{Omitted  Proofs from \S~\ref{sec:ucqneqn-equivalence}} \label{app:ucqneqn-equivalence-proofs}

\ucqneqnEquivalenceA*

\begin{proof}
For bounded distributive lattices with more than two elements under model-defining interpretations, and for $\mathrm{PosBool}(X,\overline X)$, $\mathrm{Sorp}(X,\overline X)$,
$\mathrm{Why}(X,\overline X)$, and $\mathbb B[X,\overline X]$ under provenance-tracking interpretations, the upper bounds follow from item (2) of Proposition~\ref{propogreen21} and the corresponding containment results, namely Theorems~\ref{ucqneqnmdcont}, \ref{posboolucqneqncont}, \ref{sorpucqneqncon}, \ref{whyucqneqncont}, and \ref{Bucqneqncont}. For these semirings, the matching lower bounds follow from item (1) of Proposition~\ref{propogreen21}. The same argument applies to $\mathrm{Which}(X)$ under model-defining interpretations, using Theorem~\ref{whichucqneqncont}. Hence equivalence is $\Pi^p_2$-complete for schemas with bounded arity and coNEXP-complete in general. Finally, for $\mathbb N[X,\overline X]$ and $\mathrm{Trio}(X,\overline X)$ under provenance-tracking interpretations, the upper bounds follow from item (2) of Proposition~\ref{propogreen21}, together with Theorems~\ref{N[X]ucqneqncontainment} and~\ref{Trioucqneqncontainment}. The model-defining bag equivalence bound follows from Remark~\ref{bag-equivalence-bridge}.
\end{proof}

\end{document}